\documentclass[11pt,a4paper]{article}
\pdfoutput=1
\usepackage{jheppub}
\usepackage[utf8]{inputenc}
\usepackage{amssymb, amsmath}
\usepackage[normalem]{ulem}
\usepackage{hyperref}

\usepackage{mathtools}
\usepackage{color}

\usepackage{epsfig}
\usepackage{epstopdf}
\usepackage{epsf}
\input{epsf.sty}
\usepackage{graphicx,amsmath,amssymb}
\usepackage{epsfig,multicol}
\usepackage{multirow}
\allowdisplaybreaks

\usepackage{bbm,bm,amsmath,amssymb}

\newcommand{\Su}{\color{blue}}
\newcommand{\red}{\color{red}}
\usepackage[normalem]{ulem}

\usepackage{comment}
\def\be{\begin{equation}}
	\def\ee{\end{equation}}
\def\figs/B{B}
\def\bea{\begin{eqnarray}}
	\def\eea{\end{eqnarray}}
\def\bg{\begin{eqnarray}}
	\def\nd{\end{eqnarray}}
\def\sin{{\rm sin}}

\def\ln{{\rm log}}

\def\be{\begin{equation}}
	\def\ee{\end{equation}}

\def\doi{http://doi.org}

\begin{document}
	
	\title{de Sitter Space as a Glauber-Sudarshan State: II} 	
	
	\author{Heliudson Bernardo$^{a,*}$,}
	\emailAdd{$^*$heliudson@hep.physics.mcgill.ca}
	\affiliation{$^{a}$Department of Physics, McGill University, Montr\'{e}al, QC H3A 2T8, Canada}	
	
	\author{Suddhasattwa Brahma$^{a,\dagger}$,}
	\emailAdd{$^\dagger$suddhasattwa.brahma@gmail.com}

	\author{Keshav Dasgupta$^{a,\ddagger}$,}
	\emailAdd{$^\ddagger$keshav@hep.physics.mcgill.ca}

	\author{Mir-Mehedi Faruk$^{a,\diamond}$ and}
	\emailAdd{$^\diamond$mir.faruk@mail.mcgill.ca}
	
	\author{Radu Tatar$^{b,\star}$}
	\emailAdd{$^\star$Radu.Tatar@Liverpool.ac.uk}
	\affiliation{$^{b}$Department of Mathematical Sciences, University of Liverpool,  Liverpool, L69 7ZL, UK}

	\date{today}

	\abstract{We provide further evidence to support the fact that a four-dimensional effective field theory description with de Sitter isometries in IIB string theory, overcoming the no-go and the swampland criteria, can only exist if de Sitter space is realized as a Glauber-Sudarshan state. We show here that this result is independent of the choice of de Sitter slicings. The Glauber-Sudarshan state, constructed by shifting the interacting vacuum in the M-theory uplift of the scenario, differs from a standard coherent state in QFT in the sense that the shape and size of the state changes with respect to time, implying changes in the graviton and the flux quanta. Despite this, the expectation values of the graviton and flux operators in such a state reproduce the exact de Sitter background, as long as we are within the temporal bound set by the onset of the strong coupling in the dual type IIA side, and satisfy the corresponding Schwinger-Dyson's equations in the presence of hierarchically controlled perturbative and non-perturbative quantum corrections.
		
		Additionally, we provide a detailed study of the fluxes supporting the Glauber-Sudarshan state in the M-theory uplift of the IIB scenario. We show explicitly how the Bianchi identities, anomaly cancellation and flux quantization conditions, along-with the constraints from the Schwinger-Dyson's equations, conspire together to provide the necessary temporal dependences to support such a state in full M-theory. We also discuss how our analysis points towards a surprising connection to the four-dimensional null energy condition, for a Friedman-Lemaitre-Robertson-Walker state in the IIB side, as a consistency condition for the existence of an effective field theory description in M-theory.}

	\maketitle
	
	\newpage
	
	\hskip2.5in {Lord Carnarvon: {\it Can you see anything?}}
	
	\hskip2.5in {Howard Carter: {\it Yes ..... wonderful things!}}
	
	\vskip.1in
	
	\hskip1.5in $-$ {\it Reply that Howard Carter gave to Lord Carnarvon just before}
	
	\hskip1.65in  {\it entering the main annexe of the tomb of Tut.Ankh.Amen on}
	
	\hskip1.65in {\it  Sunday, November 26, 1922 around 4:00 pm. Taken from} \cite{carter}.

\section{Introduction and summary}	
\label{sec:intro}

Just as for our protagonists in the Valley of the Kings, Luxor, Egypt almost a century ago, viewing de Sitter space as a Glauber-Sudarshan state opens up a rich plethora of wonderful things resolving, in turn, many of the conundrums raised by the no-go theorem \cite{GMN, Dasgupta:2014pma} and the swampland 
criteria \cite{swampland}. But such a viewpoint does more than simply resolve the aforementioned conundrums: it provides a much more fundamental picture of our space-time. The key point is that a four-dimensional de Sitter space, if it exists in string or M-theory, {\it cannot} be constructed as a vacuum solution: there are far too many problems that such a vacuum solution cannot resolve. 
For example, issues related to the supersymmetry breaking, renormalization of zero-point energies, existence  of a Wilsonian effective action, existence of S-matrices,  existence of a controlled laboratory for quantum computations, trans-Planckian problems, temporal dependences of the degrees of freedom, {\it et cetera}, that only a quantum state, and not a classical vacuum configuration, would be able to resolve. 

However just {\it any} quantum state is clearly not good enough as we know that our four-dimensional space-time is, to a high degree of precision, classical. Any generic quantum state would evolve in a complicated way which, at late time, would cease to resemble anything classical. Even a delta function state in the configuration space, which could in principle provide a  perfectly well-defined classical system, would cease to make sense almost immediately because of its fast spread in the configuration space. Therefore it appears that the only allowed quantum state that in-principle could reproduce the classical trajectories in the configuration space would be the coherent state. 

Viewing four-dimensional de Sitter space as a coherent state is not exactly new (see for example 
\cite{dvali}), but as far as we know, this was shown for the first time in string and M-theories in \cite{desitter2, coherbeta}\footnote{See also \cite{susskind} for an attempt to view de Sitter space as a {\it resonance}. It will be interesting to see if there is any connection between our work and \cite{susskind}.}. However viewing such a state in string or M-theory is much more challenging. There are a few reasons why this is the case. {\Su One}, the construction of a coherent state in quantum mechanics or quantum field theory requires one to displace the {\it free vacuum} by an operator (aptly called the displacement operator). However in string and in M-theory, due to the strong interactions\footnote{The IIB string theory in which we
want to realize the de Sitter space is typically strongly coupled \cite{desitter2, coherbeta}, as will also be discussed later.}, there is no simple way to construct  a free vacuum (for a given momenta ${\bf k}$).  Therefore the only allowed choice is to take an interacting vacuum that combines all the free vacua (for any given momenta ${\bf k}$) and shift it in the configuration space by some displacement operator. Unfortunately this is easier said than done: the displacement operator itself is hard to construct, as it combines the creation and annihilation operators for all momenta in a non-trivial way. Nevertheless in \cite{coherbeta} we managed to construct such an operator that does shift the interacting vacuum in the configuration space (again there are some subtleties related to the shift itself that shall be elaborated in section \ref{vac1}. For the present discussion, we will steer clear of these finer details). {\Su Two}, shifting the interacting vacuum however creates a state that {\it does not} exactly resemble the coherent state in quantum mechanics or quantum field theory. First, it is a state that can only be consistently described in a finite temporal domain governed by the trans-Planckian Censorship Conjecture (TCC) \cite{tcc}, beyond which the evolution of the state loses all its resemblance to anything classical. Secondly, the evolution of the state, 
even in the allowed temporal domain, is nothing like the usual evolution of a coherent state. Nevertheless, 
the expectation values of the graviton and flux operators would reproduce the exact de Sitter background as long as we are in the allowed temporal domain. This is a bit surprising, and the details will be elaborated in section \ref{GState}. Because of these differences, we will call the state as the {\it Glauber-Sudarshan state} instead of the usual nomenclature of the coherent state. 
{\Su Three}, the quantum effects $-$ especially the non-perturbative and non-local\footnote{Existence of the non-local quantum effects, constructed from the non-local counter-terms, does not imply any non-localities in the theory. These non-local counter-terms 
Borel sum to well-defined {\it local} contributions as elaborated in section \ref{npeffects}.}
quantum effects $-$ are necessary to construct such a state in string and M-theories. Our analysis will involve going to M-theory because, as mentioned earlier, the IIB background will typically be at $g_b = 1$ (where even S-duality doesn't help). The dual IIA background on the other hand will be {\it weakly coupled} exactly in the temporal domain governed by the TCC. In fact precisely in this domain the quantum corrections, be it perturbative, non-perturbative, non-local or even topological, can be described by  a well defined 
hierarchy between the IIA coupling $g_s$, and the eleven-dimensional Planck mass ${\rm M}_p$ as long as the degrees of freedom have temporal dependences. These temporal dependences are absolutely essential otherwise Equations of Motion (EOMs), anomaly cancellation, flux quantizations and Bianchi identities cannot be consistently described. This will be elaborated in great details in section \ref{sec2.2.5}, wherein we will also show that the results are independent of the de Sitter slicings.  

Once such Glauber-Sudarshan state is constructed in M-theory (recall that we are using the dual M-theory to study the IIB de Sitter space), it will help us explore many new directions hitherto not possible using the standard means of vacuum configurations. {\Su One}, it will
succinctly describe the physics at late time. This is of course somewhat in-built in our construction, but as we shall elaborate towards the end of section \ref{3.1} and also in section \ref{toyeft}, the result can be generalized to 
any Friedman-Lemaitre-Robertson-Walker (FLRW) type background as long as we take expanding cosmologies: the late time physics will always be weakly coupled in the IIA side (despite being strongly coupled in IIB). {\Su Two}, as we shall demonstrate in section \ref{eva}, following a series of well described dualities, one can construct a four-dimensional de Sitter state directly in M-theory that can describe the 
{\it early time} physics. Recall that the early-time physics was out of reach in the constructions of \cite{desitter2, coherbeta}, so this should open up a new window of opportunity to study the physics there.  
{\Su Three}, as we show in sections \ref{toyeft} and \ref{bernardo}, there is a surprising connection between the existence of a Effective Field Theory (EFT) in M-theory and the four-dimensional Null Energy Condition (NEC) 
in the IIB side. Even more surprisingly, this connection between EFT and NEC, does not in any way imply the existence of NEC in ten or eleven dimensions. This was proposed recently in \cite{necshort}, and here we provide a detailed proof of this statement.  {\Su Four}, it also opens up a new way of studying 
EOMs, anomaly cancellation, Bianchi identities, flux quantizations, moduli stabilizations, {\it et cetera}, in a time-dependent background. As far as we know, these avenues were never explored before in string or M-theory, and therefore it is time now to venture along these directions. As we shall elaborate in section 
\ref{sec2.2.5}, despite the complicated nature of the problem, precise quantitative analysis {\it can} be performed for each of the topics mentioned above. For example question like the meaning of {\it flux quantizations} when both the fluxes and the background are varying with respect to time, could be given a precise answer. {\Su Five}, one can also study fluctuations over our de Sitter state by constructing another state called the Agarwal-Tara state \cite{agarwal} in M-theory. Details of this construction has already appeared in \cite{coherbeta} and here we will elaborate the consequence of such a construction in section \ref{vac2}. The story is rich in details, but we will only be able to scratch the bare minimum of the subject here, and more elaborations will be kept for future publications. {\Su Six}, the existence of an Agarwal-Tara state opens up the question of {\it interactions} in the de Sitter space. Study of interactions in a curved space, which is also {\it accelerating}, has always been a challenging problem so it will be interesting to see what the new view-point would reveal. From our analysis of the time-dependent backgrounds one thing is clear: the usual techniques applied for the time-independent backgrounds are not going to be of any use for understanding the physics of scattering in a time-dependent curved background. New techniques associated with the manipulations of the graviton and the flux-added Glauber-Sudarshan state, {\it i.e.} the Agarwal-Tara state, are to be brought in. For example, computations along the lines of {\it coherent state path integrals}, where the intermediate states are now the Glauber-Sudarshan or the Agarwal-Tara states instead of the standard graviton or flux states, might be more useful here. How this may be explicitly realized in a controlled laboratory, will be elaborated elsewhere (see the first reference in \cite{desitter2} for some details on this). {\Su Seven}, so far we haven't delved deeper into the interesting topic of realizing the FLRW background as a Glauber-Sudarshan state in the dual M-theory side.  We expect the construction of this should parallel somewhat the construction of the {\it squeezed} coherent state in quantum mechanics or quantum field theory\footnote{As we discussed above, and also show in section \ref{GState}, the temporal evolution of the Glauber-Sudarshan state doesn't actually resemble the standard evolution of the coherent state in the sense that the Glauber-Sudarshan state shows a much more complicated temporal evolution that even includes some level of squeezing. What is meant here then is that the mapping of the FLRW background to a Glauber-Sudarshan state should involve a displacement operator that has at least some resemblance to the squeezing operator to the lowest order in the fields. How this is implemented in a consistent way will be discussed elsewhere as the topic is clearly beyond the scope of the present work.}. 
However subtleties aside, such a construction should pave the way to address inflationary dynamics as a dynamically evolving Glauber-Sudarshan state that leads to a late-time de Sitter space. But now the {\it early-time} physics is important and maybe the aforementioned duality chasing 
of section \ref{eva} will become useful here. Interestingly, any success along this direction will allow us to answer questions like structure formation, late-time cosmology, exit from inflation, {\it et cetera} directly from M-theory. In the long run we might even be able to connect this to the brane constructions of \cite{KKLT, KKLMMT}.

\subsection{Organization and a brief summary of the paper}

The paper is organized as follows. In section \ref{sec2.1} we study the embedding of de Sitter space in IIB and in the dual M-theory with arbitrary slicings. Recall that in \cite{desitter2} and \cite{coherbeta} we had used the {\it flat slicing} of the de Sitter space. In this section we generalize this to study all possible embedding of de Sitter space, starting with global and conformal slicing in section \ref{2.1.1}. In sections 
\ref{2.1.2} and \ref{2.1.3} we discuss the de Sitter and open slicings as well as the planar and flat slicings respectively. In all these cases we discuss how the dual IIA coupling may remain small in some given temporal domains. Interestingly, our analysis reveals the connection between these temporal domains with the temporal domains advocated by the TCC \cite{tcc}. In section \ref{2.1.4} we discuss the static patch
and point out a number of problems associated with it, one of the most prominent one being the non-existence of any weakly coupled IIA regime. Similar problems persist for 
other embeddings related to the static patch as we discuss in section \ref{2.1.5}. 

Section \ref{sec2.2} mostly deals with the quantum effects on various de Sitter patches and the conditions for the existence of Effective Field Theory (EFT) descriptions on the patches. In our earlier papers 
\cite{desitter2, coherbeta} we studied de Sitter space-time with a {\it flat slicing} and therefore the metric did not have dependences on the spatial directions, {\it i.e.} on ${\bf R}^3$ directions in IIB or ${\bf R}^2$ directions in M-theory. However now, once we go to different slicings, spatial dependences would appear which, in turn, would effect the quantum terms etc. As a bonus, our initial study on the spatial dependences leads us to discovering a condition on the temporal derivative of the dual IIA string coupling $g_s$ that would allow an EFT description to remain valid for a generic Friedman-Lemaitre-Robertson-Walker (FLRW) type universe from string/M-theory. Its connection to the Null Energy Condition (NEC) has recently appeared in \cite{necshort}, and we elaborate the story here by providing detailed proof in sections \ref{toyeft} and \ref{bernardo}. 

In section \ref{3.1} we discuss how the effects on the quantum terms may be quantified keeping in mind the specific case of de Sitter space-time with arbitrary slicings. Clearly now the number of Riemann curvature terms and the G-flux components would increase due to the ${\bf R}^2$ dependence in M-theory. We discuss their $g_s$ scalings and also the overall scaling of the quantum series \eqref{fahingsha3} in 
\eqref{botsuga}. In section \ref{3.2} we quantify precisely the validity of an EFT description for de Sitter space from the dual M-theory. We show how the relative {\it signs} in the quantum series \eqref{fahingsha3} are important for the existence of an EFT. We also study the extent to which the metric and the flux ans\"atze could be generalized without violating the EFT descriptions. The non-perturbative and non-local effects are discussed in section \ref{npeffects}. In \cite{coherbeta}, the non-local effects were discussed only to the first order in non-locality. We generalize this here to all orders in non-localities by deriving their contributions to the energy-momentum tensors. As we discuss here, existence of non-local counter-terms, which constitute these effects, do not in any way imply non-localities in the final answer. This is because when we sum the trans-series associated with the non-perturbative and non-local contributions, the final result becomes perfectly {\it local} and {\it finite}. We demonstrate this in details in section \ref{npeffects}.  

The metric and the flux dependences on the ${\bf R}^2$ directions in addition to the usual dependences on the internal six-dimensional base is further extended in section \ref{torus} to include dependences on the toroidal directions of the eight-manifold. The perturbative quantum series becomes more involved as shown in \eqref{fahingsha4}, but the $g_s$ scaling 
in \eqref{botsuga2} shows that a naive toroidal dependence would not lead to a well-defined EFT. The question that we ask here is whether one could construct a possible ans\"atze for the metric and the flux components with dependences on the eleventh direction which would still allow an EFT description in the IIB side. The answer, as we show in \eqref{makibhalu3}, is in the affirmative and it depends on a real parameter $\gamma$. Interestingly all choices of $\gamma < 5$ lead to breakdowns of EFT descriptions, so $\gamma \ge 5$ in \eqref{makibhalu3} appears to be consistent. Even more interestingly, as we show here and also in the rest of the paper, the choice $\gamma = 6$ is special: for such a choice, all the results using the metric ans\"atze \eqref{makibhalu3}  resemble the ones from the simpler ans\"atze \eqref{evader}. This resemblance even extends to the topological sector.  We also show, by explicitly computing the $g_s$ scaling in \eqref{botsuga3}, how the EFT description remains valid as long as $\gamma \ge 5$. 

Section \ref{sec2.2.5} is a detailed study of flux EOMs, anomaly cancellation, Bianchi identities and flux quantizations. In fact in this section we not only complete the story that we started in the first reference of \cite{desitter2}, but also go much beyond that by studying the behavior of each and every flux components possible. One of the crucial aspect of the flux EOMs is the determination of the ${\bf X}_8$ polynomial which is a polynomial defined in \eqref{xeight}. In section \ref{courv} we discuss all the allowed ${\bf X}_8$ polynomials for three kinds of backgrounds: the simpler background \eqref{evader}, the background that has a generic dependence on the toroidal directions, and the background \eqref{makibhalu3} that has a special dependence on the $x^{11}$ direction. As mentioned earlier, for $\gamma = 6$ in \eqref{makibhalu3}, the curvature two-forms precisely match with the ones from the metric \eqref{evader}. This is demonstrated in details in section \ref{courv} by comparing the curvature two-forms in {\bf Tables \ref{fchumbon2}} till {\bf \ref{fchumbon10}}.

In section \ref{gfluxx} we lay out all the contributions to the flux EOMs. These not only include the standard supergravity terms with the ${\bf X}_8$ polynomials derived in section \ref{courv}, but also contributions from the quantum series \eqref{fahingsha4}, the integer and fractional ${\rm M2}$ and 
$\overline{\rm M2}$-brane sources as well the non-perturbative and the non-local terms studied in section \ref{npeffects}. We show that all these contributions may be succinctly represented by a set of rank seven and eight-tensors in \eqref{mcacrisis} whose forms are shown in \eqref{olivecosta} $-$ from the perturbative sector $-$ and in \eqref{malinih} $-$ from the non-perturbative and the non-local sectors. These non-perturbative and non-local contributions typically come from the BBS \cite{BBS} and KKLT \cite{KKLT} -type instantons whose properties were derived for specific cases in \cite{coherbeta} and here we generalize the story. The $g_s$ scalings of these terms are also computed.

To study the flux EOMs, we divide the fluxes in three categories. {\Su Internal fluxes}, discussed in section \ref{intornal}, associated with all the flux components with legs along the internal eight manifold. 
{\Su ${\bf G}_{\rm 0ABC}$ fluxes}, discussed in section \ref{g0abc}, associated with flux components that have one leg along the temporal direction; and {\Su external fluxes}, discussed in section \ref{mrain},  associated with all flux components that have at least one leg along the external ${\bf R}^2$ directions (excluding the temporal direction). The internal fluxes are categorized by three set of flux components in section \ref{intornal}, although we also include brief discussion of the flux components ${\bf G}_{0ij{\rm M}}$ for completeness. The properties of all these components are given in {\bf Tables \ref{jhinuk1}} till 
{\bf \ref{jhinuk4}}. Interestingly, and as we shall observe throughout section \ref{sec2.2.5}, not only the scalings of the various rank seven and eight tensors across the tables will be different, but also the behavior of the corresponding EOMs will differ distinctly. The latter includes the $g_s$ matching of the various terms in the EOMs as they scale in distinct ways. To balance the $g_s$ scalings at the {\it lowest} orders in the moding parameters we devise a new {\red over-bracket} and {\red under-bracket} scheme shown, for example for a specific case in \eqref{lilcadu77}. This scheme not only helps us to balance the $g_s$ scalings of the various rank seven and eight tensors, but also provides the $g_s$ scalings of the perturbative and non-perturbative (including the non-local) quantum terms. At higher orders in the parameters, all the under-brackets and the over-brackets {\it merge} in a way discussed in this section.  Comparing the properties of all the four set of internal flux components, we show that they may be consistently solved with their dominant scalings given in \eqref{collateral}. Interestingly we show that the EOMs \eqref{mcacrisis} and anomaly cancellation condition \eqref{lindmonaco} allow at most {\it four} other possible choices of the dominant scalings. However once we impose flux quantization conditions, discussed in section \ref{fluxoo}, it is only the dominant scalings in \eqref{collateral} that appear to consistently fit with the other EOMs.

In section \ref{g0abc} we tackle the EOMs of the flux components of the type ${\bf G}_{\rm 0ABC}$. Their EOMs have never been studied, as in the flat-slicing (discussed in \cite{desitter2, coherbeta}) they could be put to zero in the background. For a generic slicing of the de Sitter space, it is not {\it a priori} clear that this could be consistently realized, so it makes sense to study them here. Unfortunately however, since there are too many flux components of the type ${\bf G}_{\rm 0ABC}$, they also make the perturbative quantum series much more involved with a total of {\red 60} Riemann curvature components and {\red 40} G-flux components (modulo their permutations). All of these are raised to arbitrary powers and then contracted with metric components appropriately  as shown in \eqref{fahingsha5}. We also take the special ans\"atze \eqref{makibhalu3}, and using this we show that the $g_s$ scaling of the quantum term in \eqref{botsuga4} allow for an EFT description as long as $\gamma \ge 5$ (with $\gamma = 6$ still showing the miraculous simplification).

There are eight classes of flux components now and their properties, including the $g_s$ scalings of the various rank seven and eight tensors from \eqref{mcacrisis}, are illustrated in {\bf Tables \ref{jhinuk5}} till {\bf \ref{jhinuk11}}. As before each of the flux components has a distinct behavior both in terms of the $g_s$ scalings of the rank seven and eight-tensors and the corresponding EOMs. For example, taking the flux components ${\bf G}_{\rm 0MNP}$ from {\bf Table \ref{jhinuk5}}, we see that there are two possible set of EOMs coming from how the derivatives act on the rank-seven tensors. This leads to two types of under-bracket and over-bracket diagrams captured by \eqref{lilcadu85} and \eqref{lilcadu86}. Interestingly looking at the $g_s$ scalings of the quantum terms in the first diagram \eqref{lilcadu85}, we see how and why the non-perturbative BBS \cite{BBS} -type instantons are essential to solve the system consistently. As we show in detail in this section, such behaviors are ubiquitous throughout the flux components. A careful study of all the flux components of the type ${\bf G}_{\rm 0ABC}$ lead us to propose the range of values for their dominant scalings $l_{\rm 0A}^{\rm BC}$. This is summarized in {\bf Table \ref{kat1}} and we argue that one consistent choice for the $g_s$ scalings appears to be $l_{\rm 0A}^{\rm BC} = 0$ (except for the flux components ${\bf G}_{0ij{\rm M}}$ and ${\bf G}_{0ija}$). We discuss in details the consequence stemming from such a choice and show how this would fit in what we know from the simple case with the flat-slicing studied in \cite{desitter2, coherbeta}. 

There are also other interesting new results from the detailed study in section \ref{g0abc}. For example, we manage to find the exact expressions for the flux components ${\bf G}_{0ij{\rm M}}$ and ${\bf G}_{0ija}$ in 
\eqref{kurisada} and \eqref{HiGaToK} respectively in the presence of slowly moving membranes. The former was derived in the flat-slicing case in \cite{desitter2}, and here we generalize this to incorporate all possible de Sitter slicings. Additionally we discuss how our choice of $l_{\rm 0A}^{\rm BC} = 0$ imply the flux components to be mostly {\it off-shell} and {\it not} participate at the lowest orders in the Schwinger-Dyson EOMs (except for the flux components ${\bf G}_{0{\rm N}ab}$ and ${\bf G}_{0ij{\rm M}}$). 

The final set of flux components are the {\it external fluxes} that we study in section \ref{mrain}. There are six sets of components and their properties are illustrated in {\bf Tables \ref{jhinuk16}} till {\bf \ref{jhinuk14}}. Again detailed study of these flux components reveals interesting interplays between the perturbative and the non-perturbative quantum terms and how the $g_s$ scalings are balanced via the under-bracket and the over-bracket diagrams. The result for the dominant scalings of all these components are summarized in 
{\bf Table \ref{kat2}}, and comparing the EOMs and Bianchi identities in \eqref{20lis18} we argue that one consistent choice for the dominant scalings appears to be $l_{\rm AB}^{\rm CD} = 1$. We elaborate in details how this conclusion is borne out of all our EOMs, and show that most of the external flux components, with the exception of ${\bf G}_{{\rm MN}ab}$ again {\it do not} participate at the lowest order Schwinger-Dyson equations.

The study of the Bianchi identities for the flux components ${\bf G}_{\rm 0ABC}$ and the external fluxes 
in sections \ref{g0abc} and \ref{mrain} respectively now leads us to the subtle question of flux quantizations for the {\it internal fluxes}. The subtlety arises because of the following two reasons. {\Su One}, flux quantizations when both flux components and the internal cycles are varying with time have never been studied before. In fact giving a {\it meaning} to the quantization process itself is non-trivial. {\Su Two}, the quantizations of the internal fluxes are intimately tied up with the Bianchi identities of all the other flux components, so that the system is non-trivially intertwined. Nevertheless we managed to complete the story that we started in \cite{desitter2} by precisely showing how the quantization process should be performed in section \ref{fluxoo}. We show the interesting interplay between the {\it global} and the {\it localized} fluxes, as well as the {\it dynamical} ${\rm M5}$- branes, that ultimately lead to a similar quantization rule as given by Witten in \cite{wittenfluxes} for the time-independent case. Our construction also allows us to map to the heterotic case where we could show how the anomaly cancellation condition \eqref{chetlily} does not get corrected justifying, in turn, the Adler-Bell-Jackiw \cite{ABJ} theorem even for the time-dependent case. We discuss the quantization process for each of the three flux components ${\bf G}_{\rm MNPQ}$, ${\bf G}_{{\rm MNP}a}$ and ${\bf G}_{{\rm MN}ab}$, and show how one could justify the choice \eqref{collateral} from the five allowed choices in \eqref{pughmey} for the dominant scalings of the internal flux components (including the components ${\bf G}_{0ij{\rm M}}$). 

The aforementioned computations pretty much establish the groundwork necessary to connect them to the construction of the Glauber-Sudarshan state. However before we delve into this, we take a short detour in section \ref{detourm} to discuss two different computations that  would not only answer questions related to the {\it early time} physics, but also provide new selection principles that will help us exclude many cosmologies. The latter is essentially an elaboration and proof on the condition that we discussed in 
\eqref{senrem}. In section \ref{eva} we ask the question whether it is possible to directly find a de Sitter state in M-theory instead of using the dual description from IIB. The answer turns out to be in the affirmative but with two surprises. {\Su One}, we can get the M-theory background from a certain cosmological solution in IIB, by dualizing to IIA and then uplifting to M-theory, but both the IIB and the dual IIA backgrounds are at strong couplings. Recall that the technical advantage that we got from the {\it weak} coupling limit in IIA in our earlier computations $-$ albeit being in a finite temporal domain with a strongly coupled IIB dual $-$ cannot be replicated now. Nevertheless, as we discuss here, there is still a way to study this background from a {\it different} M-theory configuration that appears from another weakly coupled IIA theory. 
This is a unique situation where a de Sitter space in M-theory is studied from another dual M-theory background. {\Su Two}, our construction reveals a way to study {\it early time} physics compared to the later time physics that we have been studying so far. It turns out, the strong coupling problem at late time reverses when we go to the early times. This is yet another advantage that opens up a door to new avenues for research for the early time physics. 

In section \ref{toyeft} we provide a detailed proof for the condition in \eqref{senrem} by asking the alternative question: what if the temporal derivative of the IIA coupling $g_s$ is given by a {\it negative} power of $g_s$ itself as in \eqref{oniston}? We keep the negative power arbitrary, namely that it can be either an integer, fractional or even an irrational number. We show that such a requirement leads to numerous problems with the existence of an EFT description. In fact, as we introduce quantum terms, including the non-perturbative ones, the only possible choice for the exponent appears to be 0. However if the temporal derivative is expressed in terms of a positive power of $g_s$ then there is not much of a constraint so long as the exponent is given by ${\mathbb{Z}\over 3}$. Our study here reveals that the temporal derivative of $g_s$ has to be a positive power of $g_s$ if we want to keep the EFT description intact. Clearly this selection principle rules out many cosmologies as we elaborate in detail in this section.

With these we are ready to connect our constructions to the Glauber-Sudarshan state. Such a connection was already established in \cite{desitter2}, and in \cite{coherbeta} we elaborated many properties of the de Sitter space when viewed as a Glauber-Sudarshan state. There were however many subtleties that were not fully explained in \cite{coherbeta}, and in section \ref{GState} we try to answer some of them. 
One of the foremost issue, as discussed in section \ref{vac1}, is the construction of such a state by displacing the vacuum state. This simple minded statement already has many subtleties because there is no {\it free} vacuum in M-theory that we could displace using some {\it displacement operator}. The only way to make sense here is to find the displacement operator that displaces the {\it interacting} vacuum by some amount in the configuration space. Two problems immediately arise even before we manage to construct the interacting vacuum by combining the free vacua. {\Su One}, how to construct the displacement operator, and {\Su two}, how much to displace in the configuration space. The former was already answered in \cite{coherbeta} as \eqref{amchase}, but the latter is more subtle. We show that the amount to be displaced in the configuration space is related to the wave-function renormalization of the Glauber-Sudarshan state by computing the precise corrections to the Glauber-Sudarshan wave-function in \eqref{privsoc}. One of the interesting outcome of the renormalization process is the temporal behavior of the graviton quanta. We show that the number of gravitons (and also the flux quanta) {\it change} as the Glauber-Sudarshan state evolves temporally, and in fact most likely loses quanta in the process. This is where our Glauber-Sudarshan state differs from the standard coherent state in quantum mechanics or quantum field theory. As the state evolves with respect to time, it gives out graviton and flux quanta, but the expectation values of the graviton and the flux operators reproduce precisely the de Sitter background as long as we are in the temporal domain governed by the TCC \cite{tcc}, the latter being related to the onset of strong coupling in the dual IIA side. 
We also make a precise prediction for the change in the number of the background quanta.

The second important question is related to the fluctuations over the de Sitter state. This was conjectured to be the Agarwal-Tara state in \cite{coherbeta}, but beyond some small computations the subject was not elaborated much in \cite{coherbeta}. In section \ref{vac2} we ask the question whether we can quantify the fluctuations when the wave-function of the Glauber-Sudarshan state gets renormalized. Such renormalization process directly effects the Agarwal-Tara state because of its relation to the Glauber-Sudarshan state via \eqref{ferrlisa}. The answer is given by computing the expectation value of the graviton operator in the Agarwal-Tara state as \eqref{articoo}. The computation is surprisingly tractable using path-integral methods and after the dust settles the final answer can be neatly packaged into \eqref{kimbas} which establishes the precise fluctuation spectra of the gravitons over the Glauber-Sudarshan state. We also briefly discuss how to connect this to fluctuations over the Bunch-Davies and the $\alpha$-vacua, leaving most of the details for future works.

Our final set of computations, in section \ref{bernardo}, now ties up one train of computations that we have followed in section \ref{toyeft} since
proposing the condition \eqref{senrem} for the existence of an EFT description. In section \ref{flrow} we discuss how the EFT constraint in \eqref{senrem} may be connected to the existence of four-dimensional NEC in the IIB side. The result is surprising due to many reasons. {\Su First}, there appears to be no reason why two completely different set of computations, one related to the EFT analysis from M-theory, and the other related to the existence of NEC in IIB, have any connection whatsoever. Yet, we show that this is not only true, but the two sets appear to be somehow intimately tied together. Such a conjecture was proposed recently in \cite{necshort} and the computations here and in section \ref{toyeft} should be viewed as a proof of its validity. {\Su Secondly}, the connection between four-dimensional EFT and four-dimensional NEC appears to have no connection with the corresponding IIB or M-theory satisfying the ten or eleven-dimensional NEC respectively.  In section \ref{iibiib} we analyze the NEC from the full ten-dimensional perspective to elucidate this point. 

In section \ref{disco} we conclude by listing some of the main results of our paper and by pointing out avenues for future works.



\section{Revisiting de Sitter with different slicings from M-theory \label{sec2.1}}

In the introduction we discussed in some details the motivation of choosing a FLRW (Friedman-Lemaitre-Roberson-Walker) metric to study the inflationary dynamics of our universe. One would like to embed this in string theory, and ask under what condition such a background may become a solution. Again, as in 
\cite{coherbeta, desitter2}, we would like to view such a background as an evolving Glauber-Sudarshan state. Unfortunately this is a rather hard problem, and in this paper we will suffice ourselves with the easier problem of viewing the four-dimensional de Sitter space as a Glauber-Sudarshan state in string theory.  
However before we elaborate the story, let us re-visit the four-dimensional de Sitter space-time with different choice of slicings. Such generalizations will not only pave the way to introduce new time-dependent backgrounds, but will also show how and why the temporal degrees of freedom are essential in realizing such backgrounds.

In \cite{coherbeta, desitter2} we studied four-dimensional de Sitter space in type IIB using exclusively the so-called {\it flat slicing}. A natural question then is what would happen if we go for other slicings of de Sitter space. Our analysis of \cite{coherbeta, desitter2} is done from a dual M-theory framework, and therefore changing the slicings would induce non-trivial changes in the dual configurations. In the following we will analyze the scenarios from all possible slicings of de Sitter space.

\subsection{Global and conformal coordinates on de Sitter \label{2.1.1}}

Let us then start our discussion with four-dimensional de Sitter space with the so called {\it global coordinates} or more appropriately the {\it conformal coordinates}.  We will use the standard prescription of viewing the four-dimensional de Sitter space as a 
hyperboloid of one sheet whose semi-major or semi-minor axes are proportional to the inverse of the cosmological constant $\Lambda$. The type IIB metric with the conformal coordinates, parametrized by 
$(t_c, x^i_c)$,  now takes the following explicit form\footnote{A question could be raised on the {\it choice} of the metric ans\"atze itself. In \cite{desitter2, coherbeta} we have explicitly demonstrated that the background EOMs, be it in the form of supergravity equations with an infinite number of perturbative and non-perturbative corrections, or in the form of the Schwinger-Dyson's equations, again in the presence of similar set of quantum terms, can be solved with such a choice of the metric ans\"atze provided the type IIA coupling $g_s < 1$. Here we will demonstrate the same with various choices of the de Sitter slicings.}:

{\footnotesize
\bg\label{morbius}
ds^2 = {1\over {\rm H}^2~ {\rm cos}^2\left(\sqrt{\Lambda} t_c\right)}\left(-dt_c^2 + g_{ij} dx_c^idx_c^i\right)
+ {\rm H}^2\Big({\rm F}_1(t_c) g_{\alpha\beta} dy^\alpha dy^\beta + {\rm F}_2(t_c) g_{mn} dy^m dy^n\Big), 
\nd}
where ${\rm H}(y)$ is the warp-factor, $g_{ij}$ is the metric of ${\bf S}^3$ (parametrized\footnote{Note the convention: In \cite{desitter2, coherbeta} the coordinates were generically written as $(t, x_i, y^m, y^\alpha)$ because of flat-slicing along space-time. Here since we take non-trivial space-time metric, we will specify a given slicing, say for example the conformal slicing, by the 4d space-time coordinates as 
$(t_c, x^i_c)$ while the internal coordinates remain $(y^m, y^\alpha)$. Additionally, the metric components will be {\it dimensionless} and therefore the dimensions will be carried by the coordinates. This remains true for any choice of the de Sitter slicing. As an example, we can express $g_{33}dx_c^3 dx_c^3$  in terms of polar coordinates as $\left({r\over r_o}\right)^2 \sin^2\theta\left(r_od\varphi\right)^2$, where $r_o >> {\rm L}_p$ is an inherent scale chosen to keep metric dimensionless, and the coordinate dimension-full, {\it i.e.}
$g_{33} = \left({r\over r_o}\right)^2 \sin^2\theta, ~x^3_c \equiv r_o\varphi$. The scale $r_o$ will become useful when we study quantum effects later on. \label{ivylab}}  
by $(\theta_1, \theta_2, \varphi)$ with $0 \le \theta_i \le \pi$ and $0 \le \varphi \le 2\pi$),  and ${\rm F}_i(t_c)$ are similar to the time-dependent functions that we had in 
\cite{desitter2}. The coordinates $(\alpha, \beta)$ and $(m, n)$ parametrize the internal manifolds ${\cal M}_2$ and ${\cal M}_4$ respectively. 
The above metric differs from the metric that we took in \cite{coherbeta, desitter2}, which was in the {\it flat slicing}, but the generic form is consistent with what we had earlier. In fact in the following we will show that the main results {\it do not} change even if we take different slicings of the de Sitter space. The type IIA coupling (by dualizing along the isometry direction $\varphi$ of ${\bf S}^3$) now takes the following form:
\bg\label{2acoup}
g_s \propto {\rm H}(y) ~{\rm cos}\left(\sqrt{\Lambda} t_c\right), \nd
which is, as in \cite{desitter2}, depends on both the internal space $(y^{m}, y^\alpha)$ as well time $t_c$. Using the standard metric for ${\bf S}^3$, the proportionality factor ${\rm H}_o$ in \eqref{2acoup} can be easily worked out, and is given by:
\bg\label{dhuka}
{\rm H}_o(\theta_1, \theta_2) \equiv g_b~ {\rm cosec}~\theta_1~{\rm cosec}~\theta_2, \nd
where $g_b$ is the constant type IIB coupling\footnote{We will still remain at the constant coupling limit of F-theory \cite{senmukh} as in \cite{desitter2, coherbeta}.} (we are ignoring the spatial pieces as we want to concentrate only on the temporal dependence). The behavior of the IIA coupling differs from the IIA coupling in \cite{desitter2}, as one would have expected, but the interesting thing is the M-theory uplift of the metric \eqref{morbius}. This takes the form:

{\footnotesize
\bg\label{evader}
ds^2 &= & g_s^{-8/3}{\rm H}^2_o\left(-dt_c^2 + g_{ij} dx_c^i dx_c^j\right) + g_s^{-2/3} {\rm H}^2(y)
\bigg({\rm F}_1(g_s/{\rm H}_1) g_{\alpha\beta} dy^\alpha dy^\beta + {\rm F}_2(g_s/{\rm H}_1) g_{mn} dy^m dy^n\bigg)\nonumber\\
&&~~~~~~~ + g_s^{4/3} g_{ab} dw^a dw^b,\nd}
where $w^{a, b} \equiv (x_c^3, x^{11})$, $g_{ij}$ is now the metric of a two-dimensional space, and 
$g_s/{\rm H}_1({\bf x}, y)$, where ${\rm H}_1 \equiv {\rm H}_o({\bf x}) {\rm H}(y)$ with ${\rm H}_o({\bf x})$ and $x \equiv ({\bf x}, t_c) = (x^i, t_c)$ from \eqref{dhuka}, is solely a function of the temporal coordinate $t_c$. The
metric \eqref{evader} takes exactly the {\it same} form as the metric in \cite{coherbeta, desitter2} (we have introduced non-trivial metric along $2+1$ space-time as well as the toroidal direction, although for the latter the metric remains diagonal because we don't want to switch on a non-trivial axion). There are however a few differences from the M-theory metric of \cite{desitter2, coherbeta}. First,  
the form of the IIA string coupling $g_s$ is the one in \eqref{2acoup}, whereas in 
\cite{coherbeta, desitter2} the IIA string coupling was $g_s \propto {\rm H}(y) \sqrt{\Lambda} t_f$. Here $t_f$ is the temporal coordinate in the flat slicing which differs from the temporal coordinate $t_c$ used here (this difference will be elaborated soon). Secondly, the unwarped space-time and the toroidal metric components, {\it i.e.} $g_{ij}$ and $g_{ab}$ are not necessarily flat anymore\footnote{It is easy to see why 
$g_{ij}$ is not flat anymore. On the other hand, for $g_{ab}$ it becomes typically a function of $(y, x^i)$, but it may also become a function of $w^a$, {\it i.e.} $g_{ab} = g_{ab}(y, x^i, w^a)$ as we shall see soon. Such dependences will introduce a non-flat metric along the toroidal direction.}.
Other than these differences, everything else remains almost the same. The string coupling \eqref{2acoup} also tells us that as long as we are in the following temporal bound:
\bg\label{tcc}
0 ~ < ~ \vert t_c \vert ~ \le ~ {\pi \over 2\sqrt{\Lambda}}, \nd
the dual IIB theory will remain under control (here we ignore the spatial warp-factor ${\rm H}(y)$). Note that this temporal bound is not only similar to the temporal bound discussed in \cite{desitter2, coherbeta} but also consistent with the trans-Planckian Cosmic Censorship (TCC) bound \cite{tcc}. Interestingly however, while the temporal bound in \cite{desitter2} makes all dynamics for $\vert t_c \vert > {1\over \sqrt{\Lambda}}$ strongly-coupled\footnote{A related question that can arise here is the following. If we go for
$t_c  > {1\over \sqrt{\Lambda}}$ the type IIA theory naturally becomes strongly coupled. However since we are analyzing the dynamics from M-theory, why should this be of any concern? The answer, as illustrated in \cite{desitter2} can be presented in two ways. One, for $g_s << 1$ we can consistently ignore quantum corrections of the form ${\rm exp}\left(-{n\over g_s^{1/3}}\right)$ for large values of $n \in \mathbb{Z}$. Two, keeping $g_s << 1$ still allows us to take eleven-dimensional {\it supergravity} with infinite number of quantum corrections as computational tools. For $g_s > 1$, there is no reason why such techniques would work in the first place. \label{moohat}}, the bound in \eqref{tcc} is by definition the maximally allowed temporal domain for the dynamics because the global temporal coordinate is by definition restricted between $ -{\pi \over 2\sqrt{\Lambda}}  
\le t_c \le {\pi \over 2\sqrt{\Lambda}}$. Of course we can always choose a different temporal coordinate 
$\tau$ by the following identification:
\bg\label{evader2}
\tau \equiv {2\over \sqrt{\Lambda}} ~{\rm tanh}^{-1}\left[{\rm tan}\left({\sqrt{\Lambda} t_c\over 2}\right)\right], \nd
such that $-\infty  \le \tau \le +\infty$, but then the IIB metric, and consequently the M-theory metric, will not take the forms \eqref{morbius} and \eqref{evader} respectively. The above coordinate transformation relates our {\it conformal coordinates} with the standard {\it global coordinates}. For us it is essential that at least the M-theory metric takes the form \eqref{evader} so that the scalings of the quantum terms, that include the perturbative and the non-perturbative terms (including the non-local ones), may be ascertained from the analysis presented in \cite{desitter2}. Choosing the temporal coordinate \eqref{evader2} will make the subsequent analysis significantly harder to track (although we expect the final results to match), so we will avoid such complications at this stage and stick with the IIB metric \eqref{morbius}. As a bonus, and as a well-known fact, the IIB metric \eqref{morbius} is also helpful in constructing the Penrose diagrams for this choice of de Sitter slicing.

\subsection{dS  slicing and open slicing of de Sitter \label{2.1.2}}

The choice of global (or conformal) coordinates, sometimes also referred to as {\it closed slicing}, is a useful way to slice the four-dimensional de Sitter space. However there exists other slicings that are equally useful. One such choice is the so-called {\it open slicing} (or sometimes as {\it dS slicing}). In the language of latter, the four-dimensional de Sitter space may be expressed as:
\bg\label{amadams}
ds^2 = dy^2 + {\rm sin}^2\left(y\sqrt{\Lambda}\right)\left[-dt^{'2} + {1\over \Lambda} {\rm sinh}^2
\left(t\sqrt{\Lambda}\right)\left(d\zeta^2 + {\rm sinh}^2\zeta~d\theta^2\right)\right], \nd
with coordinates $(y, t', \zeta, \theta)$ and a factor of $\Lambda^{-1}$ solely as convention. The metric can be brought to the open slicing by first exchanging a space-like coordinate with the time-like coordinate, then making the coordinate change $(y, \zeta, t', \theta) \to (-iT, \xi, -i\Theta, \varphi_o)$ and finally using the identification:
\bg\label{chukkam}
t_o = {1\over \sqrt{\Lambda}}~{\rm log}\left[{\rm tanh}\left({\sqrt{\Lambda} T\over 2}\right)\right], \nd
which tells us that when $T \to \infty$, the coordinate $t_o \to 0$, whereas when $T \to 0$, the coordinate
$t_o \to -\infty$. Unfortunately the coordinate $t_o$ does not cover the regime $-\infty \le T < 0$, but in the regime
$0 \le T \le \infty$, the type IIB metric, using the temporal coordinate $t_o$, takes the following form:

{\footnotesize
\bg\label{evader4}
ds^2 = {1 \over {\rm H}^2 ~{\rm sinh}^2\left(\sqrt{\Lambda}t_o\right)}\left[-dt_o^2 + 
{1\over {\Lambda}}~d\xi^2  + {1\over {\Lambda}} ~{\rm sinh}^2 \xi~d\Omega_2^2\right]
+ {\rm H}^2\Big({\rm F}_1(t) g_{\alpha\beta} dy^\alpha dy^\beta + {\rm F}_2(t) g_{mn} dy^m dy^n\Big), 
\nonumber\\ \nd} 
where $d\Omega_2^2 = \left(d\Theta^2 + {\rm sin}^2\Theta ~d\varphi_o^2\right)$ as the two-sphere. In fact 
$(\zeta, \Omega_2)$ together form the standard hyperbolic metric. To uplift the metric to M-theory, one can T-dualize along the isometry direction which, in this case, is along $\varphi_o$.  It is easy to see that the type IIA string coupling is now proportional to:
\bg\label{evader5}
g_s \propto {\rm H}(y) \sqrt{\Lambda}~\vert {\rm sinh}(\sqrt{\Lambda}t_o)\vert, \nd
with the proportionality factor given by $g_b~ {\rm cosech}~\xi~{\rm cosec}~\Theta$ which may be compared to \eqref{dhuka}.
The coupling \eqref{evader5} is consistent with the temporal coordinate $t_o$ having the natural range $-\infty \le t_o \le 0$, because we can keep the type IIA string coupling $g_s$ positive definite (the sign issue is similar to what we encountered in \cite{desitter2, coherbeta} with flat slicing).  
 However demanding $g_s < 1$, now implies the following temporal range:
\bg\label{evade6}
-{1\over \sqrt{\Lambda}}~{{\rm arcsinh}\left(1 \over \sqrt{\Lambda}\right)} ~\le ~ t_o ~ \le ~ 0, \nd
which is again well within the temporal range advocated by TCC \cite{tcc}. Interestingly, the uplifted M-theory
metric is again given by \eqref{evader} with the only change being the $g_s$ therein is to be replaced by 
$g_s$ from \eqref{evader5}. 

\subsection{Planar coordinates and flat slicing of de Sitter \label{2.1.3}}

In the same vein, one could also slice our four-dimensional de Sitter space with the so-called {\it planar coordinates}. These coordinate slicing is related to the {\it flat slicing} that we studied in \cite{desitter2, coherbeta} again by a simple coordinate transformation as we shall illustrate below. Using planar coordinates, the type IIB metric now takes the following form:

{\footnotesize
\bg\label{morbius3}
ds^2 = {1\over {\rm H}^2}\left(-dt_p^2 + {\rm exp}\left(2\sqrt{\Lambda}t_p\right)g_{ij} dx_p^idx_p^i\right)
+ {\rm H}^2\Big({\rm F}_1(t_p) g_{\alpha\beta} dy^\alpha dy^\beta + {\rm F}_2(t_p) g_{mn} dy^m dy^n\Big), 
\nd}
where $g_{ij}$ can be a non-trivial spatial metric. If we take $g_{ij} = \delta_{ij}$ and define a new coordinate mapping $(t_p, x^i_p) \to (t_f, x^i_f)$ as $x^i_p = x^i_f$ and:
\bg\label{poppy}
t_f \equiv - {{\rm exp}\left(-\sqrt{\Lambda} t_p\right)\over \sqrt{\Lambda}}, \nd
then the type IIB metric \eqref{morbius3} transform to the metric with a flat slicing from \cite{desitter2, coherbeta}. The temporal regime also works out fine because $t_f$ is now restricted between 
$-\infty \le t_f \le 0$. The M-theory uplift of such a metric has already been discussed in \cite{desitter2} which takes the form \eqref{evader} with $g_{ab} = \delta_{ab}$, including the fact that the onset of strong type IIA coupling restricts the dynamics to remain within the temporal bound dictated by TCC \cite{tcc}, {\it i.e}:
 \bg\label{flatu}
 -{1\over \sqrt{\Lambda}} ~ \le ~ t_f \le 0. \nd
 
 \subsection{Problems with the static patch of de Sitter \label{2.1.4}}
 
So far all the slicings (or coordinate choices) that we studied have temporal dependences. In fact these temporal dependences were responsible to keep the type IIA coupling $g_s$ well within the weak coupling limit. Let us now study a coordinate choice that {\it do not} have temporal dependence. This is the so-called 
{\it static patch} of de Sitter, and in this coordinate choice, parametrized by $(t_s, r_s, \theta_s, \varphi_s)$,  the type IIB metric takes the following form:

{\footnotesize
\bg\label{regina}
ds^2 = {1\over {\rm H}^2(y)}\left[-\left(1 - \Lambda r_s^2\right) dt_s^2 + {dr_s^2\over 1 - \Lambda r_s^2} + 
r_s^2 d\Omega_2^2\right]
+ {\rm H}^2(y)\Big({\rm F}_1(y, t_s) g_{\alpha\beta} dy^\alpha dy^\beta + {\rm F}_2(y, t_s) g_{mn} dy^m dy^n\Big), \nonumber\\ \nd}
where as expected, the $3+1$ dimensional space-time has no temporal dependence, and therefore it is not {\it a-priori} clear that ${\rm F}_i$'s could only be functions of $t_s$. In fact there is a possibility that they remain completely time independent and therefore we refer to the warp-factors ${\rm F}_i \equiv {\rm F}_i(t_s, y)$ more generically
compared to \eqref{morbius} and \eqref{evader4}. 

At the face value the type IIB metric \eqref{regina} already appears somewhat problematic. The radial coordinate 
$r_s < {1\over \sqrt{\Lambda}}$ and therefore we are not allowed to access the regime $r_s \ge {1\over \sqrt{\Lambda}}$. In fact the coordinate map that allows us to embed the defining one-sheeted hyperboloid in $4 + 1$ dimensional Minkowski space itself becomes undefined when we try to access the region 
$r_s \ge {1\over \sqrt{\Lambda}}$. Additionally, the non-existence of a globally time-like Killing vector is also an issue\footnote{Outside the static patch the Killing vector should be space-like, and may even change orientations.}. One might propose that in the regime $r_s > {1\over \sqrt{\Lambda}}$ we should exchange the time-like and the space-like coordinates, which would imply that any {\it spatially} dependent  flux factors that are required to support a background like \eqref{regina} should immediately develop temporal dependences once we try to access the regime $r_s \ge {1\over \sqrt{\Lambda}}$. Thus, although the temporal independence of the metric \eqref{regina} might suggest temporally independent flux factors, this is in fact {\it not} true globally  (or at least if we want to access the upper half triangular regime of the Penrose diagram). 

We can make this a bit more precise by comparing the static patch coordinates $(t_s, r_s)$ with the flat-slicing coordinates $(t_f, r_f)$ that we took in \cite{desitter2, coherbeta}. The connection between these two coordinate patches appear directly from the  
defining hyperboloid: $\eta_{\mu\nu} x^\mu x^\nu = \alpha^2 \equiv 
{1\over \Lambda}$, in $4 + 1$ dimensions with a flat Minkowski metric $\eta_{\mu\nu}$. For the flat-slicing it is easy to see that 
$x^\mu = (x^0, x^1, x^i)$ takes the form:
\bg\label{xoxi} x^0 = -\left({\alpha^2 + r_f^2\over 2}\right){1\over t_f} + {t_f\over 2}, 
~~~ x^1 = -\left({\alpha^2 - r_f^2\over 2}\right){1\over t_f} - {t_f\over 2}, ~~~ x^i = - {\alpha y^i \over t_f}, 
\nd
where $r^2_f \equiv \sum y^iy^j \delta_{ij}$ with $2 \le (i, j) \le 4$, is the radial coordinate (we take $c = 1$ so that the dimensions match). Such a choice reproduces the four-dimensional part of the metric in flat-slicing that we took in \cite{desitter2, coherbeta}. In a similar vein, the static patch coordinates may be easily expressed as:
\bg\label{xox2} x^0 = \sqrt{\alpha^2 - r^2_s}~{\rm sinh}\left({t_s\over \alpha}\right), ~~~ 
x^1 = \sqrt{\alpha^2 - r^2_s}~{\rm cosh}\left({t_s\over \alpha}\right), ~~~ x^i = r_s z^i, \nd
where $(t_s, r_s)$ denote the temporal and the radial coordinates respectively, and $\sum z^iz^j \delta_{ij} = 1$ with
$2 \le (i, j) \le 4$, 
which allows us to embed a unit ${\bf S}^2$ as shown in \eqref{regina}. From the two set of coordinate choices \eqref{xoxi} and \eqref{xox2}, it is straight forward to find a relation between the two coordinate systems (at least along the overlapping patches), which can be written in the following way:
\bg\label{overlup}
r_s = {\alpha r_f \over \vert t_f\vert }, ~~~~~~ t_s = {\alpha\over 2}~{\rm log}\left({t_f^2 - r_f^2\over \alpha^2}\right), \nd
where one may see that, as long as $r_s < \alpha$ in the static patch, $r_f < \vert t_f\vert$ in the flat-slicing. Such a choice, although consistently keeps $t_s$ real from the second relation above, is not a requirement in the flat-slicing. Therefore once we explore regimes $r_f > \alpha$, we can clearly see that the coordinate mapping does not work anymore and a whole tower of IR degrees of freedom remains out of reach in the static patch analysis.

There is yet another issue once we try to interpret the quantum theory on a static patch using Wilsonian effective action. To illustrate this, let us consider the {\it flat slicing} discussed earlier and also in \cite{desitter2, coherbeta}. We can {\it integrate} out the IR modes, up to say $\Lambda_{\rm IR}$, such that the effective dynamics may be fully 
contained in the southern diamond of the Penrose diagram. This is a sensible way to study physics in the southern diamond with energy scale $E$ having the range $\Lambda_{\rm IR} \le E \le \Lambda_{UV}$, however since the metric (and the corresponding flux factors) are inherited from the metric with a flat slicing, it'll have the corresponding temporal dependence in a natural way. Such a background would {\it differ} from 
the one in \eqref{regina} where there is a physical cut-off on the radial coordinate $r_s$. Thus if we try to strictly contain the modes in the regime $0 \le r_s < {1\over \sqrt{\Lambda}}$, instead of just integrating them out in the Wilsonian fashion, the resulting dynamics would be very different. 

A hint that there are more problems in studying the quantum theory in the static patch appears when we try to uplift the type IIB background to M-theory, basically following earlier strategies. Dualizing along the isometry direction $\varphi_s$, gives us the following form of the IIA coupling constant:
\bg\label{hannahdark}
g_s = {1 \over r_s}\Big(g_b~ {\rm H}(y)~{\rm cosec}~\theta_s\Big), \nd
with $g_b$ at the constant coupling limit of F-theory \cite{senmukh} as before. Since $1 \le {\rm cosec}~\theta_s \le \infty$ and $0 \le r_s < {1\over \sqrt{\Lambda}}$, we see that no matter what value of the warp-factor ${\rm H}(y)$ or the radial coordinate $r_s$ we choose, the type IIA coupling is generically always strongly coupled. In all the other slicings, this was saved by the appearance of temporal dependences (see for example \eqref{2acoup} and \eqref{evader5}) that restricted the temporal regime to lie in the so-called TCC bound \cite{tcc}, {\it i.e.} \eqref{tcc} and \eqref{evade6} respectively. Unfortunately there is no such saving factor now! The appearance of strong type IIA coupling, despite being in M-theory as discussed in footnote \ref{moohat}, is problematic not only because we cannot study the theory using supergravity analysis with perturbative series of corrections as in \cite{desitter2, coherbeta}, but also because, we {\it cannot decouple} any of the non-perturbative effects, that go as ${\rm exp}\left(-{n\over g_s^{1/3}}\right)$, anymore. 

One might worry if this is a problem directly in the type IIB framework. At the constant coupling scenario of 
\cite{senmukh}, and keeping {\it vanishing} axion, the type IIB coupling is generically fixed at 
$g_b = {\cal O}(1)$ (see for example \cite{DRS}), so is strongly coupled in the type IIB side also. Plus the non-existence of a Lagrangian formulation of type IIB theory poses additional problems once we try to study the infinite series of perturbative and non-perturbative quantum corrections. Of course in the {\it flat slicing} this was solved by  going to the weakly coupled limit of type IIA and restricting the dynamics in the temporal regime dictated by TCC \cite{desitter2, coherbeta}, but here we do not have such privilege. It therefore seems that the dynamics in the {\it static patch} of de Sitter are out of reach within the supergravity framework. Temporal dependence of the metric therefore appears necessary to allow for a controlled laboratory to perform computations here.

\subsection{Other patches related to the static patch \label{2.1.5}}

Other combinations of the static patch coordinates do not help either. For example we can combine the temporal and radial coordinates of the static patch, namely $t_s$ and $r_s$, in the following way:
\bg\label{claudiak}
x^\pm \equiv t_s \pm {1\over 2} ~{\rm log}\left({1 + \sqrt{\Lambda} r_s \over 1 - \sqrt{\Lambda} r_s}\right), \nd
where again we see that $r_s < {1\over \sqrt{\Lambda}}$ otherwise these coordinates become imaginary. This
coordinate system, parametrized by $(x^+, x^-, \theta_s, \varphi_s)$, is related to the Eddington-Finkelstein coordinates, and therefore some of the symmetries of the static patch are visible here too. The type IIB metric takes the following form now:
\bg\label{kathadark}
ds^2 &=& {1\over {\rm H}^2(y)}\left[-{\rm sech}^2\left({x^+ - x^-\over 2}\right) ~dx^+ dx^- + 
{\rm tanh}^2\left({x^+ - x^-\over 2}\right)~d\Omega_2^2\right] \\
&& ~~~~~~~~ +  {\rm H}^2(y)\Big({\rm F}_1(y, x^+ + x^-) g_{\alpha\beta} dy^\alpha dy^\beta 
+ {\rm F}_2(y, x^+ + x^-) g_{mn} dy^m dy^n\Big),  \nonumber \nd
where we have retained similar behavior for the internal warp-factors ${\rm F}_i \equiv {\rm F}_i(y, x^+ + x^-)$
as for the case with the static patch. $\Omega_2$ is metric of the two-sphere which means $\varphi_s$ 
is still an isometry direction. The type IIA coupling now becomes:
\bg\label{rinadark}
g_s = g_b~{\rm H}(y)~{\rm cosec}~\theta_s~{\rm coth}\left({x^+ - x^-\over 2}\right), \nd
with the IIB coupling $g_b$ at the constant coupling limit \cite{senmukh}. As in \eqref{hannahdark}, we see that generically $g_s > 1$, so the system is at strong type IIA coupling. All the earlier issues  that we faced for the static patch reappears here too, so our choice of Eddington-Finkelstein coordinates doesn't seem to alleviate the problems. In fact rewriting the $x^\pm$ coordinates as ${\rm U} \equiv e^{x^-}$  and 
${\rm V} \equiv -e^{x^+}$, and expressing the type IIB metric as:
\bg\label{kathadark2}
ds^2 &=& {1\over {\rm H}^2(y)(1 - {\rm UV})^2}\left[-4 d{\rm U} d{\rm V} + 
(1 + {\rm UV})^2 ~d\Omega_2^2\right] \\
&& ~~~~~~ +  {\rm H}^2(y)\left[{\rm F}_1\left(y, {\rm log}\left(-{{\rm U}\over {\rm V}}\right)\right) g_{\alpha\beta} dy^\alpha dy^\beta 
+ {\rm F}_2\left(y, {\rm log}\left(-{{\rm U}\over {\rm V}}\right)\right) g_{mn} dy^m dy^n\right],  \nonumber \nd
 doesn't help either. The above coordinate system is known as the {\it Kruskal coordinates}, and if we uplift 
 the IIB background to M-theory, the type IIA coupling takes the form:
 \bg\label{charlodark}
 g_s = g_b~{\rm H}(y)~{\rm cosec}~\theta_s \left({1 - {\rm UV} \over 1 + {\rm UV}}\right), \nd
 which not only restricts the dynamics to be within ${\rm UV} < 1$ $-$ lest the coupling becomes negative $-$ but is also strongly coupled because $g_s > 1$ generically. Thus the {\it Kruskal slicing} 
 has similar issues as with the {static patch} or the {Eddington-Finkelstein slicing}.

\section{Quantum effects on various de Sitter patches and EFT \label{sec2.2}}

Most of the de Sitter patches (or {\it slicings}) that we studied above have the requisite temporal dependences, except the static patch and couple other patches related to it, 
which at least allowed us to control the dynamics within the weakly coupled type IIA framework. The temporal regime where such controlled computations could be performed lied well within the temporal regime advocated by TCC \cite{tcc}. This is good, but the real question is to analyze explicitly {\it all} possible quantum terms to verify whether the predictions we made in \cite{desitter2, coherbeta} for the {\it flat slicing} of de Sitter space survive if we change the de Sitter slicings\footnote{One might raise the question that since the various slicings of de Sitter space are related to each other by coordinate transformations, the corresponding type IIB metrics should also be related to each other by coordinate transformations. While this statement is true, the implication is a bit more subtle. The de Sitter slicings that we studied in section \ref{sec2.1} are not generically {\it overlapping}, therefore the coordinate transformations will only capture the overlapping patches of the corresponding slicings, if any. On the other hand, the corresponding M-theory uplifted metrics are {\it not} related by any coordinate transformations. This is easy to see from an example. Let us consider the type IIB metric in {\it flat slicing}. We T-dualize the metric along $x^3_f$ and then uplift the configuration to M-theory as in \cite{desitter2, coherbeta}. On the other hand, if we take the type IIB metric using {\it conformal slicing}, as in \eqref{morbius}, the T-duality direction is the isometry direction $\varphi$. Since $\varphi$ and $x^3_f$ are different, the corresponding M-theory 
configurations are no longer related via coordinate transformations. An alternative way to verify this would be to note that the eleven-dimensional Ricci scalars are different for the two cases. Since all our analysis are from M-theory point of view, every uplifted metric configuration therefore needs to be treated independently without recoursing to any underlying coordinate transformations. \label{texale}}.

To proceed we will first assume that Wilsonian effective action {\it can} be written down. This by itself is subtle, and in general {\it not} true, or at least not so straight-forward,  when we have an accelerating background. The reason lies in the temporal dependence of the fluctuating frequencies that do not allow a simple integrating-out mechanism to work in the standard way \cite{tcc}. A way out of this conundrum is to view the de Sitter space itself as a Glauber-Sudarshan state over a supersymmetric Minkowski background. In such a case, the fluctuating modes over de Sitter space are linear combinations of the fluctuating modes over the Minkowski background. This has been illustrated in great details in \cite{coherbeta}, and the readers may look up the relevant details therein. 

Viewing de Sitter space as a Glauber-Sudarshan state over supersymmetric Minkowski background has an additional advantage. All the metric and flux  components that we studied in \cite{desitter2} appear as {\it expectation values} over the Glauber-Sudarshan state: the warped metric and flux components appear as the expectation values of the metric and the G-flux operators over the Glauber-Sudarshan state. In other words \cite{coherbeta}:
\bg\label{pretroxie}
\langle g_{\rm AB}\rangle_{\bar\sigma} = {\bf g}_{\rm AB}, ~~~~~ 
\langle {\rm G}_{\rm ABCD}\rangle_{\bar\sigma} = 
{\bf G}_{\rm ABCD}, \nd
where $({\rm A}, {\rm B}) \in {\bf R}^{2,1} \times {\cal M}_4 \times {\cal M}_2 \times {\mathbb{T}^2\over {\cal G}}$; and 
 the subscript ${\bar\sigma} \equiv ({\bar\alpha}, {\bar\beta})$ defines the Glauber-Sudarshan associated with the metric ($\bar\alpha$) and the flux ($\bar\beta$) components. In the language of expectation values, these metric and flux components satisfy the Schwinger-Dyson's equations. Interestingly the Schwinger-Dyson's equations take the form of the {\it supergravity} EOMs (in the presence of an infinite class of perturbative and non-perturbative quantum corrections) that was the subject of a detailed study in \cite{desitter2}.
The existence of an Effective Field Theory (EFT) description in four space-time dimensions with de Sitter isometries is then the statement that there exists an inherent {\it hierarchy} between the infinite class of perturbative and non-perturbative quantum terms.  

The point of our works \cite{desitter2, coherbeta} was to put an emphasis on the temporal dependences of the metric and the G-flux components to allow for an EFT description in four-dimensions with de Sitter isometries. In fact the temporal dependences simultaneously justified the two inter-related facts: existence of 
an hierarchy and the existence of an EFT.
Such a point of view was however stated with a specific de Sitter slicing in mind, the so-called {\it flat slicing}. The natural question is whether this remains true for all the slicings that we discussed in section  
\ref{sec2.1}. In the following we will discuss this in some details, by first starting with the {\it conformal slicing},  which is related to the {\it global slicing} of de Sitter. 

{\it A-priori} one might expect no dependence on our choice of the coordinate slicings of de Sitter space. This is of course true if we are dealing with classical EOMs, or on-shell dynamics, which are by definition {\it covariant} (see also footnote \ref{texale}). The quantum terms are however different beasts as they generically off-shell and they evolve dynamically following all possible paths. This is also the reason for the appearance of an infinite possible perturbative and non-perturbative quantum terms in the Wilsonian action at any given energy scale. Our starting point would then be to allow for the following ans\"atze for the G-flux components along the internal eight-manifold:

{\footnotesize
\bg\label{vijuan1}
\langle {\rm G}_{\rm MNPQ}\rangle_{\bar\sigma} \equiv \sum_{k > 0} {\cal G}^{(k)}_{\rm MNPQ}(y, \theta_1, \theta_2)
\left({g_s~\sin~\theta_1~\sin~\theta_2 \over {\rm H}(y)~ g_b}\right)^{2k/3} = 
\sum_{k > 0} {\cal G}^{(k)}_{\rm MNPQ}(y, \theta_1, \theta_2)\left({g_s \over {\rm H} {\rm H}_o}\right)^{2k/3}, \nd}
where $({\rm M}, {\rm N}) \in {\cal M}_4 \times {\cal M}_2 \times {{\mathbb T}^2\over {\cal G}}$, 
$k \in {\mathbb{Z}\over 2}$; and ${\rm H}_o$ as defined in \eqref{dhuka}. 
The above ans\"atze of G-flux components differs from the one taken in \cite{desitter2, coherbeta} in two crucial ways: one, the spatial part, {\rm i.e.} ${\cal G}^{(k)}_{\rm MNPQ}$, is now a function of internal coordinates 
$y \equiv (y^\alpha, y^m)$ as well as the three-dimensional spatial coordinates $(\theta_1, \theta_2)$; and two, the expansion parameter is now $\left({g_s\over {\rm H} {\rm H}_o}\right)$  instead of $\left({g_s\over {\rm H}}\right)$ ($g_b = 1$ in \cite{desitter2, coherbeta}), which captures the temporal dependences. On the other hand, the ans\"atze for the Riemann curvature is little more complicated because we expect (say along the internal eight-manifold):
\bg\label{vijuan2}
\langle {\rm R}_{\rm MNPQ}\rangle_{\bar\sigma} \equiv 
{\bf R}_{\rm MNPQ}\left(\langle {g}_{\rm RS}\rangle_{\bar\sigma}\right) + ......., \nd
where the dotted terms are the ones appearing in the path-integral with intermediate Glauber-Sudarshan states. These dotted terms {\it do} contribute to those Schwinger-Dyson's equations that relate them to the Faddeev-Popov ghosts for the metric and the flux components. Overall the picture is complicated but there does exist a set of Schwinger-Dyson's equations (amidst all of them) that replicates the {\it supergravity} equations (in the presence of quantum terms) for the flux and metric components reasonably well. All of these   
are carefully derived in section 3.3 of \cite{coherbeta} which the readers may look into for details. 

The expression for ${\bf R}_{\rm MNPQ}\left(\langle {g}_{\rm RS}\rangle_{\bar\sigma}\right)$ is not very complicated and is in fact  similar to the standard expression for the Riemann tensor with the difference being that the metric components are replaced by their expectation values. This may be illustrated as:

{\footnotesize  
\bg\label{vijuan3} 
{\bf R}_{\rm MNPQ}\left(\langle {g}_{..}\rangle_{\bar\sigma}\right) & = & 
{1\over 2} \left(\partial_{({\rm N}}\partial_{\rm P}\langle {g}_{{\rm MQ})}\rangle_{\bar\sigma} - \partial_{({\rm N}}\partial_{\rm Q}\langle {g}_{{\rm MP})}\rangle_{\bar\sigma}\right)\nonumber\\
&+& {1\over 2} \left(\partial_{({\rm P}}\langle {g}_{{\rm N}){\rm S}}\rangle_{\bar\sigma} - \partial_{\rm S} \langle {g}_{\rm NP}\rangle_{\bar\sigma}\right) \langle {g}^{\rm SL}\rangle_{\bar\sigma} \partial_{\rm M}\langle {g}_{\rm LQ}\rangle_{\bar\sigma}\nonumber\\
& - & {1\over 2} \left(\partial_{({\rm P}}\langle {g}_{{\rm M}){\rm S}}\rangle_{\bar\sigma} - \partial_{\rm S} \langle {g}_{\rm MP}\rangle_{\bar\sigma}\right) \langle {g}^{\rm SL}\rangle_{\bar\sigma} \partial_{\rm N}\langle {g}_{\rm LQ}\rangle_{\bar\sigma}\nonumber\\
&+& {1\over 4}  \left(\partial_{({\rm P}}\langle {g}_{{\rm M}){\rm S}}\rangle_{\bar\sigma} - \partial_{\rm S} \langle {g}_{\rm MP}\rangle_{\bar\sigma}\right)  \left(\partial_{\rm T} \langle {g}_{\rm NQ}\rangle_{\bar\sigma}
-\partial_{[{\rm Q}}\langle {g}_{{\rm N}]{\rm T}}\rangle_{\bar\sigma}\right) \langle {g}^{\rm ST}\rangle_{\bar\sigma} \nonumber\\
&-& {1\over 4}  \left(\partial_{({\rm P}}\langle {g}_{{\rm N}){\rm S}}\rangle_{\bar\sigma} - \partial_{\rm S} \langle {g}_{\rm NP}\rangle_{\bar\sigma}\right)  \left(\partial_{\rm T} \langle {g}_{\rm MQ}\rangle_{\bar\sigma}
-\partial_{[{\rm Q}}\langle {g}_{{\rm M}]{\rm T}}\rangle_{\bar\sigma}\right) \langle {g}^{\rm ST}\rangle_{\bar\sigma}, \nd}
where $(...)$ and $[...]$ denote symmetric and anti-symmetric terms. For example $\partial_{(a}\partial_b {\rm F}_{cd)} \equiv \partial_{a}\partial_b {\rm F}_{cd} + \partial_{c}\partial_d {\rm F}_{ab}$, with similar definitions for others. One may check that all expected properties of the Riemann tensor remain intact. 

Let us now come to our metric ans\"atze. In M-theory we expect a metric of the form \eqref{evader}, but we can generalize this even more keeping in tune with the generalized form of the G-flux components. In other words, let us consider the following expectation values of the metric components:
\bg\label{reginadark}
\langle {g}_{\rm CD}\rangle_{\bar\sigma} = \sum_{k \ge 0} g^{(k)}_{\rm CD}(y, \theta_1, \theta_2) 
\left({g_s\over {\rm H}{\rm H}_o}\right)^{a_{\rm CD} + 2k/3}, \nd
where $({\rm C}, {\rm D}) \in {\bf R}^{2, 1} \times {\cal M}_2 \times {\cal M}_4 \times {{\mathbb{T}^2\over {\cal G}}}$ and $a_{\rm CD}$ is the dominant scaling. 
This means that all the metric components are now functions\footnote{The series \eqref{reginadark} may be justified by including the series ${\rm F}_i(g_s/{\rm HH}_o)$ in the definition of the warped-metric from \eqref{evader}. This means, to allow for an exact de Sitter metric 
we should impose the condition: $g_{\rm CD}^{(k)} = 0$ for $k > 0$ when $({\rm C}, {\rm D}) \in {\bf R}^{2, 1} \times {\mathbb{T}^2\over {\cal G}}$. However in the following we will not worry too much about this as this can be imposed at any points in our computations.} of $(y, g_s)$ as well as the spatial coordinates $(\theta_1, \theta_2)$. The reason for choosing $\theta_i$ is because of our choice of the conformal slicings. We could have chosen other slicings, discussed in section \ref{sec2.1}, in which case the dependence of the metric components on the $2+1$ dimensional spatial coordinates would be different. Thus instead of taking \eqref{vijuan1} and \eqref{reginadark}, we can take the following generalized choices that would work for {\it any} de Sitter slicings:

{\footnotesize
\bg\label{theritual}
\langle {g}_{\rm CD}\rangle_{\bar\sigma} = \sum_{k \ge 0} g^{(k)}_{\rm CD}(y, x^1, x^2) 
\left({g_s\over {\rm H}{\rm H}_o}\right)^{a_{\rm CD} + 2k/3}, ~~~
\langle {\rm G}_{\rm ABCD}\rangle_{\bar\sigma}  = 
\sum_{k\ge 0} {\cal G}^{(k)}_{\rm ABCD}(y, x^1, x^2)\left({g_s \over {\rm H} {\rm H}_o}\right)^{l^{\rm CD}_{\rm AB} + 2k/3}, \nonumber\\ \nd}
where $k \in {\mathbb{Z}\over 2}$; $(x^1, x^2)$ are the two-dimensional spatial coordinates with any choice of the slicings; 
${\rm H}_o \equiv {\rm H}_o(x^1, x^2)$, although could be ${\rm H}_o(x^1, x^2, y)$, and $l^{\rm CD}_{\rm AB}$ is the dominant scaling for the G-flux components. In this language, the warped metric and the G-flux components are functions of all the space-time and the internal coordinates {\it except} the coordinates of the fiber torus 
${\mathbb{T}^2\over {\cal G}}$. We will soon speculate what happens if we relax that constraint, and make the metric and the flux components functions of {\it all} the coordinates\footnote{It is easy to see how this may come about when we change the de Sitter slicings. Consider the M-theory metric \eqref{evader} written with conformal slicing. We can generalize this and rewrite the M-theory uplifted metric, now using ${\rm H}$ and ${\rm H}_o$, in the following suggestive way:  

{\footnotesize   
\bg\label{kitagvader}
ds^2 &= & \left({g_s\over {\rm H}{\rm H}_o}\right)^{-8/3}\left(-\tilde{g}_{00} dt^2 + \tilde{g}_{ij} dx^i dx^j\right) 
+ \left({g_s\over {\rm H}{\rm H}_o}\right)^{-2/3} 
\bigg({\rm F}_1(t) \tilde{g}_{\alpha\beta} dy^\alpha dy^\beta + {\rm F}_2(t) \tilde{g}_{mn} dy^m dy^n\bigg)\nonumber\\
&&~~~~~~~ + \left({g_s\over {\rm H}{\rm H}_o}\right)^{4/3} \tilde{g}_{ab} dw^a dw^b,\nonumber \nd}
\noindent where $\tilde{g}_{\rm AB} \equiv \tilde{g}_{\rm AB}(x^i, y)$. The relation between $\tilde{g}_{\rm AB}$ and $g_{\rm AB}$ is easy to see: for the simple case here, as an example, $\tilde{g}_{\rm MN}(y, x^i) \equiv \left({{\rm H}^2(y) \over {\rm H}_o(y, x^i)}\right)^{2/3} g_{\rm MN}(y)$ with $({\rm M}, {\rm N}) \in 
{\cal M}_4 \times {\cal M}_2$; and one may similarly express the tilded metric components in terms of the un-tilded ones. To avoid too much clutter, in our analysis we will not distinguish much between the tilded and un-tilded components and simply express $g_{\rm AB} \equiv g_{\rm AB}(y, x^i)$, unless mentioned otherwise. In a similar vein, we can ask whether the G-flux scalings given in \eqref{theritual} make sense. In the flat-slicing case ${\rm H}_o = g_b \equiv 1$, but now ${\rm H}_o$ is non-trivial. To proceed, let us compare the situation with the conformal slicing. Following \eqref{xoxi} and \eqref{xox2}, we see that the conformal slicing is given by the following choices of the coordinates of the defining hyperboloid:
\bg\label{lucy}
x^0 = \alpha~{\rm tan}\left({t_c\over\alpha}\right), ~~~~~ x^i = {\alpha z^i\over {\rm cos}\left({t_c\over\alpha}\right)}, \nonumber\nd
where $1 \le i \le 4$, and $\sum z^i z^j \delta_{ij} = 1$. A direct comparison between \eqref{xoxi} and 
the above coordinate mapping is a bit subtle because here we have a unit three-sphere, whereas for the flat-slicing we had a two-sphere of radius $r_f$. Nevertheless we see that a replacement like:
\bg\label{scarjoha}
{t_f\over \alpha} \propto {g_s\over {\rm H}} ~ \rightarrow ~ {\rm cos}\left({t_c \over \alpha}\right) \propto 
{g_s \over {\rm H H}_o}, \nonumber\nd
does make sense from the coordinate mappings. Of course since the M-theory metrics are not related by coordinate transformations, such an analysis only provides indirect hints for the G-flux scalings from 
\eqref{theritual}, at least when there are overlapping patches. A more clearer hints therefore come from the scalings of the metric components, which in fact will form the basis of our subsequent analysis.  
Soon we will generalize this even further with $g_{\rm AB} \equiv g_{\rm AB}(y, x^i, w^a)$ 
and ${\rm G}_{\rm ABCD} \equiv {\rm G}_{\rm ABCD}(y, x^i, w^a)$ with $w^a \in {\mathbb{T}^2\over {\cal G}}$. \label{viruca}}.

Making all the metric and the G-flux components functions of $(x^1, x^2)$ also, compared to being functions of $(y, g_s)$ in \cite{desitter2, coherbeta}, will not only switch on new components of the curvature tensors, but will also effect some of the $g_s$ scalings of the existing curvature terms. It is easy to speculate how the existing terms, {\it i.e.} the ones appearing in \cite{desitter2, coherbeta}, change by simply following some scaling arguments, so we will pursue that first. Note that only the scalings of those curvature terms that have temporal derivatives of the metric components would be affected. Thus
If $z_1$ denotes the $\left({g_s\over {\rm H}{\rm H}_o}\right)$
scalings of those terms that have temporal derivative and $z_2$ denotes those terms that have derivatives with respect to $x^i \equiv (x^1, x^2)$, then it is easy to see that:
\bg\label{wasseypur}
z_1 \equiv {8\over 3} + {\rm x} + {\rm y} - 2, ~~~~~ z_2 \equiv {8\over 3} + {\rm x} + {\rm y}, ~~~
\left({g_s\over {\rm H}{\rm H}_o}\right)^{z_2 - z_1} =   \left({g_s\over {\rm H}{\rm H}_o}\right)^2, \nd
where ${\rm x, y}$ denote the $\left({g_s\over {\rm H}{\rm H}_o}\right)$ scalings of the two relevant metric components. The above result assures us that the scalings of the terms that appeared in \cite{desitter2, coherbeta} can at best change by ${\cal O}(g_s^2)$ which is subleading. Thus the dominant scalings of all the terms that appeared earlier in \cite{desitter2, coherbeta} should {\it not} change. 

This reassurance however relies on an important fact that can be illustrated by an example. Consider 
$\langle {\rm R}_{ijij}\rangle_{\bar\sigma}$ that scaled as $\left({g_s\over {\rm H}}\right)^{-14/3}$, when we took the {\it flat slicing} of de Sitter space-time in \cite{desitter2, coherbeta}. Once we change the slicing, {\it i.e.} allow $(x^i, x^j)$ dependences in addition to the usual $(y, g_s)$ dependences, then the expression becomes:

{\footnotesize
\bg\label{maric}
\langle {\rm R}_{ijij}\rangle_{\bar\sigma} = {\bf R}_{ijij}(y, x^i, g_s) &=& 
{\bf R}^{(1)}_{ijij}(y, x^i) \left({g_s\over {\rm H}{\rm H}_o}\right)^{-14/3} \\ 
&+ & 
{\bf R}^{(2)}_{ijij}(y, x^i) \left({g_s\over {\rm H}{\rm H}_o}\right)^{-8/3} + 
{\bf R}^{(3)}_{ijij}(y, x^i) \left({g_s\over {\rm H}{\rm H}_o}\right)^{-14/3} \left[{\partial\over \partial t}
 \left({g_s\over {\rm H}{\rm H}_o}\right)\right]^2, \nonumber \nd}
 where we have ignored the dotted terms from \eqref{vijuan2}. The reason why this is allowed is subtle and will be explained later. For the time being we will assume this to be the case.  
 
 The first term in \eqref{maric} has the right scaling with respect to $g_s$ (to avoid clutter $g_s$ scaling henceforth will mean $ \left({g_s\over {\rm H}{\rm H}_o}\right)$ scaling), {\it i.e.} $-{14\over 3}$. The second term comes from derivatives of the metric components with respect $(x^i, x^j)$. This scales as $-{8\over 3}$, which is exactly $g_s^2$ correction to the first term, implying that \eqref{wasseypur} is working. The last term is however subtle: it scales as $-{14\over 3}$, but it also has a derivative with respect to the temporal coordinate (here $t$ represents appropriate temporal coordinate associated with a given choice of de Sitter slicings).  For the various examples that we studied in section \ref{sec2.1}, we see that\footnote{For flat slicing, as studied in 
 \cite{desitter2, coherbeta}, the time derivative of $g_s$ is just a constant, which in fact confirms the dominant scaling to be $-{14\over 3}$ in \eqref{maric}. We can easily see that:
 \bg\label{rimas}
 &&{\partial\over \partial t_c}  \left({g_s\over {\rm H}{\rm H}_o}\right) = - \sqrt{\Lambda} ~{\rm sin}\left(\sqrt{\Lambda} t_c\right) = -\sqrt{\Lambda\left(1 - \left({g_s\over {\rm H}{\rm H}_o}\right)^2\right)} \nonumber\\ 
&&{\partial\over \partial t_o}  \left({g_s\over {\rm H}{\rm H}_o}\right) = +{\Lambda} ~{\rm cosh}\left(\sqrt{\Lambda} t_o\right) = + \sqrt{\Lambda^2\left(1 + \left({g_s\over {\rm H}{\rm H}_o}\right)^2\right)}, \nonumber
\nd
for the two couplings \eqref{2acoup} and \eqref{evader5} respectively. Here ${\rm H}_o$ 
takes the corresponding values for the two cases discussed in section \ref{sec2.1}. In the weak coupling limit, {\it i.e.} when  $\left({g_s\over {\rm H}{\rm H}_o}\right) << 1$, the terms inside the square-root may be expanded in perturbative series, thus confirming the generic ans\"atze \eqref{kootdiye} with 
$c_{nm} = 0$ for $m \ge 2$ and $c_{n1} = 0$ unless $n \in 2\mathbb{Z}$. \label{camil}}:
\bg\label{kootdiye}
{\partial\over \partial t}  \left({g_s\over {\rm H}{\rm H}_o}\right) = c_0 + \sum_{(n, m) \ge 1} c_{nm} 
\left({g_s\over {\rm H}{\rm H}_o}\right)^{n/m}, \nd
where $(n, m) \in (\mathbb{Z}, \mathbb{Z})$, $c_{nm}$ are constants and $c_0 \ne 0$.  Plugging \eqref{kootdiye} in \eqref{maric} one can easily infer that the dominant scaling remains $-{14\over 3}$. The condition \eqref{kootdiye} is important and useful but the condition $c_0 \ne 0$ is not essential as long as the right hand side (RHS) of \eqref{kootdiye} allows a perturbative expansion in (positive) integer or fractional powers of $g_s$. However for all the de Sitter slicings we know (except for the static patch and the ones related to the static patch), $c_0 \ne 0$ appears to be the case. 

The importance of the condition \eqref{kootdiye}, {\it i.e.} the temporal derivative of $g_s$ should always be expressed in terms of positive powers of $g_s$, can also be elucidated by another example. Let us consider the following cosmological solution in the type IIB side:

{\footnotesize
\bg\label{morbiusn}
ds^2 = {\left(\Lambda t^2\right)^n \over {\rm H}^2(y)}\left(-dt^2 + g_{ij} dx^idx^i\right)
+ {\rm H}^2(y)\Big({\rm F}_1(t) g_{\alpha\beta}(y) dy^\alpha dy^\beta + {\rm F}_2(t) g_{mn}(y) dy^m dy^n\Big), 
\nd}
where $n$ can be integer or fractional, and can also have any sign. The metric $g_{ij}$ is a function of two of the three-dimensional spatial coordinates, {\it i.e.} a function of $(x^1, x^2)$ but independent of $x^3$. The above metric represents a generic cosmological model that may not represent a four-dimensional de Sitter space for arbitrary $n$. The type IIA string coupling takes the following simple form:
\bg\label{slivia}
g_s = {g_b {\rm H}(y)\over \left(\Lambda t^2\right)^{n/2}\sqrt{g_{33}(x^1, x^2)}}, \nd
where we will assume that $g_{33}$ remains non-zero everywhere so that there are no infinite coupling points. The corresponding IIB coupling $g_b$ is at the constant coupling point \cite{senmukh} much like what we had before. As expected the IIA coupling is highly non-trivial, and plugging this in \eqref{kootdiye}, we can easily see that:
\bg\label{senrem}
\dot{g}_s ~ \propto ~ g_s^{1 + 1/n}~~~~ \Rightarrow ~~~ {1\over n} \ge -1, \nd
otherwise the dominant scaling of the relevant curvature terms, say for example ${\bf R}_{ijij}$ in \eqref{maric}, will change. Such changes in the $g_s$ scalings of the curvature terms will in turn create problems with the existence of four-dimensional Effective Field Theory (EFT) description so long as $g_s << 1$ \cite{desitter2}. This simple formalism then restricts all cosmologies with $-1 < n < 0$ in \eqref{morbiusn}. We will elaborate more on this in sections
\ref{toyeft} and \ref{bernardo}.

\subsection{Quantum terms with spatially dependent degrees of freedom \label{3.1}}

Let us now speculate how the quantum terms behave once \eqref{kootdiye} is imposed for all the de Sitter slicings. The metric and the G-flux components are defined via the ans\"atze \eqref{theritual}, which in-turn are expressed using the dominant scalings $a_{\rm CD}$ and $l^{\rm CD}_{\rm AB}$ respectively. From the metric ans\"atze \eqref{evader} (see also footnote \ref{viruca}), it is clear what $a_{\rm CD}$ should be. Question is, what values do $l^{\rm CD}_{\rm AB}$ take? In \cite{desitter2, coherbeta}, where 
${\rm H}_o = g_b$, 
$l^{\rm CD}_{\rm AB}$ took the following form:
\bg\label{immaQ}
l_{12}^{0m}  = l_{12}^{0\alpha} = -4, ~~~~~ l^{\rm PQ}_{\rm MN}  = 1, \nd
where $({\rm M}, {\rm N}) \in {\cal M}_4 \times {\cal M}_2 \times {\mathbb{T}^2\over {\cal G}}$; $m \in 
{\cal M}_4$ and $\alpha \in {\cal M}_2$.  The situation at hand is bit more subtle now because of the 
$(x^i, x^j)$ dependences. Clearly there should be many new G-flux components of the form 
${\bf G}_{{\rm MNP}i}$ and ${\bf G}_{{\rm MN}ij}$. There is also the possibilities of ${\bf G}_{0ija}$ without 
involving derivatives along $w^a$ direction (in fact similar possibilities could exist for all ${\bf G}_{0ij{\rm M}}$ components). On the other hand, G-flux components like ${\bf G}_{0{\rm MNP}}$ could be made to vanish by switching on quantum terms, via:
\bg\label{chupke}
{\bf G}_{0{\rm MNP}} = \partial_{[0} {\bf C}_{{\rm MNP}]} + c_1 \big(\hat{\mathbb{Y}}_4\big)_{0{\rm MNP}} 
+ c_2 \left(\ast_{11}\mathbb{Y}_7\right)_{0{\rm MNP}} = 0, \nd
where $\hat{\mathbb{Y}}_4$ and $\mathbb{Y}_7$ are respectively four and seven-forms topological quantum terms derived in sections 3.2.7 and 4.2 of \cite{desitter2}. Note that we do not make ${\bf G}_{0ij{\rm M}} = 0$ using the quantum terms because, in the presence of dynamical M2-branes, non-zero ${\bf G}_{0ij{\rm M}}$ components are switched on \cite{desitter2}\footnote{We will discuss the case for 
${\bf G}_{0ija}$ later.}. Combining together, we have the following form of the quantum terms:
\bg\label{fahingsha3}
\mathbb{Q}_{\rm T}^{(\{l_i\}, n_i)} &= & \left[{\bf g}^{-1}\right] \prod_{i = 0}^3 \left[\partial\right]^{n_i} 
\prod_{{\rm k} = 1}^{41} \left({\bf R}_{\rm A_k B_k C_k D_k}\right)^{l_{\rm k}} \prod_{{\rm r} = 42}^{67} 
\left({\bf G}_{\rm A_r B_r C_r D_r}\right)^{l_{\rm r}}\\
& = & {\bf g}^{m_i m'_i}.... {\bf g}^{j_k j'_k} 
\{\partial_m^{n_1}\} \{\partial_\alpha^{n_2}\} \{\partial_i^{n_3}\}\{\partial_0^{n_0}\}
\left({\bf R}_{mnpq}\right)^{l_1} \left({\bf R}_{abab}\right)^{l_2}\left({\bf R}_{pqab}\right)^{l_3}\left({\bf R}_{\alpha a b \beta}\right)^{l_4} \nonumber\\
&\times& \left({\bf R}_{\alpha\beta mn}\right)^{l_5}\left({\bf R}_{\alpha\beta\alpha\beta}\right)^{l_6}
\left({\bf R}_{ijij}\right)^{l_7}\left({\bf R}_{ijmn}\right)^{l_8}\left({\bf R}_{iajb}\right)^{l_9}
\left({\bf R}_{i\alpha j \beta}\right)^{l_{10}}\left({\bf R}_{0mnp}\right)^{l_{11}}
\nonumber\\
& \times & \left({\bf R}_{0m0n}\right)^{l_{12}}\left({\bf R}_{0i0j}\right)^{l_{13}}\left({\bf R}_{0a0b}\right)^{l_{14}}\left({\bf R}_{0\alpha 0\beta}\right)^{l_{15}}
\left({\bf R}_{0\alpha\beta m}\right)^{l_{16}}\left({\bf R}_{0abm}\right)^{l_{17}}\left({\bf R}_{0ijm}\right)^{l_{18}}
\nonumber\\
& \times & \left({\bf R}_{mnp\alpha}\right)^{l_{19}}\left({\bf R}_{m\alpha ab}\right)^{l_{20}}
\left({\bf R}_{m\alpha\alpha\beta}\right)^{l_{21}}\left({\bf R}_{m\alpha ij}\right)^{l_{22}}
\left({\bf R}_{0mn \alpha}\right)^{l_{23}}\left({\bf R}_{0m0\alpha}\right)^{l_{24}}
\left({\bf R}_{0\alpha\beta\alpha}\right)^{l_{25}}
\nonumber\\
&\times& \left({\bf R}_{0ab \alpha}\right)^{l_{26}}\left({\bf R}_{0ij\alpha}\right)^{l_{27}}
\left({\bf R}_{mnpi}\right)^{l_{28}}\left({\bf R}_{mni0}\right)^{l_{29}}
\left({\bf R}_{mn i\alpha}\right)^{l_{30}}\left({\bf R}_{0m0i}\right)^{l_{31}}
\left({\bf R}_{mijk}\right)^{l_{32}}\nonumber\\
&\times& \left({\bf R}_{m\beta i \alpha}\right)^{l_{33}}\left({\bf R}_{abmi}\right)^{l_{34}}
\left({\bf R}_{ijk0}\right)^{l_{35}}\left({\bf R}_{\alpha 0i0}\right)^{l_{36}}
\left({\bf R}_{\alpha\beta i 0}\right)^{l_{37}}\left({\bf R}_{ab0i}\right)^{l_{38}}
\left({\bf R}_{\alpha ijk}\right)^{l_{39}}\nonumber\\
&\times& \left({\bf R}_{ab i \alpha}\right)^{l_{40}}\left({\bf R}_{\alpha\beta i \alpha}\right)^{l_{41}}
\left({\bf G}_{mnpq}\right)^{l_{42}}\left({\bf G}_{mnp\alpha}\right)^{l_{43}}
\left({\bf G}_{mnpa}\right)^{l_{44}}\left({\bf G}_{mn\alpha\beta}\right)^{l_{45}}
\left({\bf G}_{mn\alpha a}\right)^{l_{46}}\nonumber\\
&\times&\left({\bf G}_{m\alpha\beta a}\right)^{l_{47}}\left({\bf G}_{0ijm}\right)^{l_{48}} 
\left({\bf G}_{0ij\alpha}\right)^{l_{49}}
\left({\bf G}_{mnab}\right)^{l_{50}}\left({\bf G}_{ab\alpha\beta}\right)^{l_{51}}
\left({\bf G}_{m\alpha ab}\right)^{l_{52}} \left({\bf G}_{mnpi}\right)^{l_{53}} \nonumber\\
&\times&\left({\bf G}_{m\alpha\beta i}\right)^{l_{54}}\left({\bf G}_{mn\alpha i}\right)^{l_{55}} 
\left({\bf G}_{mnai}\right)^{l_{56}}
\left({\bf G}_{mabi}\right)^{l_{57}}\left({\bf G}_{a\alpha\beta i}\right)^{l_{58}}
\left({\bf G}_{\alpha ab i}\right)^{l_{59}} \left({\bf G}_{ma\alpha i}\right)^{60} \nonumber\\
&\times&\left({\bf G}_{mn ij}\right)^{l_{61}}\left({\bf G}_{m\alpha ij}\right)^{l_{62}} 
\left({\bf G}_{\alpha\beta ij}\right)^{l_{63}}
\left({\bf G}_{maij}\right)^{l_{64}}\left({\bf G}_{\alpha a ij}\right)^{l_{65}}
\left({\bf G}_{ab ij}\right)^{l_{66}} \left({\bf G}_{oija}\right)^{l_{67}}, \nonumber \nd
for any choices of $(l_i, n_i)$. Note that there are at most 41 independent choices of the Riemann tensors and 26 possible choices of G-flux components (modulo their permutations). This should be compared to what we had for the flat-slicing in \cite{desitter2, coherbeta} namely,  27 choices of Riemann tensors and 11 choices of G-flux components (again, modulo their permutations).

\begin{table}[tb]  
 \begin{center}
\renewcommand{\arraystretch}{1.5}
\begin{tabular}{|c||c||c||c|}\hline Riemann tensors  & Components & 
$g_s$ expansions \\ \hline\hline
${\bf R}_{\rm MNPQ}$ &  ${\bf R}_{mnpq}, {\bf R}_{mnp\alpha}, {\bf R}_{mn\alpha\beta}, {\bf R}_{m\alpha\alpha\beta}, {\bf R}_{\alpha\beta\alpha\beta}$ &  ${}^{\Sigma}_{{}_{k \ge 0}} {\rm R}^{(k)}_{\rm MNPQ} 
\left({g_s\over {\rm H}{\rm H}_o}\right)^{2(k-1)/3}$ \\ \hline 
${\bf R}_{{\rm MN}ab}$ &  ${\bf R}_{mnab}, {\bf R}_{m\alpha ab}, 
{\bf R}_{\alpha\beta ab}$ &  ${}^{\Sigma}_{{}_{k \ge 0}} {\rm R}^{(k)}_{{\rm MN}ab} 
\left({g_s\over {\rm H}{\rm H}_o}\right)^{2(k + 2)/3}$ \\ \hline 
${\bf R}_{abab}$ &  ${\bf R}_{abab}$ & ${}^{\Sigma}_{{}_{k \ge 0}} {\rm R}^{(k)}_{abab} 
\left({g_s\over {\rm H}{\rm H}_o}\right)^{2(k + 5)/3}$ \\ \hline 
${\bf R}_{{\rm MNP}0}$ &  ${\bf R}_{mnp0}, {\bf R}_{mn\alpha 0}, {\bf R}_{m\alpha\beta 0}, 
{\bf R}_{0\alpha\alpha\beta}$ & ${}^{\Sigma}_{{}_{k \ge 0}} {\rm R}^{(k)}_{{\rm MNP}0} 
\left({g_s\over {\rm H}{\rm H}_o}\right)^{(2k - 5)/3}$ \\ \hline 
${\bf R}_{{\rm MNP}i}$ &  ${\bf R}_{mnpi}, {\bf R}_{mn\alpha i}, {\bf R}_{m\alpha\beta i}, 
{\bf R}_{i\alpha\alpha\beta}$ & ${}^{\Sigma}_{{}_{k \ge 0}} {\rm R}^{(k)}_{{\rm MNP}i} 
\left({g_s\over {\rm H}{\rm H}_o}\right)^{2(k - 1)/3}$ \\ \hline 
${\bf R}_{{\rm MN}i0}$ &  ${\bf R}_{mni0}, {\bf R}_{\alpha\beta i0}$
 &  ${}^{\Sigma}_{{}_{k \ge 0}} {\rm R}^{(k)}_{{\rm MN}i0} 
\left({g_s\over {\rm H}{\rm H}_o}\right)^{(2k - 5)/3}$ \\ \hline 
${\bf R}_{{\rm MN}\mu\nu}$ &  ${\bf R}_{mnij}, {\bf R}_{m\alpha ij}, {\bf R}_{\alpha\beta ij}, {\bf R}_{0m0n}, {\bf R}_{0\alpha 0\beta}, {\bf R}_{0m 0\alpha}$
 &  ${}^{\Sigma}_{{}_{k \ge 0}} {\rm R}^{(k)}_{{\rm MN}\mu\nu} 
\left({g_s\over {\rm H}{\rm H}_o}\right)^{2(k - 4)/3}$ \\ \hline 
${\bf R}_{{\rm M}0ij}$ &  ${\bf R}_{m0ij}, {\bf R}_{\alpha 0ij}$
 & ${}^{\Sigma}_{{}_{k \ge 0}} {\rm R}^{(k)}_{{\rm M}0ij} 
\left({g_s\over {\rm H}{\rm H}_o}\right)^{(2k - 11)/3}$ \\ \hline  
${\bf R}_{{\rm M}\mu\nu\rho}$ &  ${\bf R}_{m0i0}, {\bf R}_{mijk}, {\bf R}_{\alpha 0 i0}, {\bf R}_{\alpha ijk}$
 &  ${}^{\Sigma}_{{}_{k \ge 0}} {\rm R}^{(k)}_{{\rm M}\mu\nu\rho} 
\left({g_s\over {\rm H}{\rm H}_o}\right)^{2(k - 4)/3}$ \\ \hline 
${\bf R}_{\mu\nu\rho\sigma}$ &  ${\bf R}_{ijk0}$
 & ${}^{\Sigma}_{{}_{k \ge 0}} {\rm R}^{(k)}_{\mu\nu\rho\sigma} 
\left({g_s\over {\rm H}{\rm H}_o}\right)^{(2k - 11)/3}$ \\ \hline   
${\bf R}_{\mu\nu\mu\nu}$ &  ${\bf R}_{ijij}, {\bf R}_{0i0j}$
 & ${}^{\Sigma}_{{}_{k \ge 0}} {\rm R}^{(k)}_{\mu\nu\mu\nu} 
\left({g_s\over {\rm H}{\rm H}_o}\right)^{2(k - 7)/3}$ \\ \hline   
${\bf R}_{0{\rm M} ab}$ &  ${\bf R}_{0mab}, {\bf R}_{0\alpha ab}$
 &  ${}^{\Sigma}_{{}_{k \ge 0}} {\rm R}^{(k)}_{0{\rm M}ab} 
\left({g_s\over {\rm H}{\rm H}_o}\right)^{(2k +1)/3}$ \\ \hline   
${\bf R}_{{\rm M}ab i}$ &  ${\bf R}_{m ab i}, {\bf R}_{\alpha ab i}$
 &  ${}^{\Sigma}_{{}_{k \ge 0}} {\rm R}^{(k)}_{{\rm M}abi} 
\left({g_s\over {\rm H}{\rm H}_o}\right)^{2(k + 2)/3}$ \\ \hline 
${\bf R}_{ab \mu 0}$ &  ${\bf R}_{abi 0}$
 &  ${}^{\Sigma}_{{}_{k \ge 0}} {\rm R}^{(k)}_{ab\mu 0} 
\left({g_s\over {\rm H}{\rm H}_o}\right)^{(2k + 1)/3}$ \\ \hline    
${\bf R}_{\mu\nu ab}$ &  ${\bf R}_{abij}, {\bf R}_{0a 0 b}$
 &  ${}^{\Sigma}_{{}_{k \ge 0}} {\rm R}^{(k)}_{\mu\nu ab} 
\left({g_s\over {\rm H}{\rm H}_o}\right)^{2(k - 1)/3}$ \\ \hline    
 \end{tabular}
\renewcommand{\arraystretch}{1}
\end{center}
 \caption[]{The ${g_s\over {\rm H H}_o}$ expansions of the components of the curvature tensors associated with the M-theory metric \eqref{evader}. The warp-factor ${\rm H}(y)$ is the universal warp-factor appearing in \cite{desitter2, coherbeta},  whereas ${\rm H}_o \equiv {\rm H}_o(x, y)$ depends on the choice of the de Sitter slicings, as defined in section \ref{sec2.1}. The components of the Riemann tensors are defined in the 
usual way:  $(m, n) \in {\cal M}_4$,  
 $(\alpha, \beta) \in {\cal M}_2$, $(a, b) \in {\mathbb{T}^2\over {\cal G}}$ and 
 $(\mu, \nu) \in {\bf R}^{2, 1}$; with $x \equiv (x^i, x^j)$ and $y^m \in {\cal M}_4 \times {\cal M}_2$.  The modes of the curvature tensor are defined as ${\rm R}^{(k)}_{\rm A_1A_2 A_3 A_4}  = 
 {\rm R}^{(k)}_{\rm A_1A_2A_3A_4}(x, y)$ where ${\rm A}_i \in {\bf R}^{2, 1} \times {\cal M}_4 \times {\cal M}_2 \times {\mathbb{T}^2\over {\cal G}}$  and 
 $k \in {\mathbb{Z}\over 2}$.
These curvature tensors form the essential ingredients of the quantum terms \eqref{fahingsha3}.} 
  \label{firzasut}
 \end{table}

To see how the scalings of the G-flux components would work in the generic case, let us start with 
${\bf G}_{{\rm MNP}i}$. Typically these would appear from the spatial derivatives of the three-form fields
${\bf C}_{\rm MNP}$, plus curvature contributions to satisfy the Bianchi identities (see section 4.2 of 
\cite{desitter2}). They would therefore follow similar ans\"atze as in \eqref{theritual} with dominant scaling 
$l_{\rm MN}^{{\rm P}i}$. On the other hand, we can view ${\bf G}_{{\rm MN}ij}$ as {\it localized} fluxes of the form:
\bg\label{sessence}
{\bf G}_{{\rm MN}ij}(g_s, y, x) = \sum_{k\ge 0} \left({\cal G}^{(k)}_{{\rm MN}ij}(x, y) + 
{\cal F}^{(k)}_{ij}(x) \Omega^{(k)}_{\rm MN}(y)\right) 
\left({g_s\over {\rm H}{\rm H}_o}\right)^{l_{\rm MN}^{ij} + 2k/3}, \nd
where $x \equiv (x^i, x^j)$ and $y \equiv (y^m, y^\alpha, w^a)$ represent the coordinate dependences; and 
$\Omega^{(k)}_{\rm MN}(y)$ are the {\it localized} two-forms on the internal eight-manifold 
${\cal M}_4 \times {\cal M}_2 \times {\mathbb{T}^2\over {\cal G}}$. We have left a non-localized piece 
${\cal G}^{(k)}_{{\rm MN}ij}(x, y)$ for completeness, which may or may not be a constant. The gauge field components ${\cal F}^{(k)}_{ij}(x)$ will appear in the type IIB side either on the ${\rm D}3, \overline{{\rm D}3}$
or the $(p, q)$ seven-branes (in the full F-theory formalism). These gauge fields, which were absent in the 
flat-slicing case because of the absence of the $(x^i, x^j)$ dependence, now appear with a dominant scaling $l_{\rm MN}^{ij}$. Such localized fluxes are necessary to allow for a de Sitter space in the IIB side with different slicings.

Let us now discuss how the quantum series \eqref{fahingsha3} would scale with respect to $g_s$. The scaling of the Riemann curvature terms are given in {\bf Table \ref{firzasut}}, where $k \ge 0$ so that the dominant scaling of any curvature component is always the $k$ independent numerical factor appearing there. Similarly, the dominant scaling of any G-flux component appears from $l_{\rm AB}^{\rm CD}$ where 
$({\rm A, B})$ scans the full eleven-dimensional space-time. Let $\theta_{nl}$ be the dominant scaling of the quantum term \eqref{fahingsha3}, whose expression, using the parameters appearing in \eqref{fahingsha3}, may be written as:
\bg\label{botsuga}
\theta_{nl} &= & {2\over 3} \sum_{i = 1}^{27} l_i + {5\over 3} \sum_{j = 28}^{41} l_j + {1\over 3}\left(n_1 + n_2 + 4n_3\right) + \left({1\over 3} + \left[{n\over m}\right]\right)n_0 
+ \left(l_{0a}^{ij} + {4\over 3}\right)l_{67}\\
&+& \left(l_{mn}^{pq} + {4\over 3}\right)l_{42} + \left(l_{mn}^{p\alpha} + {4\over 3}\right)l_{43} +
\left(l_{mn}^{pa} + {1\over 3}\right)l_{44} + \left(l_{mn}^{\alpha\beta} + {4\over 3}\right)l_{45} 
+ \left(l_{mn}^{\alpha a} + {1\over 3}\right)l_{46}\nonumber\\
&+& \left(l_{ma}^{\alpha\beta} + {1\over 3}\right)l_{47} + \left(l_{ij}^{0m} + {13\over 3}\right)l_{48} 
+ \left(l_{ij}^{0\alpha} + {13\over 3}\right)l_{49} + \left(l_{mn}^{ab} {\red - {2\over 3}}\right){\red l_{50}} 
+ \left(l_{\alpha\beta}^{ab} {\red - {2\over 3}}\right){\red l_{51}} \nonumber\\
&+& \left(l_{m\alpha}^{ab} {\red - {2\over 3}}\right){\red l_{52}} + \left(l_{mn}^{pi} + {7\over 3}\right)l_{53} 
+ \left(l_{m\alpha}^{\beta i} + {7\over 3}\right)l_{54} + \left(l_{mn}^{\alpha i} + {7\over 3}\right)l_{55} 
+ \left(l_{mn}^{ai} + {4\over 3}\right)l_{56} \nonumber\\
&+& \left(l_{ab}^{mi} + {1\over 3}\right)l_{57} + \left(l_{\alpha\beta}^{ai} + {4\over 3}\right)l_{58} 
+ \left(l_{ab}^{\alpha i} + {1\over 3}\right)l_{59} + \left(l_{m\alpha}^{ai} + {4\over 3}\right)l_{60} 
+ \left(l_{mn}^{ij} + {10\over 3}\right)l_{61} \nonumber\\
&+& \left(l_{m\alpha}^{ij} + {10\over 3}\right)l_{62} + \left(l_{\alpha\beta}^{ij} + {10\over 3}\right)l_{63} 
+ \left(l_{ma}^{ij} + {7\over 3}\right)l_{64} + \left(l_{\alpha a}^{ij} + {7\over 3}\right)l_{65} 
+ \left(l_{ab}^{ij} + {4\over 3}\right)l_{66}, \nonumber \nd
where $\left[{n\over m}\right]$ is a positive ratio defined in \eqref{kootdiye}; and as in \cite{desitter2, coherbeta} we see that \eqref{botsuga} is mostly positive except at three places
marked in {\red red} associated with the G-flux components ${\bf G}_{{\rm MN}ab}$ with 
$({\rm M, N}) \in {\cal M}_4 \times {\cal M}_2$. Despite equivalence with \cite{desitter2, coherbeta} there are however a few differences. First, the number of terms is significantly larger from what we had earlier. There are now 41 curvature terms (modulo their permutations), out of which the first 27 of them all scale as 
$\left({g_s\over {\rm HH}_o}\right)^{2/3}$ whereas the remaining 14 scale as $\left({g_s\over {\rm HH}_o}\right)^{5/3}$. Interestingly the $g_s$ scalings still remain positive definite. Secondly, the coefficient of $n_0$ is not just ${1\over 3}$: there is an extra factor coming from 
\eqref{kootdiye}. As emphasized earlier, such a factor is essential to restore the four-dimensional EFT description in the type IIB side. Here we see why: a relative {\it minus} sign for the coefficient of $n_0$ will have disastrous consequence. On the other hand, the relative minus signs for the coefficients of $(l_{50}, l_{51}, l_{52})$
can only be flipped if we allow non-zero $g_s$ scalings for the corresponding G-flux components. Combining them together, we see that {\it at least} if we impose the following conditions:
\bg\label{botspivee}
\left[{n\over m}\right] ~ \ge 0, ~~~~ l_{\rm MN}^{ab} > {2\over 3}, \nd
where $({\rm M, N})$ span, as before, the six-dimensional base of the eight-manifold, there appears to be 
some possibility for the existence of a four-dimensional EFT in the dual IIB side. On the other hand, if 
$l_{\rm MN}^{ab} = 0$, $\theta_{nl}$ will have relative minus sign, we will lose the $g_s$ hierarchy in the system.

The loss of $g_s$ hierarchy {\it does not} immediately imply a loss of ${\rm M}_p$ hierarchy. One could still have operators with the same $\left({g_s\over {\rm HH}_o}\right)^{\theta_{nl}}$ scalings for a given choice of $n_i$ and $l_i$ in \eqref{fahingsha3}, but typically they would be suppressed by different powers of ${\rm M}_p$. For example, the operator $\mathbb{Q}_{\rm T}^{(\{l_i\}, n_i)}$ in \eqref{fahingsha3} can be expressed in terms of the $g_s$ and ${\rm M}_p$ scalings in the following way:
\bg\label{magiQ}
\mathbb{Q}_{\rm T}^{(\{l_i\}, n_i)}(x, y;  g_s, {\rm M}_p) \equiv 
{\bf{Q}}_{\rm T}^{(\{l_i\}, n_i)}(x, y) \left({g_s\over {\rm HH}_o}\right)^{\theta_{nl}} \times 
{1\over {\rm M}^{\sigma_{nl}}_p}, \nd
where $\theta_{nl}$ is the same scaling that appears in \eqref{botsuga}, and $\sigma_{nl}$ is the ${\rm M}_p$ scaling of the quantum term \eqref{fahingsha3}. A naive derivative countings of the curvature and G-flux components, including the spatial and temporal derivatives, in the quantum term \eqref{fahingsha3} will tell us that:
\bg\label{harpno}
 \sigma_{nl} = \sum_{i = 0}^3 n_i + 2\sum_{j = 1}^{41} l_j + \sum_{p = 42}^{67} l_p, \nd
 which contains all positive definite integers. If this was only the case, then clearly there will always be some level of ${\rm M}_p$ hierarchy in the system and the operators can thus be distinguished. However, as in \cite{coherbeta}, a careful look reveals hidden subtleties. The first one comes from looking at the G-flux components ${\bf G}_{{\rm MN}ab}$. Such flux components can come from the following choices of the three-form potentials: ${\bf C}_{{\rm MN}b}$ with derivative acting along $x^a \equiv x^3$ direction, or
 ${\bf C}_{{\rm MN}a}$ with derivative acting along $w^b \equiv w^{11}$ direction, or 
${\bf C}_{{\rm M}ab}$ with derivative acting along $y^{\rm M} = (y^m, y^\alpha)$ directions. None are helpful: the first choice will clash with our T-duality rule, the second choice will create problems with a EFT description in the IIB side (this will be clarified a bit later when we discuss the quantum terms in \eqref{fahingsha4}) and the third choice will create a cross-term metric component ${\bf g}_{{\rm M}3}$ in the dual IIB side. These are of course the same issues that we encountered in \cite{desitter2}, and the resolution therein  was to view ${\bf G}_{{\rm MN}ab}$ as {\it localized} fluxes in the same vein as \eqref{sessence}. 

The localized fluxes are defined using localized forms, $\Omega_{\rm MN}(y)$ and $\Omega_{ab}(y)$, in the way they appear in \eqref{sessence}. The way we have written them, however, doesn't reveal the hidden 
$g_s$ and ${\rm M}_p$ dependences. These dependences are in {\it addition} to the standard $g_s$ and ${\rm M}_p$ dependences expected from the four-forms. These {\it hidden} dependences appear from the following definition for $\Omega_{\rm AB}$:
\bg\label{samaraw}
\Omega_{\rm AB}(y; g_s, {\rm M}_p) &\equiv & \sum_{l, k = 0}^{\infty} {\cal B}^{(\small{\rm AB})}_{lk} 
{\rm exp}\left[-y^{2l} {\rm M}_p^{2l} - \left({\bf g}_{\rm MN}y^{\rm M} y^{\rm N}\right)^k {\rm M}^{2k}_p\right] \epsilon_{\rm AB}
\nonumber\\
&= & 
\sum_{l, k = 0}^{\infty} {\cal B}^{({\rm AB})}_{lk} 
{\rm exp}\left[-y^{2l} {\rm M}_p^{2l} - \left({{g}_{\rm MN}y^{\rm M} y^{\rm N}\over g^{2/3}_s}\right)^k {\rm M}^{2k}_p\right]
\epsilon_{\rm AB}, \nd 
where ${\cal B}^{({\rm AB})}_{lk}$ are just constants but can have $g_s$ dependences (although no ${\rm M}_p$ dependences), and for large values of $l$ and $k$ we can easily see that, as long as ${\rm M}_p \to \infty$ (or even ${\rm M}_p >> 1$), they are decoupled. Note that the first term is written without recoursing to any underlying metric ({\it i.e.} the metric choice is a flat one: $y^2 \equiv \eta_{\rm MN} y^{\rm M} y^{\rm N}$) whereas the second term has a metric factor. There is however an issue with the second term: it is non-perturbative in $g_s$ and therefore for $g_s << 1$ this would completely decouple\footnote{One could argue that for $g_s$ going to zero as $g_s = \epsilon$, where $\epsilon \to 0$, we see that the gaussian function \eqref{samaraw} is typically peaked near $y = \epsilon^\gamma$ with $\gamma > {1\over 3}$. Therefore near 
$y^{\rm M} = 0 + \epsilon^\gamma$ we could have a localized normalizable  form. This is not true (and also erroneous) because of a few reasons. {\Su One}, the unwarped metric $g_{\rm MN}(y)$ is typically a function of $y^{\rm M}$ and so the naive scaling cannot be right. {\Su Two}, for $y^{\rm M} > 0$ the gaussian piece always vanishes. {\Su Three}, since we have consistently kept the non-perturbative contributions of the form ${\rm exp}\left[-{b(y, x)\over g_s^{2/3}}\right]$ to the flux components zero, we cannot introduce such a contribution now (although, as discussed in \cite{desitter2, coherbeta}, a non-perturbative effect of the form 
${\rm exp}\left[-{c(y, x)\over g_s^2}\right]$ to the energy momentum tensor does become essential). {\Su Four}, demanding the normalization scheme of the form:
\bg\label{fainplay} {\rm M}_p^6\int d^6 y \sqrt{-{\bf g}_6} ~{\bf g}^{\rm MP} {\bf g}^{\rm NQ} ~\Omega_{\rm MN} 
\Omega_{\rm PQ} = 1, \nonumber\nd 
with $({\rm M, P}) \in {\cal M}_4 \times {\cal M}_2$, we see that ${\cal B}^{\rm MN}_{l0} \propto g_s^{1/3}$ whereas ${\cal B}^{\rm MN}_{0k} \propto g_s^{-2/3}$.
The latter appearance of an inverse $g_s$ factor is alarming, although wholly expected from our normalization scheme. The D3 or D7 gauge fields in the two cases would scale as $g_s^\sigma$ where 
$\sigma > {1\over 3}$ and $\sigma > {4\over 3}$ respectively to allow for \eqref{botspivee}; and {\Su finally}, such a choice of the gaussian might clash with the existence of a four-dimensional EFT description with de Sitter isometries in the IIB side. This will become clearer as we proceed. \label{playground}}. This means only the first gaussian piece with ${\cal B}_{l0}^{\rm AB} \ne 0$ contributes to the two-form. Such a form will lead to gauge fields ${\cal F}_{\rm MN}(y)$ and ${\cal F}_{ij}(x)$ on either or both the three and the seven-branes in the IIB side. Note that their $g_s$ dependence will come mostly from the generic dependence as in \eqref{sessence}.  

\subsection{Existence of an EFT description with de Sitter isometries \label{3.2}}

Let us now see how this works for a generic choice of the energy-momentum tensor $\mathbb{T}_{\rm AB}$.
The energy-momentum tensor gets contributions from both the perturbative and the non-perturbative quantum terms (see \cite{desitter2, coherbeta}), but here we will discuss the contributions from the perturbative terms of the form \eqref{fahingsha3}. The energy-momentum tensor defined as:
\bg\label{faining}
\mathbb{T}_{\rm AB}(x, y; g_s, {\rm M}_p) &\equiv & \sum_{\{l_i\}, n_j} \left(a {\bf g}_{\rm AB} 
\mathbb{Q}^{(\{l_i\}, n_j)}_{\rm T} + {\partial \mathbb{Q}^{(\{l_i\}, n_j)}_{\rm T} \over \partial {\bf g}^{\rm AB}}\right) {b\over {\rm M}_p^{\sigma_{\{l_i\}, n_j}}}\nonumber\\
& = & \sum_{n, l} \left({g_s\over {\rm HH}_o}\right)^{a_{\rm AB}+ \theta_{nl}}
{1\over {\rm M}_p^{\sigma_{nl} + \gamma_{nl}}}
~{\bf T}_{\rm AB}(x, y), \nd
where $(a, b)$ are constants independent of $(g_s, {\rm M}_p)$; 
$({\rm A, B}) \in {\bf R}^{2, 1} \times {\cal M}_4 \times {\cal M}_2 \times {\mathbb{T}^2\over {\cal G}}$;  and $\theta_{nl}$ and $\sigma_{nl}$ are defined in \eqref{botsuga} and \eqref{harpno} respectively. We can take $a \equiv 1$, but $b$ will depend on the precise way $\mathbb{Q}^{(\{l_i\}, n_j)}_{\rm T}$ appears in our M-theory Lagrangian. Note also the appearance of $\left({g_s\over {\rm HH}_o}\right)^{a_{\rm AB}}$ in the $g_s$ scaling above which is associated with the metric scalings in \eqref{theritual}. Thus $a_{\rm AB}$ takes three possible values:
\bg\label{sillyP}
a_{\rm AB} \equiv \left(-{8\over 3}, ~-{2\over 3}, ~+{4\over 3}\right), \nd
depending on whether we are looking at space-time ${\bf R}^{2, 1}$, the internal space ${\cal M}_4 \times {\cal M}_2$ or the toroidal direction ${\mathbb{T}^2\over {\cal G}}$ respectively, as seen from the M-theory metric \eqref{evader}. The other difference is the appearance of $\gamma_{nl}$ in addition to $\sigma_{nl}$, defined in \eqref{harpno}, for the ${\rm M}_p$ scalings of the quantum terms. Such a factor appears when we take the {\it hidden} ${\rm M}_p$ dependence from the localized form 
 $\Omega_{\rm AB}$ with ${\cal B}^{\rm AB}_{0k} = 0$ in \eqref{samaraw}. This way, as expected, only the 
 ${\rm M}_p$ scalings are effected (see also footnote \ref{playground}). 
 
 The other quantities $(\theta_{nl}, \sigma_{nl}, \gamma_{nl})$ can be defined more efficiently with the aid of a few notations. We define $\mathbb{N}_i$ in the following suggestive way:

{\footnotesize
 \bg\label{lilrabe}
 &&\mathbb{N}_1 = \sum_{i = 1}^{27}l_i, ~~~~~ \mathbb{N}_2 = \sum_{j = 28}^{41}l_j, ~~~~~ 
 \mathbb{N}_3 =  \sum_{k = 48}^{49} l_k  \nonumber\\
 && \mathbb{N}_4 =  \sum_{p = 61}^{63} l_p, ~~~~~ \mathbb{N}_5 =  \sum_{q = 53}^{55} l_q   + 
 \sum_{r = 64}^{65} l_r,  ~~~~ \mathbb{N}_6 =  \sum_{s = 44}^{46} l_s   +  l_{57} + l_{59} \nonumber\\
 && \mathbb{N}_7 = \sum_{u = 42}^{43} l_{u} + l_{45} + l_{56} + l_{58} + l_{60} + \sum_{v = 66}^{67} l_{v}, ~~~~ 
 \mathbb{N}_8 =  \sum_{t = 50}^{52} l_t, ~~~~ \mathbb{N}_9 =  \sum_{1 = 1}^{2} n_i, \nd}
where $l_i$ are the powers of the curvature and the G-flux components appearing in the quantum term 
\eqref{fahingsha3}.   Similarly $n_1$ and $n_2$ denote the derivatives along ${\cal M}_4 \times {\cal M}_2$ here. The coefficients \eqref{lilrabe} can then be used directly to express the scaling $\theta_{nl}$ in the {\it absence} of time-dependent fluxes, {\it i.e.} when $l_{\rm AB}^{\rm CD} = 0$ in \eqref{botsuga}. In fact we can show that all the three scaling coefficients, {\it i.e.} $\theta_{nl}, \sigma_{nl}$ and $\gamma_{nl}$ can be re-written as:

{\footnotesize
\bg\label{evehew}
 && \sigma_{nl} = n_0 + n_3 + 2 \sum_{1 = 1}^2 \mathbb{N}_i + \sum_{j = 3}^9 \mathbb{N}_j, ~~~~~~
 \gamma_{nl} = {\red -2 \mathbb{N}_9} \nonumber\\
 && 3 \theta_{nl} = 2\mathbb{N}_1 + 5\mathbb{N}_2 + 13\mathbb{N}_3 + 10\mathbb{N}_4 + 
 7\mathbb{N}_5 + \mathbb{N}_6 + 4\mathbb{N}_7 {\red -2 \mathbb{N}_8} + \mathbb{N}_9 + 4n_3 + 
 \left(1 + 3\left[{n\over m}\right]\right)n_0, \nd}
with $n_0$ and $n_3$ are the coefficients associated with the temporal derivative and derivatives along the spatial, {\it i.e.} $(x^1, x^2)$ directions, respectively. Note the appearance of the relative {\red minus} signs for both $g_s$ and ${\rm M}_p$ scalings. This is important, as the ${\rm M}_p$ scaling is related to the combination $\sigma_{nl} + \gamma_{nl}$, so a relative minus sign there shows that for a given power of 
${\rm M}_p$ there would be in principle infinite number of possible terms. An example here would help clarify the scenario. Let us assume that we want to find how many operators are possible with the $(g_s, {\rm M}_p)$ scalings of the form $\left({g_s\over {\rm HH}_o}\right)^{a_1} {1\over {\rm M}_p^{b_1}}$. 
For concreteness, we can take the energy-momentum tensor associated with space-time ${\bf R}^{2, 1}$.
Combining the results of \eqref{sillyP}, \eqref{lilrabe} and \eqref{evehew} together then reproduces the following conditions on the various coefficients of \eqref{lilrabe}:

{\footnotesize
\bg\label{kilmill}
&&n_0 + n_3 + 2\sum_{i = 1}^2 \mathbb{N}_i + \sum_{j = 3}^8 \mathbb{N}_j {\red - \mathbb{N}_9} = b_1\\
&& 2\mathbb{N}_1 + 5\mathbb{N}_2  + 13 \mathbb{N}_3 + 10 \mathbb{N}_4 + 7 \mathbb{N}_5 
+ \mathbb{N}_6 + 4 \mathbb{N}_7 {\red -2 \mathbb{N}_8} + \mathbb{N}_9 + 4n_3 + \left(1+ 3\left[{n\over m}\right] \right)n_0 = 8 + 3a_1, \nonumber \nd}
where the relative minus signs are shown in {\red red}. For a given value of $b_1$ in ${\rm M}_p^{-b_1}$, we see that the first equation allows an infinite number of solutions. Similarly, for a given choice of 
$a_1$ in $\left({g_s\over {\rm HH}_o}\right)^{a_1}$ there are also an infinite number of solutions to the second equation. Thus the {\it overlap} between two infinite sets  is also an infinite set. This may be quantified by eliminating $\mathbb{N}_9$ using the first equation in \eqref{kilmill} to 
produce\footnote{Consider this in the following way. The first equation in \eqref{kilmill} is an equation for a plane in eleven dimensions (this has nothing to do with the fact that we are in eleven dimensions!). The second equation in \eqref{kilmill} is also an equation for a plane in eleven dimensions. For various choices of $(a_1, b_1)$ these two planes would overlap on another plane in ten (or eleven) dimensions. However since we are looking at integer values of $(n_i, \mathbb{N}_j)$, only integer points on the three planes would make sense here. Thus if and only if there are relative minus signs in the equations of {\it all} the three planes then there would be the possibility for an infinite number of integer points in the overlapping plane. When either or both the equations in \eqref{kilmill} have no relative minus signs, the intersecting plane argument is not helpful and there would be some hierarchy in the system with the possibility of an EFT description in the dual IIB side. \label{aprpolyps}}:

{\footnotesize
\bg\label{hereyes}
4\mathbb{N}_1 + 7\mathbb{N}_2 + 14 \mathbb{N}_3 +11 \mathbb{N}_4 + 8\mathbb{N}_5
+ 2\mathbb{N}_6+ 5 \mathbb{N}_7 {\red - \mathbb{N}_8} + \left(2 + \left[{n\over m}\right]\right)n_0 + 5n_3 = 8 +  3a_1 + b_1, \nd}
where we see that for a given choice of $(a_1, b_1)$ there are indeed an infinite possible 
solutions\footnote{There is yet another constraint that we kept under the rug so far, and it appears from imposing the Schwinger-Dyson's equations to the expectation values. The localized form of the G-flux components ${\bf G}_{{\rm MN}ab}$, from imposing \eqref{samaraw}, implies a similar behavior for the metric components, {\it i.e.}:
\bg\label{tutya}
{\bf g}_{\rm MN}(x, y; g_s, {\rm M}_p) = \left({g_s\over {\rm HH}_o}\right)^{-2/3} \sum_{k = 0}^\infty a_k^{(mn)}
g^{(k)}_{\rm MN}(x, y) ~{\rm exp}\left(-b^{(mn)}_k y^{2k}{\rm M}^{2k}_p\right), \nonumber \nd
where $(a^{(mn)}_k, b^{(mn)}_k)$ are constants, and we have restricted ourselves to only define the dominant metric components along $({\rm M, N}) \in {\cal M}_4 \times {\cal M}_2$ this way. Such a choice of the metric components, in turn, would effect the Riemann tensors. Since the Riemann tensors come with two derivatives, this will change $2(\mathbb{N}_1 + \mathbb{N}_2)$ to $(2 - 4k)(\mathbb{N}_1 + \mathbb{N}_2)$
in the first equation of \eqref{kilmill}, converting \eqref{hereyes}, in the simpler case with $k = 1$, to: 
\bg\label{hereyes2}
{\red 0\mathbb{N}_1} + 3\mathbb{N}_2 + 14 \mathbb{N}_3 +11 \mathbb{N}_4 + 8\mathbb{N}_5
+ 2\mathbb{N}_6+ 5 \mathbb{N}_7 {\red - \mathbb{N}_8} + \left(2 + \left[{n\over m}\right]\right)n_0 + 5n_3 = 8 +  3a_1 + b_1. \nonumber \nd
This has similar issues as \eqref{hereyes}, due to the vanishing of the $\mathbb{N}_1$ coefficient, implying that the situation might worsen in the time-independent case. Note that multiplying the second equation in \eqref{kilmill} by an arbitrary factor doesn't help. One might resort to the weaker form of the breakdown 
\eqref{hereyes} by demanding that the scale appearing in \eqref{samaraw} to be much {\it smaller} then ${\rm M}_p$. This would mean $\Omega_{\rm AB} \propto {\rm exp}\left[-y^{2l}{\rm M}^{2l}\right]\epsilon_{\rm AB}$, where ${\rm M} \equiv c {\rm M}_p$, and $c << 1$. Since $y > {\rm M}^{-1}$, the exponential piece is convergent, and with $\mathbb{N}_9$ derivative action we could 
get the two equations \eqref{kilmill}. There is however a subtlety that one needs to take here: all the metric and the flux components are functions of the dimensionless coordinates $x/R_x$ and $y/R_y$, where 
$R_i >> {\rm L}_p$ are some low-energy scales (see footnote \ref{ivylab}). Clearly ${\rm M}$ needs to be much larger than $R^{-1}_i$ for this to make sense, otherwise the derivative actions will introduce $R_i^{-1}$ scale dependences in our analysis. On the other hand a choice like ${\rm M}_p >> {\rm M} >> R^{-1}_i$ would work in the limit when 
$y \sim R_y$ and $x \sim R_x$. 
Although such a choice in \eqref{samaraw} allows us to use \eqref{kilmill},
this unfortunately doesn't help as we saw earlier, and therefore from either viewpoints the problem persists. 
In fact a more careful analysis, presented later, will show that an arbitrary choice of an intermediate scale is {\it not} essential to imply EFT breakdown. \label{mar1}} for the
$(\mathbb{N}_i, n_j)$ coefficients precisely because of the relative {\red minus} sign for $\mathbb{N}_8$.  From the quantum term \eqref{fahingsha3} and \eqref{lilrabe}, we see that $\mathbb{N}_8$ is related to the 
G-flux components ${\bf G}_{{\rm MN}ab}$. Thus it is this breakdown of $g_s$ hierarchy that eventually also breaks the ${\rm M}_p$ hierarchy in the system, eliminating any chance for a well defined EFT in four-dimensions with de Sitter isometries in the IIB side. Note that replacing $\mathbb{N}_9$ by $(2k - 1)\mathbb{N}_9$ by choosing higher powers of $y^{2k} {\rm M}_p^{2k}$ in the localized two-form \eqref{samaraw} would not change the result because \eqref{hereyes} will now be replaced by:

{\footnotesize
\bg\label{hereyes2}
&&4k\mathbb{N}_1 + (10 k - 3)\mathbb{N}_2 + (26k - 12) \mathbb{N}_3 +(20 k - 9) \mathbb{N}_4 
+ (14k - 6) \mathbb{N}_5
+ 2k \mathbb{N}_6 \\
&+ & (8k - 3) \mathbb{N}_7 {\red - (4k - 3)\mathbb{N}_8} + \left(2k + (2k -1)\left[{n\over m}\right]\right)n_0 + (8k - 3)n_3 = (2k - 1) (8 +  3a_1) + b_1, \nonumber \nd}
where, as before, most of the terms are positive definite except for one (shown in {\red red}) 
because $k > 1$.
Clearly, once we {\it flip} the sign of the $\mathbb{N}_8$ term by switching on time-dependent G-flux components ${\bf G}_{{\rm MN}ab}$, {\it i.e.} keeping $l_{\rm MN}^{ab} > {2\over 3}$, the second equation in \eqref{kilmill} will have a {\it finite} number of solutions for a given value of $a_1$. The first equation in \eqref{kilmill} continues to have an infinite number of solutions for a given value of $b_1$, but only a finite set of solutions can solve both these equations consistently. Taking the EOM constraint from imposing the Schwinger-Dyson's equations do not change the outcome. For example using the metric behavior as in the footnote \ref{mar1}, the constraint \eqref{hereyes}, in the absence of time-dependent G-flux components, takes the following form:

{\footnotesize
\bg\label{marjomey}
&&{\red 0\mathbb{N}_1} + (6k - 3)\mathbb{N}_2 + (26k - 12) \mathbb{N}_3 +(20 k - 9) \mathbb{N}_4 
+ (14k - 6) \mathbb{N}_5
+ 2k \mathbb{N}_6 \\
&+ & (8k - 3) \mathbb{N}_7 {\red - (4k - 3)\mathbb{N}_8} + \left(2k + (2k -1)\left[{n\over m}\right]\right)n_0 + (8k - 3)n_3 = (2k - 1) (8 +  3a_1) + b_1, \nonumber \nd}
which may be compared to \eqref{hereyes2}. Expectedly, as pointed out in footnote \ref{mar1}, this has similar issues as \eqref{hereyes2}. We could have also taken different powers of ${\rm M}_p$ in the exponential parts for the metric and the G-flux components, but the result would remain unchanged (we will deal with this generic case a little later). In the time-dependent case the sign of the $\mathbb{N}_8$ coefficient becomes positive, but the absence of the $\mathbb{N}_1$ factor doesn't create a problem because this will be fixed by the second equation in \eqref{kilmill}.
This results in a well defined set of operators for a given values of $(a_1, b_1)$ with time-dependent G-flux components ${\bf G}_{{\rm MN}ab}$.  

The above analysis at least shows why time-dependent G-flux components ${\bf G}_{{\rm MN}ab}$ are essential to allow for an EFT description in the IIB side. The EOM constraints do not change the results,  although it appears that the other components could in principle remain time-{\it independent}. This is {\it not} true because we haven't analyzed the flux EOMs. In our earlier analysis, done with a flat-slicing in \cite{desitter2, coherbeta}, the flux EOMs and anomaly cancellations told us that {\it all}  G-flux components should become time-{\it dependent}. We will discuss this a bit later, but before going into it, let us ask what happens if we had taken ${\cal B}_{0k} \ne 0$ in the choice of the two-form \eqref{samaraw}. Since this is time-dependent, the corresponding G-flux components 
${\bf G}_{{\rm MN}ab}$ would automatically develop some time-dependences. We will however keep 
$l_{\rm AB}^{\rm CD} = 0$ to see if this mild temporal dependences help us in any way. In this limit, the two equations in \eqref{kilmill} change to:

{\footnotesize
\bg\label{kilmill2}
&&n_0 + n_3 + 2\sum_{i = 1}^2 \mathbb{N}_i + \sum_{j = 3}^8 \mathbb{N}_j {\red - \mathbb{N}_9} = b_1\\
&& 2\mathbb{N}_1 + 5\mathbb{N}_2  + 13 \mathbb{N}_3 + 10 \mathbb{N}_4 + 7 \mathbb{N}_5 
+ \mathbb{N}_6 + 4 \mathbb{N}_7 {\red -2 \mathbb{N}_8 - \mathbb{N}_9} + 4n_3 + \left(1+ \left[{n\over m}\right] \right)n_0 = 8 + 3a_1. \nonumber \nd}
This unfortunately allows a more stronger breakdown of the four-dimensional EFT in the IIB side with $k = 1$ in \eqref{samaraw}. Once we impose additional time-dependences by switching on  $l_{\rm AB}^{\rm CD} = 0$ in the G-flux components, including the ones associated with 
$\mathbb{N}_8$ {\it i.e.} ${\bf G}_{{\rm MN}ab}$, the coefficient of the $\mathbb{N}_8$ factor would flip sign. However the $\mathbb{N}_9$ factor would remain unchanged. Here one could argue that, since the coefficients of both the $\mathbb{N}_9$ factors in \eqref{kilmill2} are $-1$ each, we could eliminate $\mathbb{N}_9$ from both these equations, and therefore $(\mathbb{N}_i, n_j)$ can have finite number of solutions. While this is generically true, the fact that the two-form in \eqref{samaraw} with 
${\cal B}^{\rm AB}_{0k} \ne 0$ vanishes as ${\rm exp}\left(-{1\over g_s^{2/3}}\right)$ for $g_s \to 0$, tells us that this choice of two-form is not useful. Thus it appears that the two-form \eqref{samaraw} cannot lead to a finite number of solutions in the absence of time-dependent G-flux components.

Let us see how this may work using an explicit example when no time-dependent G-flux components are switched on. The dangerous G-flux components are the ones with indices ${\bf G}_{{\rm MN}ab}$, and to match-up with the corresponding Riemann curvature components\footnote{We will not worry too much on the constraints coming from imposing the Schwinger-Dyson's equations. This is possible because, while our metric choice in the presence of \eqref{samaraw} develops additional exponential piece 
as ${\bf g}_{\rm AB} \propto g_s^{a_{\rm AB}} \tilde{g}_{\rm AB} e^{-y^2 {\rm M}^2_p}$ where $a_{\rm AB}$ is as in \eqref{sillyP}, we can choose to ignore it because of the following reasoning. The derivatives could be arranged to act on the  $\tilde{g}_{\rm AB}$ part in the Riemann curvature tensors, but on the exponential piece of the G-flux components 
${\bf G}_{{\rm MN}ab} \propto {\cal F}_{\rm MN} \Omega_{ab}$ if we take the gauge field ${\cal F}_{\rm MN}$ as a {\it slowly} varying function and the un-warped part of the metric to dominate. In other words, we are imposing the following conditions on the metric and the gauge field components:
\bg\label{mca53}
\partial_{\rm C}~ \tilde{g}_{\rm AB} >> 2 \eta_{{\rm CD}} ~
\tilde{g}_{\rm AB} ~y^{\rm D} {\rm M}_p^2, ~~~~~ \partial_{\rm C} {\cal F}_{\rm AB} \approx 0. \nonumber
\nd
Such conditions are not hard to impose in the Schwinger-Dyson's equations but since (as we show below) such choices of time-independent G-flux components {\it do not} lead to a well-defined EFT in four dimensions, we do not have to worry too much about the exact realizations of these aforementioned conditions when we classify the operators below. \label{mcabday}}, we will take the triplet $(\mathbb{N}_1, \mathbb{N}_8, \mathbb{N}_9)$ from \eqref{lilrabe}, keeping other $\mathbb{N}_i$ as well as $(n_0, n_3)$ vanishing. The $\mathbb{N}_9$ components are necessary to take care of the derivatives as described earlier, and $\mathbb{N}_1$ and $\mathbb{N}_8$ are each described using the sub-triplets 
$(l_3, l_4, l_{20})$ and $(l_{50}, l_{51}, l_{52})$ respectively (see \eqref{fahingsha3} and \eqref{lilrabe} for details). For further concreteness, we will consider the quantum terms contributing to 
$\mathbb{T}_{\mu\nu}$ and to order ${g_s^0\over {\rm M}_p^2}$, {\it i.e.} to zeroth order in $g_s$ and 
second order in ${\rm M}_p$. The equations governing the $\mathbb{N}_i$ can be easily read from 
\eqref{kilmill2} and take the form:
\bg\label{jenamal}
&&2\mathbb{N}_1 + \mathbb{N}_8 - \mathbb{N}_9 = b_1 \equiv 2 \nonumber\\
&&2\mathbb{N}_1 -2 \mathbb{N}_8 + \mathbb{N}_9 = 8 + 3a_1 \equiv 8, \nd
where we have chosen $(a_1, b_1) = (0, 2)$ to allow for quantum terms with ${g_s^0\over {\rm M}^2_p}$. Note that it is easy to generalize this to arbitrary $(a_1, b_1)$. A simple analysis of the above set of equations gives us:
\bg\label{capitani}
\left(\mathbb{N}_1, ~ \mathbb{N}_8, ~\mathbb{N}_9\right) ~ = ~ (k, ~4k - 10, ~6k - 12), ~~~~ k \ge 3, \nd
where the bound on $k$ keeps all $\mathbb{N}_i$ positive definite. Negative $\mathbb{N}_9$ would imply 
{\it non-local} operators which, as discussed in \cite{desitter2, coherbeta}, are very important for the consistency and existence of an EFT in the dual IIB side.  Here however we will only concentrate on the {\it local} and perturbative operators for simplicity. Two specific local operators contributing to the energy-momentum tensor $\mathbb{T}_{\mu\nu}$ may be written as:

{\footnotesize
\bg\label{zetazone}
&& {\bf g}_{\mu\nu} {\bf R}^3 \partial^6 {\bf G}^{2} \equiv 
{\bf g}_{\mu\nu} {\bf g}^{a_1 b_5} {\bf g}^{m_1 n_5} \prod_{i = 1}^4 {\bf g}^{b_i a_{i + 1}} 
\prod_{j = 1}^4 {\bf g}^{n_j m_{j + 1}} \prod_{l = 1}^3 {\bf R}_{m_l n_l a_l b_l} \prod_{p = 6}^{11} 
\partial_{m_p}
\partial_{n_p} \left(\prod_{s = 4}^5 {\bf G}_{m_s n_s a_s b_s} \right)\nonumber\\
&& {\bf g}_{\mu\nu} {\bf R}^4 \partial^{12} {\bf G}^{6} \equiv 
{\bf g}_{\mu\nu} {\bf g}^{a_1 b_{10}} {\bf g}^{m_1 n_{10}} \prod_{i = 1}^9 {\bf g}^{b_i a_{i + 1}} 
\prod_{j = 1}^9 {\bf g}^{n_j m_{j + 1}} \prod_{l = 1}^4 {\bf R}_{m_l n_l a_l b_l} \prod_{p = 11}^{22} 
\partial_{m_p}
\partial_{n_p} \left(\prod_{s = 5}^{10} {\bf G}_{m_s n_s a_s b_s}\right), \nonumber\\ \nd}
where $(m_i, n_i) \in {\cal M}_4$ and $(a_j, b_j) \in {\mathbb{T}^2\over {\cal G}}$, and the derivatives act 
on the G-flux components in a way described earlier. Note that there are multiple possible rearrangements of the various metric and flux factors possible, but here we only concentrate on a specific ordering of the various fields. Therefore, considering the two choices in \eqref{zetazone}, we see that a sub-class of operators contributing to the energy-momentum tensor $\mathbb{T}_{\mu\nu}$ may be written as the following series\footnote{A question could be asked here, and even earlier when we had introduced the quantum terms \eqref{fahingsha3}, regarding the {\it existence} of all these operators in M-theory. For example, how do we know that all these operators appear in the low energy Wilsonian action? The answer is that, in the absence of the knowledge of the full M-theory (or the corresponding type IIA) action, we cannot predict the precise values of the coefficients $c_k$ appearing in \eqref{sadakala}, or to discern whether some of the $c_k$'s vanish. The point however is not this and what is important is the following. In the absence of the precise form of the M-theory action, question is how much {\it genericity} we can impose on the choice of the operators. In other words, what constraints on the EFT  may be imposed once a generic form of the Wilsonian action is taken into account. If, in the end, our exercise can discern the way to avoid EFT breakdowns in 
four-dimensions,  we can be sure that  once we know the precise values  of $c_k$'s in \eqref{sadakala} the final answer will remain unchanged, albeit quantified better.}:
\bg\label{sadakala}
\mathbb{T}_{\mu\nu}(x, y; g_s, {\rm M}_p) \equiv {g_s^0 \over {\rm M}_p^2} \left(\sum_{k = 3}^\infty 
c_k ~{\bf g}_{\mu\nu} {\bf R}^k \partial^{6k - 12} {\bf G}^{4k - 10} + .... \right) + {\cal O}\left({g_s^{a} 
\over {\rm M}_p^b}\right), \nd
with $a > 0, b \ge 0$ and $c_k$ are {\it constant} coefficients that are independent of either $g_s$ or ${\rm M}_p$ but can take any signs. Note that the sub-class in \eqref{sadakala} is already infinite dimensional, but there exists an even bigger class of operators (also infinite dimensional, and shown here by the dotted terms) if we take all $\mathbb{N}_i$, $n_0$  and $n_3$ in \eqref{kilmill2} into account. 

In the time-independent case clearly we have lost both $g_s$ and ${\rm M}_p$ hierarchies, as evident from the infinite number of operators. However one might try to ask whether there could exist a different level of hierarchies, say in the terms inside the bracket in \eqref{sadakala}. The metric and the G-flux terms in \eqref{sadakala} or in \eqref{zetazone} appear from the expectation values over a Glauber-Sudarshan state, as in \eqref{pretroxie}, and therefore their functional forms are known. Typically they could be written in terms of the space-time and internal coordinates $x^i$ and $y^{\rm M}$ respectively, but since all the fields here are dimensionless, they typically appear as functions of ${x\over R_x}$ and ${y \over R_y}$ where $R_i$ are some scales to measure distances (see also footnotes \ref{ivylab} and \ref{mar1}). These scales cannot be short-distance scales as we have already integrated out the high-energy modes, so can only be IR scales that are many order smaller than 
${\rm M}_p$ (or many order larger than ${\rm L}_p$, the Planck length). As such one might try to construct a hierarchy using these scales. However this fails precisely when $x \sim R_x$ and $y \sim R_y$, showing that any arbitrary choice of scales cannot save the day, and the only saving grace appears from switching on temporal dependences to the G-flux components. This is further reinforced from the fact that the conclusion remains unchanged even if we switch on $k$ dependences from say {\bf Table \ref{firzasut}} and 
metric and flux ans\"atze from \eqref{reginadark}, because:
\bg\label{booang}
\theta_{nl} ~ \to ~ \theta_{nl} + {2\over 3} \sum_{i = 1}^{67} l_i k_i, \nd
where $k_i \equiv k \ge 0$, {\it i.e.} the lowest values for all $k_i$ are zero so in the $g_s$ expansions they cannot change the dominant scalings $\theta_{nl}$ from \eqref{botsuga}. The other factor $l_i$, defined in \eqref{fahingsha3}, are positive definite integers.

One might also be concerned by the specific choice of the localized function \eqref{samaraw}. What if we had a different choice of the localized function which may not dependent on the UV scale ${\rm M}_p$? For example we could have a localized function expressed in terms of the dimensionless coordinate $y/R_y$, which would not have introduced relative minus signs in the first lines of \eqref{evehew} and 
\eqref{kilmill}. This is of course a valid possibility and here we can see that, in the absence of time-dependent G-flux components, the breakdown happens in a slightly different way as shown in \cite{evan} earlier, namely:
\bg\label{taniarma}  
\mathbb{T}_{\mu\nu}(x, y; g_s, {\rm M}_p) = {\bf g}_{\mu\nu}
\sum_{a_1 = 0}^\infty \left({g_s\over {\rm HH}_o}\right)^{a_1} 
\left(\sum_{l = 0}^\infty b_l(x, y){\rm M}^l_p  + \sum_{k = 0}^\infty {c_k(x, y)\over {\rm M}^k_p}\right) + ...., \nd 
where we have taken the energy-momentum tensor along $2+1$ dimensional space-time, and the dotted terms are additional contributions coming from the derivative term in \eqref{faining}. The coefficients 
$c_k(x, y)$ and $b_l(x, y)$ are functions that can be derived from \eqref{fahingsha3}, the former directly from 
\eqref{fahingsha3} and the latter indirectly from the non-perturbative (and hence non-local) quantum terms
(see \cite{desitter2, coherbeta}).
Note that in \eqref{taniarma} we have lost the $g_s$ hierarchy in the usual way: the relative minus sign in say the second equation of \eqref{kilmill} allows an infinite number of solutions for a given value of $a_1$.  On the other hand, all these terms are suppressed by different powers of ${\rm M}_p$. However there are also non-local counter-terms that scale as positive powers of ${\rm M}_p$ (their contributions are {\it finite} as shown in \cite{desitter2, coherbeta}). Thus when ${\rm M}_p \to \infty$ we cannot ignore the positive powers of ${\rm M}_p$. Similarly, when ${\rm M}_p \to 0$, we cannot ignore the inverse powers of 
${\rm M}_p$. In this sense we have lost the ${\rm M}_p$ hierarchy altogether, implying again that a loss of $g_s$ hierarchy ultimately governs the loss of ${\rm M}_p$ hierarchy also. Therefore it appears that no matter how we want to count the operator contributions to the energy-momentum tensor, in the absence of temporal dependences of the G-flux components, finiteness or alternatively the existence of an EFT description cannot be achieved.

Before moving ahead, let us clarify another subtlety that could arise once time dependences are switched on. Some aspect of this have been described in footnotes \ref{mar1} and \ref{mcabday}, and has to do with the existence of the localized form for the G-flux components ${\bf G}_{{\rm MN}ab}$. Imposing mild temporal dependences using the exponential piece do not help as we saw in \eqref{kilmill2}, so question is what metric and flux ans\"atze we should take in the time-dependent case. A possible ans\"atze for the relevant G-flux components
may be written as:

{\footnotesize
\bg\label{tanyarbon}
{\bf G}_{{\rm MN}ab}(x, y; g_s, {\rm M}_p) \equiv \left({g_s\over {\rm HH}_o}\right)^{l_{\rm MN}^{ab}} \sum_{k = 0}^\infty c_k~
{\cal F}^{(k)}_{\rm MN}(x, y) ~{\rm exp}\left[-d_k\left({g_s\over {\rm HH}_o}\right)^{\xi k/3} y^{2(k+1)} {\rm M}^{2(k+1)}_p\right]\epsilon_{ab}, \nd} 
where we see that there is an additional $g_s$ dependence in the exponential piece with $(c_k, d_k)$ as constants, and $\xi$ is a constant. In the limit $g_s$ goes to zero as $g_s = \epsilon^\alpha$ and ${\rm M}_p$ goes to infinity as ${\rm M}_p = \epsilon^{-\beta}$, such that $\alpha > 12\beta/\xi$, one may perturbatively expand the exponential piece. The power of $g_s$ in the perturbative expansion follows consistently with the generic ans\"atze for the G-flux components from \eqref{theritual}. One may then envision the following ans\"atze for the metric components:

{\footnotesize
\bg\label{makibhalu}
{\bf g}_{\rm AB}(x, y; g_s, {\rm M}_p) \equiv \left({g_s\over {\rm HH}_o}\right)^{a_{\rm AB}} 
\sum_{k = 0}^\infty a^{(ab)}_k \widetilde{g}^{(k)}_{\rm AB}(x, y)~{\rm exp}\left[-b^{(ab)}_k\left({g_s\over {\rm HH}_o}\right)^{\xi k/3}
y^{2k} {\rm M}^{2k}_p\right], \nd}
with constant $(a^{(ab)}_k, b^{(ab)}_k)$, and $a_{\rm AB}$ as in \eqref{sillyP}. The exponential piece is clearly sub-dominant\footnote{In fact both \eqref{tanyarbon} and \eqref{makibhalu} will have additional sub-dominant series in 
${g_s\over {\rm HH}_o}$: the former coming from \eqref{sessence} and the latter coming from ${\rm F}_i(g_s/{\rm HH}_o)$ in the metric ans\"atze \eqref{evader} and \eqref{theritual}. Therefore for every order in ${g_s\over {\rm HH}_o}$ there should be a series 
expressed using ${\rm exp}(-y^{2k}{\rm M}^{2k}_p)$ using equivalent parameters like $(c_k, d_k)$ and $(a^{(ab)}_k, b^{(ab)}_k)$ for \eqref{tanyarbon} and \eqref{makibhalu} respectively in the following way: 
\bg\label{bic707}
&& {\bf g}_{\rm AB}(x, y; g_s, {\rm M}_p) \equiv \sum_{l = 0}^\infty \left({g_s\over {\rm HH}_o}\right)^{a_{\rm AB} + {2l\over 3}} 
\sum_{k = 0}^\infty a^{(ab)}_{kl} \widetilde{g}^{(k, l)}_{\rm AB}(x, y)~{\rm exp}\left[-b^{(ab)}_{kl}\left({g_s\over {\rm HH}_o}\right)^{\xi k/3}
y^{2k} {\rm M}^{2k}_p\right]\nonumber\\
&& {\bf G}_{{\rm MN}ab}(x, y; g_s, {\rm M}_p) \equiv \sum_{p = 0}^\infty\left({g_s\over {\rm HH}_o}\right)^{l_{\rm MN}^{ab} + {2p\over 3}} \sum_{k = 1}^\infty c_{kp}~
{\cal F}^{(k, p)}_{\rm MN}(x, y) ~{\rm exp}\left[-d_{kp}\left({g_s\over {\rm HH}_o}\right)^{\xi k/3} y^{2(k+1)} {\rm M}^{2(k+1)}_p\right]\epsilon_{ab},\nonumber \nd
where the series in \eqref{tanyarbon} and \eqref{makibhalu} are understood to 
be defined for $p = l = 0$. To avoid these complications, we will restrict ourselves with the series for the dominant $g_s$ expansions unless mentioned otherwise. \label{margincall}} and we expect for the metric components corresponding to ${\bf R}^{2, 1} \times 
{\mathbb{T}^2\over {\cal G}}$, the exponential correction piece to vanish (so that the four-dimensional de Sitter metric remains unchanged). In the following, we will not worry too much about this constraint, and count the operators as though $({\rm A, B}) \in {\bf R}^{2, 1} \times {\cal M}_4 \times {\cal M}_2 \times 
{\mathbb{T}^2\over {\cal G}}$, but there are other subtleties that we will have to consider. {\Su First}, in the scaling of the Riemann curvatures, as given in {\bf Table \ref{firzasut}}, the derivatives acted on the 
$\widetilde{g}_{\rm AB}$ part as no exponential piece was taken. Now of course there is a possibility that the derivatives act on the exponential piece too. Fortunately, the $g_s$ scalings of the Riemann curvature tensors only get sub-dominant contributions from the exponential piece, so the results of {\bf Table \ref{firzasut}} remain unchanged. This is good, but the exponential piece could contribute to the ${\rm M}_p$ scalings (albeit suppressed by the $g_s$). In the limit $g_s$ goes to zero {\it faster} than ${\rm M}_p$ goes to infinity, this is not much of an issue. The issue, if any, then appears when the exponential piece dominates. {\Su Secondly}, we could repeat the same story for the G-flux components \eqref{tanyarbon}, but there is not much of a concern here because the derivatives are controlled by $\mathbb{N}_9$ factor in \eqref{lilrabe}, and we could always study those operators where they act on the exponential piece\footnote{Recall that the degrees of freedom are actually the world-volume gauge fields ${\cal F}_{\rm AB}(x, y)$ and they would appear in the Schwinger-Dyson's equations.}. Thus the distinction is between the {\it inherent} derivatives, defining the Riemann curvature tensors, versus the {\it external} derivatives, governed by say $\mathbb{N}_9$ here, that can act on the exponential terms. {\Su Thirdly}, we see that the metric components in 
\eqref{makibhalu} as well as the ones in footnote \ref{mar1}, are defined by  a series with $k \ge 0$, whereas the G-flux components are defined by  a series with $k \ge 1$ (or with $k \ge 0$ but with $k+1$ modings). In the Schwinger-Dyson's equations, this would imply the appearance of pieces of the curvature tensors {\it without} exponential parts as well as pieces of the curvature tensors accompanied by the exponential parts. On the other hand, the G-flux contributions, specific to \eqref{tanyarbon}, are always accompanied by the exponential pieces. A natural question is whether we can allow $a_k^{(ab)} = 0$, for $k \ge 1$. The answer is surprisingly yes, because the form of the G-flux components in say \eqref{tanyarbon}, or even in \eqref{sessence}\footnote{In a similar vein, the localized two-form $\Omega^{(k)}_{\rm MN}$ can be defined in an identical way to accommodate the form in \eqref{tanyarbon} because of the way the series appears in \eqref{sessence}. For example we can write it as:
\bg\label{khalame}
\Omega^{(k)}_{\rm MN} \equiv {\rm exp}\left[-d_{k} \left({g_s\over {\rm HH}_o}\right)^{\beta k}y^{2(k+1)}{\rm M}^{2(k+1)}_p\right]\epsilon_{\rm MN}, \nonumber \nd 
where $k \ge 0$ now, and $\beta$ is a constant factor inserted so that the perturbative expansion keeps 
$l^{ij}_{\rm MN}$ as the dominant scaling. One could also take $\beta = 0$, so that the $g_s$ is completely determined by $l^{ij}_{\rm MN} + {2k\over 3}$ in \eqref{sessence}. Either definitions give consistent answers so we don't have to be particularly careful which definition to use as long as the dominant scaling remains unchanged.}, is particularly reminiscent of the sum of the complete set of states for a higher-dimensional harmonic oscillator where the gauge fields 
${\cal F}^{(k)}_{\rm MN}(x, y)$ can be replaced by the corresponding Hermite polynomials of the form
$\mathbb{H}^{(k)}(x, y) \epsilon_{\rm MN}$. This means any arbitrary functional form for ${\bf G}_{{\rm MN}ab}$ (or ${\bf G}_{{\rm MN} ij}$) may be constructed satisfying the Schwinger-Dyson's equations with appropriate choices of $(c_k, d_k)$, thus eliminating the pressure on the metric components to simulate an accompanying exponential factor (see also footnote \ref{marjobrief}).  

One might however worry that such a choice of the metric ans\"atze may be too restrictive. In the following let us give 
an example to help us to see how this may work more quantitatively even when we consider 
the generic form \eqref{makibhalu} for the metric components (from here going to the simpler case would be easier).
Again the relevant degrees of freedom are the curvature tensors ${\bf R}_{{\rm MN}ab}$ governed by $\mathbb{N}_1$, the G-flux components 
${\bf G}_{{\rm MN}ab}$ governed by $\mathbb{N}_8$, and the derivatives 
governed by $\mathbb{N}_9$  in \eqref{lilrabe}. The only metric components contributing to ${\bf R}_{{\rm MN}ab}$ are ${\bf g}_{{\rm MN}}$ and ${\bf g}_{ab}$. If we want the exponential part in \eqref{makibhalu} to dominate $-$ which is opposite to what we took in footnote \ref{mcabday} $-$ we can take ${\cal M}_4 \times {\cal M}_2$ to be $\mathbb{T}^4 \times \mathbb{T}^2$ locally (globally they need to be non-trivial otherwise the Euler characteristics would vanish). This means 
both $\widetilde{g}_{\rm MN}$ and $\widetilde{g}_{ab}$ can be taken to be locally {\it flat}. This way the dominant ${\rm M}_p$ scalings appear from the two derivatives acting on the exponential part.  Putting everything together, we see that the result now changes for the energy-momentum tensor $\mathbb{T}_{\mu\nu}$ from \eqref{jenamal} to the following:
\bg\label{mameyjodi}
&&~\mathbb{N}_8 - (4k_1 - 2)\mathbb{N}_1 - (2k_2 - 1)\mathbb{N}_9 = b_1\nonumber\\
&&\left(3l_{\rm MN}^{ab} - 2\right)\mathbb{N}_8 + 2(k_1 + 1)\mathbb{N}_1 + (k_2 + 1)\mathbb{N}_9 = 
8 + 3a_1, \nd
where $k_1$ is the $k$-scaling of the metric in \eqref{makibhalu} and $k_2$ is the $k$-scalings of either the metric or the G-flux components \eqref{tanyarbon}, or both depending on how the $\mathbb{N}_9$ derivatives act\footnote{We have also assumed that $\partial_y{\cal F}^{(k)}_{\rm MN} << {\rm M}_p$ so the 
derivative action on the exponential term dominates in \eqref{tanyarbon}.}. One can easily show that as long as $l_{\rm MN}^{ab} > {2\over 3}$, the second equation in \eqref{mameyjodi} has finite number of solutions for a given $a_1$. Thus, although the first equation in 
\eqref{mameyjodi} has an infinite number of solution for any $b_1$, the overlap between the two equations has only a finite number of solutions. Therefore, once time-dependent G-flux components are switched on, {\it i.e.} 
taking $l_{\rm MN}^{ab} > {2\over 3}$, there are only finite number of operators possible at order 
$\left({g_s \over {\rm HH}_o}\right)^{a_1} \times {1\over {\rm M}^{b_1}_p}$. This should also be obvious by combining the two equations in \eqref{mameyjodi} to get the following condition:
\bg\label{tanholojeni}
6\mathbb{N}_1 + 3\mathbb{N}_9 + \left(6l_{\rm MN}^{ab} - 3\right)\mathbb{N}_8 = 16 + 6a_1 + b_1, \nd
which clearly has a finite number of solutions, and is also independent of $(k_1, k_2)$. Interestingly, one may easily see that relaxing the flatness conditions of the metric components $\widetilde{g}_{\rm MN}$ and $\widetilde{g}_{ab}$ doesn't change the above conclusion. In fact even if we consider the exponential function to be defined with respect to $y/R_y$, where 
$R_y$ is an IR scale instead of ${\rm M}_p$, the result remains unchanged, {\it i.e.} there continues to be a finite number of operators contributing to the energy-momentum tensor $\mathbb{T}_{\mu\nu}$ once time-dependent G-flux components  are switched on.

\subsection{Contributions from the non-perturbative quantum effects \label{npeffects}}

There are also non-perturbative contributions coming from wrapped instantons and other effects. The instanton ones get further contributions from the non-local counter-terms that was discussed in much details in \cite{coherbeta}. However in \cite{coherbeta} we only took the contributions from the {\it lowest} order in non-localities by summing over the trans-series. Questions can be raised regarding the generic effects one might see once higher order non-localites are taken into account.  In the following we will discuss under what conditions they might contribute finite values. 

The non-local counter-terms, by definitions, should not be visible at low energies, which means in the classical ({\it i.e.} the supergravity part of the) M-theory action, their effects should be minimal. That this is true was shown in the first reference of \cite{desitter2}. However, since we are going beyond the standard classical effects, one could ask how the generic non-localities manifest themselves from loop effects. In the limit $g_s << 1$, and to lowest orders in the non-localities, these effects could be quantified in the form of  
the BBS \cite{BBS}  and KKLT \cite{KKLT} type instantons. Note that these effects are in addition to the standard BBS and KKLT instantons' contributions.

Let us analyze these effects when we allow non-localities to $q$-th order. The non-localities are defined using functions $\mathbb{F}^{(q)}(y_{q+1} - y_q)$ for two points $(y_{q+1}, y_q)$ in the internal eight-dimensional manifold. However now we have metric and G-flux components to have dependences on all the eleven-dimensional space, or at least have dependences on the coordinates of ${\bf R}^{2, 1} \times {\cal M}_4 \times {\cal M}_2$ as we have seen so far. We can then in principle demand that the non-local functions 
also develop such dependences, but then this will involve introducing non-localities along the non-compact space-time directions leading to a more involved scenario. To avoid such complications we will restrict ourselves to non-localities that involve only the coordinates of the internal eight-manifold, although subtleties could appear once we have dependences on $w^a \equiv (x^3, w^{11})$ directions. We will not worry too much about the latter, and express the non-perturbative contributions as:
\bg\label{brazmukh}
{\bf S}_{\rm np}^{(q)} = {\rm M}_p^{11} \int d^3x d^8 y \sqrt{-{\bf g}_{11}(x, y)} ~\sum_k c_k\Big[{\rm exp}
\left(-k \big\vert\mathbb{G}^{(q)}(x, y)\big\vert\right) - 1\Big], \nd
where the superscript $q$ denotes the order of non-localities that we consider here, and the sum is over all integer $k$. Such a form was derived in the second reference of \cite{coherbeta} which the readers may look up for more details.  Note the appearance of $\vert\mathbb{G}^{(q)}(x, y)\vert$ instead of just  $\mathbb{G}^{(q)}(x, y)$. Such a choice facilitates convergence,  but one can argue this directly when we sum the trans-series. For example, if $\mathbb{G}^{(q)}(x, y)$ is a negative definite function, then one can choose the coefficients $d_n$ in eq. (3.58) and (3.59) of the second reference of \cite{coherbeta} such that there is an overall minus sign in the exponential of \eqref{brazmukh}. This means we can always allow an exponentially decaying function in \eqref{brazmukh} implying $\mathbb{G}^{(q)}(x, y)$ appears as 
$\vert\mathbb{G}^{(q)}(x, y)\vert$ there. The function $\mathbb{G}^{(q)}(x, y)$ contains all information about the non-localities, and may be expressed as:

{\footnotesize
\bg\label{brazmukh2}
\mathbb{G}^{(q)}(x, y) \equiv \int d^8y_q \sqrt{{\bf g}_8(y_q, x)}~\mathbb{F}^{(q)}(y - y_q) \prod_{r = 1}^{q-1} d^8 y_r 
\sqrt{{\bf g}_8(y_r, x)}\left({\mathbb{F}^{(r)}(y_{r+1} - y_r)\mathbb{Q}_{\rm T}^{(\{l_i\}, n_i)}(y_1, x)\over 
{\rm M}_p^{\sigma(\{l_i\}, n_i) - 8q}}\right), \nonumber\\ \nd}
where $\mathbb{Q}_{\rm T}^{(\{l_i\}, n_i)}(y_1, x)$ is the quantum series given in \eqref{fahingsha3} and 
$\sigma(\{l_i\}, n_i)$ is the typical ${\rm M}_p$ dimension of the quantum series that could be related to 
\eqref{harpno} and \eqref{evehew}. We will come back to this a little bit later. We can also sum over 
$(\{l_i\}, n_i)$, instead of just $k$ in \eqref{brazmukh}, to contain all possible quantum terms, but this will simply make the above series more complicated without revealing any new physics. We will avoid such complications at this stage.

We can now ask how much does \eqref{brazmukh} contributes to the energy-momentum tensor. More importantly, since we expect the contributions to the energy-momentum tensor to involve the non-locality functions $\mathbb{F}^{(q)}(y_{q+1} - y_q)$, are these contributions finite? The energy-momentum tensor becomes:

{\footnotesize
\bg\label{brazmukh3}
\mathbb{T}^{(q)}_{\rm AB} &\equiv &  {-2{\rm M}_p^{11}\over \sqrt{-{\bf g}_{11}(z_1, z_2)}} {\delta \over 
\delta {\bf g}^{\rm AB}(z_1, z_2)}\left(\int d^3x d^8 y \sqrt{-{\bf g}_{11}(x, y)} ~\sum_k c_k\Big[{\rm exp}
\left(-k\big\vert \mathbb{G}^{(q)}(x, y)\big\vert\right) - 1\Big]\right) \nonumber\\
& = & \sum_k c_k~{\bf g}_{\rm AB}(z_1, z_2) ~\Big[{\rm exp}
\left(-k\big\vert \mathbb{G}^{(q)}(z_1, z_2)\big\vert\right) - 1\Big] 
+  \sum_k {2k c_k \over {\rm M}_p^{\sigma(\{l_i\}, n_i) - 8q}}\int d^8 y d^8 y_q
 ~\mathbb{F}^{(q)}(y - y_q)\nonumber\\
&\times& \sum_{s = 1}^q \delta^{8}(y_s - z_2)  \prod_{r = 1}^{q - 1} d^8 y_r \mathbb{F}^{(r)}(y_{r+1} - y_r){\delta\over \delta {\bf g}^{\rm AB}(z_1, z_2)} 
 \left(\sqrt{{\bf g}_8(y_q, z_1) {\bf g}_8(y_r, z_1)}\mathbb{Q}_{\rm T}^{(\{l_i\}, n_i)}(y_1, z_1)\right)
 \nonumber\\ 
&\times & {\rm exp} \left(-k\big\vert \mathbb{G}^{(q)}(z_1, y)\big\vert\right) \sqrt{{{\bf g}_{11}(z_1, y) \over {\bf g}_{11}(z_1, z_2)}},
  \nd}
where $z_1 \in {\bf R}^{2, 1}$ and $z_2 \in {\cal M}_4 \times {\cal M}_2 \times {\mathbb{T}^2\over {\cal G}}$, 
${\bf g}_{\rm AB}$ is the full warped metric, 
$\mathbb{G}^{(q)}(x, y)$ is as defined in \eqref{brazmukh2} and we have assumed that 
$\mathbb{F}^{(0)}(y_1 - y_0) \equiv 0$. To see how each of the terms contribute, we have to find how they scale with respect to $g_s$ and ${\rm M}_p$. If we assume that the non-locality functions have no 
explicit $g_s$ or ${\rm M}_p$ dependences, then it is easy to see that $\mathbb{G}^{(q)}(x, y)$ scales as
$\left({g_s \over {\rm HH}_o}\right)^{\theta_{nl} - 2q/3}$. Combining the ${\rm M}_p$ scalings of every term, we see that the energy-momentum tensor \eqref{brazmukh3} has the following $(g_s, {\rm M}_p)$ scaling:
\bg\label{orihitt}
\mathbb{T}^{(q)}_{\rm AB} &= & \left({g_s \over {\rm HH}_o}\right)^{a_{\rm AB}} \left[{\rm exp}
\left(-{\rm M}_p^{8q - \sigma(\{l_i\}, n_i)} \left({g_s \over {\rm HH}_o}\right)^{\theta_{nl} - 2q/3}\right) - 1\right]
\\
&+ & {\rm M}_p^{8q - \sigma(\{l_i\}, n_i)} \left({g_s \over {\rm HH}_o}\right)^{a_{\rm AB} 
+ \theta_{nl} - 2q/3}
~{\rm exp}\left(-{\rm M}_p^{8q - \sigma(\{l_i\}, n_i)} \left({g_s \over {\rm HH}_o}\right)^{\theta_{nl} - 2q/3}\right), 
 \nonumber \nd
where $a_{\rm AB}$ is defined in \eqref{sillyP}. When the fluxes are time-independent, $\theta_{nl}$ is given by the second relation in \eqref{evehew}, and as such will have relative minus sign(s). With time-dependent G-flux components, this takes the form \eqref{botsuga}. Thus if we are looking for contributions of order 
$\left({g_s\over {\rm HH}_o}\right)^a {\rm M}_p^{|b|}$, then they can be extracted from \eqref{fahingsha3} 
with:
\bg\label{annamie}
\theta_{nl} = a - a_{\rm AB} + {2q\over 3}, ~~~~~ \sigma(\{l_i\}, n_i) = 8q - |b|, \nd
at $q$-th order in non-locality. In the Schwinger-Dyson's equations, the most negative $a$ could be is when 
$a = -2$, which is related to the space-time equations. For this case $a_{\rm AB} = -{8\over 3}$, implying that 
$\theta_{nl}$ in \eqref{annamie} to be positive definite. When $a_{\rm AB} = +{4\over 3}$, $a = +2$, implying that $\theta_{nl}$ remains positive definite again. Thus if there are no relative minus signs in 
$\theta_{nl}$ $-$ which is the case when $l_{\rm AB}^{\rm CD} > {2\over 3}$ in \eqref{botsuga} $-$ the first equation in \eqref{annamie} has finite number of solutions for {\it any} order $q$ of non-locality. In that case it doesn't really matter whether $\sigma(\{l_i\}, n_i)$ has any relative minus signs: there would still be finite number of quantum terms contributing to the energy-momentum tensor at any order in $q$. Interestingly, the contributions from the second term in \eqref{brazmukh3} become {\it finite} and {\it local} because of the nested integral structure. This may be seen first from:

{\footnotesize
\bg\label{romy}
\int d^3x d^8 y ~\sqrt{-{\bf g}_{11}(x, y)} ~\mathbb{F}^{(q)}(y - y_q) ~{\rm exp}
\bigg(-k\big\vert \mathbb{G}^{(q)}(x, y)\big\vert\bigg) \equiv \left({g_s\over {\rm HH}_o}\right)^{-14/3}{\cal F}^{(q)}(y_q; g_s, 
{\rm M}_p, k), \nd}
which gives a finite and local answer because the non-locality has been integrated away. If the higher order non-localities can be ignored, then this is all there is to the analysis, and we don't have to know the precise functional form for the function $\mathbb{F}^{(1)}(y_2 - z_2)$. In the case we keep higher order non-locality function, we see that the exponential pieces in \eqref{brazmukh3} typically scale as:

{\footnotesize
\bg\label{tabula}
\Bigg[1 + {\rm M}_p^{|b|}\left({g_s\over {\rm HH}_o}\right)^{a - a_{\rm AB}}\prod_{r = 1}^q \mathbb{F}^{(r)}(y_{r+1} - y_r)
\Bigg]{\rm exp}\left[-k {\rm M}_p^{|b|} \left({g_s\over {\rm HH}_o}\right)^{a - a_{\rm AB}}
\prod_{r = 1}^q \mathbb{F}^{(r)}(y_{r+1} - y_r)\right], \nd}
where $y_{q+1} \equiv y$. We note that for fixed ${\rm M}_p, g_s$ and $\mathbb{F}^{(r)}$, the exponential term is suppressed by $k$ because $a - a_{\rm AB} > 0$ as well as $|b| > 0$. Therefore sum over $k$ gives rise to a convergent series. Similarly for ${\rm M}_p \to \infty$, the exponential part goes to zero faster than any polynomial powers, when we keep $g_s, k$ and $\mathbb{F}^{(r)}$ fixed. In a similar vein, keeping 
${\rm M}_p, k$ and $\mathbb{F}^{(r)}$ fixed, the exponential part goes to identity when $g_s \to 0$. This would then decouple the first term with ${\bf g}_{\rm AB}$ in \eqref{brazmukh3}, but keep the second term intact (albeit suppressed by powers of $g_s$ as one would expect). Finally in the limit for fixed $g_s, k$ and ${\rm M}_p$, when the products of the non-locality functions $\mathbb{F}^{(r)}$ with $1 \le r \le q$ become smaller for large values of $q$, both the terms in \eqref{brazmukh3} vanishes. This limiting choice is expected because higher order non-localities cannot effect the low-energy results. It also resolves one puzzle we could have had in \eqref{annamie}, namely, the relative {\it minus} sign in $\theta_{nl} - {2q\over 3}$ breaking the $g_s$ hierarchy even when $\theta_{nl}$ itself doesn't have relative minus signs. Clearly, the decoupling of quantum terms with arbitrary higher orders in $q$, saves the day here. Of course when 
$\theta_{nl}$ has relative minus signs, due to the absence of time-dependent G-flux components, this still doesn't help and we lose $g_s$ hierarchy altogether. 

The above discussion relies on the fact that the non-locality function does not have any explicit dependence on $g_s$ or ${\rm M}_p$. However one could ask if the implicit dependence on $g_s$ and ${\rm M}_p$ could change anything. To answer this, we will have to take a specific form of the non-locality function. One such example is constructed in the first reference of \cite{desitter2}, and may be quoted here as:
\bg\label{povedaE}
\mathbb{F}^{(r)}(y_{r+1} - y_r) = \sum_{\{n\}} {\cal C}^{(r)}_{\{n\}} 
\mathbb{H}_{\{n\}}\left({y_{r+1} - y_r\over \mathbb{V}_8^{1/8}}\right)
{\rm exp}\left[-{(y_{r+1} - y_r)^2\over \mathbb{V}_8^{1/4}}\right], \nd
where $1 \le r \le q$ with $y_{q+1} \equiv y$, ${\cal C}^{(r)}_{\{n\}}$ are constants and $\mathbb{H}_{\{n\}}\left({y_{r+1} - y_r\over \mathbb{V}_8^{1/8}}\right)$ are the eight-dimensional Hermite polynomials (constructed by solving the eigenstates of a Simple Harmonic Oscillator (SHO) in eight Euclidean dimensions\footnote{The eight-dimensional Hermite polynomials are defined using products of Hermite polynomials along the eight directions, much like what we have for the eigenstates of a three-dimensional harmonic oscillator. Therefore:
$$\mathbb{H}_{\{n\}}\left({y - y'\over \mathbb{V}_8^{1/8}}\right) \equiv \prod_{i = 1}^8 
\mathbb{H}_{n_i}\left({y_i - y'_i\over \mathbb{R}_i}\right)$$
\noindent $\{n\} \equiv (n_1, n_2, ....., n_8)$, and the arguments in the Hermite polynomials are made dimensionless using the scale of the internal eight manifold $\mathbb{R}_i$, such that $\mathbb{V}_8 \equiv \mathbb{R}_1...\mathbb{R}_8$. If $\zeta$ denotes a dimensionless coordinate, then these polynomials are normalized using the standard procedure: 
$\int d\zeta\mathbb{H}_{n}(\zeta) \mathbb{H}_{m}(\zeta) e^{-\zeta^2} = 2^m m! \sqrt{\pi} \delta_{nm}$, which in turn may be used to express any function in the internal eight-manifold.  For example, we can identify the gauge fields in \eqref{tanyarbon} with the eight-dimensional Hermite polynomials as ${\cal F}^{(k)}_{\rm MN}(x, y) \equiv 
\mathbb{H}_{\{k\}}(x, y) \epsilon_{\rm MN}$ with $(x, y)$ made dimensionless using some IR scales much like $(R_x, R_y)$ used earlier. Note that such a choice can justify the $k = 0$ case in our metric 
ans\"atze of \eqref{makibhalu} for appropriate values of $c_k$ in \eqref{tanyarbon}. \label{marjobrief}}). As such for specific choices of the coefficients ${\cal C}^{(r)}_{\{n\}}$, any desired functional form for $\mathbb{F}^{(r)}(y_{r+1} - y_r)$ may be constructed.  The volume $\mathbb{V}_8$ is defined as 
the {\it difference} between the volume of the internal eight-manifold viewed as a Glauber-Sudarshan 
state\footnote{One subtlety arises here compared to what we had in \cite{desitter2}. Since we are using generic slicing of four-dimensional de Sitter space, the volume of the internal eight-manifold viewed as a Glauber-Sudarshan state will be a function two-dimensional spatial coordinate $x^i$ as well as of the expected $(y, g_s)$ (assuming no dependence on $w^a$). Thus to avoid the $x^i$ dependence in \eqref{povedaE} we can either integrate over all $x^i$ or keep $x^i$ to some fixed slice. See also footnote 
\ref{hana00}.} and the 
volume of the solitonic eight-manifold. It is easy to see that the non-locality function decouples in the usual supergravity limit, and typically in the limit $g_s \to 0$, but does provide non-trivial contribution in the limit when the toroidal volume ${\mathbb{T}^2\over {\cal G}}$ as well as $g_s$ vanish (see footnote 49 in the first reference of \cite{desitter2}).

The coefficients ${\cal C}^{(r)}_{\{n\}}$ are generically chosen so that the $\mathbb{F}^{(r)}(y_{r+1} - y_r)$ 
functions remain as 
close deformations of gaussian functions in a way that we regain locality in the limit the deformed gaussian functions become delta functions. This means the superscript $r$ in \eqref{povedaE}  implies multiplying 
$r$ gaussian-like functions in \eqref{brazmukh2}. As $r$ increases, the gaussian-like functions would decrease exponentially, making $\mathbb{G}^{(r)}(x, y)$ in \eqref{brazmukh2} as well as the action 
${\bf S}^{(r)}_{\rm np}$ in \eqref{brazmukh} to vanish,  implying no contributions from higher order non-localities. Thus the relative minus sign in the $g_s$ scaling of the form 
$\left({g_s \over {\rm HH}_o}\right)^{\theta_{nl} - 2r/3}$ in say \eqref{orihitt} poses no problems in constructing EFT description in the IIB side as long as $\theta_{nl}$ by itself is devoid of any relative minus signs.

The decoupling of higher order non-localities is definitely an attractive (and possibly an expected) feature of the construction, but it does pose some subtleties in defining the energy-momentum tensor in the time-independent case. For example, one could raise the question that $b_l$ in \eqref{taniarma}, for large values of $l$, should 
decouple and therefore there should appear some level of ${\rm M}_p$ hierarchy in the system even if there isn't a $g_s$ hierarchy. Shouldn't this be a way out to constructing an EFT description in the time-independent case?

In the following we will argue that the decoupling of higher order non-localities {\it cannot} help us regain 
four-dimensional EFT description as long as there is no $g_s$ hierarchy.
Of course we should keep in mind that the construction \eqref{taniarma} is in the limit where the localized function, for example 
\eqref{samaraw}, {\it does not} depend on the UV scale ${\rm M}_p$. When the localized function \eqref{samaraw} does depend on the UV scale ${\rm M}_p$, the analysis is pretty clear (see for example the second reference of \cite{desitter2}), and the existence or non-existence of four-dimensional EFT description can be precisely quantified.  For the former case we start by expressing the warped volume
$\mathbb{V}_8$ as ${\bf V}_8^{(1)} - {\bf V}_8^{(2)}$ where  ${\bf V}_8^{(1)}$ is the volume of the eight-manifold Glauber-Sudarshan state and ${\bf V}_8^{(2)}$ is the corresponding volume of the supersymmetric vacuum configuration (see footnote 49 in the first reference of \cite{desitter2}). However, now due to our choice of de Sitter slicings, ${\bf V}_8^{(1)}$ can be expressed as an average volume in the following way:
\bg\label{the12}
{\bf V}_8^{(1)} \equiv {\int d^8 y d^2 {\bf x} \sqrt{{\bf g}_2(x, y) {\bf g}_8(x, y)} \over \int d^2 {\bf x} \sqrt{{\bf g}_2(x, y)}}
= \left({g_s \over {\rm HH}_o}\right)^{-2/3} {\rm M}_p^{-8} {\rm V}_8^{(1)}, \nd
where $x \equiv (x^0, {\bf x})$ and ${\rm V}_8^{(1)}$ is the dimensionless eight-volume. On the other hand, 
the volume of the eight-manifold for the supersymmetric vacuum is ${\rm M}_p^{-8} {\rm V}_8^{(2)}$ with dimensionless ${\rm V}_8^{(2)}$. Thus it appears, both in the exponential and in the Hermite polynomials, there are $\left({g_s\over {\rm HH}_o}\right)^{1/6} {\rm M}_p^2$ dependences. Taking the simplest case where ${\cal C}^{(r)}_{\{n\}} = 0$ for $n > 0$ in \eqref{povedaE}, we see that:
\bg\label{equinox}
\mathbb{F}^{(r)}(y_{r+1} - y_r) = 
{\cal C}^{(r)}_{\{0\}} ~{\rm exp}\left[-{{\rm M}_p^2 (y_{r+1} - y_r)^2 \over \left({\rm V}_8^{(1)}
- {\widetilde{\rm V}}_8^{(2)}\right)^{1/4}}\left({g_s\over {\rm HH}_o}\right)^{1/6}\right], \nd
as the non-locality function to order $r$, where $\widetilde{\rm V}_8^{(2)} \equiv \left({g_s\over {\rm HH}_o}\right)^{2/3}{\rm V}_8^{(2)}$ and $1 \le r \le q$ with $y_{q+1} \equiv y$. This clearly vanishes for ${\rm M}_p \to \infty$, and becomes trivial when $g_s \to 0$ faster than ${\rm M}_p \to \infty$. In the usual time independent supergravity limit where ${\rm M}_p \to \infty$ we expect
${\rm V}_8^{(1)} = {\rm V}_8^{(2)}$, because ${g_s\over {\rm HH}_o} \equiv 1$, and therefore 
\eqref{equinox} vanishes. Non-trivial contributions appear in the limit $g_s \to 0$ 
and ${\rm M}_p \to \infty$ when ${\rm M}_p^2 \left({g_s\over {\rm HH}_o}\right)^{1/6} =$ constant. This is the limit where we expect non-local effects to show up in our scenario, at least for the simple choice of \eqref{equinox}. Keeping this in mind, we can rewrite \eqref{brazmukh2} in the following way:

{\footnotesize
\bg\label{brazmukh5}
\mathbb{G}^{(q)}(x, y) \equiv -\prod_{r = 1}^{q-1}\int d^8y_q d^8y_r~\partial^{(n_i)}\Bigg[\sqrt{{\bf g}_8(y_q, x){\bf g}_8(y_r, x)}~\mathbb{F}^{(q)}(y - y_q)~\mathbb{F}^{(r)}(y_{r+1} - y_r)\Bigg]{{\bf Q}_{\rm T}^{(\{l_i\})}(y_1, x)\over 
{\rm M}_p^{\sigma(\{l_i\}, n_i) - 8q}}, \nonumber\\ \nd}
where we have defined the quantum series \eqref{fahingsha3} as $\mathbb{Q}_{\rm T}^{(\{l_i\}, n_i)}(y_1, x) 
\equiv \partial^{(n_i)}{\bf Q}_{\rm T}^{(\{l_i\})}(y_1, x)$ with the superscripts denoting appropriate contractions with inverse metric components. The derivative action is arranged to be along the internal eight-manifold, which is possible because the internal eight-manifold has no singularities or 
boundaries\footnote{The orbifold structure
coming from ${\mathbb{T}^2\over {\cal G}}$, which keeps the Euler characteristics from vanishing, does not introduce any singularities on the eight-manifold at the quantum level. It is not too hard to demonstrate the appropriate blow-ups that resolve the classical singularities.}  (this provides the 
overall minus sign in \eqref{brazmukh5}). 

Let us now analyze the situation when the G-flux components are time-independent and consider the simpler form \eqref{equinox} instead of \eqref{povedaE}. We will also assume no dependence on the toroidal fibre directions (this will be studied in section \ref{torus}), and therefore the quantum series 
is as given in \eqref{fahingsha3}. Using the same parameters as in \eqref{lilrabe}, we see that $\sigma_{nl}$
does shift by a factor $\gamma_{nl}$ as in \eqref{evehew} due to the derivative action from \eqref{brazmukh5}. However now $3\theta_{nl}$ also changes by $+{\mathbb{N}_9\over 2}$ in \eqref{evehew}.
The analysis then changes slightly from what we had in \eqref{kilmill}, and resembles somewhat the relations \eqref{annamie}, with the exception that both $\theta_{nl}$ and $\sigma(\{l_i\}, n_i)$ are being shifted by appropriate amounts determined by $\mathbb{N}_9$. Putting everything together, to
$\left({g_s\over {\rm HH}_o}\right)^{a_1} {\rm M}_p^{+b_1}$ order in expansion and $q$-th order in non-localities, we get the following two relations:

{\footnotesize
\bg\label{kilmill3}
&&n_0 + n_3 + 2\sum_{i = 1}^2 \mathbb{N}_i + \sum_{j = 3}^8 \mathbb{N}_j {\red - \mathbb{N}_9} 
{\red -8q} + b_1 = 0\\
&& 2\mathbb{N}_1 + 5\mathbb{N}_2  + 13 \mathbb{N}_3 + 10 \mathbb{N}_4 + 7 \mathbb{N}_5 
+ \mathbb{N}_6 + 4 \mathbb{N}_7 {\red -2 \mathbb{N}_8} + {3\over 2} \mathbb{N}_9 + 4n_3 {\red -2q}
+ \left(1+ \left[{n\over m}\right] \right)n_0 = 8 + 3a_1, \nonumber \nd}
where the relative minus signs are again shown in {\red red}. The above should be compared to the perturbative case discussed earlier in \eqref{kilmill}. Note that the ${\red -q}$ factor appearing in \eqref{kilmill3} is harmless: higher order non-localities automatically decouple 
because of our choice of \eqref{povedaE} or \eqref{equinox}, so again the other two relative minus signs 
create problems. For a given choice of $a_1$ in the $g_s$ expansion, the second equation in \eqref{kilmill3} allows an infinite number of solutions for $(n_i, \mathbb{N}_j)$. Similarly, for a given value of $+b_1$ in the ${\rm M}_p$ expansion, there are again an infinite number of solutions for $(n_i, \mathbb{N}_j)$. Together we see that the situation, in the {\it absence} of time-dependent fluxes, is worse: not only there is a breakdown of $(g_s, {\rm M}_p)$ hierarchies due to \eqref{taniarma}, but now to any order in 
$\left({g_s\over {\rm HH}_o}\right)^{a_1} {\rm M}_p^{+b_1}$ there are an infinite number of terms. Once we switch on time-dependent G-flux components, the relative minus sign associated with $\mathbb{N}_8$ 
in the second equation of \eqref{kilmill3} goes away (the relative minus sign due to $q$ still remains harmless due to decoupling of higher order non-localities), and therefore there are {\it finite} number of 
terms contributing. Each of these terms will have different ${\rm M}_p$ scalings, so to any order in 
$\left({g_s\over {\rm HH}_o}\right)^{a_1} {\rm M}_p^{+b_1}$ there are only finite number of operators contributing to the energy-momentum tensor. 

The above analysis hopefully demonstrates convincingly that, no matter how we define the scale of our theory, there is no EFT description with four-dimensional de Sitter isometries in the IIB side when G-flux components are time-independent. The result does not depend on our choice of de Sitter slicings, so resorting to the simplest slicing, for example the {\it flat-slicing} we took in \cite{coherbeta, desitter2}, suffices. However one might ask what happens when we take the following choice for 
$\mathbb{F}^{(r)}(y_{r+1} - y_r)$ in \eqref{povedaE}, namely:
\bg\label{hana12}
\mathbb{F}^{(r)}(y_{r+1} - y_r) = {{\cal C}^{(r)}_{\{0\}}~\delta^{8 - d\delta_{qr}}(y_{r+1} - y_r) \over 
{\rm M}_p^{8 - d\delta_{qr}} \sqrt{{\bf g}_{8 - d\delta_{qr}}(y_r, x^0, \bar{\bf x})}}, \nd
with $1 \le r \le q$ and $y_{q+1} \equiv y$.
We define the delta function to satisfy $\int d^8 y' \delta^8(y - y') \equiv 1$, the Kronecker delta 
$\delta_{qr}$ vanishes unless $r = q$ and ${\cal C}^{(r)}_{\{0\}}$ is a constant that appeared earlier. This implies that the delta function has inverse length dimensions. Since $\mathbb{F}^{(r)}(y_{r+1} - y_r)$ is dimensionless, it would approach the fully localized eight-dimensional delta function when $d = 0$ in \eqref{hana12}. This result also follows directly by shrinking the size of the gaussian function in \eqref{equinox} to zero. Additionally we note that:
\bg\label{friepalmers}
{\bf g}_8(y, x^0, {\bf x}) = \left({g_s\over {\rm H}(y){\rm H}_o(y, {\bf x})}\right)^{-4/3}\left({{\rm H(y)}^2\over 
{\rm H}_o(y, {\bf x})}\right)^{16/3} 
{\rm F}_1^2(x^0) {\rm F}_2^4(x^0)~ {\bf g}_8(y), \nd
from where ${\bf g}_8(y, x^0, \bar{\bf x})$ is defined by fixing ${\bf x}$ to some fixed slice
$\bar{\bf x}$. This only changes ${\rm H}_o^{-4}(y, {\bf x})$ to ${\rm H}_o^{-4}(y, \bar{\bf x})$ in \eqref{friepalmers} so helps us to uniquely fix the slice\footnote{The fact that ${\bf x}$ dependence comes from 
only ${\rm H}_o(y, {\bf x})$ helps us determine where the fixed slice may be placed. The point is, if we want the average value to {\it coincide} with the one got from the fixed slice, then it is easy to infer:
$${\int d^2{\bf x} ~{\rm H}_o^{-16/3}(y, {\bf x}) \over \int d^2{\bf x} ~{\rm H}_o^{-4/3}(y, {\bf x})} \equiv
{1\over {\rm H}^4_o(y, \bar{\bf x})}$$
\noindent where we see that the actual spatial metric does not enter, and we can easily determine $\bar{\bf x}$ from above. The above choice is consistent with the way we have defined the volume of the eight-manifold by averaging in \eqref{the12} and thus defines the 
non-locality function $\mathbb{F}^{(r)}(y_{r+1} - y_r)$ in a unique way in the limit \eqref{hana12}. \label{hana00}}. In fact
in the limit where $\mathbb{F}^{(r)}(y_{r+1} - y_r)$ approaches a delta function, it is irrelevant to take $r > 1$, so we can stick with the simplest case with $r = 1$, {\it i.e.} to the first order in non-locality. For such a case, plugging \eqref{hana12}, \eqref{friepalmers} into \eqref{brazmukh}, with $d = 0$, we get:

{\footnotesize
\bg\label{brazmukh6}
{\bf S}^{(1)}_{\rm np} = {\rm M}_p^{11} \int d^3x d^8y \sqrt{-{\bf g}_{11}(x, y)} ~\sum_{k, \{l_i\}, n_i} c_k 
\left[{\rm exp}\left(-k {\cal C}_{\{0\}}^{(1)}\bigg\vert \mathbb{Q}_{\rm T}^{(\{l_i\}, n_i)}(y, x) 
~{{\rm H}^4_o(y, \bar{\bf x})\over {\rm H}^4_o(y, {\bf x})}\bigg\vert\right) - 1\right], \nd}
which, not surprisingly, only contributes {\it perturbatively} to the energy momentum tensor. This is consistent with the fact that product of two series of \eqref{fahingsha3} is already contained in the series because the exponent $(\{l_i\}, n_i)$ simply changes to $(\{l_i + l_j\}, n_i + n_j)$. Thus exponentiation in \eqref{brazmukh6} doesn't change the perturbative nature (thus also removing the $-c_k$ pieces from 
\eqref{brazmukh6}).  On the other hand, keeping $d > 0$ in 
\eqref{hana12} does create a genuine non-perturbative series because\footnote{This may be exemplified by the following. Let $q = 4$ in \eqref{brazmukh2}. We can easily see that for $d > 0$ in \eqref{hana12} makes $\mathbb{F}^{(r)}(y_{r+1} - y_r)$ for $1 \le r \le 3$ 
proportional to $\delta^8(y_{r+1} - y_r)$, {\it i.e.}
identifies the non-local function with a localized eight-dimensional delta function (although we are not required to be on any {\it slice}) whereas it makes $\mathbb{F}^{(4)}(y - y_4)$ proportional to $\delta^{8-d}(y - y_4)$. This removes most of the nested integrals and change 
$\mathbb{Q}^{(\{l_i\}, n_i)}_{\rm T}(y_1, x)$ to $\mathbb{Q}^{(\{l_i\}, n_i)}_{\rm T}(y_4, x)$. At the last stage, 
using $\mathbb{F}^{(4)}(y - y_4) \propto \delta^{8-d}(y - y_4)$, finally converts the nested integral structure of $\mathbb{G}^{(4)}(x, y)$ in \eqref{brazmukh2} to:
\bg\label{ryankipaa}
\mathbb{G}^{(4)}(y, x) \equiv \int^y d^d y_4 \sqrt{{\bf g}_d(y_4, x)}~\mathbb{Q}^{(\{l_i\}, n_i)}_{\rm T}(y_4, x),
\nonumber \nd
with the final integral domain being defined by $y$, where $y \in {\cal M}_4 \times {\cal M}_2 \times 
{\mathbb{T}^2\over {\cal G}}$. There are of course other ways to go from the non-local non-perturbative action \eqref{brazmukh} to the purely non-perturbative one \eqref{brazmukh7}, but we will stick with the simple choice of 
\eqref{hana12} to avoid over-complicating the operation. \label{kibaloryan}}:

{\footnotesize
\bg\label{brazmukh7}
{\bf S}_{\rm np} = {\rm M}_p^{11} \int d^3x d^8y \sqrt{-{\bf g}_{11}(x, y)} ~\sum_{k, \{l_i\}, n_i} c_k 
\left[{\rm exp}\left(-k \bigg\vert \int^y d^d y' \sqrt{{\bf g}_d(y', x)} ~\mathbb{Q}_{\rm T}^{(\{l_i\}, n_i)}(y', x)\bigg\vert\right) - 1\right], \nonumber\\ \nd}
from where one may extract the instanton effects by appropriately choosing $d$. For example, with $d = 6$ we get the M5-brane instantons and with $d = 3$ we get the M2-brane instantons. Note that choosing $d = 8$ in \eqref{hana12} converts $\mathbb{F}^{(q)}(y - y_q) = 1$, thus producing the corresponding {\it de-localized} instanton effects. One may easily check that the convergence and the scaling remain unchanged from what we discussed earlier with non-trivial $\mathbb{F}^{(r)}(y_{r+1} - y_r)$. In the case with time-dependent G-flux components, \eqref{brazmukh7} will continue to provide {\it finite} contributions to the energy-momentum tensors.

\subsection{Quantum terms with toroidal fibre dependences and EFT \label{torus}}

Our analysis above justified the use of time-dependent G-flux components to allow for a four-dimensional EFT description in the IIB side. There were two important requirements. One, the derivative constraint 
\eqref{kootdiye}, and two, the definition of the localized form using ${\cal B}_{lk}^{\rm AB} = 0$ for $k > 0$ in \eqref{samaraw}, although the latter could in principle be relaxed. 
Using these two conditions, metric and G-flux components were allowed to depend on all internal and space-time coordinates of the form $(x^0, x^i , y^m, y^\alpha)$ for any choice of the de Sitter slicings. Question is what happens if we allow the metric and the flux components to also depend on 
$w^a = (x^3, w^{11})$. Clearly the number of Riemann curvature components will increase from 41 independent components (modulo their permutations) to  60 components (again, modulo their permutations). The number of G-flux components will remain 26 as complete anti-symmetrization of their indices put an upper bound on the allowed number of components. The quantum term will then take the following form:
\bg\label{fahingsha4}
\mathbb{Q}_{\rm T}^{(\{l_i\}, n_i)} &= & \left[{\bf g}^{-1}\right] \prod_{i = 0}^4 \left[\partial\right]^{n_i} 
\prod_{{\rm k} = 1}^{60} \left({\bf R}_{\rm A_k B_k C_k D_k}\right)^{l_{\rm k}} \prod_{{\rm r} = 61}^{86} 
\left({\bf G}_{\rm A_r B_r C_r D_r}\right)^{l_{\rm r}}\nonumber\\
& = & {\bf g}^{m_i m'_i}.... {\bf g}^{j_k j'_k} 
\{\partial_m^{n_1}\} \{\partial_\alpha^{n_2}\} \{\partial_a^{n_3}\}\{\partial_i^{n_4}\}\{\partial_0^{n_0}\}
\left({\bf R}_{a0b0}\right)^{l_1} \left({\bf R}_{abab}\right)^{l_2}\left({\bf R}_{pqab}\right)^{l_3}\left({\bf R}_{\alpha a b \beta}\right)^{l_4} \nonumber\\
&\times& \left({\bf R}_{abij}\right)^{l_5}\left({\bf R}_{\alpha\beta\alpha\beta}\right)^{l_6}
\left({\bf R}_{ijij}\right)^{l_7}\left({\bf R}_{ijmn}\right)^{l_8}\left({\bf R}_{\alpha\beta mn}\right)^{l_9}
\left({\bf R}_{i\alpha j \beta}\right)^{l_{10}}\left({\bf R}_{0\alpha 0\beta}\right)^{l_{11}}
\nonumber\\
& \times & \left({\bf R}_{0m0n}\right)^{l_{12}}\left({\bf R}_{0i0j}\right)^{l_{13}}\left({\bf R}_{mnpq}\right)^{l_{14}}\left({\bf R}_{0mnp}\right)^{l_{15}}
\left({\bf R}_{0\alpha\beta m}\right)^{l_{16}}\left({\bf R}_{0abm}\right)^{l_{17}}\left({\bf R}_{0ijm}\right)^{l_{18}}
\nonumber\\
& \times & \left({\bf R}_{mnp\alpha}\right)^{l_{19}}\left({\bf R}_{m\alpha ab}\right)^{l_{20}}
\left({\bf R}_{m\alpha\alpha\beta}\right)^{l_{21}}\left({\bf R}_{m\alpha ij}\right)^{l_{22}}
\left({\bf R}_{0mn \alpha}\right)^{l_{23}}\left({\bf R}_{0m0\alpha}\right)^{l_{24}}
\left({\bf R}_{0\alpha\beta\alpha}\right)^{l_{25}}
\nonumber\\
&\times& \left({\bf R}_{0ab \alpha}\right)^{l_{26}}\left({\bf R}_{0ij\alpha}\right)^{l_{27}}
\left({\bf R}_{mnai}\right)^{l_{28}}\left({\bf R}_{\alpha\beta ai}\right)^{l_{29}}
\left({\bf R}_{a0i0}\right)^{l_{30}}\left({\bf R}_{aijk}\right)^{l_{31}}
\left({\bf R}_{abai}\right)^{l_{32}}\nonumber\\
&\times& \left({\bf R}_{m\beta i \alpha}\right)^{l_{33}}\left({\bf R}_{abmi}\right)^{l_{34}}
\left({\bf R}_{ijk0}\right)^{l_{35}}\left({\bf R}_{\alpha 0i0}\right)^{l_{36}}
\left({\bf R}_{\alpha\beta i 0}\right)^{l_{37}}\left({\bf R}_{ab0i}\right)^{l_{38}}
\left({\bf R}_{\alpha ijk}\right)^{l_{39}}\nonumber\\
&\times& \left({\bf R}_{ab i \alpha}\right)^{l_{40}}\left({\bf R}_{\alpha\beta i \alpha}\right)^{l_{41}}
\left({\bf R}_{mni\alpha}\right)^{l_{42}}\left({\bf R}_{mni0}\right)^{l_{43}}
\left({\bf R}_{mnpi}\right)^{l_{44}}\left({\bf R}_{0m0i}\right)^{l_{45}}
\left({\bf R}_{mijk}\right)^{l_{46}}\nonumber\\
&\times& \left({\bf R}_{maij}\right)^{l_{47}}\left({\bf R}_{ma\alpha\beta}\right)^{l_{48}}
\left({\bf R}_{maba}\right)^{l_{49}}\left({\bf R}_{aij0}\right)^{l_{50}}
\left({\bf R}_{mnpa}\right)^{l_{51}}\left({\bf R}_{a\alpha\beta 0}\right)^{l_{52}}
\left({\bf R}_{a 0\alpha 0}\right)^{l_{53}}\nonumber\\
&\times& \left({\bf R}_{ab a0}\right)^{l_{54}}\left({\bf R}_{mna\alpha}\right)^{l_{55}}
\left({\bf R}_{a\alpha ij}\right)^{l_{56}}\left({\bf R}_{a\alpha\alpha\beta}\right)^{l_{57}}
\left({\bf R}_{ab a\alpha}\right)^{l_{58}}\left({\bf R}_{m0a0}\right)^{l_{59}} 
\left({\bf R}_{mna0}\right)^{l_{60}}\nonumber\\
&\times & \left({\bf G}_{mnpq}\right)^{l_{61}} \left({\bf G}_{mnp\alpha}\right)^{l_{62}}
\left({\bf G}_{mnpa}\right)^{l_{63}}\left({\bf G}_{mn\alpha\beta}\right)^{l_{64}}
\left({\bf G}_{mn\alpha a}\right)^{l_{65}}
\left({\bf G}_{m\alpha\beta a}\right)^{l_{66}}\left({\bf G}_{0ijm}\right)^{l_{67}}\nonumber\\ 
&\times & \left({\bf G}_{0ij\alpha}\right)^{l_{68}}
\left({\bf G}_{mnab}\right)^{l_{69}}\left({\bf G}_{ab\alpha\beta}\right)^{l_{70}}
\left({\bf G}_{m\alpha ab}\right)^{l_{71}} \left({\bf G}_{mnpi}\right)^{l_{72}}
\left({\bf G}_{m\alpha\beta i}\right)^{l_{73}}\left({\bf G}_{mn\alpha i}\right)^{l_{74}} \nonumber\\
&\times &\left({\bf G}_{mnai}\right)^{l_{75}}
\left({\bf G}_{mabi}\right)^{l_{76}}\left({\bf G}_{a\alpha\beta i}\right)^{l_{77}}
\left({\bf G}_{\alpha ab i}\right)^{l_{78}} \left({\bf G}_{ma\alpha i}\right)^{79} 
\left({\bf G}_{mn ij}\right)^{l_{80}}\left({\bf G}_{m\alpha ij}\right)^{l_{81}} \nonumber\\
&\times& \left({\bf G}_{\alpha\beta ij}\right)^{l_{82}}
\left({\bf G}_{maij}\right)^{l_{83}}\left({\bf G}_{\alpha a ij}\right)^{l_{84}}
\left({\bf G}_{ab ij}\right)^{l_{85}} \left({\bf G}_{0ija}\right)^{l_{86}},
\nd
which should be compared to \eqref{fahingsha3}. To proceed, we will take similar ans\"atze for the metric and the G-flux components as in \eqref{makibhalu} and \eqref{tanyarbon} respectively except that 
$\widetilde{g}^{(k)}_{\rm AB}(x, y)$ in \eqref{makibhalu} is to be replaced by $\widetilde{g}^{(k)}_{\rm AB}(x, y, w^a)$ and ${\cal F}^{(k)}_{\rm MN}(x, y)$ in \eqref{tanyarbon} is to be replaced by ${\cal F}^{(k)}_{\rm MN}(x, y, w^a)$. With this, the ${g_s\over {\rm HH}_o}$ scalings of the additional metric components, numbered from 
$l_{42}$ to $l_{60}$ above, are given in {\bf Table \ref{sutariaF}}. Interestingly, the dominant scalings of the 
Riemann curvature components, appearing in {\bf Table \ref{firzasut}}, do not change even if we incorporate the $w^a$ dependences except for the six classes of curvature components ${\bf R}_{ab\mu\nu}, 
{\bf R}_{\rm MNPQ}, {\bf R}_{\mu\nu\rho\sigma}, {\bf R}_{{\rm MN}\mu\nu}, 
{\bf R}_{{\rm MN}ab}$ and ${\bf R}_{abab}$. The change in the dominant scalings of these terms are shown by asterisks in {\bf Table \ref{sutariaF}}. 

What is important now is to see how the quantum terms in \eqref{fahingsha3} scale with respect to $g_s$. Of course demanding $x^3$ dependence goes against our T-duality rules which took us to M-theory in the first place.  Secondly, demanding $w^{11}$ dependence of the quantum terms in \eqref{fahingsha4} somehow implies that the dual IIB theory {\it knows} about the eleven-dimensional dependence. This may not be an issue if we consider the full F-theory framework.  We will however not worry about these subtleties at this stage because, as we shall show below and as has also been anticipated in \cite{coherbeta, desitter2}, the problem lies elsewhere. This becomes clearer once we work out the $g_s$ scalings of the quantum terms of 
\eqref{fahingsha4}, which may now be expressed in the following way: 
\bg\label{botsuga2}
\theta_{nl} &= & {\red -{4\over 3}} \sum_{i = 1}^{14} {\red l_i} + {2\over 3} \sum_{j = 15}^{32} l_j 
+ {5\over 3} \sum_{k = 33}^{46} l_k  {\red -{1\over 3}} \sum_{p = 47}^{60} {\red l_p} +
{1\over 3}\left(n_1 + n_2 + 4n_4\right) {\red -{2n_3\over 3}} + \left({1\over 3} + \left[{n\over m}\right]\right)n_0 
\nonumber\\
&+& \left(l_{mn}^{pq} + {4\over 3}\right)l_{61} + \left(l_{mn}^{p\alpha} + {4\over 3}\right)l_{62} +
\left(l_{mn}^{pa} + {1\over 3}\right)l_{63} + \left(l_{mn}^{\alpha\beta} + {4\over 3}\right)l_{64} 
+ \left(l_{mn}^{\alpha a} + {1\over 3}\right)l_{65}\nonumber\\
&+& \left(l_{ma}^{\alpha\beta} + {1\over 3}\right)l_{66} + \left(l_{ij}^{0m} + {13\over 3}\right)l_{67} 
+ \left(l_{ij}^{0\alpha} + {13\over 3}\right)l_{68} + \left(l_{mn}^{ab} {\red - {2\over 3}}\right){\red l_{69}} 
+ \left(l_{\alpha\beta}^{ab} {\red - {2\over 3}}\right){\red l_{70}} \nonumber\\
&+& \left(l_{m\alpha}^{ab} {\red - {2\over 3}}\right){\red l_{71}} + \left(l_{mn}^{pi} + {7\over 3}\right)l_{72} 
+ \left(l_{m\alpha}^{\beta i} + {7\over 3}\right)l_{73} + \left(l_{mn}^{\alpha i} + {7\over 3}\right)l_{74} 
+ \left(l_{mn}^{ai} + {4\over 3}\right)l_{75} \nonumber\\
&+& \left(l_{ab}^{mi} + {1\over 3}\right)l_{76} + \left(l_{\alpha\beta}^{ai} + {4\over 3}\right)l_{77} 
+ \left(l_{ab}^{\alpha i} + {1\over 3}\right)l_{78} + \left(l_{m\alpha}^{ai} + {4\over 3}\right)l_{79} 
+ \left(l_{mn}^{ij} + {10\over 3}\right)l_{80} \nonumber\\
&+& \left(l_{m\alpha}^{ij} + {10\over 3}\right)l_{81} + \left(l_{\alpha\beta}^{ij} + {10\over 3}\right)l_{82} 
+ \left(l_{ma}^{ij} + {7\over 3}\right)l_{83} + \left(l_{\alpha a}^{ij} + {7\over 3}\right)l_{84} 
+ \left(l_{ab}^{ij} + {4\over 3}\right)l_{85}, \nonumber\\
& + & \left(l_{0a}^{ij} + {4\over 3}\right)l_{86}, \nd
where we now see that there are too many relative {\red minus} signs compared to what we had in 
\eqref{botsuga}. This is alarming because, while the minus signs associated with $(l_{69}, l_{70}, l_{71})$ can be flipped by switching on $l_{\rm MN}^{ab} > {2\over 3}$, the other minus signs associated with the curvature tensors as well as the derivatives along the $w^a$ directions cannot be easily flipped to positive. Thus at least at face value it appears that for both time-independent as well as time-dependent G-flux components, the relative minus signs in $\theta_{nl}$ will {\it not} allow a finite number of operator contributions 
to the energy-momentum tensor at any order in $\left({g_s\over {\rm HH}_o}\right)^{a} 
\times {1\over {\rm M}^{\pm b}_p}$. As a result there appears to be no four-dimensional EFT description possible in dual IIB side if we allow dependence on the toroidal directions.  

The above conclusion is based on our choice of M-theory metric as in say \eqref{evader} from where we can compute the $g_s$ and ${\rm M}_p$ scalings of \eqref{fahingsha4}. Allowing localized G-flux components 
${\bf G}_{{\rm MN} ab}$ we have seen how relative minus signs appear in the ${\rm M}_p$ scalings. However we also know that there could be constraints coming from the underlying Schwinger-Dyson's equations. Such constraints typically change the metric and the flux components to \eqref{makibhalu} and 
\eqref{tanyarbon} respectively (with appropriate $w^a$ dependences as discussed above). Could we change the ans\"atze of the metric and the G-flux components further so that  we can flip some (or all) of the minus signs in \eqref{botsuga2} for the time-dependent case?

\begin{table}[tb]  
 \begin{center}
\renewcommand{\arraystretch}{1.5}
\begin{tabular}{|c||c||c||c|}\hline Riemann tensors  & Components & 
$g_s$ expansions \\ \hline\hline
${\bf R}_{{\rm MNP}a}$ &  ${\bf R}_{mnpa}, {\bf R}_{mna\alpha}, {\bf R}_{m\alpha\beta a}, {\bf R}_{a\alpha\alpha\beta}$ &  ${}^{\Sigma}_{{}_{k \ge 0}} {\rm R}^{(k)}_{{\rm MNP}a} 
\left({g_s\over {\rm H}{\rm H}_o}\right)^{2(k-1)/3}$ \\ \hline 
${\bf R}_{{\rm MN}a0}$ &  ${\bf R}_{mna0}, {\bf R}_{\alpha\beta a0}$ &  
${}^{\Sigma}_{{}_{k \ge 0}} {\rm R}^{(k)}_{{\rm MN}a0} 
\left({g_s\over {\rm H}{\rm H}_o}\right)^{(2k - 5)/3}$ \\ \hline 
${\bf R}_{{\rm MN}ai}$ &  ${\bf R}_{mnai}, {\bf R}_{\alpha\beta ai}$ & 
${}^{\Sigma}_{{}_{k \ge 0}} {\rm R}^{(k)}_{{\rm MN}ai} 
\left({g_s\over {\rm H}{\rm H}_o}\right)^{2(k - 1)/3}$ \\ \hline 
${\bf R}_{{\rm M}a\mu\nu}$ &  ${\bf R}_{m0a0}, {\bf R}_{maij}, {\bf R}_{\alpha 0a 0}, 
{\bf R}_{\alpha aij}$ & ${}^{\Sigma}_{{}_{k \ge 0}} {\rm R}^{(k)}_{{\rm M}a\mu\nu} 
\left({g_s\over {\rm H}{\rm H}_o}\right)^{2(k - 4)/3}$ \\ \hline 
${\bf R}_{{\rm M}abc}$ &  ${\bf R}_{mabc}, {\bf R}_{\alpha abc}$ & 
${}^{\Sigma}_{{}_{k \ge 0}} {\rm R}^{(k)}_{{\rm M}abc} 
\left({g_s\over {\rm H}{\rm H}_o}\right)^{2(k + 2)/3}$ \\ \hline 
${\bf R}_{a\mu\nu\rho}$ &  ${\bf R}_{a0i0}, {\bf R}_{aijk}$
 &  ${}^{\Sigma}_{{}_{k \ge 0}} {\rm R}^{(k)}_{a\mu\nu\rho} 
\left({g_s\over {\rm H}{\rm H}_o}\right)^{2(k - 4)/3}$ \\ \hline 
${\bf R}_{a\mu\nu 0}$ &  ${\bf R}_{aij0}$
 &  ${}^{\Sigma}_{{}_{k \ge 0}} {\rm R}^{(k)}_{a\mu\nu 0} 
\left({g_s\over {\rm H}{\rm H}_o}\right)^{(2k - 11)/3}$ \\ \hline 
${\bf R}_{abc0}$ &  ${\bf R}_{aba0}$
 & ${}^{\Sigma}_{{}_{k \ge 0}} {\rm R}^{(k)}_{abc0} 
\left({g_s\over {\rm H}{\rm H}_o}\right)^{(2k + 1)/3}$ \\ \hline  
${\bf R}_{abci}$ &  ${\bf R}_{abai}$
 &  ${}^{\Sigma}_{{}_{k \ge 0}} {\rm R}^{(k)}_{abci} 
\left({g_s\over {\rm H}{\rm H}_o}\right)^{2(k + 2)/3}$ \\ \hline 
${}^\ast{\bf R}_{{\rm MN}\mu\nu}$ &  ${\bf R}_{m0n0}, {\bf R}_{\alpha 0 \beta 0}, {\bf R}_{\alpha\beta ij}, 
{\bf R}_{mnij}$
 & ${}^{\Sigma}_{{}_{k \ge 0}} {\rm R}^{(k)}_{{\rm MN}\mu\nu} 
\left({g_s\over {\rm H}}\right)^{2(k - 7)/3}$ \\ \hline   
${}^\ast{\bf R}_{\mu\nu\rho\sigma}$ &  ${\bf R}_{i0j0}, {\bf R}_{ijkl}$
 & ${}^{\Sigma}_{{}_{k \ge 0}} {\rm R}^{(k)}_{\mu\nu\rho\sigma} 
\left({g_s\over {\rm H}}\right)^{2(k - 10)/3}$ \\ \hline   
${}^\ast{\bf R}_{ab\mu\nu}$ &  ${\bf R}_{a0b0}, {\bf R}_{abij}$
 & ${}^{\Sigma}_{{}_{k \ge 0}} {\rm R}^{(k)}_{ab\mu\nu} 
\left({g_s\over {\rm H}}\right)^{2(k - 4)/3}$ \\ \hline   
${}^\ast{\bf R}_{\rm MNPQ}$ &  ${\bf R}_{mnpq}, {\bf R}_{mn\alpha\beta}, {\bf R}_{\alpha\beta\alpha\beta}$
 & ${}^{\Sigma}_{{}_{k \ge 0}} {\rm R}^{(k)}_{\rm MNPQ} 
\left({g_s\over {\rm H}}\right)^{2(k - 4)/3}$ \\ \hline   
${}^\ast{\bf R}_{{\rm MN}ab}$ &  ${\bf R}_{mnab}, {\bf R}_{\alpha\beta ab}$
 & ${}^{\Sigma}_{{}_{k \ge 0}} {\rm R}^{(k)}_{{\rm MN} ab} 
\left({g_s\over {\rm H}{\rm H}_o}\right)^{2(k - 1)/3}$ \\ \hline   
${}^\ast{\bf R}_{abcd}$ &  ${\bf R}_{abab}$
 &  ${}^{\Sigma}_{{}_{k \ge 0}} {\rm R}^{(k)}_{abcd} 
\left({g_s\over {\rm H}{\rm H}_o}\right)^{2(k +2)/3}$ \\ \hline 
 \end{tabular}
\renewcommand{\arraystretch}{1}
\end{center}
 \caption[]{The ${g_s\over {\rm HH}_o}$ expansions for the {\it additional} Riemann curvature components 
 arising from the $w^a \equiv (x^3, w^{11})$  dependences of the metric components.
 Here as before, 
  $(m, n) \in {\cal M}_4$,  
 $(\alpha, \beta) \in {\cal M}_2$, $(a, b) \in {\mathbb{T}^2\over {\cal G}}$ and 
 $(\mu, \nu) \in \mathbb{R}^{2, 1}$. The curvature tensors ${\rm R}^{(k)}_{\rm N_1N_2 N_3 N_4}  = 
 {\rm R}^{(k)}_{\rm N_1N_2 N_3 N_4}(x, y, w^a)$ populate the quantum terms 
 \eqref{fahingsha4}. Note that there are six set of curvature components, given by $\ast$ here, that scale differently from the ones in {\bf Table \ref{firzasut}} because of their $w^a$ dependences.}
  \label{sutariaF}
 \end{table}

Clearly the metric and the flux choices in \eqref{makibhalu} and \eqref{tanyarbon}, with the addition of the $w^a$ dependence, respectively {\it cannot} help because the $w^a$ derivatives in the definitions of the Riemann curvatures, or the external derivatives in the quantum terms \eqref{fahingsha5}, cannot introduce
extra powers of $g_s$ to change the scalings in \eqref{botsuga2}. In fact the $y^{\rm M}$ dependent exponential factors in either \eqref{tanyarbon} or \eqref{makibhalu} cannot change the dominant scaling, even if we replace them by $w^a$ dependences as long as we have $w^a$ dependences in 
$\widetilde{g}^{(k)}_{\rm AB}(x, y, w^a)$ and ${\cal F}^{(k)}_{\rm MN}(x, y, w^a)$.  What about a different choice of the metric components by allowing dependence on the $w^a \equiv (x^3, w^{11})$ coordinates on the exponential factors only? For example, a choice of the form:

{\footnotesize
\bg\label{makibhalu2}
{\bf g}_{\rm CD}(x, y, w^a; g_s, {\rm M}_p) \equiv \left({g_s\over {\rm HH}_o}\right)^{a_{\rm CD}} 
\sum_{k = 0}^\infty a^{(cd)}_k \widetilde{g}^{(k)}_{\rm CD}(x, y)~{\rm exp}\left[-b^{(cd)}_k\left({g_s\over {\rm HH}_o}\right)^{\gamma k/3}
(w^a)^{2k} {\rm M}^{2k}_p\right], \nd}
where $({\rm C, D}) \in {\bf R}^{2, 1} \times {\cal M}_4 \times {\cal M}_2 \times {\mathbb{T}^2\over {\cal G}}$, with constants $(a^{(cd)}_k, b^{(cd)}_k)$, and $a_{\rm CD}$ as in \eqref{sillyP}, might be much better suited. This is almost like what we had in \eqref{makibhalu} except that the $w^a$ dependence comes solely from the exponential piece with the sub-dominant $e^{-y^2{\rm M}_p^2}$ as well as ${\rm F}_i(g_s/{\rm HH}_o)$ absent, although their presence will not change anything (see footnote \ref{margincall}). Note that the $g_s$ power in the exponential piece is written as $\gamma$, and since curvatures involve two derivatives, we expect $\gamma > 2$. There are however few issues with the metric choice 
\eqref{makibhalu2}. {\Su First}, $x^3$ dependence of the metric components in \eqref{makibhalu2} clashes with the T-duality rules that we used to lift the IIB configuration 
to M-theory. {\Su Secondly}, if $({\rm C, D}) \in {\bf R}^{2, 1} \times {\bf S}^1_{(3)}$, where  ${\bf S}^1_{(3)}$
is parametrized locally by $x^3$, the metric components do not dualize to an exact de Sitter space-time in the IIB side unless $x^{11} = 0$. In fact in the limit $g_s$ and $x^{11}$ go to zero  as $(g_s, x^{11}) = (\epsilon, \epsilon)$, and ${\rm M}_p$ goes to infinity as 
${\rm M}_p = \epsilon^{-(2 + \gamma/3)}$, the spatial metric deviates significantly away from de Sitter space-time in the dual IIB side, although in the limit $x^{11}$ goes to zero as $x^{11} = \epsilon^{1 + z}$ with a constant $z$, the deviation from an exact de Sitter space-time is minimal. Interestingly in this limit there {\it can} be curvature contributions that can offset the ${\red - {4\over 3}}$ scalings in \eqref{botsuga2} if we take $\gamma > 4$ in \eqref{makibhalu2}. However such terms are suppressed by powers of $x^{11}$. As an example, let us consider the curvature term ${\bf R}_{\rm MNPQ}$ defined in the following 
way\footnote{Recall that $\langle {\rm R}_{\rm MNPQ}\rangle_{\bar\sigma} = {\bf R}_{\rm MNPQ}(y, x, w^a, g_s)$ as in \eqref{maric}.}:

{\footnotesize
\bg\label{maric3}
{\bf R}_{\rm MNPQ}(y, x, w^a, g_s) &=&
{\bf R}^{(1)}_{\rm MNPQ}(y, x, w^a) \left({g_s\over {\rm H}{\rm H}_o}\right)^{-2/3} 
+ c~{\partial {\bf g}_{(mn} \over \partial w^a} {\partial {\bf g}_{pq)} \over \partial w^b}{\bf g}^{ab} \\ 
&+& 
{\bf R}^{(2)}_{\rm MNPQ}(y, x, w^a) \left({g_s\over {\rm H}{\rm H}_o}\right)^{+4/3} + 
{\bf R}^{(3)}_{\rm MNPQ}(y, x^i) \left({g_s\over {\rm H}{\rm H}_o}\right)^{-2/3} \left[{\partial\over \partial t}
 \left({g_s\over {\rm H}{\rm H}_o}\right)\right]^2, \nonumber \nd}
where $(.., ..)$ denotes all possible symmetric or anti-symmetric permutations of the indices, $c$ is a constant, and in the second line we allow the condition \eqref{kootdiye} (see footnote \ref{camil}). The first and the third term appears from derivatives along ${\cal M}_4 \times {\cal M}_2$ and ${\bf R}^2$ respectively. On the other hand, the second term, in the usual case, scales as $\left({g_s\over {\rm HH}_o}\right)^{-8/3}$, which produces the ${\red -{4l_{14}\over 3}}$ factor in \eqref{botsuga2}. Question is, whether there is a simple way to change this scaling.

The form of the second term in \eqref{maric3} gives us a hint: if the $w^a$ derivative could bring down extra powers of $g_s$ then clearly the negative scalings from the Riemann curvature terms in \eqref{botsuga2} could be changed. However, as discussed above, we cannot allow $x^3$ derivatives if we want to preserve the underlying T-duality map. Our discussion above seems to allow $\gamma \ge 4$ in \eqref{makibhalu2}. 
Is there a lower bound for $\gamma$? In the following let us see whether we can find one. 

Since $k \ge 0$ for the metric components in \eqref{makibhalu2}, the dominant scalings do not change and at least in the limit discussed above we recover the de Sitter space-time in the dual IIB side. 
Note also that the $x^{11}$ derivatives will directly act on the exponential pieces for both the flux and the metric components. Such actions will contribute $+{ 4kl_i\over 3}$ to at least all the unstarred Riemann curvatures from {\bf Table \ref{sutariaF}} and at least $+{4kn_3\over 3}$ to the $n_3$ derivatives when we take $\gamma \ge  4$ in \eqref{makibhalu2}. These could in principle 
change the negative scalings in \eqref{botsuga2}. 

Interestingly with the choice of $\gamma \ge 4$ in \eqref{makibhalu2}, the second term in \eqref{maric3}
scales as $\left({g_s\over {\rm HH}_o}\right)^{0+}$ instead of $\left({g_s\over {\rm HH}_o}\right)^{-8/3}$. This means the {\it dominant} scalings of all components of ${\bf R}_{\rm MNPQ}$ now become 
$\left({g_s\over {\rm HH}_o}\right)^{-2/3}$, thus contributing $+{2\over 3}(l_6, l_9, l_{14})$ to 
\eqref{botsuga2}.  On the other hand if $\gamma \ge 2$ in \eqref{makibhalu2} the scaling would have been quite different: the second term in \eqref{maric3} would have scaled as $\left({g_s\over {\rm HH}_o}\right)^{-4/3}$, which would have become the dominant scalings for all the components associated with ${\bf R}_{\rm MNPQ}$, thus contributing ${\red 0(l_6, l_9, l_{14})}$ to \eqref{botsuga2}. The zero scalings of these components mean that they are time-neutral, and therefore any powers of these components do not contribute to the $g_s$ scalings of $\theta_{nl}$ in \eqref{botsuga2}. Existence of time-neutral series is not good as was shown in \cite{evan}, and therefore $\gamma = 2$ does not appear to be a viable possibility here. What about $\gamma = 3$? Can this be realized in \eqref{makibhalu2}? To answer this, let us consider another component of the Riemann tensor of the form:

{\footnotesize
\bg\label{maric4}
{\bf R}_{abab}(y, x, w^a, g_s) &=&
{\bf R}^{(1)}_{abab}(y, x, w^a) \left({g_s\over {\rm H}{\rm H}_o}\right)^{+10/3} 
+ {\bf R}^{(2)}_{abab}(y, x^i) \left({g_s\over {\rm H}{\rm H}_o}\right)^{+10/3} \left[{\partial\over \partial t}
 \left({g_s\over {\rm H}{\rm H}_o}\right)\right]^2 \\ 
&+& 
{\bf R}^{(3)}_{abab}(y, x, w^a) \left({g_s\over {\rm H}{\rm H}_o}\right)^{+16/3} 
+ c_1~{\partial {\bf g}_{(ab} \over \partial w^c} {\partial {\bf g}_{ab)} \over \partial w^d}{\bf g}^{cd} 
+ c_2\left({\partial^2 {\bf g}_{aa}\over \partial w^b \partial w^b} - {\partial^2 {\bf g}_{bb}\over \partial w^a \partial w^a}\right), \nonumber\nd}
where the first term appears from derivatives along ${\cal M}_4 \times {\cal M}_2$, the second term appears from temporal derivatives and applying the condition \eqref{kootdiye}, the third term appears from derivatives along the spatial directions ${\bf R}^2$, and the remaining two terms with constant coefficients $(c_1, c_2)$ appear from the derivatives along the toroidal directions ${\mathbb{T}^2\over {\cal G}}$ (we will not worry about the T-duality constraint here). Note, in the absence of the dependence on $w^a$, the dominant scaling is $\left({g_s\over {\rm HH}_o}\right)^{10/3}$, assuming 
\eqref{kootdiye}. Once $w^a$ dependence is switched on, with a choice of the metric given by 
\eqref{makibhalu} with $\widetilde{g}^{(k)}_{\rm AB}$ function of $(x, y, w^a)$, the dominant scaling changes to 
$\left({g_s\over {\rm HH}_o}\right)^{4/3}$, thus contributing ${\red -{4\over 3}}$ to \eqref{botsuga2}. On the other hand, if we use the metric ans\"atze \eqref{makibhalu2}, the dominant $g_s$ scaling of the last two terms in \eqref{maric4} become $\left({g_s\over {\rm HH}_o}\right)^{2(2 + \gamma k)/3}$ and 
$\left({g_s\over {\rm HH}_o}\right)^{(4 + \gamma k)/3}$ respectively. Their respective contributions to 
\eqref{botsuga2} would be ${2\over 3}(\gamma k - 2)$ or ${1\over 3}(\gamma k - 4)$ depending on which of them is the dominating one. Taking $k = 1$
for the first contributing term in the series, and $\gamma = 3$, we see that the two values are $+{2\over 3}$ and ${\red -{2\over 3}}$ respectively, implying that the contribution to \eqref{botsuga2} will be 
${\red -{2l_2\over 3}}$. Unfortunately $\gamma = 4$, which worked well for the other curvature components, now contributes as ${\red 0l_2}$ to \eqref{botsuga2} leading to time-neutral series. This clearly shows that both $\gamma = 3$ and 
$\gamma = 4$ are unacceptable
and therefore our only choice in \eqref{makibhalu2} appears to be $\gamma \ge 5$ (see also the detailed scalings of the curvature tensors in {\bf Table \ref{fchumbon}}).
Taking all these into account, then allows us to consider the following ans\"atze for the 
{\it dominant} scalings of the metric and the G-flux components (see also footnote \ref{margincall}):

{\footnotesize
\bg\label{makibhalu3}
&&{\bf g}_{\rm CD}(x, y, x^{11}; g_s, {\rm M}_p) \equiv \left({g_s\over {\rm HH}_o}\right)^{a_{\rm CD}} 
\sum_{k = 0}^\infty a^{(cd)}_k \widetilde{g}^{(k)}_{\rm CD}(x, y)~{\rm exp}\left[-b^{(cd)}_k\left({g_s\over {\rm HH}_o}\right)^{\gamma k/3}
x^{2k}_{11} {\rm M}^{2k}_p\right]\\
&&{\bf G}_{{\rm ABCD}}(x, y, x^{11}; g_s, {\rm M}_p) \equiv \left({g_s\over {\rm HH}_o}\right)^{l_{\rm AB}^{\rm CD}} \sum_{k = 1}^\infty c_k~
{\cal G}^{(k)}_{\rm ABCD}(x, y) ~{\rm exp}\left[-d_k\left({g_s\over {\rm HH}_o}\right)^{\gamma k/3} x^{2k}_{11} {\rm M}^{2k}_p\right], \nonumber \nd}
with $a_{\rm CD}$ and $l_{\rm AB}^{\rm CD}$ as in \eqref{sillyP} and \eqref{theritual} respectively; and we have chosen a conservative lower bound of ${\Su \gamma = 5}$ although generically we should allow $\gamma \ge 5$ (in fact later we will argue that it's $\gamma = 6$ that makes more sense here. However to arrive at this conclusion requires a few steps of reasoning which will be elaborated soon). With the choice of $\gamma \ge 5$ all the Riemann curvature terms and the derivatives along $x^{11}$ (parametrized by $n_3$ in \eqref{fahingsha4}) contribute positive scalings to \eqref{botsuga2}. With time-dependent G-flux components, {\it i.e.} the flux choice in \eqref{makibhalu3}, there appears a possibility of a four-dimensional an EFT description in the dual IIB side.

\begin{table}[tb]  
 \begin{center}
\renewcommand{\arraystretch}{1.5}
\begin{tabular}{|c||c||c|}\hline Riemann tensors  & Dominant $g_s$ scaling & 
Contributions to \eqref{botsuga3} \\ \hline\hline
${\bf R}_{mnpa}, {\bf R}_{mna\alpha}, {\bf R}_{m\alpha\beta a}, {\bf R}_{a\alpha\alpha\beta}$ &  
${\gamma k\over 3} - {2\over 3}$ & ${\gamma k\over 3} - {1 \over 3}$ \\ \hline 
 ${\bf R}_{mna0}, {\bf R}_{\alpha\beta a0}$ &  
 ${\gamma k\over 3} - {5\over 3}$ & ${\gamma k\over 3} - {1 \over 3}$  
 \\ \hline 
${\bf R}_{mnai}, {\bf R}_{\alpha\beta ai}$ & ${\gamma k\over 3} - {2\over 3}$ & ${\gamma k\over 3} + {2 \over 3}$\\ \hline 
${\bf R}_{m0a0}, {\bf R}_{maij}, {\bf R}_{\alpha 0a 0}, 
{\bf R}_{\alpha aij}$ & ${\gamma k\over 3} - {8\over 3}$ & ${\gamma k\over 3} - {1 \over 3}$
\\ \hline 
${\bf R}_{mabc}, {\bf R}_{\alpha abc}$ & ${\gamma k\over 3} + {4 \over 3}$ & ${\gamma k\over 3} - {1 \over 3}$
 \\ \hline 
${\bf R}_{a0i0}, {\bf R}_{aijk}$
 &  ${\gamma k\over 3} - {8 \over 3}$ & ${\gamma k\over 3} + {2 \over 3}$\\ \hline
${\bf R}_{aij0}$
 &  ${\gamma k\over 3} - {11 \over 3}$ & ${\gamma k\over 3} - {1 \over 3}$\\ \hline 
${\bf R}_{aba0}$
 & ${\gamma k\over 3} + {1 \over 3}$ & ${\gamma k\over 3} - {1 \over 3}$\\ \hline 
 ${\bf R}_{abai}$
 & ${\gamma k\over 3} + {4\over 3}$ & ${\gamma k\over 3} + {2 \over 3}$\\ \hline 
 ${\bf R}_{m0n0}, {\bf R}_{\alpha 0 \beta 0}, {\bf R}_{\alpha\beta ij}, 
{\bf R}_{mnij}$
 & ${\rm dom}\left({2\gamma k\over 3} - {14\over 3}, -{8\over 3}\right)$ & 
 ${\rm dom}\left({2\gamma k\over 3} - {4\over 3}, {2 \over 3}\right)$ \\ \hline
${\bf R}_{i0j0}, {\bf R}_{ijkl}$
 & ${\rm dom}\left({2\gamma k\over 3} - {20 \over 3}, -{14 \over 3}\right)$ & 
 ${\rm dom}\left({2\gamma k\over 3} - {4\over 3}, {2 \over 3}\right)$ \\ \hline
 ${\bf R}_{a0b0}, {\bf R}_{abij}$
 & ${\rm dom}\left({\gamma k\over 3} - {8 \over 3}, -{2 \over 3}\right)$ & 
 ${\rm dom}\left({\gamma k\over 3} - {4\over 3}, {2 \over 3}\right)$ \\ \hline 
 ${\bf R}_{mnpq}, {\bf R}_{mn\alpha\beta}, {\bf R}_{\alpha\beta\alpha\beta}$
 & ${\rm dom}\left({2\gamma k\over 3} - {8 \over 3}, -{2 \over 3}\right)$ & 
 ${\rm dom}\left({2\gamma k\over 3} - {4\over 3}, {2 \over 3}\right)$ \\ \hline 
${\bf R}_{mnab}, {\bf R}_{\alpha\beta ab}$
 & ${\rm dom}\left({\gamma k\over 3} - {2 \over 3}, {4 \over 3}\right)$ & 
 ${\rm dom}\left({\gamma k\over 3} - {4\over 3}, {2 \over 3}\right)$ \\ \hline 
 ${\bf R}_{abab}$
 & ${\rm dom}\left({\gamma k\over 3} + {4\over 3}, {10 \over 3}\right)$ & 
 ${\rm dom}\left({\gamma k\over 3} - {4\over 3}, {2 \over 3}\right)$ \\ \hline 
 \end{tabular}
\renewcommand{\arraystretch}{1}
\end{center}
 \caption[]{The ${g_s\over {\rm HH}_o}$ expansions for the {additional} Riemann curvature components 
 arising from the $x^{11}$  dependences of the metric components and using the metric ans\"atze as in 
 \eqref{makibhalu3} with $k \ge 1$. The third column denotes their contributions to \eqref{botsuga3}, where
 ${\rm dom}(a, b)$ choses the dominant one between $\left({g_s\over {\rm HH}_o}\right)^a$ and 
 $\left({g_s\over {\rm HH}_o}\right)^b$. If $a > 0, b> 0$, then the dominant one is the smaller of the two, whereas if $a < 0, b < 0$ ({\it i.e.} when both are negative), the dominant one is the larger of the two. 
Note that the scalings in {\bf Table \ref{firzasut}} remain unchanged even with the metric choice \eqref{makibhalu3}.}
  \label{fchumbon}
 \end{table}

It is interesting that while most terms in {\bf Table \ref{fchumbon}} works with $\gamma \ge 3$, the component \eqref{maric4} and two other set of curvature tensors scale as ${1\over 3}(\gamma k - 4)$ instead of ${2 \over 3}(\gamma k - 2)$. This is because of the derivative action which, in turn, also effects the scaling of the $n_3$ term in \eqref{botsuga2}. Taking all these into account, we now see that the $g_s$ scaling of the quantum term in \eqref{fahingsha4} with the background ans\"atze \eqref{makibhalu3}, changes from \eqref{botsuga2} to the following:
\bg\label{botsuga3}
\theta_{nl} &= & \left({\gamma k \over 3} {\red -{4\over 3}}\right) \sum_{r = 1 }^{5} {\red l_r} +{2\over 3} \sum_{i = 6}^{27} l_i  + \left({\gamma k \over 3} + {2\over 3}\right) \sum_{j = 28}^{32} l_j  
+ {5\over 3} \sum_{k = 33}^{46} l_k  + \left({\gamma k \over 3} {\red -{1\over 3}}\right) 
\sum_{p = 47}^{60} {\red l_p} 
\nonumber\\
&+& \left(l_{mn}^{pq} + {4\over 3}\right)l_{61} + \left(l_{mn}^{p\alpha} + {4\over 3}\right)l_{62} +
\left(l_{mn}^{pa} + {1\over 3}\right)l_{63} + \left(l_{mn}^{\alpha\beta} + {4\over 3}\right)l_{64} 
+ \left(l_{mn}^{\alpha a} + {1\over 3}\right)l_{65}\nonumber\\
&+& \left(l_{ma}^{\alpha\beta} + {1\over 3}\right)l_{66} + \left(l_{ij}^{0m} + {13\over 3}\right)l_{67} 
+ \left(l_{ij}^{0\alpha} + {13\over 3}\right)l_{68} + \left(l_{mn}^{ab} {\red - {2\over 3}}\right){\red l_{69}} 
+ \left(l_{\alpha\beta}^{ab} {\red - {2\over 3}}\right){\red l_{70}} \nonumber\\
&+& \left(l_{m\alpha}^{ab} {\red - {2\over 3}}\right){\red l_{71}} + \left(l_{mn}^{pi} + {7\over 3}\right)l_{72} 
+ \left(l_{m\alpha}^{\beta i} + {7\over 3}\right)l_{73} + \left(l_{mn}^{\alpha i} + {7\over 3}\right)l_{74} 
+ \left(l_{mn}^{ai} + {4\over 3}\right)l_{75} \nonumber\\
&+& \left(l_{ab}^{mi} + {1\over 3}\right)l_{76} + \left(l_{\alpha\beta}^{ai} + {4\over 3}\right)l_{77} 
+ \left(l_{ab}^{\alpha i} + {1\over 3}\right)l_{78} + \left(l_{m\alpha}^{ai} + {4\over 3}\right)l_{79} 
+ \left(l_{mn}^{ij} + {10\over 3}\right)l_{80} \nonumber\\
&+& \left(l_{m\alpha}^{ij} + {10\over 3}\right)l_{81} + \left(l_{\alpha\beta}^{ij} + {10\over 3}\right)l_{82} 
+ \left(l_{ma}^{ij} + {7\over 3}\right)l_{83} + \left(l_{\alpha a}^{ij} + {7\over 3}\right)l_{84} 
+ \left(l_{ab}^{ij} + {4\over 3}\right)l_{85}, \nonumber\\
& + & \left(l_{0a}^{ij} + {4\over 3}\right)l_{86} + \left({\gamma k\over 6} {\red -{2\over 3}}\right) {\red n_3} + {1\over 3}\left(n_1 + n_2 + 4n_4\right) + \left({1\over 3} + \left[{n\over m}\right]\right)n_0, \nd
where we note that the 60 curvature tensors have varied set of scalings. With $\gamma = 4$, the vanishing of both the coefficients of $n_3$ and $l_r$ in \eqref{botsuga3} implies that the time-neutral operators could also be made ${\rm M}_p$ neutral\footnote{As an example, we can use $n_3$ derivative actions, and the corresponding collection of operators from \eqref{fahingsha4}, to write the following operator:
\bg\label{dormezvous}
{\cal O} \equiv \lim_{x^{11} \to 0} \sum_{\{l_i\}, n_3} {{\cal C}_{n_3l_1...l_5} \over {\rm M}_p^{2\sigma(\{l_i\}, n_3)}}~
 \{\partial^{2n_3}_{11}\} \left({\bf R}_{a0b0}\right)^{l_1} \left({\bf R}_{abab}\right)^{l_2}\left({\bf R}_{mnab}\right)^{l_3}\left({\bf R}_{\alpha a b \beta}\right)^{l_4} \left({\bf R}_{abij}\right)^{l_5}{\bf g}^{a_i b_i}....
 {\bf g}^{m_jn_j}, \nonumber \nd
where ${\cal C}_{n_3l_1...l_5}$ are constant coefficients, the inverse metic components are used for complete contractions,  and $\sigma = n_3 + \sum_{i = 1}^5 l_i$ with 
$(n_3, l_i) \in (\mathbb{Z}, \mathbb{Z})$. We may note two things: one, that the above operator scales as 
$\left({g_s\over {\rm HH}_o}\right)^0 \times {1\over {\rm M}_p^0}$. It is easy to see why it is time-neutral from
\eqref{botsuga3}. The ${\rm M}_p$ neutrality is seen from the derivatives along $x^{11}$ appearing from both the external derivative action as well as the derivatives inside the definition of the curvature tensors. Two, even in the limit of vanishing $x^{11}$ the operator remains non-zero.}.
Clearly this is unacceptable for a healthy EFT. For $\gamma > 4$ there is in-fact a subset of curvature tensors, from the set that scales as ${2l_i\over 3}$, which would in principle scale as 
${\rm dom}\left({2\gamma k\over 3} - {4\over 3}, {2 \over 3}\right)$ as evident from {\bf Table \ref{fchumbon}}. 
For $\gamma \ge 5$, the dominant scalings of these tensors are all ${2\over 3}$, which is what appears in 
\eqref{botsuga3}. Therefore it seems that even with toroidal fibre dependence, if we allow the background ans\"atze of the form \eqref{makibhalu3}, an EFT description may be allowed in the dual IIB side. Question however is how robust is this conclusion that relies crucially on the background choice \eqref{makibhalu3}. 
Analysis of this would require us to explore the flux EOMs as well as conditions for global anomaly cancellations. This is the subject to which we turn next.

\section{Flux equations of motion, anomaly cancellations and healthy EFT \label{sec2.2.5}}

For generic choice of the metric dependences on the toroidal direction ${\mathbb{T}^2\over {\cal G}}$, as we saw above, the $g_s$ scaling of the quantum terms \eqref{fahingsha4} is expressed as \eqref{botsuga2}.  This clearly has neither $g_s$ nor ${\rm M}_p$ hierarchies even in the presence of time-dependent G-flux components
(in fact it is not even necessary to express the G-flux components ${\bf G}_{{\rm MN}ab}$ as localized fluxes to see this\footnote{A question can be raised at this point on the usefulness of the ans\"atze \eqref{makibhalu3} for the G-flux components ${\bf G}_{{\rm MN}ab}$. Recall that the motivation for taking a localized form for the G-flux components ${\bf G}_{{\rm MN}ab}$ was to avoid problems with Buscher's duality as well as with the existence of an EFT in four-dimensions in the IIB side (see discussions after 
\eqref{harpno}). However now that we do know EFT can exist with $w^a$ dependence, can we allow non-localized three-form field components ${\bf C}_{{\rm MN}3}(x, y, x^{11})$? The answer is still {\it no} because such a three-form would imply a generic form of the metric components ${\bf g}_{AB}(x, y, x^{11})$ with no exponential suppression (in other words, the $w^a$ is not necessarily in any exponential factor as in \eqref{makibhalu3}). Scalings of the curvature tensors would then imply the quantum terms \eqref{fahingsha4} to scale as \eqref{botsuga2} and {\it not} as \eqref{botsuga3}. This would imply non-existence of an EFT description in the dual IIB side. Therefore it appears that the only way an EFT would exist in the dual side if the G-flux components ${\bf G}_{{\rm MN}ab}$ are viewed as {\it localized} fluxes.}). However for a special choice of the background, as given as \eqref{makibhalu3}, the $g_s$ scaling goes as \eqref{botsuga3} which, for $\gamma \ge 5$, allows an EFT description in the dual IIB side.
Unfortunately, most of the curvature tensors, that contribute for the metric choice of \eqref{makibhalu3}, come proportional to $x^{11}$ or $x^2_{11}$ as may be easily seen from {\bf Table \ref{fchumbon}}. All of these would vanish in the limit $x^{11} \to 0$ except for the three set of Riemann tensors:
\bg\label{lizolsen}
{\bf R}_{{\rm MN}ab}, ~~~{\bf R}_{abab}, ~~~ {\bf R}_{ab\mu\nu}, \nd
although in the last set, the component ${\bf R}_{abi0}$ would exist whether or not the metric has any dependence on the toroidal directions (see {\bf Table \ref{firzasut}}). In fact the non-vanishing pieces of \eqref{lizolsen} scale differently as 
${g_s\over {\rm HH}_o}, \left({g_s\over {\rm HH}_o}\right)^3$ and $\left({g_s\over {\rm HH}_o}\right)^{-1}$
respectively, but contribute equally as $+{1\over 3}l_r$ to \eqref{botsuga3}. Interestingly, and as a side note, in this limit only 
{\it even} number of $n_3$ derivatives in \eqref{fahingsha3} would survive.

\begin{table}[tb]  
 \begin{center}
\renewcommand{\arraystretch}{1.5}
\begin{tabular}{|c||c||c||c|}\hline Contributions to ${\bf R}^{a_0b_0}_{[{\rm MN}]}$  & Using
\eqref{makibhalu3} & Using generic ${\mathbb{T}^2\over {\cal G}}$ dep & Using
\eqref{evader} \\ \hline\hline
${\bf R}_{{\rm MNP}a}$ &  ${\gamma k\over 3} - 1$ & $-1$ & ..... \\ \hline
${\bf R}_{{\rm MN}a0}$ &  ${\gamma k\over 3} - 1$ & $-1$ & ..... \\ \hline
${\bf R}_{{\rm MN}ai}$ &  ${\gamma k\over 3} $ & $~~0$ & ..... \\ \hline 
${\bf R}_{{\rm MN}\mu\nu}$ &  ${\rm dom}\left({2\gamma k\over 3} - 2, 0\right)$ & $-2$ & 0 \\ \hline
${\bf R}_{{\rm MNPQ}}$ &  ${\rm dom}\left({2\gamma k\over 3} - 2, 0\right)$ & $-2$ & 0 \\ \hline
${\bf R}_{{\rm MN}ab}$ &  ${\rm dom}\left({\gamma k\over 3} - 2, 0\right)$ & $-2$ & 0 \\ \hline
${\bf R}_{{\rm MNP}0}$ &  $~~~0$ & $0$ & 0 \\ \hline
${\bf R}_{{\rm MNP}i}$ &  $~~~1$ & $1$ & 1 \\ \hline
${\bf R}_{{\rm MN}i0}$ &  $~~~1$ & $1$ & 1 \\ \hline
\end{tabular}
\renewcommand{\arraystretch}{1}
\end{center}
 \caption[]{Comparing the $g_s$ scalings of the two-form ${\bf R}^{a_0b_0}_{\rm MN} dy^{\rm M} \wedge dy^{\rm N}$ from the metric choices \eqref{makibhalu3}, generic toroidal fibre dependence, and \eqref{evader}. The first column has all the Riemann tensors contributing to the two-form via the contraction 
 ${\bf R}_{[{\rm MN}][{\rm CD}]} e^{a_0{\rm C}}e^{b_0{\rm D}} \equiv {\bf R}^{a_0b_0}_{[{\rm MN}]}$ where $e^{a_0{\rm M}}$ is the eleven-dimensional vielbein. The second and the fourth columns, using the metric choices \eqref{makibhalu3} and 
 \eqref{evader} respectively, allow four-dimensional EFT description to be valid in the IIB side, whereas the third column does not allow  an EFT description. The dotted terms in the fourth column represent the non-existence of the corresponding Riemann tensors.}
  \label{fchumbon2}
 \end{table}

The more important question now is whether we gain any mileage from making the metric and the flux components to have toroidal fibre dependences. In \cite{coherbeta} and \cite{heliudson} we argued how using a flat slicing of de Sitter space {\it prohibits} any G-flux components to be time-independent. Such a restriction is {\it not} imposed by the $g_s$ scaling of the quantum series: for example \eqref{botsuga}, 
\eqref{botsuga3}, and even \eqref{botsuga2}, can allow all G-flux components except ${\bf G}_{{\rm MN}ab}$ to remain in principle time-{\it independent}. The constraint actually comes from anomaly cancellations and flux EOMs. In \cite{desitter2, coherbeta, heliudson} we performed detailed computations for the case with a flat-slicing to show the non-existence of time-independent G-flux components. With generic slicing of de Sitter space, or with eleven-dimensional fibre dependences, this should be revisited to see whether some components of G-flux could remain time-independent. For example, could it be possible to keep flux components like ${\bf G}_{{\rm MNP}a}$ to remain time-independent? Such flux components map to 
RR and NS three-form fluxes, $({\bf F}_3)_{\rm MNP}$ and $({\bf H}_3)_{\rm MNP}$ respectively
 in the IIB side. If this were possible, then maybe we can justify a KKLT \cite{KKLT} like Glauber-Sudarshan state in our construction. Unfortunately, as we shall argue below, all computations seem to indicate the existence of only time-{\it dependent} G-flux components. The time-independent cases lead to breakdowns of EFTs in the IIB side. 

\begin{table}[tb]  
 \begin{center}
\renewcommand{\arraystretch}{1.5}
\begin{tabular}{|c||c||c||c|}\hline Contributions to ${\bf R}^{a_0b_0}_{[ab]}$  & Using
\eqref{makibhalu3} & Using generic ${\mathbb{T}^2\over {\cal G}}$ dep & Using
\eqref{evader} \\ \hline\hline
${\bf R}_{abai}$ &  ${\gamma k\over 3} + 2$ & $2$ & ..... \\ \hline
${\bf R}_{aba0}$ &  ${\gamma k\over 3} + 1$ & $1$ & ..... \\ \hline
${\bf R}_{ab{\rm M}c}$ &  ${\gamma k\over 3} + 1 $ & $1$ & ..... \\ \hline 
${\bf R}_{abab}$ &  ${\rm dom}\left({\gamma k\over 3}, 2\right)$ & $0$ & 2 \\ \hline
${\bf R}_{ab{\rm MN}}$ &  ${\rm dom}\left({\gamma k\over 3}, 2\right)$ & $0$ & 2 \\ \hline
${\bf R}_{ab\mu\nu}$ &  ${\rm dom}\left({\gamma k\over 3}, 2\right)$ & $0$ & 2 \\ \hline
${\bf R}_{ab{\rm M}0}$ &  $2$ & $2$ & $2$ \\ \hline
${\bf R}_{ab{\rm M}i}$ &  $3$ & $3$ & $3$ \\ \hline
${\bf R}_{ab\mu 0}$ &  $3$ & $3$ & $3$ \\ \hline
\end{tabular}
\renewcommand{\arraystretch}{1}
\end{center}
 \caption[]{Comparing the $g_s$ scalings of the two-form ${\bf R}^{a_0b_0}_{ab} dw^a \wedge dw^b$ from the metric choices \eqref{makibhalu3}, generic toroidal fibre dependence, and \eqref{evader}. As before, the two forms are constructed by contracting the corresponding Riemann tensors using eleven-dimensional vielbeins. The third column again represents the $g_s$ scalings that do not lead to a well defined EFT description in the dual IIB side.}
  \label{fchumbon3}
 \end{table}

\subsection{Computing the curvature forms and polynomials \label{courv}}

An easy way to verify the existence of G-flux components with or without time-dependences is to work out the ${\bf X}_8$ polynomial. This polynomial enters crucially in both the flux EOMs as well as in the anomaly cancellation equations \cite{BB}, and decides the fate of the fluxes on the compact eight-manifold. It may be expressed, using the curvature two-forms, in the following way:
\bg\label{xeight}
{\bf X}_8 \equiv  {1\over 3\cdot 2^9 \cdot \pi^4} \left({\rm tr} ~\mathbb{R}^4 
- {1\over 4} \left({\rm tr}~\mathbb{R}^2\right)^2\right), \nd
where $\mathbb{R}$ is the curvature two-form to be defined momentarily. Note that, compared to the Calabi-Yau four-fold case, the integral of ${\bf X}_8$ over the eight-manifold can neither be time-independent, nor represent the Euler characteristics. On the other hand, it does form a crucial part of the flux EOMs as mentioned earlier. The trace is defined over the holonomy matrices $\mathbb{M}_{a_0b_0}$, whose detailed properties are discussed in \cite{desitter2}, so we will avoid elaborating them here. The curvature two-form then can be expressed as:
\bg\label{elfadogg}
\mathbb{R} = \left({\bf R}^{a_0 b_0}_{[{\rm MN}]}~dy^{\rm M} \wedge dy^{\rm N} + 
{\bf R}^{a_0 b_0}_{[ab]}~dw^{a} \wedge dw^{b} + {\bf R}^{a_0 b_0}_{[{\rm M}a]}~dy^{\rm M} \wedge 
dw^{a}\right)\mathbb{M}_{a_0b_0}, \nd
where the curvature tensors contributing to ${\bf R}^{a_0 b_0}_{[{\rm MN}]}$ for the metric choices 
\eqref{makibhalu3}, \eqref{evader} and generic ${\mathbb{T}^2\over {\cal G}}$ are given in 
{\bf Table \ref{fchumbon2}}. Similar contributions to ${\bf R}^{a_0 b_0}_{[ab]}$ and 
${\bf R}^{a_0 b_0}_{[{\rm M}a]}$ are given in {\bf Tables \ref{fchumbon3}} and 
{\bf \ref{fchumbon4}} respectively.

\begin{table}[tb]  
 \begin{center}
\renewcommand{\arraystretch}{1.5}
\begin{tabular}{|c||c||c||c|}\hline Contributions to ${\bf R}^{a_0b_0}_{[{\rm M}a]}$  & Using
\eqref{makibhalu3} & Using generic ${\mathbb{T}^2\over {\cal G}}$ dep & Using
\eqref{evader} \\ \hline\hline 
${\bf R}_{{\rm M}abc}$ &  ${\gamma k\over 3}$ & $~~0$ & ..... \\ \hline
${\bf R}_{{\rm M}a\mu\nu}$ &  ${\gamma k\over 3} $ & $~~0$ & ..... \\ \hline
${\bf R}_{{\rm M}a{\rm N}i}$ &  ${\gamma k\over 3} + 1 $ & $~~1$ & ..... \\ \hline 
${\bf R}_{{\rm M}a{\rm N}0}$ &  ${\gamma k\over 3}$ & $~~0$ & ..... \\ \hline
${\bf R}_{{\rm M}a{\rm NP}}$ &  ${\gamma k\over 3}$ & $~~0$ & ..... \\ \hline
${\bf R}_{{\rm M}a{\rm N}b}$ &  ${\rm dom}\left({\gamma k\over 3} -1, 1\right)$ & $-1$ & 1 \\ \hline
${\bf R}_{{\rm M}ab0}$ &  $1$ & $~~1$ & 1 \\ \hline
${\bf R}_{{\rm M}abi}$ &  $2$ & $~~2$ & $2$ \\ \hline
\end{tabular}
\renewcommand{\arraystretch}{1}
\end{center}
 \caption[]{Comparing the $g_s$ scalings of the two-form ${\bf R}^{a_0b_0}_{{\rm M}a} dy^{\rm M} \wedge dw^a$ from the metric choices \eqref{makibhalu3}, generic toroidal fibre dependence, and \eqref{evader}. 
 These cross-term forms are constructed using the eleven-dimensional vielbeins in a way described earlier.
  The third column still represents the $g_s$ scalings that do not lead to a well defined EFT description in the dual IIB side, although we note that there are many Riemann tensors absent in the fourth column.}
  \label{fchumbon4}
 \end{table}

The $g_s$ scalings of the curvature two-forms are important because if there are $g_s$ independent contributions to ${\bf X}_8$ \eqref{xeight} that would imply that certain components of the fluxes may be allowed to remain time-{\it independent}. Using the metric choice \eqref{makibhalu3}, the two-form 
${\bf R}^{a_0 b_0}_{[{\rm MN}]}$ may be expressed as:

{\footnotesize
\bg\label{case}
{\bf R}^{a_0 b_0}_{[{\rm MN}]}(x, y, w^a; k, g_s)\mathbb{M}_{a_0b_0} & =& 
{\bf R}^{(1)}_{[{\rm MN}]}(x, y, w^a) + {\bf R}^{(2)}_{[{\rm MN}]}(x, y, w^a)\left({g_s\over {\rm HH}_o}\right)^{{\gamma k\over 3}} \\
& + & {\bf R}^{(3)}_{[{\rm MN}]}(x, y, w^a)\left({g_s\over {\rm HH}_o}\right) + 
{\bf R}^{(4)}_{[{\rm MN}]}(x, y, w^a)\left({g_s\over {\rm HH}_o}\right)^{{\rm dom}\left({2\gamma k\over 3} -2, 0\right)} \nonumber\\
& + & {\bf R}^{(5)}_{[{\rm MN}]}(x, y, w^a)\left({g_s\over {\rm HH}_o}\right)^{{\rm dom}\left({\gamma k\over 3} -2, 0\right)} 
 + {\bf R}^{(6)}_{[{\rm MN}]}(x, y, w^a)\left({g_s\over {\rm HH}_o}\right)^{{\gamma k\over 3} - 1}\nonumber \nd}
where we suppress the internal ${\rm M}_p$ dependence that appears from the exponential term in 
\eqref{makibhalu3}. Taking $k = 1$ and $\gamma = 5$, we see that all terms have positive $g_s$ powers except ${\bf R}^{(5)}_{[{\rm MN}]}(x, y, w^a)$ which scales as $\left({g_s\over {\rm HH}_o}\right)^{-{1/3}}$. The three components, ${\bf R}^{(1)}_{[{\rm MN}]}(x, y, w^a)$, ${\bf R}^{(4)}_{[{\rm MN}]}(x, y, w^a)$ and 
${\bf R}^{(5)}_{[{\rm MN}]}(x, y, w^a)$, scale as $\left({g_s\over {\rm HH}_o}\right)^0$ because this is the dominant scaling when $\gamma = 5$. In a similar vein the two-form ${\bf R}^{a_0 b_0}_{[ab]}$ may be defined in the following way:

\begin{table}[tb]  
 \begin{center}
\renewcommand{\arraystretch}{1.5}
\begin{tabular}{|c||c||c||c|}\hline Contributions to ${\bf R}^{a_0b_0}_{[ij]}$  & Using
\eqref{makibhalu3} & Using generic ${\mathbb{T}^2\over {\cal G}}$ dep & Using
\eqref{evader} \\ \hline\hline
${\bf R}_{ij{\rm M}a}$ &  ${\gamma k\over 3} - 3$ & $-3$ & ..... \\ \hline
${\bf R}_{ijka}$ &  ${\gamma k\over 3} - 2$ & $-2$ & ..... \\ \hline
${\bf R}_{ija0}$ &  ${\gamma k\over 3} -3 $ & $-3$ & ..... \\ \hline 
${\bf R}_{ij{\rm MN}}$ &  ${\rm dom}\left({2\gamma k\over 3} - 4, -2\right)$ & $-4$ & $-2$ \\ \hline
${\bf R}_{ij{\rm M}0}$ &  $-2$ & $-2$ & $-2$ \\ \hline
${\bf R}_{ijk{\rm M}}$ &  $-1$ & $-1$ & $-1$ \\ \hline
${\bf R}_{ijk0}$ &  $-1$ & $-1$ & $-1$ \\ \hline
${\bf R}_{ijij}$ &  ${\rm dom}\left({2\gamma k\over 3} - 4, -2\right)$ & $-4$ & $-2$ \\ \hline
${\bf R}_{ijab}$ &  ${\rm dom}\left({\gamma k\over 3} - 4, -2\right)$ & $-4$ & $-2$ \\ \hline
\end{tabular}
\renewcommand{\arraystretch}{1}
\end{center}
 \caption[]{Comparing the $g_s$ scalings of the two-form ${\bf R}^{a_0b_0}_{ij} dx^{i} \wedge dx^j$ from the metric choices \eqref{makibhalu3}, generic toroidal fibre dependence, and \eqref{evader}. As before, the dotted terms in the fourth column represent the non-existence of the corresponding Riemann tensors.}
  \label{fchumbon5}
 \end{table}

{\footnotesize
\bg\label{brynhildur}
{\bf R}^{a_0 b_0}_{[ab]}(x, y, w^a; k, g_s)\mathbb{M}_{a_0b_0} & =& 
{\bf R}^{(1)}_{[ab]}(x, y, w^a)\left({g_s\over {\rm HH}_o}\right)^{{\gamma k\over 3} + 2} 
 +  {\bf R}^{(2)}_{[ab]}(x, y, w^a)\left({g_s\over {\rm HH}_o}\right)^{{\rm dom}\left({\gamma k\over 3}, 2\right)}\\
&+& {\bf R}^{(3)}_{[ab]}(x, y, w^a)\left({g_s\over {\rm HH}_o}\right)^2 
+ {\bf R}^{(4)}_{[ab]}(x, y, w^a)\left({g_s\over {\rm HH}_o}\right)^3
+  {\bf R}^{(5)}_{[ab]}(x, y, w^a)\left({g_s\over {\rm HH}_o}\right)^{{\gamma k\over 3} + 1}
\nonumber, \nd}
where $\left({g_s\over {\rm HH}_o}\right)^{5/3}$ is the dominant scaling now for $\gamma = 5$ and $k = 1$. Interestingly, all scalings are positive definite, and there is no zero scaling now compared to what we had in \eqref{case}. In fact a similar story appears for ${\bf R}^{a_0 b_0}_{[{\rm M}a]}$, which may be expressed as:

{\footnotesize  
\bg\label{hannalara}
{\bf R}^{a_0 b_0}_{[{\rm M}a]}(x, y, w^a; k, g_s)\mathbb{M}_{a_0b_0} & =& 
{\bf R}^{(1)}_{[{\rm M}a]}(x, y, w^a)\left({g_s\over {\rm HH}_o}\right)^{{\gamma k\over 3} + 1} 
 +  {\bf R}^{(2)}_{[{\rm M}a]}(x, y, w^a)\left({g_s\over {\rm HH}_o}\right)^{{\rm dom}\left({\gamma k\over 3} -1, 1\right)}\\
&+& {\bf R}^{(3)}_{[{\rm M}a]}(x, y, w^a)\left({g_s\over {\rm HH}_o}\right) 
+ {\bf R}^{(4)}_{[{\rm M}a]}(x, y, w^a)\left({g_s\over {\rm HH}_o}\right)^2
+  {\bf R}^{(5)}_{[{\rm M}a]}(x, y, w^a)\left({g_s\over {\rm HH}_o}\right)^{{\gamma k\over 3}}
\nonumber, \nd}
with the dominant scaling now being $\left({g_s\over {\rm HH}_o}\right)^{2/3}$, and we see no components with zero scaling. This means the only way \eqref{xeight} can have a $g_s$ independent term if the negative scalings of ${\bf R}^{a_0 b_0}_{[{\rm MN}]}$ somehow cancel the positive $g_s$ scalings of 
${\bf R}^{a_0 b_0}_{[ab]}$ and ${\bf R}^{a_0 b_0}_{[{\rm M}a]}$. Could this happen here?

The dominant scalings of the three curvature forms in \eqref{case}, \eqref{brynhildur} and \eqref{hannalara}
tell us that we can remove the $k$ dependences of the three two-forms and express the generic curvature two-form \eqref{elfadogg} in the following suggestive 
way\footnote{We have suppressed one intermediate step that would take us from \eqref{case}, \eqref{brynhildur} and \eqref{hannalara} to \eqref{elfadogg2}.  A more appropriate way to 
express \eqref{elfadogg2} would be to observe, for example, that ${\bf R}_{[ab]} \equiv {\bf R}^{a_ob_o}_{[ab]}\mathbb{M}_{a_ob_o}$ scales in the following way:
\bg\label{ferrern}
{\bf R}_{[ab]}(x, y, w^a; k, g_s) = \sum_{l = 0}^{p_k} {\bf R}^{(l, k)}_{[ab]}(x, y, w^a) \left({g_s\over {\rm HH}_o}\right)^{{1\over 3}(l + 5)k}, 
\nonumber \nd
for every choice of $k$ in \eqref{makibhalu3} and $p_1 = 6, p_2 = 11$ etc. Once we sum over all $k \ge 1$, we can easily see that the dominant scale remains $\left({g_s\over {\rm HH}_o}\right)^{5\over 3}$ and the series takes the generic form as in \eqref{elfadogg2}. Similar arguments could be given for the other two two-forms
${\bf R}_{[{\rm MN}]}$ and ${\bf R}_{[{\rm M}a]}$. \label{elish}}:
\bg\label{elfadogg2}
\mathbb{R}(x, y, w^a; g_s) &=& \sum_{l = 0}^\infty {\bf R}^{(l)}_{[{\rm MN}]}(x, y, w^a)\left({g_s\over {\rm HH}_o}\right)^{l-1\over 3} dy^{\rm M} \wedge dy^{\rm N}\\
& + & \sum_{l = 0}^\infty
\left[{\bf R}^{(l)}_{[ab]}\left({g_s\over {\rm HH}_o}\right)^{l+5\over 3} dw^{a} \wedge dw^{b} 
+ {\bf R}^{(l)}_{[{\rm M}a]}\left({g_s\over {\rm HH}_o}\right)^{l+2\over 3}~dy^{\rm M} \wedge 
dw^{a}\right], \nonumber \nd
which could then be inserted in the definition of ${\bf X}_8$. In terms of the individual two-forms, there are two ways of doing this: in one case we take three ${\bf R}_{[{\rm MN}]}$ and one ${\bf R}_{[ab]}$ two-forms, and in the other case we can use two ${\bf R}_{[{\rm MN}]}$ and two ${\bf R}_{[{\rm M}a]}$ two-forms to compute \eqref{xeight}. In both cases the answer turns out to be the same and is given by:
\bg\label{thetrial}
{\bf X}_8(x, y, w^a; g_s) = \sum_{\{l_i\}} \widetilde{\bf X}_{(8, c)}^{(l_1,..., l_4)}(x, y, w^a) 
\left({g_s\over {\rm HH}_o}\right)^{{1\over 3}(l_1 + l_2 + l_3 + l_4 + 2)}, \nd 
which unfortunately does not have a $g_s$ independent piece because $l_i \ge 0$. This is consistent with a similar conclusion that we had for the case with flat-slicing in \cite{desitter2, heliudson}, so it's no surprise that we get the same story here too. In fact, using the metric ans\"atze \eqref{evader} and the last columns in the {\bf Tables \ref{fchumbon2}}, {\bf \ref{fchumbon3}} and {\bf \ref{fchumbon4}},  it is easy to see that the curvature two-form \eqref{elfadogg2} changes to:
\bg\label{elfadogg3}
\mathbb{R}(x, y; g_s) &=& \sum_{l = 0}^\infty {\bf R}^{(l)}_{[{\rm MN}]}(x, y)\left({g_s\over {\rm HH}_o}\right)^{l\over 3} dy^{\rm M} \wedge dy^{\rm N}\\
& + & \sum_{l = 0}^\infty
\left[{\bf R}^{(l)}_{[ab]}(x, y)\left({g_s\over {\rm HH}_o}\right)^{l+ 6\over 3} dw^{a} \wedge dw^{b} 
+ {\bf R}^{(l)}_{[{\rm M}a]}(x, y)\left({g_s\over {\rm HH}_o}\right)^{l+ 3\over 3}~dy^{\rm M} \wedge 
dw^{a}\right], \nonumber \nd

\begin{table}[tb]  
 \begin{center}
\renewcommand{\arraystretch}{1.5}
\begin{tabular}{|c||c||c||c|}\hline Contributions to ${\bf R}^{a_0b_0}_{[i0]}$  & Using
\eqref{makibhalu3} & Using generic ${\mathbb{T}^2\over {\cal G}}$ dep & Using
\eqref{evader} \\ \hline\hline
${\bf R}_{i0a0}$ &  ${\gamma k\over 3} - 2$ & $-2$ & ..... \\ \hline
${\bf R}_{i0aj}$ &  ${\gamma k\over 3} - 3$ & $-3$ & ..... \\ \hline
${\bf R}_{i0ab}$ &  $-1 $ & $-1$ & $-1$ \\ \hline 
${\bf R}_{i0j0}$ &  ${\rm dom}\left({2\gamma k\over 3} - 4, -2\right)$ & $-4$ & $-2$ \\ \hline
${\bf R}_{i0{\rm MN}}$ &  $-1$ & $-1$ & $-1$ \\ \hline
${\bf R}_{i0{\rm M}j}$ &  $-2$ & $-2$ & $-2$ \\ \hline
${\bf R}_{i0{\rm M}0}$ &  $-1$ & $-1$ & $-1$ \\ \hline
${\bf R}_{i0jk}$ &  $-1$ & $-1$ & $-1$ \\ \hline
\end{tabular}
\renewcommand{\arraystretch}{1}
\end{center}
 \caption[]{Comparing the $g_s$ scalings of the two-form ${\bf R}^{a_0b_0}_{i0} dx^{i} \wedge dx^0$ from the metric choices \eqref{makibhalu3}, generic toroidal fibre dependence, and \eqref{evader}. Comparing this with {\bf Table \ref{fchumbon5}} we see that the individual components behave differently from having fully spatial indices $(i, j)$ to having mixed indices $(i, 0)$. This will have important consequences for the flux EOMs and anomaly cancellations.}
  \label{fchumbon6}
 \end{table}

\noindent where the only difference from \eqref{elfadogg2} is the $g_s$ scalings: the dominant scalings are shown up-front, and the sub-dominant ones follow in the usual way. The infinite series appears from summing over all $k \ge 1$ in \eqref{makibhalu3} as outlined in footnote \ref{elish}. Once we plug in \eqref{elfadogg3} into \eqref{xeight}, we get the following structure of the eight-form ${\bf X}_8$:
\bg\label{thetrial2}
{\bf X}_8(x, y; g_s) = \sum_{\{l_i\}} \widetilde{\bf X}_{(8, a)}^{(l_1,..., l_4)}(x, y) 
\left({g_s\over {\rm HH}_o}\right)^{{1\over 3}(l_1 + l_2 + l_3 + l_4 + 6)}, \nd 
which is in fact the {\it same} scaling that we saw for the case with flat-slicing in \cite{desitter2, heliudson}. 
Unfortunately again the eight-form has no $g_s$ independent term. Note that both the eight-forms \eqref{thetrial} and \eqref{thetrial2} are no longer topological or related to the Euler characteristics of the internal eight-manifold. Thus it appears, for the metric choices \eqref{evader} and \eqref{makibhalu3}, ${\bf X}_8$ 
remains time-dependent. What happens for the case when the metric has a generic dependence on the toroidal direction?

The generic dependence on the toroidal direction may be quantified by allowing a $w^a$ dependence to the metric components $\widetilde{g}^{(k)}_{\rm CD} \equiv \widetilde{g}^{(k)}_{\rm CD}(x, y, w^a)$ in say 
\eqref{makibhalu3}. In such a case, we can look up the third columns in {\bf Tables \ref{fchumbon2}}, 
{\bf \ref{fchumbon3}} and {\bf \ref{fchumbon4}}. The curvature two-form can now be expressed as:
\bg\label{elfadogg4}
\mathbb{R}(x, y, w^a; g_s) &=& \sum_{l = 0}^\infty {\bf R}^{(l)}_{[{\rm MN}]}(x, y, w^a)\left({g_s\over {\rm HH}_o}\right)^{l - 6\over 3} dy^{\rm M} \wedge dy^{\rm N}\\
& + & \sum_{l = 0}^\infty
\left[{\bf R}^{(l)}_{[ab]}\left({g_s\over {\rm HH}_o}\right)^{l\over 3} dw^{a} \wedge dw^{b} 
+ {\bf R}^{(l)}_{[{\rm M}a]}\left({g_s\over {\rm HH}_o}\right)^{l - 3\over 3}~dy^{\rm M} \wedge 
dw^{a}\right], \nonumber \nd
where we see that there are two places where the dominant scalings become {\it negative}: for 
${\bf R}_{[{\rm MN}]}$ and for ${\bf R}_{[{\rm M}a]}$. We can plug in \eqref{elfadogg4} in \eqref{xeight} to compute the eight-form in two different ways. The results are the same, and ${\bf X}_8$ takes the following form:
\bg\label{thetrial3}
{\bf X}_8(x, y, w^a; g_s) = \sum_{\{l_i\}} \widetilde{\bf X}_{(8, b)}^{(l_1,..., l_4)}(x, y, w^a) 
\left({g_s\over {\rm HH}_o}\right)^{{1\over 3}(l_1 + l_2 + l_3 + l_4 {\red - 18})}, \nd 
which, because of the {\red minus} sign, does have a $g_s$ independent term for all $l_i$ satisfying 
${\red l_1 + l_2 + l_3 + l_4 = 18}$. According to the flux EOM studied in \cite{desitter2, heliudson}, such a scenario should lead to time-{\it independent} ${\bf H}_3$ and ${\bf F}_3$ three-form fluxes in the dual IIB side. Unfortunately, as shown in \eqref{botsuga2}, such a system {\it does not} have an EFT description even if we switch on time-dependent G-flux components ${\bf G}_{{\rm MN}ab}$ (keeping ${\bf G}_{{\rm MNP}a}$ time-independent).

\begin{table}[tb]  
 \begin{center}
\renewcommand{\arraystretch}{1.5}
\begin{tabular}{|c||c||c||c|}\hline Contributions to ${\bf R}^{a_0b_0}_{[{\rm M}i]}$  & Using
\eqref{makibhalu3} & Using generic ${\mathbb{T}^2\over {\cal G}}$ dep & Using
\eqref{evader} \\ \hline\hline
${\bf R}_{{\rm M}i{\rm N}a}$ &  ${\gamma k\over 3} - 1$ & $- 1$ & ..... \\ \hline
${\bf R}_{{\rm M}iaj}$ &  ${\gamma k\over 3} - 2$ & $- 2$ & ..... \\ \hline
${\bf R}_{{\rm M}i{\rm NP}}$ &  $0$ & $~~0$ & $~~0$ \\ \hline 
${\bf R}_{{\rm M}i{\rm N}j}$ &  ${\rm dom}\left({2\gamma k\over 3} - 3, -1\right)$ & $-3$ & $-1$ \\ \hline
${\bf R}_{{\rm M}i{\rm N}0}$ &  $0$ & $~~0$ & $~~0$ \\ \hline
${\bf R}_{{\rm M}ij0}$ &  $-1$ & $-1$ & $-1$ \\ \hline
${\bf R}_{{\rm M}ijk}$ &  $0$ & $~~0$ & $~~0$ \\ \hline
${\bf R}_{{\rm M}iab}$ &  $0$ & $~~0$ & $~~0$ \\ \hline
\end{tabular}
\renewcommand{\arraystretch}{1}
\end{center}
 \caption[]{Comparing the $g_s$ scalings of the two-form ${\bf R}^{a_0b_0}_{{\rm M}i} dy^{\rm M} \wedge dx^i$ from the metric choices \eqref{makibhalu3}, generic toroidal fibre dependence, and \eqref{evader}. 
 These two-forms are useful ingredients of the eight-form \eqref{xeight} defined partly along spatial and 
 partly along the internal directions. They also have important consequences for the flux EOMs and anomaly cancellations.}
  \label{fchumbon7}
 \end{table}
 
 \subsection{G-flux equations of motion and consistency conditions \label{gfluxx}}

Our above conclusion, about the non-existence of time-independent G-flux components (for the cases that allow an EFT description in the dual IIB side), relies on analyzing {\it all} the flux EOMs. Here, due to the toroidal fibre dependence, the story is more generic than what we encountered in \cite{heliudson, desitter2}, but fortunately the precise EOMs are no different. Recall that the flux EOMs may be succinctly presented as:
\bg\label{elenag}
d\ast_{11} {\bf G}_4 = b_1 {\bf G}_4 \wedge {\bf G}_4 + b_2 \mathbb{Y}_8 + b_3 d\ast_{11} \mathbb{Y}_4 + 
(n_b - \bar{n}_b) {\bf \Lambda}_8, \nd
where $\ast_{11}$ is the warped Hodge duality operator, $b_i$ are the coefficients that only depend on 
${\rm M}_p$, $\mathbb{Y}_4$ is the quantum terms that may be extracted from \eqref{fahingsha4} \cite{desitter2, heliudson}, 
$(n_b, \bar{n}_b)$ denote the number of space-filling M2 and $\overline{\rm M2}$ branes respectively that have localized form ${\bf \Lambda}_8$; and $\mathbb{Y}_8$ is an eight-form constructed out of curvature and 
G-flux forms. 
Thus the curvature part of $\mathbb{Y}_8$ is precisely the ${\bf X}_8$ form discussed above. 
In writing \eqref{elenag} we have ignored the non-perturbative contributions. This will be inserted in a bit later.
If we now only consider the ${\bf X}_8$ piece of $\mathbb{Y}_8$, then \eqref{elenag} leads to the familiar 
constraint:
\bg\label{lindmonaco}
b_1\int {\bf G}_4 \wedge {\bf G}_4 + \vert n_b - \bar{n}_b\vert = -b_2 \int {\bf X}_8, \nd
where the absolute sign is there to choose only the positive difference by keeping $n_b > \bar{n}_b$. This choice is made to avoid potential conflict with the anti-brane back-reaction issues pointed out in 
\cite{beno}, although in our case we can either keep $n_b \ge \bar{n}_b$ or remove the branes and the anti-branes altogether. The flexibility in our choice appears from our ability to choose a different source for supersymmetry breaking $-$ non-self-dual G-flux components $-$ thus removing any dependence on
anti-branes. In such a case, we see from \eqref{thetrial} and \eqref{thetrial2} that the RHS of \eqref{lindmonaco} is always time-dependent\footnote{There is a subtlety that we will have to consider here related to the dependence of the metric and the G-flux components on the spatial directions 
$x^i$. One way to make sense of \eqref{lindmonaco} is to take an {\it average} over the ${\bf R}^2$ directions for the two terms with coefficients $b_1$ and $b_2$. Since ${\bf \Lambda}_8$ is a localized form on the eight-manifold, the averaging will not affect the $\vert n_b - \bar{n}_b\vert$ factor. Alternatively we can fix the spatial coordinate over a slice and define the integral \eqref{lindmonaco}.  See also footnote 
\ref{hana00}. A third alternative would be to take $n_b = \bar{n}_b$, and balance the $x^i$ dependence of the two remaining terms in \eqref{lindmonaco}.}. 
This means the left hand side (LHS) of \eqref{lindmonaco}, which include products of G-flux components, cannot be kept time-independent. It is easy to see why this may be the case: Since ${\bf G}_{{\rm MN}ab}$ components need to be time-dependent to avoid EFT breakdowns (see \eqref{botsuga} and \eqref{botsuga3}), \eqref{lindmonaco} will imply ${\bf G}_{\rm MNPQ}$ to become time-dependent also, at least when $n_b = \bar{n}_b$ or $n_b = \bar{n}_b = 0$. For example, using the analysis of flux EOMs from \cite{desitter2, heliudson}, 
and the ans\"atze from \eqref{theritual}, it is easy to see from \eqref{lindmonaco} that:
\bg\label{gmvero}
{2\over 3}(k_1 + k_2) + l_{\rm MN}^{\rm PQ} + l_{\rm RS}^{ab} = l + {p\over 3}, \nd
where $(k_1, k_2)$ are the $g_s$ scalings of the flux components ${\bf G}_{\rm MNPQ}$ and 
${\bf G}_{{\rm RS}ab}$ respectively from \eqref{theritual}, $p = 2$ for \eqref{thetrial} and $p = 6$ for \eqref{thetrial2}, $l \equiv {1\over 3}(l_1 + l_2 + l_3 + l_4)$ is the scaling appearing in either \eqref{thetrial} or \eqref{thetrial2}, and 
$l_{\rm AB}^{\rm CD}$ is the original scaling of the flux components defined earlier. Since we expect 
$l_{\rm RS}^{ab} \ge 1$ to avoid issues with the existence of EFT, and $(k_i, l) \ge (0, 0)$, it is easy to see that $\vert l_{\rm MN}^{\rm PQ}\vert \ge \big\vert{p - 3\over 3}\big\vert$, implying that the {\it smallest} scaling of  
${\bf G}_{\rm MNPQ}$ is $\left({g_s\over {\rm HH}_o}\right)^{p - 3\over 3}$. For \eqref{thetrial} the scaling is 
$\left({g_s\over {\rm HH}_o}\right)^{-1/3}$, whereas for \eqref{thetrial2} it is 
$\left({g_s\over {\rm HH}_o}\right)^{+1}$.  However the $-{1\over 3}$ scaling of the flux components ${\bf G}_{\rm MNPQ}$ may be alarming, and we will come back to this issue a bit later. 
What happens if we make $l_{\rm MN}^{\rm PQ} = 0$?  Allowing  this will lead to, by choosing $l_{\rm RS}^{ab} = 1 + {n\over 3}$ with $n \in \mathbb{Z}$:
\bg\label{meaador}
k_1 + k_2 = {3\over 2}\left[l - \left({n+ 3 - p\over 3}\right)\right], \nd
which would imply, for $l = {n+1\over 3}$ in \eqref{thetrial} and $l = {n-3\over 3}$ in \eqref{thetrial2}, we can keep $k_1 = k_2 = 0$. The relative minus sign for the second case ({\it i.e.} for \eqref{thetrial2}) now implies that  $n \ge 4$ if we want to keep non-zero G-flux component ${\cal G}^{(0)}_{\rm MNPQ}$. In the same vein, for the case \eqref{thetrial}, $n \ge 0$ to allow the aforementioned conditions.  In the two cases, {\it i.e.} for \eqref{thetrial} and \eqref{thetrial2}, the ${\cal G}^{(0)}_{{\rm MN}ab}$ components scale at least as 
$\left({g_s\over {\rm HH}_o}\right)^{+1}$ and $\left({g_s\over {\rm HH}_o}\right)^{7/3}$ respectively. Although such high scaling of G-flux components for the latter case is consistent with the EFT, it can now no longer appear to the lowest orders in the flux or the Einstein's EOMs. This means that the supersymmetry breaking condition\footnote{Recall that we derived the supersymmetry breaking condition by comparing the Einstein's equations for space-time ${\bf R}^{2, 1}$ and the flux equations for ${\bf G}_{0ij{\rm M}}$. In both cases it was crucial that at least the G-flux components ${\bf G}_{{\rm MN}ab}$ appear. However if the lowest components of ${\bf G}_{{\rm MN}ab}$, {\it i.e.} ${\cal G}^{(0)}_{{\rm MN}ab}$, have $g_s$ scalings {\it bigger} than $+1$ they do not appear in either of the flux or the Einstein's EOMs. Unfortunately the scaling requirements also prohibit other {\it internal} flux components to appear, as we will see soon.}, from the non self-duality of the G-flux components ${\cal G}^{(0)}_{{\rm MN}ab}$ (see \cite{desitter2, heliudson}), cannot appear easily and we have to resort to anti-branes as the source for this. The latter has numerous issues (see \cite{beno}), so it appears that keeping ${\bf G}_{\rm MNPQ}$ time-independent is problematic, at least for the case \eqref{thetrial2}. We will soon see that a similar issue plagues the former case, {\it i.e.} the case 
\eqref{thetrial}, too.  
Similarly, for $a \equiv (3, 11)$, either or both the G-flux components 
${\bf G}_{{\rm MNP}a}$ will have to be time-dependent. This is because, using the cue from \eqref{gmvero}, we expect $l_{\rm MN}^{{\rm P}a}$ to satisfy:
\bg\label{gmvero2}
l_{\rm MN}^{{\rm P}a} + l_{\rm RS}^{{\rm Q}b} ~\ge ~ {p\over 3}, \nd 
as making them zero would lead to $k_1 + k_2 = {3\over 2}\left(l + {p\over 3}\right)$ 
contradicting\footnote{For example, matching the powers of $g_s$ in the flux EOM would imply
${\cal G}^{(0)}_{{\rm MNP}a} {\cal G}^{(0)}_{{\rm RSQ}b} = 0$. This cannot lead to any non-trivial time-independent components.} again the fact that $(k_i, l) \ge (0, 0)$.
Since ${\bf G}_{{\rm MNP}a}$ components dualize to ${\bf H}_3$ and ${\bf F}_3$ three-form fluxes, 
${\bf G}_{{\rm MN}ab}$ components dualize to seven-brane world-volume gauge fluxes, and ${\bf G}_{\rm MNPQ}$ components dualize to five-form fluxes in the IIB side, our analysis show that none of these flux components can be 
time-independent, at least when we use \eqref{lindmonaco}.  
Question is, what happens when we cannot impose the condition \eqref{lindmonaco}, {\it i.e.} when we analyze the case corresponding to non-compact directions? Addtionally, what happens $n_b > \bar{n}_b$ with 
$0 \le \bar{n}_b \le 1$ (the latter to avoid conflict with \cite{beno})?

\begin{table}[tb]  
 \begin{center}
\renewcommand{\arraystretch}{1.5}
\begin{tabular}{|c||c||c||c|}\hline Contributions to ${\bf R}^{a_0b_0}_{[{\rm M}0]}$  & Using
\eqref{makibhalu3} & Using generic ${\mathbb{T}^2\over {\cal G}}$ dep & Using
\eqref{evader} \\ \hline\hline
${\bf R}_{{\rm M}0{\rm N}a}$ &  ${\gamma k\over 3} - 2$ & $- 2$ & ..... \\ \hline
${\bf R}_{{\rm M}0a\mu}$ &  ${\gamma k\over 3} - 2$ & $- 2$ & ..... \\ \hline
${\bf R}_{{\rm M}0{\rm NP}}$ &  $-1$ & $-1$ & $-1$ \\ \hline 
${\bf R}_{{\rm M}0{\rm N}0}$ &  ${\rm dom}\left({2\gamma k\over 3} - 3, -1\right)$ & $-3$ & $-1$ \\ \hline
${\bf R}_{{\rm M}0{\rm N}i}$ &  $0$ & $~~0$ & $~~0$ \\ \hline
${\bf R}_{{\rm M}0ij}$ &  $-1$ & $-1$ & $-1$ \\ \hline
${\bf R}_{{\rm M}0j0}$ &  $0$ & $~~0$ & $~~0$ \\ \hline
${\bf R}_{{\rm M}0ab}$ &  $-1$ & $-1$ & $-1$ \\ \hline
\end{tabular}
\renewcommand{\arraystretch}{1}
\end{center}
 \caption[]{Comparing the $g_s$ scalings of the two-form ${\bf R}^{a_0b_0}_{{\rm M}0} dy^{\rm M} \wedge dx^0$ from the metric choices \eqref{makibhalu3}, generic toroidal fibre dependence, and \eqref{evader}. 
 The contributions to the three columns differ from the respective contributions in {\bf Table \ref{fchumbon7}}
 because the temporal derivatives create extra powers of ${g_s\over {\rm HH}_o}$ and we are using 
 \eqref{kootdiye} as mentioned earlier.}
  \label{fchumbon8}
 \end{table}

This is where, as emphasized in \cite{heliudson}, other flux EOMs may become important. Recall that the 
flux EOM that actually reproduces the anomaly cancellation condition \eqref{lindmonaco}, appears from looking at the EOM for the G-flux components ${\bf G}_{0ij{\rm A}}$, where ${\rm A} \in {\bf R}^{2, 1} \times 
{\cal M}_4 \times {\cal M}_2 \times {\mathbb{T}^2\over {\cal G}}$, although the antisymmetry will only restrict 
${\rm A}$ to lie within the eight-manifold.  In the presence of space-filling branes, other flux EOMs become relevant, which in turn necessitates the determination of the generic form of the polynomial \eqref{xeight}. 
The generic flux EOMs, for the various embeddings of de Sitter space-time, have not been discussed in details anywhere (in \cite{desitter2, heliudson} we studied the case for {\it flat slicing} only). Thus it is time now to study them when we allow dependences on {\it all} eleven-dimensional coordinates. As one might expect, and because of the proliferation of the number of G-flux components, the scenario at hand is little more involved than what we studied in \cite{desitter2, heliudson} but can nevertheless be presented in the following {\it tensorial} language:
\bg\label{mcacrisis}
\partial_{{\rm N}_k}\left(\mathbb{T}_7^{(f)}\right)_{\rm N_1.....N_7} &= &  b_1~
\partial_{{\rm N}_k}\left(\mathbb{T}_7^{(q)}\right)_{\rm N_1.....N_7} + 
b_2 \left(\mathbb{T}_8^{(k)}\right)_{{\rm N_1.....N_7}{\rm N}_k} \\
&+& b_3 \left(\mathbb{X}_8^{(k)}\right)_{{\rm N_1.....N_7}{\rm N}_k} + 
b_4 \left({\bf \Lambda}_8\right)_{{\rm N_1.....N_7}{\rm N}_k} + 
b_5~ {\overbracket[1pt][7pt]{\mathbb{T}_0 \partial_{{\rm N}_k}\Big(\mathbb{T}}}{}_7^{(np)}\Big)_{\rm N_1.....N_7},\nonumber\nd
which is expressed using rank zero, seven and eight tensors. The rank seven tensors are classified by superscripts $f, q$ and  $np$ which stand for flux, quantum and non-perturbative respectively. The rank eight tensors, namely $\mathbb{T}_8^{(k)}$ and $\mathbb{X}_8^{(k)}$, have superscripts $k$ which match-up with the derivative actions $\partial_{{\rm N}_k}$ that act on the rank seven tensors. The various choices of $k$ represent the various choices of the three-form field strengths for a given choice of a four-form G-flux components. The rank eight tensor ${\bf \Lambda}_8$ is associated with M2 and $\overline{\rm M2}$ branes, and is typically a localized tensor. Since the sources we take are space-filling\footnote{This is not an essential requirement when we take a generic slicing of de Sitter space in the IIB side. For the flat-slicing case in \cite{desitter2, coherbeta} this requirement was imposed to avoid the appearance of four-dimensional isometry breaking factors in IIB. Here this requirement is imposed to simply avoid non-trivial
contributions from branes and anti-branes to the Schwinger-Dyson's equations. Inclusion of these effects is left as an exercise for the reader.}, for two-branes 
wrapped on cycles of the eight-manifold, we expect ${\bf \Lambda}_8 = 0$. We can also replace the two-branes by {\it fractional} two-branes (and the corresponding anti-branes), but the condition on 
${\bf \Lambda}_8$ remains the same.  
Putting everything together, the tensors appearing in \eqref{mcacrisis} are defined in the following way:
\bg\label{olivecosta}
&&\left(\mathbb{T}_7^{(q)}\right)_{\rm N_1 ...... N_7} = \sqrt{-{\bf g}_{11}} \left(\mathbb{Y}_4\right)^{\rm ABCD}~
\epsilon_{{\rm ABCDN_1 N_2.... N_7}}\nonumber\\
&&\left(\mathbb{T}_7^{(f)}\right)_{\rm N_1 ...... N_7} = \sqrt{-{\bf g}_{11}} ~{\bf G}^{\rm ABCD}~
\epsilon_{{\rm ABCDN_1 N_2.... N_7}} \nonumber\\ 
&&\left(\mathbb{T}_8^{(k)}\right)_{{\rm N_1 ...... N_7}{\rm N}_k} = {\bf G}_{[{\rm N_1 N_2 N_3 N_4}} 
{\bf G}_{{\rm N_5 N_6 N_7}{\rm N}_k]} \nonumber\\
&&\left(\mathbb{X}_8^{(k)}\right)_{{\rm N_1 ...... N_7}{\rm N}_k} = 
{1\over 3\cdot 2^9 \cdot \pi^4} \left({\rm tr} ~\mathbb{R}_{\rm tot}^4 
- {1\over 4} \left({\rm tr}~\mathbb{R}_{\rm tot}^2\right)^2\right)_{{\rm N_1 N_2.....N_7}{\rm N}_k}, \nd
where we have used the warped eleven-dimensional metric and the raising or lowering of the indices are performed using the same warped metric. The subscript $k$ in ${\rm N}_k$ is defined as follows. When 
$k \equiv (m, \alpha) \in {\cal M}_4 \times {\cal M}_2$, the partial derivatives act along the sub-manifold 
${\cal M}_4 \times {\cal M}_2$; when $k \equiv (i, j) \in {\bf R}^2$, the derivative action act along the spatial directions ${\bf R}^2$; and when $k \equiv (a, b) \in {\mathbb{T}^2\over {\cal G}}$ the partial derivatives act along the toroidal sub-manifold ${\mathbb{T}^2\over {\cal G}}$. The quantum piece $\mathbb{Y}_4$, that appeared in \eqref{elenag}, is defined in \cite{desitter2, heliudson} and can be extracted from \eqref{fahingsha4} as mentioned earlier. The remaining tensors in \eqref{mcacrisis}, including the ones with the over-bracket structure, are defined as follows.

The generic form for $\mathbb{X}_8$ appearing in \eqref{mcacrisis} differs from
\eqref{xeight} by the presence of $\mathbb{R}_{\rm tot}$. It turns out, there are {\it three} possible ways to define $\mathbb{R}_{\rm tot}$ depending on whether the metric components depend only on the $(x, y)$ coordinates, or generically on all the eleven-dimensions, {\it i.e.}
${\bf g}_{\rm CD} \equiv {\bf g}_{\rm CD}(x, y)$ or ${\bf g}_{\rm CD} \equiv {\bf g}_{\rm CD}(x, y, w^a)$
respectively; or take the special eleven-dimensional dependence as in 
\eqref{makibhalu3}. For the last case, {\it i.e.} the metric with $\gamma$ dependence as in 
\eqref{makibhalu3}
it will be advisable at this stage to rewrite 
\eqref{elfadogg2} in the following suggestive way:

{\footnotesize
\bg\label{elfadogg22}
\mathbb{R}(x, y, w^a; g_s) &=& \sum_{l = 0}^\infty {\bf R}^{(l)}_{[{\rm MN}]}(x, y, w^a)\left({g_s\over {\rm HH}_o}\right)^{{l\over 3} + {\rm dom}\left({\gamma\over 3} -2, 0\right)}~ dy^{\rm M} \wedge dy^{\rm N}\\
& + & \sum_{l = 0}^\infty
\left[{\bf R}^{(l)}_{[ab]}\left({g_s\over {\rm HH}_o}\right)^{{l\over 3} + {\rm dom}\left({\gamma\over 3}, 2\right)} ~dw^{a} \wedge dw^{b} 
+ {\bf R}^{(l)}_{[{\rm M}a]}\left({g_s\over {\rm HH}_o}\right)^{{l\over 3} + {\rm dom}\left({\gamma\over 3}-1, 1\right)}~dy^{\rm M} \wedge 
dw^{a}\right], \nonumber \nd}
where we expect $\gamma \ge 5$ for consistency, as described above. We are however no longer restricted  to only the internal three components any more, as the flux EOMs \eqref{mcacrisis} 
 will require us to compute all possible curvature two-forms allowed in our case. The $g_s$ scalings of all the other allowed two-forms are given in {\bf Tables \ref{fchumbon5}}, 
{\bf \ref{fchumbon6}}, {\bf \ref{fchumbon7}}, {\bf \ref{fchumbon8}}, {\bf \ref{fchumbon9}} and 
{\bf \ref{fchumbon10}}.  Looking at the first columns in {\bf Tables \ref{fchumbon5}} to 
{\bf \ref{fchumbon10}}, and taking the conservative choice of $\gamma \ge 5$, we find the following additional curvature two-form:

{\footnotesize
\bg\label{elfadogg5}
\mathbb{R}^{(e)}(x, y, w^a; g_s) &=&  \sum_{l = 0}^\infty
\left[{\bf R}^{(l)}_{[{\rm M}i]}~ dy^{\rm M} \wedge dx^{i} 
+ {\bf R}^{(l)}_{[{\rm M}0]}~dy^{\rm M} \wedge 
dx^{0}\right] \left({g_s\over {\rm HH}_o}\right)^{l-3\over 3}\\
& + & \sum_{l = 0}^\infty
\left[{\bf R}^{(l)}_{[ai]}~ dw^{a} \wedge dx^{i} 
+ {\bf R}^{(l)}_{[a0]}~dw^{a} \wedge 
dx^{0}\right]\left({g_s\over {\rm HH}_o}\right)^{{l\over 3} + {\rm dom}\left({\gamma\over 3} -2, 0\right)}\nonumber\\ 
&+& \sum_{l = 0}^\infty
\left[{\bf R}^{(l)}_{[ij]}\left({g_s\over {\rm HH}_o}\right)^{{l\over 3} + {\rm dom}\left({\gamma\over 3} -4, -2\right)} dx^{i} \wedge dx^{j} 
+ {\bf R}^{(l)}_{[i0]}\left({g_s\over {\rm HH}_o}\right)^{l - 6\over 3}~dx^{i} \wedge 
dx^{0}\right], \nonumber \nd}
where as before, we show the dominant scalings up front, and rest of the sub-dominant ones follow when we sum over $l$. For $\gamma = 5$, the fifth term scales as $\left({g_s\over {\rm HH}_o}\right)^{l - 7\over 3}$
whereas the third and the fourth terms scale as $\left({g_s\over {\rm HH}_o}\right)^{l -1\over 3}$. We can now combine the two set of curvature two-forms $\mathbb{R}$ and $\mathbb{R}^{(e)}$ and define 
$\mathbb{R}_{\rm tot}$ as:
\bg\label{candace} 
\mathbb{R}_{\rm tot} \equiv \mathbb{R} + \mathbb{R}^{(e)}, \nd
which is the one that appears in \eqref{olivecosta}. In a similar vein we can define $\mathbb{R}_{\rm tot}$
for the case when the metric has only $(x, y)$ dependences by replacing \eqref{elfadogg3} by:

{\footnotesize
\bg\label{elfadogg34}
\mathbb{R}_{\rm tot}(x, y; g_s) &=& \sum_{l = 0}^\infty {\bf R}^{(l)}_{[{\rm MN}]}(x, y)\left({g_s\over {\rm HH}_o}\right)^{l\over 3} dy^{\rm M} \wedge dy^{\rm N}\\
& + & \sum_{l = 0}^\infty
\left[{\bf R}^{(l)}_{[ai]}(x, y)~dw^{a} \wedge dx^{i} 
+ {\bf R}^{(l)}_{[a0]}(x, y)~dw^{a} \wedge 
dx^{0}\right]\nonumber\\ 
& + & \sum_{l = 0}^\infty
\left[{\bf R}^{(l)}_{[ij]}(x, y)~dx^{i} \wedge dx^{j} 
+ {\bf R}^{(l)}_{[0i]}(x, y)~dx^{0} \wedge 
dx^{i}\right] \left({g_s\over {\rm HH}_o}\right)^{l- 6\over 3}\nonumber\\
& + & \sum_{l = 0}^\infty
\left[{\bf R}^{(l)}_{[{\rm M}i]}(x, y)dy^{\rm M} \wedge dx^{i} 
+ {\bf R}^{(l)}_{[{\rm M}0]}(x, y)~dy^{\rm M} \wedge 
dx^{0}\right] \left({g_s\over {\rm HH}_o}\right)^{l- 3\over 3}\nonumber\\
& + & \sum_{l = 0}^\infty
\left[{\bf R}^{(l)}_{[ab]}(x, y)\left({g_s\over {\rm HH}_o}\right)^{l+ 6\over 3} dw^{a} \wedge dw^{b} 
+ {\bf R}^{(l)}_{[{\rm M}a]}(x, y)\left({g_s\over {\rm HH}_o}\right)^{l+ 3\over 3}~dy^{\rm M} \wedge 
dw^{a}\right], \nonumber \nd}
where other than ${\bf R}_{[{\rm MN}]}$, we have two additional components, ${\bf R}_{[ai]}$ and 
${\bf R}_{[a0]}$, that remain $g_s$ independent. On the other hand, if we demand a generic dependence on the eleven-dimensions, {\it i.e.} not the special case \eqref{makibhalu3}, the total curvature form becomes:

{\footnotesize
\bg\label{elfadogg35}
\mathbb{R}_{\rm tot}(x, y, w^a; g_s) &=& \sum_{l = 0}^\infty {\bf R}^{(l)}_{[{\rm MN}]}(x, y, w^a)\left({g_s\over {\rm HH}_o}\right)^{l - 6\over 3} dy^{\rm M} \wedge dy^{\rm N}\\
& + & \sum_{l = 0}^\infty
\left[{\bf R}^{(l)}_{[ij]}(x, y, w^a)~dx^{i} \wedge dx^{j} 
+ {\bf R}^{(l)}_{[0i]}(x, y, w^a)~dx^{0} \wedge 
dx^{i}\right] \left({g_s\over {\rm HH}_o}\right)^{l- 12\over 3}\nonumber\\
& + & \sum_{l = 0}^\infty
\left[{\bf R}^{(l)}_{[ai]}(x, y, w^a)~dw^{a} \wedge dx^{i} 
+ {\bf R}^{(l)}_{[a0]}(x, y, w^a)~dw^{a} \wedge 
dx^{0}\right]\left({g_s\over {\rm HH}_o}\right)^{l- 6\over 3}\nonumber\\ 
& + & \sum_{l = 0}^\infty
\left[{\bf R}^{(l)}_{[{\rm M}i]}(x, y, w^a)dy^{\rm M} \wedge dx^{i} 
+ {\bf R}^{(l)}_{[{\rm M}0]}(x, y, w^a)~dy^{\rm M} \wedge 
dx^{0}\right] \left({g_s\over {\rm HH}_o}\right)^{l- 9\over 3}\nonumber\\
& + & \sum_{l = 0}^\infty
\left[{\bf R}^{(l)}_{[ab]}(x, y, w^a)~dw^{a} \wedge dw^{b} 
+ {\bf R}^{(l)}_{[{\rm M}a]}(x, y, w^a)\left({g_s\over {\rm HH}_o}\right)^{l - 1\over 3}~dy^{\rm M} \wedge 
dw^{a}\right], \nonumber \nd}
where only ${\bf R}_{[ab]}$ remains $g_s$ independent. The above expression is very different from 
\eqref{elfadogg22} and \eqref{elfadogg5}. Interestingly however, \eqref{elfadogg22} and \eqref{elfadogg5}
become exactly equivalent to \eqref{elfadogg34} when ${\Su \gamma = 6}$. This may be a bit surprising because 
\eqref{elfadogg34} is only for the case when the dependence extend to space-time directions {\it i.e.} 
\eqref{evader} whereas \eqref{elfadogg22} and \eqref{elfadogg5} is for the special case \eqref{makibhalu3} where the dependence also extends to incorporate $w^a$ directions. When $\gamma = 5$, 
\eqref{elfadogg22} and \eqref{elfadogg5} reduce to:

 {\footnotesize
\bg\label{elfadogg36}
\mathbb{R}_{\rm tot}(x, y, w^a; g_s) &=& \sum_{l = 0}^\infty {\bf R}^{(l)}_{[{\rm MN}]}(x, y, w^a)\left({g_s\over {\rm HH}_o}\right)^{l - 1 \over 3} dy^{\rm M} \wedge dy^{\rm N}\\
& + & \sum_{l = 0}^\infty
\left[{\bf R}^{(l)}_{[ai]}(x, y, w^a)~dw^{a} \wedge dx^{i} 
+ {\bf R}^{(l)}_{[a0]}(x, y)~dw^{a} \wedge 
dx^{0}\right]\left({g_s\over {\rm HH}_o}\right)^{l- 1 \over 3}\nonumber\\ 
& + & \sum_{l = 0}^\infty
\left[{\bf R}^{(l)}_{[{\rm M}i]}(x, y, w^a)dy^{\rm M} \wedge dx^{i} 
+ {\bf R}^{(l)}_{[{\rm M}0]}(x, y)~dy^{\rm M} \wedge 
dx^{0}\right] \left({g_s\over {\rm HH}_o}\right)^{l- 3\over 3}\nonumber\\
& + & \sum_{l = 0}^\infty
\left[{\bf R}^{(l)}_{[ab]}(x, y)\left({g_s\over {\rm HH}_o}\right)^{l+ 5\over 3} dw^{a} \wedge dw^{b} 
+ {\bf R}^{(l)}_{[{\rm M}a]}(x, y)\left({g_s\over {\rm HH}_o}\right)^{l+ 2\over 3}~dy^{\rm M} \wedge 
dw^{a}\right] \nonumber\\
& + & \sum_{l = 0}^\infty
\left[{\bf R}^{(l)}_{[ij]}(x, y, w^a) \left({g_s\over {\rm HH}_o}\right)^{l- 7\over 3}dx^{i} \wedge dx^{j} 
+ {\bf R}^{(l)}_{[0i]}(x, y, w^a)\left({g_s\over {\rm HH}_o}\right)^{l- 6\over 3}dx^{0} \wedge 
dx^{i}\right], \nonumber \nd}
where none of the terms are $g_s$ independent. The above collects all possible contributions to 
$\mathbb{R}_{\rm tot}$ with metric choices \eqref{evader}, \eqref{makibhalu3} and with generic dependences on the toroidal directions. It is now time to elaborate on the last chain from the flux EOM 
\eqref{mcacrisis}, {\it i.e.} the non-perturbative contributions. 

Such a contribution may be extracted from the non-perturbative action \eqref{brazmukh} and 
\eqref{brazmukh2}, which includes the contributions from the non-local counter-terms, or directly from 
\eqref{brazmukh7}, which includes a more standard non-perturbative interactions. We will dwell with the former, {\it i.e.} \eqref{brazmukh}, as it is easy to go to the simpler case \eqref{brazmukh7}. The non-perturbative contribution that we are looking at may be succinctly presented as:

{\footnotesize
\bg\label{olivecandace}
{\delta {\bf S}_{\rm np}^{(q)} \over 
\delta {\bf C}_{\rm ABC}(z', x')} &= & {\rm M}_p^{11} \int d^3x d^8 z \sqrt{-{\bf g}_{11}(x, z)} 
~\sum_k c_k~{\delta \over 
\delta {\bf C}_{\rm ABC}(z', x')} \Big[{\rm exp}
\left(-k \big\vert\mathbb{G}^{(q)}(x, z)\big\vert\right) - 1\Big]\\
& = & -{\rm M}_p^{8q - \sigma(\{l_i\}, n_i)}\sum_k k~c_k\prod_{r = 2}^{q-1}\int d^8 z~ d^8 z_q ~d^8 z_r \sqrt{-{\bf g}_{11}(x', z)} 
 ~\mathbb{F}^{(q)}(z - z_q)~ \mathbb{F}^{(r)}(z_{r+1} - z_r)\nonumber\\
&\times&  
{\partial \over \partial W^{\rm N}}\left[\left(\mathbb{Y}_4\right)^{\rm ABCN}(z', x') 
\sqrt{{\bf g}_8(z_q, x'){\bf g}_8(z_r, x') {\bf g}_8(z', x')} \mathbb{F}^{(1)}(z_2 - z')\right]
{\rm exp}
\left(-k \big\vert\mathbb{G}^{(q)}(x', z)\big\vert\right), \nonumber \nd}
where the overall minus sign is from the modulus structure of \eqref{brazmukh} and $W \equiv (z', x')$ with 
$z = (y, w^a)$. The above result is more general than the one presented in \cite{heliudson} as \eqref{olivecandace} works for
any order $q$ in non-localities. The RHS of \eqref{olivecandace} is a function of $(z', x')$ and its 
tensorial structure is of the form:
\bg\label{satte}
{\overbracket[1pt][7pt]{\mathbb{S} ~\partial_{\rm N} \mathbb{T}}}{}^{\rm ABCN} \equiv
\int d^8 z_2~ \mathbb{S}(x', z_2)~ {\partial \over \partial W^{{\rm N}}} \mathbb{T}^{\rm ABCN}(z', x', z_2), \nd
depending on how the partial derivatives act on the quantum terms $\mathbb{Y}_4$. In 
\eqref{satte}
both $\mathbb{S}(x')$ and $\mathbb{T}^{\rm ABCN}(z, x')$ may be easily read up from 
\eqref{olivecandace}, and we can see that the non-perturbative contributions for the 
${\bf C}_{012}$ EOM is a total derivative with respect to $z'$ (although it does depend on $x'$). 
The total derivative structure means that, integrating over the eight-manifold ${\cal M}_4 \times {\cal M}_2 
\times {\mathbb{T}^2\over {\cal G}}$, the contribution vanishes, thus keeping the anomaly equation 
\eqref{lindmonaco} unchanged. Locally however it does contribute, as one might have expected. 

There is one subtlety that we have not discussed which would connect \eqref{olivecandace} with 
\eqref{satte}. As it stands \eqref{satte} is true only when ${\bf g}_8(z, x) = {\bf g}_8^{(1)}(z) {\bf g}_8^{(2)}(x)$. For our case, and looking at footnote \ref{camil}, this appears to be true although in general this may not be. 
Assuming the case where the determinant of the eight-dimensional metric can be split in the aforementioned way, 
the rank zero tensor $\mathbb{T}_0$ in \eqref{mcacrisis} can now be identified with 
${\mathbb{S}(x', z_2)\over \sqrt{{\bf g}^{(1)}_8(z_2)}}$ in \eqref{satte}. Note that by construction it can depend on the space-time as well as the internal directions and when the metric and the flux components are functions of the internal coordinates only $-$ an example would be the oft-studied 
{\it flat slicing} of de Sitter $-$ $\mathbb{T}_0$ is just a function of the internal coordinates. The rank seven tensor 
$\mathbb{T}_7^{np}$ tensor on the other hand is slightly different from $\mathbb{T}^{\rm ABCN}$ in 
\eqref{satte}; and we can express both $\mathbb{T}_7$ and $\mathbb{T}_0$ from \eqref{mcacrisis}, using \eqref{olivecandace}, in the following way:

{\footnotesize
\bg\label{malinih}
&& {\overbracket[1pt][7pt]{\mathbb{T}_0 ~\partial \mathbb{T}}}{}^{(np)}_7 \equiv \int d^8z_2 \sqrt{{\bf g}^{(1)}_8(z_2)}~
\mathbb{T}_0(x', z_2) \partial_{[{\rm N}_k} \left(\mathbb{T}_7^{np}\right)_{{\rm N_1.....N_7}]}(x', z', z_2)\\
&& \mathbb{T}_0(x', z_2) = \prod_{r = 3}^{q-1}\int d^8 z~ d^8 z_q ~d^8 z_r \sqrt{-{\bf g}_{11}(x', z)
~{\bf g}^{(1)}_8(z_q)~{\bf g}^{(1)}_8(z_r)} \nonumber\\
&& ~~~~~~~~~~~~~~~~~ \times \sum_k k~c_k 
 ~\mathbb{F}^{(q)}(z - z_q)~ \mathbb{F}^{(r)}(z_{r+1} - z_r)~
\mathbb{F}^{(2)}(z_3 - z_2)~{\rm exp}
\left(-k \big\vert\mathbb{G}^{(q)}(x', z)\big\vert\right)\nonumber\\
&&\left(\mathbb{T}_7^{np}\right)_{\rm N_1.....N_7}(x', z', z_2) \equiv 
\left(\mathbb{Y}_4\right)^{\rm ABCD}(z', x') 
{\bf g}^{(2)q/2}_8(x')~{\bf g}^{(1)1/2}_8(z')~ \mathbb{F}^{(1)}(z_2 - z')~
\epsilon_{{\rm A..DN_1...N_7}}, \nonumber \nd}
where the anti-symmetry is put in for completeness. Note the appearance of the first non-locality function
$\mathbb{F}^{(1)}(z_2 - z')$ in the definitions of the tensors. The {\it contraction} of the zero and seven rank tensors is defined via this non-locality function, although the rank zero tensor has other non-locality functions with nested integral structure.

\begin{table}[tb]  
 \begin{center}
\renewcommand{\arraystretch}{1.5}
\begin{tabular}{|c||c||c||c|}\hline Contributions to ${\bf R}^{a_0b_0}_{[ai]}$  & Using
\eqref{makibhalu3} & Using generic ${\mathbb{T}^2\over {\cal G}}$ dep & Using
\eqref{evader} \\ \hline\hline
${\bf R}_{ai{\rm MN}}$ &  ${\gamma k\over 3}$ & $~~0$ & ..... \\ \hline
${\bf R}_{ai{\rm M}j}$ &  ${\gamma k\over 3} -1$ & $-1$ & ..... \\ \hline
${\bf R}_{aijk}$ &  ${\gamma k\over 3}$ & $~~0$ & ..... \\ \hline
${\bf R}_{aij0}$ &  ${\gamma k\over 3} -1$ & $-1$ & ..... \\ \hline
${\bf R}_{aibc}$ &  ${\gamma k\over 3} $ & $~~0$ & ..... \\ \hline
${\bf R}_{aibj}$ &  ${\rm dom}\left({\gamma k\over 3} - 2, 0\right)$ & $-2$ & $~~0$ \\ \hline
${\bf R}_{aib0}$ &  $1$ & $~~1$ & $~~1$ \\ \hline
${\bf R}_{ai{\rm M}b}$ &  $1$ & $~~1$ & $~~1$ \\ \hline
\end{tabular}
\renewcommand{\arraystretch}{1}
\end{center}
 \caption[]{Comparing the $g_s$ scalings of the two-form ${\bf R}^{a_0b_0}_{ai} dw^a \wedge dx^i$ from the metric choices \eqref{makibhalu3}, generic toroidal fibre dependence, and \eqref{evader}. 
The contributions follow similar structure as in {\bf Table \ref{fchumbon7}} although the last column has 
only few contributions. In fact most contributions appear when we impose dependence on the toroidal
directions.}
  \label{fchumbon9}
 \end{table}

To see the contributions of the standard non-perturbative effects we can either use the non-perturbative action \eqref{brazmukh7} in \eqref{olivecandace}, or use the limiting case from \eqref{hana12} in a way described in footnote \ref{kibaloryan}. If we follow the latter procedure $-$ and  
since the integral variable in \eqref{olivecandace} is $z_2$ $-$ we have to first 
use $\mathbb{F}^{(1)}(z_2 - z') = 
\mathbb{F}^{(1)}(z' - z_2)$ to replace the non-locality function there and then apply the limiting condition from \eqref{hana12}. In either case, {\it i.e.} using \eqref{brazmukh7} or \eqref{hana12} in 
\eqref{olivecandace}, the result is:
\bg\label{jesbiel}
{\overbracket[1pt][7pt]{\mathbb{T}_0 ~\partial \mathbb{T}}}{}^{(np)}_7 &\equiv & 
-\sum_k k c_k \int d^8z \sqrt{-{\bf g}_{11}(z, x')}~{\rm exp}\left(-k \big\vert\mathbb{G}^{(q)}(x', z)\big\vert\right)
\nonumber\\
&\times &
\partial_{{\rm N}_k}\left(\sqrt{{\bf g}_d(x', z')}~\left(\mathbb{Y}_4\right)^{\rm ABCD}(x', z')\right)
\epsilon_{\rm ABCD
N_1N_2......N_7}, \nd
for a $d$ dimensional instanton gas,
where $\mathbb{G}^{(q)}(x', z)$ is defined as in \eqref{brazmukh7} (see also footnote \ref{kibaloryan}).
Note that, since $z$ is integrated over, the terms outside the total derivative become functions of $x'$ only whereas the terms inside the derivative are functions of $(x', z')$. This means, when 
${\rm N}_k \in {\cal M}_4 \times {\cal M}_2 \times {\mathbb{T}^2\over {\cal G}}$, the RHS of 
\eqref{jesbiel} is a {\it total derivative}.  Since this situation is specific for the EOM for the three-form 
${\bf C}_{012}$, \eqref{jesbiel} doesn't contribute {\it globally}, and therefore cannot change the anomaly cancellation condition \eqref{lindmonaco}, as anticipated earlier. After the dust settles, the $g_s$ scaling of the non-local non-perturbative contributions to the flux EOMs \eqref{malinih} becomes:
\bg\label{jesbielbon}
\left({g_s\over {\rm HH}_o}\right)^{\theta_{nl} - l_{\rm AB}^{\rm CD} -{14\over 3} -{2q\over 3} - {2k_g\over 3}}
{\rm exp}\left[-k\left({g_s\over {\rm HH}_o}\right)^{\theta_{nl} - {2q\over 3}}\right], \nd
for the generic case \eqref{olivecandace} and we have used the G-flux scalings with $l_{\rm AB}^{\rm CD}$ and $k_g$
as given in \eqref{theritual}. If we restrict the integrals in \eqref{olivecandace} to ${\cal M}_4 \times {\cal M}_2$, which would be related to the localized BBS \cite{BBS} instantons, ${2q\over 3}$ appearing in \eqref{jesbielbon}
should be replaced by $2q$, so that the exponential suppression becomes 
${\rm exp}\left[-k\left({g_s\over {\rm HH}_o}\right)^{\theta_{nl}-2q}\right]$.  For the KKLT instantons 
\cite{KKLT}, the ${2q\over 3}$ part in \eqref{jesbielbon} gets cancelled away \cite{coherbeta}. 
The coefficients of the exponential pieces, for the ${\bf C}_{012}$ EOMs then scale as: 
$\left({g_s\over {\rm HH}_o}\right)^{\theta_{nl} -{2\over 3}(3q+1)}$  and 
$\left({g_s\over {\rm HH}_o}\right)^{\theta_{nl} -{2\over 3}}$ respectively for $k_g = 0$. From here we see that, to order $\left({g_s\over {\rm HH}_o}\right)^{s\over 3}$ for $s \in \mathbb{Z}$, the non-local quantum terms that contribute to the flux EOMs \eqref{mcacrisis} are classified by:
\bg\label{natferrer}
\theta_{nl} = l_{\rm AB}^{\rm CD} + {14\over 3} + {2 \over 3}(s + k_g) 
- \left({2d_2\over 3} - {d_1\over 3}\right)q, \nd
where $\theta_{nl}$ is the one in \eqref{botsuga3} as we are taking the eleven-dimensional dependence 
with $\gamma \ge 5$ from \eqref{makibhalu3}; and 
$(d_1, d_2) = (6, 0)$ for the localized BBS \cite{BBS} instantons with $(d_1, d_2) = (4, 2)$ for the localized KKLT \cite{KKLT} instantons. Higher values of $(k_g, q)$ for a fixed value of $s$ are 
{\it suppressed} by the corresponding powers of ${\rm M}_p$ which in fact produces the necessary hierarchy in the system.
On the other hand, the $g_s$ scalings for the standard localized BBS and KKLT instantons from \eqref{jesbiel} go as:

\begin{table}[tb]  
 \begin{center}
\renewcommand{\arraystretch}{1.5}
\begin{tabular}{|c||c||c||c|}\hline Contributions to ${\bf R}^{a_0b_0}_{[a0]}$  & Using
\eqref{makibhalu3} & Using generic ${\mathbb{T}^2\over {\cal G}}$ dep & Using
\eqref{evader} \\ \hline\hline
${\bf R}_{a0{\rm MN}}$ &  ${\gamma k\over 3} - 1$ & $-1$ & ..... \\ \hline
${\bf R}_{a0{\rm M}0}$ &  ${\gamma k\over 3} -1$ & $-1$ & ..... \\ \hline
${\bf R}_{a0ij}$ &  ${\gamma k\over 3} -1$ & $-1$ & ..... \\ \hline
${\bf R}_{a0i0}$ &  ${\gamma k\over 3}$ & $~~0$ & ..... \\ \hline
${\bf R}_{a0bc}$ &  ${\gamma k\over 3} -1 $ & $-1$ & ..... \\ \hline
${\bf R}_{a0b0}$ &  ${\rm dom}\left({\gamma k\over 3} - 2, 0\right)$ & $-2$ & $~~0$ \\ \hline
${\bf R}_{a0bi}$ &  $1$ & $~~1$ & $~~1$ \\ \hline
${\bf R}_{a0{\rm M}b}$ &  $0$ & $~~0$ & $~~0$ \\ \hline
\end{tabular}
\renewcommand{\arraystretch}{1}
\end{center}
 \caption[]{Comparing the $g_s$ scalings of the two-form ${\bf R}^{a_0b_0}_{a0} dw^a \wedge dx^0$ from the metric choices \eqref{makibhalu3}, generic toroidal fibre dependence, and \eqref{evader}. 
 Again two factors play roles here: one, the temporal dependence follow results from {\bf Table \ref{fchumbon8}}; and two, the usage of \eqref{kootdiye}. The importance of the latter has been emphasized 
 before, and more details will be elaborated later.}
  \label{fchumbon10}
 \end{table}

{\footnotesize
\bg\label{beverwickmey}
\left({g_s\over {\rm HH}_o}\right)^{\theta_{nl} - l_{\rm AB}^{\rm CD} -{20\over 3} - {2k_g\over 3}}
{\rm exp}\left[-k\left({g_s\over {\rm HH}_o}\right)^{\theta_{nl} - 2}\right],~~~~
\left({g_s\over {\rm HH}_o}\right)^{\theta_{nl} - l_{\rm AB}^{\rm CD} -{14\over 3} - {2k_g\over 3}}
{\rm exp}\left[-k\left({g_s\over {\rm HH}_o}\right)^{\theta_{nl}}\right], \nonumber\\ \nd}
respectively. This means, for the ${\bf C}_{012}$ EOMs, the coefficients in front of the exponential factors 
in \eqref{beverwickmey} scale as $\left({g_s\over {\rm HH}_o}\right)^{\theta_{nl} - {8\over 3}}$ and 
$\left({g_s\over {\rm HH}_o}\right)^{\theta_{nl} - {2\over 3}}$ respectively for $k_g = 0$. From here, the contributions of the non-perturbative terms to the flux EOMs \eqref{mcacrisis} at order 
$\left({g_s\over {\rm HH}_o}\right)^{s\over 3}$ for $s \in \mathbb{Z}$, are classified by:
\bg\label{natferrer2}
\theta_{nl} = l_{\rm AB}^{\rm CD} + {14\over 3} + {2 \over 3}(s + k_g) 
- {2d_2\over 3} + {d_1\over 3}, \nd
where as before the choice of $(d_1, d_2)$ as $(6, 0)$ and $(4, 2)$ provide the contributions from the localized BBS and KKLT instantons respectively. Again, for a given value of $s \in \mathbb{Z}$, the higher order terms 
coming from \eqref{natferrer2} are hierarchically suppressed by powers of  ${\rm M}_p$.

\subsection{Internal fluxes for metric with toroidal dependence \label{intornal}}

With this we have defined all the tensors appearing in \eqref{mcacrisis}. It is now time to study all the flux EOMs for all the three cases: (a) where the metric is of the form \eqref{evader}, (b) where the metric allows the special toroidal dependence \eqref{makibhalu3}, and (c) where the toroidal fibre dependence is generic.
Let us see how this would work in the present case. Imagine we want to find the EOM for the three-form 
${\bf C}_{0ij}$ where $(i, j) \in {\bf R}^2$. This would require us to find all the seven and the eight form tensors in \eqref{mcacrisis}, including the 
${\bf X}_8$ form oriented along the eight manifold. This is summarized in {\bf Table \ref{jhinuk1}}. Note that the ${\bf X}_8$ form has already been computed in \eqref{thetrial}, \eqref{thetrial3} and \eqref{thetrial2}  
respectively for the aforementioned three cases (we identify ${\bf X}_8^{(m,\alpha)}$ from {\bf Table \ref{jhinuk1}} with the ${\bf X}_8$ polynomials above). Since there are only two possible permutations of indices for the ${\bf X}_8$ form, $f^{(i)}_1(\gamma)$, which are the two possible values for $i = 1, 2$, can be written in the following way:
\bg\label{cora1}
&&f^{(1)}_1(\gamma) = 3~{\rm dom}\left({\gamma\over 3} - 2, 0\right) + {\rm dom}\left({\gamma\over 3}, 2\right)
\nonumber\\ 
&&f^{(2)}_1(\gamma) = 2~{\rm dom}\left({\gamma\over 3} - 2, 0\right) + 2~{\rm dom}\left({\gamma\over 3} -1, 1\right), \nd
where both leads to $+{2\over 3}$ when $\gamma = 5$ with $l \equiv {1\over 3}(l_1 + l_2 + l_3 + l_4)$ 
in {\bf Table \ref{jhinuk1}}. 
As we shall see below, for many of the ${\bf X}_8$ polynomial, this would be the case: there are multiple forms for $f_n^{(i)}(\gamma)$ for the same {\it dominant} value when $\gamma = 5$.
The other eight-form tensor ${\bf \Lambda}_8$ is a localized delta function over the eight-manifold and has no $g_s$ scaling. The $g_s$ scalings of the various rank seven and eight tensors can now be collected from {\bf Table \ref{jhinuk1}} to satisfy\footnote{The set of relations in \eqref{cora2} does not prohibit intermediate cancellations of terms. For example, with $k_1 = k_{n_i} = 0$, we can allow $l_{0i}^{j{\rm M}} = -4$ to {\it balance} with the $b_4$ term in \eqref{mcacrisis} that scales as $\left({g_s\over {\rm HH}_o}\right)^0$. In fact in \cite{desitter2} this identification was used to determine the warp-factor ${\rm H}(y)$ in terms of M2 and $\overline{\rm M2}$ branes. \label{redcut}}:

{\footnotesize
\bg\label{cora2}
l_{0i}^{j{\rm M}} + 4 + {2\over 3}\left(k_1 + k_{n_1} 
+ k_{n_2}\right) = \theta_{nl}(k_2) - l_{0i}^{j{\rm M}} - {14\over 3}  
- {2k_2\over 3} = {2\over 3}\left(k_4 + k_5\right) + l_{[{\rm MN}}^{[{\rm PQ}} \oplus l_{{\rm RS}]}^{ab]}, \nd}
where the sum $\oplus$ symbolizes the possible flux components that would appear in the definition of the rank eight tensor $\mathbb{T}_8^{(1)}$. For example here the rank eight tensor will be represented by 
${\bf G}_{\rm MNPQ} {\bf G}_{{\rm RS}ab}$ and ${\bf G}_{{\rm MNP}a} {\bf G}_{{\rm QRS}b}$ which are captured by the anti-symmetry in {\bf Table \ref{jhinuk1}} and thus represented by 
$l_{\rm MN}^{\rm PQ} + l_{\rm RS}^{ab}$  and $l_{\rm MN}^{{\rm P}a} + l_{\rm QR}^{{\rm S}b}$ respectively. The other parameters appearing in \eqref{cora2} are defined as follows. $k_{n_i}$ are the modings that appear in the definition of ${\rm F}_i \equiv \sum_{k_{n_i}}{\rm F}_i^{(k_{n_i})} \left({g_s\over {\rm HH}_o}\right)^{k_{n_i}}$; 
$(k_1, k_2)$ are the G-flux modings from \eqref{theritual} for the flux components 
${\bf G}_{0ij{\rm M}}$ (we use two different modings to distinguish their appearance as flux components and as in the quantum series $\mathbb{Y}_4^{0ij{\rm M}}$); and $(k_4, k_5)$ are the corresponding scalings for 
${\bf G}_{\rm MNPQ}$ and ${\bf G}_{{\rm RS}ab}$ respectively, or  ${\bf G}_{{\rm MNP}a}$ and 
${\bf G}_{{\rm QRS}b}$ respectively (depending on which set is considered in \eqref{cora2}. In both cases, the relations \eqref{cora2} should be further identified to either $l + 2$, $l - 6$ or $l + f_1^{(1, 2)}(\gamma)$ depending on the metric choice \eqref{evader}, generic eleven-dimensional metric or the special dependence \eqref{makibhalu3} respectively. For $\gamma = 5$ and $l_{0i}^{j{\rm M}} = -4$, the consequences of \eqref{cora2} have already been discussed in \eqref{gmvero}, \eqref{meaador} and \eqref{gmvero2}, wherefrom we conclude that $\left(l_{\rm MN}^{\rm PQ}, l_{\rm RS}^{ab}, l_{\rm MN}^{{\rm P}a}, 
l_{\rm QR}^{{\rm S}b}\right) > (0, 0, 0, 0)$, so at least no time-independent G-flux components related to these scalings are allowed. Schematically therefore, to lowest orders\footnote{By {\it lowest orders} we will always mean lowest orders in 
${g_s\over {\rm HH}_o}$ {and} {\it zeroth} orders in $(k_i, k_{n_i}, l)$ unless mentioned otherwise. Needless to say, {\it higher orders} would mean either higher orders in ${g_s\over {\rm HH}_o}$ or higher orders in $(k_i, k_{n_i}, l)$ or both, depending on how the quantum terms are accommodated in. For 
example in 
\eqref{lilcadu77}, $\theta_2 = {8\over 3}$ captures zeroth orders in $(k_i, k_{n_i}, l)$ and second orders in ${g_s\over {\rm HH}_o}$, which happens to be one of the two lowest orders here. We can go to higher orders in multiple ways: second orders in 
${g_s\over {\rm HH}_o}$ but higher orders in $(k_i, k_{n_i}, l)$, or zeroth orders in $(k_i, k_{n_i}, l)$ but higher orders in ${g_s\over {\rm HH}_o}$, or a mixture of both. In general, a relation like \eqref{cora2} for each of the following cases will be the guiding principle when we want to go to higher orders. 
To avoid these complications we will only study the lowest orders cases here. \label{honikjul3}} and taking the metric \eqref{evader} the balancing of various terms in the EOM \eqref{mcacrisis} happens in the following way:
\bg\label{lilcadu77} 
d_{({\rm M})}\ast\overbracket[1pt][7pt]{{\bf G}_4 + b_1 d_{({\rm M})}\ast\Big(\mathbb{Y}_4}(\overbracket[1pt][7pt]{\theta_1) + \mathbb{Y}_4\underbracket[1pt][7pt]{(\theta_2) + ..\Big) - b_2\Big(\mathbb{T}_8}\underbracket[1pt][7pt]{{}^{(m, \alpha)}\Big) - b_3 \Big( \mathbb{X}_8}_{\theta_2 \ge {8\over 3}}{}^{(m,\alpha)}\Big) - b_4{\bf \Lambda}}^{\theta_1 \ge {2\over 3}}{}_8 = 0, \nonumber\\ \nd
where the over-bracket and under-bracket are used to identify various parts of the EOMs that scale in the same way with respect to ${g_s\over {\rm HH}_o}$; ${\rm M} \equiv (m, \alpha) \in {\cal M}_4 \times {\cal M}_2$; and $\theta_i$ are the scaling of the quantum terms as in \eqref{botsuga3} such that $\theta_1 \ge {2\over 3}$ and $\theta_2 \ge {8\over 3}$. The zeroth order identification to $b_4$ term is what fixes the warp-factor \cite{desitter2}, and once we go to higher orders, the quantum terms take over. In fact, 
the dotted terms in the quantum series will start manifesting themselves once we go to higher orders in $\left(k_i, k_{n_i}, l, {g_s\over {\rm HH}_o}\right)$  in such a way that eventually {\it all} terms in \eqref{lilcadu77} will be related to each other via \eqref{cora2}\footnote{There is some subtlety that needs elaboration here. As we saw in the schematic diagram \eqref{lilcadu77}, 
when $k_i = k_{n_i} = l = 0$ there can be hierarchies between the ${g_s\over {\rm HH}_o}$ powers at the lowest orders. In fact we are dealing with hierarchies between $\left({g_s\over {\rm HH}_o}\right)^0$ and 
$\left({g_s\over {\rm HH}_o}\right)^{+2}$ at the lowest orders (see {\bf Table \ref{jhinuk1}}). This means the quantum terms appearing for the underbracket case in \eqref{lilcadu77}, {\it i.e.} $\mathbb{Y}_4(\theta_2)$, could in-principle allow $(k_1, k_{n_1}, k_{n_2}) \ge \left({1\over 2}, {1\over 2}, {1\over 2}\right)$ even for $k_2 = k_4 = k_5 = 0$. This is of course where the over-bracket merges in with the under-bracket 
in \eqref{lilcadu77} to fully realize \eqref{cora2}. In the ensuing analysis it should be understood that such realizations of {\it merging} would happen for all the cases to be studied here. However we will not separately discuss them for individual cases and, to avoid making the analysis too complicated, we will stick with the lowest orders where
$k_i = k_{n_i} = l = 0$. Such a simplification has its advantage. For example, it is easy to see from the fact that $\theta_1 \ge {2\over 3}$, the lowest orders quantum terms simply renormalize the 
flux factors. Thus the warp-factor is indeed determined by the distribution of M2 and $\overline{\rm M2}$ branes. See also section 4.2.3 in the first reference of \cite{desitter2}. \label{hemramal}}. 
Note that the non-local non-perturbative terms cannot change the conclusion because they scale as $\left({g_s\over {\rm HH}_o}\right)^p$ where:
\bg\label{Badangti}
p \equiv \theta_{nl}(k_2) - l_{0i}^{j{\rm M}} - {14\over 3} - {2q\over 3} - {2k_2\over 3} - 
k\left({g_s\over {\rm HH}_o}\right)^{\theta_{nl}(k_2) - {2q\over 3}}
{\rm log}^{-1} \left({g_s\over {\rm HH}_o}\right), \nd
which is similar to the rank seven quantum term $\mathbb{T}^{(q)}_7$ appearing in {\bf Table \ref{jhinuk1}}
except for the sub-dominant terms which (a) for large $q$, {\it i.e.} for large non-locality, they decouple and (b) go to zero as ${{\rm x}\over {\rm log}~{\rm x}}$ for ${\rm x} \to 0$ if we identify ${\rm x} \equiv {g_s\over {\rm HH}_o}$. In the above discussion, we have also identified $k_2$ as the $g_s$ scaling of the G-flux components 
${\bf G}_{0ij{\rm M}}$.  Replacing ${2q\over 3}$ by $2$ in \eqref{Badangti} will incorporate the effects of the BBS instantons from \eqref{beverwickmey}. They too cannot change the aforementioned conclusion. In the following we will see if this continues to hold generically.

\begin{table}[tb]  
 \begin{center}
\renewcommand{\arraystretch}{1.5}
\begin{tabular}{|c||c||c|}\hline ${\bf G}_{0ij{\rm M}}$ tensors & form & ${g_s\over {\rm HH}_o}$ scaling \\ \hline\hline
$\left(\mathbb{T}_7^{(f)}\right)_{{\rm NPQRS}ab}$ & $\sqrt{-{\bf g}_{11}} {\bf G}^{0ij{\rm M}} {\rm F}_1^{(n_1)}
{\rm F}_2^{(n_2)} \epsilon_{0ij{\rm MNPQRS}ab}$ & $l_{0i}^{j{\rm M}} + 4 + {2\over 3}\left(k_1 + k_{n_1} 
+ k_{n_2}\right)$ \\ \hline 
$\left(\mathbb{T}_7^{(q)}\right)_{{\rm NPQRS}ab}$ & $\sqrt{-{\bf g}_{11}} \mathbb{Y}_4^{0ij{\rm M}} 
\epsilon_{0ij{\rm MNPQRS}ab}$ & $\theta_{nl}(k_2) - l_{0i}^{j{\rm M}} - {14\over 3}  
- {2k_2\over 3}$ \\ \hline 
$\left(\mathbb{T}_8^{(m, \alpha)}\right)_{{\rm MNPQRS}ab}$ & ${\bf G}_{[{\rm MNPQ}} {\bf G}_{{\rm RS}ab]}$ & ${2\over 3}\left(k_4 + k_5\right) + l_{[{\rm MN}}^{[{\rm PQ}} \oplus l_{{\rm RS}]}^{ab]}$ 
\\ \hline 
$\left(\mathbb{X}_8^{(m, \alpha)}\right)_{{\rm MNPQRS}ab}$ & \eqref{thetrial2}, \eqref{thetrial3}, \eqref{cora1}
& $l + 2, l - 6, l + f^{(i)}_1(\gamma)$ \\ \hline
$\left({\bf \Lambda}_8\right)_{{\rm MNPQRS}ab}$ & $\delta^8(z - z')$
& $0$ \\ \hline
\end{tabular}
\renewcommand{\arraystretch}{1}
\end{center}
 \caption[]{Comparing the $g_s$ scalings of the various tensors that contribute to the EOM for the G-flux components ${\bf G}_{0ij{\rm M}}$ in \eqref{mcacrisis}. The other factors appearing above are defined as follows: 
 $\theta_{nl}(k_2)$ is defined with $k_2$ G-flux modings in \eqref{botsuga} for the case \eqref{thetrial2}, 
 in \eqref{botsuga2} for the case \eqref{thetrial3}; and in \eqref{botsuga3} for the case \eqref{thetrial}. For the G-flux components we use the ans\"atze \eqref{theritual} so $(k_4, k_5)$ represent the flux modings, and 
 $l_{\rm AB}^{\rm CD}$ the inherent scalings. $k_{n_i}$ represents the moding for the internal metric coefficients ${\rm F}_1$ and ${\rm F}_2$ in say \eqref{evader}; $f_1^{(i)}$ are given in \eqref{cora1}; and the superscript $(m, \alpha)$ are defined just after \eqref{olivecosta}.} 
  \label{jhinuk1}
 \end{table}

The next interesting case is for the three form ${\bf C}_{\rm MNP}(x, y, w^a; g_s)$ which would correspond to the EOMs for the G-flux components ${\bf G}_{\rm MNPQ}, {\bf G}_{{\rm MNP}i}$ and 
${\bf G}_{{\rm MNP}a}$. Recall that we are using the gauge condition \eqref{chupke} so components like 
${\bf G}_{0{\rm MNP}}$ do not appear, except ${\bf G}_{0ij{\rm M}}$. The other related component is 
${\bf G}_{0ija}$, which we can define in the same way as ${\bf G}_{0ij{\rm M}}$ but this will make the warp-factor ${\rm H} \equiv {\rm H}(y, w^a)$ leading to a breakdown of EFT as we saw earlier. Thus we will keep 
${\rm H} = {\rm H}(y)$ and make ${\bf G}_{0ija} = 0$. (See discussions in the next sub-section \ref{g0abc}.) The $g_s$ dependence of the ${\bf C}_{\rm MNP}$ field will determine the various $g_s$ dependence of the rank seven and eight tensors in the corresponding G-flux EOMs.  For the present case we will first concentrate on the ${\bf X}_8$ form which will be oriented
along ${\bf R}^{2, 1} \times {\cal C}_3 \times {\mathbb{T}^2\over {\cal G}}$, where ${\cal C}_3$ is a three-cycle in ${\cal M}_4 \times {\cal M}_2$. The latter is possible because both 
${\cal M}_4$ as well as ${\cal M}_2$ are non-K\"ahler manifolds, and therefore allowed to have odd-cycles. 
The eight-form \eqref{xeight} however now should include curvature two-forms \eqref{candace} that are computed away from the eight-manifold. As mentioned in {\bf Table \ref{jhinuk1}}
this eight-form, and not the one in \eqref{xeight}, will become essential to study the EOM for the flux component ${\bf G}_{\rm MNPQ}$. The eight-form now includes the following  indices and their permutations:
\bg\label{lilcadu}
\overbracket[1pt][7pt]{{\rm Q}~\underbracket[1pt][7pt]{{\rm R}~{\rm S}}~a} ~\overbracket[1pt][7pt]{b~\underbracket[1pt][7pt]{0~i} ~j} ~~~ 
+~~~ 34 ~{\rm permutations}, \nd
where it is understood that a pair of alphabets determine the curvature two form indices and the trace over holonomy matrices are taken in the end. When the metric depends on space-time coordinates 
as in \eqref{evader}, the scaling of ${\bf X}_8$ is slightly different from \eqref{thetrial2} and is given by:
\bg\label{thetrial4}
{\bf X}_8(x, y; g_s) = \sum_{\{l_i\}} \widetilde{\bf X}_{(8, 2a)}^{(l_1, ...., l_4)}(x, y) 
\left({g_s\over {\rm HH}_o}\right)^{{1\over 3}(l_1 + l_2 + l_3 + l_4 - 3)}, \nd
which should be identified with the corresponding ${\bf X}_8^{(m, \alpha)}$ in {\bf  Table \ref{jhinuk2}}. Interestingly, and as one would expect, all the 35 permutations of the alphabets in \eqref{lilcadu} leads to the {\it same} scaling of $l - 1$, where $l = {1\over 3}(l_1 + ... + l_4)$. The story repeats, when the metric has generic dependence on $w^a$, but with a different scaling:
\bg\label{thetrial5}
{\bf X}_8(x, y, w^a; g_s) = \sum_{\{l_i\}} \widetilde{\bf X}_{(8, 2b)}^{(l_1, ...., l_4)}(x, y, w^a) 
\left({g_s\over {\rm HH}_o}\right)^{{1\over 3}(l_1 + l_2 + l_3 + l_4 - 27)}, \nd
compared to what we had earlier in \eqref{thetrial3}. Such a large negative value is alarming, because it tells us that for ${\red l_1 + .... + l_4 = 27}$, ${\bf X}_8$ can become time-independent. One might be equally concerned by the negative value for the $g_s$ scaling in \eqref{thetrial4}, but note that the G-flux EOM for 
${\bf G}_{\rm MNPQ}$ includes the product of ${\bf G}_{{\rm RS}ab} {\bf G}_{0ij{\rm M}}$, that involves negative $g_s$ scalings because we expect ${\bf G}_{0ij{\rm M}}$ to scale at least as 
$\left({g_s\over {\rm HH}_o}\right)^{-4}$. This requirement stems from slowly moving membranes as shown in \cite{desitter2}. If we take the back-reactions of the moving membrane, the dominant scaling of $-4$ does not change as was also shown in \cite{desitter2}. Interestingly, and as noticed earlier, all the 35 permutations of the alphabets in \eqref{lilcadu}, scale in exactly the same way as $l - 9$, which is what appears in {\bf Table \ref{jhinuk2}} when we identify ${\bf X}_8^{(m, \alpha)}$ in the table with 
\eqref{thetrial5}. 
Finally, when we take the special toroidal dependence as in \eqref{makibhalu3}, the {\it dominant} $g_s$ scaling becomes:
\bg\label{thetrial6}
{\bf X}_8(x, y, w^a; g_s) = \sum_{\{l_i\}} \widetilde{\bf X}_{(8, 2c)}^{(l_1, ...., l_4)}(x, y, w^a) 
\left({g_s\over {\rm HH}_o}\right)^{{1\over 3}(l_1 + l_2 + l_3 + l_4 - 7)}, \nd
when $\gamma = 5$ in \eqref{makibhalu3}. However not all $g_s$ scalings of the 35 alphabets in 
\eqref{lilcadu} are the same. If we denote the $g_s$ scaling of any pair of alphabets in \eqref{lilcadu} as
$\left({g_s\over {\rm HH}_o}\right)^{-|p|}$, then ${4\over 3} \le p \le {7\over 3}$. Interestingly only {\it one} combination of the curvature forms provide the necessary dominant scaling for $\gamma = 5$. For example:
\bg\label{bbox}
{\rm tr}\left({\bf R}_{[{\rm QR}]}~ {\bf R}_{[{\rm S}a]} ~{\bf R}_{[b0]} ~{\bf R}_{[ij]}\right) ~dy^{\rm R} \wedge ..... \wedge dx^j,
\nd
contributing to ${\rm tr}~ \mathbb{R}^4_{\rm tot}$ would have a $g_s$ scaling of $-{7\over 3}$, where the trace is over the holonomy matrices. In terms of $\gamma$, this implies that there is only one possible value for $f_2(\gamma)$ in {\bf Table \ref{jhinuk2}} given by:
\bg\label{cora3}
f_2(\gamma) = 2~{\rm dom}\left({\gamma\over 3} - 2, 0\right) + {\rm dom}\left({\gamma\over 3} - 1, 1\right)
+ {\rm dom}\left({\gamma\over 3} - 4, -2\right), \nd
that allows the dominant scaling. One can easily check that when $\gamma = 6$, \eqref{cora3} gives us 
$f_2(6) = -1$ leading to the ${\bf X}_8$ scaling of $l - 1$. This is similar to the ${\bf X}_8$ scaling for the metric choice \eqref{evader}. In \eqref{cora1}, we see that $f_1^{(1)}(6) = f_1^{(2)}(6) = 2$ , which also matches with the result from the metric \eqref{evader}. In the following we will see whether this continues to be the case.

\begin{table}[tb]  
 \begin{center}
\renewcommand{\arraystretch}{1.5}
\begin{tabular}{|c||c||c|}\hline ${\bf G}_{{\rm MNPQ}}$ tensors & form & ${g_s\over {\rm HH}_o}$ scaling \\ \hline\hline
$\left(\mathbb{T}_7^{(f)}\right)_{{\rm RS}ab0ij}$ & $\sqrt{-{\bf g}_{11}} {\bf G}^{{\rm MNPQ}} {\rm F}_1^{(n_1)}
{\rm F}_2^{(n_2)} \epsilon_{{\rm MNPQRS}ab0ij}$ & $l_{\rm MN}^{{\rm PQ}} - 2 + {2\over 3}\left(k_1 + k_{n_1} 
+ k_{n_2}\right)$ \\ \hline 
$\left(\mathbb{T}_7^{(q)}\right)_{{\rm RS}ab0ij}$ & $\sqrt{-{\bf g}_{11}} \mathbb{Y}_4^{{\rm MNPQ}} 
\epsilon_{{\rm MNPQRS}ab0ij}$ & $\theta_{nl}(k_2) - l_{\rm MN}^{{\rm PQ}} - {14\over 3}  
- {2k_2\over 3}$ \\ \hline 
$\left(\mathbb{T}_8^{(m, \alpha)}\right)_{{\rm MRS}ab0ij}$ & ${\bf G}_{[{\rm RS}ab} {\bf G}_{0ij{\rm M}]}$ & ${2\over 3}\left(k_4 + k_5\right) + l_{[{\rm RS}}^{[{ab}} \oplus l_{{0i}]}^{j{\rm M}]}$ 
\\ \hline 
$\left(\mathbb{X}_8^{(m, \alpha)}\right)_{{\rm MRS}ab0ij}$ & \eqref{thetrial4}, \eqref{thetrial5}, 
\eqref{cora3}
& $l - 1, l - 9, l + f_2(\gamma)$ \\ \hline
\end{tabular}
\renewcommand{\arraystretch}{1}
\end{center}
 \caption[]{Comparing the $g_s$ scalings of the various tensors that contribute to the EOM for the G-flux components ${\bf G}_{{\rm MNPQ}}$ in \eqref{mcacrisis}. The other factors appearing above are defined as follows: 
 $\theta_{nl}(k_2)$ is defined with $k_2$ G-flux modings in \eqref{botsuga} for the metric choice \eqref{evader}, 
 in \eqref{botsuga2} for the generic dependence of the metric on the toroidal direction 
 ${\mathbb{T}^2\over {\cal G}}$; and in \eqref{botsuga3} for the special case of \eqref{makibhalu3}. For the G-flux components we use the ans\"atze \eqref{theritual} so the flux modings follow this with the inherent scalings given by
 $l_{\rm AB}^{\rm CD}$. Finally the inherent scalings. $k_{n_i}$ represents the moding for the internal metric coefficients ${\rm F}_1$ and ${\rm F}_2$ in say \eqref{evader}; and $f_2(\gamma)$ is given in \eqref{cora3}.} 
  \label{jhinuk2}
 \end{table}

The G-flux EOM for the flux components ${\bf G}_{\rm MNPQ}$ from \eqref{mcacrisis} leads to various constraints as may be extracted from {\bf Table \ref{jhinuk2}}. As in \eqref{cora2}, we expect:

{\footnotesize
\bg\label{cora4}
l_{\rm MN}^{{\rm PQ}} - 2 + {2\over 3}\left(k_1 + k_{n_1} 
+ k_{n_2}\right) = \theta_{nl}(k_2) - l_{\rm MN}^{{\rm PQ}} - {14\over 3}  
- {2k_2\over 3} = {2\over 3}\left(k_4 + k_5\right) + l_{[{\rm RS}}^{[{ab}} \oplus l_{{0i}]}^{j{\rm M}]}, \nd}
where most of the parameters appearing in \eqref{cora4} are defined below \eqref{cora2}. The only difference is that the $k_i$ modings now correspond to the flux components appearing in 
{\bf Table \ref{jhinuk2}}. The various relations in \eqref{cora4} are not complete till we identify them with 
either $l - 1$ for the metric choice \eqref{evader}, $l - 9$ for the metric with generic dependence on the toroidal directions; and $l + f_2(\gamma)$ for the special choice \eqref{makibhalu3}. Whichever identifications we make, we should remember that the raising and lowering of the flux or the metric indices have to be performed with the corresponding metric choice.  Failure to do so will lead to the wrong EOMs. 

Let us see how various terms in \eqref{cora4} may be balanced. As pointed out in footnote \ref{redcut}, we can allow intermediate cancellations of the terms in \eqref{mcacrisis}. For example if we take 
$l_{\rm MN}^{\rm PQ} = 1$ from \eqref{cora2}, we see that it can be balanced, 
by the ${\bf X}_8$ polynomial \eqref{thetrial4} for $l = 0$ with the metric choice \eqref{evader}. This means, to lowest orders, and for the same metric choice \eqref{evader}, we can allow:
\bg\label{vanand1}
\theta_{nl} - l_{\rm MN}^{{\rm PQ}} - {14\over 3} = l_{{\rm RS}}^{{ab}} +  l_{{0i}}^{j{\rm M}}, \nd
 giving us $\theta_{nl} = {8\over 3}$ from \eqref{botsuga} for $l_{{\rm RS}}^{{ab}} = 1$ and   
 $l_{{0i}}^{j{\rm M}} = -4$. As in \eqref{lilcadu77}, the $g_s$ scalings of the other terms  in \eqref{mcacrisis} may now be balanced in the following schematic way:
 \bg\label{lilcadu78}
d_{({\rm M})}\ast\overbracket[1pt][7pt]{{\bf G}_4 + b_1 d_{({\rm M})}\ast\Big(\mathbb{Y}_4}(\overbracket[1pt][7pt]{\theta_1) + \mathbb{Y}_4\underbracket[1pt][7pt]{(\theta_2) + ..\Big) - b_2\Big(\mathbb{T}_8}_{\theta_2 \ge {8\over 3}}{}^{(m, \alpha)}\Big) - b_3 \Big( \mathbb{X}_8}^{\theta_1 \ge {14\over 3}}{}^{(m,\alpha)}\Big) = 0, \nonumber\\ \nd 
where $(\theta_1, \theta_2)  \ge \left({14\over 3}, {8\over 3}\right)$ from \eqref{botsuga3}; and again the dotted terms start manifesting themselves as we go to higher orders in 
$\left(k_i, k_{n_i}, l, {g_s\over {\rm HH}_o}\right)$, eventually reproducing \eqref{cora4}.
When ${\bf C}_{\rm MNP}$ also has dependence on the toroidal direction, the balancing of the first term in {\bf Table \ref{jhinuk2}} with ${\bf X}_8$ polynomial has to involve $f_2(\gamma)$. When $\gamma = 5$, we see that this balance becomes hard because $f_2(5) = -{7\over 3}$. However when $\gamma = 6$, $f_2(6) = -1$, and the balance works perfectly. Does this mean that $\gamma = 6$ is the right value chosen by the background? To justify this we will need more data, and in the following we will analyze other flux EOMs and see what value of $\gamma$ allows a consistent background.

Before moving ahead, let us point out a subtlety with $\theta_2 = {8\over 3}$ in the under-bracket choice of \eqref{lilcadu78}. Since the contribution to $\theta_2$ comes exclusively from 
${7\over 3}\left(l_{61} + l_{62} + l_{64}\right)$ in \eqref{botsuga3}, the choice of $\theta_2 = {8\over 3}$ puts a tighter constraint on the quantum terms, and to lower orders there may not be any relevant candidate if we only look at {\it local} operators. In this case we expect 
the sum of the product of the fluxes to vanish, at least to lowest orders. Since ${\bf G}_{0ij{\rm M}}$ involves derivative on the warp-factor ${\rm H}(y)$, the algebraic constraint will involve a non-trivial differential equation for the warp-factor. Once we incorporate the effects of non-perturbative instantons from 
\eqref{jesbiel}, especially the effects of the BBS \cite{BBS} instantons from \eqref{beverwickmey}, then 
\eqref{vanand1} changes to:
\bg\label{vanand2}
\theta_{nl} - l_{\rm MN}^{{\rm PQ}} - {20\over 3} = l_{{\rm RS}}^{{ab}} +  l_{{0i}}^{j{\rm M}}, \nd
giving us $\theta_{nl} \equiv \theta_2 = {14\over 3}$. This provides a consistent way to balance 
$\mathbb{Y}_4(\theta_2)$ with $\mathbb{T}_8^{(m, \alpha)}$. Furthermore, 
incorporating non-localities, especially the non-local and non-perturbative operators, the $g_s$ scaling involves an extra factor of ${2q\over 3}$ in \eqref{vanand1} (see \eqref{jesbielbon}). This can in principle increase the values of both $\theta_1$ and $\theta_2$ in \eqref{lilcadu78} as long as $q$ is not too large (as large $q$ non-localities are suppressed). Including these thus appears to provide a well-defined quantum system.
 
The choice of ${\bf C}_{\rm MNP} = {\bf C}_{\rm MNP}(x, y, {\Su w^a}; g_s)$ now opens up the possibilities of G-flux components ${\bf G}_{{\rm MNP}a}$, the three-form ${\bf C}_{{\rm NP}a}$ as well as the eight-form ${\bf X}_8 \equiv \left({\bf X}_8\right)_{{\rm MQRS}b0ij}$. The latter then allows the following permutations of alphabets:
\bg\label{lilcadu2}
\overbracket[1pt][7pt]{{\rm M}~\underbracket[1pt][7pt]{{\rm Q}~{\rm R}}~{\rm S}} ~\overbracket[1pt][7pt]{b~\underbracket[1pt][7pt]{0~i} ~j} ~~~ 
+~~~ 31 ~{\rm permutations}, \nd
which should be compared to \eqref{lilcadu} earlier.  For the case when the metric has the form \eqref{evader}, the scaling of the ${\bf X}_8$ polynomial again differs from what we had earlier. The result now is:
\bg\label{thetrial7}
{\bf X}_8(x, y; g_s) = \sum_{\{l_i\}} \widetilde{\bf X}_{(8, 3a)}^{(l_1, ...., l_4)}(x, y) 
\left({g_s\over {\rm HH}_o}\right)^{{1\over 3}(l_1 + l_2 + l_3 + l_4 - 6)}, \nd
which remains the same for all the 32 permutations of the alphabets in \eqref{lilcadu2}. On the other hand, if we allow the metric to generically depend on all the eleven-dimensional coordinates, the result takes the form:
\bg\label{thetrial8}
{\bf X}_8(x, y, w^a; g_s) = \sum_{\{l_i\}} \widetilde{\bf X}_{(8, 3b)}^{(l_1, ...., l_4)}(x, y, w^a) 
\left({g_s\over {\rm HH}_o}\right)^{{1\over 3}(l_1 + l_2 + l_3 + l_4 - 30)}, \nd
which is a bigger number than what we encountered in \eqref{thetrial5}. Such a big number should be a concern when we try to match the various terms in the flux EOM \eqref{mcacrisis}, but since the generic toroidal dependence do not lead to a well-defined EFT, this is not much of a concern now. On the other hand, allowing the toroidal dependence to be the special one in \eqref{makibhalu3}, and taking 
$\gamma = 5$ therein gives us the following dominant scaling:
\bg\label{thetrial9}
{\bf X}_8(x, y, w^a; g_s) = \sum_{\{l_i\}} \widetilde{\bf X}_{(8, 3c)}^{(l_1, ...., l_4)}(x, y, w^a) 
\left({g_s\over {\rm HH}_o}\right)^{{1\over 3}(l_1 + l_2 + l_3 + l_4 - 10)}, \nd
which appears from only {\it one} specific combination of the alphabets in \eqref{lilcadu2}. Other permutations of the alphabets in \eqref{lilcadu} produce $g_s$ scaling as 
$\left({g_s\over {\rm HH}_o}\right)^{-|p|}$ with ${7\over 3} \le p \le {10\over 3}$. For example, the following trace of the curvature forms:
\bg\label{bbox2}
{\rm tr}\left({\bf R}_{[{\rm PQ}]}~ {\bf R}_{[{\rm RS}]} ~{\bf R}_{[b0]} ~{\bf R}_{[ij]}\right) ~dy^{\rm P} \wedge ..... \wedge dx^j,
\nd
contributing to ${\rm tr}~\mathbb{R}^4_{\rm tot}$ scales as $-{10\over 3}$ when $\gamma = 5$ in 
\eqref{makibhalu3}. If we take generic $\gamma$, the aforementioned dominant scaling appears from 
the following value for $f_3(\gamma)$ in {\bf Table \ref{jhinuk3}}:
\bg\label{cora5}
f_3(\gamma) = 3~{\rm dom}\left({\gamma\over 3} - 2, 0\right) + {\rm dom}\left({\gamma\over 3} - 4, -2\right), \nd
which when $\gamma = 6$ takes the form $f_3(6) = -2$, the same as what we have when we allow the metric ans\"atze \eqref{evader}. One may want to compare this with what we had earlier in \eqref{cora3}, as this will be important when we study the EOM constraints from \eqref{mcacrisis} and 
{\bf Table \ref{jhinuk3}}.

\begin{table}[tb]  
 \begin{center}
\renewcommand{\arraystretch}{1.5}
\begin{tabular}{|c||c||c|}\hline ${\bf G}_{{\rm MNP}a}$ tensors & form & ${g_s\over {\rm HH}_o}$ scaling \\ \hline\hline
$\left(\mathbb{T}_7^{(f)}\right)_{{\rm QRS}b0ij}$ & $\sqrt{-{\bf g}_{11}} {\bf G}^{{\rm MNP}a} {\rm F}_1^{(n_1)}
{\rm F}_2^{(n_2)} \epsilon_{{\rm MNP}a{\rm QRS}b0ij}$ & $l_{\rm MN}^{{\rm P}a} - 4 + {2\over 3}\left(k_1 + k_{n_1} 
+ k_{n_2}\right)$ \\ \hline 
$\left(\mathbb{T}_7^{(q)}\right)_{{\rm QRS}b0ij}$ & $\sqrt{-{\bf g}_{11}} \mathbb{Y}_4^{{\rm MNP}a} 
\epsilon_{{\rm MNP}a{\rm QRS}b0ij}$ & $\theta_{nl}(k_2) - l_{\rm MN}^{{\rm P}a} - {14\over 3}  
- {2k_2\over 3}$ \\ \hline 
$\left(\mathbb{T}_8^{(a, b)}\right)_{{\rm QRS}ab0ij}$ & ${\bf G}_{[{\rm QR}ab} {\bf G}_{0ij{\rm S}]}$ & ${2\over 3}\left(k_4 + k_5\right) + l_{[{\rm QR}}^{[{ab}} \oplus l_{{0i}]}^{j{\rm S}]}$ 
\\ \hline 
$\left(\mathbb{T}_8^{(m, \alpha)}\right)_{{\rm MQRS}b0ij}$ & ${\bf G}_{[{\rm MQR}b} {\bf G}_{0ij{\rm S}]}$ & ${2\over 3}\left(k_4 + k_5\right) + l_{[{\rm MQ}}^{[{{\rm R}b}} \oplus l_{{0i}]}^{j{\rm S}]}$ 
\\ \hline 
$\left(\mathbb{X}_8^{(a, b)}\right)_{{\rm QRS}ab0ij}$ & \eqref{thetrial4}, \eqref{thetrial5}, 
\eqref{cora3}
& $l - 1, l - 9, l + f_2(\gamma)$ \\ \hline
$\left(\mathbb{X}_8^{(m, \alpha)}\right)_{{\rm MQRS}b0ij}$ & \eqref{thetrial7}, \eqref{thetrial8}, 
\eqref{cora5}
& $l - 2, l - 10, l + f_3(\gamma)$ \\ \hline
\end{tabular}
\renewcommand{\arraystretch}{1}
\end{center}
 \caption[]{Comparing the $g_s$ scalings of the various tensors that contribute to the EOM for the G-flux components ${\bf G}_{{\rm MNP}a}$ in \eqref{mcacrisis}. Note that now, all the eight forms have {\it two} possible choices stemming from how the derivatives act on the seven-forms in \eqref{mcacrisis}.} 
  \label{jhinuk3}
 \end{table}

The story now proceeds in somewhat similar way although with a crucial difference: there would be two different choices for all the rank eight tensors depending on how the derivatives act on the rank seven tensors in {\bf Table \ref{jhinuk3}} and in \eqref{mcacrisis}. The two different derivatives actions are along the toroidal directions ${\mathbb{T}^2\over {\cal G}}$ and along the six-dimensional base 
${\cal M}_4 \times {\cal M}_2$. Let us start with the latter case first. Looking at {\bf Table \ref{jhinuk3}}, we see that one possible way to balance all the terms in \eqref{mcacrisis} is to impose:

{\footnotesize
\bg\label{cora6}
l_{\rm MN}^{{\rm P}a} - 4 + {2\over 3}\left(k_1 + k_{n_1} 
+ k_{n_2}\right) = \theta_{nl}(k_2) - l_{\rm MN}^{{\rm P}a} - {14\over 3}  
- {2k_2\over 3} = {2\over 3}\left(k_4 + k_5\right) + l_{[{\rm MQ}}^{[{\rm R}{b}} \oplus l_{{0i}]}^{j{\rm S}]}, \nd}
which should further be identified to $l -2$ when the metric is of the form \eqref{evader}, or to 
$l - 10$ when the metric depends generically on all the eleven-directions, or to 
$l + f_3(\gamma)$ when the metric has the special form of \eqref{makibhalu3}. Here $l \equiv 
{1\over 3}(l_1 + ... + l_4)$. To the lowest orders, when $k_i = k_{n_i} = l = 0$, taking 
$l_{\rm MN}^{{\rm P}a} = 1$, the balance 
$l_{\rm MN}^{{\rm P}a} - 4 = l_{{\rm MQ}}^{[{\rm R}{b}} + l_{{0i}]}^{j{\rm S}}$ may be easily achieved because of our earlier choice of
$l_{{0i}]}^{j{\rm S}} = -4$. The remaining two terms in \eqref{mcacrisis}\footnote{Recall that $b_4 = 0$ in \eqref{mcacrisis}, and we are ignoring the non-perturbative and non-local contributions for the time being.} balance when:
\bg\label{cora62}
\theta_{nl} = {8\over 3} + l_{\rm MN}^{{\rm P}a} + l + {2k_2\over 3}, \nd
in \eqref{botsuga}; 
where $k_2$ is the moding for the G-flux components ${\bf G}_{{\rm MNP}a}$. Note that \eqref{cora62} implies that $\theta_{nl} \ge {11\over 3}$ for this to make sense which necessarily involves curvatures, fluxes and their derivatives in \eqref{fahingsha3}. If we take the special choice of the metric \eqref{makibhalu2} (or 
\eqref{makibhalu3} with $\gamma = 5$), then $\theta_{nl} \ge f_3(\gamma) + {17\over 3}$. This boils down to the same lower bound when $f_3(6) = -2$ as one would expect. The schematic balancing of the $g_s$ scaling of various terms in \eqref{mcacrisis} can be expressed as:
 \bg\label{lilcadu79}
d_{({\rm M})}\ast\overbracket[1pt][7pt]{{\bf G}_4 + b_1 d_{({\rm M})}\ast\Big(\mathbb{Y}_4}(\overbracket[1pt][7pt]{\theta_1) + \mathbb{Y}_4(\theta_2) + \mathbb{Y}_4\underbracket[1pt][7pt]{(\theta_3) + ..\Big) - b_3\Big(\mathbb{X}_8}_{\theta_3 \ge {11\over 3}}{}^{(m, \alpha)}\Big) - b_2 \Big( \mathbb{T}_8}^{\theta_1 \ge {8\over 3}}{}^{(m,\alpha)}\Big) = 0, \nonumber\\ \nd 
with $\theta_i$ being the quantum scaling from \eqref{botsuga3}. Note that $\mathbb{Y}_4(\theta_2)$ is not identified with anything at the lowest orders, and as such have vanishing leading contributions to \eqref{lilcadu79}, at least when the derivatives on the rank seven tensors act along the internal six directions. However when the derivatives act along the toroidal directions, the story is different as we see in the following.

When the derivatives act along the toroidal directions, the scenario is more subtle. A quick look at 
{\bf Table \ref{jhinuk3}} tells us that the balancing of various terms in \eqref{mcacrisis} has to involve other ingredients if we want to keep:
\bg\label{collateral}
l_{0i}^{j{\rm M}} = -4, ~~~ l_{\rm MN}^{\rm PQ} = l_{\rm MN}^{{\rm P}a} = l_{\rm MN}^{ab} = 1, \nd
where $({\rm M, N}) \in {\cal M}_4 \times {\cal M}_2, (a, b) \in {\mathbb{T}^2\over {\cal G}}$ and 
$(i, j) \in {\bf R}^2$. Question is, where can the extra contributions appear in {\bf Table \ref{jhinuk3}}?
The answer lies in \eqref{mcacrisis}: when the derivatives along the toroidal directions act on the rank seven tensors, with the metric and the G-flux components taking the special form \eqref{makibhalu3} with arbitrary 
$\gamma$, then extra factors of $\left({g_s\over {\rm HH}_o}\right)^{\gamma k'/3}$ appear from the exponential terms. The balancing now involves one possibility of the form:

{\footnotesize
\bg\label{cora7}
l_{\rm MN}^{{\rm P}a} - 4 + {2\over 3}\left(k_1 + k_{n_1} + {\gamma k'\over 2}
+ k_{n_2}\right) = \theta_{nl}(k_2) - l_{\rm MN}^{{\rm P}a} - {14\over 3}  
- {2k_2\over 3} + {\gamma k'\over 3} = {2\over 3}\left(k_4 + k_5\right) + l_{[{\rm MQ}}^{[{\rm R}{b}} \oplus l_{{0i}]}^{j{\rm S}]}, \nonumber\\ \nd}
where $k_i$ are the $g_s$ moding that appears from \eqref{theritual}, and $k'$ is the additional moding appearing from the derivative action $\partial_a$ on the exponential piece in \eqref{makibhalu3}. Since 
$k_i \in {\mathbb{Z}\over 2}$ and $k' \ge 1$, to lowest orders the first row in {\bf Table \ref{jhinuk3}} scales as
$l_{\rm MN}^{{\rm P}a} - 4 + {\gamma\over 3}$; whereas the second row scales as $\theta_{nl} - l_{\rm MN}^{{\rm P}a} + {\gamma - 14\over 3}$. This means when $\gamma = 6$, we can easily balance the various terms of \eqref{mcacrisis} in the following way:
\bg\label{ferrpach}
l_{\rm MN}^{{\rm P}a} - 4 + {\gamma\over 3}  = -1, ~~~~ \theta_{nl} = l_{{\rm QR}}^{{ab}} + 
l_{{\rm MN}}^{{\rm P}{a}} + l_{0i}^{j{\rm S}} + {14 -\gamma \over 3}, \nd
where the $-1$ factor on the RHS of the first equation appears from the fifth row of {\bf Table \ref{jhinuk3}}
with $f_2(\gamma) = -1$ for $\gamma = 6$. Similarly, following \eqref{collateral}, $\theta_{nl} \ge {2\over 3}$, 
 so can only equate the flux product from the third row of {\bf Table \ref{jhinuk3}} with curvatures and fluxes 
 when $\theta_{nl} = {2p\over 3}$ with $p > 2$
 as one may infer from  \eqref{botsuga3}. Clearly when $\gamma = 5$, the aforementioned balancing gets harder to
 perform. The schematic balancing of the various $g_s$ factors now becomes:
  \bg\label{lilcadu80}
d_{(a)}\ast\overbracket[1pt][7pt]{{\bf G}_4 + b_1 d_{(a)}\ast\Big(\mathbb{Y}_4}(\overbracket[1pt][7pt]{\theta_1) + \mathbb{Y}_4\underbracket[1pt][7pt]{(\theta_2) + \mathbb{Y}_4(\theta_3) +  ..\Big) - b_2\Big(\mathbb{T}_8}_{\theta_2 \ge {2\over 3}}{}^{(a, b)}\Big) - b_3 \Big( \mathbb{X}_8}^{\theta_1 \ge {8\over 3}}{}^{(a, b)}\Big) = 0, \nonumber\\ \nd 
where now the balancing at the lowest orders involves $\mathbb{Y}_4(\theta_2)$ instead of 
$\mathbb{Y}_4(\theta_3)$. Unfortunately the relevant term contributing to $\theta_2$ is 
${4\over 3}\left(l_{63} + l_{65} + l_{66}\right)$ as one may infer from \eqref{botsuga3} (or \eqref{botsuga4} discussed later). For $\theta_2 = {2\over 3}$ there are no contributing terms, so the under-bracket  will only equate the sum of the product of fluxes to zero. The situation is similar to what we encountered earlier 
in the under-bracket choice of \eqref{lilcadu78}. The resolution therein was the same: allow the sum of the products of the fluxes to vanish. As pointed out there, this involves a non-trivial differential equation for the warp-factor. As we go to higher orders, the full identification as in \eqref{cora7} will become manifest as higher order quantum terms start participating. Once we incorporate the effects of the BBS \cite{BBS} instantons from \eqref{beverwickmey}, $\theta_2$ will change to ${8\over 3}$, thus preventing any reasons that may hinder consistent dynamics  in the system. Additionally,
incorporating non-local and non-perturbative effects, both $\theta_1$ and $\theta_2$ change to ${2\over 3}(4 + q)$ and 
${2\over 3}(q + 1)$ respectively where $q$ is the degree of non-locality (see \eqref{jesbielbon}). This way the system can be solved consistently as long as $q$ is not too large.

\begin{table}[tb]  
 \begin{center}
\renewcommand{\arraystretch}{1.5}
\begin{tabular}{|c||c||c|}\hline ${\bf G}_{{\rm MN}ab}$ tensors & form & ${g_s\over {\rm HH}_o}$ scaling \\ \hline\hline
$\left(\mathbb{T}_7^{(f)}\right)_{{\rm PQRS}0ij}$ & $\sqrt{-{\bf g}_{11}} {\bf G}^{{\rm MN}ab} {\rm F}_1^{(n_1)}
{\rm F}_2^{(n_2)} \epsilon_{{\rm MN}ab{\rm PQRS}0ij}$ & $l_{\rm MN}^{ab} - 6 + {2\over 3}\left(k_1 + k_{n_1} 
+ k_{n_2}\right)$ \\ \hline 
$\left(\mathbb{T}_7^{(q)}\right)_{{\rm PQRS}0ij}$ & $\sqrt{-{\bf g}_{11}} \mathbb{Y}_4^{{\rm MN}ab} 
\epsilon_{{\rm MN}ab{\rm PQRS}0ij}$ & $\theta_{nl}(k_2) - l_{\rm MN}^{ab} - {14\over 3}  
- {2k_2\over 3}$ \\ \hline 
$\left(\mathbb{T}_8^{(a, b)}\right)_{{\rm PQRS}b0ij}$ & ${\bf G}_{[{\rm PQR}b} {\bf G}_{0ij{\rm S}]}$ & ${2\over 3}\left(k_4 + k_5\right) + l_{[{\rm PQ}}^{[{\rm R}{b}} \oplus l_{{0i}]}^{j{\rm S}]}$ 
\\ \hline 
$\left(\mathbb{T}_8^{(m, \alpha)}\right)_{{\rm MPQRS}0ij}$ & ${\bf G}_{[{\rm MPQR}} {\bf G}_{0ij{\rm S}]}$ & ${2\over 3}\left(k_4 + k_5\right) + l_{[{\rm MP}}^{[{{\rm QR}}} \oplus l_{{0i}]}^{j{\rm S}]}$ 
\\ \hline 
$\left(\mathbb{X}_8^{(a, b)}\right)_{{\rm PQRS}b0ij}$ & \eqref{thetrial7}, \eqref{thetrial8}, 
\eqref{cora5}
& $l - 2, l - 10, l + f_3(\gamma)$ \\ \hline
$\left(\mathbb{X}_8^{(m, \alpha)}\right)_{{\rm MPQRS}0ij}$ & \eqref{thetrial10}, \eqref{thetrial11}, 
\eqref{cora8}
& $l - 3, l - 11, l + f_4(\gamma)$ \\ \hline
\end{tabular}
\renewcommand{\arraystretch}{1}
\end{center}
 \caption[]{Comparing the $g_s$ scalings of the various tensors that contribute to the EOM for the G-flux components ${\bf G}_{{\rm MN}ab}$ in \eqref{mcacrisis}. Note that now, all the eight forms again have {\it two} possible choices stemming from how the derivatives act on the seven-forms in \eqref{mcacrisis} as well as on the behavior of the three form fields ${\bf C}_{{\rm MN}a}$ and ${\bf C}_{{\rm N}ab}$.} 
  \label{jhinuk4}
 \end{table}
 
The dependence of the three-form ${\bf C}_{{\rm MN}a}$ on the toroidal direction now opens up the possibility of the three-form ${\bf C}_{{\rm N}ab} = {\bf C}_{{\rm N}ab}(x, y, w^a; g_s)$. The presence of these two then leads to the EOM for the G-flux components ${\bf G}_{{\rm MN}ab}$ whose various tensor contributions to 
\eqref{mcacrisis} appears in {\bf Table \ref{jhinuk4}}. Existence of ${\bf C}_{{\rm N}ab}$ three-form field then switches on the corresponding eight-form $\left({\bf X}_8\right)_{{\rm MPQRS}0ij}$, that involves 15 permutations of the alphabets in the following way:
\bg\label{lilcadu3}
\overbracket[1pt][7pt]{{\rm M}~\underbracket[1pt][7pt]{{\rm P}~{\rm Q}}~{\rm R}} ~\overbracket[1pt][7pt]{{\rm S}~\underbracket[1pt][7pt]{0~i} ~j} ~~~ 
+~~~ 14 ~{\rm permutations}, \nd
leading to various scaling possibilities that we describe in the following. First of all, when the metric takes the simplest dependence of the form \eqref{evader}, all the 15 permutations of the alphabets in \eqref{lilcadu3} leads to the same $g_s$ scaling of $-3$. This means the corresponding ${\bf X}_8$ polynomial takes the form:
\bg\label{thetrial10}
{\bf X}_8(x, y; g_s) = \sum_{\{l_i\}} \widetilde{\bf X}_{(8, 4a)}^{(l_1, ...., l_4)}(x, y) 
\left({g_s\over {\rm HH}_o}\right)^{{1\over 3}(l_1 + l_2 + l_3 + l_4 - 9)}, \nd 
which should be compared to \eqref{thetrial7} that also appears in {\bf Table \ref{jhinuk4}}. On the other hand, demanding a generic dependence of the metric and the flux components on all the eleven-dimensional coordinates, produces a $g_s$ scaling of $-11$ that remains the same for all the 15 permutations of the alphabets in \eqref{lilcadu3}. This leads to: 
\bg\label{thetrial11}
{\bf X}_8(x, y, w^a; g_s) = \sum_{\{l_i\}} \widetilde{\bf X}_{(8, 4b)}^{(l_1, ...., l_4)}(x, y, w^a) 
\left({g_s\over {\rm HH}_o}\right)^{{1\over 3}(l_1 + l_2 + l_3 + l_4 - 33)}, \nd
 with a further increase of the $g_s$ scaling from \eqref{thetrial8}. Reassuringly however, the non-existence of the EFT description from \eqref{botsuga2}, helps us to avoid worrying about balancing the $g_s$ dependences of the various terms of \eqref{mcacrisis}. Finally, when we choose the metric with special dependence on the toroidal direction as in \eqref{makibhalu3}, the {\it dominant} $g_s$ scaling becomes
 $- 4$ leading us to: 
 \bg\label{thetrial12}
{\bf X}_8(x, y, w^a; g_s) = \sum_{\{l_i\}} \widetilde{\bf X}_{(8, 4c)}^{(l_1, ...., l_4)}(x, y, w^a) 
\left({g_s\over {\rm HH}_o}\right)^{{1\over 3}(l_1 + l_2 + l_3 + l_4 - 12)}, \nd
 for $\gamma = 5$. In fact the dominant scaling is shared by 3 permutations of the alphabets in 
 \eqref{lilcadu3}, all taking the value:
 \bg\label{cora8}
 f_4(\gamma) = 2~{\rm dom}\left({\gamma\over 3} - 2, 0\right) + {\rm dom}\left({\gamma\over 3} - 4, -2\right) - 1, \nd
 which becomes $f_3(\gamma) = -3$ when $\gamma = 6$, thus matching with the $g_s$ scaling for the metric \eqref{evader}. If we go beyond the dominant scaling, then the typical $g_s$ scaling of the various perturbations in \eqref{lilcadu3} go as $\left({g_s\over {\rm HH}_o}\right)^{-|p|}$ with ${10\over 3} \le p \le
 {12\over 3}$. Thus an example of the dominant contribution to ${\rm tr}~\mathbb{R}^4_{\rm tot}$ would be:
\bg\label{bbox3}
{\rm tr}\left({\bf R}_{[{\rm MP}]}~ {\bf R}_{[{\rm QR}]} ~{\bf R}_{[{\rm S}0]} ~{\bf R}_{[ij]}\right) ~dy^{\rm M} \wedge ..... \wedge dx^j.
\nd
With these we are ready to analyze the various terms in the EOM for the G-flux components 
${\bf G}_{{\rm MN}ab}$ in \eqref{mcacrisis}. Since we will be following the generic scalings of the G-flux components as in \eqref{theritual}, the fact that the G-flux components ${\bf G}_{{\rm MN}ab}$ take the 
{\it localized} form \eqref{tanyarbon} will not be relevant in the following analysis. The localized form 
\eqref{tanyarbon} does however place some constraints on the gauge field components 
${\cal F}^{(k)}_{\rm MN}$, but we will not separately analyze them here.
 
Our first case would be to study the various scalings of the rank seven and eight tensors from 
{\bf Table \ref{jhinuk4}} when the derivatives act along the base ${\cal M}_4 \times {\cal M}_2$ directions. Looking at rows 1, 2 and 4 in {\bf Table \ref{jhinuk4}}, we see that one possible balancing of various terms in \eqref{mcacrisis} can happen when:

{\footnotesize
\bg\label{cora9}
l_{\rm MN}^{ab} - 6 + {2\over 3}\left(k_1 + k_{n_1} 
+ k_{n_2}\right) = \theta_{nl}(k_2) - l_{\rm MN}^{ab} - {14\over 3}  
- {2k_2\over 3} = {2\over 3}\left(k_4 + k_5\right) + l_{[{\rm MP}}^{[{\rm QR}} \oplus l_{{0i}]}^{j{\rm S}]}, \nd}
which could further be equated to either $l - 3$ when the metric takes the form \eqref{evader}, or to 
$l - 11$ when the metric has generic dependence on all the eleven-dimensional coordinates, or to 
$l + f_4(\gamma)$, with $f_4(\gamma)$ as in \eqref{cora8}, when the metric and the fluxes take the special form of \eqref{makibhalu3}. Assuming $l_{\rm MN}^{ab} = 1$ from \eqref{collateral}, we see that the product of the fluxes in row 4 of {\bf Table \ref{jhinuk4}}, scales exactly as the ${\bf X}_8$ polynomial in 
\eqref{thetrial10}. 

Consider now the case when the derivatives along the base ${\cal M}_4 \times {\cal M}_2$ act on the rank seven-tensors in \eqref{mcacrisis}. Looking at the flux ans\"atze in \eqref{makibhalu} (or the detailed form in footnote \ref{margincall}), we notice that the derivatives cannot bring down factors of $g_s$ even if 
$\xi \ne 0$ there (at least when we consider dominant contributions). This means we will have to balance the terms from the first and the second row in {\bf Table \ref{jhinuk4}}. This can happen when:
\bg\label{lechalet}
\theta_{nl} = 2l_{\rm MN}^{ab} - {4\over 3} + {2\over 3}(k_1 + k_2 + k_{n_1} + k_{n_2}), \nd
in \eqref{botsuga3} (or in \eqref{botsuga}). When $k_i = k_{n_i} = 0$, this implies 
$\theta_{nl} = 2l_{\rm MN}^{ab} - {4\over 3}$, so $l_{\rm MN}^{ab} > {2\over 3}$ for positivity of $\theta_{nl}$, 
{\it i.e.} for $\theta_{nl} > 0$. Happily, this is also the condition one infers from either \eqref{botsuga} or \eqref{botsuga3} for the validity of the EFT description. Thus $l_{\rm MN}^{ab} = 1$, passes successfully another consistency check.  The schematic mapping of the various $g_s$ factors in the flux EOM \eqref{mcacrisis} now happens in the following way: 
\bg\label{lilcadu81}
d_{({\rm M})}\ast\overbracket[1pt][7pt]{{\bf G}_4 + b_1 d_{({\rm M})}\ast\Big(\mathbb{Y}_4}^{\theta_1 \ge {2\over 3}}(\theta_1) + \mathbb{Y}_4(\theta_2) + \mathbb{Y}_4\underbracket[1pt][7pt]{(\theta_3)  +  ..\Big) - b_2\Big(\mathbb{T}_8}\underbracket[1pt][7pt]{{}^{(m, \alpha)}\Big) - b_3 \Big( \mathbb{X}_8}_{\theta_3 \ge {8\over 3}}{}^{(m,\alpha)}\Big) = 0, \nonumber\\ \nd 
where $\mathbb{Y}_4(\theta_2)$ remains untouched at the lowest orders. However for $\theta_{nl} \equiv \theta_1 = {2\over 3}$, the quantum term 
$\mathbb{Y}_4^{{\rm MN}ab} \propto {\bf G}^{{\rm NN}ab}$, so the matching doesn't actually provide a dynamical equation for the corresponding flux component. Such an equation appears from 
$\theta_3 = {8\over 3}$ in the identification \eqref{lilcadu81} above. One may alternatively choose 
$l_{\rm MN}^{ab} = 2$, which is only possible when $l_{\rm MN}^{PQ} = 0$ otherwise anomaly cancellation condition will be violated\footnote{Other two choices are: 
$\left(l_{\rm MN}^{ab}, l_{\rm MN}^{\rm PQ}\right) = \left({4\over 3}, {2\over 3}\right)$ and $\left({5\over 3}, {1\over 3}\right)$. \label{mar_c}}. Such a choice leaves other scalings unchanged, giving rise to:
\bg\label{automoon}
l_{0i}^{jm} = -4, ~~~l_{\rm MN}^{{\rm P}a} = 1, ~~~ l_{\rm MN}^{ab} = 2, ~~~ l_{\rm MN}^{\rm PQ} = 0, \nd
where the first one is fixed by moving membranes, the second one is fixed by anomaly cancellation condition \eqref{lindmonaco}, and the third is fixed by constraints from EFT \eqref{botsuga3}, anomaly cancellation condition \eqref{lindmonaco} and EOM \eqref{mcacrisis}. Finally the fourth one gets fixed automatically from the anomaly cancellation condition once we demand
$l_{\rm MN}^{ab} = 2$, as alluded to above. Interestingly, the choice \eqref{automoon} leads to the choice 
$\theta_{nl} \equiv \theta_1 = {8\over 3}$ for the dynamical evolutions for {\it all} the G-flux components:
${\bf G}_{\rm MNPQ}, {\bf G}_{{\rm MNP}a}$ and ${\bf G}_{{\rm MN}ab}$. 

On the other hand, if we stick with the dominant scalings \eqref{collateral}, then the dynamical evolution of the flux components ${\bf G}_{{\rm MN}ab}$ appears in the same way as above, albeit in a slightly different placement of the quantum terms,
when we study the derivatives' action along the toroidal direction in \eqref{mcacrisis}. Looking at \eqref{makibhalu3}, the toroidal derivative action brings down an extra factor of $\left({g_s\over {\rm HH}_o}\right)^{\gamma\over 3}$ that contributes to the $g_s$ scaling in the first row of {\bf Table \ref{jhinuk4}}. This scales exactly as the 
terms in the third row of {\bf Table \ref{jhinuk4}} when $\gamma = 6$. Similarly the toroidal derivatives on the quantum terms in the second row, scales exactly as the ${\bf X}_8$ polynomial \eqref{cora5} $-$ which in turn is the same as in \eqref{thetrial7} $-$ with $\gamma = 6$ in \eqref{makibhalu3} when:
\bg\label{adelchalet}
\theta_{nl} = l_{\rm MN}^{ab} + l + {2\over 3}(1 + k_2). \nd
For $l = k_2 = 0$, this gives $\theta_{nl} = {5\over 3}$ with $l_{\rm MN}^{ab} = 1$,  thus providing the necessary dynamical equation involving curvature, fluxes and their derivatives. The schematic arrangement of the various $g_s$ scalings in the EOM \eqref{mcacrisis} now becomes:
\bg\label{lilcadu82}
d_{(a)}\ast\overbracket[1pt][7pt]{{\bf G}_4 + b_1 d_{(a)}\ast\Big(\mathbb{Y}_4}
\overbracket[1pt][7pt]{(\theta_1) + \mathbb{Y}_4\underbracket[1pt][7pt]{(\theta_2) + \mathbb{Y}_4(\theta_3)  +  ..\Big) - b_3\Big(\mathbb{X}_8}_{\theta_2 \ge {5\over 3}}{}^{(a, b)}\Big) 
- b_2 \Big(\mathbb{T}_8}^{\theta_1 \ge {2\over 3}}{}^{(a, b)}\Big) = 0, \nonumber\\ \nd 
where we now expect the dynamical evolution to be governed by $\theta_2 \ge {5\over 3}$. With the choice 
\eqref{automoon}, the placement of the over- and under-brackets would be slightly different, but for both cases the quantum series will be controlled by $\theta_{nl} = {8\over 3}$.

Our above analysis justifies that the scaling choice \eqref{collateral} (or even \eqref{automoon}) does lead to consistent matching the $g_s$ powers of all the internal flux equations. One might however wonder if there are other possibilities of $g_s$ scalings that could equally match all the possible flux equations. Question is what other possible matching could be envisioned here? One possibility is to match the product of the fluxes, {\it i.e.} the rank eight tensors $\mathbb{T}_8$ with the corresponding ${\bf X}_8$ polynomial, which is another rank eight tensor at the lowest orders. Similarly the derivatives of the rank seven tensors could be identified equivalently, again at the lowest orders. Schematically this would mean the following identifications of the $g_s$ scalings in \eqref{mcacrisis}:
\bg\label{lilcadu83}
d_{({\rm M}, a)}\ast\overbracket[1pt][7pt]{{\bf G}_4 + b_1 d_{({\rm M}, a)}\ast\Big(\mathbb{Y}_4}^{\theta_1}(\theta_1) + \mathbb{Y}_4\underbracket[1pt][7pt]{(\theta_{{\rm M}, a})  +  ..\Big) - b_2\Big(\mathbb{T}_8}\underbracket[1pt][7pt]{{}^{({\rm M}, a)}\Big) - b_3 \Big( \mathbb{X}_8}_{\theta_{{\rm M}, a}}{}^{({\rm M}, a)}\Big) = 0, \nonumber\\ \nd 
where the subscript on $d$ denotes derivatives along the internal six directions ${\cal M}_4 \times {\cal M}_2$ or derivatives along the toroidal directions ${\mathbb{T}^2\over {\cal G}}$. The remain sub- and superscripts have one-to-one correspondence with the corresponding derivative actions. 

The above schematic construction then instructs us how to match the ${g_s\over {\rm HH}_o}$ scalings of the various rank seven and eight tensors. 
Imagine now we want to follow the under-bracket identification from \eqref{lilcadu83}. For this to happen, 
it would require us to first assign the dominant scalings of the G-flux components as:
\bg\label{renaissance}
l_{\rm MN}^{\rm PQ} = x_1, ~~~ l_{\rm RS}^{ab} = x_2, ~~~ l_{\rm MN}^{{\rm P}a} = x_3, ~~~ 
l_{0i}^{j{\rm M}} = x_4, \nd
where $x_i$ are the factors that we shall determine from balancing the rank eight tensors. From 
\eqref{mcacrisis} we know that the forms of the rank eight tensors do depend on how the derivatives act on the rank seven tensors, and therefore the balancing the rank eight tensors would depend whether the derivatives act along the base ${\cal M}_4 \times {\cal M}_2$ or the toroidal ${\mathbb{T}^2\over {\cal G}}$ directions. In the first case, since we do not distinguish between the two toroidal directions, 
$l_{\rm MN}^{{\rm P}a} = l_{\rm MN}^{{\rm P}b} \equiv x_3 = 1$, from the anomaly cancellation condition. This then leads to an {\it over-determined} set of equations that may be expressed in the following matrix form:
\bg\label{renaisstagra}
\begin{pmatrix}
1 & ~ & 1 & ~ & 0 & ~ &  0 \\  
0 & ~ & 1 & ~ & 0 & ~ & 1 \\  
0 & ~&  0 & ~ & 1 & ~ & 1 \\  
1 & ~ & 0 & ~ & 0 & ~ & 1
\end{pmatrix}
\begin{pmatrix}
  x_1\\
  x_2\\
  x_3\\
  x_4 \end{pmatrix} = \begin{pmatrix} ~~2 \\ -1 \\ -2 \\ -3 \end{pmatrix},
\nd
which can nevertheless be solved consistently and leads to $(x_1, x_2, x_3, x_4) = (0, 2, 1, -3)$, where we see that $x_3 = 1$ reappears from the matrix equation. On the other hand, if we demand the derivatives act along the toroidal directions on the rank seven tensors, we get the following matrix equation:
\bg\label{renaisstagra2}
\begin{pmatrix}
1 & ~ & 1 & ~ & 0 & ~ &  0 \\  
0 & ~ & 1 & ~ & 0 & ~ & 1 \\  
0 & ~&  0 & ~ & 1 & ~ & 1 \\  
0 & ~ & 0 & ~ & 1 & ~ & 0
\end{pmatrix}
\begin{pmatrix}
  x_1\\
  x_2\\
  x_3\\
  x_4 \end{pmatrix} = \begin{pmatrix} ~~2 \\ -1 \\ -2 \\ ~~1 \end{pmatrix},
\nd
which is not over-determined because the anomaly cancelation condition is already incorporated in 
the last row of the bigger matrix. Somewhat surprisingly, the above matrix equation reproduces exactly the same answers for $x_i$, namely, $(x_1, x_2, x_3, x_4) = (0, 2, 1, -3)$, showing that the system appears to be self-consistent with:
\bg\label{renaissance2}
l_{\rm MN}^{\rm PQ} = 0, ~~~ l_{\rm RS}^{ab} = 2, ~~~ l_{\rm MN}^{{\rm P}a} = 1, ~~~ 
l_{0i}^{j{\rm M}} = -3, \nd
which may now be compared to \eqref{collateral}. The above equation again confirms that the three-form RR and NS fluxes in the dual IIB side {\it cannot} be time-independent because $l_{\rm MN}^{{\rm P}a} 
= l_{\rm MN}^{{\rm P}b} = 1$, although the G-flux components ${\bf G}_{\rm MNPQ}$ appears to be time-independent. While the condition \eqref{renaissance2} does not conflict with either 
\eqref{botsuga} or \eqref{botsuga2}, thus allowing an EFT description\footnote{Interestingly, if we now balance all the rank seven tensors associated with flux components and the quantum terms from \eqref{fahingsha4}, including their derivatives, these quantum terms  appear to scale in the same way as 
$\theta_{nl} = {8\over 3}$ in \eqref{botsuga3}. Similar behavior also stems from the identification \eqref{automoon}.}, there is however something off about it 
when we compare it to both \eqref{collateral} and \eqref{automoon}: the $g_s$ scaling of the flux components ${\bf G}_{0ij{\rm M}}$ now scales as 
$\left({g_s\over {\rm HH}_o}\right)^{-3}$  instead of $\left({g_s\over {\rm HH}_o}\right)^{-4}$ (although the remain scalings are similar to \eqref{automoon}). Recall that the latter scaling of ${\red -4}$ was derived from the back-reaction effects  of the space-filling two-branes (both integer and fractional as well as branes and anti-branes as shown in section 4.2.4 in the first reference of 
\cite{desitter2}). These branes would automatically appear from the localized G-flux components 
${\bf G}_{{\rm MN}ab}$, so it will be inconsistent to ignore them 
here\footnote{The localized fluxes 
${\bf G}_{{\rm MN}ab}$ lead to seven-branes in the IIB side. Instantons on the seven-branes, when they become {\it small}, lead to bound states of three-branes which dualize to two-branes in M-theory. These are the two-branes whose back-reaction effects contribute to  $\left({g_s\over {\rm HH}_o}\right)^{-4}$ scalings for the G-flux components ${\bf G}_{0ij{\rm M}}$ \cite{Dasgupta:2014pma, desitter2}. \label{KoTo}}. Therefore although \eqref{renaissance2} appears to be a new class of solution for the internal flux configuration in the system, there are reason why \eqref{collateral} (and maybe even \eqref{automoon}, although flux quantization discussed in section \ref{fluxoo} does put some constraint on this choice) is still the preferred one at every point in the moduli space of the system.

\subsection{Flux equations for ${\bf G}_{0{\rm ABC}}$ components and EFT \label{g0abc}}

Our discussion in the previous section has at least taught us two things: one, for $\gamma = 6$ 
in \eqref{makibhalu3} the results of curvature etc. resemble the ones from the metric \eqref{evader}; and two, 
the quantum terms from \eqref{fahingsha4} can in principle {\it balance} all the $g_s$ scalings of the 
rank seven and eight tensors in the flux EOM \eqref{mcacrisis}. The reason for such a leverage is simple. The quantum terms \eqref{fahingsha4} may be alternatively expressed as \cite{desitter2}:

{\footnotesize
\bg\label{suhaag}
\int d^{11}z \sqrt{-{\bf g}_{11}}~\sum_{\{l_i\}, n_i}~\mathbb{Q}_{\rm T}^{(\{l_i\}, n_i)}(x, y, w^a; g_s) \equiv
\int d^{11}z \sqrt{-{\bf g}_{11}}~\sum_{i = 1}^\infty ~{\bf G}_{\rm ABCD} \mathbb{Y}_4^{\rm ABCD}(\theta_i), \nd}
where $z \equiv (x, y, w^a; t)$ with $t$ related to $g_s$ in the way described earlier; and $\theta_i$ are the values that $\theta_{nl}$ from \eqref{botsuga3} can take. Note for any given value of $\theta_{nl} = \theta_i$, there is a series of values due to various modings of the fluxes and curvature components. This means, for example, a higher moding of $\theta_k$ could match with  a lower moding of $\theta_m$ for $\theta_m > 
\theta_k$. This way, at higher orders in 
$\left(k_i, k_{n_i}, l, {g_s\over {\rm HH}_o}\right)$, 
a large set of quantum terms from 
\eqref{fahingsha4} can start participating (see footnote \ref{honikjul3}). With this in mind, a generic strategy at lowest orders to balance all EOMs for 
G-fluxes of the form ${\bf G}_{\rm 0ABC}$ may be schematically expressed in the following way:
\bg\label{lilcadu84}
d_{(z)}\ast\overbracket[1pt][7pt]{{\bf G}_4 + b_1 d_{(z)}\ast\Big(\mathbb{Y}_4}^{\theta_1}(\theta_1) +
\mathbb{Y}_4\overbracket[1pt][7pt]{(\theta_{z, 2}) + \mathbb{Y}_4\underbracket[1pt][7pt]{(\theta_{z, 3})  +  ..\Big)- b_2\Big(\mathbb{T}_8}_{\theta_{z, 3}}{}^{(z)}\Big) -b_3 \Big(\mathbb{X}_8}^{\theta_{z, 2}}{}^{(z)}\Big) = 0 \nonumber\\ \nd
where $z \equiv (0, {\rm M}, a)$ depending on what directions the derivatives act. Note that the above $g_s$ identifications are {\it always} true\footnote{One may easily verify that the first identification with $\theta_1$ is trivially true at the lowest orders in ${g_s\over {\rm HH}_o}$ from the definition \eqref{suhaag}. In other words, the first identification will imply 
$\theta_1 = 2\left(l_{\rm AB}^{\rm CD} + b\right)$ to the lowest orders where $b$ can be read up from \eqref{botsuga4} given below. As we go to higher orders in $\left(k_i, k_{n_i}, l, {g_s\over {\rm HH}_o}\right)$, the difference between the ${\bf G}_4$ and $\mathbb{Y}_4(\theta_1)$ starts becoming significant.}, but an extra knowledge of the actual scalings of all the G-flux components could help us ascertain whether $\theta_1 = \theta_{z, 3}$ or $\theta_1 = \theta_{z, 2}$, or even
$\theta_1 = \theta_{z, 2} = \theta_{z, 3}$. Eventually of course we expect:
\bg\label{hgtok}
\theta_1(\{l_1\}, n_1; \{k_1\}) = \theta_{z, 2}(\{l_2\}, n_2; \{k_2\}) = \theta_{z, 3}(\{l_3\}, n_3; \{k_3\}), \nd
where $(\{l_3\}, n_3) > (\{l_2\}, n_2) > (\{l_1\}, n_1)$ and $\{k_1\} > \{k_2\} > \{k_3\}$, with $\{k_i\}$ forming the set of modings with the choice $(\{l_i\}, n_i)$ in \eqref{fahingsha4}. In the following therefore, we will start by imposing the equivalences \eqref{lilcadu83}, and then try to infer whether further identifications between 
the set of triplets $(\theta_1, \theta_{z, 2}, \theta_{z, 3})$ are possible.  

Before moving ahead let us point out a possible caveat when we try to identify the $g_s$ scalings of the 
remaining G-flux components.
Since ${\bf G}_4 \ne d{\bf C}_3$, one shouldn't infer the scalings of ${\bf C}_3$ three-form fields from the knowledge of the scalings of the corresponding four-form fluxes. The exception appears to be the G-flux components ${\bf G}_{0ij{\rm M}}$, which scale as $\left({g_s\over {\rm HH}_o}\right)^{-4}$. Question is, what about components like ${\bf G}_{0ija}$ which we kept zero in our earlier discussions?

In this section we will go for more detailed study of components like ${\bf G}_{0{\rm ABC}}$ where 
$({\rm A, B}) \in {\bf R}^2 \times {\cal M}_4 \times {\cal M}_2 \times {\mathbb{T}^2\over {\cal G}}$. Earlier 
in \eqref{chupke} we argued how quantum terms could keep ${\bf G}_{0{\rm MNP}} = 0$, and in the study of the quantum terms \eqref{fahingsha3} and \eqref{fahingsha4}, we did keep other related components zero to simplify the ensuing analysis. There is however no strong reason why the condition \eqref{chupke} when extended to other components may not lead to an over-determined set of conditions. Since we don't know the exact form of $\mathbb{Y}_4$ and $\mathbb{Y}_7$, it would seem safer to keep components like 
${\bf G}_{0{\rm ABC}}$ non-zero and work out their precise EOMs. In fact there is another deeper reason for entertaining these components: they would contribute to both internal ({\it i.e.} fluxes on the internal eight-manifold) and external (fluxes with at least one leg along ${\bf R}^2$) flux EOMs. Only after we have laid out 
{\it all} the flux EOMs, we can try to see how much simplification can be allowed without trivializing the system.

One immediate consequence of keeping components like ${\bf G}_{0{\rm ABC}}$ non-zero is the 
increase in the number of components contributing to \eqref{fahingsha4}. We will discuss this below and also elaborate on the scaling behavior. Our metric choice will be the one with special dependence on the 
toroidal direction, {\it i.e.} \eqref{makibhalu3}, and for this case \eqref{fahingsha4} changes to:
\bg\label{fahingsha5}
\mathbb{Q}_{\rm T}^{(\{l_i\}, n_i)} &= & \left[{\bf g}^{-1}\right] \prod_{i = 0}^4 \left[\partial\right]^{n_i} 
\prod_{{\rm k} = 1}^{60} \left({\bf R}_{\rm A_k B_k C_k D_k}\right)^{l_{\rm k}} \prod_{{\rm r} = 61}^{100} 
\left({\bf G}_{\rm A_r B_r C_r D_r}\right)^{l_{\rm r}}\nonumber\\
& = & {\bf g}^{m_i m'_i}.... {\bf g}^{j_k j'_k} 
\{\partial_m^{n_1}\} \{\partial_\alpha^{n_2}\} \{\partial_a^{n_3}\}\{\partial_i^{n_4}\}\{\partial_0^{n_0}\}
\left({\bf R}_{a0b0}\right)^{l_1} \left({\bf R}_{abab}\right)^{l_2}\left({\bf R}_{pqab}\right)^{l_3}\left({\bf R}_{\alpha a b \beta}\right)^{l_4} \nonumber\\
&\times& \left({\bf R}_{abij}\right)^{l_5}\left({\bf R}_{\alpha\beta\alpha\beta}\right)^{l_6}
\left({\bf R}_{ijij}\right)^{l_7}\left({\bf R}_{ijmn}\right)^{l_8}\left({\bf R}_{\alpha\beta mn}\right)^{l_9}
\left({\bf R}_{i\alpha j \beta}\right)^{l_{10}}\left({\bf R}_{0\alpha 0\beta}\right)^{l_{11}}
\nonumber\\
& \times & \left({\bf R}_{0m0n}\right)^{l_{12}}\left({\bf R}_{0i0j}\right)^{l_{13}}\left({\bf R}_{mnpq}\right)^{l_{14}}\left({\bf R}_{0mnp}\right)^{l_{15}}
\left({\bf R}_{0\alpha\beta m}\right)^{l_{16}}\left({\bf R}_{0abm}\right)^{l_{17}}\left({\bf R}_{0ijm}\right)^{l_{18}}
\nonumber\\
& \times & \left({\bf R}_{mnp\alpha}\right)^{l_{19}}\left({\bf R}_{m\alpha ab}\right)^{l_{20}}
\left({\bf R}_{m\alpha\alpha\beta}\right)^{l_{21}}\left({\bf R}_{m\alpha ij}\right)^{l_{22}}
\left({\bf R}_{0mn \alpha}\right)^{l_{23}}\left({\bf R}_{0m0\alpha}\right)^{l_{24}}
\left({\bf R}_{0\alpha\beta\alpha}\right)^{l_{25}}
\nonumber\\
&\times& \left({\bf R}_{0ab \alpha}\right)^{l_{26}}\left({\bf R}_{0ij\alpha}\right)^{l_{27}}
\left({\bf R}_{mnai}\right)^{l_{28}}\left({\bf R}_{\alpha\beta ai}\right)^{l_{29}}
\left({\bf R}_{a0i0}\right)^{l_{30}}\left({\bf R}_{aijk}\right)^{l_{31}}
\left({\bf R}_{abai}\right)^{l_{32}}\nonumber\\
&\times& \left({\bf R}_{m\beta i \alpha}\right)^{l_{33}}\left({\bf R}_{abmi}\right)^{l_{34}}
\left({\bf R}_{ijk0}\right)^{l_{35}}\left({\bf R}_{\alpha 0i0}\right)^{l_{36}}
\left({\bf R}_{\alpha\beta i 0}\right)^{l_{37}}\left({\bf R}_{ab0i}\right)^{l_{38}}
\left({\bf R}_{\alpha ijk}\right)^{l_{39}}\nonumber\\
&\times& \left({\bf R}_{ab i \alpha}\right)^{l_{40}}\left({\bf R}_{\alpha\beta i \alpha}\right)^{l_{41}}
\left({\bf R}_{mni\alpha}\right)^{l_{42}}\left({\bf R}_{mni0}\right)^{l_{43}}
\left({\bf R}_{mnpi}\right)^{l_{44}}\left({\bf R}_{0m0i}\right)^{l_{45}}
\left({\bf R}_{mijk}\right)^{l_{46}}\nonumber\\
&\times& \left({\bf R}_{maij}\right)^{l_{47}}\left({\bf R}_{ma\alpha\beta}\right)^{l_{48}}
\left({\bf R}_{maba}\right)^{l_{49}}\left({\bf R}_{aij0}\right)^{l_{50}}
\left({\bf R}_{mnpa}\right)^{l_{51}}\left({\bf R}_{a\alpha\beta 0}\right)^{l_{52}}
\left({\bf R}_{a 0\alpha 0}\right)^{l_{53}}\nonumber\\
&\times& \left({\bf R}_{ab a0}\right)^{l_{54}}\left({\bf R}_{mna\alpha}\right)^{l_{55}}
\left({\bf R}_{a\alpha ij}\right)^{l_{56}}\left({\bf R}_{a\alpha\alpha\beta}\right)^{l_{57}}
\left({\bf R}_{ab a\alpha}\right)^{l_{58}}\left({\bf R}_{m0a0}\right)^{l_{59}} 
\left({\bf R}_{mna0}\right)^{l_{60}}\nonumber\\
&\times & \left({\bf G}_{mnpq}\right)^{l_{61}} \left({\bf G}_{mnp\alpha}\right)^{l_{62}}
\left({\bf G}_{mnpa}\right)^{l_{63}}\left({\bf G}_{mn\alpha\beta}\right)^{l_{64}}
\left({\bf G}_{mn\alpha a}\right)^{l_{65}}
\left({\bf G}_{m\alpha\beta a}\right)^{l_{66}}\left({\bf G}_{0ijm}\right)^{l_{67}}\nonumber\\ 
&\times & \left({\bf G}_{0ij\alpha}\right)^{l_{68}}
\left({\bf G}_{mnab}\right)^{l_{69}}\left({\bf G}_{ab\alpha\beta}\right)^{l_{70}}
\left({\bf G}_{m\alpha ab}\right)^{l_{71}} \left({\bf G}_{mnpi}\right)^{l_{72}}
\left({\bf G}_{m\alpha\beta i}\right)^{l_{73}}\left({\bf G}_{mn\alpha i}\right)^{l_{74}} \nonumber\\
&\times &\left({\bf G}_{mnai}\right)^{l_{75}}
\left({\bf G}_{mabi}\right)^{l_{76}}\left({\bf G}_{a\alpha\beta i}\right)^{l_{77}}
\left({\bf G}_{\alpha ab i}\right)^{l_{78}} \left({\bf G}_{ma\alpha i}\right)^{79} 
\left({\bf G}_{mn ij}\right)^{l_{80}}\left({\bf G}_{m\alpha ij}\right)^{l_{81}} \nonumber\\
&\times& \left({\bf G}_{\alpha\beta ij}\right)^{l_{82}}
\left({\bf G}_{maij}\right)^{l_{83}}\left({\bf G}_{\alpha a ij}\right)^{l_{84}}
\left({\bf G}_{ab ij}\right)^{l_{85}} \left({\bf G}_{0ija}\right)^{l_{86}}
\left({\bf G}_{0mnp}\right)^{l_{87}}
\left({\bf G}_{0mn\alpha}\right)^{l_{88}} \nonumber\\
&\times& \left({\bf G}_{0m\alpha\beta}\right)^{l_{89}}\left({\bf G}_{0mab}\right)^{l_{90}}
\left({\bf G}_{0\alpha ab}\right)^{l_{91}}\left({\bf G}_{0mna}\right)^{l_{92}}
\left({\bf G}_{0m\alpha a}\right)^{l_{93}} \left({\bf G}_{0\alpha\beta a}\right)^{l_{94}}
\left({\bf G}_{0mni}\right)^{l_{95}}\nonumber\\
&\times& \left({\bf G}_{0m\alpha i}\right)^{l_{96}} \left({\bf G}_{0\alpha\beta i}\right)^{l_{97}}
\left({\bf G}_{0mia}\right)^{l_{98}}
\left({\bf G}_{0\alpha ia}\right)^{l_{99}}\left({\bf G}_{0ab i}\right)^{l_{100}}, \nd
with 100 possible distinct terms (60 curvatures and 40 fluxes), modulo their permutations. Question naturally arises whether this proliferation still preserves the necessary EFT when $\gamma \ge 5$ in \eqref{makibhalu3}. For the quantum series \eqref{fahingsha4}, the $g_s$ scaling in \eqref{botsuga3} showed us under what conditions 
$g_s$ hierarchy could be preserved. Does this happen here too? To see this, it is instructive to first work out the modifications to \eqref{botsuga3}. This may be quantified as:
\bg\label{botsuga4}
\theta_{nl} &= & \left({\gamma k \over 3} {\red -{4\over 3}}\right) \sum_{r = 1 }^{5} {\red l_r} +{2\over 3} \sum_{i = 6}^{27} l_i  + \left({\gamma k \over 3} + {2\over 3}\right) \sum_{j = 28}^{32} l_j  
+ {5\over 3} \sum_{k = 33}^{46} l_k  + \left({\gamma k \over 3} {\red -{1\over 3}}\right) 
\sum_{p = 47}^{60} {\red l_p} 
\nonumber\\
&+& \left(l_{mn}^{pq} + {4\over 3}\right)l_{61} + \left(l_{mn}^{p\alpha} + {4\over 3}\right)l_{62} +
\left(l_{mn}^{pa} + {1\over 3}\right)l_{63} + \left(l_{mn}^{\alpha\beta} + {4\over 3}\right)l_{64} 
+ \left(l_{mn}^{\alpha a} + {1\over 3}\right)l_{65}\nonumber\\
&+& \left(l_{ma}^{\alpha\beta} + {1\over 3}\right)l_{66} + \left(l_{ij}^{0m} + {13\over 3}\right)l_{67} 
+ \left(l_{ij}^{0\alpha} + {13\over 3}\right)l_{68} + \left(l_{mn}^{ab} {\red - {2\over 3}}\right){\red l_{69}} 
+ \left(l_{\alpha\beta}^{ab} {\red - {2\over 3}}\right){\red l_{70}} \nonumber\\
&+& \left(l_{m\alpha}^{ab} {\red - {2\over 3}}\right){\red l_{71}} + \left(l_{mn}^{pi} + {7\over 3}\right)l_{72} 
+ \left(l_{m\alpha}^{\beta i} + {7\over 3}\right)l_{73} + \left(l_{mn}^{\alpha i} + {7\over 3}\right)l_{74} 
+ \left(l_{mn}^{ai} + {4\over 3}\right)l_{75} \nonumber\\
&+& \left(l_{ab}^{mi} + {1\over 3}\right)l_{76} + \left(l_{\alpha\beta}^{ai} + {4\over 3}\right)l_{77} 
+ \left(l_{ab}^{\alpha i} + {1\over 3}\right)l_{78} + \left(l_{m\alpha}^{ai} + {4\over 3}\right)l_{79} 
+ \left(l_{mn}^{ij} + {10\over 3}\right)l_{80} \nonumber\\
&+& \left(l_{m\alpha}^{ij} + {10\over 3}\right)l_{81} + \left(l_{\alpha\beta}^{ij} + {10\over 3}\right)l_{82} 
+ \left(l_{ma}^{ij} + {7\over 3}\right)l_{83} + \left(l_{\alpha a}^{ij} + {7\over 3}\right)l_{84} 
+ \left(l_{ab}^{ij} + {4\over 3}\right)l_{85}, \nonumber\\
& + & \left(l_{0a}^{ij} + {10\over 3}\right)l_{86} 
+ \left(l_{0m}^{np} + {7\over 3}\right)l_{87} 
+ \left(l_{0m}^{n \alpha} + {7\over 3}\right)l_{88} + \left(l_{0m}^{\alpha\beta} + {7\over 3}\right)l_{89} 
+ \left(l_{0m}^{ab} + {1\over 3}\right)l_{90} \nonumber\\
&+& \left(l_{0\alpha}^{ab} + {1\over 3}\right)l_{91} + \left(l_{0m}^{na} + {4\over 3}\right)l_{92} 
+ \left(l_{0m}^{\alpha a} + {4\over 3}\right)l_{93} + \left(l_{0\alpha}^{\beta a} + {4\over 3}\right)l_{94} 
+ \left(l_{0m}^{ni} + {10\over 3}\right)l_{95} \nonumber\\
&+& \left(l_{0m}^{\alpha i} + {10\over 3}\right)l_{96} + \left(l_{0\alpha}^{\beta i} + {10\over 3}\right)l_{97} 
+ \left(l_{0m}^{ia} + {7\over 3}\right)l_{98} + \left(l_{0\alpha}^{ia} + {7\over 3}\right)l_{99} 
+ \left(l_{0a}^{bi} + {4\over 3}\right)l_{100}\nonumber\\
&+ & \left({\gamma k\over 6} {\red -{2\over 3}}\right) {\red n_3} + {1\over 3}\left(n_1 + n_2 + 4n_4\right) + \left({1\over 3} + \left[{n\over m}\right]\right)n_0, \nd
where we see that the extra proliferation does not create additional relative minus signs. However one would also need to justify that the dominant scalings $l_{0{\rm A}}^{\rm BC}$ of the G-flux components 
${\bf G}_{0{\rm ABC}}$ do not themselves become negative definite and violate the bound posed in 
\eqref{botsuga4}. In fact EFT constraints put the following bounds on the dominant scalings:

{\footnotesize
\bg\label{perfum}
l_{0{\rm M}}^{\rm NP} \ge -2, ~~~l_{0{\rm M}}^{ab} \ge 0, ~~~ l_{0{\rm N}}^{{\rm P}a} \ge -1, ~~~
l_{0{\rm M}}^{{\rm N}i} \ge -3, ~~~ l_{0{\rm N}}^{ia} \ge -2, ~~~ l_{0{a}}^{bi} \ge -1, ~~~
l_{0{i}}^{{j}a} \ge -3, \nd}
which may be easily inferred from \eqref{botsuga4}. Note that \eqref{perfum} {\it does not} imply that the actual dominant scalings take the lower bounds given above. In fact similar bounds from EFT may also be proposed for the remaining terms in \eqref{fahingsha5}:

{\footnotesize
\bg\label{perfum2}
l_{{\rm MN}}^{ia} \ge -1, ~~~l_{{\rm M}a}^{ij} \ge -2, ~~~ l_{{\rm M}i}^{ab} \ge 0, ~~~
l_{ab}^{ij} \ge -1, ~~~ l_{{\rm MN}}^{{\rm P}i} \ge -2, ~~~ l_{{\rm MN}}^{ij} \ge -3 \nd}
which are related to the so-called {\it external} fluxes and whose dynamics will be discussed in the next sub-section. It is easy to infer that the {\it internal} fluxes are bounded from below as $l_{\rm MN}^{\rm PQ} \ge -1, l_{\rm MN}^{{\rm P}a} \ge 0$ and $l_{\rm MN}^{ab} \ge 1$. Clearly our choice of $+1$ for each of them in \eqref{collateral} is consistent with the EFT bound. Interestingly the other choices from 
\eqref{automoon}, \eqref{renaissance2} and footnote \ref{mar_c} also appear to be consistent with the EFT bound. However out of these five allowed choices:

{\footnotesize 
\bg\label{pughmey}
\left(l_{0i}^{j{\rm M}}, l_{\rm MN}^{{\rm P}a}, l_{\rm MN}^{ab}, l_{\rm MN}^{{\rm PQ}}\right) = 
(-4, 1, 1, 1), ~~ (-4, 1, 2, 0), ~~\left(-4, 1, {4\over 3}, {2\over 3}\right), ~~\left(-4, 1, {5\over 3}, {1\over 3}\right),
~~(-3, 1, 2, 0),  \nonumber\\ \nd}
we will only take the first one to fix the scalings of the remaining G-flux components by comparing their 
EOMs against their dominant scalings.    
According to \eqref{perfum} and \eqref{perfum2}, the dominant scalings could take any values above or equal to the lower bounds. As we shall see below, this is exactly what will turn out from the corresponding flux EOMs, so this gives yet another motivation to work them out here.

Let us make couple more comments before we explicitly elucidate every flux 
components. {\Su One}, in \cite{desitter2}, the dominant scalings were all taken to be positive definite because the negative scalings could be Borel summed to an exponentially decaying factor at late time. In the following, for illustrative purpose, we will consider negative $g_s$ scalings\footnote{Despite the fact that these flux components would blow-up at late time where $g_s \to 0$.} as long as they are within the EFT bounds \eqref{perfum} and \eqref{perfum2} unless mentioned otherwise. How the trans-series behave when the negative $g_s$ scalings are summed over will be discussed later. {\Su Two}, in the flux EOMs from \eqref{mcacrisis} there will be a subset of equations where the temporal derivatives act on the rank seven tensors. As discussed in \eqref{senrem}, and will also be elaborated later in section \ref{toyeft}, the temporal derivative of $g_s$ should only be expressed as positive power of $g_s$. Following footnote 
\ref{camil}, we will take any additional powers of $g_s$ coming from the temporal derivatives to be {\it zero}. This can be easily rectified, but for all the de Sitter slicings studied in section \ref{sec2.1} this seems unnecessary so we will stick with the simplest case here.

\vskip.2in

\noindent {\it Case 1: ${\bf G}_{0{\rm MNP}}$ components}

\vskip.2in

\noindent Let us start by first studying the EOMs for the flux components ${\bf G}_{0{\rm MNP}}$ where 
$({\rm M, N}) \in {\cal M}_4 \times {\cal M}_2$. The various rank seven and eight tensors contributing to the EOMs \eqref{mcacrisis} are given in {\bf Table \ref{jhinuk5}}. A quick matching between rows 1 and 2 of 
{\bf Table \ref{jhinuk5}} reveals that: 
\bg\label{jadushman}
\theta_{nl} = 2\left(l_{\rm 0M}^{\rm NP} + {7\over 3}\right), \nd
which is exactly how the components ${\bf G}_{\rm 0MNP}$ contribute to \eqref{botsuga4}. The factor of 2 is the values that $\left(l_{87}, l_{88}, l_{89}\right)$ from \eqref{fahingsha5} would take to provide the kinetic terms. Additionally, 
as we know, the contributions to 
\eqref{mcacrisis} typically come from the types of derivatives acting on the rank seven tensors. Here we only have derivative along the temporal and the internal six directions corresponding to the three-form fields
${\bf C}_{\rm MNP}$ and ${\bf C}_{0{\rm MN}}$ respectively. These three-forms are also responsible for switching on  ${\bf X}_8^{(0, 0)}$ and ${\bf X}_8^{(m, \alpha)}$ respectively. The structure of 
${\bf X}_8^{(0, 0)}$ already appears in \eqref{thetrial4}, \eqref{thetrial5} and in \eqref{cora3} corresponding to the metric choices \eqref{evader}, generic toroidal dependent metric and \eqref{makibhalu3} respectively. For the case \eqref{makibhalu3}, and as we discussed in much detail in the previous subsection, the result for ${\bf X}_8$ reduces to the one for the metrc \eqref{evader} when $\gamma = 6$ in \eqref{makibhalu3}. In the following therefore we will only concentrate on this specific case of $\gamma = 6$ unless mentioned otherwise.

The ${\bf X}_8 \equiv {\bf X}_8^{(0, 0)}$ form related to the three-form field ${\bf C}_{\rm MNP}$, computed in 
\eqref{thetrial4} for the metric \eqref{evader} (or the metric \eqref{makibhalu3} with $\gamma = 6$) and \eqref{thetrial5} for the metric with generic dependence on all the eleven directions,  differs from the other
${\bf X}_8 \equiv {\bf X}_8^{(m, \alpha)}$ form. This difference may be quantified by following arrangements of the alphabets:
\bg\label{lilcadu00}
\overbracket[1pt][7pt]{{\rm P}~\underbracket[1pt][7pt]{{\rm Q}~{\rm R}}~{\rm S}} ~\overbracket[1pt][7pt]{{a}~\underbracket[1pt][7pt]{b~i} ~j} ~~~ 
+~~~~{\rm permutations}, \nd
underlying the rank eight nature of the eight-forms for the two metric choices \eqref{evader} and the generic one. 
For the former case, {\it i.e.} when the metric allows the special dependence on the toroidal directions as in 
\eqref{makibhalu3} with $\gamma= 6$ (or the metric \eqref{evader}), the ${\bf X}_8$ form becomes:
 \bg\label{thetrial20}
{\bf X}_8(x, y, w^a; g_s) = \sum_{\{l_i\}} \widetilde{\bf X}_{(8, 5a)}^{(l_1, ...., l_4)}(x, y, w^a) 
\left({g_s\over {\rm HH}_o}\right)^{{1\over 3}(l_1 + l_2 + l_3 + l_4)}, \nd
which is now related to the three-form ${\bf C}_{0{\rm MN}}$. Note that 
the dominant scaling is 0. This should be compared to the other ${\bf X}_8$ computed earlier, all of which had non-trivial dominant $g_s$ dependences. In fact \eqref{thetrial20} is the first appearance of 
the eight-form whose dominant scaling is time-independent. On the other hand, taking the metric to generically depend on all the eleven-dimensional coordinates, the eight-form becomes:
\bg\label{thetrial21}
{\bf X}_8(x, y, w^a; g_s) = \sum_{\{l_i\}} \widetilde{\bf X}_{(8, 5b)}^{(l_1, ...., l_4)}(x, y, w^a) 
\left({g_s\over {\rm HH}_o}\right)^{{1\over 3}(l_1 + l_2 + l_3 + l_4 - 8)}, \nd
which has a dominant scaling of $-8$. Both the eight-forms \eqref{thetrial20} and \eqref{thetrial21} contribute to the EOMs for ${\bf G}_{0{\rm MNP}}$ as may be seen from {\bf Table \ref{jhinuk5}}, and which we illustrate in the following.

\begin{table}[tb]  
 \begin{center}
\renewcommand{\arraystretch}{1.5}
\begin{tabular}{|c||c||c|}\hline ${\bf G}_{0{\rm MNP}}$ tensors & form & ${g_s\over {\rm HH}_o}$ scaling \\ \hline\hline
$\left(\mathbb{T}_7^{(f)}\right)_{{\rm QRS}abij}$ & $\sqrt{-{\bf g}_{11}} {\bf G}^{0{\rm MNP}} {\rm F}_1^{(n_1)}
{\rm F}_2^{(n_2)} \epsilon_{0{\rm MNP}{\rm QRS}abij}$ & $l_{0{\rm M}}^{\rm NP} + {2\over 3}\left(k_1 + k_{n_1} 
+ k_{n_2}\right)$ \\ \hline 
$\left(\mathbb{T}_7^{(q)}\right)_{{\rm QRS}abij}$ & $\sqrt{-{\bf g}_{11}} \mathbb{Y}_4^{0{\rm MNP}} 
\epsilon_{0{\rm MNP}{\rm QRS}abij}$ & $\theta_{nl}(k_2) - l_{0{\rm M}}^{\rm NP} - {14\over 3}  
- {2k_2\over 3}$ \\ \hline 
$\left(\mathbb{T}_8^{(0, 0)}\right)_{{\rm QRS}ab0ij}$ & ${\bf G}_{[{\rm QRS}a} {\bf G}_{b0ij]}$ & ${2\over 3}\left(k_4 + k_5\right) + l_{[{\rm QR}}^{[{\rm S}a} \oplus l_{{b0}]}^{ij]}$ 
\\ \hline 
$\left(\mathbb{T}_8^{(m, \alpha)}\right)_{{\rm PQRS}abij}$ & ${\bf G}_{[{\rm PQRS}} {\bf G}_{abij]}$ & ${2\over 3}\left(k_4 + k_5\right) + l_{[{\rm PQ}}^{[{{\rm RS}}} \oplus l_{ab]}^{ij]}$ 
\\ \hline 
$\left(\mathbb{X}_8^{(0, 0)}\right)_{{\rm QRS}ab0ij}$ & \eqref{thetrial4}, \eqref{thetrial5} 
& $l - 1, ~~l - 9$  \\ \hline
$\left(\mathbb{X}_8^{(m, \alpha)}\right)_{{\rm PQRS}abij}$ & \eqref{thetrial20}, \eqref{thetrial21}
& $l , ~~l - 8$ \\ \hline
\end{tabular}
\renewcommand{\arraystretch}{1}
\end{center}
 \caption[]{Comparing the $g_s$ scalings of the various tensors that contribute to the EOM for the G-flux components ${\bf G}_{0{\rm MN}P}$ in \eqref{mcacrisis}. Note that now we take $\gamma = 6$ in 
 \eqref{makibhalu3} so the results of the curvature etc. will get identified to the corresponding results from the metric \eqref{evader}. We will also show the results for the metric with generic dependence on the toroidal direction. The three forms contributing to the EOMs are now ${\bf C}_{\rm MNP}$ and 
 ${\bf C}_{0{\rm MN}}$.} 
  \label{jhinuk5}
 \end{table}

To start, let us compare the rows 1, 2, 3 and 5 in {\bf Table \ref{jhinuk5}} associated with ${\bf C}_{\rm MNP}$. The subtlety here is that the EOM 
will involve {\it temporal} derivatives. This means the dominant scalings of rows 1 and 2 should change by 
${\red -1}$, implying the dominant scalings of these rows become ${\Su l_{\rm 0M}^{\rm NP} - 1}$ and 
${\Su \theta_{nl} - l_{\rm 0M}^{\rm NP} - {17\over 3}}$ respectively. On the other hand, the third row involves the following product of the four-forms:
\bg\label{perfum3}
{\bf G}_{[{\rm QRS}a} {\bf G}_{b0ij]} = {1\over n}\left({\bf G}_{{\rm QR}ab} {\bf G}_{0ij{\rm S}} \pm {\bf G}_{{\rm QRS}a} {\bf G}_{0ijb} \pm {\bf G}_{0{\rm QRS}} {\bf G}_{abij} + ....\right), \nd
where $n$ takes care of the permutations and $\pm$ sign takes care of the antisymmetries. From 
\eqref{perfum} and \eqref{perfum2} the dominant scaling is $-4$ as is carried by ${\bf G}_{0ij{\rm M}}$ (see also {\bf Table \ref{jhinuk1}}). Thus the product of G-fluxes in row 3 can atmost have a dominant scaling of
$-3$. Alternatively, defining ${\red {\rm x} \equiv {g_s\over {\rm HH}_o}}$, we can express \eqref{perfum3} as:
\bg\label{perfumo}
{\rm log}_{\rm x}\Big({\bf G}_{[{\rm QRS}a} {\bf G}_{b0ij]}\Big) \ge -3, \nd
which would be a concise way to express the $g_s$ bound of the product of G-flux components {\it i.e.} on the rank eight tensors $\left(\mathbb{T}^{(k)}_8\right)_{{\rm N}_1...{\rm N}_7{\rm N}_k}$.
In general we expect from {\bf Table \ref{jhinuk5}}:

{\footnotesize
\bg\label{cora20}
l_{\rm 0M}^{\rm NP} - 1 + {2\over 3}\left(k_1 + k_{n_1} 
+ k_{n_2}\right) = \theta_{nl}(k_2) - l_{\rm MN}^{ab} - {17\over 3}  
- {2k_2\over 3} = {2\over 3}\left(k_4 + k_5\right) + l_{[{\rm QR}}^{[{\rm S}a} \oplus l_{{b0}]}^{ij]}, \nd}
where $k_i$ are the G-flux modings and $k_{n_i}$ are the ${\rm F}_i$ modings.
Now comparing it to the ${\bf X}_8$ scaling of $l - 1$ from \eqref{thetrial4}, it implies that 
$l_{0{\rm M}}^{\rm NP}$ can have a dominant scaling of either $-2$ or $0$. Since both are within the EFT bound from \eqref{perfum}, we will have to look further to fix the dominant scaling\footnote{We are ignoring the fact that $-2$ scaling would make the flux components to blow-up at late time. In the following, and as mentioned earlier, we will not worry too much about the negative $g_s$ scalings as long as they are within the EFT bounds. We will come back to this towards the end of section \ref{mrain}.}.

Looking further means comparing the rows 1, 2, 4 and 6 associated with the three-form 
${\bf C}_{0{\rm MN}}$. Since this involved derivatives with respect to the internal six directions, {\it i.e.} along ${\cal M}_4 \times {\cal M}_2$, the dominant scalings from rows 1 and 2 now remain what they are in 
{\bf Table \ref{jhinuk5}}. However the product of fluxes from row 4 will now involve:
\bg\label{perfum4}
{\bf G}_{[{\rm PQRS}} {\bf G}_{abij]} = {1\over n}\left({\bf G}_{{\rm PQ}ab} {\bf G}_{{\rm RS}ij} 
\pm {\bf G}_{{\rm PQR}a} {\bf G}_{{\rm S}bij} \pm {\bf G}_{{\rm PQR}i} {\bf G}_{{\rm S}abj} + ....\right), \nd
which can at most have a dominant scaling of $-2$ from \eqref{perfum2}. This may be easily seen by choosing ${\bf G}_{{\rm PQ}ij}$ with the lower bound $-3$ or ${\bf G}_{{\rm MNP}i}$ with the lower bound $-2$. The remaining G-flux components then provide the lower bound of $-2$. In other words:
\bg\label{perfumo2}
{\rm log}_{\rm x}\Big({\bf G}_{[{\rm PQRS}} {\bf G}_{abij]}\Big) \ge -2. \nd
The two lower bounds \eqref{perfumo} and \eqref{perfumo2}, much like the lower bounds in \eqref{perfum}
and \eqref{perfum2}, do not tell us what the actual scaling is. This means we have to look at all possible EOMs carefully. In general, and as in 
\eqref{cora20}, we expect from {\bf Table \ref{jhinuk5}}:

{\footnotesize
\bg\label{cora21}
l_{\rm 0M}^{\rm NP} + {2\over 3}\left(k_1 + k_{n_1} 
+ k_{n_2}\right) = \theta_{nl}(k_2) - l_{\rm MN}^{ab} - {14\over 3}  
- {2k_2\over 3} = {2\over 3}\left(k_4 + k_5\right) + l_{[{\rm PQ}}^{[{\rm RS}} \oplus l_{{ab}]}^{ij]}, \nd}
which should then be equated to the ${\bf X}_8$ form from \eqref{thetrial20}. The equalities proposed in 
\eqref{cora20} and \eqref{cora21} are possible for large values of $(k_i, k_{n_i})$ but for $k_i = k_{n_i} = 0$
this doesn't look possible. What is the most efficient way to work out the $g_s$ scalings of the various G-flux components when $k_i = k_{n_i} = 0$? One way would be to compare all the possible EOMs in the system and from there work out the $g_s$ scalings. This is in general hard because of the sheer number of possible components as well as the mixing between components. The other way, maybe slightly more efficient, would be to make an educated guess for the $g_s$ scalings and see whether this satisfies {\it at least} all the flux EOMs. The downside of such an approach is the {\it adhoc} nature of the analysis, plus the absence of a compelling rule to discern the various scalings.
 In the following therefore we will follow a mixture of both the procedures, namely, write all the EOMs and from there determine what are {\it all} the allowed possibilities for a given G-flux components. Once we have a scan of all the choices, we will try to find the ones that fit the full data. This is undoubtedly a hard exercise, but unfortunately there is no simpler way to solve the problem.

Looking at {\bf Table \ref{jhinuk5}}, we see that the most dominant scaling is the one carried by the flux components ${\bf G}_{0ij{\rm M}}$. Since this is fixed to $-4$ (see \eqref{collateral} or \eqref{pughmey}), the temporal derivative action will tell us to equate $l_{\rm 0M}^{\rm NP} -1$ with the $g_s$ scaling of either the product of the fluxes or the ${\bf X}_8$ polynomial. This will give: 
\bg\label{angeld1}
l_{\rm 0M}^{\rm NP} = -2~~~ {\rm or} ~~~ l_{\rm 0M}^{\rm NP} = 0, \nd 
respectively. Once we take $l_{0{\rm M}}^{\rm NP} = 0$, then it is easy to see that 
there is a match between first, second and the fifth rows as well as a match between second and the third rows
in {\bf Table \ref{jhinuk5}} provided:
\bg\label{hostag}
l \equiv {2\over 3}(k_1 + k_{n_1} + k_{n_2}),~~ \theta_1 = {14\over 3} + {2k_2\over 3} + l, ~~ \theta_{2} = {8\over 3} + {2\over 3}(k_2 + k_4 + k_5), \nd
where $(k_1, k_2)$ are the modings of the flux components ${\bf G}_{\rm 0MNP}$, 
$(k_{n_1}, k_{n_2})$ are the modings of $({\rm F}_1, {\rm F}_2)$, and $(k_4, k_5)$ are the modings of the dominant flux components ${\bf G}_{{\rm QR}ab}$ and ${\bf G}_{0ij{\rm M}}$ respectively. 
These two conditions will govern the EOM with temporal derivatives on the rank seven tensors. On the other hand, when we take $l_{\rm 0M}^{\rm NP} = -2$, the matching is between the first, second and the third rows, as well as between second and fifth rows in {\bf Table \ref{jhinuk5}}. This happens for:
\bg\label{hostag2}
\theta_1 = {2\over 3}(1 + k_1 + k_2 + k_{n_1} + k_{n_2}), ~~~ \theta_2 = {8\over 3} + {2k_2 \over 3} + l, \nd
where now $k_4 + k_5 = k_1 + k_{n_1} + k_{n_2}$. In the schematic construction of \eqref{lilcadu84}, 
the identification in \eqref{hostag} will imply $\theta_1 = \theta_{0, 2} = {14\over 3}$ and $\theta_{0, 3} = 
\theta_2 = {8\over 3}$ in \eqref{botsuga4} when $k_i = k_{n_i} = 0$. This leads to the following diagram at the lowest orders:
\bg\label{lilcadu85}
d_{(0)}\ast\overbracket[1pt][7pt]{{\bf G}_4 + b_1 d_{(0)}\ast\Big(\mathbb{Y}_4}
\overbracket[1pt][7pt]{(\theta_1) + \mathbb{Y}_4\underbracket[1pt][7pt]{(\theta_2) + \mathbb{Y}_4(\theta_3)  +  ..\Big) - b_2\Big(\mathbb{T}_8}_{\theta_2 \ge {8 \over 3}}{}^{(0, 0)}\Big) 
- b_3 \Big(\mathbb{X}_8}^{\theta_1 \ge {14 \over 3}}{}^{(0, 0)}\Big) = 0, \nonumber\\ \nd  
where $\mathbb{Y}_4(\theta_3)$ and beyond only start participating once we go to higher orders in the parameters $\left(k_i, k_{n_i}, l, {g_s\over {\rm HH}_o}\right)$.
Eventually, for sufficiently higher order, we expect the complete identifications of the various tensor scalings from \eqref{cora20} to be realized. On the other hand, for \eqref{hostag2}, the diagram at lowest orders becomes:
\bg\label{lilcadu86}
d_{(0)}\ast\overbracket[1pt][7pt]{{\bf G}_4 + b_1 d_{(0)}\ast\Big(\mathbb{Y}_4}
\overbracket[1pt][7pt]{(\theta_1) + \mathbb{Y}_4\underbracket[1pt][7pt]{(\theta_2) + \mathbb{Y}_4(\theta_3)  +  ..\Big) - b_3\Big(\mathbb{X}_8}_{\theta_2 \ge {8 \over 3}}{}^{(0, 0)}\Big) 
- b_2 \Big(\mathbb{T}_8}^{\theta_1 \ge {2 \over 3}}{}^{(0, 0)}\Big) = 0, \nonumber\\ \nd  
which means, compared to \eqref{lilcadu84}, $\theta_1 = \theta_{0, 3} = {2\over 3}$ and $\theta_2 = 
\theta_{0, 2} = {2\over 3}$ in \eqref{botsuga4} when $k_i = k_{n_i} = 0$. The identifications \eqref{cora20} will again be realized at higher orders in
 $\left(k_i, k_{n_i}, l, {g_s\over {\rm HH}_o}\right)$, much like what we had before. 

There are however few caveats associated with the identifications \eqref{lilcadu85} and \eqref{lilcadu86} that need to be pointed out now. {\Su One}, 
the choice of $l_{\rm 0M}^{\rm NP} = 0$, {\it i.e.} zero scaling of ${\bf G}_{\rm 0MNP}$, may look consistent with the $+1$ scaling of 
${\bf G}_{\rm MNPQ}$, with the latter implying a similar $+1$ scaling for the field ${\bf C}_{\rm MNP}$. This reasoning is not {\it always} true because ${\bf G}_4$ differs from $d{\bf C}_3$ by the quantum terms to avoid conflicting with the Bianchi identity. In fact it is these quantum terms whose dominant scaling governs the scaling of the G-flux components. {\Su Two}, the choice $l_{\rm 0M}^{\rm NP} = 0$ appears to clash with our original identification of $l_{\rm 0M}^{\rm NP} - 1$ with the various rank eight tensors and the quantum terms, since the vanishing of  $l_{\rm 0M}^{\rm NP}$ implies that the temporal derivative of the G-flux components 
${\bf G}_{\rm 0MNP}$ itself is zero. However if we view ${\bf G}_{\rm 0MNP}$ to scale as 
$\left({g_s\over {\rm HH}_o}\right)^\epsilon$ when $\epsilon \to 0$ for $l_{\rm 0M}^{\rm NP} \to 0$, then the derivative action does change $\epsilon$ to $\epsilon - 1$, but there will be an overall $\epsilon$ coefficient that will render the temporal derivative to zero. Interestingly, the fact that 
$d_{(0)} \ast \mathbb{Y}_4(\theta_1)$ is still identified to $\mathbb{X}_8^{(0, 0)}$, continues to provide a dynamical evolution of the remaining flux components. However since the contribution to $\theta_2$ comes from \eqref{botsuga4} as ${7\over 3}(l_{87} + l_{88} + l_{89})$, when $\theta_2 = {8\over 3}$, the equations may get highly constrained unless we incorporate non-perturbative  and/or non-local effects. Including non-perturbative BBS instanton effects from \eqref{beverwickmey}, will change $\theta_2$ from ${8\over 3}$ to 
${14\over 3}$, and provide next order corrections to $\theta_1$. Similarly, incorporating non-localities will have similar effects in changing $\theta_1$ and $\theta_2$ by the addition of ${2q\over 3}$ factors, where $q$ is the degree of non-locality (see \eqref{jesbielbon}). {\Su Three}, when $l_{\rm 0M}^{\rm NP} = -2$, the identifications \eqref{lilcadu86} tells us that $\theta_1 \ge {2\over 3}$. To lowest orders, and from 
\eqref{botsuga4}, $l_{87} = l_{88} = l_{89} = 2$ so $\mathbb{Y}_4^{\rm 0MNP} \propto {\bf G}^{\rm 0MNP}$, therefore the quantum term only renormalizes the corresponding flux components, leading to an algebraic relation between various components of fluxes. Thus the scenario is the same as with $l_{\rm 0M}^{\rm NP} = 0$ and $\theta_1 = {14\over 3}$. However now, since the contribution to $\theta_2$ comes from 
${1\over 3}(l_{87} + l_{88} + l_{89})$, $\theta_2 = {8\over 3}$ may actually provide non-trivial relations between fluxes and curvature forms, even in the absence of non-local and non-perturbative effects.

In the second case, 
when the derivatives along ${\cal M}_4 \times {\cal M}_2$ act on the rank seven tensors, one can match 
$l_{\rm 0M}^{\rm NP}$ with the $g_s$ scaling of either the product of the fluxes or the ${\bf X}_8$ polynomial.  The former is given in \eqref{perfumo2} and the latter in \eqref{thetrial20}, leading to:
\bg\label{angeld2}
l_{\rm 0M}^{\rm NP} \ge -2~~~ {\rm and/or} ~~~ l_{\rm 0M}^{\rm NP} = 0, \nd 
respectively. This may now be compared to \eqref{angeld1} which had {\it two} different choices whereas here there is a possibility of an overlap at $l_{\rm 0M}^{\rm NP} = 0$ because the first relation in 
\eqref{angeld2} only provides a lower bound (which in turn is consistent with the EFT bound from \eqref{perfum}). For the case when $l_{\rm 0M}^{\rm NP} = 0$, 
the matching happens between the first, second and the sixth rows, and between the second and the fourth rows of {\bf Table \ref{jhinuk5}}. The schematic diagram for this case is exactly as in \eqref{lilcadu85} with 
$(\theta_1, \theta_2) \ge \left({14\over 3}, {8\over 3}\right)$ in \eqref{botsuga4} with the following notational changes: 
\bg\label{sevignic}
d_{(0)} \to d_{({\rm M})}, ~~~~~ \mathbb{T}^{(0, 0)}_8 \to \mathbb{T}^{(m, \alpha)}_8, ~~~~~ 
\mathbb{X}^{(0, 0)}_8 \to \mathbb{X}^{(m, \alpha)}_8. \nd
For the other case when 
$l_{\rm 0M}^{\rm NP} \ge -2$, the matching is between first, second and fourth rows and between second and sixth rows of {\bf Table \ref{jhinuk5}}. The schematic diagram is the same as in \eqref{lilcadu86}
with $(\theta_1, \theta_2) \ge \left({2 \over 3}, {8\over 3}\right)$ in \eqref{botsuga4} with the notational changes \eqref{sevignic}, implying that the concerns we had earlier, reappears here too. This 
seems to imply that although $l_{\rm 0M}^{\rm NP} \ge -2$ may lead to consistent dynamics (modulo the blow-up at late time), the choice 
$l_{\rm 0M}^{\rm NP} = 0$ may be a better fit in the presence of non-local and non-perturbative effects.

\vskip.2in

\noindent {\it Case 2: ${\bf G}_{{\rm 0NP}a}$ components}

\vskip.2in

\noindent Our next case with ${\bf G}_{{\rm 0NP}a}$ components, all three-form fluxes contributing to the EOMs are of the kind ${\bf C}_{{\rm NP}a}$ with derivative along the temporal direction, ${\bf C}_{\rm 0NP}$ with derivatives along the toroidal directions ${\mathbb{T}^2\over {\cal G}}$; and ${\bf C}_{{\rm 0P}a}$ with derivatives along the internal six directions ${\cal M}_4 \times {\cal M}_2$. These three-forms are not only useful to write the EOMs, but also to express the corresponding eight-forms ${\bf X}_8$. For example, the ${\bf X}_8$ forms associated with ${\bf C}_{\rm 0NP}$ has already been worked out in \eqref{thetrial20} for the metric choice \eqref{evader}, and in \eqref{thetrial21} for the metric with generic dependence on all the eleven directions. Similarly, the ${\bf X}_8$ forms associated with ${\bf C}_{{\rm NP}a}$ appear in \eqref{thetrial7} and \eqref{thetrial8} for the aforementioned two metric choices respectively. For the metric choice \eqref{makibhalu3}, the ${\bf X}_8$ form resembles the one with the metric choice \eqref{evader} (at least with respect to the $g_s$ scaling) when we take $\gamma = 6$ in \eqref{makibhalu3}. Since this is the case we will henceforth concentrate on, we will not worry about the generic $\gamma$ dependent metric 
\eqref{makibhalu3}. 

For the three-form ${\bf C}_{{\rm 0P}a}$, the ${\bf X}_8$ form involves four curvature two-forms in various 
possible permutations much like what we had for example in \eqref{lilcadu00}. For us here, this will involve the following arrangements of alphabets:
\bg\label{lilcadu0N}
\overbracket[1pt][7pt]{{\rm Q}~\underbracket[1pt][7pt]{{\rm R}~{\rm S}}~{\rm T}} ~\overbracket[1pt][7pt]{{\rm M}~\underbracket[1pt][7pt]{b~i} ~j} ~~~ 
+~~~~{\rm permutations}, \nd
underlying the rank eight nature of the eight-forms for the two metric choices \eqref{makibhalu3} with $\gamma = 6$ and the generic one. 
For the former case, {\it i.e.} when the metric allows special dependence on the toroidal directions as in 
\eqref{makibhalu3} with $\gamma = 6$, the ${\bf X}_8$ form becomes:
 \bg\label{thetrial22}
{\bf X}_8(x, y, w^a; g_s) = \sum_{\{l_i\}} \widetilde{\bf X}_{(8, 6a)}^{(l_1, ...., l_4)}(x, y, w^a) 
\left({g_s\over {\rm HH}_o}\right)^{{1\over 3}(l_1 + l_2 + l_3 + l_4 - 3)}, \nd
which may be compared to \eqref{thetrial20}. Note that now
the dominant scaling is $-1$.  On the other hand, taking the metric to generically depend on all the eleven-dimensional coordinates, the eight-form becomes:
\bg\label{thetrial23}
{\bf X}_8(x, y, w^a; g_s) = \sum_{\{l_i\}} \widetilde{\bf X}_{(8, 6b)}^{(l_1, ...., l_4)}(x, y, w^a) 
\left({g_s\over {\rm HH}_o}\right)^{{1\over 3}(l_1 + l_2 + l_3 + l_4 - 27)}, \nd
which has a dominant scaling of $-9$. Both the eight-forms \eqref{thetrial22} and \eqref{thetrial23} contribute to the EOMs for ${\bf G}_{0{\rm MP}a}$ as may be seen from {\bf Table \ref{jhinuk6}}, and which we shall elaborate in the following.

Compared to the situation in {\bf Table \ref{jhinuk5}} for the G-flux components ${\bf G}_{\rm 0MNP}$, there are three different class of EOMs associated with three possible ways in which the derivatives may act on the rank seven tensors in {\bf Table \ref{jhinuk6}}: the temporal, the toroidal and the internal six directions. Let us start with the case with the temporal derivative. First it is easy to see that the product of the fluxes are bounded from below as:
\bg\label{jules1}
{\rm log}_{\rm x}\left({\bf G}_{[{\rm QRST}} {\bf G}_{b0ij]}\right) \ge -4, \nd
where ${\rm x}$ is defined just above \eqref{perfumo}. The bound appears from the EFT constraints 
\eqref{perfum} and \eqref{perfum2} as we saw earlier. From the first row in {\bf Table \ref{jhinuk6}} we see that $l_{\rm 0N}^{{\rm P}a} - 3$ (where the extra $-1$ comes from the temporal derivative) can be matched with the $g_s$ scalings of either row 3 or row 6. Since we are taking the metric \eqref{makibhalu3} with $\gamma = 6$, or the metric \eqref{evader}, the latter comes directly from \eqref{thetrial7}. This immediately gives us:
\bg\label{lonpa}
l_{\rm 0N}^{{\rm P}a} \ge -1, ~~~ {\rm and/or} ~~~ l_{\rm 0N}^{{\rm P}a} = 1, \nd
respectively, as the allowed two cases. Depending on what the actual $g_s$ scalings of the flux components turn out, the two cases could actually be overlapping. Taking $l_{\rm 0N}^{{\rm P}a} \ge -1$, now leads to exactly the same schematic diagram as in \eqref{lilcadu86} with $(\theta_1, \theta_2) = 
\left({2\over 3}, {8\over 3}\right)$ in \eqref{botsuga4}. The value of $\theta_1$, which appears from matching rows 1 and 2 in {\bf Table \ref{jhinuk6}}, is consistent with how these components contribute to \eqref{botsuga4}, namely:
\bg\label{renaroy}
\theta_{nl} \equiv \theta_1 = 2\left(l_{\rm 0N}^{{\rm P}a} + {4\over 3}\right), \nd
with the factor 2 represents the values $\left(l_{92}, l_{93}, l_{94}\right)$ would take to produce the kinetic terms from \eqref{fahingsha5}. When $l_{\rm 0N}^{{\rm P}a} = 1$, the schematic diagram resembles \eqref{lilcadu85} with $(\theta_1, \theta_2) = \left({14\over 3}, {8\over 3}\right)$. Again, since the contributions to $\theta_2$ comes from ${7\over 3}\left(l_{92} + l_{93} + l_{94}\right)$, $\theta_2 = 
{8\over 3}$ can become over-constraining unless we incorporate the effects of the BBS instantons 
from \eqref{beverwickmey} changing $\theta_2$ to ${14\over 3}$, or the effects of the $q$-th order non-localities changing $\theta_2$ generically to ${2\over 3}(4 + q)$ (or to $2\left(q + {4\over 3}\right)$ if we incorporate BBS type non-local non-perturbative effects). 
On the other hand, we could also take $l_{\rm 0N}^{{\rm P}a} = 0$. In that case the diagram will become:
\bg\label{lilcadu87}
d_{(0)}\ast\overbracket[1pt][7pt]{{\bf G}_4 + b_1 d_{(0)}\ast\Big(\mathbb{Y}_4}^{\theta_1 \ge {8\over 3}}(\theta_1) +
\mathbb{Y}_4\overbracket[1pt][7pt]{(\theta_{0, 2}) + \mathbb{Y}_4\underbracket[1pt][7pt]{(\theta_{0, 3})  +  ..\Big)- b_2\Big(\mathbb{T}_8}_{\theta_{0, 3} \ge {5\over 3}}{}^{(0, 0)}\Big) -b_3 \Big(\mathbb{X}_8}^{\theta_{0, 2} \ge {11\over 3}}{}^{(0, 0)}\Big) = 0, \nonumber\\ \nd
with three possible quantum series participating unless $\theta_1 = \theta_{0, 2}$ or $\theta_1 = \theta_{0, 3}$. Again $\theta_{0, 3}$ can change from ${5\over 3}$ to ${11\over 3}$ with the incorporations of BBS instantons from \eqref{beverwickmey}. Eventually, for large enough $g_s$ scalings, we expect locally:

{\footnotesize
\bg\label{hitlib}
l_{\rm 0N}^{{\rm P}a} - 3 + {2\over 3}(k_1 + k_{n_1} + k_{n_2}) = \theta_{nl} - l_{\rm 0N}^{{\rm P}a} - {17\over 3}  - {2k_2\over 3} = {2\over 3}(k_4 + k_5) + l_{[{\rm QR}}^{[{\rm ST}} \oplus l_{b0]}^{ij]}, \nd}
which should further be identified to $l - 2$ appearing from \eqref{thetrial7} for the metric \eqref{makibhalu3} with $\gamma = 6$, or with $l -10$ coming from the metric with generic dependence on all the eleven-dimensional coordinates. The latter doesn't lead to an EFT description so we will refrain from elaborating this case further.

When the derivatives on the rank seven tensors in \eqref{mcacrisis} act along the toroidal directions, the story is different from what we had earlier. First, the product of the fluxes are now bounded from below as:
\bg\label{jules2}
{\rm log}_{\rm x}\left({\bf G}_{[{\rm QRST}} {\bf G}_{abij]}\right) \ge -2, \nd
instead of $-4$ in \eqref{jules1} earlier. There are other differences also: $l_{\rm 0N}^{{\rm P}a}$, instead of 
$l_{\rm 0N}^{{\rm P}a} - 2$, is now identified with the $g_s$ scalings of either the product of fluxes in row 4, or with the ${\bf X}_8$ polynomial is row 7 of {\bf Table \ref{jhinuk6}}. The latter comes from \eqref{thetrial20} as we saw before, and putting everything together now leads to the following two possibilities:
\bg\label{lonpa2}
l_{\rm 0N}^{{\rm P}a} \ge -2 ~~~ {\rm and/or} ~~~ l_{\rm 0N}^{{\rm P}a} = 0, \nd
compared to \eqref{lonpa} earlier. They can again be overlapping, but this can only be ascertained once we know the precise scalings of all the flux components. Unfortunately, comparing to \eqref{perfum} we see that 
$l_{\rm 0N}^{{\rm P}a} \ge -2$ violates the EFT bound, so all scaling between $-2 \le l_{\rm 0N}^{{\rm P}a} < -1$ are eliminated. The schematic diagram for $l_{\rm 0N}^{{\rm P}a} = 0$ now becomes:

{\footnotesize
\bg\label{lilcadu88}
d_{(a)}\ast\overbracket[1pt][7pt]{{\bf G}_4 + b_1 d_{(0)}\ast\Big(\mathbb{Y}_4}
\overbracket[1pt][7pt]{(\theta_1) + \mathbb{Y}_4\underbracket[1pt][7pt]{(\theta_2) + \mathbb{Y}_4(\theta_3)  +  ..\Big) - b_2\Big(\mathbb{T}_8}_{\theta_2 \ge {2 \over 3}}{}^{(a, b)}\Big) 
- b_3 \Big(\mathbb{X}_8}^{\theta_1 \ge {8 \over 3}}{}^{(a, b)}\Big) = 0, \nd}  
where $\theta_1 = {8\over 3}$ is again consistently {\it double} the effective scaling with which 
${\bf G}_{{\rm 0NP}a}$ enters \eqref{fahingsha5} or \eqref{botsuga4}. However $\theta_2 = {2\over 3}$ is alarming as the minimum $g_s$ scaling should be ${4\over 3}$ for this to make sense. Clearly non-perturbative effects are again necessary and with the addition of BBS instantons from \eqref{beverwickmey}
would convert $\theta_2$ to  ${8\over 3}$. Additional non-local effects would change this further in the same vein as we saw earlier. Taking $l_{\rm 0N}^{{\rm P}a} = -1$,  leads to a diagram of the form:
\bg\label{lilcadu89}
d_{(a)}\ast\overbracket[1pt][7pt]{{\bf G}_4 + b_1 d_{(a)}\ast\Big(\mathbb{Y}_4}^{\theta_1 \ge {2\over 3}}(\theta_1) +
\mathbb{Y}_4\overbracket[1pt][7pt]{(\theta_{a, 2}) + \mathbb{Y}_4\underbracket[1pt][7pt]{(\theta_{a, 3})  +  ..\Big)- b_2\Big(\mathbb{T}_8}_{{\red \theta_{a, 3} >  -{1\over 3}}}{}^{(a, b)}\Big) -b_3 \Big(\mathbb{X}_8}^{\theta_{a, 2} \ge {5 \over 3}}{}^{(a, b)}\Big) = 0, \nonumber\\ \nd
where, since the effective scaling of ${\bf G}_{{\rm 0NP}a}$ participating in \eqref{fahingsha5} and \eqref{botsuga4} is ${1\over 3}$, both $\theta_1 = {2\over 3}$ and $\theta_{a, 2} = {5\over 3}$ fit well with the quantum dynamics. However ${\red \theta_{a, 3} = -{1\over 3}}$ can be alarming because it apparently breaks the $g_s$ hierarchy in the system, leading to an eventual breakdown of EFT. Fortunately, 
$\theta_{a, 3} = \pm {1\over 3}$ cannot be accommodated in the system because there are no quantum terms contributing to the dynamics for such low values (recall that the minimum scaling required here is 
${2\over 3}$). Thus non-perturbative, with the addition of BBS instantons, $\theta_{a, 3} = {5\over 3}$ from 
\eqref{beverwickmey}, and we could have consistent dynamics. On the other hand at higher orders in $\left(k_i, k_{n_i}, l, {g_s\over {\rm HH}_o}\right)$ 
we expect, at least perturbatively:

\begin{table}[tb]  
 \begin{center}
\renewcommand{\arraystretch}{1.5}
\begin{tabular}{|c||c||c|}\hline ${\bf G}_{0{\rm NP}a}$ tensors & form & ${g_s\over {\rm HH}_o}$ scaling \\ \hline\hline
$\left(\mathbb{T}_7^{(f)}\right)_{{\rm QRST}bij}$ & $\sqrt{-{\bf g}_{11}} {\bf G}^{0{\rm NP}a} {\rm F}_1^{(n_1)}
{\rm F}_2^{(n_2)} \epsilon_{0{\rm NP}a{\rm QRST}bij}$ & $l_{0{\rm N}}^{{\rm P}a} - 2 + {2\over 3}\left(k_1 + k_{n_1} 
+ k_{n_2}\right)$ \\ \hline 
$\left(\mathbb{T}_7^{(q)}\right)_{{\rm QRST}bij}$ & $\sqrt{-{\bf g}_{11}} \mathbb{Y}_4^{0{\rm NP}a} 
\epsilon_{0{\rm NP}a{\rm QRST}bij}$ & $\theta_{nl}(k_2) - l_{\rm 0N}^{{\rm P}a} - {14\over 3}  
- {2k_2\over 3}$ \\ \hline 
$\left(\mathbb{T}_8^{(0)}\right)_{{\rm QRST}b0ij}$ & ${\bf G}_{[{\rm QRST}} {\bf G}_{b0ij]}$ & ${2\over 3}\left(k_4 + k_5\right) + l_{[{\rm QR}}^{[{\rm ST}} \oplus l_{{b0}]}^{ij]}$ 
\\ \hline 
$\left(\mathbb{T}_8^{(a, b)}\right)_{{\rm QRST}abij}$ & ${\bf G}_{[{\rm QRST}} {\bf G}_{abij]}$ & ${2\over 3}\left(k_4 + k_5\right) + l_{[{\rm QR}}^{[{{\rm ST}}} \oplus l_{{a}b]}^{ij]}$ 
\\ \hline 
$\left(\mathbb{T}_8^{(m, \alpha)}\right)_{{\rm MQRST}bij}$ & ${\bf G}_{[{\rm MQRS}} {\bf G}_{b{\rm T}ij]}$ & ${2\over 3}\left(k_4 + k_5\right) + l_{[{\rm MQ}}^{[{{\rm RS}}} \oplus l_{b{\rm T}]}^{ij]}$ 
\\ \hline 
$\left(\mathbb{X}_8^{(0)}\right)_{{\rm QRST}b0ij}$ & \eqref{thetrial7}, \eqref{thetrial8}
& $l - 2, l - 10$ \\ \hline
$\left(\mathbb{X}_8^{(a, b)}\right)_{{\rm QRST}abij}$ & \eqref{thetrial20}, \eqref{thetrial21} 
& $l, l - 8$ \\ \hline
$\left(\mathbb{X}_8^{(m, \alpha)}\right)_{{\rm MQRST}bij}$ & \eqref{thetrial22}, \eqref{thetrial23}
& $l - 1, l - 9$ \\ \hline
\end{tabular}
\renewcommand{\arraystretch}{1}
\end{center}
 \caption[]{Comparing the $g_s$ scalings of the various tensors that contribute to the EOM for the G-flux components ${\bf G}_{0{\rm NP}a}$ in \eqref{mcacrisis}. The three-form fluxes contributing to the EOMs are now ${\bf C}_{{\rm NP}a}, {\bf C}_{0{\rm NP}}$ and ${\bf C}_{0{\rm P}a}$ with derivatives acting along the temporal, toroidal and the internal six-directions respectively.} 
  \label{jhinuk6}
 \end{table}

{\footnotesize
\bg\label{hitlib2}
l_{\rm 0N}^{{\rm P}a} + {2\over 3}(k_1 + k_{n_1} + k_{n_2}) = \theta_{nl} - l_{\rm 0N}^{{\rm P}a} - {8 \over 3}  - {2k_2\over 3} = {2\over 3}(k_4 + k_5) + l_{[{\rm QR}}^{[{\rm ST}} \oplus l_{ab]}^{ij]} = l, \nd} 
where $l \in {\mathbb{Z}\over 3}$ for the metric choice \eqref{makibhalu3} with $\gamma = 6$.
When the derivatives on the rank seven tensors in \eqref{mcacrisis} act along ${\cal M}_4 \times {\cal M}_2$ directions, the story is again different from the two earlier cases. From {\bf Table \ref{jhinuk6}} we now see that $l_{\rm 0N}^{{\rm P}a} - 2$ can be identified with either the quantum terms, or the product of fluxes or the ${\bf X}_8$ polynomial. The $g_s$ scaling of the product of the fluxes is now bounded from below as:
\bg\label{lalonde}
{\rm log}_{\rm x}\left({\bf G}_{[{\rm MQRS}} {\bf G}_{b{\rm T}ij]}\right) \ge -3, \nd
which can be easily ascertained from the bounds \eqref{perfum} and \eqref{perfum2} and 
${\rm x} = {g_s\over {\rm HH}_o}$. Similarly the $g_s$ scalings of the relevant ${\bf X}_8$ polynomial can be read up from \eqref{thetrial22} for the metric \eqref{makibhalu3} with $\gamma = 6$ (or from \eqref{evader}), and from \eqref{thetrial23} for the metric with generic dependence on all the eleven dimensions. Comparing 
$l_{\rm 0N}^{{\rm P}a} - 2$ with the $g_s$ scalings of the flux products and the ${\bf X}_8$ polynomial lead to:
\bg\label{lonpa3}
l_{\rm 0N}^{{\rm P}a}  \ge -1, ~~~ {\rm and/or}~~~ l_{\rm 0N}^{{\rm P}a} = 1, \nd
respectively, where both are within the EFT bound \eqref{perfum} and are also overlapping, similar to \eqref{lonpa}. The case with 
$l_{\rm 0N}^{{\rm P}a}$ was dealt above when the derivatives on the rank seven tensors in 
\eqref{mcacrisis} act along the toroidal directions. For this case, taking $l_{\rm 0N}^{{\rm P}a} = -1$ leads to the following schematic diagram:
\bg\label{lilcadu90}
d_{({\rm M})}\ast\overbracket[1pt][7pt]{{\bf G}_4 + b_1 d_{({\rm M})}\ast\Big(\mathbb{Y}_4}
\overbracket[1pt][7pt]{(\theta_1) + \mathbb{Y}_4\underbracket[1pt][7pt]{(\theta_2) + \mathbb{Y}_4(\theta_3)  +  ..\Big) - b_3\Big(\mathbb{X}_8}_{\theta_2 \ge {8 \over 3}}{}^{(m, \alpha)}\Big) 
- b_2 \Big(\mathbb{T}_8}^{\theta_1 \ge {2 \over 3}}{}^{(m, \alpha)}\Big) = 0, \nonumber\\ \nd  
where $\theta_1 = {2\over3}$ is as expected double the effective scaling of ${1\over 3}$ for the flux components ${\bf G}_{{\rm 0NP}a}$. Similarly $\theta_2 = {8\over 3}$ perfectly provides adequate number of quantum terms from \eqref{fahingsha5} to balance the corresponding ${\bf X}_8$ polynomial that scales as 
$\left({g_s\over {\rm HH}_o}\right)^{l - 1}$ where $l \in {\mathbb{Z}\over 3}$. At higher orders in $\left(k_i, k_{n_i}, l, {g_s\over {\rm HH}_o}\right)$ we expect, similar to \eqref{hitlib2}, all terms would balance in the following way:

{\footnotesize
\bg\label{hitlib3}
l_{\rm 0N}^{{\rm P}a} - 2 + {2\over 3}(k_1 + k_{n_1} + k_{n_2}) = \theta_{nl} - l_{\rm 0N}^{{\rm P}a} - {14\over 3}  - {2k_2\over 3} = {2\over 3}(k_4 + k_5) + l_{[{\rm MQ}}^{[{\rm RS}} \oplus l_{b{\rm T}]}^{ij]} = l-1, \nd}
where $l \in {\mathbb{Z}\over 3}$ for the metric choice \eqref{makibhalu3} with $\gamma = 6$. Finally when 
$l_{\rm 0N}^{{\rm P}a} = 0$, the schematic diagram is similar to what we had in \eqref{lilcadu87}, except for a few important notational changes:

{\footnotesize
\bg\label{lilcadu91} 
d_{({\rm M})}\ast\overbracket[1pt][7pt]{{\bf G}_4 + b_1 d_{({\rm M})}\ast\Big(\mathbb{Y}_4}^{\theta_1 \ge {8\over 3}}(\theta_1) +
\mathbb{Y}_4\overbracket[1pt][7pt]{(\theta_{{\rm M}, 2}) + \mathbb{Y}_4\underbracket[1pt][7pt]{(\theta_{{\rm M}, 3})  +  ..\Big)- b_2\Big(\mathbb{T}_8}_{\theta_{{\rm M}, 3} \ge {5\over 3}}{}^{(m, \alpha)}\Big) -b_3 \Big(\mathbb{X}_8}^{\theta_{{\rm M}, 2} \ge {11\over 3}}{}^{(m, \alpha)}\Big) = 0, \nd}
showing that quantum terms participating for the temporal derivative case in \eqref{lilcadu87} may be related to the quantum terms participating for the case \eqref{lilcadu91} when the derivatives act along the internal six directions ${\cal M}_4 \times {\cal M}_2$. Similar notational changes may be incorporated for 
$l_{\rm 0N}^{{\rm P}a}$ appropriately from {\bf Table \ref{jhinuk6}} showing consistent behavior of 
${\bf G}_{{\rm 0NP}a}$ components for the whole range $-1 \le l_{\rm 0N}^{{\rm P}a} \le 1$. 

\begin{table}[tb]  
 \begin{center}
\renewcommand{\arraystretch}{1.5}
\begin{tabular}{|c||c||c|}\hline ${\bf G}_{0{\rm N}ab}$ tensors & form & ${g_s\over {\rm HH}_o}$ scaling \\ \hline\hline
$\left(\mathbb{T}_7^{(f)}\right)_{{\rm MPQRS}ij}$ & $\sqrt{-{\bf g}_{11}} {\bf G}^{0{\rm N}ab} {\rm F}_1^{(n_1)}
{\rm F}_2^{(n_2)} \epsilon_{0{\rm N}ab{\rm MPQRS}ij}$ & $l_{\rm 0N}^{ab} - 4 + {2\over 3}\left(k_1 + k_{n_1} 
+ k_{n_2}\right)$ \\ \hline 
$\left(\mathbb{T}_7^{(q)}\right)_{{\rm MPQRS}ij}$ & $\sqrt{-{\bf g}_{11}} \mathbb{Y}_4^{0{\rm N}ab} 
\epsilon_{0{\rm N}ab{\rm MPQRS}ij}$ & $\theta_{nl}(k_2) - l_{\rm 0N}^{ab} - {14\over 3}  
- {2k_2\over 3}$ \\ \hline 
$\left(\mathbb{T}_8^{(0)}\right)_{{\rm MPQRS}0ij}$ & ${\bf G}_{[{\rm MPQR}} {\bf G}_{0ij{\rm S}]}$ & ${2\over 3}\left(k_4 + k_5\right) + l_{[{\rm MP}}^{[{\rm QR}} \oplus l_{{0i}]}^{j{\rm S}]}$ 
\\ \hline 
$\left(\mathbb{T}_8^{(m, \alpha)}\right)_{{\rm MNPQRS}ij}$ & ${\bf G}_{[{\rm MNPQ}} {\bf G}_{{\rm S}bij]}$ & ${2\over 3}\left(k_4 + k_5\right) + l_{[{\rm MN}}^{[{{\rm PQ}}} \oplus l_{{\rm S}b]}^{ij]}$ 
\\ \hline 
$\left(\mathbb{T}_8^{(a, b)}\right)_{{\rm MPQRS}bij}$ & ${\bf G}_{[{\rm MPQR}b} {\bf G}_{{\rm S}bij]}$ & ${2\over 3}\left(k_4 + k_5\right) + l_{[{\rm MP}}^{[{{\rm QR}}} \oplus l_{{\rm S}b]}^{ij]}$ 
\\ \hline 
$\left(\mathbb{X}_8^{(0)}\right)_{{\rm MPQRS}0ij}$ & \eqref{thetrial10}, \eqref{thetrial11} 
& $l - 3, l - 11$ \\ \hline
$\left(\mathbb{X}_8^{(m, \alpha)}\right)_{{\rm MNPQRS}ij}$ & \eqref{meggy1}, \eqref{meggy2}
& $l - 2, l - 10$ \\ \hline
$\left(\mathbb{X}_8^{(a, b)}\right)_{{\rm MPQRS}bij}$ & \eqref{thetrial22}, \eqref{thetrial23}
& $l - 1, l - 9$ \\ \hline
\end{tabular}
\renewcommand{\arraystretch}{1}
\end{center}
 \caption[]{Comparing the $g_s$ scalings of the various tensors that contribute to the EOM for the G-flux components ${\bf G}_{0{\rm N}ab}$ in \eqref{mcacrisis}. As before, we have kept $\gamma = 6$ in 
 \eqref{makibhalu3} and considered the three-forms ${\bf C}_{{\rm N}ab}, {\bf C}_{0ab}$ and 
 ${\bf C}_{0{\rm N}a}$ with derivatives acting along the temporal, the internal six  and the toroidal directions respectively to determine the EOM.} 
  \label{jhinuk7}
 \end{table}

\vskip.2in

\noindent {\it Case 3: ${\bf G}_{{\rm 0N}ab}$ components}

\vskip.2in

\noindent We start by making a consistency check from the $g_s$ scalings of the various rank seven and eight tensors for the flux components ${\bf G}_{{\rm 0N}ab}$ from {\bf Table \ref{jhinuk7}}: equating the 
$g_s$ scalings from rows 1 and 2 gives us:
\bg\label{gonab}
\theta_{nl} = 2\left(l_{\rm 0N}^{ab} + {1\over 3}\right), \nd
which is precisely how these components contribute to \eqref{botsuga4}.  The factor of $2$ is the value that either $l_{90}$ or $l_{91}$ from \eqref{fahingsha5} would take in the first place to provide the lowest order kinetic terms. The contributing three-forms are now: ${\bf C}_{{\rm N}ab}$ with temporal derivative, 
${\bf C}_{0ab}$ with derivatives along ${\cal M}_4 \times {\cal M}_2$ directions, and ${\bf C}_{{\rm 0N}a}$ with derivatives along the toroidal directions. The $g_s$ dependence of the ${\bf X}_8$ form associated with the three-forms ${\bf C}_{{\rm N}ab}$ and ${\bf C}_{{\rm 0N}a}$ have already been worked out in 
\eqref{thetrial10} and \eqref{thetrial22} respectively for the metric \eqref{makibhalu3} with $\gamma = 6$ (or the metric \eqref{evader}; and in \eqref{thetrial11} and \eqref{thetrial23} respectively for the metric with generic dependence on all the eleven dimensions. For the three-form ${\bf C}_{0ab}$, the ${\bf X}_8$ polynomial takes the following form:
\bg\label{meggy1}
{\bf X}_8(x, y, w^a; g_s) = \sum_{\{l_i\}} \widetilde{\bf X}_{(8, 7a)}^{(l_1, ...., l_4)}(x, y, w^a) 
\left({g_s\over {\rm HH}_o}\right)^{{1\over 3}(l_1 + l_2 + l_3 + l_4 - 6)}, \nd
where one may remove the toroidal dependence if one wants to take the metric \eqref{evader}. Expectedly the $g_s$ dependence does not change. The above behavior of the polynomial appears from the following arrangements of alphabets in the wedge product of the curvature two-forms:
\bg\label{lilcadu88}
\overbracket[1pt][7pt]{{\rm M}~\underbracket[1pt][7pt]{{\rm N}~{\rm P}}~{\rm Q}} ~\overbracket[1pt][7pt]{{\rm R}~\underbracket[1pt][7pt]{{\rm S}~i} ~j} ~~~ 
+~~~~{\rm permutations}, \nd
which may be compared to the other arrangements of the curvature two-forms. Once we take the metric with generic dependence on all the eleven dimensions, the ${\bf X}_8$ form changes from \eqref{meggy1} to the following:
\bg\label{meggy2}
{\bf X}_8(x, y, w^a; g_s) = \sum_{\{l_i\}} \widetilde{\bf X}_{(8, 7b)}^{(l_1, ...., l_4)}(x, y, w^a) 
\left({g_s\over {\rm HH}_o}\right)^{{1\over 3}(l_1 + l_2 + l_3 + l_4 - 30)}, \nd
which would clearly alter the behavior of the quantum terms entering the dynamics. Fortunately the {\it absence} of an EFT description in this case allows us to not worry about the dynamics. All the scalings of the ${\bf X}_8$ polynomials relevant for the EOMs of ${\bf G}_{{\rm 0N}ab}$ appears in 
{\bf Table \ref{jhinuk7}}, which we shall analyze in the following.

Our starting point, as before, would be the ones with the temporal derivatives on the rank seven tensors 
in {\bf Table \ref{jhinuk7}}. From {\bf Table \ref{jhinuk7}}, we now see that $l_{\rm 0N}^{ab} - 5$ can now be balanced with the $g_s$ scalings of either the quantum terms, the product of the fluxes or the ${\bf X}_8$ polynomials. The product of the fluxes, on the other hand, are bounded from below as (we define 
${\rm x} \equiv {g_s\over {\rm HH}_o}$):
\bg\label{gonab2}
{\rm log}_{\rm x}\left({\bf G}_{[{\rm MPQR}} {\bf G}_{0ij{\rm S}]}\right) \ge -5, \nd
which may easily be inferred from \eqref{perfum} and \eqref{perfum2}. Note that the above bound is {\it smaller} than the actual scalings of ${\bf G}_{{\rm MPQR}}$ and  ${\bf G}_{0ij{\rm S}]}$, which are respectively $+1$ and $-4$, because of the presence of other components in the product series (here for example they are ${\bf G}_{{\rm 0MPQ}}$ and ${\bf G}_{{\rm RS}ij}$ which are bounded by $-2$ and $-3$ respectively from \eqref{perfum} and \eqref{perfum2}). Therefore when we compare $l_{0N}^{ab} - 5$ with the $g_s$ scalings from \eqref{gonab2} and \eqref{thetrial10}, we find that:
\bg\label{gonab3}
l_{\rm 0N}^{ab} \ge 0, ~~~{\rm and/or}~~~ l_{\rm 0N}^{ab} = 2, \nd
which are well within the EFT bound set in \eqref{perfum}. Thus it appears the whole range 
$0 \le l_{\rm 0N}^{ab} \le 2$ should be allowed. When $l_{\rm 0N}^{ab} = 0$, the schematic diagram balancing the various tensors from \eqref{mcacrisis} from {\bf Table \ref{jhinuk7}} may be expressed as:

{\footnotesize
\bg\label{lilcadu92}
d_{({\rm 0})}\ast\overbracket[1pt][7pt]{{\bf G}_4 + b_1 d_{({\rm 0})}\ast\Big(\mathbb{Y}_4}
\overbracket[1pt][7pt]{(\theta_1) + \mathbb{Y}_4\underbracket[1pt][7pt]{(\theta_2) + \mathbb{Y}_4(\theta_3)  +  ..\Big) - b_3\Big(\mathbb{X}_8}_{\theta_2 \ge {8 \over 3}}{}^{(0, 0)}\Big) 
- b_2 \Big(\mathbb{T}_8}^{\theta_1 \ge {2 \over 3}}{}^{(0, 0)}\Big) = 0, \nd}  
where $\theta_1 = {2\over 3}$ expectedly fits with \eqref{gonab}, and $\theta_2 = {8\over 3}$ now provides 
adequate number of quantum terms from \eqref{fahingsha5} to balance with the ${\bf X}_8$ polynomial 
\eqref{thetrial10}, at least at lowest orders in $g_s$ and zeroth orders in 
$(k_i, k_{n_i})$. At higher orders in $\left(k_i, k_{n_i}, l, {g_s\over {\rm HH}_o}\right)$, we would expect:

{\footnotesize
\bg\label{megmojer}
l_{\rm 0N}^{ab} - 5 + {2\over 3}(k_1 + k_{n_1} + k_{n_2}) = \theta_{nl} - l_{\rm 0N}^{ab} - {17\over 3} 
-{2k_2\over 3} = {2\over 3}(k_4 + k_5) + l_{[{\rm MP}}^{[{\rm QR}} \oplus l_{{0i}]}^{j{\rm S}]}, \nd}
which should be equated to either $l - 3$ for the special metric \eqref{makibhalu3} with $\gamma = 6$ (or the metric \eqref{evader}), or with $l - 11$ for the metric with generic dependence on all the eleven dimensions.  Note that the identifications in \eqref{lilcadu92} do not require any non-perturbative effects (which are sub-leading here), but this will not be the case when $l_{\rm 0N}^{ab} = 2$. The schematic diagram for this case becomes:
\bg\label{lilcadu93}
d_{({\rm 0})}\ast\overbracket[1pt][7pt]{{\bf G}_4 + b_1 d_{({\rm 0})}\ast\Big(\mathbb{Y}_4}
\overbracket[1pt][7pt]{(\theta_1) + \mathbb{Y}_4(\theta_2) + \mathbb{Y}_4\underbracket[1pt][7pt]{(\theta_3)  +  ..\Big) - b_2\Big(\mathbb{T}_8}_{\theta_3 \ge {8 \over 3}}{}^{(0, 0)}\Big) 
- b_3 \Big(\mathbb{X}_8}^{\theta_1 \ge {14 \over 3}}{}^{(0, 0)}\Big) = 0, \nonumber\\ \nd  
where, while $\theta_1 = {14\over 3}$ is perfectly consistent with \eqref{gonab}, $\theta_3 = {8\over 3}$ doesn't have perturbative terms to contribute to the balancing with the flux products. Non-perturbative effects, like BBS instantons from \eqref{beverwickmey} are then essential and incorporating them would change $\theta_3$ to ${14\over 3}$ which would now consistently balance with the flux products. Eventually
the balancing will involve the full identifications from \eqref{megmojer} with $l - 3$ for 
$l \in {\mathbb{Z}\over 3}$. Additional non-local effects would start contributing at sub-leading orders.
 
When the derivatives, acting on the rank seven tensors from \eqref{mcacrisis}, are along the base 
${\cal M}_4 \times {\cal M}_2$ then the balancing of the various terms take slightly different structure as may be seem from 
{\bf Table \ref{jhinuk7}}. We now have to balance $l_{\rm 0N}^{ab} - 4$ with the $g_s$ scalings from either the quantum terms, product of the fluxes or the ${\bf X}_8$ polynomials, all of which in turn take different values from what we had earlier. For example, the $g_s$ scalings of the product of the fluxes are now bounded from below by:
\bg\label{gonab3}
{\rm log}_{\rm x}\left({\bf G}_{[{\rm MNPQ}} {\bf G}_{{\rm RS}ij]}\right) \ge -4, \nd
instead of \eqref{gonab2} earlier (for example here the participating flux components are 
${\bf G}_{{\rm MNP}i}$ and ${\bf G}_{{\rm QRS}j}$, both of which are bounded from below by $-2$ as may be  easily seen from \eqref{perfum2}). The ${\bf X}_8$ polynomial also scales differently, and is given by 
\eqref{meggy1} instead of \eqref{thetrial10} earlier. Interestingly, comparing $l_{\rm 0N}^{ab} - 4$ with 
\eqref{gonab3} or with \eqref{meggy1} now reproduces the {\it same} range for $l_{\rm 0N}^{ab}$
that we had in \eqref{gonab3} earlier, namely $ 0 \le l_{\rm 0N}^{ab} \le 2$. The schematic diagrams are also similar from what we had in \eqref{lilcadu92} and \eqref{lilcadu93}: when 
$l_{\rm 0N}^{ab} = 0$, we have $\theta_1 = {2\over 3}$ and $\theta_2 = {8\over 3}$; and when 
$l_{\rm 0N}^{ab} = 2$, we have $\theta_1 = {14\over 3}$ and $\theta_3 = {8\over 3}$, implying the necessity of the contributions of the BBS instantons from \eqref{beverwickmey}. The only changes to 
\eqref{lilcadu92} and \eqref{lilcadu93} are the notational changes as in \eqref{sevignic}. At higher orders in $\left(k_i, k_{n_i}, l, {g_s\over {\rm HH}_o}\right)$ we expect:

{\footnotesize
\bg\label{megmojer2}
l_{\rm 0N}^{ab} - 4 + {2\over 3}(k_1 + k_{n_1} + k_{n_2}) = \theta_{nl} - l_{\rm 0N}^{ab} - {14\over 3} 
-{2k_2\over 3} = {2\over 3}(k_4 + k_5) + l_{[{\rm MN}}^{[{\rm PQ}} \oplus l_{{\rm RS}]}^{{ij}]}, \nd}
which should be equated to $l - 2$ now for the metric choice \eqref{makibhalu3} with $\gamma = 6$, or with 
$l - 10$ for the metric with generic dependence on all the eleven dimensions, where $l \in
{\mathbb{Z}\over 3}$. 

It would be interesting to speculate how the schematic diagram would change if we take an intermediate value for $l_{\rm 0N}^{ab}$ in the range $0 \le l_{\rm 0N}^{ab} \le 2$, for example what would happen if we take $l_{\rm 0N}^{ab} = 1$. To be specific we take the derivatives to act along the 
${\cal M}_4 \times {\cal M}_2$ directions. The schematic diagram now becomes:

{\footnotesize
\bg\label{lilcadu94} 
d_{({\rm M})}\ast\overbracket[1pt][7pt]{{\bf G}_4 + b_1 d_{({\rm M})}\ast\Big(\mathbb{Y}_4}^{\theta_1 \ge {8\over 3}}(\theta_1) +
\mathbb{Y}_4\overbracket[1pt][7pt]{(\theta_{{\rm M}, 2}) + \mathbb{Y}_4\underbracket[1pt][7pt]{(\theta_{{\rm M}, 3})  +  ..\Big)- b_2\Big(\mathbb{T}_8}_{\theta_{{\rm M}, 3} \ge {5\over 3}}{}^{(m, \alpha)}\Big) -b_3 \Big(\mathbb{X}_8}^{\theta_{{\rm M}, 2} \ge {11\over 3}}{}^{(m, \alpha)}\Big) = 0, \nonumber\\ \nd}
similar to what we had in \eqref{lilcadu91} for $l_{\rm 0N}^{{\rm P}a}$, including the addition of BBS instantons from \eqref{beverwickmey} to enhance $\theta_{{\rm M}, 3} = {5\over 3}$ to
$\theta_{{\rm M}, 3} = {11 \over 3}$. The similarities of the schematc diagrams over various flux components are  a sign that consistent dynamics may be achieved with perturbative and non-perturbative quantum corrections over the appropriate ranges of $l_{\rm 0A}^{\rm BC}$. 

Finally, when the derivatives act along the toroidal directions, the story is again different from the two earlier cases. This time $l_{\rm 0N}^{ab} - 2$ needs to be balanced with the $g_s$ scalings of the quantum terms, the flux products or the ${\bf X}_8$ polynomial. The $g_s$ scalings of the flux products are now bounded from below by:
\bg\label{gonab4}
{\rm log}_{\rm x}\left({\bf G}_{[{\rm MPQR}} {\bf G}_{{\rm S}bij]}\right) \ge -3, \nd
corresponding to say, ${\bf G}_{{\rm MPQ}i}$ and ${\bf G}_{{\rm RS}bj}$ whose $g_s$ scalings are bounded by $-2$ and $-1$ respectively. Once we incorporate the $g_s$ scaling of the ${\bf X}_8$ polynomial from
\eqref{thetrial22}, the range of values for $l_{\rm 0N}^{ab}$ becomes:
\bg\label{gonab0}
l_{\rm 0N}^{ab} \ge -1, ~~~{\rm and/or}~~~ l_{\rm 0N}^{ab} = 1. \nd
Unfortunately the EFT constraint from \eqref{perfum} rules out the values of $l_{\rm 0N}^{ab}$ in the range
$-1 \le l_{\rm 0N}^{ab} < 0$, which means the allowed range is $0 \le l_{\rm 0N}^{ab} \le 1$. For 
$l_{\rm 0N}^{ab} = 0$, the schematic diagram becomes:

{\footnotesize
\bg\label{lilcadu944} 
d_{(a)}\ast\overbracket[1pt][7pt]{{\bf G}_4 + b_1 d_{({a})}\ast\Big(\mathbb{Y}_4}^{\theta_1 \ge {2\over 3}}(\theta_1) +
\mathbb{Y}_4\overbracket[1pt][7pt]{(\theta_{a, 2}) 
+ \mathbb{Y}^{\rm BBS}_4\underbracket[1pt][7pt]{(\theta_{{a}, 3})  +  ..\Big)- b_2\Big(\mathbb{T}_8}_{\theta_{a, 3} \ge {\red -{1\over 3}} + 2}{}^{(a, b)}\Big) -b_3 \Big(\mathbb{X}_8}^{\theta_{a, 2} \ge {5 \over 3}}{}^{(a, b)}\Big) = 0, \nd}
which tells us that while perturbative series contribute as $\theta_1 = {2\over 3}$ and $\theta_{a, 2} = {5\over 3}$, it is the non-perturbative BBS instantons from \eqref{beverwickmey} that can convert the $g_s$ scaling
of ${\red -{1\over 3}}$ to $+{5\over 3}$ to balance the flux products whose $g_s$ scalings are, in turn, bounded from below by $-3$. At higher orders in 
$\left(k_i, k_{n_i}, l, {g_s\over {\rm HH}_o}\right)$ we expect, at least perturbatively:

{\footnotesize
\bg\label{megmojer3}
l_{\rm 0N}^{ab} - 2 + {2\over 3}(k_1 + k_{n_1} + k_{n_2}) = \theta_{nl} - l_{\rm 0N}^{ab} - {8 \over 3} 
-{2k_2\over 3} = {2\over 3}(k_4 + k_5) + l_{[{\rm MP}}^{[{\rm QR}} \oplus l_{{\rm S}b]}^{{ij}]} 
= \Big\{\begin{matrix}~l-1\\ ~l-9\end{matrix}~ , \nd}
where the top and the bottom choices are associated with the metric \eqref{makibhalu3} for $\gamma =6$, and the metric with generic dependence on all the eleven dimensions respectively. When $l_{\rm 0N}^{ab} = 1$, the quantum terms balancing the ${\bf X}_8$ polynomial and the flux products are classified by 
$\theta_1 =  \theta_2 = {8\over 3}$ respectively except that the latter come from the BBS instantons 
\eqref{beverwickmey}. Thus it appears that consistent dynamics may be achieved within the range  
$0 \le l_{\rm 0N}^{ab} \le 2$ for the flux components ${\bf G}_{{\rm 0N}ab}$.

 \begin{table}[tb]  
 \begin{center}
\renewcommand{\arraystretch}{1.5}
\begin{tabular}{|c||c||c|}\hline ${\bf G}_{0{\rm N}ia}$ tensors & form & ${g_s\over {\rm HH}_o}$ scaling \\ \hline\hline
$\left(\mathbb{T}_7^{(f)}\right)_{{\rm MPQRS}bj}$ & $\sqrt{-{\bf g}_{11}} {\bf G}^{0{\rm N}ia} {\rm F}_1^{(n_1)}
{\rm F}_2^{(n_2)} \epsilon_{0{\rm N}ia{\rm MPQRS}bj}$ & $l_{\rm 0N}^{ia}  + {2\over 3}\left(k_1 + k_{n_1} 
+ k_{n_2}\right)$ \\ \hline 
$\left(\mathbb{T}_7^{(q)}\right)_{{\rm MPQRS}bj}$ & $\sqrt{-{\bf g}_{11}} \mathbb{Y}_4^{0{\rm N}ia} 
\epsilon_{0{\rm N}ia{\rm MPQRS}bj}$ & $\theta_{nl}(k_2) - l_{\rm 0N}^{ia} - {14\over 3}  
- {2k_2\over 3}$ \\ \hline 
$\left(\mathbb{T}_8^{(0, 0)}\right)_{{\rm MPQRS}b0j}$ & ${\bf G}_{[{\rm MPQR}} {\bf G}_{{\rm S}b0j]}$ & ${2\over 3}\left(k_4 + k_5\right) + l_{[{\rm MP}}^{[{\rm QR}} \oplus l_{{{\rm S}b}]}^{0j]}$ 
\\ \hline 
$\left(\mathbb{T}_8^{(i, j)}\right)_{{\rm MPQRS}bij}$ & ${\bf G}_{[{\rm MPQR}} {\bf G}_{{\rm S}bij]}$ & ${2\over 3}\left(k_4 + k_5\right) + l_{[{\rm MP}}^{[{{\rm QR}}} \oplus l_{{\rm S}b]}^{ij]}$ 
\\ \hline 
$\left(\mathbb{T}_8^{(m, \alpha)}\right)_{{\rm MNPQRS}bj}$ & ${\bf G}_{[{\rm MNPQ}} {\bf G}_{{\rm RS}bj]}$ & ${2\over 3}\left(k_4 + k_5\right) + l_{[{\rm MN}}^{[{{\rm PQ}}} \oplus l_{{\rm RS}]}^{bj]}$ 
\\ \hline 
$\left(\mathbb{T}_8^{(a, b)}\right)_{{\rm MPQRS}abj}$ & ${\bf G}_{[{\rm MPQR}} {\bf G}_{{\rm S}abj]}$ & ${2\over 3}\left(k_4 + k_5\right) + l_{[{\rm MP}}^{[{{\rm QR}}} \oplus l_{{\rm S}j]}^{ab]}$ 
\\ \hline 
$\left(\mathbb{X}_8^{(0, 0)}\right)_{{\rm MPQRS}b0j}$ & \eqref{meggy3}, \eqref{meggy4}
& $l - 1, l - 9$ \\ \hline
$\left(\mathbb{X}_8^{(i, j)}\right)_{{\rm MPQRS}bij}$ & \eqref{thetrial22}, \eqref{thetrial23}
& $l - 1, l - 9$ \\ \hline
$\left(\mathbb{X}_8^{(m, \alpha)}\right)_{{\rm MNPQRS}bj}$ & \eqref{meggy5}, \eqref{meggy6}
& $l, l - 8$ \\ \hline
$\left(\mathbb{X}_8^{(a, b)}\right)_{{\rm MPQRS}abj}$ & \eqref{meggy7}, \eqref{meggy8}
& $l + 1, l - 7$ \\ \hline
\end{tabular}
\renewcommand{\arraystretch}{1}
\end{center}
 \caption[]{Comparing the $g_s$ scalings of the various tensors that contribute to the EOM for the G-flux components ${\bf G}_{0{\rm N}ia}$ in \eqref{mcacrisis}. 
 Note that now, all the eight forms as well as the EOM have {\it four} possible choices stemming from how the derivatives act on the seven-forms in \eqref{mcacrisis} as well as on the behavior of the three form fields ${\bf C}_{{\rm N}ia}, {\bf C}_{0{\rm N}a}, {\bf C}_{0i{\rm N}}$ and ${\bf C}_{0ia}$.} 
  \label{jhinuk8}
 \end{table}

\vskip.2in

\noindent {\it Case 4: ${\bf G}_{{\rm 0N}ia}, {\bf G}_{{\rm 0MN}i}$ and ${\bf G}_{0abi}$ components}

\vskip.2in

\noindent Our next case deals with three flux components ${\bf G}_{{\rm 0N}ia}, {\bf G}_{{\rm 0MN}i}$ and ${\bf G}_{0abi}$ that have one leg along the spatial ${\bf R}^2$ directions. All these three flux components can be viewed as {\it localized} flux components, much like what we had in \eqref{sessence} but now having dependence on the toroidal directions as:

{\footnotesize
\bg\label{sessence3}
{\bf G}_{0i\mathbb{CD}}(x, y, w^a; g_s) = \sum_{k\ge 0} \left({\cal G}^{(k)}_{0i\mathbb{CD}}(x, y, w^a) + 
{\cal F}^{(k)}_{0i}(x) \Omega^{(k)}_{\mathbb{CD}}(y, w^a)\right) 
\left({g_s\over {\rm H}{\rm H}_o}\right)^{l_{\mathbb{CD}}^{0i} + 2k/3}, \nd}
where we have added a non-localized piece just for completeness, $(\mathbb{C, D}) \in {\cal M}_4 \times 
{\cal M}_2 \times {\mathbb{T}^2\over {\cal G}}$, and $\Omega_{\mathbb{CD}}(y, w^a)$ is a localized two-form in the internal eight-manifold. The gauge fields ${\cal F}_{0i}^{(k)}$ appear on the integer and fractional three-branes and also on the $(p, q)$ seven-branes (if we take the full F-theory framework). We can even allow gauge fields to have dependence on the internal directions, but for the time being we keep it simple by allowing the coordinate dependences as in \eqref{sessence3}. We can also make the following consistency check: Equating the first two rows of  {\bf Tables \ref{jhinuk8}}, 
{\bf \ref{jhinuk9}} and {\bf \ref{jhinuk12}}, we get:
\bg\label{gonia0}
\theta^{(1)}_{nl} = 2\left(l_{\rm 0N}^{ia} + {7\over 3}\right), ~~ \theta^{(2)}_{nl} = 2\left(l_{\rm 0M}^{{\rm N}i} + {10 \over 3}\right), ~~ \theta^{(3)}_{nl} = 2\left(l_{0a}^{bi} + {4 \over 3}\right), \nd
which are precisely the scalings by which these components contribute in \eqref{botsuga4}. The factors of 2 in each of them are the minimal values that $\left(l_{98}, l_{99}\right)$ for ${\bf G}_{{\rm 0N}ia}$, 
$\left(l_{95}, l_{96}, l_{97}\right)$ for ${\bf G}_{{\rm 0MN}i}$, and $\left(l_{100}\right)$ for 
${\bf G}_{{0a}bi}$ can take to form the corresponding kinetic terms in \eqref{fahingsha5}. 

Let us now start by taking the flux components ${\bf G}_{{\rm 0N}ia}$. The three-forms contributing to the EOMs are ${\bf C}_{{\rm N}ia}$ with temporal derivative, ${\bf C}_{{\rm 0N}a}$ with derivatives along the spatial directions ${\bf R}^2$, ${\bf C}_{0ia}$ with derivatives along the internal six directions ${\cal M}_4 \times {\cal M}_2$, and ${\bf C}_{0i{\rm N}}$ with derivatives along the toroidal directions. Out of these four three-forms, only the first two are related to the gauge fields ${\cal A}_i$ and ${\cal A}_0$ respectively, whereas the other two are related to ${\cal F}_{0i}$ via localized one-forms in the internal eight-manifold. 
These one-forms are not closed, and they would contribute to the dynamics in the usual way, but we will not worry too much about the structure of these forms. Our aim would to be use the three-forms to derive the EOMs as in \eqref{mcacrisis} so that, in the end, the dynamics is captured by the rank seven and eight tensors as given in {\bf Table \ref{jhinuk8}}.  

Corresponding to these four three-forms, there would be four different kinds of ${\bf X}_8$ polynomials. For example, corresponding to ${\bf C}_{{\rm N}ia}$ the ${\bf X}_8$ polynomial involves the following permutation of the indices of the product of the curvature two-forms:
\bg\label{lilcadu99}
\overbracket[1pt][7pt]{{\rm M}~\underbracket[1pt][7pt]{{\rm P}~{\rm Q}}~{\rm R}} ~\overbracket[1pt][7pt]{{b}~\underbracket[1pt][7pt]{{\rm S}~0} ~j} ~~~ 
+~~~~{\rm permutations}, \nd
where the permutations involve at least 23 different rearrangements of the alphabets in 
\eqref{lilcadu99}. For the metric choice \eqref{makibhalu3} with $\gamma = 6$ (or even for the metric choice 
\eqref{evader}), all the above permutations lead to only {\it one} dominant scaling for the eight polynomial. This is given by:
\bg\label{meggy3}
{\bf X}_8(x, y, w^a; g_s) = \sum_{\{l_i\}} \widetilde{\bf X}_{(8, 8a)}^{(l_1, ...., l_4)}(x, y, w^a) 
\left({g_s\over {\rm HH}_o}\right)^{{1\over 3}(l_1 + l_2 + l_3 + l_4 - 3)}, \nd
where $l_1 + .... + l_4 \equiv 3l$ with $l \in {\mathbb{Z}\over 3}$. Interestingly, if we take any other values for $\gamma \ge 5$, then different permutations in \eqref{lilcadu99} lead to different values. On the other hand, if we take the metric with generic dependence on all the eleven dimensions, all the permutations in \eqref{lilcadu99} again lead to a unique dominant scaling for ${\bf X}_8$ as:
\bg\label{meggy4}
{\bf X}_8(x, y, w^a; g_s) = \sum_{\{l_i\}} \widetilde{\bf X}_{(8, 8b)}^{(l_1, ...., l_4)}(x, y, w^a) 
\left({g_s\over {\rm HH}_o}\right)^{{1\over 3}(l_1 + l_2 + l_3 + l_4 - 27)}, \nd
where again such a large negative value might be concerning, but since this case doesn't lead to a well defined EFT, we don't have to worry too much about it. In a similar vein, associated to the three-form
${\bf C}_{{\rm 0N}a}$, the ${\bf X}_8$ polynomials have already been worked out in 
\eqref{thetrial22},  and \eqref{thetrial23} for the metric \eqref{makibhalu3} with $\gamma = 6$, and the metric with generic dependence on all the eleven dimensions respectively. It is interesting to note that the $g_s$ scalings of the corresponding ${\bf X}_8$ polynomials match precisely with the $g_s$ scalings of the 
${\bf X}_8$ polynomials in \eqref{meggy3} and \eqref{meggy4} respectively but the other coordinate dependences differ. Similarly, if we take the three-form components ${\bf C}_{0ia}$, the ${\bf X}_8$ polynomial associated with the metric \eqref{makibhalu3} with $\gamma = 6$ becomes:
\bg\label{meggy5}
{\bf X}_8(x, y, w^a; g_s) = \sum_{\{l_i\}} \widetilde{\bf X}_{(8, 9a)}^{(l_1, ...., l_4)}(x, y, w^a) 
\left({g_s\over {\rm HH}_o}\right)^{{1\over 3}(l_1 + l_2 + l_3 + l_4)}, \nd
which allows a dominant $g_s$ scaling of 0, much like what we had in \eqref{thetrial20}. In fact the similarity
also extends to \eqref{thetrial21}, which allows a dominant scaling of $-8$ and here it happens with:
\bg\label{meggy6}
{\bf X}_8(x, y, w^a; g_s) = \sum_{\{l_i\}} \widetilde{\bf X}_{(8, 9b)}^{(l_1, ...., l_4)}(x, y, w^a) 
\left({g_s\over {\rm HH}_o}\right)^{{1\over 3}(l_1 + l_2 + l_3 + l_4 - 24)}, \nd
although the coordinate dependences would differ as one might have expected. Both of the ${\bf X}_8$ polynomials appear from the following permutations of the indices of the curvature two-forms:
\bg\label{lilcadu100}
\overbracket[1pt][7pt]{{\rm M}~\underbracket[1pt][7pt]{{\rm N}~{\rm P}}~{\rm Q}} ~\overbracket[1pt][7pt]{{\rm R}~\underbracket[1pt][7pt]{{\rm S}~b} ~j} ~~~ 
+~~~~{\rm permutations}, \nd
where as before, all possible permutations reproduce the {\it same} dominant $g_s$ scalings for the two choices of metric ans\"atze as long as $\gamma = 6$ in \eqref{makibhalu3}. Finally, once we take the three-form ${\bf C}_{0i{\rm N}}$, the ${\bf X}_8$ polynomials would appear from the following permutations of the indices of the corresponding four curvature two-forms:
\bg\label{lilcadu101}
\overbracket[1pt][7pt]{{\rm M}~\underbracket[1pt][7pt]{{\rm P}~{\rm Q}}~{\rm R}} ~\overbracket[1pt][7pt]{{\rm S}~\underbracket[1pt][7pt]{a~b} ~j} ~~~ 
+~~~~{\rm permutations}, \nd
where the values of the curvature two-forms may be computed either for the metric \eqref{makibhalu3} with $\gamma = 6$, or with the metric that depends generically on all the eleven dimensions. Taking the former case leads to the following expression for the ${\bf X}_8$ polynomial:
\bg\label{meggy7}
{\bf X}_8(x, y, w^a; g_s) = \sum_{\{l_i\}} \widetilde{\bf X}_{(8, 10a)}^{(l_1, ...., l_4)}(x, y, w^a) 
\left({g_s\over {\rm HH}_o}\right)^{{1\over 3}(l_1 + l_2 + l_3 + l_4 + 3)}, \nd
where again all the permutations in \eqref{lilcadu101} leads to the same dominant scaling of $+1$. Exactly similar criterion, but with a different dominant scaling, is met when we take the metric to generically depend on all of the eleven dimensions. The ${\bf X}_8$ polynomial now becomes:
\bg\label{meggy8}
{\bf X}_8(x, y, w^a; g_s) = \sum_{\{l_i\}} \widetilde{\bf X}_{(8, 10b)}^{(l_1, ...., l_4)}(x, y, w^a) 
\left({g_s\over {\rm HH}_o}\right)^{{1\over 3}(l_1 + l_2 + l_3 + l_4 - 21)}, \nd
with a dominant scaling of $-7$. All the above choices of the eight-forms are still not enough to determine the EOMs corresponding to the flux components ${\bf G}_{{\rm 0N}ia}$. We need to determine the lower 
bounds on the product of the fluxes as they appear in rows 3 to 6 in {\bf Table \ref{jhinuk8}}. Using the EFT bounds on the individual components of fluxes in \eqref{perfum} and \eqref{perfum2}, the lower bounds on the four set of the product of the fluxes appear to be:
\bg\label{lilmadi}
&&{\rm log}_{\rm x}\left({\bf G}_{[{\rm MPQR}} {\bf G}_{{\rm S}b0j]}\right) \ge -3, ~~~~ 
{\rm log}_{\rm x}\left({\bf G}_{[{\rm MPQR}} {\bf G}_{{\rm S}bij]}\right) \ge -3\nonumber\\
&& {\rm log}_{\rm x}\left({\bf G}_{[{\rm MNPQ}} {\bf G}_{{\rm RS}bj]}\right) \ge -1, ~~~~ 
{\rm log}_{\rm x}\left({\bf G}_{[{\rm MPQR}} {\bf G}_{{\rm S}abj]}\right) \ge -1, \nd
where ${\rm x} = {g_s\over {\rm HH}_o}$. With these bounds, and with the list of ${\bf X}_8$ forms determined above, we are now ready to tackle the EOMs corresponding to the flux components 
${\bf G}_{{\rm 0N}ia}$

Tackling the flux EOMs involve analyzing and comparing the $g_s$ scalings of the various rank seven and eight tensors appearing in {\bf Table \ref{jhinuk8}}. This is subtle because, according to \eqref{mcacrisis}, the behaviors of these tensors do depend on how the derivatives act of the rank seven tensors. There are four kinds of derivatives that could act of the rank seven tensors, namely the tensors appearing in rows 
1 and 2 of {\bf Table \ref{jhinuk8}}: temporal derivative, derivatives along the spatial ${\bf R}^2$ directions, derivatives along the ${\cal M}_4 \times {\cal M}_2$ directions and the derivatives along the toroidal ${\mathbb{T}^2\over {\cal G}}$ directions. The $g_s$ scalings of the rank seven tensor $\mathbb{T}_7^{(f)}$ are also different depending on how the derivatives act. For example, the $g_s$ scalings are $l_{\rm 0N}^{ia} - 1$, 
$l_{\rm 0N}^{ia}$, $l_{\rm 0N}^{ia}$ and $l_{\rm 0N}^{ia} + 2$ respectively, corresponding to how the derivatives act. If we now balance these $g_s$ scalings with rows 3 and 7, with rows 4 and 8, with rows 5 and 9 and with rows 6 and 10 in {\bf Table \ref{jhinuk8}} respectively for the four possible actions of the derivatives on the rank seven tensors, then $l_{\rm 0N}^{ia}$ appears to have the following four possible ranges of values:
\bg\label{harmrein}
-2 \le l_{\rm 0N}^{ia} \le 0, ~~~~ -3 \le l_{\rm 0N}^{ia} \le -1, ~~~~ -1 \le l_{\rm 0N}^{ia} \le 0, ~~~~
-3 \le l_{\rm 0N}^{ia} \le -1, \nd
respectively. Unfortunately these ranges cannot be all consistent with EFT. In fact looking at 
\eqref{perfum} we see that the range of values of $l_{\rm 0N}^{ia}$ lying between $-3$ and $-2$, {\it i.e.}
$-3 \le l_{\rm 0N}^{ia} < -2$ are inconsistent with EFT as may be easily extracted from \eqref{botsuga4}. Thus the consistent range appears to be $-2 \le l_{\rm 0N}^{ia} \le 0$. For the case with temporal derivative, this implies the following identifications between the $g_s$ scalings of the various rank seven and eight tensors:

{\footnotesize
\bg\label{megmojer4}
l_{\rm 0N}^{ia} - 1 + {2\over 3}(k_1 + k_{n_1} + k_{n_2}) = \theta_{nl} - l_{\rm 0N}^{ia} - {17 \over 3} 
-{2k_2\over 3} = {2\over 3}(k_4 + k_5) + l_{[{\rm MP}}^{[{\rm QR}} \oplus l_{{\rm S}a]}^{{bj}]} 
= \Big\{\begin{matrix}~l-1\\ ~l-9\end{matrix}~ , \nd}
which we expect to hold at least when we go to higher orders in $\left(k_i, k_{n_i}, l, {g_s\over {\rm HH}_o}\right)$. Here $l \in {\mathbb{Z}\over 3}$, and the two values correspond to the metric choices \eqref{makibhalu3} with $\gamma = 6$, and the one with generic dependence on all the eleven dimensions respectively. At lowest orders the full identifications 
\eqref{megmojer4} cannot hold, and the balancing of the $g_s$ dependences of the various tensors 
of \eqref{mcacrisis} leads to the same schematic diagram for $l_{\rm 0N}^{ia} = -2$ as we had in \eqref{lilcadu92} with 
$(\theta_1, \theta_2) = \left({2\over 3}, {8\over 3}\right)$. The value of $\theta_1 = {2\over 3}$ is expectedly consistent with $\theta_{nl}^{(1)}$ in \eqref{gonia0} for $l_{\rm 0N}^{ia} = -2$.  When 
$l_{\rm 0N}^{ia} = 0$, the schematic diagram resembles \eqref{lilcadu93} with 
$(\theta_1, \theta_2) = \left({14\over 3}, {8\over 3}\right)$ which necessitates the inclusion of BBS instantons from \eqref{beverwickmey} as $\theta_2 = {8\over 3}$ doesn't have perturbative terms to balance the $g_s$ scalings of the flux products. 

When the derivatives act along the spatial ${\bf R}^2$ directions, clearly it is only the quantum terms that can balance the $g_s$ scalings of the all the rank seven and eight tensors in {\bf Table \ref{jhinuk8}} at lowest orders. For $l_{\rm 0N}^{ia} = -2$, the schematic diagram is somewhat similar to \eqref{lilcadu944} but with a few notational changes:

{\footnotesize
\bg\label{lilcadu95} 
d_{(i)}\ast\overbracket[1pt][7pt]{{\bf G}_4 + b_1 d_{({i})}\ast\Big(\mathbb{Y}_4}^{\theta_1 \ge {2\over 3}}(\theta_1) +
\mathbb{Y}_4\overbracket[1pt][7pt]{(\theta_{i, 2}) 
+ \mathbb{Y}^{\rm BBS}_4\underbracket[1pt][7pt]{(\theta_{{i}, 3})  +  ..\Big)- b_2\Big(\mathbb{T}_8}_{\theta_{i, 3} \ge {\red -{1\over 3}} + 2}{}^{(i, j)}\Big) -b_3 \Big(\mathbb{X}_8}^{\theta_{i, 2} \ge {5 \over 3}}{}^{(i, j)}\Big) = 0, \nd}
where the negative value in $\theta_{i, 3}$ tells us that there are {\it no} perturbative terms from 
\eqref{fahingsha5} that can ever contribute even to the lowest orders. For $l_{\rm 0N}^{ia} = 0$, the diagram is similar to \eqref{lilcadu95} but now $\theta_1 = {14\over 3}, \theta_{i, 2} = \theta_{i, 3} = {11\over 3}$. Once we go to higher orders in $\left(k_i, k_{n_i}, l, {g_s\over {\rm HH}_o}\right)$, we expect:

{\footnotesize
\bg\label{megmojer5}
l_{\rm 0N}^{ia} + {2\over 3}(k_1 + k_{n_1} + k_{n_2}) = \theta_{nl} - l_{\rm 0N}^{ia} - {14 \over 3} 
-{2k_2\over 3} = {2\over 3}(k_4 + k_5) + l_{[{\rm MP}}^{[{\rm QR}} \oplus l_{{\rm S}a]}^{{0b}]} 
= \Big\{\begin{matrix}~l-1\\ ~l-9\end{matrix}~ , \nd}
with the notations similar to \eqref{megmojer4}. The story becomes a bit different when the derivatives act along the internal six-manifold ${\cal M}_4 \times {\cal M}_2$ on the rank seven tensors in 
{\bf Table \ref{jhinuk8}}. For $l_{\rm 0N}^{ia} = -2$, the schematic diagram is similar to what we had in 
\eqref{lilcadu94}, but with $\theta_1 = {2\over 3}$ (consistent with \eqref{gonia0}), 
$\theta_{{\rm M}, 2} = {8\over 3}$ and $\theta_{{\rm M}, 3} = {5\over 3}$. The key difference from 
\eqref{lilcadu94} is that now all the quantum terms, at the lowest orders, are given by the perturbative series in \eqref{fahingsha5} with the sub-leading contributions from the instantons (local and non-local ones). When $l_{\rm 0N}^{ia} = 0$, the schematic diagram is similar to \eqref{lilcadu78}
with $(\theta_1, \theta_2) = \left({14\over 3}, {11\over 3}\right)$. The choice of $\theta_2 = {11\over 3}$ naively doesn't necessitates the introduction of BBS instantons from \eqref{beverwickmey} compared to \eqref{lilcadu78}. However,
the fact that the lowest possible scaling is ${7\over 3}$, introducing local and non-local instantons may become necessary to alleviate any tensions one might have in dealing with only perturbative operators from 
\eqref{fahingsha5}. At higher orders in $\left(k_i, k_{n_i}, l, {g_s\over {\rm HH}_o}\right)$, all terms in \eqref{mcacrisis} may be balanced via:

{\footnotesize
\bg\label{megmojer6}
l_{\rm 0N}^{ia} + {2\over 3}(k_1 + k_{n_1} + k_{n_2}) = \theta_{nl} - l_{\rm 0N}^{ia} - {14 \over 3} 
-{2k_2\over 3} = {2\over 3}(k_4 + k_5) + l_{[{\rm MN}}^{[{\rm PQ}} \oplus l_{{\rm RS}]}^{{bj}]} 
= \Big\{\begin{matrix}~l\\ ~l- 8\end{matrix}~ . \nd}
Finally, when the derivatives act along the toroidal directions, the story again differs from the three cases above. For $l_{\rm 0N}^{ia} = -2$, the schematic diagram resembles the one we had in \eqref{lilcadu944}, with similar choices of the quantum terms, including the BBS instantons from \eqref{beverwickmey}. For 
$l_{\rm 0N}^{ia} = 0$, the schematic diagram again resembles \eqref{lilcadu944} but with the following choices of the quantum terms: $\theta_1 = {14\over 3}, \theta_{a, 2} = \theta_{a, 3} = {11\over 3}$, with the latter coming from the BBS instantons. Interestingly, since the lowest order scaling for the flux components ${\bf G}_{{\rm 0N}ia}$ is ${7\over 3}$, the choice of  $\theta_{a, 2} = \theta_{a, 3} = {11\over 3}$ may still require higher order BBS instantons' contributions to alleviate any tension while balancing the quantum terms with the flux products and the ${\bf X}_8$ polynomial. Eventually, when we go sufficiently higher orders in $\left(k_i, k_{n_i}, l, {g_s\over {\rm HH}_o}\right)$, the balancing of the various terms in {\bf Table \ref{jhinuk8}} may occur {\it perturbatively} via the following identifications:

{\footnotesize
\bg\label{megmojer7}
l_{\rm 0N}^{ia} + 2 + {2\over 3}(k_1 + k_{n_1} + k_{n_2}) = \theta_{nl} - l_{\rm 0N}^{ia} - {8 \over 3} 
-{2k_2\over 3} = {2\over 3}(k_4 + k_5) + l_{[{\rm MP}}^{[{\rm QR}} \oplus l_{{\rm S}j]}^{{ab}]} 
= \Big\{\begin{matrix}~l + 1\\ ~l- 7\end{matrix}~ , \nd}
 with sub-leading non-perturbative (and non-local) contributions. Therefore, putting everything together, there appears to be consistent quantum dynamics for the flux components ${\bf G}_{{\rm0N}ia}$ when the dominant scaling lies in the range $-2 \le l_{\rm 0N}^{ia} \le 0$. 
 
 \begin{table}[tb]  
 \begin{center}
\renewcommand{\arraystretch}{1.5}
\begin{tabular}{|c||c||c|}\hline ${\bf G}_{0{\rm MN}i}$ tensors & form & ${g_s\over {\rm HH}_o}$ scaling \\ \hline\hline
$\left(\mathbb{T}_7^{(f)}\right)_{{\rm PQRS}jab}$ & $\sqrt{-{\bf g}_{11}} {\bf G}^{0{\rm MN}i} {\rm F}_1^{(n_1)}
{\rm F}_2^{(n_2)} \epsilon_{0{\rm MN}i{\rm PQRS}jab}$ & $l_{\rm 0M}^{{\rm N}i} + 2 + {2\over 3}\left(k_1 + k_{n_1} 
+ k_{n_2}\right)$ \\ \hline 
$\left(\mathbb{T}_7^{(q)}\right)_{{\rm PQRS}jab}$ & $\sqrt{-{\bf g}_{11}} \mathbb{Y}_4^{0{\rm MN}i} 
\epsilon_{0{\rm MN}i{\rm PQRS}jab}$ & $\theta_{nl}(k_2) - l_{\rm 0M}^{{\rm N}i} - {14\over 3}  
- {2k_2\over 3}$ \\ \hline 
$\left(\mathbb{T}_8^{(0, 0)}\right)_{{\rm PQRS}ab0j}$ & ${\bf G}_{[{\rm PQRS}} {\bf G}_{ab0j]}$ & ${2\over 3}\left(k_4 + k_5\right) + l_{[{\rm PQ}}^{[{\rm RS}} \oplus l_{{ab}]}^{0j]}$ 
\\ \hline 
$\left(\mathbb{T}_8^{(i, j)}\right)_{{\rm PQRS}ijab}$ & ${\bf G}_{[{\rm PQRS}} {\bf G}_{ijab]}$ & ${2\over 3}\left(k_4 + k_5\right) + l_{[{\rm PQ}}^{[{{\rm RS}}} \oplus l_{ij]}^{ab]}$ 
\\ \hline 
$\left(\mathbb{T}_8^{(m, \alpha)}\right)_{{\rm MPQRS}jab}$ & ${\bf G}_{[{\rm MPQR}} {\bf G}_{{\rm S}jab]}$ & ${2\over 3}\left(k_4 + k_5\right) + l_{[{\rm MP}}^{[{{\rm QR}}} \oplus l_{{\rm S}j]}^{ab]}$ 
\\ \hline 
$\left(\mathbb{X}_8^{(0, 0)}\right)_{{\rm PQRS}ab0j}$ & \eqref{meggy9}, \eqref{meggy10}
& $l, l - 8$ \\ \hline
$\left(\mathbb{X}_8^{(i, j)}\right)_{{\rm PQRS}ijab}$ & \eqref{thetrial20}, \eqref{thetrial21}
& $l, l - 8$ \\ \hline
$\left(\mathbb{X}_8^{(m, \alpha)}\right)_{{\rm MPQRS}jab}$ & \eqref{meggy7}, \eqref{meggy8}
& $l + 1, l - 7$ \\ \hline
\end{tabular}
\renewcommand{\arraystretch}{1}
\end{center}
 \caption[]{Comparing the $g_s$ scalings of the various tensors that contribute to the EOM for the G-flux components ${\bf G}_{0{\rm MN}i}$ in \eqref{mcacrisis}. The contributing three-forms are now
 ${\bf C}_{{\rm MN}i}, {\bf C}_{0{\rm MN}}$ and ${\bf C}_{0{\rm N}i}$ with derivatives along temporal, spatial and the internal six directions respectively.} 
  \label{jhinuk9}
 \end{table}

Our next interesting case is with the flux components ${\bf G}_{{\rm 0MN}i}$ whose EOMs are controlled by the three-form fluxes: ${\bf C}_{{\rm MN}i}, {\bf C}_{\rm 0MN}$ and ${\bf C}_{{\rm 0N}i}$. The first two can be associated with the gauge field ${\cal A}_i$ and ${\cal A}_0$ respectively defined using the localized two-form $\Omega_{\rm MN}$, whereas the last one may be related to the field strength ${\cal F}_{0i}$ defined using a localized one-form. All of these should lead us to the localized form \eqref{sessence3}. On the other hand, the three-forms are also related to the ${\bf X}_8$ polynomials as we discussed earlier. The 
${\bf X}_8$ polynomials associated to ${\bf C}_{\rm 0MN}$ and ${\bf C}_{{\rm 0N}i}$ have already been derived in \eqref{thetrial20} and \eqref{meggy7} for the metric choices \eqref{makibhalu3} with
$\gamma = 6$ and in \eqref{thetrial21} and \eqref{meggy8} for the metric choices 
 with generic dependences on all the eleven dimensions respectively. For the three-form 
 ${\bf C}_{{\rm MN}i}$, the ${\bf X}_8$ polynomial becomes:
 \bg\label{meggy9}
{\bf X}_8(x, y, w^a; g_s) = \sum_{\{l_i\}} \widetilde{\bf X}_{(8, 11a)}^{(l_1, ...., l_4)}(x, y, w^a) 
\left({g_s\over {\rm HH}_o}\right)^{{1\over 3}(l_1 + l_2 + l_3 + l_4)}, \nd 
for the metric choice \eqref{makibhalu3} with $\gamma = 6$. Notice the similarity of the $g_s$ scaling with the one from \eqref{thetrial20}, although their functional forms differ. In fact similar identifications between the $g_s$ scalings may be made for the following ${\bf X}_8$ polynomial:
\bg\label{meggy10}
{\bf X}_8(x, y, w^a; g_s) = \sum_{\{l_i\}} \widetilde{\bf X}_{(8, 11b)}^{(l_1, ...., l_4)}(x, y, w^a) 
\left({g_s\over {\rm HH}_o}\right)^{{1\over 3}(l_1 + l_2 + l_3 + l_4  - 24)}, \nd
and the one from \eqref{thetrial21} although, again, their functional forms differ. \eqref{meggy10} is determined for the metric with generic dependence on all the eleven dimensions, and both \eqref{meggy10} and \eqref{meggy9} appear from the following permutations of the indices of the four curvature two-forms:
\bg\label{lilcadu102}
\overbracket[1pt][7pt]{{\rm P}~\underbracket[1pt][7pt]{{\rm Q}~{\rm R}}~{\rm S}} ~\overbracket[1pt][7pt]{{\rm 0}~\underbracket[1pt][7pt]{a~b} ~j} ~~~ 
+~~~~{\rm permutations}, \nd
with the dominant scalings remaining the same over all the possible permutations of the indices in 
\eqref{lilcadu102}. The only thing now remaining is to find the lower bounds on the products of fluxes in 
{\bf Table \ref{jhinuk9}}. This may be quantified as (${\rm x} = {g_s\over {\rm HH}_o}$):

{\footnotesize
\bg\label{brejame}
{\rm log}_{\rm x}\left({\bf G}_{[{\rm PQRS}} {\bf G}_{ab0j]}\right) \ge -2, ~~~  
{\rm log}_{\rm x}\left({\bf G}_{[{\rm PQRS}} {\bf G}_{abij]}\right) \ge -2, ~~~
{\rm log}_{\rm x}\left({\bf G}_{[{\rm MPQR}} {\bf G}_{{\rm S}jab]}\right) \ge -1. \nd}  
 With these we are ready to work out the EOMs from {\bf Table \ref{jhinuk9}}. The first step would be to match the $g_s$ scalings of the rank seven tensors with the rank eight tensors, namely the flux products and the ${\bf X}_8$ polynomials. More appropriately however it is the $g_s$ scalings of the derivative actions on the rank seven tensors that need to be identified with the rank eight tensors. The derivative actions change the $g_s$ scalings slightly such that they become $l_{\rm 0M}^{{\rm N}i} + 1$ and $l_{\rm 0M}^{{\rm N}i} + 2$ 
 respectively for the temporal derivative and the derivatives along ${\bf R}^2$ or along ${\cal M}_4 \times 
 {\cal M}_2$ respectively. Equating them to the rank eight tensors, now lead to the following ranges of choices:
 \bg\label{greenjan}
 -3 \le l_{\rm 0M}^{{\rm N}i} \le -1, ~~~~  -4 \le l_{\rm 0M}^{{\rm N}i} \le -2, ~~~~ 
  -3 \le l_{\rm 0M}^{{\rm N}i} \le -1, \nd
  respectively for the derivatives along the temporal, spatial and the internal six directions respectively. Unfortunately EFT considerations from \eqref{perfum} rules out the range of values 
  $-4 \le l_{\rm 0M}^{{\rm N}i} < -3$, so it appears that the valid range consistent with EFT will be 
  $ -3 \le l_{\rm 0M}^{{\rm N}i} \le -1$. Question is, whether this is consistent with EOMs. This is what we turn to next.

Let us start with the temporal derivative. For $l_{\rm 0M}^{{\rm N}i} = -3$, the schematic diagram resembles the one in \eqref{lilcadu92} with the same values for $\theta_1$ and $\theta_2$ (the former is consistent with 
$\theta_{nl}^{(2)}$ in \eqref{gonia0}). When $l_{\rm 0M}^{{\rm N}i} = -1$, the schematic diagram now resembles \eqref{lilcadu93}, again with the same values of $(\theta_1, \theta_3) = \left({14\over 3}, {8\over 3}\right)$ (the former is again consistent with \eqref{gonia0}). Since $\theta_3 = {8\over 3}$ does not allow 
perturbative terms from \eqref{fahingsha5} to participate, we can introduce the effects of the BBS instantons from \eqref{beverwickmey}. This will convert $\theta_3$ to ${14\over 3}$, which is much like what we had in 
\eqref{lilcadu93}. Of course at higher orders in $\left(k_i, k_{n_i}, l, {g_s\over {\rm HH}_o}\right)$, the balancing between the various tensors appears to be:

{\footnotesize
\bg\label{megmojer7}
l_{\rm 0M}^{{\rm N}i} + 1 + {2\over 3}(k_1 + k_{n_1} + k_{n_2}) = \theta_{nl} - l_{\rm 0M}^{{\rm N}i} - {17 \over 3} 
-{2k_2\over 3} = {2\over 3}(k_4 + k_5) + l_{[{\rm PQ}}^{[{\rm RS}} \oplus l_{0j]}^{{ab}]} 
= \Big\{\begin{matrix}~l\\ ~l- 8\end{matrix}~ , \nd} 
with sub-leading contributions from the non-perturbative (and non-local) terms. When the derivatives act along the spatial ${\bf R}^2$ directions, taking $l_{\rm 0M}^{{\rm N}i} = -3$, leads to the same schematic diagram as in \eqref{lilcadu95} including identical values for $\theta_1, \theta_{i, 2}$ and $\theta_{i, 3}$. On the other hand, taking $l_{\rm 0M}^{{\rm N}i} = -1$ also leads to \eqref{lilcadu95}, but now $\theta_1 = {14\over 3}, \theta_{i, 2} = \theta_{i, 3} = {11\over 3}$ with the latter appearing from the contributions of the BBS instantons. At higher orders in $\left(k_i, k_{n_i}, l, {g_s\over {\rm HH}_o}\right)$ the identifications between the various rank seven and eight tensors become:

{\footnotesize
\bg\label{megmojer8}
l_{\rm 0M}^{{\rm N}i} + 2 + {2\over 3}(k_1 + k_{n_1} + k_{n_2}) = \theta_{nl} - l_{\rm 0M}^{{\rm N}i} - {14 \over 3} 
-{2k_2\over 3} = {2\over 3}(k_4 + k_5) + l_{[{\rm PQ}}^{[{\rm RS}} \oplus l_{ij]}^{{ab}]} 
= \Big\{\begin{matrix}~l\\ ~l- 8\end{matrix}~ , \nd} 
with other factors remaining same as in \eqref{megmojer7}, including any changes incurred from the non-perturbative or non-local terms. Finally, when the derivatives act along the internal six directions, the schematic diagram for $l_{\rm 0M}^{{\rm N}i} = -3$ is similar to the one in \eqref{lilcadu90} with identical values for $\theta_1$ and $\theta_2$. These values change to $\theta_1 = {14\over 3}$ and $\theta_2 = {8\over 3}$ when $l_{\rm 0M}^{{\rm N}i} = -1$ in \eqref{lilcadu90}. Once the effects of the BBS instantons are 
incorporated in from \eqref{beverwickmey}, we expect $\theta_2 = {14\over 3}$, which is big enough to accommodate quantum terms to balance the flux products from the fifth row in {\bf Table \ref{jhinuk9}}. Putting everything together, it therefore appears that consistent quantum dynamics may be ascertained in the dominant scaling range $-3 \le l_{\rm 0M}^{{\rm N}i} \le -1$ for the flux components 
${\bf G}_{{\rm 0MN}i}$.

 \begin{table}[tb]  
 \begin{center}
\renewcommand{\arraystretch}{1.5}
\begin{tabular}{|c||c||c|}\hline ${\bf G}_{0abi}$ tensors & form & ${g_s\over {\rm HH}_o}$ scaling \\ \hline\hline
$\left(\mathbb{T}_7^{(f)}\right)_{{\rm MNPQRS}j}$ & $\sqrt{-{\bf g}_{11}} {\bf G}^{0abi} {\rm F}_1^{(n_1)}
{\rm F}_2^{(n_2)} \epsilon_{0abi{\rm MNPQRS}j}$ & $l_{0a}^{bi} - 2 + {2\over 3}\left(k_1 + k_{n_1} 
+ k_{n_2}\right)$ \\ \hline 
$\left(\mathbb{T}_7^{(q)}\right)_{{\rm MNPQRS}j}$ & $\sqrt{-{\bf g}_{11}} \mathbb{Y}_4^{0abi} 
\epsilon_{0abi{\rm MNPQRS}j}$ & $\theta_{nl}(k_2) - l_{0a}^{bi} - {14\over 3}  
- {2k_2\over 3}$ \\ \hline 
$\left(\mathbb{T}_8^{(0, 0)}\right)_{{\rm MNPQRS}0j}$ & ${\bf G}_{[{\rm MNPQ}} {\bf G}_{{\rm RS}0j]}$ & ${2\over 3}\left(k_4 + k_5\right) + l_{[{\rm MN}}^{[{\rm PQ}} \oplus l_{{\rm RS}]}^{0j]}$ 
\\ \hline 
$\left(\mathbb{T}_8^{(i, j)}\right)_{{\rm MNPQRS}ij}$ & ${\bf G}_{[{\rm MNPQ}} {\bf G}_{{\rm RS}ij]}$ & ${2\over 3}\left(k_4 + k_5\right) + l_{[{\rm MN}}^{[{{\rm PQ}}} \oplus l_{{\rm RS}]}^{ij]}$ 
\\ \hline 
$\left(\mathbb{T}_8^{(a, b)}\right)_{{\rm MNPQRS}aj}$ & ${\bf G}_{[{\rm MNPQ}} {\bf G}_{{\rm RS}aj]}$ & ${2\over 3}\left(k_4 + k_5\right) + l_{[{\rm MN}}^{[{{\rm PQ}}} \oplus l_{{\rm RS}]}^{aj]}$ 
\\ \hline 
$\left(\mathbb{X}_8^{(0, 0)}\right)_{{\rm MNPQRS}0j}$ & \eqref{meggy11}, \eqref{meggy12} 
& $l - 2, l - 10$ \\ \hline
$\left(\mathbb{X}_8^{(i, j)}\right)_{{\rm MNPQRS}ij}$ & \eqref{meggy1}, \eqref{meggy2} 
& $l - 2, l - 10$ \\ \hline
$\left(\mathbb{X}_8^{(a, b)}\right)_{{\rm MNPQRS}aj}$ & \eqref{meggy5}, \eqref{meggy6} 
& $l, l - 8$ \\ \hline
\end{tabular}
\renewcommand{\arraystretch}{1}
\end{center}
 \caption[]{Comparing the $g_s$ scalings of the various tensors that contribute to the EOM for the G-flux components ${\bf G}_{0abi}$ in \eqref{mcacrisis}. The contributing three-form fields are now 
 ${\bf C}_{abi}$ with derivative along the temporal direction, ${\bf C}_{0ab}$ with derivatives along the spatial directions, 
 and ${\bf C}_{0ai}$ with derivatives along the toroidal directions.} 
  \label{jhinuk12}
 \end{table}

Our final case is the flux components ${\bf G}_{0abi}$ whose dynamics are controlled by the three-forms
${\bf C}_{abi}$ with temporal derivative, ${\bf C}_{0ab}$ with derivatives along ${\bf R}^2$ and 
${\bf C}_{0ai}$ with derivatives along the toroidal directions. The ${\bf X}_8$ polynomials associated with 
${\bf C}_{0ab}$ have already been computed in \eqref{meggy1} and \eqref{meggy2} for the metric 
\eqref{makibhalu3} with $\gamma = 6$ and for the metric with generic dependence on all the eleven dimensions respectively. Similarly the ${\bf X}_8$ polynomials associated with ${\bf C}_{0ai}$ have also been computed in \eqref{meggy5} and \eqref{meggy6} respectively for the two above-mentioned metric choices. The ${\bf X}_8$ polynomials associated with ${\bf C}_{abi}$ may be computed by looking at the following permutations of the indices of the curvature two-forms:
\bg\label{lilcadu103}
\overbracket[1pt][7pt]{{\rm M}~\underbracket[1pt][7pt]{{\rm N}~{\rm P}}~{\rm Q}} ~\overbracket[1pt][7pt]{{\rm R}~\underbracket[1pt][7pt]{{\rm S}~0} ~j} ~~~ 
+~~~~{\rm permutations}. \nd
When we take the special metric choice \eqref{makibhalu3} with $\gamma = 6$, all the permutations in 
\eqref{lilcadu103} leads to the same dominant scaling $-2$, much like what we had for the other cases. The ${\bf X}_8$ polynomial then becomes:
\bg\label{meggy11}
{\bf X}_8(x, y, w^a; g_s) = \sum_{\{l_i\}} \widetilde{\bf X}_{(8, 12a)}^{(l_1, ...., l_4)}(x, y, w^a) 
\left({g_s\over {\rm HH}_o}\right)^{{1\over 3}(l_1 + l_2 + l_3 + l_4  - 6)}, \nd
with the dominant scaling resembling the one from \eqref{meggy1}, although, as expected, their functional forms differ. Similarly, once we take the metric with generic dependence on all the eleven dimensions, the dominant scale becomes $-10$, which is uniform over all the permutations in \eqref{lilcadu103}:
\bg\label{meggy12}
{\bf X}_8(x, y, w^a; g_s) = \sum_{\{l_i\}} \widetilde{\bf X}_{(8, 12b)}^{(l_1, ...., l_4)}(x, y, w^a) 
\left({g_s\over {\rm HH}_o}\right)^{{1\over 3}(l_1 + l_2 + l_3 + l_4 - 30)}, \nd
where one may notice the scaling resemblance to \eqref{meggy2}. All the ${\bf X}_8$ polynomials, including their $g_s$ scalings, fill up rows 6 to 8 in {\bf Table \ref{jhinuk12}}. To complete the story, we now need the lower bounds on the flux products appearing in rows 3 to 5 in {\bf Table \ref{jhinuk12}}. For these three cases the lower bounds become:

{\footnotesize
\bg\label{kuria}
{\rm log}_{\rm x}\left({\bf G}_{[{\rm MNPQ}} {\bf G}_{{\rm RS}0j]}\right) \ge -4, ~~~
{\rm log}_{\rm x}\left({\bf G}_{[{\rm MNPQ}} {\bf G}_{{\rm RS}ij]}\right) \ge -4, ~~~
{\rm log}_{\rm x}\left({\bf G}_{[{\rm MNPQ}} {\bf G}_{{\rm RS}aj]}\right) \ge -1,\nd}
respectively, where ${\rm x} = {g_s\over {\rm HH}_o}$. With these at hand, we are now ready to analyze all the EOMs associated with the flux components ${\bf G}_{0abi}$. First, the bound on $l_{0a}^{bi}$ may be easily ascertained by comparing the $g_s$ scalings in row 1 with the $g_s$ scalings of the flux products in rows 3 to 5 and with the ${\bf X}_8$ polynomials in rows 6 to 8 from {\bf Table \ref{jhinuk12}}. A before, the subtlety appears from how the derivatives act on the rank seven tensors here. For example, the $g_s$ scalings of row 1 will be $l_{0a}^{bi} - 3$ for temporal derivative, $l_{0a}^{bi} -2$ for derivatives along 
spatial ${\bf R}^2$ directions, and $l_{0a}^{bi}$ for derivatives along the toroidal ${\mathbb{T}^2\over {\cal G}}$ directions. Taking this into account, the bounds on $l_{0a}^{bi}$ appear to be:
\bg\label{kurtuki}
-1 \le l_{0a}^{bi} \le 1, ~~~~ -2 \le l_{0a}^{bi} \le 0, ~~~~ -2 \le l_{0a}^{bi} \le -1, \nd
for the three cases respectively. Unfortunately, not all the range of values can be consistent with EFT, and comparing with \eqref{perfum} one may easily see that $l_{0a}^{bi}$ lying between the range $-2$ and $-1$ are inconsistent with EFT, {\it i.e.} $-2 \le l_{0a}^{bi} < -1$ should be discarded. Interestingly, EFT considerations appears to rule out the whole range of values for $l_{0a}^{bi}$ when we compare row 1 with 
row 5 and row 8 in  {\bf Table \ref{jhinuk12}} for the case in which the derivatives are along the toroidal directions. Does that mean that there is {\it no} consistent way to express the EOM for this case? The answer is clearly {\it no} because the aforementioned comparison presumes that the quantum term from row 2 can only participate at the lowest orders. This is clearly not the case, as we saw in many of the other cases earlier, and we can always balance the rank eight tensors with various pieces of the quantum terms stemming from either \eqref{fahingsha5} or from the local and non-local BBS instantons 
\eqref{beverwickmey} (and other non-perturbative contributions). This means the allowed range can be taken to be $-1 \le l_{0a}^{bi} \le 1$, and the question is whether this is also consistent with EOMs. In the following we will verify this against the EOMs.

Starting with the temporal derivative the schematic diagram now resembles \eqref{lilcadu92} with the same values for $(\theta_1, \theta_2)$ for $l_{0a}^{bi} = -1$. When $l_{0a}^{bi} = 1$, the schematic diagram resembles the one in \eqref{lilcadu93}, again with the same values for $(\theta_1, \theta_2)$ therein, thus necessitating the contributions from the BBS instantons which would boost up $\theta_2$ to 
${14\over 3}$. At higher orders in $\left(k_i, k_{n_i}, l, {g_s\over {\rm HH}_o}\right)$, the various tensors in \eqref{mcacrisis} from 
{\bf Table \ref{jhinuk12}} may be identified pertubatively as:

{\footnotesize
\bg\label{megmojer9}
l_{0a}^{bi} -3 + {2\over 3}(k_1 + k_{n_1} + k_{n_2}) = \theta_{nl} - l_{0a}^{bi} - {17 \over 3} 
-{2k_2\over 3} = {2\over 3}(k_4 + k_5) + l_{[{\rm MN}}^{[{\rm PQ}} \oplus l_{{\rm RS}]}^{{0j}]} 
= \Big\{\begin{matrix}~l - 2\\ ~l - 10\end{matrix}~ , \nd} 
with $l \in {\mathbb{Z}\over 3}$ and the two choices are respectively for the two aforementioned metrics, \eqref{makibhalu3} with $\gamma = 6$ and the generic eleven-dimensional coordinate dependent metric.
When the derivatives act along the spatial ${\bf R}^2$ directions, for $l_{0a}^{bi} = -1$, the schematic diagram resembles the one in \eqref{lilcadu95} with the same choices of quantum terms as depicted therein. For $l_{0a}^{bi} = 1$, the diagram is similar to \eqref{lilcadu95} but the quantum contributions come from $\theta_1 = {14\over 3}, \theta_{i, 2} = \theta_{i, 3} = {11\over 3}$ with the latter getting contributions from the BBS instantons \eqref{beverwickmey}. Again, for higher orders in $\left(k_i, k_{n_i}, l, {g_s\over {\rm HH}_o}\right)$ the identifications of the various rank seven and eight tensors would entail:

{\footnotesize
\bg\label{megmojer9}
l_{0a}^{bi} - 2 + {2\over 3}(k_1 + k_{n_1} + k_{n_2}) = \theta_{nl} - l_{0a}^{bi} - {14 \over 3} 
-{2k_2\over 3} = {2\over 3}(k_4 + k_5) + l_{[{\rm MN}}^{[{\rm PQ}} \oplus l_{{\rm RS}]}^{{ij}]} 
= \Big\{\begin{matrix}~l - 2\\ ~l - 10\end{matrix}~ , \nd} 
perturbatively with sub-leading non-perturbative and non-local contributions. Finally, when the derivatives act along the toroidal ${\mathbb{T}^2\over {\cal G}}$ directions, for the case $l_{0a}^{bi} = -1$, the schematic diagram resembles \eqref{lilcadu82} but with $\theta_2 = {\red -{1\over 3}}$. Introducing BBS instantons would change this to $+{5\over 3}$. Of course since the lowest scaling carried by ${\bf G}_{0abi}$ is 
${1\over 3}$, the value of $\theta_2 = {5\over 3}$ may still be not enough to balance with the ${\bf X}_8$ polynomial, so the next order may become important. When $l_{0a}^{bi} = 1$, the schematic diagram 
resembles somewhat \eqref{lilcadu89}, but with the following differences:

{\footnotesize
\bg\label{lilcadux}
d_{(a)}\ast\overbracket[1pt][7pt]{{\bf G}_4 + b_1 d_{(a)}\ast\Big(\mathbb{Y}_4}^{\theta_1 \ge {14 \over 3}}(\theta_1) +
\mathbb{Y}^{\rm BBS}_4\overbracket[1pt][7pt]{(\theta_{a, 2}) + \mathbb{Y}^{\rm BBS}_4\underbracket[1pt][7pt]{(\theta_{a, 3})  +  ..\Big)- b_2\Big(\mathbb{T}_8}_{\theta_{a, 3} \ge  {8\over 3} + 2}{}^{(a, b)}\Big) -b_3 \Big(\mathbb{X}_8}^{\theta_{a, 2} \ge {5 \over 3} + 2}{}^{(a, b)}\Big) = 0, \nonumber\\ \nd}
where we now see that, since the lowest scaling contributed by ${\bf G}_{0abi}$ is ${7\over 3}$,  there are no perturbative contributions balancing the quantum terms with the flux products or the ${\bf X}_8$ polynomial. Non-perturbative contributions become essential here, although for $\theta_{a, 2} = {11\over 3}$
one might also have to incorporate non-local contributions for consistency. On the other hand, for sufficiently higher orders in $\left(k_i, k_{n_i}, l, {g_s\over {\rm HH}_o}\right)$ we expect at least perturbatively we can balance the various tensors in 
{\bf  Table \ref{jhinuk12}} via:
\bg\label{megmojer10}
l_{0a}^{bi}  + {2\over 3}(k_1 + k_{n_1} + k_{n_2}) = \theta_{nl} - l_{0a}^{bi} - {8 \over 3} 
-{2k_2\over 3} = {2\over 3}(k_4 + k_5) + l_{[{\rm MN}}^{[{\rm PQ}} \oplus l_{{\rm RS}]}^{{aj}]} 
= \Big\{\begin{matrix}~l \\ ~l - 8\end{matrix}~ , \nonumber\\ \nd 
with the non-pertubative (and non-local) terms contributing at the sub-leading order. This is where the merging of the over-brackets and the under-brackets in\eqref{lilcadux} can happen. 
Putting everything together, it appears that consistency with the quantum dynamics can again be met for the flux components 
${\bf G}_{0abi}$ in the range $-1 \le l_{0a}^{bi} \le 1$.

 \begin{table}[tb]  
 \begin{center}
\renewcommand{\arraystretch}{1.5}
\begin{tabular}{|c||c||c|}\hline ${\bf G}_{0{\rm M}ij}$ tensors & form & ${g_s\over {\rm HH}_o}$ scaling \\ \hline\hline
$\left(\mathbb{T}_7^{(f)}\right)_{{\rm NPQRS}ab}$ & $\sqrt{-{\bf g}_{11}} {\bf G}^{0ij{\rm M}} {\rm F}_1^{(n_1)}
{\rm F}_2^{(n_2)} \epsilon_{0ij{\rm M}{\rm NPQRS}ab}$ & $l_{0i}^{j{\rm M}} + 4 + {2\over 3}\left(k_1 + k_{n_1} 
+ k_{n_2}\right)$ \\ \hline 
$\left(\mathbb{T}_7^{(q)}\right)_{{\rm NPQRS}ab}$ & $\sqrt{-{\bf g}_{11}} \mathbb{Y}_4^{0ij{\rm M}} 
\epsilon_{0ij{\rm M}{\rm NPQRS}ab}$ & $\theta_{nl}(k_2) - l_{0i}^{j{\rm M}} - {14\over 3}  
- {2k_2\over 3}$ \\ \hline 
$\left(\mathbb{T}_8^{(0, 0)}\right)_{{\rm NPQRS}0ab}$ & ${\bf G}_{[{\rm NPQR}} {\bf G}_{{\rm S}0ab]}$ & ${2\over 3}\left(k_4 + k_5\right) + l_{[{\rm NP}}^{[{\rm QR}} \oplus l_{{0{\rm S}}]}^{ab]}$ 
\\ \hline 
$\left(\mathbb{T}_8^{(i, j)}\right)_{{\rm NMPQRS}abi}$ & ${\bf G}_{[{\rm NPQR}} {\bf G}_{{\rm S}abi]}$ & ${2\over 3}\left(k_4 + k_5\right) + l_{[{\rm NP}}^{[{{\rm QR}}} \oplus l_{{\rm S}a]}^{bi]}$ 
\\ \hline 
$\left(\mathbb{T}_8^{(m, \alpha)}\right)_{{\rm MNPQRS}ab}$ & ${\bf G}_{[{\rm MNPQ}} {\bf G}_{{\rm RS}ab]}$ & ${2\over 3}\left(k_4 + k_5\right) + l_{[{\rm MN}}^{[{{\rm PQ}}} \oplus l_{{\rm RS}]}^{ab]}$ 
\\ \hline 
$\left(\mathbb{X}_8^{(0, 0)}\right)_{{\rm NPQRS}0ab}$ & \eqref{meggy13}, \eqref{meggy14}
& $l + 1, l - 7$ \\ \hline
$\left(\mathbb{X}_8^{(i, j)}\right)_{{\rm NPQRS}abi}$ & \eqref{meggy7}, \eqref{meggy8} 
& $l + 1, l - 7$ \\ \hline
$\left(\mathbb{X}_8^{(m, \alpha)}\right)_{{\rm MNPQRS}ab}$ & \eqref{thetrial2}, \eqref{thetrial3}
& $l + 2, l - 6$ \\ \hline
$\left({\bf \Lambda}_8\right)_{{\rm MNPQRS}ab}$ & $\delta^8(z - z')$
& $0$ \\ \hline
\end{tabular}
\renewcommand{\arraystretch}{1}
\end{center}
 \caption[]{Comparing the $g_s$ scalings of the various tensors that contribute to the EOM for the G-flux components ${\bf G}_{0ij{\rm M}}$ in \eqref{mcacrisis}. Note that this is an elaborated version of the results in {\bf Table \ref{jhinuk1}} by including the effects from the three-forms ${\bf C}_{{\rm M}ij}, {\bf C}_{0{\rm M}i}$ and ${\bf C}_{0ij}$ with derivatives along the temporal, spatial and the internal six directions respectively.} 
  \label{jhinuk10}
 \end{table}

\vskip.2in

\noindent {\it Case 5: ${\bf G}_{0ij{\rm M}}$ and ${\bf G}_{0ija}$ components}

\vskip.2in

\noindent We start by the flux components ${\bf G}_{0ij{\rm M}}$ that we already dealt in some details in {\bf Table \ref{jhinuk1}}. However therein we only considered the EOM coming from the three-form field 
${\bf C}_{0ij}$, and did not study the remaining EOMs from the three-form fields ${\bf C}_{ij{\rm M}}$ and 
${\bf C}_{0i{\rm M}}$ with derivatives along the temporal and the spatial ${\bf R}^2$ directions. In the same vein, we should also consider the flux components ${\bf G}_{0ija}$ whose EOMs originate from the three-form fields ${\bf C}_{ija}$, with temporal derivative, ${\bf C}_{0ij}$, with derivatives along the toroidal directions and ${\bf C}_{0ia}$, with derivatives along the spatial directions. First however let us get some of the consistency checks out of the way. Comparing rows 1 and 2 in {\bf Tables \ref{jhinuk10}} and 
{\bf \ref{jhinuk11}}, it is easy to see that the flux components ${\bf G}_{0ij{\rm M}}$ and 
${\bf G}_{0ija}$ contribute as:
\bg\label{gomji0}
\theta_{nl}^{(1)} = 2\left(l_{0i}^{j{\rm M}} + {13\over 3}\right), ~~~~~
\theta_{nl}^{(2)} = 2\left(l_{0i}^{j{\rm M}} + {10\over 3}\right), \nd
respectively, which matches with the scalings in \eqref{botsuga4} when $(l_{67}, l_{68})$ and 
$l_{86}$ take values 2 at the lowest orders to provide the corresponding kinetic terms. The three-forms associated with the two set of flux components are associated with the ${\bf X}_8$ polynomials. For 
${\bf G}_{0ij{\rm M}}$, the ${\bf X}_8$ polynomials related to ${\bf C}_{0ij}$ and ${\bf C}_{0i{\rm M}}$ have already been derived respectively in \eqref{thetrial2} and \eqref{meggy7} for the metric \eqref{makibhalu3} with $\gamma = 6$ and in \eqref{thetrial3} and \eqref{meggy8} for the metric with generic eleven-dimensional dependence. Associated with the three-form ${\bf C}_{ij{\rm M}}$, the ${\bf X}_8$ polynomials 
are related to the following permutations of the indices of the four curvature two-forms:
\bg\label{lilcadu106x}
\overbracket[1pt][7pt]{{\rm N}~\underbracket[1pt][7pt]{{\rm R}~{\rm P}}~{\rm Q}} ~\overbracket[1pt][7pt]{{\rm S}~\underbracket[1pt][7pt]{{\rm 0}~a} ~b} ~~~ 
+~~~~{\rm permutations}, \nd
with the two input metric ans\"atze, one, \eqref{makibhalu3} with $\gamma = 6$, and two, generic metric with eleven-dimensional dependences. For the first case, all permutations lead to the same dominant $g_s$ scaling of:
\bg\label{meggy13}
{\bf X}_8(x, y, w^a; g_s) = \sum_{\{l_i\}} \widetilde{\bf X}_{(8, 13a)}^{(l_1, ...., l_4)}(x, y, w^a) 
\left({g_s\over {\rm HH}_o}\right)^{{1\over 3}(l_1 + l_2 + l_3 + l_4 + 3)}, \nd
which is a positive integer. Interestingly, if we had taken $\gamma = 5$ in \eqref{makibhalu3}, this would not have been the case, and the dominant scale would be $-{1\over 3}$ captured by three of the fourteen permutations of the alphabets in \eqref{lilcadu106x} (the remaining are 0). For the second case, {\it i.e.} when the metric allows a generic dependence on all the eleven-dimensional coordinates, the polynomial becomes:
\bg\label{meggy14}
{\bf X}_8(x, y, w^a; g_s) = \sum_{\{l_i\}} \widetilde{\bf X}_{(8, 13b)}^{(l_1, ...., l_4)}(x, y, w^a) 
\left({g_s\over {\rm HH}_o}\right)^{{1\over 3}(l_1 + l_2 + l_3 + l_4 - 21)}, \nd
where one may note the similarities between the $g_s$ scalings from \eqref{meggy7} and 
\eqref{meggy8} although their functional forms differ.  To complete the story, one may also work out the bounds on the products of the flux components from rows 3 to 5 in {\bf Table \ref{jhinuk10}}. They are given by:

{\footnotesize
\bg\label{kuriaT}
{\rm log}_{\rm x}\left({\bf G}_{[{\rm NPQR}} {\bf G}_{{\rm S}0ab]}\right) \ge -1, ~~~
{\rm log}_{\rm x}\left({\bf G}_{[{\rm NPQR}} {\bf G}_{{\rm S}abi]}\right) \ge -1, ~~~
{\rm log}_{\rm x}\left({\bf G}_{[{\rm MNPQ}} {\bf G}_{{\rm RS}ab]}\right) = 2,\nd}
where note the {\it equality} in the last case. This stems from our earlier choice \eqref{collateral} which, in fact, also fixes $l_{0i}^{j{\rm M}} = -4$. Can this continue to hold once we take other EOMs associated with the flux components ${\bf G}_{0ij{\rm M}}$? To see this we will have to find the allowed range of choices for 
$l_{0i}^{j{\rm M}}$. 

However before going into this, let us ask the reason for the choice $l_{0i}^{j{\rm M}} = -4$ in the first place. 
As shown in great details in the first reference of \cite{desitter2} (see section 4.2.4 therein), the choice of 
$l_{0i}^{j{\rm M}} = -4$ stems from the presence of dynamical membranes in the system. In fact this fixes the form of ${\bf G}_{0ij{\rm M}}$ to be \cite{desitter2}:
\bg\label{kurisada}
{\bf G}_{0ij{\rm M}}(x, y, w^a; g_s) &= & \epsilon_{0ij}\left[\bar{c}_1\partial_{\rm M}\left({\rm HH}_o\right)^{-4} + 
{\bar{c}_2\over \left({\rm HH}_o\right)^4} ~\widetilde{g}^{\mu\nu} ~\partial_{\rm M} \widetilde{g}_{\mu\nu}\right] 
\left({g_s\over {\rm HH}_o}\right)^{-4} \nonumber\\
&&~~~~~~~~~~~ +  
\sum_{k \in {\mathbb{Z}\over 2}} {\cal G}^{(k)}_{0ij{\rm M}}(x, y, w^a)
\left({g_s\over {\rm HH}_o}\right)^{{2\over 3}(k - 6)}, \nd
where $(\bar{c}_1, \bar{c}_2)$ are numerical constants; ${\rm H} = {\rm H}(y)$ and ${\rm H}_o = {\rm H}_o(x, y)$ are the warp-factors; and $\widetilde{g}_{\mu\nu}$ is the un-warped ({\it i.e.} $g_s$ independent) metric along the space-time ${\bf R}^{2, 1}$ as we took in 
\eqref{makibhalu3} with $\gamma = 6$. Note the piece that depends on the derivative of the un-warped metric. Such a piece is absent for the {\it flat-slicing} that we took in \cite{desitter2}, but appears when we take the de Sitter space with generic slicings. For the metric with generic dependence on all the eleven dimensional coordinates, 
both ${\rm H}$ and ${\rm H}_o$ can develop dependences on $w^a$. This would lead to a breakdown of EFT, so we won't elaborate it further. Note that the first term of \eqref{kurisada} 
seems to allow a three-form ${\bf C}_{0ij} = \epsilon_{0ij}~ g_s^{-4}$, if we assume sufficiently slowly 
moving membranes such that \eqref{kurisada} may be approximated to 
${\bf G}_{0ij{\rm M}} = {1\over 4!} \partial_{[{\rm M}} {\bf C}_{0ij]}$. In such a case both temporal and spatial derivatives on the rank seven tensors from {\bf Table \ref{jhinuk10}} should contribute to the EOMs. What happens if they are accounted for?

To see the effects of the other EOMs coming from the three derivative actions $-$ temporal, spatial and internal six-dimensional $-$ on the rank seven tensors 
in {\bf Table \ref{jhinuk10}}, we will first ignore the fact that $l_{0i}^{j{\rm M}} = -4$. This means comparing 
the $g_s$ scalings $l_{0i}^{j{\rm M}} + 3$, $l_{0i}^{j{\rm M}} + 4$ and $l_{0i}^{j{\rm M}} + 4$ with rows 
$(3, 6)$, rows $(4, 7)$ and rows $(5, 8, 9)$ from {\bf Table \ref{jhinuk10}} respectively. This leads to the following range of values for $l_{0i}^{j{\rm M}}$:
\bg\label{kalakuria}
-4 \le l_{0i}^{j{\rm M}} \le -2, ~~~~~ -5 \le l_{0i}^{j{\rm M}} \le -3, ~~~~~ l_{0i}^{j{\rm M}} = (-2, -4), \nd
where the lower bound $l_{0i}^{j{\rm M}} = -5$ clashes with EFT bound from \eqref{perfum} and therefore the whole range $-5 \le l_{0i}^{j{\rm M}} < -4$ is eliminated. Thus the valid range, consistent with EFT, appears to be 
$-4 \le l_{0i}^{j{\rm M}} \le -2$ which matches well with the values $l_{0i}^{j{\rm M}} = -2$ or 
$l_{0i}^{j{\rm M}} = -4$ coming from the derivatives along ${\cal M}_4 \times {\cal M}_2$ on the rank seven tensors. 

The last term in \eqref{kalakuria} deserves some comments. The choice $l_{0i}^{j{\rm M}} = -4$ appears when we identify row 1 with row 9 in {\bf Table \ref{jhinuk10}}, {\it i.e.} identify the flux derivatives along the internal six directions with the membrane sources. On the other hand, $l_{0i}^{j{\rm M}} = -2$ appears from identifying row 1 with either flux products (row 5) or the ${\bf X}_8$ polynomial (row 8) in 
{\bf Table \ref{jhinuk10}}. We also see the appearance of $l_{0i}^{j{\rm M}} = -3$, when we take the spatial derivatives. Such a choice appears when we take no branes in the system, as predicted earlier in 
\eqref{renaissance2}. 

There is however a subtlety when we look at \eqref{lilcadu83}. The lowest order $g_s$ identifications therein involve $\mathbb{Y}_4\left(\theta_{{\rm M}, a}\right)$ matching with the $g_s$ scalings of {\it both}
$\mathbb{T}_8^{({\rm M}, a)}$ and $\mathbb{X}_8^{({\rm M}, a)}$. This doesn't appear to be the case when we take the EOMs associated with derivatives along the spatial ${\bf R}^2$ directions. However if we take 
$l_{0i}^{j{\rm M}} = -3$ for the case with derivatives along the internal six directions, the schematic diagram resembles \eqref{lilcadu83} with $\theta_1 = {8\over 3}$ and $\theta_{\rm M} = {11\over 3}$ in 
\eqref{botsuga4}. All of these happens in the absence of branes, but the subtlety pointed out in footnote \ref{KoTo} still remains. Thus it appears that the realization of $l_{0i}^{j{\rm M}} = -3$ may be a bit tricky here. 

Let us move forward by asking what does the choice $l_{0i}^{j{\rm M}} = -4$ entails for the other EOMs. The EOM with temporal derivative allows the range $-4 \le l_{0i}^{j{\rm M}} \le -2$ as we saw in \eqref{kalakuria}.
When $l_{0i}^{j{\rm M}} = -4$, the schematic diagram here resembles \eqref{lilcadu92} with the same choices for $(\theta_1, \theta_2)$. Similarly when $l_{0i}^{j{\rm M}} = -2$, the diagram resembles \eqref{lilcadu93} with same values for 
$(\theta_1, \theta_2)$ therein implying the necessity of adding BBS instantons from \eqref{beverwickmey}
thus boosting $\theta_2$ to ${14\over 3}$. At higher orders in $\left(k_i, k_{n_i}, l, {g_s\over {\rm HH}_o}\right)$ we expect a relation like:

{\footnotesize
\bg\label{megmojer100}
l_{0i}^{j{\rm M}}  + 3 +  {2\over 3}(k_1 + k_{n_1} + k_{n_2}) = \theta_{nl} - l_{0i}^{j{\rm M}} - {17 \over 3} 
-{2k_2\over 3} = {2\over 3}(k_4 + k_5) + l_{[{\rm NP}}^{[{\rm QR}} \oplus l_{{\rm S}0]}^{{ab}]} 
= \Big\{\begin{matrix}~l + 1 \\ ~l - 7\end{matrix}~ , \nd} 
perturbatively, much like what we had earlier. When the derivatives act along the internal six directions, we 
expect \eqref{cora2} to be valid between the $g_s$ scalings of the various tensors in {\bf Table \ref{jhinuk10}}. In the absence of branes, there is a possibility of allowing $l_{0i}^{j{\rm M}} = -2$  (modulo the subtlety pointed out in footnote \ref{KoTo}). For this case the schematic diagram becomes:
\bg\label{lilcadu911}
d_{({\rm M})}\ast\overbracket[1pt][7pt]{{\bf G}_4 + b_1 d_{({\rm M})}\ast\Big(\mathbb{Y}_4}
\overbracket[1pt][7pt]{(\theta_1) + \mathbb{Y}_4(\theta_2)  +  ..\Big) - b_2\Big(\mathbb{T}_8}^{\theta_1 \ge {14 \over 3}}\overbracket[1pt][7pt]{{}^{(m, \alpha)}\Big) 
- b_3 \Big(\mathbb{X}_8}{}^{(m, \alpha)}\Big) = 0, \nonumber\\ \nd  
which would mean that, even to the lowest orders, \eqref{cora2} could be perturbatively realized (once we further identify it to $l + 2$ where $l \in {\mathbb{Z}\over 3}$ for the metric \eqref{makibhalu3} with 
$\gamma = 6$). For the case with spatial derivatives, taking $l_{0i}^{j{\rm M}} = -4$ gives vanishing $g_s$ scaling of the flux derivatives ({\it i.e.} derivatives on the rank seven tensors in row 1 of 
{\bf Table \ref{jhinuk10}}), but since we are not allowing membranes to wrap ${\bf S}^1 \in 
{\cal M}_4 \times {\cal M}_2$, the schematic diagram will have to resemble \eqref{lilcadu95} with the input of
BBS instantons in the same way therein. Perturbatively however, at higher orders in $\left(k_i, k_{n_i}, l, {g_s\over {\rm HH}_o}\right)$, the identifications become:

{\footnotesize
\bg\label{megmojer101}
l_{0i}^{j{\rm M}}  +  4 +  {2\over 3}(k_1 + k_{n_1} + k_{n_2}) = \theta_{nl} - l_{0i}^{j{\rm M}} - {14 \over 3} 
-{2k_2\over 3} = {2\over 3}(k_4 + k_5) + l_{[{\rm NP}}^{[{\rm QR}} \oplus l_{{\rm S}a]}^{{bi}]} 
= \Big\{\begin{matrix}~l + 1 \\ ~l - 7\end{matrix}~, \nd} 
which may be compared to \eqref{cora2} and \eqref{megmojer100}. 
After the dust settles, it appears that our original choice of 
$l_{0i}^{j{\rm M}} = -4$ still works consistently even in the presence of additional EOMs for the flux components ${\bf G}_{0ij{\rm M}}$. The fact that 
negative $g_s$ scaling doesn't create a problem here at late time because it is proportional to the volume form which also scales as 
$\left({g_s\over {\rm HH}_o}\right)^{-4}$.

 \begin{table}[tb]  
 \begin{center}
\renewcommand{\arraystretch}{1.5}
\begin{tabular}{|c||c||c|}\hline ${\bf G}_{0ija}$ tensors & form & ${g_s\over {\rm HH}_o}$ scaling \\ \hline\hline
$\left(\mathbb{T}_7^{(f)}\right)_{{\rm MNPQRS}b}$ & $\sqrt{-{\bf g}_{11}} {\bf G}^{0ija} {\rm F}_1^{(n_1)}
{\rm F}_2^{(n_2)} \epsilon_{0ija{\rm MNPQRS}b}$ & $l_{0i}^{ja} + 2 + {2\over 3}\left(k_1 + k_{n_1} 
+ k_{n_2}\right)$ \\ \hline 
$\left(\mathbb{T}_7^{(q)}\right)_{{\rm MNPQRS}b}$ & $\sqrt{-{\bf g}_{11}} \mathbb{Y}_4^{0ija} 
\epsilon_{0ija{\rm MNPQRS}b}$ & $\theta_{nl}(k_2) - l_{0i}^{ja} - {14\over 3}  
- {2k_2\over 3}$ \\ \hline 
$\left(\mathbb{T}_8^{(0, 0)}\right)_{{\rm MNPQRS}0b}$ & ${\bf G}_{[{\rm MNPQ}} {\bf G}_{{\rm RS}0b]}$ & ${2\over 3}\left(k_4 + k_5\right) + l_{[{\rm MN}}^{[{\rm PQ}} \oplus l_{{\rm RS}]}^{0b]}$ 
\\ \hline 
$\left(\mathbb{T}_8^{(a, b)}\right)_{{\rm MNPQRS}ab}$ & ${\bf G}_{[{\rm MNPQ}} {\bf G}_{{\rm RS}ab]}$ & ${2\over 3}\left(k_4 + k_5\right) + l_{[{\rm MN}}^{[{{\rm PQ}}} \oplus l_{{\rm RS}]}^{ab]}$ 
\\ \hline 
$\left(\mathbb{T}_8^{(i, j)}\right)_{{\rm MNPQRS}bi}$ & ${\bf G}_{[{\rm MNPQ}} {\bf G}_{{\rm RS}bi]}$ & ${2\over 3}\left(k_4 + k_5\right) + l_{[{\rm MN}}^{[{{\rm PQ}}} \oplus l_{{\rm RS}]}^{bi]}$ 
\\ \hline 
$\left(\mathbb{X}_8^{(0, 0)}\right)_{{\rm MNPQRS}0b}$ & \eqref{meggy15}, \eqref{meggy16} 
& $l, l - 8$ \\ \hline
$\left(\mathbb{X}_8^{(a, b)}\right)_{{\rm MNPQRS}ab}$ & \eqref{thetrial2}, \eqref{thetrial3} 
& $l + 2, l - 6$ \\ \hline
$\left(\mathbb{X}_8^{(i, j)}\right)_{{\rm MNPQRS}bi}$ & \eqref{meggy5}, \eqref{meggy6} 
& $l, l - 8$ \\ \hline
\end{tabular}
\renewcommand{\arraystretch}{1}
\end{center}
 \caption[]{Comparing the $g_s$ scalings of the various tensors that contribute to the EOM for the G-flux components ${\bf G}_{0ia}$ in \eqref{mcacrisis}. Despite similarities to the flux components 
 ${\bf G}_{0ij{\rm M}}$, the behavior will turn out to be quite different. The three-form fluxes contributing to the
 eight-forms as well as the EOMs are
  ${\bf C}_{ija}$ with derivative acting along the temporal direction, ${\bf C}_{0ij}$ with derivatives acting along the toroidal directions
 and ${\bf C}_{0ia}$ with derivatives acting the spatial directions.} 
  \label{jhinuk11}
 \end{table}

The story however gets a bit more subtle for the flux components ${\bf G}_{0ija}$. Clearly since both the warp factors ${\rm H}$ and ${\rm H}_o$ {\it cannot} be functions of $w^a$ (to avoid breakdown of EFT), the flux components ${\bf G}_{0ija}$ cannot take the form \eqref{kurisada}. Furthermore EFT bound from 
\eqref{perfum} will tell us that such flux components cannot scale as 
$\left({g_s \over {\rm HH}_o}\right)^{-4}$.

What could be the simplest way to derive the dominant scaling for ${\bf G}_{0ija}$ components? One way would be to analyze all the EOMs from {\bf Table \ref{jhinuk11}}, as we have done for the other flux components before. Another, maybe a bit more efficient, way would be to look at the membrane dynamics and determine the form for the flux components ${\bf G}_{0ija}$, much like the way we derived 
\eqref{kurisada}. However as cautioned above, the scaling {\it cannot} be 
$\left({g_s\over {\rm HH}_o}\right)^{-4}$, so we need to proceed carefully. The analysis of the slowly moving {\rm M2} branes, as derived in section 4.2.4 in the first reference of \cite{desitter2} leads to the following possible form for the dominant scaling of ${\bf G}_{0ija}$:
\bg\label{HiGaToK}
{\bf G}_{0ija} = -{3\over 2}\sqrt{-\gamma_{(2)}}~{\bf g}^{\mu\nu} ~\partial_a {\bf g}_{\mu\nu}~\epsilon_{0ij}
 =  {\bar{c}_3(w^a)\over \left({\rm HH}_o\right)^4} \left({g_s\over {\rm HH}_o}\right)^{-4 + {\gamma\over 3}} + ....., 
\nd  
where the dotted terms capture some of the sub-dominant powers of $g_s$, and 
$\bar{c}_3(w^a)$ is a function of the toroidal coordinates, especially $x^{11}$. The complete form of the sub-dominant pieces can be worked out once we take dynamical membranes into account, but we will not do so here. Instead let us examine \eqref{HiGaToK} carefully. The $\gamma$ appearing\footnote{Not to be confused with $\gamma_{(2)}$ which is the determinant of the pull-back metric on the membrane \cite{desitter2}.} in \eqref{HiGaToK} stems from our metric choice \eqref{makibhalu3}, so when $\gamma = 6$, the {\it dominant} scaling for the flux components ${\bf G}_{0ija}$ appears to be $\left({g_s\over {\rm HH}_o}\right)^{-2}$, which is well within the EFT bound from \eqref{perfum}. There is however a subtlety here: the choice 
\eqref{HiGaToK} relies on the space-time metric to take the form \eqref{makibhalu3}. While the choice of the metric \eqref{makibhalu3}, {\it i.e.} ${\bf g}_{\rm AB}$,  when $({\rm A, B}) \in {\cal M}_4 \times {\cal M}_2$ makes sense, an extension to the space-time and the toroidal directions would imply a {\it deviation} from exact de Sitter space-time in the dual IIB side, unless we take the limit $x^{11} \to 0$ in the end. Unfortunately this makes $c_3(x^{11}) \to 0$, so produces vanishing flux components. 

The outcome of the above reasonings is that space filling membranes may not be a good source for the 
${\bf G}_{0ija}$ components, so our only choice appears to be an analysis of the EOMs from 
{\bf Table \ref{jhinuk11}} as alluded to earlier. The three-forms responsible for the EOMs are 
${\bf C}_{ija}$, with temporal derivative, ${\bf C}_{0ij}$, with derivatives along the toroidal directions, and 
${\bf C}_{0ia}$, with derivatives along the spatial directions. Similarly the ${\bf X}_8$ polynomials associated with ${\bf C}_{0ij}$ and ${\bf C}_{0ia}$ have been worked out in \eqref{thetrial2} and 
\eqref{meggy5} respectively for the metric choice \eqref{makibhalu3} with $\gamma = 6$; and in 
\eqref{thetrial3} and \eqref{meggy6} respectively for the metric with generic dependence on all the eleven dimensional coordinates. For the three-form ${\bf C}_{ija}$, the ${\bf X}_8$ polynomial becomes:
\bg\label{meggy15}
{\bf X}_8(x, y, w^a; g_s) = \sum_{\{l_i\}} \widetilde{\bf X}_{(8, 14a)}^{(l_1, ...., l_4)}(x, y, w^a) 
\left({g_s\over {\rm HH}_o}\right)^{{1\over 3}(l_1 + l_2 + l_3 + l_4)}, \nd
for the metric choice \eqref{makibhalu3} with $\gamma = 6$. Note the similarity of the $g_s$ scaling with the one for \eqref{meggy5}. This similarity continues to hold even when we take the generic metric, and the 
${\bf X}_8$ polynomial becomes:
\bg\label{meggy16}
{\bf X}_8(x, y, w^a; g_s) = \sum_{\{l_i\}} \widetilde{\bf X}_{(8, 14b)}^{(l_1, ...., l_4)}(x, y, w^a) 
\left({g_s\over {\rm HH}_o}\right)^{{1\over 3}(l_1 + l_2 + l_3 + l_4 - 24)}, \nd
although the functional forms differ. If we had taken $\gamma = 5$ in \eqref{makibhalu3}, the dominant scaling would have been $-{4\over 3}$ from one of the permutations of the indices of the four curvature two-forms in the following:
\bg\label{lilcadu107x}
\overbracket[1pt][7pt]{{\rm M}~\underbracket[1pt][7pt]{{\rm N}~{\rm P}}~{\rm Q}} ~\overbracket[1pt][7pt]{{\rm R}~\underbracket[1pt][7pt]{{\rm S}~0} ~b} ~~~ 
+~~~~{\rm permutations}, \nd
with all the remaining ones giving $-1$ as the $g_s$ scaling. Expectedly, for both \eqref{meggy15} and 
\eqref{meggy16}, the arrangements of the curvature forms across the permutations, all lead to the same dominant scalings of 0 and $-8$ respectively. The lower bounds on the flux products from rows 3 to 5 in
{\bf Table \ref{jhinuk11}} now become:

{\footnotesize
\bg\label{ktoki}
{\rm log}_{\rm x}\left({\bf G}_{[{\rm MNPQ}} {\bf G}_{{\rm RS}0b]}\right) \ge -1, ~~~
{\rm log}_{\rm x}\left({\bf G}_{[{\rm MNPQ}} {\bf G}_{{\rm RS}ab]}\right) = 2, ~~~
{\rm log}_{\rm x}\left({\bf G}_{[{\rm MNPQ}} {\bf G}_{{\rm RS}bi]}\right) \ge -1,\nd}
where now the equality in the middle term appears from our choice \eqref{collateral}. Again, as before, 
we need to compare the $g_s$ scalings of the flux products and the ${\bf X}_8$ polynomials with the flux derivatives from row 1 of {\bf Table \ref{jhinuk11}}. The latter takes values $l_{0i}^{ja} + 1, l_{0i}^{ja} + 4$ and $l_{0i}^{ja} + 2$ for the temporal, toroidal and spatial derivatives respectively, producing the following range of choices:
\bg\label{duilibmey}
-2 \le l_{0i}^{ja} \le -1, ~~~~~ l_{0i}^{ja} = -2, ~~~~~ -3 \le l_{0i}^{ja} \le -2, \nd
which when combined together gives us $l_{0i}^{ja}$ lying between $-3$ and $-1$, {\it i.e.} 
$-3 \le l_{0i}^{ja} \le -1$. Note that the case with derivatives along the toroidal directions, reproduces the dominant scaling of $l_{0i}^{ja} = -2$ which should be compared to \eqref{HiGaToK}. The schematic diagram for this case becomes:
\bg\label{lilcadu912}
d_{(a)}\ast\overbracket[1pt][7pt]{{\bf G}_4 + b_1 d_{(a)}\ast\Big(\mathbb{Y}_4}
\overbracket[1pt][7pt]{(\theta_1) + \mathbb{Y}_4(\theta_2)  +  ..\Big) - b_2\Big(\mathbb{T}_8}^{\theta_1 \ge {8 \over 3}}\overbracket[1pt][7pt]{{}^{(a, b)}\Big) 
- b_3 \Big(\mathbb{X}_8}{}^{(a, b)}\Big) = 0, \nonumber\\ \nd  
with the choice of $\theta_1 = {8\over 3}$ consistent with \eqref{gomji0}. This equality of the $g_s$ scalings of the various tensors in \eqref{lilcadu912} at the lowest orders implies that the equality would continue to hold at higher orders in $\left(k_i, k_{n_i}, l, {g_s\over {\rm HH}_o}\right)$ provided:
\bg\label{megmojer102}
{2\over 3}(k_1 + k_{n_1} + k_{n_2}) + 2 = \theta_{nl} -{2\over 3}(1 + k_2) = {2\over 3}(k_4 + k_5) + 2
= l + 2, \nd
where $l \in {\mathbb{Z}\over 3}$ and $\theta_{nl}$ can be fixed from \eqref{botsuga4}. Interestingly, while \eqref{lilcadu911} and \eqref{lilcadu912} are realized for $l_{0i}^{j{\rm M}} = l_{0i}^{ja} = -2$, both appears to be riddled with subtleties. In general, for $-3 \le l_{0i}^{ja} \le -1$, we expect to some higher orders in $\left(k_i, k_{n_i}, l, {g_s\over {\rm HH}_o}\right)$, the following matching conditions:

{\footnotesize
\bg\label{megmojer103}
l_{0i}^{ja}  +  2 + {\red {\rm X}_{\partial}^r} +  {2\over 3}(k_1 + k_{n_1} + k_{n_2}) = \theta_{nl} 
- l_{0i}^{ja} - {14 \over 3} -{2k_2\over 3}  + {\red {\rm X}_{\partial}^r} = {2\over 3}(k_4 + k_5) + l_{[{\rm MN}}^{[{\rm PQ}} \oplus 
l_{{\rm RS}]}^{{b{\red r}}]} 
= \Big\{\begin{matrix}~l + {\red \alpha_r} \\ ~l + {\red \beta_r} \end{matrix}~, \nonumber\\ \nd} 
where ${\red r} = (0, i, a)$ depending on the temporal, spatial and toroidal derivatives respectively; with the above matching conditions to hold {\it pertubatively} so that we can read up $\theta_{nl}$ from \eqref{botsuga4} from the pertubative quantum terms in \eqref{fahingsha5}. The other parameters appearing in 
\eqref{megmojer103} are defined as follows:

{\footnotesize
\bg\label{megkev}
\left({\red {\rm X}_{\partial}^0, {\rm X}_{\partial}^i, {\rm X}_{\partial}^a}\right) = (-1, 0, 2), ~~
 \left({\red \alpha_0, \beta_0}\right) = (0, -8),~~ \left({\red \alpha_i, \beta_i}\right) = (0, -8), ~~
\left({\red \alpha_a, \beta_a}\right) = (2, -6), \nd}
which may be easily read up from {\bf  Table \ref{jhinuk11}}. The consistency of \eqref{megkev} may be seen from taking the limiting values of $l_{0i}^{ja}$. For example when $l_{0i}^{ja} = -3$ and ${\red r} = 0$, the schematic diagram resembles \eqref{lilcadu87}, but with $\left(\theta_2, \theta_{0,2}, \theta_{0,3}\right) = 
\left({2\over 3}, {8\over 3}, {5\over 3}\right)$. When $l_{0i}^{ja} = -1$, the diagram resembles 
\eqref{lilcadu93} but with $\theta_3 = {11\over 3}$. One can similarly work out the other cases. Thus there appears to be well defined dynamics for the flux components ${\bf G}_{0ija}$ in the range 
$-3 \le l_{0i}^{ja} \le -1$, however for $l_{0i}^{ja} = -2$ one can keep ${\bf G}_{0ija} = 0$ to avoid conflicting with EFT.

\begin{table}[tb]  
 \begin{center}
\renewcommand{\arraystretch}{1.5}
\begin{tabular}{|c||c||c||c|}\hline ${\bf G}_{0{\rm ABC}}$ tensors & contributing 3-forms & EFT bounds & allowed range of choices \\ \hline\hline
${\bf G}_{0{\rm MNP}}$ & ${\bf C}_{\rm MNP}, ~{\bf C}_{0{\rm MN}}$ & $-2$ & $-2 \le l_{\rm 0M}^{\rm NP} \le 0$ \\ \hline 
${\bf G}_{0{\rm NP}a}$ & ${\bf C}_{{\rm NP}a}, ~{\bf C}_{0{\rm NP}}, ~ {\bf C}_{{\rm 0P}a}$ & $-1$ & 
$-1 \le l_{\rm 0N}^{{\rm P}a} \le 1$ \\ \hline 
${\bf G}_{0{\rm N}ab}$ & ${\bf C}_{{\rm N}ab}, ~{\bf C}_{0{ab}}, ~ {\bf C}_{{\rm 0N}a}$ & $~~0$ & 
$0 \le l_{\rm 0N}^{ab} \le 2$ \\ \hline 
${\bf G}_{0{\rm MN}i}$ & ${\bf C}_{{\rm MN}i}, ~{\bf C}_{0{\rm MN}}, ~ {\bf C}_{{\rm 0N}i}$ & $-3$ & 
$-3 \le l_{\rm 0M}^{{\rm N}i} \le -1$ \\ \hline 
${\bf G}_{0{\rm N}ia}$ & ${\bf C}_{{\rm N}ia}, ~{\bf C}_{0{\rm N}a}, ~ {\bf C}_{{\rm 0}ia}~{\bf C}_{0i{\rm N}}$ & $-2$ & 
$-2 \le l_{\rm 0N}^{ia} \le 0$ \\ \hline 
${\bf G}_{0ij{\rm M}}$ & ${\bf C}_{ij{\rm M}}, ~{\bf C}_{0{\rm M}i}, ~ {\bf C}_{{\rm 0}ij}$ & $-4$ & 
$l_{0i}^{j{\rm M}} = -4$ \\ \hline 
${\bf G}_{0{ij}a}$ & ${\bf C}_{{ij}a}, ~{\bf C}_{0{ij}}, ~ {\bf C}_{{\rm 0}ia}$ & $-3$ & 
$-3 \le l_{0i}^{ja} \le -1$ \\ \hline 
${\bf G}_{0abi}$ & ${\bf C}_{abi}, ~{\bf C}_{0ab}, ~ {\bf C}_{0ai}$ & $-1$ & 
$-1 \le l_{0a}^{bi} \le 1$ \\ \hline 
\end{tabular}
\renewcommand{\arraystretch}{1}
\end{center}
 \caption[]{Comparing all the {\it allowed} $g_s$ scalings, $\left({g_s\over {\rm HH}_o}\right)^{l_{\rm 0A}^{\rm BC}}$, of the various tensors that contribute to the EOMs for the G-flux components ${\bf G}_{0{\rm ABC}}$ in \eqref{mcacrisis}, where $l_{\rm 0A}^{\rm BC}$ are the dominant scalings. We have also shown the EFT bounds that appear from \eqref{botsuga4} for the various components, as well as the three-form tensors associated with these components. These three-forms are essential to construct the structure of the associated ${\bf X}_8$ polynomials. In the range of choices in the $g_s$ scalings,  although the lower bounds are fixed, the upper bounds are flexible as elaborated in the text.} 
  \label{kat1}
 \end{table}

\vskip.2in

\noindent {\it Case Summary: ${\bf G}_{\rm 0ABC}$ flux equations and their behaviors}

\vskip.2in

\noindent In this section we studied in detail the behavior of the flux components 
${\bf G}_{\rm 0ABC}$. The result of our analysis is summarized in {\bf Table \ref{kat1}}. While the lower limits in the various range of choices are strictly bounded by EFT considerations, the upper limits are somewhat flexible. If we fix the form of ${\bf G}_{0ij{\rm M}}$ from \eqref{kurisada} and take ${\bf G}_{0ija} = 0$, then
we see from {\bf Table \ref{kat1}} that almost all of the dominant scalings $l_{\rm 0A}^{\rm BC}$ passes through $l_{\rm 0A}^{\rm BC} = 0$, except the flux components ${\bf G}_{{\rm 0MN}i}$ (which is not strict because of the flexibility of the upper bound). One choice then would be to fix all the dominant scalings, except $l_{0i}^{j{\rm M}}$ and $l_{0i}^{ja}$, to be zero, {\it i.e.} keep $l_{\rm 0A}^{\rm BC} = 0$. This attractive choice however suffers from the following issue: viewing the de Sitter space as a Glauber-Sudarshan state, the choice $l_{\rm 0A}^{\rm BC} = 0$ would imply that these flux components can only appear {\it off-shell}. 
A way to avoid the appearance of such off-shell states is to make the $g_s$ independent pieces of 
${\bf G}_{\rm 0ABC}$ to vanish. That this is possible can be seen from {\bf Tables \ref{jhinuk5}} to 
{\bf \ref{jhinuk11}}: the quantum terms can balance the ${\bf X}_8$ polynomials and the flux products, without invoking flux derivatives at this order, thus satisfying the EOMs consistently as we saw in our above analyses. 

Interestingly, keeping $l_{\rm 0A}^{\rm BC} = 0$ implies that some flux components would contribute at higher orders in $\left(k_i, k_{n_i}, l, {g_s\over {\rm HH}_o}\right)$, as may be seen from \eqref{botsuga4}. For example ${\bf G}_{\rm 0MNP}$ would contribute as $\left({g_s\over {\rm HH}_o}\right)^{7/3}$,  ${\bf G}_{{\rm 0MN}i}$ would contribute as 
$\left({g_s\over {\rm HH}_o}\right)^{10/3}$, and so on. They therefore cannot influence the Schwinger-Dyson's equations associated with the metric components. On the other hand, if their scalings are given by the lower bounds in {\bf Table \ref{kat1}}, then {\it all} flux components of the type ${\bf G}_{\rm 0ABC}$, including 
${\bf G}_{0ij{\rm M}}$, would contribute as $\left({g_s\over {\rm HH}_o}\right)^{1\over 3}$ (we still keep 
${\bf G}_{0ija} = 0$). The contributions of these to the space-time equations, {\it i.e.} the Schwinger-Dyson's equations for ${\bf g}_{\mu\nu}$, would be $-{8\over 3} + 2\times {1\over 3} = -2$ which is also how the flux components ${\bf G}_{{\rm MN}ab}$ contributed in \cite{desitter2}. From the quantum side, such contributions can only come from instanton effects as we discussed in \cite{coherbeta}. 
Because of the no-go conditions \cite{GMN, Dasgupta:2014pma}, it is only these non-perturbative quantum terms that could make an effect to this cause. One might then question whether there is any strong reason to keep the $\left({g_s\over {\rm HH}_o}\right)^p$ contributions from the fluxes with $p$ {\it small}. Can we not go with any $p \in {\mathbb{Z}\over 3}$, instead of $p = {\cal O}\left({1\over 3}\right)$?  The answer is {\it no}
because if $p >> {1\over 3}$, then \eqref{fahingsha5} and \eqref{botsuga4} will tell us that there may not be 
flux related contributions non-perturbatively either to the flux EOMs or to the Einstein's equations (both of these appearing here as consequence of the Schwinger-Dyson's equations). Therefore keeping $p$ small is advisable and useful. 

Our above discussions provide strong motivations to keep the flux components time-dependent. On the other hand, if we want to keep $l_{\rm 0A}^{\rm BC} = 0$ (except for $l_{0i}^{j{\rm M}}$ and 
$l_{0i}^{j{a}}$), then the only flux components that would contribute as 
$\left({g_s\over {\rm HH}_o}\right)^{1\over 3}$ are ${\bf G}_{{\rm 0M}ab}$, as may be easily seen from
\eqref{fahingsha5} and \eqref{botsuga4} (see the coefficients of $l_{90}$ and $l_{91}$ in \eqref{botsuga4}). Interestingly, these flux components go hand-in-hand with the flux components ${\bf G}_{{\rm MN}ab}$ that also scale as $\left({g_s\over {\rm HH}_o}\right)^{1\over 3}$ (see the coefficients of $l_{69}, l_{70}$ and 
$l_{71}$ once we insert \eqref{collateral} in \eqref{botsuga4})\footnote{This is not surprising. The flux components ${\bf G}_{{\rm MN}ab}$ are localized fluxes, so appear as field strengths ${\cal F}^{(k)}_{\rm MN}$ on the IIB branes (see \eqref{sessence}). The gauge fields then come from the three-form 
${\bf C}_{{\rm M}ab}$, which would scale as $\left({g_s\over {\rm HH}_o}\right)^{+1}$ from \eqref{collateral}. As such, this automatically switches on ${\bf G}_{{\rm 0M}ab}$ components scaling as 
$\left({g_s\over {\rm HH}_o}\right)^{0}$. The subtlety is that ${\bf G}_4 \ne d{\bf C}_3$ because of the Bianchi identity involving quantum terms, so the above argument is only true if ${\bf C}_3$ has the required dominant scaling, or share the same dominant scaling as the quantum terms. \label{Noha}}. We can then keep the time-independent parts of the other flux components zero, as alluded to earlier. In fact quantum terms can conspire to keep components like ${\bf G}_{\rm 0MNP}$ zero as in \eqref{chupke} (non-zero ${\bf G}_{\rm 0MNP}$ flux components contribute only to order $\left({g_s\over {\rm HH}_o}\right)^{-{8\over 3} + 2\times {7\over 3}} =
\left({g_s\over {\rm HH}_o}\right)^2$ from \eqref{botsuga4}
 to the space-time Einstein's equations when $l_{\rm 0M}^{\rm NP} = 0$. This is a far cry from the lowest order $\left({g_s\over {\rm HH}_o}\right)^{-2}$ contributions necessary to balance the space-time Einstein's equations).

Let us try to quantify some of the above comments. As mentioned many times, ${\bf G}_4$ is not just 
$d{\bf C}_3$ because of the non-trivial Bianchi identity. The Bianchi identity, derived around eq. (4.139) in the first reference of \cite{desitter2}, combines the quantum terms $\hat{\mathbb{Y}}_4$ and $\mathbb{Y}_7$ in a self-consistent way to reproduce flux quantization etc. for the time-dependent background (see details in section \ref{fluxoo} and also in \cite{desitter2}). For the present case, let us define ${\bf G}_4$ using ${\bf C}_3$ and the quantum terms
 (which we shall denote as $\mathbb{Q}_i(\hat{\mathbb{Y}_4}, \mathbb{Y}_7)$) for the flux components 
 ${\bf G}_{\rm 0ABC}$ as ${\bf G}_4 = d{\bf C}_3 + \mathbb{Q}$. In terms of components we can write this as:
 \bg\label{nomads}
 {\bf G}_{\rm 0ABC} &= & {1\over 4!} ~ \partial_{[0} \left[\sum_{k \in {\mathbb{Z} \over 2}} {\bf C}^{(k)}_{{\rm ABC}]}(x, y, w^a) \left({g_s\over {\rm HH}_o}\right)^{l_{\rm AB}^{\rm C} + {2k\over 3}}\right]  \\
 &+ & 
 \sum_{k \in {\mathbb{Z} \over 2}} \left[\mathbb{Q}^{(p, k)}_{\rm 0ABC}(x, y, w^a) 
 \left({g_s\over {\rm HH}_o}\right)^{\tilde{l}_{{\rm 0A}}^{\rm ~BC} + {2k\over 3}} +
 \mathbb{Q}^{(np, k)}_{\rm 0ABC}(x, y, w^a) 
 \left({g_s\over {\rm HH}_o}\right)^{\hat{l}_{{\rm 0A}}^{\rm ~BC} + {2k\over 3}}\right], \nonumber \nd
where $(p, np)$ denote the perturbative (that include \eqref{fahingsha5} and topological) and non-perturbative (including the non-local) quantum terms respectively. The dominant scalings of the three terms on the RHS of \eqref{nomads} are 
$(l_{\rm AB}^{\rm C}, \tilde{l}_{\rm 0A}^{\rm ~BC}, \hat{l}_{\rm 0A}^{\rm ~BC})$ and they have to be related to 
$l_{\rm 0A}^{\rm BC}$, the dominant scaling of ${\bf G}_{\rm 0ABC}$ from \eqref{makibhalu3}. As discussed in the first reference of 
\cite{desitter2}, when the components of the G-flux lie in the eight-dimensional internal space,  the dominant scaling of the G-flux components match exactly with the dominant scalings of the quantum terms. These identifications are {\it essential} to satisfy flux quantizations and Bianchi identities. Here, although there are no flux quantization rules, Bianchi identities, that give rise to \eqref{nomads} in the first place, need to be satisfied. This means at least $l_{\rm 0A}^{\rm BC}$ have to match up with 
$(\tilde{l}_{\rm 0A}^{\rm ~BC}, \hat{l}_{\rm 0A}^{\rm ~BC})$, although the $g_s$ scalings of the three-form fields can match up at higher orders\footnote{With temporal derivatives, the matching will involve 
$l_{\rm AB}^{\rm C} - 1$ as expected \label{fern}.}. In the following let us speculate what happens when the 
dominant scalings of the three-form fields match with the dominant scalings of the corresponding G-flux components (with due consideration to footnote \ref{fern}). For this we have to study the Bianchi identities 
associated with the various G-flux components and their connections to the corresponding three-form 
fields. 
 Looking at {\bf Table \ref{kat1}} we see that
 ${\bf C}_{\rm MNP}$, for example, has multiple components because $({\rm M, N}) \in {\cal M}_4 \times 
 {\cal M}_2$. We can denote them, symbolically, as:
 \bg\label{boxmaric}
 {\bf C}_{\rm MNP} = \left({\rm X}_1^a, {\rm X}_1^b, {\rm X}_1^c, .....\right) \equiv {\rm X}_1(x, y, w^a; g_s), \nd
 such that ${\rm X}_1$ denote a {\it set} of functions with various tensor indices restricted to a give subspace. The subspaces could be ${\cal M}_4 \times {\cal M}_2$, or ${\mathbb{T}^2\over {\cal G}}$, or 
 ${\bf R}^2$, depending on which components we consider from {\bf Table \ref{kat1}}. Let ${\rm X}_i$, with 
 $i = 1,..., 14$ denote the functions appearing in {\bf Table \ref{lokeys}}. The Bianchi identities then will lead to the following set of equations:
 \bg\label{chloefern}
&&\partial_0 {\rm X}_1 + \left[\partial_{\rm P} {\rm X}_2\right] + \mathbb{Q}_1 = {\rm G}_1\nonumber\\
&&\partial_0 {\rm X}_3 + \partial_a {\rm X}_2 + \left[\partial_{\rm N} {\rm X}_4\right] + \mathbb{Q}_2 = {\rm G}_2\nonumber\\
&&\partial_0 {\rm X}_5 + \partial_{\rm N} {\rm X}_6 + \left[\partial_{b} {\rm X}_4\right] + \mathbb{Q}_3 = {\rm G}_3\nonumber\\
 &&\partial_0 {\rm X}_7 + \partial_i {\rm X}_2 + \left[\partial_{\rm M} {\rm X}_8\right] + \mathbb{Q}_4 = {\rm G}_4\nonumber\\
 &&\partial_0 {\rm X}_{14} + \partial_i {\rm X}_6 + \left[\partial_{b} {\rm X}_{10}\right] + \mathbb{Q}_8 = {\rm G}_8\nonumber\\ 
 &&\partial_0 {\rm X}_{11} + \partial_{\rm M} {\rm X}_{12} + \left[\partial_{j} {\rm X}_8\right] + \mathbb{Q}_6 = {\rm G}_6\nonumber\\
&&\partial_0 {\rm X}_{13} + \partial_a {\rm X}_{12} + \left[\partial_{j} {\rm X}_{10}\right] + \mathbb{Q}_7 = {\rm G}_7\nonumber\\
&&\partial_0 {\rm X}_9 + \partial_i {\rm X}_4 + \partial_{\rm N} {\rm X}_{10} + \partial_{a} {\rm X}_8 
 + \mathbb{Q}_5 = {\rm G}_5, \nd
where ${\rm G}_i$ are the set of G-flux components\footnote{The precise forms of ${\rm G}_i$, or the flux components ${\bf G}_{\rm 0ABC}$, will be spelled out in sub-section \ref{fluxoo}. For the time being it will suffice to note that they scale in the same way with respect to ${g_s\over {\rm HH}_o}$.} according to their order of appearance in {\bf Table \ref{kat1}}, $\mathbb{Q}_i$ are the $(p, np)$ quantum terms from \eqref{nomads} (albeit written in terms of {\it dual variables}, {\it i.e.} seven-forms instead of four-forms, as in \cite{desitter2}\footnote{Recall that they scale in the same way as \eqref{fahingsha5}.}), and $[....]$ denote partial anti-symmetrization, for example: 
$\left[\partial_{\rm P} {\rm X}_2\right] \equiv \partial_{[{\rm P}}{\bf C}_{|0|{\rm MN}]}$, with $|q|$ denoting components neutral under anti-symmetrization. We are also not careful about the relative signs, because 
${\rm X}_i$ denote set of functions as in \eqref{boxmaric} so the components can be arranged appropriately
to allow relative plus signs.

Let us start by analyzing the first equation in the 
set of equations in \eqref{chloefern}. We should begin by noting that, since $({\rm X}_1, {\rm X}_2)$ are themselves set of flux components, this is not just a single equation but involves at least 20 different equations. The first term ${\rm X}_1$ scales as $\left({g_s\over {\rm HH}_o}\right)^{+1}$ from 
\eqref{collateral}, therefore according to footnote \ref{fern}, this would scale as 
$\left({g_s\over {\rm HH}_o}\right)^{0}$, which we will henceforth write as $0$. (In fact, to avoid clutter, we will henceforth denote the $g_s$ scaling $\left({g_s\over {\rm HH}_o}\right)^{\pm p}$ by $\pm p$.) However since the EFT (or the EOM) bound from {\bf Table \ref{kat1}} for the flux components ${\rm G}_1$ is $-2$, there is a possibility that 
${\rm X}_2$ scales as $-2$. If this is true, then in the second equation, ${\rm G}_2$  will at least scale as $-2$ contradicting\footnote{There is a subtlety here: in the second equation ${\rm X}_2$ appears with derivatives along the toroidal directions. From our choice of the metric \eqref{makibhalu3} one might think that it should bring down some factors of ${\gamma\over 3}$. This is not quite true because when 
${\rm X}_3 = {\rm X}_4 = \mathbb{Q}_2 = 0$, the second equation can be integrated to give:
\bg\label{lilmadd}
{\bf C}_{\rm 0NP}(x, y, x^{11}; g_s) = {\sqrt{\pi}\over 2}\left({g_s\over {\rm HH}_o}\right)^{l_{\rm 0N}^{{\rm P}a} - {\gamma\over 6}}
\sum_{k = 1}^\infty c_k~{\cal G}^{(k)}_{{\rm 0NP}a}(x, y)~{\bf Erf}\left[\sqrt{d_1}~{\rm M}_p x^{11} 
\left({g_s\over {\rm HH}_o}\right)^{\gamma\over 6}\right], \nonumber \nd
where ${\bf Erf}(a x^{11})$ is the error function, and we have restricted the exponential piece from \eqref{makibhalu3} to only quadratic factors. In the limit $(g_s, {\rm M}_p x^{11}) \to (0, 0)$, the error function can be expanded as a Taylor series, and from there it is easy to infer the scaling of ${\bf C}_{\rm 0NP}$ to be 
$l_{\rm 0N}^{{\rm P}a}$ {\it without} the $\gamma$ factor. This implies that the dominant scaling of ${\rm X}_2$ doesn't change by the derivative action as long as we do not involve temporal derivative. The other limit of ${\rm M}_p x^{11} \to \infty$ is not very useful for us here. \label{harmonimad}}
the EFT (or EOM) bound of $-1$. This means ${\rm X}_2$ can at most scale as $-1$, but nothing smaller implying that ${\rm G}_1$ can at most scale as $-1$. On the other hand, the $-1$ scaling of ${\rm G}_1$, or the flux components ${\bf G}_{\rm 0MNP}$, means that it can only contribute as 
$+{4\over 3}$ to the quantum terms \eqref{fahingsha5} (see coefficients ($l_{87}, l_{88}, l_{89}$) from 
\eqref{botsuga4}). The $+{4\over 3}$ scaling of the G-flux components ${\bf G}_{\rm 0MNP}$ then implies that they {\it cannot} contribute to the lowest order Einstein's equations and therefore {\it cannot} influence the lowest order metric components. This conclusion does not change even if we allow all parts of the first equation in \eqref{chloefern} to scale in the same way, namely as $0$. 

In the second equation of \eqref{chloefern}, which is in fact a collection of at least 30 equations, the first term
$\partial_0 {\rm X}_3$ would scale as zero because of our choice \eqref{collateral}. The second term can scale as $-1$ or $0$ depending on how the first equation scales. The third term can at most scale as $-1$ because of the EFT bound in {\bf Table \ref{kat1}}. However this would create a contradiction because it will make the third equation in \eqref{chloefern} to scale at least as $-1$, thus violating the EFT bound of $0$. To avoid such issues, ${\rm X}_4$ can only scale as $0$ or higher, implying further that 
${\bf G}_{{\rm 0NP}a}$ can scale as ${\rm X}_2$ or $\mathbb{Q}_2$ if their scalings are smaller than $0$, otherwise ${\bf G}_{{\rm 0NP}a}$ will scale as $0$. Since we assumed ${\rm X}_2$ to scale as $0$ from the first equation in \eqref{chloefern}, and G-flux components share the same dominant scalings as the corresponding three-form fields, means ${\bf G}_{{\rm 0NP}a}$ components will also scale as $0$. Looking at the coefficients of $(l_{91}, l_{92}, l_{93})$ from \eqref{botsuga4}, we see that ${\bf G}_{{\rm 0NP}a}$ components contribute as $+{4\over 3}$, and therefore {\it cannot} participate at the lowest order Einstein's equations.

In the third equation of \eqref{chloefern}, which is a collection of 6 equations, both ${\rm X}_5$ and 
${\rm X}_4$ scale as 1 and 0 respectively and therefore ${\rm X}_6$ can only scale as 0 or higher. Since 0 is also the EFT bound, we see that ${\bf G}_{{\rm 0N}ab}$ scaling as $0$, {\it can} contribute to the lowest order Einstein's equations as alluded to earlier (see also footnote \ref{Noha}).

In the fourth equation of \eqref{chloefern}, both ${\rm X}_7$ and ${\rm X}_8$ can scale as $-3$ or higher because of the EFT bound, but then the latter will contradict the EFT bound of $-2$ for ${\bf G}_{{\rm 0N}ia}$ (unless the toroidal derivatives bring down extra powers of $g_s$). Near $(g_s, {\rm M}_p x^{11}) \to (0, 0)$ this is not possible (see footnote \ref{harmonimad}), and assuming the dominant scalings to remain the same across the various three-form fields contributing to 
${\rm G}_4$ in \eqref{chloefern}, the flux components 
${\bf G}_{{\rm 0MN}i}$ would scale as $0$. One may similarly argue the scalings of the other G-flux components, and show that the scalings of $0$ for each of them (except for ${\bf G}_{0ij{\rm M}}$) work under the assumed conditions\footnote{There are still a few subtleties that need elaborations here. {\Su First}, in the sixth equation of \eqref{chloefern}, ${\rm X}_{12}$ would scale as $-4$, but ${\rm X}_{11}$ can scale as 
$-3$ or higher (similarly ${\rm X}_8$ can scale as $-2$ or higher). According to \eqref{collateral}, the dominant scaling of ${\rm G}_6$, {\it i.e.} ${\bf G}_{0ij{\rm M}}$, should be $-4$.  From the exact expression in \eqref{kurisada} we see that this is indeed the case (although the knowledge of the un-warped space-time metric $\widetilde{g}_{\mu\nu}$ is essential). However the expression \eqref{kurisada} also suggests
{\it vanishing} ${\bf C}_{ij{\rm M}}$ and ${\bf C}_{0i{\rm M}}$. This would seem to consistently fit with our procedure of fixing the scalings of the G-flux components from the corresponding three-form fields: since 
both ${\bf C}_{ij{\rm M}}$ and ${\bf C}_{0i{\rm M}}$, {\it i.e.} ${\rm X}_{11}$ and ${\rm X}_8$ respectively, do not share the same dominant scaling of ${\rm X}_{12}$, they should be put to zero.  {\Su Secondly}, the choice of the dominant scaling $-4$ for ${\rm X}_{12}$, {\it i.e.} ${\bf C}_{0ij}$, would seem to contradict the EFT bound of $-3$ for ${\rm G}_7$, {\it i.e.} ${\bf G}_{0ija}$, in the seventh equation of \eqref{chloefern} in light of the footnote \ref{harmonimad}.  This is fortunately saved by the fact that the expression for 
${\bf G}_{0ija}$ is given by \eqref{HiGaToK} and therefore differs from the general expression \eqref{makibhalu3}. Since ${\rm X}_{10}$ can scale as $-1$ (see the equation for ${\rm G}_8$), it doesn't share the dominant scaling of $-2$, so should be put to zero. On the other hand, ${\rm X}_{13}$ could scale as $-1$, but since it does not appear in the exact expression \eqref{HiGaToK}, we should also put this to zero.
After the dust settles, the $-2$ scaling of ${\bf G}_{0ija}$ would prohibit it to contribute to the lowest order Einstein's equations, plus it vanishes when $x^{11} \to 0$. {\Su Finally}, comparing the third and the fifth equation in \eqref{chloefern}, ${\rm X}_6$ scale as $0$ or higher, and since ${\rm X}_{10}$ vanishes, $\partial_0 {\rm X}_{14}$ can in principle scale as $-1$ or higher, or as $0$ or higher if we  want to balance it with the scaling of ${\rm X}_6$.  The choice of $-1$ scaling would mean that ${\rm X}_{14}$ scales as 
${\rm log}\left({g_s\over {\rm HH}_o}\right)$ which blows up in the limit $g_s \to 0$. Thus the latter option is better as it also fits with our general procedure. \label{lisjol}}. Unfortunately this means none of the flux components, except 
${\bf G}_{{\rm 0N}ab}$ and ${\bf G}_{0ij{\rm M}}$, can participate to the lowest order Einstein's equations. 

To summarize, if the dominant scalings of the flux components ${\bf G}_{\rm 0ABC}$ are kept zero, then most of the flux components (except 
${\bf G}_{{\rm 0N}ab}$ and ${\bf G}_{0ij{\rm M}}$) do not contribute to the lowest order Einstein's equations for the space-time or the internal space. In that case one can keep the time-independent parts of the flux components zero without violating any of the flux EOMs. If the dominant scalings  are given by the EFT bounds ({\it i.e.} the third column in {\bf Table \ref{kat1}}) then almost all flux components (except again
${\bf G}_{{\rm 0N}ab}$) are time-dependent and contribute equally as 
$\left({g_s\over {\rm HH}_o}\right)^{1\over 3}$ to \eqref{fahingsha5}. If the dominant scalings $l_{\rm 0A}^{\rm BC} \ge {1\over 3}$, then {\it none} of the flux components contribute to the Einstein's equations at the lowest orders.
Away from these limits, the ${\bf G}_{\rm 0ABC}$ flux components  are generically time-dependent, and can be allowed as a part of the Glauber-Sudarshan state at {\it higher order} in $g_s$ without violating the de Sitter isometries (for a generic embedding of de Sitter space in the dual IIB side).

\subsection{External fluxes and additional consistency conditions \label{mrain}}

The lesson that we learnt from the previous section is that most of the flux components of the type 
${\bf G}_{\rm 0ABC}$ generically do not contribute at the lowest orders to the Einstein's equations with the exception of ${\bf G}_{0ij{\rm M}}$, as long as we are away from the EFT bounds given in {\bf Table \ref{kat1}}, or when we allow $l_{\rm 0A}^{\rm BC} = 0$. The latter condition does allow 
the flux components ${\bf G}_{{\rm 0N}ab}$, but we can always use the Bianchi identity to kill off the $g_s$ independent parts of them. This is of course what we would have expected, and it is satisfying to see that the EOMs can be consistently solved for all of these flux components. The story however does not end here as there are additional flux components that have at least one leg along the spatial ${\bf R}^2$ directions. We will call them the {\it external fluxes}, and in the following elaborate their properties carefully. Needless to say, all the external flux components will have their own distinct properties governed by the rank seven and eight tensors from \eqref{mcacrisis}, much like what we saw earlier.

\begin{table}[tb]  
 \begin{center}
\renewcommand{\arraystretch}{1.5}
\begin{tabular}{|c||c||c|}\hline ${\bf G}_{{\rm MN}ij}$ tensors & form & ${g_s\over {\rm HH}_o}$ scaling \\ \hline\hline
$\left(\mathbb{T}_7^{(f)}\right)_{{\rm PQRS}0ab}$ & $\sqrt{-{\bf g}_{11}} {\bf G}^{{\rm MN}ij} {\rm F}_1^{(n_1)}
{\rm F}_2^{(n_2)} \epsilon_{{\rm MN}ij{\rm PQRS}0ab}$ & $l_{\rm MN}^{ij} + 2 + {2\over 3}\left(k_1 + k_{n_1} 
+ k_{n_2}\right)$ \\ \hline 
$\left(\mathbb{T}_7^{(q)}\right)_{{\rm PQRS}0ab}$ & $\sqrt{-{\bf g}_{11}} \mathbb{Y}_4^{{\rm MN}ij} 
\epsilon_{{\rm MN}ij{\rm PQRS}0ab}$ & $\theta_{nl}(k_2) - l_{\rm MN}^{ij} - {14\over 3}  
- {2k_2\over 3}$ \\ \hline 
$\left(\mathbb{T}_8^{(i, j)}\right)_{{\rm PQRS}0abi}$ & ${\bf G}_{[{\rm PQRS}} {\bf G}_{0abi]}$ & ${2\over 3}\left(k_4 + k_5\right) + l_{[{\rm PQ}}^{[{\rm RS}} \oplus l_{{0i}]}^{ab]}$ 
\\ \hline 
$\left(\mathbb{T}_8^{(m, \alpha)}\right)_{{\rm MPQRS}0ab}$ & ${\bf G}_{[{\rm MPQR}} {\bf G}_{{\rm S}0ab]}$ & ${2\over 3}\left(k_4 + k_5\right) + l_{[{\rm MP}}^{[{{\rm QR}}} \oplus l_{{\rm S}0]}^{ab]}$ 
\\ \hline 
$\left(\mathbb{X}_8^{(i, j)}\right)_{{\rm PQRS}0abi}$ & \eqref{meggy9}, \eqref{meggy10} 
& $l, l - 8$ \\ \hline
$\left(\mathbb{X}_8^{(m, \alpha)}\right)_{{\rm MPQRS}0ab}$ & \eqref{meggy13}, \eqref{meggy14} 
& $l + 1, l - 7$ \\ \hline
\end{tabular}
\renewcommand{\arraystretch}{1}
\end{center}
 \caption[]{Comparing the $g_s$ scalings of the various tensors that contribute to the EOM for the G-flux components ${\bf G}_{{\rm MN}ij}$ in \eqref{mcacrisis}. Again, comparing it to \eqref{sessence44} we can view this as localized fluxes with a possibility of delocalized parts. In either case, the structure of the EOMs stem 
 from the three form fields ${\bf C}_{{\rm MN}i}$ and ${\bf C}_{{\rm M}ij}$ with derivatives along the spatial and the internal six directions respectively.} 
  \label{jhinuk16}
 \end{table}

\vskip.2in

\noindent {\it Case 1: ${\bf G}_{{\rm MN}ij}, {\bf G}_{{\rm M}aij}$ and ${\bf G}_{abij}$ components}

\vskip.2in

\noindent Our story starts with the flux components of the form ${\bf G}_{ij\mathbb{CD}}$, where 
$\left(\mathbb{C}, \mathbb{D}\right) \in {\cal M}_4 \times {\cal M}_2 \times {\mathbb{T}^2\over {\cal G}}$. From their indices one can see that these flux components go hand-in-hand with the ones 
studied in case 4 earlier, namely the ${\bf G}_{0i\mathbb{CD}}$ components. In fact we can even conjecture their structure to take similar {\it localized} form as in \eqref{sessence3}, namely: 

{\footnotesize
\bg\label{sessence44}
{\bf G}_{ij\mathbb{CD}}(x, y, w^a; g_s) = \sum_{k\ge 0} \left({\cal G}^{(k)}_{ij\mathbb{CD}}(x, y, w^a) + 
{\cal F}^{(k)}_{ij}(x) \Omega^{(k)}_{\mathbb{CD}}(y, w^a)\right) 
\left({g_s\over {\rm H}{\rm H}_o}\right)^{l_{\mathbb{CD}}^{0i} + 2k/3}, \nd}
where $\Omega_{\mathbb{CD}}$ is the same localized two-form in the internal eight-manifold that we encountered in \eqref{sessence3} (see also \eqref{sessence}). The above form \eqref{sessence44} provides the remaining gauge fluxes ${\cal F}^{(k)}_{ij}$ on the IIB branes, thus satisfactorily completing the $U(1)$ 
gauge structure on the dual IIB branes once we incorporate ${\cal F}_{0i}^{(k)}$ components from 
\eqref{sessence3}. The non-abelian enhancement of the gauge theory can come from wrapped M2-branes on the vanishing two-cycles in the internal eight-manifold. There is an interesting story there but 
any elaborations will take us too far away from the present topic, so we shall leave it for future 
works\footnote{It suffices to add here that any non-abelian gauge dynamics on the dual side {\it may not} contribute at the lowest orders, and would therefore not influence the metric configuration coming from the lowest order Schwinger-Dyson's equations. This will be elaborated later.}. At the abelian level, these flux components contribute as:
\bg\label{bicLLL}
\theta^{(1)}_{nl} = 2\left(l_{\rm MN}^{ij} + {10\over 3}\right), ~~~ 
\theta^{(2)}_{nl} = 2\left(l_{{\rm M}a}^{ij} + {7 \over 3}\right), ~~~ 
\theta^{(3)}_{nl} = 2\left(l_{ab}^{ij} + {4 \over 3}\right), \nd
to the kinetic terms as may be seen by equations rows 1 and 2 in {\bf Tables \ref{jhinuk16}},
{\bf \ref{jhinuk13}} and {\bf \ref{jhinuk15}} respectively. This is also consistent with \eqref{botsuga4} as is evident by looking at the coefficients of $(l_{80}, l_{81}, l_{82}), (l_{83}, l_{84})$ and $l_{85}$, corresponding to the three cases respectively. 

Let us then start with the flux components ${\bf G}_{{\rm MN}ij}$ whose dynamics are controlled by the three-form fields ${\bf C}_{{\rm MN}i}$, with spatial derivatives, and ${\bf C}_{{\rm N}ij}$, with derivatives along ${\cal M}_4 \times {\cal M}_2$ directions.  The ${\bf X}_8$ polynomials associated with the three forms are \eqref{meggy9} and \eqref{meggy13} respectively for the metric \eqref{makibhalu3} with $\gamma = 6$; and \eqref{meggy10} and \eqref{meggy14} respectively for the metric with generic dependence on all the eleven dimensions. The various rank seven and eight tensors contributing to the flux EOMs are summarized in {\bf Table \ref{jhinuk16}}, and the product of fluxes in rows 3 and 4 are bounded from below by:
\bg\label{ktokii}
{\rm log}_{\rm x}\left({\bf G}_{[{\rm PQRS}} {\bf G}_{0abi]}\right) \ge -2, ~~~~~
{\rm log}_{\rm x}\left({\bf G}_{[{\rm MPQR}} {\bf G}_{{\rm S0}ab]}\right) \ge -1, \nd
which may now be compared with the flux derivatives from row 1 and the ${\bf X}_8$ polynomials from 
rows 5 and 6 to determine the possible range of choices for the dominant scaling $l_{\rm MN}^{ij}$. Since the flux components do not involve derivatives along the toroidal directions, the scaling from row 1 is 
always $l_{\rm MN}^{ij} + 2$, giving us the following range of choices:
\bg\label{eveleb}
-4 \le l_{\rm MN}^{ij} \le -2, ~~~~~~ -3 \le l_{\rm MN}^{ij} \le -1. \nd
Unfortunately the EFT constraints from \eqref{perfum2} rules out all the dominant scalings lying within the range $-4 \le l_{\rm MN}^{ij} < -3$, and therefore the allowed range for $l_{\rm MN}^{ij}$ appears to be 
$-3 \le l_{\rm MN}^{ij} \le - 1$ (the upper bound is not strict as we saw earlier).  In this range, and at some higher orders in $\left(k_i, k_{n_i}, l, {g_s\over {\rm HH}_o}\right)$ we expect, at least perturbatively, the following conditions to hold:

{\footnotesize
\bg\label{megmojer106}
l_{\rm MN}^{ij}  +  2 + {\red {\rm X}_{\partial}^r} +  {2\over 3}(k_1 + k_{n_1} + k_{n_2}) = \theta_{nl} 
- l_{\rm MN}^{ij} - {14 \over 3} -{2k_2\over 3}  + {\red {\rm X}_{\partial}^r} = {2\over 3}(k_4 + k_5) + l_{[{\rm PQ}}^{[{\rm RS}} \oplus 
l_{0a]}^{{b{\red r}}]} 
= \Big\{\begin{matrix}~l + {\red \alpha_r} \\ ~l + {\red \beta_r} \end{matrix}~, \nonumber\\ \nd} 
where ${\red r} = (i, {\rm M})$ signifying how the derivatives along the spatial and the internal six directions act on the rank seven tensors in {\bf Table \ref{jhinuk16}}. The other quantities appearing in 
\eqref{megmojer106} are defined as follows:
\bg\label{tagmontses}
\left({\red {\rm X}_{\partial}^i}, {\red {\rm X}_{\partial}^{\rm M}}\right) = (0, 0), ~~~~~ 
\left({\red \alpha_i, \beta_i}\right) = (0, -8), ~~~~~ \left({\red \alpha_{\rm M}, \beta_{\rm M}}\right) = (1, -7), \nd
with the latter parts coming from the values of the ${\bf X}_8$ polynomials given in \eqref{meggy9}, 
\eqref{meggy10}, as well as from \eqref{meggy13} and \eqref{meggy14} in the way described earlier. At lowest orders, the balance is more subtle than what we have in \eqref{megmojer106}. For example, when
$l_{\rm MN}^{ij} = -3$, the schematic diagram resembles \eqref{lilcadu95} with the same values for 
$(\theta_1, \theta_{i, 2}, \theta_{i, 3})$ when the derivatives act along the spatial ${\bf R}^2$ directions on the rank seven tensors in {\bf Table \ref{jhinuk16}}. On the other hand, when $l_{\rm MN}^{ij} = +1$, the schematic diagram will again be of the form \eqref{lilcadu95}, but now 
$(\theta_1, \theta_{i, 2}, \theta_{i, 3}) = \left({26\over 3}, {17\over 3}, {11\over 3} + 2\right)$, where the BBS instanton contributions fron \eqref{beverwickmey} is shown by the addition of $+2$ for $\theta_{i, 3}$. The perturbative corrections can come from terms like 
${\bf G}_{{\rm MN}ij}\left({\bf R}^{{\rm MN CD}}{\bf R}^{ij}_{~~{\rm CD}} + ...\right)$ that scale as $+{17\over 3}$, as long as we restrict the curvature terms to the set of the first 27 terms in \eqref{fahingsha5} (and take $\gamma = 6$ in \eqref{makibhalu3}). When the derivatives act along the internal six directions, {\it i.e.} along 
${\cal M}_4 \times {\cal M}_2$, the schematic diagram for $l_{\rm MN}^{ij} = -3$ resembles the one from 
\eqref{lilcadu90} with the same values for $(\theta_1, \theta_2)$ therein, with sub-leading instanton contributions.  With $l_{\rm MN}^{ij} = +1$, the diagram resembles \eqref{lilcadu91} but with 
$(\theta_1, \theta_{{\rm M}, 2}, \theta_{{\rm M}, 3}) = \left({26\over 3}, {20\over 3}, {14\over 3}\right)$, implying the necessity of adding BBS instantons from \eqref{beverwickmey} to enhance 
$\theta_{{\rm M}, 3}$ from $+{14\over 3}$ to $+{20\over 3}$. Finally, with $l_{\rm MN}^{ij} = -1$, the schematic diagram resembles \eqref{lilcadu78} with the same values for $(\theta_1, \theta_2)$ therein.

 \begin{table}[tb]  
 \begin{center}
\renewcommand{\arraystretch}{1.5}
\begin{tabular}{|c||c||c|}\hline ${\bf G}_{{\rm M}aij}$ tensors & form & ${g_s\over {\rm HH}_o}$ scaling \\ \hline\hline
$\left(\mathbb{T}_7^{(f)}\right)_{{\rm NPQRS}0b}$ & $\sqrt{-{\bf g}_{11}} {\bf G}^{{\rm M}ija} {\rm F}_1^{(n_1)}
{\rm F}_2^{(n_2)} \epsilon_{{\rm M}ija{\rm NPQRS}0b}$ & $l_{{\rm M}i}^{ja} + {2\over 3}\left(k_1 + k_{n_1} 
+ k_{n_2}\right)$ \\ \hline 
$\left(\mathbb{T}_7^{(q)}\right)_{{\rm NPQRS}0b}$ & $\sqrt{-{\bf g}_{11}} \mathbb{Y}_4^{{\rm M}ija} 
\epsilon_{{\rm M}ija{\rm NPQRS}0b}$ & $\theta_{nl}(k_2) - l_{{\rm M}i}^{ja} - {14\over 3}  
- {2k_2\over 3}$ \\ \hline 
$\left(\mathbb{T}_8^{(i, j)}\right)_{{\rm NPQRS}0bi}$ & ${\bf G}_{[{\rm NPQR}} {\bf G}_{{\rm S}0bi]}$ & ${2\over 3}\left(k_4 + k_5\right) + l_{[{\rm NP}}^{[{\rm QR}} \oplus l_{{\rm S}0]}^{bi]}$ 
\\ \hline 
$\left(\mathbb{T}_8^{(a, b)}\right)_{{\rm NPQRS}0ab}$ & ${\bf G}_{[{\rm NPQR}} {\bf G}_{{\rm S}0ab]}$ & ${2\over 3}\left(k_4 + k_5\right) + l_{[{\rm NP}}^{[{{\rm QR}}} \oplus l_{{\rm S}0]}^{ab]}$ 
\\ \hline 
$\left(\mathbb{T}_8^{(m, \alpha)}\right)_{{\rm MNPQRS}0b}$ & ${\bf G}_{[{\rm MNPQ}} {\bf G}_{{\rm RS}0b]}$ & ${2\over 3}\left(k_4 + k_5\right) + l_{[{\rm MN}}^{[{{\rm PQ}}} \oplus l_{{\rm RS}]}^{0b]}$ 
\\ \hline 
$\left(\mathbb{X}_8^{(i, j)}\right)_{{\rm NPQRS}0bi}$ & \eqref{meggy3}, \eqref{meggy4} 
& $l - 1, l - 9$ \\ \hline
$\left(\mathbb{X}_8^{(a, b)}\right)_{{\rm NPQRS}0ab}$ & \eqref{meggy13}, \eqref{meggy14}
& $l + 1, l - 7$ \\ \hline
$\left(\mathbb{X}_8^{(m, \alpha)}\right)_{{\rm MNPQRS}0b}$ & \eqref{meggy15}, \eqref{meggy16}
& $l, l - 8$ \\ \hline
\end{tabular}
\renewcommand{\arraystretch}{1}
\end{center}
 \caption[]{Comparing the $g_s$ scalings of the various tensors that contribute to the EOM for the G-flux components ${\bf G}_{{\rm M}ija}$ in \eqref{mcacrisis}. As expected there are no temporal derivatives now, and the three three-form contributing to the EOMs are ${\bf C}_{{\rm M}ja}$ with derivatives along the spatial directions, ${\bf C}_{{\rm M}ij}$ with derivatives acting along the toroidal directions, and 
 ${\bf C}_{ija}$ with derivatives acting along the internal six directions. We are again taking $\gamma = 6$
 in \eqref{makibhalu3}, so there is no third column for the ${\bf X}_8$ polynomial({\it i.e.} it coincides with the result from the metric choice \eqref{evader}.} 
  \label{jhinuk13}
 \end{table}

The story for the flux components ${\bf G}_{{\rm M}aij}$ is little more involved than the previous one because there are now three possible derivative actions on the three-forms, namely, spatial derivatives on 
${\bf C}_{{\rm M}ja}$, toroidal derivatives on ${\bf C}_{{\rm M}ij}$ and internal six-dimensional derivatives on
${\bf C}_{ija}$, as evident from {\bf Table \ref{jhinuk13}}. The product of the fluxes, appearing in rows 3 to 5 are bounded from below by:

{\footnotesize
\bg\label{ktokii2}
{\rm log}_{\rm x}\left({\bf G}_{[{\rm NPQR}} {\bf G}_{{\rm S}0bi]}\right) \ge -3, ~~~~
{\rm log}_{\rm x}\left({\bf G}_{[{\rm NPQR}} {\bf G}_{{\rm S0}ab]}\right) \ge -1, ~~~~
{\rm log}_{\rm x}\left({\bf G}_{[{\rm MNPQ}} {\bf G}_{{\rm RS0}b]}\right) \ge -1, \nd}
and the ${\bf X}_8$ polynomials associated with the three derivatives' actions are given by 
\eqref{meggy3}, \eqref{meggy13} and \eqref{meggy15} respectively for the metric \eqref{makibhalu3} with 
$\gamma = 6$ (or the metric \eqref{evader} in the simpler case); and by \eqref{meggy4}, \eqref{meggy13} and \eqref{meggy15} for the metric with generic dependence on all the eleven dimensional coordinates. Combining these together we see the following matching conditions appear for the various rank seven and
eight tensors:

{\footnotesize
\bg\label{megmojer107}
l_{{\rm M}i}^{ja}  + {\red {\rm X}_{\partial}^r} +  {2\over 3}(k_1 + k_{n_1} + k_{n_2}) = \theta_{nl} 
- l_{{\rm M}i}^{ja} - {14 \over 3} -{2k_2\over 3}  + {\red {\rm X}_{\partial}^r} = {2\over 3}(k_4 + k_5) + l_{[{\rm NP}}^{[{\rm QR}} \oplus 
l_{{\rm S}0]}^{{b{\red r}}]} 
= \Big\{\begin{matrix}~l + {\red \alpha_r} \\ ~l + {\red \beta_r} \end{matrix}~, \nonumber\\ \nd} 
 at the perturbative level and at some higher orders in $\left(k_i, k_{n_i}, l, {g_s\over {\rm HH}_o}\right)$. The non-perturbative effects are typically sub-leading at this order. We have also defined ${\red r} = (i, a, {\rm M})$ in \eqref{megmojer107}, and the other quantities appearing therein are defined as follows:

{\footnotesize
\bg\label{tagmontses2}
\left({\red {\rm X}_{\partial}^i}, {\red {\rm X}_{\partial}^{a}}, {\red {\rm X}_{\partial}^{\rm M}}\right) = (0, 2, 0), ~~~
\left({\red \alpha_i, \beta_i}\right) = (-1, -9), ~~~\left({\red \alpha_a, \beta_a}\right) = (1, -7), ~~~ \left({\red \alpha_{\rm M}, \beta_{\rm M}}\right) = (0, -8), \nd}
the latter being associated with the various ${\bf X}_8$ polynomials discussed above and in 
{\bf Table \ref{jhinuk13}}. These ${\bf X}_8$ polynomials and the lower bounds on the flux products from
\eqref{ktokii2}, when inserted in \eqref{megmojer107} reproduce the following range of values for 
$l_{{\rm M}i}^{ja}$:
\bg\label{cjulesc}
-3 \le l_{{\rm M}i}^{ja} \le -1, ~~~~~~ -1 \le l_{{\rm M}i}^{ja} \le 0, \nd
where the former coming from the two cases where the spatial and the toroidal derivatives act on the 
rank seven tensors in {\bf Table \ref{jhinuk13}}, and the latter coming from the case where the six-dimensional derivatives act on the rank seven tensors. Unfortunately the lower bound in the  former case is inconsistent with the EFT bound from \eqref{perfum2}, and therefore the range 
$-3 \le l_{{\rm M}i}^{ja} < -2$ should be eliminated. Thus the allowed range appears to be 
$-2 \le l_{{\rm M}i}^{ja} \le 0$ although the upper bound is not strict.  A higher value for the upper bound implies, from \eqref{bicLLL}, that the flux components ${\bf G}_{{\rm M}aij}$ can only contribute at higher orders in $\left(k_i, k_{n_i}, {g_s\over {\rm HH}_o}\right)$ to the Schwinger-Dyson's equations for the metric components.  The lower bound of 
$-2$, on the other hand, implies that ${\bf G}_{{\rm M}aij}$ would contribute as $+{1\over 3}$ to the quantum series \eqref{fahingsha5}, and therefore would appear in the lowest order Schwinger-Dyson's equations. 

Let us see how the various quantum terms scale in the EOMs. When the derivatives along the spatial directions act on the rank seven tensors, the choice of $l_{{\rm M}i}^{ja} = -2$, would lead to a schematic diagram of the form \eqref{lilcadu95} with the same values for $(\theta_1, \theta_{i, 2}, \theta_{i, 3})$. For 
$l_{{\rm M}i}^{ja} = 0$, the schematic diagram \eqref{lilcadu} appears with 
$(\theta_1, \theta_{i, 2}, \theta_{i, 3}) = \left({14\over 3}, {11\over 3}, {5\over 3} + 2\right)$, where the addition of $+2$ implies BBS instanton contributions from \eqref{beverwickmey}. When we take the derivatives to be along the toroidal directions, the dominant scaling has an additional shift by $+2$  as shown in 
\eqref{tagmontses2}, which means, for $l_{{\rm M}i}^{ja} = -2$, the schematic diagram resembles 
\eqref{lilcadu944} with the same values for $(\theta_1, \theta_{a, 2}, \theta_{a, 3})$ therein. For 
$l_{{\rm M}i}^{ja} = 0$, the schematic diagram \eqref{lilcadu944} appears with 
$(\theta_1, \theta_{a, 2}, \theta_{a, 3}) = \left({14\over 3}, {11\over 3}, {5\over 3} + 2\right)$, where $+2$ again signals the presence of BBS instantons. Finally, when the derivatives act along the internal six directions, the choice of $l_{{\rm M}i}^{ja} = -2$ resembles the schematic diagram \eqref{lilcadu94}, but with 
$(\theta_1, \theta_{{\rm M}, 2}, \theta_{{\rm M}, 3}) = \left({2\over 3}, {8\over 3}, {5\over 3}\right)$, implying that now the BBS instantons can only have sub-leading effects. For $l_{{\rm M}i}^{ja} = 0$, the schematic diagram resembles \eqref{lilcadu78} but with $(\theta_1, \theta_2) = \left({14\over 3}, {11\over 3}\right)$, again with sub-leading instanton effects. 

 \begin{table}[tb]  
 \begin{center}
\renewcommand{\arraystretch}{1.5}
\begin{tabular}{|c||c||c|}\hline ${\bf G}_{abij}$ tensors & form & ${g_s\over {\rm HH}_o}$ scaling \\ \hline\hline
$\left(\mathbb{T}_7^{(f)}\right)_{{\rm MNPQRS}0}$ & $\sqrt{-{\bf g}_{11}} {\bf G}^{abij} {\rm F}_1^{(n_1)}
{\rm F}_2^{(n_2)} \epsilon_{abij{\rm MNPQRS}0}$ & $l_{ab}^{ij} - 2 + {2\over 3}\left(k_1 + k_{n_1} 
+ k_{n_2}\right)$ \\ \hline 
$\left(\mathbb{T}_7^{(q)}\right)_{{\rm MNPQRS}0}$ & $\sqrt{-{\bf g}_{11}} \mathbb{Y}_4^{abij} 
\epsilon_{abij{\rm MNPQRS}0}$ & $\theta_{nl}(k_2) - l_{ab}^{ij} - {14\over 3}  
- {2k_2\over 3}$ \\ \hline 
$\left(\mathbb{T}_8^{(i, j)}\right)_{{\rm MNPQRS}0j}$ & ${\bf G}_{[{\rm MNPQ}} {\bf G}_{{\rm RS}0j]}$ & ${2\over 3}\left(k_4 + k_5\right) + l_{[{\rm MN}}^{[{\rm PQ}} \oplus l_{{\rm RS}]}^{0j]}$  
\\ \hline 
$\left(\mathbb{T}_8^{(a, b)}\right)_{{\rm MNPQRS}0a}$ & ${\bf G}_{[{\rm MNPQ}} {\bf G}_{{\rm RS}0a]}$ & ${2\over 3}\left(k_4 + k_5\right) + l_{[{\rm MN}}^{[{{\rm PQ}}} \oplus l_{{\rm RS}]}^{0a]}$ 
\\ \hline 
$\left(\mathbb{X}_8^{(i, j)}\right)_{{\rm MNPQRS}0j}$ & \eqref{meggy11}, \eqref{meggy12}, 
& $l - 2, l - 10$ \\ \hline
$\left(\mathbb{X}_8^{(a, b)}\right)_{{\rm MNPQRS}0a}$ & \eqref{meggy15}, \eqref{meggy16} 
& $l, l - 8$ \\ \hline 
\end{tabular}
\renewcommand{\arraystretch}{1}
\end{center}
 \caption[]{Comparing the $g_s$ scalings of the various tensors that contribute to the EOM for the G-flux components ${\bf G}_{abij}$ in \eqref{mcacrisis}. Note that the form of the G-flux components would imply a localized structure of the type \eqref{sessence}. Generically, we can view the flux components as coming from the three-forms ${\bf C}_{abi}$ with derivatives along the spatial directions, and ${\bf C}_{ija}$ with derivatives along the toroidal directions.  In terms of localized forms, the former could be thought of giving rise to a gauge field ${\bf A}_i$ and the latter giving rise to a gauge field ${\bf A}_a$.}
  \label{jhinuk15}
 \end{table} 
 
The last set of flux components are of the form ${\bf G}_{abij}$ whose dynamics are captured by rank seven and eight tensors listed in {\bf Table \ref{jhinuk15}}. There are two set of three-forms contributing here: 
${\bf C}_{abi}$, with derivatives along the spatial direction, and ${\bf C}_{ija}$, with derivatives along the toroidal directions. The corresponding ${\bf X}_8$ polynomials are \eqref{meggy11} and \eqref{meggy15} for the metric \eqref{makibhalu3} with $\gamma = 6$ and \eqref{meggy12} and \eqref{meggy16} with metric that generically depends on the eleven dimensions, respectively. The flux products from rows 3 and 4 in 
{\bf Table \ref{jhinuk15}} are bounded from below by:
\bg\label{ktokii3}
{\rm log}_{\rm x}\left({\bf G}_{[{\rm MNPQ}} {\bf G}_{{\rm RS}0j]}\right) \ge -4, ~~~~
{\rm log}_{\rm x}\left({\bf G}_{[{\rm MNPQ}} {\bf G}_{{\rm RS}0a]}\right) \ge -1, \nd
where the dominant scaling of $-4$ appears for example from the combination of flux components
${\bf G}_{\rm 0MNP} {\bf G}_{{\rm RSQ}j}$ that are bounded from below by $-2$ (see 
\eqref{perfum} and \eqref{perfum2}). With these, once we compare all the rows of {\bf Table \ref{jhinuk15}}, we find the following balancing conditions:

{\footnotesize
\bg\label{megmojer108}
l_{ab}^{ij} -2 + {\red {\rm X}_{\partial}^r} +  {2\over 3}(k_1 + k_{n_1} + k_{n_2}) = \theta_{nl} 
- l_{ab}^{ij} - {14 \over 3} -{2k_2\over 3}  + {\red {\rm X}_{\partial}^r} = {2\over 3}(k_4 + k_5) + l_{[{\rm MN}}^{[{\rm PQ}} \oplus 
l_{{\rm RS}]}^{{0{\red r}}]} 
= \Big\{\begin{matrix}~l + {\red \alpha_r} \\ ~l + {\red \beta_r} \end{matrix}~, \nonumber\\ \nd} 
which would appear at higher orders in $\left(k_i, k_{n_i}, l, {g_s\over {\rm HH}_o}\right)$ as we had for the other cases before. The parameter ${\red r}$ now scans ${\red r} = (i, a)$, which are related to the two derivatives' actions, and the other quantities appearing in \eqref{megmojer108} are defined in the following way:
\bg\label{tagmontses3}
\left({\red {\rm X}_{\partial}^i}, {\red {\rm X}_{\partial}^{a}}\right) = (0, 2), ~~~
\left({\red \alpha_i, \beta_i}\right) = (-2, -10), ~~~\left({\red \alpha_a, \beta_a}\right) = (0, -8), \nd
with the latter values coming from the ${\bf X}_8$ polynomials as can be inferred from 
{\bf Table \ref{jhinuk15}}. With these at hand, we now have all the data to predict the behavior of the EOMs for the flux components ${\bf G}_{abij}$. First, let us check the allowed range of choices for $l_{ab}^{ij}$. 
Plugging \eqref{ktokii3} in \eqref{megmojer108}, we find that:
\bg\label{BBmaa}
-2 \le l_{ab}^{ij} \le 0, ~~~~~~ -1 \le l_{ab}^{ij} \le 0, \nd
for the two possible derivative actions respectively. Unfortunately EFT constraints from \eqref{perfum2} rules out $l_{ab}^{ij}$ lying in the range $-2 \le l_{ab}^{ij} < -1$, and therefore the allowed range seems to be 
$-1 \le l_{ab}^{ij} \le 0$, at least at lowest orders. Once we go to higher orders in $\left(k_i, k_{n_i}, l, {g_s\over {\rm HH}_o}\right)$, the upper bound can be easily extended to higher values, although the lower bound is fixed from the EFT constraints.
For $l_{ab}^{ij} = -1$, with the derivatives along the spatial directions acting on the rank seven tensors, the schematic diagram becomes:
\bg\label{lilcadu9o2}
d_{(i)}\ast\overbracket[1pt][7pt]{{\bf G}_4 + b_1 d_{(i)}\ast\Big(\mathbb{Y}_4}
\overbracket[1pt][7pt]{(\theta_1) + \mathbb{Y}_4\underbracket[1pt][7pt]{(\theta_2) + \mathbb{Y}_4(\theta_3)  +  ..\Big) - b_3\Big(\mathbb{X}_8}_{\theta_2 \ge {5 \over 3}}{}^{(i, j)}\Big) 
- b_2 \Big(\mathbb{T}_8}^{\theta_1 \ge {2 \over 3}}{}^{(i, j)}\Big) = 0, \nonumber\\ \nd  
with sub-leading non-local and non-perturbative contributions. For $l_{ab}^{ij} = +1$,  the schematic diagram resembles \eqref{lilcadu95} but with 
$(\theta_1, \theta_{i, 2}, \theta_{i.3}) = \left({14\over 3}, {11\over 3}, {5\over 3} + 2\right)$. The factor of $+2$ comes from the BBS instantons. When the derivatives act along the toroidal directions due considerations have to be taken when scalings are concerned because of the ${\red {\rm X}^a_\partial}$ factor in 
\eqref{megmojer108}. For $l_{ab}^{ij} = -1$, the schematic diagram resembles \eqref{lilcadu82} with the same values for $(\theta_1, \theta_2)$. For $l_{ab}^{ij} = +1$, the schematic diagram resembles
\eqref{lilcadu944} but with $(\theta_1, \theta_{a, 2}, \theta_{a, 3}) = \left({14\over 3}, {11\over 3}, {8\over 3} 
+ 2\right)$, with $+2$ coming from the BBS instantons. Combining everything together, we see that consistent dynamics for the three set of flux components, namely ${\bf G}_{{\rm MN}ij}, {\bf G}_{{\rm M}aij}$ and ${\bf G}_{abij}$, may be worked out in the presence of perturbative and non-perturbative quantum corrections.

 \begin{table}[tb]  
 \begin{center}
\renewcommand{\arraystretch}{1.5}
\begin{tabular}{|c||c||c|}\hline ${\bf G}_{{\rm MNP}i}$ tensors & form & ${g_s\over {\rm HH}_o}$ scaling \\ \hline\hline
$\left(\mathbb{T}_7^{(f)}\right)_{{\rm QRS}0jab}$ & $\sqrt{-{\bf g}_{11}} {\bf G}^{{\rm MNP}i} {\rm F}_1^{(n_1)}
{\rm F}_2^{(n_2)} \epsilon_{{\rm MNP}a{\rm QRS}0jab}$ & $l_{\rm MN}^{{\rm P}i} + {2\over 3}\left(k_1 + k_{n_1} 
+ k_{n_2}\right)$ \\ \hline 
$\left(\mathbb{T}_7^{(q)}\right)_{{\rm QRS}0jab}$ & $\sqrt{-{\bf g}_{11}} \mathbb{Y}_4^{{\rm MNP}i} 
\epsilon_{{\rm MNP}i{\rm QRS}0jab}$ & $\theta_{nl}(k_2) - l_{\rm MN}^{{\rm P}i} - {14\over 3}  
- {2k_2\over 3}$ \\ \hline 
$\left(\mathbb{T}_8^{(i, j)}\right)_{{\rm QRS}0ijab}$ & ${\bf G}_{[{\rm QRS}0} {\bf G}_{abij]}$ & ${2\over 3}\left(k_4 + k_5\right) + l_{[{\rm QR}}^{[{\rm S}0} \oplus l_{{ab}]}^{ij]}$ 
\\ \hline 
$\left(\mathbb{T}_8^{(m, \alpha)}\right)_{{\rm MQRS}0jab}$ & ${\bf G}_{[{\rm MQRS}} {\bf G}_{0jab]}$ & ${2\over 3}\left(k_4 + k_5\right) + l_{[{\rm MQ}}^{[{{\rm RS}}} \oplus l_{0j]}^{ab]}$ 
\\ \hline 
$\left(\mathbb{X}_8^{(i, j)}\right)_{{\rm QRS}0ijab}$ & \eqref{thetrial4}, \eqref{thetrial5} 
& $l - 1, l - 9$ \\ \hline
$\left(\mathbb{X}_8^{(m, \alpha)}\right)_{{\rm MQRS}0jab}$ & \eqref{meggy9}, \eqref{meggy10} 
& $l, l - 8$ \\ \hline
\end{tabular}
\renewcommand{\arraystretch}{1}
\end{center}
 \caption[]{Comparing the $g_s$ scalings of the various tensors that contribute to the EOM for the G-flux components ${\bf G}_{{\rm MNP}i}$ in \eqref{mcacrisis}. The contributing  three form fields are now
 ${\bf C}_{{\rm MNP}}$ and  ${\bf C}_{{\rm NP}i}$ with derivatives along spatial and the internal six directions 
 respectively. One should compare the data here with the ones from {\bf Table \ref{jhinuk5}}.} 
  \label{jhinuk17}
 \end{table}

\vskip.2in

\noindent {\it Case 2: ${\bf G}_{{\rm MNP}i}, {\bf G}_{{\rm MN}ai}$ and ${\bf G}_{{\rm M}abi}$ components}

\vskip.2in

\noindent Our last three set of flux components, all have one leg along the spatial ${\bf R}^2$ directions. 
They should be compared to the three flux components, namely 
${\bf G}_{0{\rm MNP}}, {\bf G}_{{\rm 0NP}a}$ and ${\bf G}_{{\rm 0N}ab}$, 
studied earlier where all have one leg along the temporal direction (we will call them the temporal counterparts). Comparing the data in 
{\bf Tables \ref{jhinuk17}, \ref{jhinuk18}} and {\bf \ref{jhinuk14}} with the data in {\bf Tables \ref{jhinuk5}, 
\ref{jhinuk6}} and {\bf \ref{jhinuk7}} respectively, we see interesting similarities despite the fact that the tensorial components do not necessarily match. For example the $g_s$ scalings of the ${\bf X}_8$ polynomials in 
{\bf Tables \ref{jhinuk17}, \ref{jhinuk18}} and {\bf \ref{jhinuk14}} match with their temporal counterparts exactly. Similarly the $g_s$ scalings of the rank seven tensors  also have interesting similarities, although in all these cases we expect the functional forms (as well as the tensors indices) to differ in appropriate ways. The similarities between the $g_s$ scalings between these two cases are not very surprising because of the exchange of spatial and temporal coordinates. The fact that the $g_s$ scalings of ${\bf g}_{00}$ and 
${\bf g}_{ij}$ are exactly $\left({g_s\over {\rm HH}_o}\right)^{-{8\over 3}}$ but their functional forms differ (for arbitrary slicing of de Sitter space), is responsible for creating the necessary similarities (and the corresponding differences).

The {\it differences} are important. One set of difference comes from the functional forms as discussed above, but another crucial set of difference comes from the fact that the temporal derivatives on the rank seven tensors {\it change} the corresponding $g_s$ scalings by factors of $-1$. However replacing the temporal derivatives with the spatial ones keep the $g_s$ scalings of the rank seven tensors unchanged. This means, in the schematic diagrams, the quantum terms would appear {\it differently}. To see this more concretely, let us consider the flux components ${\bf G}_{{\rm MNP}i}$ from {\bf Table \ref{jhinuk17}}. One may easily check that the flux products are bounded from below by $-3$ and $-2$ from rows 3 and 4 respectively in {\bf Table \ref{jhinuk17}}. This is similar to what we had in {\bf Table \ref{jhinuk5}} earlier. However the balancing the $g_s$ scaling of the various tensors now involve:

{\footnotesize
\bg\label{megmojer109}
l_{\rm MN}^{{\rm P}i} + {\red {\rm X}_{\partial}^r} +  {2\over 3}(k_1 + k_{n_1} + k_{n_2}) = \theta_{nl} 
- l_{\rm MN}^{{\rm P}i} - {14 \over 3} -{2k_2\over 3}  + {\red {\rm X}_{\partial}^r} = {2\over 3}(k_4 + k_5) 
+ l_{[{\rm QR}}^{[{\rm S0}} \oplus 
l_{{ab}]}^{{j{\red r}}]}  
= \Big\{\begin{matrix}~l + {\red \alpha_r} \\ ~l + {\red \beta_r} \end{matrix}~, \nonumber\\ \nd} 
which should be compared to \eqref{cora20} and \eqref{cora21}. For the present case, there is no extra factor of $-1$ as may be seen from the following definition of the parameters appearing in \eqref{megmojer109}:
\bg\label{tagmontses4}
\left({\red {\rm X}_{\partial}^i}, {\red {\rm X}_{\partial}^{\rm M}}\right) = (0, 0), ~~~
\left({\red \alpha_i, \beta_i}\right) = (-1, -9), ~~~\left({\red \alpha_{\rm M}, \beta_{\rm M}}\right) = (0, -8), \nd
where the $(0, 0)$ factor signifies the key difference. The other parameters appearing above are expectedly similar to what we had earlier, and now comparing the flux derivative (from row 1 of 
{\bf Table \ref{jhinuk17}}) with the flux products and the ${\bf X}_8$ polynomials we find that:
\bg\label{evnot}
-3 \le l_{\rm MN}^{{\rm P}i} \le -1, ~~~~~~ -2 \le l_{\rm MN}^{{\rm P}i} \le 0. \nd
The lower bound of $-3$ violates the EFT bound from \eqref{perfum2}, so it seems the allowed range is 
$-2 \le l_{\rm MN}^{{\rm P}i} \le 0$. This is much like what we had earlier, but the choice therein did not involve an excursion to the dis-allowed regions. Nevertheless, the regime of interest matches with what we had earlier although it does allow us to 
go beyond $0$, as the upper bound on $l_{\rm MN}^{{\rm P}i}$ is not strict. Note that earlier we argued, from 
say {\bf Table \ref{kat1}}, EFT considerations restrict $l_{\rm 0M}^{\rm NP}$ to $l_{\rm 0M}^{\rm NP} \ge -1$ thus dis-allowing it to contribute to the lowest order Einstein's equations. Does this happen here too? To see this we will have to study the other flux components. We will come back to this later.

Due to the small difference in \eqref{megmojer109}, the quantum terms would scale differently from what we had for the flux components ${\bf G}_{\rm 0MNP}$. As an example, let us consider $l_{\rm MN}^{{\rm P}i} = -2$ with derivatives along the spatial directions acting on the rank seven tensors. One may easily see that the schematic diagram {\it differs} from what we had in \eqref{lilcadu86} even if we change the subscript $(0)$ to $(i)$. The resemblance here is more to \eqref{lilcadu95}, and now the contributions of the non-perturbative BBS instantons become crucial to make sense of the EOMs. When $l_{\rm MN}^{{\rm P}i} = +1$, 
the schematic diagram again resembles \eqref{lilcadu95} but with 
$(\theta_1, \theta_{i, 2}, \theta_{i, 3}) = \left({20\over 3}, {14\over 3}, {8\over 3} + 2\right)$, where $+2$ signifies the contributions from the BBS instantons. One may check that the scalings of the quantum terms differ from the previous case for $l_{\rm 0M}^{\rm N} = 1$. Interestingly, when the derivatives act along the internal six directions, the schematic diagrams are identical (as an expected consequence of the similarities of the flux products and the ${\bf X}_8$ polynomials).

 \begin{table}[tb]  
 \begin{center}
\renewcommand{\arraystretch}{1.5}
\begin{tabular}{|c||c||c|}\hline ${\bf G}_{{\rm MN}ai}$ tensors & form & ${g_s\over {\rm HH}_o}$ scaling \\ \hline\hline
$\left(\mathbb{T}_7^{(f)}\right)_{{\rm PQRS}b0j}$ & $\sqrt{-{\bf g}_{11}} {\bf G}^{{\rm MN}ia} {\rm F}_1^{(n_1)}
{\rm F}_2^{(n_2)} \epsilon_{{\rm MN}ia{\rm PQRS}b0j}$ & $l_{\rm MN}^{ia} - 2 + {2\over 3}\left(k_1 + k_{n_1} 
+ k_{n_2}\right)$ \\ \hline 
$\left(\mathbb{T}_7^{(q)}\right)_{{\rm PQRS}b0j}$ & $\sqrt{-{\bf g}_{11}} \mathbb{Y}_4^{{\rm MN}ia} 
\epsilon_{{\rm MN}ia{\rm PQRS}b0j}$ & $\theta_{nl}(k_2) - l_{\rm MN}^{ia} - {14\over 3}  
- {2k_2\over 3}$ \\ \hline 
$\left(\mathbb{T}_8^{(i, j)}\right)_{{\rm PQRS}b0ij}$ & ${\bf G}_{[{\rm PQR}b} {\bf G}_{0ij{\rm S}]}$ & ${2\over 3}\left(k_4 + k_5\right) + l_{[{\rm PQ}}^{[{{\rm R}b}} \oplus l_{0i]}^{j{\rm S}]}$ 
\\ \hline 
$\left(\mathbb{T}_8^{(a, b)}\right)_{{\rm PQRS}ab0j}$ & ${\bf G}_{[{\rm PQRS}} {\bf G}_{ab0j]}$ & ${2\over 3}\left(k_4 + k_5\right) + l_{[{\rm PQ}}^{[{\rm RS}} \oplus l_{{ab}]}^{0j]}$ 
\\ \hline 
$\left(\mathbb{T}_8^{(m, \alpha)}\right)_{{\rm MPQRS}b0j}$ & ${\bf G}_{[{\rm MPQR}} {\bf G}_{{\rm S}b0j]}$ & ${2\over 3}\left(k_4 + k_5\right) + l_{[{\rm MP}}^{[{{\rm QR}}} \oplus l_{{\rm S}b]}^{0j]}$ 
\\ \hline 
$\left(\mathbb{X}_8^{(i, j)}\right)_{{\rm PQRS}b0ij}$ & \eqref{thetrial7}, \eqref{thetrial8}
& $l - 2, l - 10$ \\ \hline
$\left(\mathbb{X}_8^{(a, b)}\right)_{{\rm PQRS}ab0j}$ & \eqref{meggy9}, \eqref{meggy10} 
& $l, l - 8$ \\ \hline
$\left(\mathbb{X}_8^{(m, \alpha)}\right)_{{\rm MPQRS}b0j}$ & \eqref{meggy3}, \eqref{meggy4} 
& $l - 1, l - 9$ \\ \hline
\end{tabular}
\renewcommand{\arraystretch}{1}
\end{center}
 \caption[]{Comparing the $g_s$ scalings of the various tensors that contribute to the EOM for the G-flux components ${\bf G}_{{\rm MN}ai}$ in \eqref{mcacrisis}. Note that now, all the eight forms  have {\it three} possible choices stemming from how the derivatives act on the seven-forms in \eqref{mcacrisis} as well as on the behavior of the three form fields ${\bf C}_{{\rm MN}a}, {\bf C}_{{\rm MN}i}$ 
 and ${\bf C}_{{\rm N}ai}$. One should also compare the data here with the ones in {\bf Table \ref{jhinuk6}}.} 
  \label{jhinuk18}
 \end{table}

For the second case with the flux components ${\bf G}_{{\rm MN}ai}$ one should compare 
{\bf Table \ref{jhinuk18}} and {\bf Table \ref{jhinuk6}}, with the latter corresponding to the flux components 
${\bf G}_{{\rm 0NP}a}$. Again there is a set of similarities and differences. The {\it similarities}, ignoring the differences in the tensor indices, appear in the scalings of the rank seven tensors, the scalings of the
${\bf X}_8$ polynomials as well as on the lower bounds of the flux products appearing in rows 3 to 5 in 
{\bf Table \ref{jhinuk18}}. The {\it differences} appear at two places: {\Su one}, when we balance the various 
$g_s$ scalings of the rank seven and eight tensors, we find that:
\bg\label{tagmontses5}
\left({\red {\rm X}_{\partial}^0}, {\red {\rm X}_{\partial}^{a}}, {\red {\rm X}_{\partial}^{\rm M}}\right) = (-1, 2, 0) 
~~ \to ~~ 
\left({\red {\rm X}_{\partial}^i}, {\red {\rm X}_{\partial}^{a}}, {\red {\rm X}_{\partial}^{\rm M}}\right) = (0, 2, 0), \nd
where the former is for the flux components ${\bf G}_{{\rm 0NP}a}$, while the latter is for the flux components
${\bf G}_{{\rm MN}ai}$. {\Su Two}, this change of $-1$ to $0$, changes the behavior of the quantum terms in the 
EOMs corresponding to the two flux components associated with the temporal and spatial derivatives respectively. {\Su Three}, comparing the flux derivatives in row 1 of {\bf Table \ref{jhinuk18}} with the flux products and the ${\bf X}_8$ polynomials, produces the following range of values for $l_{\rm MN}^{ia}$:
\bg\label{notieve}
-2 \le l_{\rm MN}^{ia} \le 0, ~~~~~~ -1 \le l_{\rm MN}^{ia} \le 1, \nd
with the former coming from the spatial and the toroidal derivatives and the latter coming from the derivatives along the internal six directions. Clearly the range $-2 \le l_{\rm MN}^{ia} < -1$ is eliminated from the EFT considerations (see \eqref{perfum2}), and the allowed range appears to be 
$-1 \le l_{\rm MN}^{ia} \le 1$, much like what we had for the temporal counterpart. The upper bound is again un-restricted and we will discuss later whether such flux components can participate at lowest orders. 

The difference in \eqref{tagmontses5} can lead to a different choice of the quantum terms. To see this, let us consider $l_{\rm MN}^{ai} = 0$ with derivatives along the spatial directions. The schematic diagram expectedly differs from \eqref{lilcadu87} for $l_{\rm 0N}^{{\rm P}a}$ and resembles 
\eqref{lilcadu95}. The quantum contributions are also different: instead of 
$(\theta_1, \theta_{0, 2}, \theta_{0, 3}) = \left({8\over 3}, {11\over 3}, {5\over 3} + 2\right)$, we have
$(\theta_1, \theta_{i, 2}, \theta_{i, 3}) = \left({8\over 3}, {8\over 3}, {2 \over 3} + 2\right)$, with the addition of $+2$ for both cases signify the contributions from the BBS instantons. For $l_{\rm MN}^{ai} = 1$, the diagram resembles \eqref{lilcadu95} with 
$(\theta_1, \theta_{i, 2}, \theta_{i, 3}) = \left({14 \over 3}, {11\over 3}, {5 \over 3} + 2\right)$, with $+2$ again signifying the contributions from the BBS instantons. One may easily check that 
$(\theta_1, \theta_{0, 2}, \theta_{0, 3}) = \left({14 \over 3}, {14 \over 3}, {8 \over 3} + 2\right)$
for $l_{\rm 0N}^{{\rm P}a} = +1$, so the quantum terms contribute differently for the two cases. When the derivatives act along the toroidal and the internal six directions, the story is similar to what we had for the temporal counterparts.

 \begin{table}[tb]  
 \begin{center}
\renewcommand{\arraystretch}{1.5}
\begin{tabular}{|c||c||c|}\hline ${\bf G}_{{\rm M}abi}$ tensors & form & ${g_s\over {\rm HH}_o}$ scaling \\ \hline\hline
$\left(\mathbb{T}_7^{(f)}\right)_{{\rm NPQRS}0j}$ & $\sqrt{-{\bf g}_{11}} {\bf G}^{{\rm M}abi} {\rm F}_1^{(n_1)}
{\rm F}_2^{(n_2)} \epsilon_{{\rm M}abi{\rm NPQRS}0j}$ & $l_{{\rm M}i}^{ab} - 4 + {2\over 3}\left(k_1 + k_{n_1} 
+ k_{n_2}\right)$ \\ \hline 
$\left(\mathbb{T}_7^{(q)}\right)_{{\rm NPQRS}0j}$ & $\sqrt{-{\bf g}_{11}} \mathbb{Y}_4^{{\rm M}abi} 
\epsilon_{{\rm M}abi{\rm NPQRS}0j}$ & $\theta_{nl}(k_2) - l_{{\rm M}i}^{ab} - {14\over 3}  
- {2k_2\over 3}$ \\ \hline 
$\left(\mathbb{T}_8^{(i, j)}\right)_{{\rm NPQRS}0ij}$ & ${\bf G}_{[{\rm NPQR}} {\bf G}_{{\rm S}0ij]}$ & ${2\over 3}\left(k_4 + k_5\right) + l_{[{\rm NP}}^{[{\rm QR}} \oplus l_{{\rm S} 0]}^{ij]}$ 
\\ \hline 
$\left(\mathbb{T}_8^{(m, \alpha)}\right)_{{\rm MNPQRS}0j}$ & ${\bf G}_{[{\rm MNPQ}} {\bf G}_{{\rm RS}0j]}$ & ${2\over 3}\left(k_4 + k_5\right) + l_{[{\rm MN}}^{[{{\rm PQ}}} \oplus l_{{\rm RS}]}^{0j]}$ 
\\ \hline 
$\left(\mathbb{T}_8^{(a, b)}\right)_{{\rm NPQRS}0ja}$ & ${\bf G}_{[{\rm NPQR}} {\bf G}_{{\rm S}0ja]}$ & ${2\over 3}\left(k_4 + k_5\right) + l_{[{\rm NP}}^{[{{\rm QR}}} \oplus l_{{\rm S}0]}^{ja]}$ 
\\ \hline 
$\left(\mathbb{X}_8^{(i, j)}\right)_{{\rm NPQRS}0ij}$ & \eqref{thetrial10}, \eqref{thetrial11} 
& $l - 3, l - 11$ \\ \hline
$\left(\mathbb{X}_8^{(m, \alpha)}\right)_{{\rm MNPQRS}0j}$ & \eqref{meggy11}, \eqref{meggy12} 
& $l - 2, l - 10$ \\ \hline
$\left(\mathbb{X}_8^{(a, b)}\right)_{{\rm NPQRS}0ja}$ & \eqref{meggy3}, \eqref{meggy4} 
& $l - 1, l - 9$ \\ \hline
\end{tabular}
\renewcommand{\arraystretch}{1}
\end{center}
 \caption[]{Comparing the $g_s$ scalings of the various tensors that contribute to the EOM for the G-flux components ${\bf G}_{{\rm M}abi}$ in \eqref{mcacrisis}. The contributing three-forms are now 
 ${\bf C}_{{\rm M}ab}$ with derivatives along the spatial directions, ${\bf C}_{abi}$ with derivatives along the internal six directions, and ${\bf C}_{{\rm S} ai}$ with derivatives along the toroidal directions. As before, one may compare this with {\bf Table \ref{jhinuk7}}.}
  \label{jhinuk14}
 \end{table}

Our final set of flux components ${\bf G}_{{\rm M}abi}$, whose details appear in {\bf Table \ref{jhinuk14}}
should be compared to the ones in {\bf Table \ref{jhinuk7}} for flux components ${\bf G}_{{\rm 0N}ab}$. 
The similarities between these two set of flux components are evident from their respective tables: flux products and the ${\bf X}_8$ polynomials come with the same $g_s$ scalings despite slight differences in their tensorial structure. The main {\it differences} stem from what we had in \eqref{tagmontses5} leading to different choices of the quantum terms etc. leading to the elimination of all dominant scalings lying in the range $-1 \le l^{ab}_{i{\rm M}} < 0$ and to allow $0 \le l^{ab}_{i{\rm M}} < 2$. When derivatives act along the spatial directions, one may easily work out the differences in the choices of the quantum terms for the various choices of $l^{ab}_{i{\rm M}}$ and $l_{\rm 0N}^{ab}$ respectively. After the dust settles, it appears that consistent dynamics can be elucidated with appropriate choices of the quantum terms for 
$l_{i{\rm M}}^{ab}$ lying within the allowed range.

\vskip.2in

\noindent {\it Case summary: External flux equations and their behaviors}

\vskip.2in

\noindent There are two universal behaviors that arise from the study of the properties of the flux components from {\bf Tables \ref{jhinuk1}} to {\bf \ref{jhinuk14}}. {\Su One}, by equating rows 1 and 2 in each of the 18 tables, we can reproduce the contributions of the G-flux components to the quantum series \eqref{fahingsha5} in \eqref{botsuga4}. {\Su Two}, the $g_s$ scalings of the ${\bf X}_8$ polynomials for the two metric configurations, the special one \eqref{makibhalu3} with $\gamma = 6$ (or \eqref{evader}) and the general one with arbitrary dependence on all eleven dimensions, always differ by $+8$, {\it i.e.}:
\bg\label{curvkat}
{\rm log}_{\rm x}\left({{\bf X}_8^{({\rm spe})}\over {\bf X}_8^{({\rm gen})}}\right) = + 8, \nd
where the super-scripts denote special (spe) and general (gen) associated with the two metric choices respectively, and ${\rm x} = {g_s\over {\rm HH}_o}$. In writing \eqref{curvkat}, we have ignored the functional forms of the ${\bf X}_8$ polynomials. They should of course be put in, but the $g_s$ scalings would still show the universal behavior of \eqref{curvkat}.  

The results of our analysis of all the flux components have been collected in {\bf Table \ref{kat2}}, which should now be compared with {\bf Table \ref{kat1}}. For the flux components of the form 
${\bf G}_{\rm 0ABC}$, by analyzing the flux EOMs as well as the Bianchi identities \eqref{chloefern}, it was found that generically they tend to scale as $\left({g_s\over {\rm HH}_o}\right)^0$ except for the flux components ${\bf G}_{0ij{\rm M}}$ and ${\bf G}_{0ija}$ (the latter typically vanishing when $x^{11} \to 0$). This means {\it none} of the flux components, except ${\bf G}_{0ij{\rm M}}$ and ${\bf G}_{{\rm 0N}ab}$, contribute to the lowest order Schwinger-Dyson's equations for the metric components. What happens 
here? To see this we have to write the Bianchi identities associated to the spatial flux components. The three-form fields are necessary, and so are the quantum terms,  and since the former have already been defined in say \eqref{boxmaric} as collections of various set of functions (with appropriate tensorial structure), we can use them here too. The Bianchi identities, looking at {\bf Table \ref{kat2}}, take the following form (see {\bf Table \ref{lokeys}} for the definitions of ${\rm G}_i$ and ${\rm X}_i$):
\bg\label{20lis18} 
 &&\left[\partial_{\rm Q} {\rm X}_1\right] + \mathbb{Q}_9 = {\rm G}_9\nonumber\\
&&\partial_i {\rm X}_1 + \left[\partial_{\rm M} {\rm X}_7\right] + \mathbb{Q}_{12} = {\rm G}_{12}\nonumber\\
&&\partial_a {\rm X}_1 + \left[\partial_{\rm M} {\rm X}_3\right]  + \mathbb{Q}_{10} = {\rm G}_{10}\nonumber\\
&&\left[\partial_b {\rm X}_3\right] + \left[\partial_{\rm M} {\rm X}_5\right] + \mathbb{Q}_{11} = 
{\rm G}_{11}\nonumber\\
&&\left[\partial_j {\rm X}_{7}\right] + \left[\partial_{\rm M} {\rm X}_{11}\right] + \mathbb{Q}_{15} = {\rm G}_{15}\nonumber\\
&&\left[\partial_j {\rm X}_{14}\right] + \left[\partial_b {\rm X}_{13}\right] + \mathbb{Q}_{17} = {\rm G}_{17}\nonumber\\
&&\partial_i {\rm X}_{5} + \partial_{\rm M} {\rm X}_{14} + \partial_{b} {\rm X}_9 + \mathbb{Q}_{14} = {\rm G}_{14}\nonumber\\
  &&\partial_i {\rm X}_{3} + \partial_a {\rm X}_7 + \left[\partial_{\rm M} {\rm X}_{9}\right] + \mathbb{Q}_{13} = {\rm G}_{13}\nonumber\\ 
 &&\left[\partial_i {\rm X}_9\right] + \partial_a {\rm X}_{11} + \partial_{\rm M} {\rm X}_{13} 
 + \mathbb{Q}_{16} = {\rm G}_{16}, \nd
where ${\rm X}_i$ denote a set of flux components as defined in \eqref{boxmaric}; $\mathbb{Q}_9$ to 
$\mathbb{Q}_{17}$ denote quantum terms (both perturbative and non-perturbative ones) but expressed using dual flux variables; and ${\rm G}_i$ are the flux terms that are related to ${\bf G}_{\rm ABCD}$ tensors from {\bf Table \ref{kat2}}. As before the precise connection between ${\rm G}_i$ and the four-form flux components will be specified in the next sub-section, but they scale in the same way. 

 \begin{table}[tb]  
 \begin{center}
\renewcommand{\arraystretch}{1.5}
\begin{tabular}{|c||c||c||c||c||c||c||c|}\hline ${\Su {\rm G}_i}$ & ${\Su{\bf G}_{\rm ABCD}}$ & ${\Su{\rm G}_i}$& ${\Su{\bf G}_{\rm ABCD}}$ & ${\Su{\rm X}_i}$ & ${\Su{\bf C}_{\rm ABC}}$ & ${\Su{\rm X}_i}$ & ${\Su{\bf C}_{\rm ABC}}$ \\ \hline\hline
${\rm G}_1$ & ${\bf G}_{\rm 0MNP}$ & ${\rm G}_{10}$ & ${\bf G}_{{\rm MNP}a}$ & ${\rm X}_1$ & 
${\bf C}_{\rm MNP}$ & ${\rm X}_{10}$ & ${\bf C}_{0ia}$ \\ \hline
${\rm G}_2$ & ${\bf G}_{{\rm 0NP}a}$ &${\rm G}_{11}$ & ${\bf G}_{{\rm MN}ab}$ 
& ${\rm X}_2$ & ${\bf C}_{\rm 0MN}$ & ${\rm X}_{11}$ & ${\bf C}_{ij{\rm M}}$ \\ \hline
${\rm G}_3$ & ${\bf G}_{{\rm 0N}ab}$ & ${\rm G}_{12}$ & ${\bf G}_{{\rm MNP}i}$ & ${\rm X}_3$ & ${\bf C}_{{\rm NP}a}$ & ${\rm X}_{12}$ & ${\bf C}_{0ij}$ \\ \hline
${\rm G}_4$ & ${\bf G}_{{\rm 0MN}i}$ & ${\rm G}_{13}$ & ${\bf G}_{{\rm MN}ai}$ & ${\rm X}_4$ & ${\bf C}_{{\rm 0P}a}$ & ${\rm X}_{13}$ & ${\bf C}_{ija}$ \\ \hline
${\rm G}_5$ & ${\bf G}_{{\rm 0N}ia}$ & ${\rm G}_{14}$ & ${\bf G}_{{\rm M}abi}$ & ${\rm X}_5$ & ${\bf C}_{{\rm N}ab}$ & ${\rm X}_{14}$ & ${\bf C}_{abi}$ \\ \hline
${\rm G}_6$ & ${\bf G}_{0ij{\rm M}}$ & ${\rm G}_{15}$ & ${\bf G}_{{\rm MN}ij}$ & ${\rm X}_6$ & ${\bf C}_{0ab}$ & $....$ & $.....$ \\ \hline
${\rm G}_7$ & ${\bf G}_{0ija}$ & ${\rm G}_{16}$ & ${\bf G}_{{\rm M}aij}$ & ${\rm X}_7$ & ${\bf C}_{{\rm MN}i}$ & $....$ & $......$ \\ \hline 
${\rm G}_8$ & ${\bf G}_{0abi}$ & ${\rm G}_{17}$ & ${\bf G}_{abij}$  & ${\rm X}_8$ & ${\bf C}_{{\rm 0N}i}$ 
& $....$ & $.....$ \\ \hline
${\rm G}_9$ & ${\bf G}_{\rm MNPQ}$ & $....$ & $.....$ & ${\rm X}_9$ & ${\bf C}_{{\rm N}ia}$ & $....$ & $.....$\\ \hline 
\end{tabular}
\renewcommand{\arraystretch}{1}
\end{center}
 \caption[]{Identifications of the ${\rm G}_i$ and the ${\rm X}_i$ tensors to the four and three-form flux components  ${\bf G}_{\rm ABCD}$ and ${\bf C}_{\rm ABC}$ respectively. These flux components appear in the Bianchi identities 
\eqref{chloefern} and \eqref{20lis18}. Note that $({\rm A, B}) \in {\bf R}^{2, 1} \times {\cal M}_4 \times {\cal M}_2 \times {\mathbb{T}^2\over {\cal G}}$.}
  \label{lokeys}
 \end{table}

Let us start by looking at the equation for ${\rm G}_{12}$.  Since ${\rm X}_1$ scales as 
$\left({g_s\over {\rm HH}_o}\right)^{+1}$, which we will write as $+1$, ${\rm X}_7$ can scale at least as 
$-2$ without violating the EFT bound from {\bf Table \ref{kat2}}. This will however violate the EFT bound for 
${\rm G}_{13}$, so ${\rm X}_7$ can scale as $-1$ or higher. Since the $g_s$ scalings determine whether flux components can contribute to the Schwinger-Dyson's equations for the metric components, the $-1$ scaling of ${\rm X}_7$ will prohibit ${\rm G}_{12}$ to participate in the lowest order equations. Similarly, looking at the equation for ${\rm G}_{13}$, we see that ${\rm X}_9$ can scale at least as $-1$, since ${\rm X}_3$ scales as $+1$. This will however clash with the EFT bound for ${\rm G}_{14}$, unless ${\rm X}_9$ scales at least as $0$. This means ${\rm G}_{13}$ can participate at the lowest orders as long as ${\rm X}_7$ scales as $-1$. Unfortunately, from our analysis of {\bf Table \ref{kat1}}, ${\rm X}_7$ can scale as $+1$ or higher, implying that ${\rm G}_{13}$ can scale as $0$ or higher. In either case, the scaling is too high to participate to the lowest order equations. In fact our previous analysis also showed that both ${\rm X}_9$ and ${\rm X}_{14}$ can scale as $+1$ or higher, implying that both ${\rm G}_{13}$ and ${\rm G}_{14}$ would scale as $+1$ and therefore would not contribute at the lowest orders. For the last three cases, {\it i.e.} for ${\rm G}_{15}$ to ${\rm G}_{17}$, since ${\rm X}_{11}$ and ${\rm X}_{14}$ vanish  
(see footnote \ref{lisjol}), ${\rm G}_{15}$ would scale as $+1$ thus prohibiting it to participate at the lowest orders. The EFT constraint also rules out ${\rm G}_{16}$ to participate as it would appear to scale as 
$-1$ or higher. Interestingly, this $-1$ bound would have allowed ${\it G}_{17}$ to participate, but unfortunately from earlier considerations ${\rm X}_{13}$ scales as $+1$, thus ruling out its participation at lowest orders. Therefore to conclude, {\it all} internal ({\it i.e.} ${\rm G}_9$ to ${\rm G}_{11}$) and external 
({\it i.e.} ${\rm G}_{12}$ to ${\rm G}_{17}$) fluxes scale as $+1$, and other than ${\rm G}_{11}$, none of them contribute to the lowest order Einstein's equations.

Let us end this section by answering couple of questions. The {\Su first} one was already asked at the begining, namely, what happens when we take both positive and negative $g_s$ scalings for the G-flux components? To answer this, let us modify the ans\"atze for the G-flux components from \eqref{theritual} to the following:
\bg\label{monrain}
{\bf G}_{\rm ABCD}(x, y, w^a; g_s) = \sum_{k, l} {\cal G}^{(k, l)}_{\rm ABCD}(x, y, w^a) 
\left({g_s\over {\rm HH}_o}\right)^{l_{\rm AB}^{\rm CD} + 2k/3} 
\left({g_s\over {\rm HH}_o}\right)^{-2l/3}, \nd
where $(k, l) \in \left({\mathbb{Z}\over 2}, {\mathbb{Z}\over 2}\right)$ and 
$({\rm A, B}) \in {\bf R}^{2, 1} \times {\cal M}_4 \times {\cal M}_2 \times 
{\mathbb{T}^2\over {\cal G}}$. The above ans\"atze collects both positive and negative powers of $g_s$, but unfortunately it is not well defined 
at late time. In fact as $g_s \to 0$, the flux components appear to 
blow-up\footnote{The readers might ask whether this is of any concern since most of the metric components in M-theory also blow-up at late time. The answer lies in the IIB side. The six-dimensional internal metric therein behaves perfectly well at late time despite the fact that the corresponding M-theory metric blows-up at late time (the M-theory Lagrangian however remains finite at late time). The magic lies in the duality itself: dualizing from M-theory to IIB removes all the offending factors from the metric to make the IIB metric well-defined. On the other hand, the story for the fluxes is different. The M-theory fluxes {\it retain} their behavior when they are dualized to IIB (albeit with change of nomenclature). This means, from IIB point of view, we expect the fluxes to be well-behaved at late time so as to consistently fit with the well-behaved metric components. The only differences are the ${\bf G}_{0ij{\rm M}}$ and ${\bf G}_{0ija}$ flux components that dualize to five-form fluxes in IIB and blow-up at late time. This is however perfectly consistent from IIB point of view because the $3+1$ dimensional space-time metric itself blows up at late time (because of the de Sitter nature). \label{casin125}}.  
The only way to make sense of the above ans\"atze is to sum the series inverse in $g_s$ in the following way:
\bg\label{monrain2}
\sum_{l\in {\mathbb{Z}\over 2}} c_l \left({g_s\over {\rm HH}_o}\right)^{-2l/3} 
= \sum_{n,m = 0,1}^\infty b_{nm} 
~{\rm exp}\left[-n\left({g_s\over {\rm HH}_o}\right)^{-m/3}\right], \nd
where both $(c_l, b_{nm})$ are constants. The RHS is now well defined and vanishes when $n \ne 0$ (for $n = 0$, we expect $\sum b_{0m}$ to be a finite integer). For some details on the derivation of \eqref{monrain2}, one may look up 
\cite{coherbeta}. It is now easy to see that at late time, and when 
$l_{\rm AB}^{\rm CD} \ge 0$ (except for the flux components ${\bf G}_{0ij{\rm M}}$ and ${\bf G}_{0ija}$), the series \eqref{monrain2} may be approximated
by the first term $b_{n1}$ (as other terms will go to zero faster). This means,
\eqref{monrain} may be re-written at late time as: 

{\footnotesize
\bg\label{monrain3}
{\bf G}_{\rm ABCD}(x, y, w^a; g_s) = \sum_{k, n} {\cal G}^{(k, n)}_{\rm ABCD}(x, y, w^a) 
\left({g_s\over {\rm HH}_o}\right)^{l_{\rm AB}^{\rm CD} + 2k/3} 
{\rm exp}\left[-n\left({g_s\over {\rm HH}_o}\right)^{-1/3}\right], \nd}
which matches exactly with the flux ans\"atze that we had in \cite{desitter2}, and for $n = 0$, it is clearly \eqref{theritual}. Both are useful way to 
quantify the flux components at late time, but the analysis leading to 
\eqref{monrain3} depends crucially on the dominant scaling being positive or 
zero. A negative dominant scaling means that the moding $l$ in \eqref{monrain}
is replaced by $l + {2\over 3}\vert l_{\rm AB}^{\rm CD}\vert$ for such a case there is no simple way to perform the sum in \eqref{monrain2}. Therefore, although negative dominant scalings lying within the EFT bounds \eqref{perfum} and 
\eqref{perfum2} may contribute finitely to the EOMs, there is no simple way to 
sum the trans-series to get finite answers for the flux components at late time. Therefore keeping $l_{\rm AB}^{\rm CD} \ge 0$ appears to be necessary (except of course for the flux components ${\bf G}_{0ij{\rm M}}$ and 
${\bf G}_{0ija}$). 

The next question is an interesting one and is related to an alternative viewpoint regarding our construction, namely the choice of the metric 
ans\"atze \eqref{makibhalu3} or \eqref{evader}. How is this related to a metric 
ans\"atze that includes some of the {\it compensators} studied in 
\cite{andrew}? The answer is that, the compensators in \cite{andrew} are related to the {\it solitonic} configuration $-$ and not to \eqref{makibhalu3} or \eqref{evader} $-$ and also only when we want to analyze the moduli motion. The solitonic background, {\it i.e.} eq. (2.1) in the second reference of \cite{coherbeta}, however remains as it is and the compensators arise once we study moduli dynamics. This is much like what we usually encounter for the flux-less Calabi-Yau case: the background is a product space and the moduli motions are captured by cross-terms in the internal space that lead to the Lichnerowicz equation whose solutions are the K\"ahler and the complex structure moduli. Once we have a warped solitonic background, 
one way to deal with the moduli is to introduce the compensators.  
For concreteness consider introducing the compensators as 
${\bf g}^{(0)}_{\mu {\rm M}}$ directly in IIB from start, where $\mu \in {\bf R}^{3, 1}$, with ${\rm M} \in {\cal M}_4 \times {\cal M}_2$ and the super-script are for the solitonic background. There could be cross-terms for the fluxes, but we will not worry about them right now. We can also quantify the compensators as ${\bf g}^{(0)}_{im}$ and ${\bf g}^{(0)}_{jn}$ with $m \ne n$. T-dualizing once along $y^m$ and then once more along $y^n$, we can convert the two metric cross-terms into ${\bf B}_{\rm NS}$ fluxes: ${\bf B}^{(0)}_{im}$ and ${\bf B}^{(0)}_{jn}$. We are now in the same situation as before, and so uplifting to M-theory by first T-dualizing along $x^3$ and then lifting along $w^b \equiv x^{11}$ will give us a metric configuration similar to eq. (2.1) in the second reference of 
\cite{coherbeta} with G-flux components ${\bf G}^{(0)}_{mpib}$ and ${\bf G}^{(0)}_{nqjb}$ along-with other flux 
components\footnote{Note that, at the solitonic level, (a) if we had odd number of metric cross-terms, we could still T-dualize even number of times along the local one-cycles in the internal space; (b) if we also had NS flux cross-terms along-with the metric cross-terms, we could first S-dualize to convert the NS fluxes to the RR fluxes and then perform even number of T-dualities; and (c) if we have both NS and RR fluxes, then in the presence of a D-brane, the NS flux can be gauge transformed to world-volume gauge-fields. After which we can follow the aforementioned duality chasing. Therefore to summarize: generically, it appears that introducing the compensators to study the moduli motion at the solitonic level is equivalent to introducing extra flux components. None of these discussions affect the solitonic background, {\it i.e.} eq. (2.1) in the second reference of \cite{coherbeta}, or the construction of Glauber-Sudarshan or the Agarwal-Tara states, discussed in \cite{coherbeta} and in section \ref{GState}.}. We see that the compensators are useful to study the {\it moduli}, but says nothing about the 
original background (for which we still need to solve the supergravity EOMs). This means the procedure of constructing the Glauber-Sudarshan state {\it et cetera} are unaffected by the aforementioned analysis implying the necessity of introducing (a) temporal dependence of the fluxes and (b) the non-perturbative and non-local interactions. Once the moduli are stabilized at the solitonic level, expectation values of the metric and the flux operators in the Glauber-Sudarshan state will give us the {\it dynamical} moduli in the de Sitter background (this is of course a part of the {\it dynamical moduli stabilization} discussed in \cite{desitter2} and \cite{coherbeta}).


\subsection{Flux quantizations, Bianchi identities and five-branes \label{fluxoo}}

There are a few issues that we have kept under the rug so far, and have to do with flux quantizations and 
heterotic anomaly cancellation. The latter might look surprising and out-of-context, but there is a deeper reason to connect the M-theory compactification to heterotic theory. We will come back to this below and also in section \ref{eva}, but it suffices to say at this point that both these issues have some bearings on the Bianchi identities 
\eqref{chloefern} and \eqref{20lis18}. In the process we will also be able to define the flux factors 
${\rm G}_1$ to ${\rm G}_{14}$ appearing in the two set of Bianchi identities
more precisely. 

The Bianchi identities discussed here appear from the EOMs of the dual seven-forms as shown in 
\cite{desitter2}. As such they may be derived from the following M-theory action written in terms of the {\it dual} variables:
\bg\label{sdmer}
\mathbb{S}_{11} = c_1\int {\bf G}_7 \wedge \ast {\bf G}_7 + c_2 \int {\bf G}_7 \wedge \hat{\mathbb{Y}}_4 
+ c_3 \int {\bf G}_7 \wedge \ast \mathbb{Y}_7 + {\rm N}_5 \int {\bf C}_6 \wedge {\bf \Lambda}_5, \nd
where $c_i$ are constants that depend on ${\rm M}_p$ (similar to the $b_i$ constants in \eqref{elenag}), 
${\rm N}_5$ is the number of M5-branes defined using localized form ${\bf \Lambda}_5$ (much like the localized form ${\bf \Lambda}_8$ for the M2-branes in \eqref{elenag});  $\hat{\mathbb{Y}}_4$ is a topological four-form that can be related to $\mathbb{Y}_8$ polynomial; 
and $\mathbb{Y}_7$ enclose the perturbative quantum terms, such that 
${\bf G}_7 \wedge \ast \mathbb{Y}_7$ is defined in the same vein as \eqref{suhaag}. In fact this would exactly be \eqref{fahingsha5} written using the dual seven-forms keeping the curvature terms therein untouched (see discussions in sections 3.2.7 and 4.2 in the first reference of \cite{desitter2}). The six-form EOM would then lead us to the required Bianchi identity:

 \begin{table}[tb]  
 \begin{center}
\renewcommand{\arraystretch}{1.5}
\begin{tabular}{|c||c||c||c|}\hline ${\bf G}_{{\rm ABCD}}$ tensors & contributing 3-forms & EFT bounds & allowed range \\ \hline\hline
${\bf G}_{{\rm MNPQ}}$ & ${\bf C}_{\rm MNP}$ & $-1$ & $+1$ \\ \hline 
${\bf G}_{{\rm MNP}a}$ & ${\bf C}_{{\rm MNP}}, ~{\bf C}_{{\rm NP}a}$ & $~~0$ & $+1$ \\ \hline 
${\bf G}_{{\rm MN}ab}$ & ${\bf C}_{{\rm MN}a}, ~{\bf C}_{{\rm N}{ab}}$ & $+1$ & $+1$ \\ \hline 
${\bf G}_{{\rm MNP}i}$ & ${\bf C}_{\rm MNP}, ~{\bf C}_{{\rm NP}i}$ & $-2$ & $-2 \le l_{\rm MN}^{{\rm P}i} \le 0^+$ \\ \hline 
${\bf G}_{{\rm MN}ai}$ & ${\bf C}_{{\rm MN}a}, ~{\bf C}_{{\rm MN}i}, ~ {\bf C}_{{\rm N}ia}$ & $-1$ & $-1 \le 
l_{\rm MN}^{{a}i} \le 1^+$ \\ \hline 
${\bf G}_{{\rm M}abi}$ & ${\bf C}_{{\rm M}ab}, ~{\bf C}_{abi}, ~ {\bf C}_{{\rm M}ai}$ & $~~0$ & $0 \le 
l_{i{\rm M}}^{ab} \le 2^+$ \\ \hline 
${\bf G}_{{\rm MN}ij}$ & ${\bf C}_{{\rm MN}i}, ~{\bf C}_{{\rm N}ij}$ & $-3$ & $-3 \le l_{\rm MN}^{ij} \le -1^+$ \\ \hline 
${\bf G}_{{\rm M}aij}$ & ${\bf C}_{{\rm M}ja}, ~{\bf C}_{{\rm M}ij}, ~ {\bf C}_{ija}$ & $-2$ & $-2 \le 
l_{{\rm M}a}^{ij} \le 0^+$ \\ \hline 
${\bf G}_{ab{ij}}$ & ${\bf C}_{{abi}}, ~{\bf C}_{{ij}a}$ & $-1$ & $-1 \le l_{ab}^{ij} \le 0^+$ \\ \hline 
\end{tabular}
\renewcommand{\arraystretch}{1}
\end{center}
 \caption[]{Comparing the $g_s$ scalings of the various tensors that contribute to the EOM for the G-flux components ${\bf G}_{{\rm ABCD}}$ in \eqref{mcacrisis} with $({\rm A, B}) \in {\bf R}^2 \times {\cal M}_4 \times {\cal M}_2 \times {\mathbb{T}^2\over {\cal G}}$. As before, we have shown the EFT bounds that appear from \eqref{botsuga4} for the various components, as well as the three-form tensors associated with these components. These three-forms are essential to construct the structure of the associated ${\bf X}_8$ polynomials. The superscript $+$ signs for the upper bounds tell us that they can be {\it higher} than what appears from matching the rank eight tensors with the rank seven quantum terms in {\bf Tables \ref{jhinuk16}} to 
 {\bf \ref{jhinuk14}}. One should also compare the results here with the ones in {\bf Table \ref{kat1}}.}
  \label{kat2}
 \end{table}

\bg\label{vacage}
d{\bf G}_4 = {1\over c_1}\Big(\hat{\rm N}_5  {\bf \Lambda}_5 - c_2~d\hat{\mathbb{Y}}_4 
- c_3 ~d\ast \mathbb{Y}_7\Big), \nd 
where $c_i$ are the same constants that appeared in \eqref{sdmer}, and we have used $\hat{\rm N}_5$ instead of ${\rm N}_5$. This has to do with the presence of {\it dynamical} M5-branes. Dynamical M5 branes can arise in two possible ways. {\Su One}, when M5-branes are physically moving along an internal sub-space of a {\it static} eight-manifold, and {\Su two}, when the M5-branes are stationary but the size of the internal eight-manifold is changing. For the second case we can easily see, relative to some point in the internal space, the M5-branes appear to {\it move}. Of course we could also have a combination of both, much like the dynamical M2-branes studied in section 4.2.4 in the first reference of \cite{desitter2}. Such dynamical motion would appear to make ${\rm N}_5$ time-dependent, {\it i.e.} $g_s$ dependent\footnote{This may be motivated from the familiar example from electrodynamics. Consider ${\rm N}_0$ point charges. They appear in the $0 + 1$ dimensional Lagrangian as ${\rm N}_0 \int \mathbb{A} = {\rm N}_0 \int \left({\rm A}_0 + {\bf A} \cdot {\bf v}\right) dx^0$, where ${\bf v}$ is the velocity vector. In general there is an acceleration and therefore ${\bf v} \equiv {\bf v}(x^0)$. In $3+1$ dimensions, such an action may be rewritten as: 
\bg\label{cheque}
\mathbb{S}_{\rm EM} = {\rm N}_0 \int d^4 x {\rm A}_0\left(1 + {{\bf A}\cdot {\bf v}\over {\rm A}_0}\right) 
\delta^3\left({\bf x} - {\bf w}\right) \equiv \int dx^0~{\rm A}_0 \hat{\rm N}_0, \nonumber \nd
where $\hat{\rm N}_0$ captures the dynamical behavior of the point charges. In a curved space we can make the usual replacements: $d^4 x \to \sqrt{\bf g} ~d^4 x, ~~ \delta^3\left({\bf x} - {\bf w}\right) \to
{\delta^3\left({\bf x} - {\bf w}\right)\over \sqrt{\bf g}}$, but the result remains unchanged. In general 
$\hat{\rm N}_0 = \hat{\rm N}_0({\bf x}, x^0)$, but if $\vert {\bf A} \vert  = {\rm A}_0$, then 
$\hat{\rm N}_0 = \hat{\rm N}_0(x^0)$ since ${\bf v} = {\bf v}(x^0)$. For us this is the dynamical behavior we are aiming for. The case with the M5-branes on an internal space that is time-dependent, similar considerations should lead $\hat{\rm N}_5$ to capture the temporal dependence. \label{22061:37}}. Let us then propose:
\bg\label{benson}
\hat{\rm N}_5 \equiv \sum_{k \in {\mathbb{Z}\over 2}} \hat{\rm N}_5^{(k)} 
\left({g_s\over {\rm HH}_o}\right)^{l + {2k\over 3}}, \nd
where at least $\hat{\rm N}_5^{(0)} \in \mathbb{Z}$, which can be justified from our above considerations; and $l$ is the dominant scaling that depends on how the M5-branes are oriented in the internal eight-manifold. Note that there is a small leeway here: $\hat{\rm N}_5$ doesn't strictly have to be $g_s$ dependent. It can have dependence on the orthogonal directions in the eight-manifold so long as such dependences may be splitted into a product of $g_s$ dependence as in \eqref{benson}, and an internal space 
dependence. We can then absorb the internal space dependence in the localized form ${\bf \Lambda}_5$ (since this is exactly a function of the internal space coordinates {\it orthogonal} to the 
M5-branes). Such splitting has been the main theme of how the metric and flux components appear, so it should be no surprise that it continues to hold for the case with the dynamical M5-branes. We can also make the above discussion a bit more quantitative. Consider the ${\rm N}_5$ term in the action 
\eqref{sdmer}. This may be expressed as:

{\footnotesize
\bg\label{lillmadi}
{\rm N}_5 \int_{{\bf \Sigma}_{11}} {\bf C}_6 \wedge {\bf \Lambda}_5 = \sum_{k_1 \in {\mathbb{Z}\over 2}} {\rm N}_5 
\int_{{\bf \Sigma}_{11}} {\bf C}_{6, ||} f^{(k_1)}(y_\perp, w^a_\perp) 
\left({g_s\over {\rm HH}_o}\right)^{2k_1\over 3}\wedge {\bf \Lambda}_5(y_\perp, w^a_\perp)
= \int_{{\bf \Sigma}_6} \hat{\rm N}_5 ~{\bf C}_{6, ||}, \nonumber\\ \nd}
where ${\bf C}_{6, ||}$ is the six-form parallel to the world-volume of the M5-branes, and 
$f^{(k_1)}(y_\perp, w^a_\perp)$ is the off-shoot of the dynamical behavior of the M5-branes much like what we have in footnote \ref{22061:37} for the charged point particle case. The orthogonal coordinates on the eight-manifold are denoted by $(y_\perp, w^a_\perp)$. Integrating $f^{(k_1)}(y_\perp, w^a_\perp)$ with the localized form ${\bf \Lambda}_5(y_\perp, w^a_\perp)$ over an internal five-cycle can create additional 
${g_s\over {\rm HH}_o}$ factors as $(y_\perp, w^a_\perp)$ dependences are integrated out. Such 
${g_s}$ factors together with the already-existing $g_s$ factors in \eqref{lillmadi} conspire to give the 
$\hat{\rm N}_5$ factor from \eqref{benson} (this would also explain how the dominant scaling of $l$ in 
\eqref{benson} might arise). On the other hand, if all terms on the RHS of \eqref{vacage} are 
globally defined, then integrating $d{\bf G}_4$ over a five-manifold {\it without} boundary will give vanishing 
$\hat{\rm N}_5$. To study flux quantization, we want $\hat{\rm N}_5$ to be non-zero, implying a five-manifold with boundary. Therefore, 
integrating 
\eqref{vacage} over a five-manifold ${\bf \Sigma}_5$ such that it has a four-dimensional boundary
 ${\bf \Sigma}_4$, {\it i.e.} 
$\partial {\bf \Sigma}_5 = {\bf \Sigma}_4$, gives the following flux quantization condition\footnote{Recall that \eqref{vacage} is derived from 
${\delta \mathbb{S}_{11} \over \delta{\bf C}_{6, ||}} = 0$, {\it i.e.} from the EOM for the six-form ${\bf C}_{6, ||}$ parallel to the space-filling M5-branes.}:
\bg\label{beccamey}
c_1\int_{{\bf \Sigma}_4} {\bf G}_4 = \hat{\rm N}_5 - c_2 \int_{{\bf \Sigma}_4} \hat{\mathbb{Y}}_4
- c_3 \int_{{\bf \Sigma}_4} \ast \mathbb{Y}_7, \nd
where $c_i$ are the constants that appear in the action \eqref{sdmer}. The above relation provides a consistent way to study flux quantization, and in the absence of time-dependences and the 
$\ast \mathbb{Y}_7$ piece, \eqref{beccamey} does reproduce Witten's flux quantization condition 
\cite{wittenfluxes}. However once time-dependences and quantum corrections are switched on,
the quantization condition seems to {\it deviate} from \cite{wittenfluxes}. Additionally, non-perturbative and non-local corrections to \eqref{vacage} would make the results deviate further from \cite{wittenfluxes}. This cannot be right, so we must be interpreting things wrongly here. Question is, where are we making an error?
Note that we cannot reinterpret $\ast {\bf G}_7$ as a {\it different} four-form ${\bf G}'_4$, and then combine it with $\ast \mathbb{Y}_7$ to redefine ${\bf G}_4 \equiv {\bf G}'_4 + {c_3\over c_1}\ast\mathbb{Y}_7$ as this would violate the electric-magnetic dualities in M-theory. Further issue arises when ${\mathbb{T}^2\over {\cal G}} = 
{{\bf S}^1\over \mathbb{Z}_2} \times {\bf S}^1$ locally. In this limit the M-theory background dualizes to 
$E_8 \times E_8$ heterotic theory on a circle \cite{horava}, with the four-form fluxes dualizing to the heterotic three-form\footnote{Not all components of the four-forms in M-theory dualize to the heterotic 
three-forms. If we identify ${\cal G}$ as a $\mathbb{Z}_2$ action, then:
\bg\label{leslaine} 
{\mathbb{T}^2\over {\cal G}} ~ \to ~ {{\bf S}^1_b \times {\bf S}^1_a \over \mathbb{Z}_2} ~ \to ~ 
{{\bf S}^1_b \over \mathbb{Z}_2} \times {\bf S}^1_a, \nonumber \nd
when we are away from the fixed points of $\mathbb{Z}_2$ (this is what we meant by {\it local} action earlier), and we take $(a, b) = (3, 11)$. In the IIA language this is an orientifold operation and therefore eliminates all three-form fields that have no legs along ${\bf S}^1_b$ direction \cite{horava}. In other words, the surviving three-form fields are ${\bf C}_{{\rm MN}b}$ and ${\bf C}_{{\rm M}ab}$. In terms of G-flux components, they are ${\bf G}_{{\rm MNP}b}$ and ${\bf G}_{{\rm MN}ab}$ respectively. The former dualizes to the heterotic three-forms ${\cal H}_{\rm MNP}$ and the latter to the $U(1)$ gauge fields. The G-flux components ${\bf G}_{\rm MNPQ}$ are thereby eliminated.  Of course, away from the orientifold point, the M-theory fixed points are blown-up, and all the G-flux components survive. This is where ${\mathbb{T}^2 \over {\cal G}}$ becomes a smooth two-manifold, albeit non-K\"ahler. \label{hetdual}}
${\cal H}$. Heterotic anomaly cancellation constrains ${\cal H}$ to satisfy:
\bg\label{chetlily}
d{\cal H} = {\rm tr}~\mathbb{R} \wedge \mathbb{R} - {1\over 30}~{\rm tr}~\mathbb{F} \wedge \mathbb{F}, \nd
where the integral over the last term related to the $E_8$ gauge field $\mathbb{F}$ counts the number of small instantons, or alternatively the heterotic five-branes\footnote{It's a bit more subtle: curvature corrections on type IIB O7/D7 or on type I O9/D9 tell us that the type I five-branes are also counted by first Pontryagin class of the tangent bundle. These type I five-branes S-dualize to the heterotic five-branes, so the curvature form, including the first Chern class of the vector bundle, also contribute here. For the $E_8$ heterotic case, these two contributions may be accounted a bit differently \cite{horava}, but the conclusion remains the same. \label{mca66k}}. 
These five-branes are of course directly related to the M5-branes in the M-theory side. 
On the other hand, if we take \eqref{vacage}, the duality to heterotic theory will imply corrections to the anomaly cancellation condition \eqref{chetlily}. This definitely cannot be right,  
so the resolution to the conundrum should lie elsewhere.

Looking at Witten's flux quantization condition in \cite{wittenfluxes} gives us a hint. The condition therein is expressed completely in terms of topological forms, namely the Chern classes. On the other hand, both 
$\ast \mathbb{Y}_7$ as well as the non-perturbative (or non-local) corrections are {\it non-topological}. Additionally, because of the explicit appearance of the metric factors in them, they may not always be globally defined over the eight-manifold. The {\it local} nature of $\ast \mathbb{Y}_4$ implies that it can easily be balanced by localized fluxes, whereas the global part of the G-fluxes can be balanced by 
$\hat{\rm N}_5$ and $\hat{\mathbb{Y}}_4$. The division of the flux components into global and localized forms is of course the heart of our construction (see for example \eqref{sessence}, \eqref{tanyarbon}, \eqref{sessence3}, and \eqref{sessence44}), so we can express the relevant flux components in the following suggestive way:

{\footnotesize
\bg\label{sessence70}
{\bf G}_{\mathbb{ABCD}}({\bf x}, y, w^a; g_s) & = & {\bf G}^{\rm global}_{\mathbb{ABCD}}({\bf x}, y, w^a; g_s) + 
{\bf G}^{\rm local}_{\mathbb{ABCD}}({\bf x}, y, w^a; g_s)\nonumber\\
& = & \sum_{k \in {\mathbb{Z}\over 2}} \Big({\cal G}^{(k)}_{\mathbb{ABCD}}({\bf x}, y, w^a) 
+ {\cal F}^{(k)}_{\mathbb{AB}}({\bf x}, y, w^a) \Omega^{(k)}_{\mathbb{CD}}(y, w^a)\Big)
\left({g_s\over {\rm HH}_o}\right)^{l_{\mathbb{AB}}^{\mathbb{CD}} + {2k\over 3}}, \nd}
where $(\mathbb{A, B, C, D}) \in {\cal M}_4 \times {\cal M}_2 \times {\mathbb{T}^2\over {\cal G}}$. We are therefore dealing with the three flux components ${\bf G}_{{\rm MNPQ}}, {\bf G}_{{\rm MN}ab}$ 
and ${\bf G}_{{\rm MNP}a}$ as they are the only ones defined over the compact eight-manifold (other flux components have at least one leg along the spatial ${\bf R}^{2, 1}$ directions and are therefore not required to be quantized). For ${\bf G}_{{\rm MN}ab}$, \eqref{sessence70} is natural, but now we see that the other two set of fluxes ${\bf G}_{\rm MNPQ}$ and ${\bf G}_{{\rm MNP}a}$ also have localized 
pieces.
In retrospect this is not surprising: the existence of localized two-form $\Omega_{\mathbb{CD}}$ which enabled us to have localized fluxes in \eqref{sessence3} and \eqref{sessence44} also necessitates the existence of these localized pieces. 
Our flux quantization scheme can then be expressed by the following schematic diagram:
\bg\label{lilcaduxox}
c_1\int_{{\bf \Sigma}_4} \overbracket[1pt][7pt]{{\bf G}^{\rm global}_4 + c_1\int_{{\bf \Sigma}_4} \underbracket[1pt][7pt]{{\bf G}^{\rm local}_4 = 
-c_3\int_{{\bf \Sigma}_4} \ast\mathbb{Y}}_{\theta_{nl}}{}_7 + \hat{\rm N}}\overbracket[1pt][7pt]{{}_5 - c_2 \int_{{\bf \Sigma}_4} \hat{\mathbb{Y}}}{}_4, \nd
implying that the localized fluxes may be completely balanced by the quantum terms $\ast\mathbb{Y}_7$, whereas the integral of the global fluxes over appropriate four-cycles are properly quantized. Note that it {\it almost} clarifies the conundrum we had earlier regarding the heterotic dual: the global flux components dualize to the heterotic side to the heterotic three-form ${\cal H}$,  the $\hat{\mathbb{Y}}_4$ term dualizes to the curvature polynomial and $\hat{\rm N}_5$ dualizes to the instanton class in \eqref{chetlily}. Combining everything together we have:
\bg\label{kellerson1}
\int_{{\bf \Sigma}_4}{\bf G}^{\rm global}_4 + {c_2\over c_1}\int_{{\bf \Sigma}_4} \hat{\mathbb{Y}}_4 
= {\hat{\rm N}_5\over c_1}, \nd
as our flux quantization rule. This is exactly what we expect from \cite{wittenfluxes} but now there is a difference: both LHS and RHS of \eqref{kellerson1} have $g_s$ dependence. In fact this would imply that the heterotic anomaly cancellation condition should also develop appropriate $g_s$ dependence in the right way\footnote{This doesn't mean that the anomaly cancellation condition \eqref{chetlily} has 
${\cal O}(g_s)$ corrections. Rather it means that, ${\cal H}$ may be expressed in powers of $g_s$ much like how we expressed the G-fluxes. In a similar vein, the curvature polynomial and the instanton term should also be expressed in powers of $g_s$. We will demonstrate this soon for individual cases.}. In the following we will verify whether this $g_s$ dependence actually comes out from our earlier analysis or not. But before going into this, let us write down the constraints on the localized fluxes:
\bg\label{kellerson2}
c_1\int_{{\bf \Sigma}_4}{\bf G}^{\rm local}_4 = 
-{c_3}\int_{{\bf \Sigma}_4} \ast\mathbb{Y}_7 + {\rm nonperturbative~corrections}, \nd
which would make sense if and only if at every order in $g_s$ this equality can be established. Does this happen here? The answer is happily yes as was shown rigorously in the first reference of 
\cite{desitter2} (see section 4.2.1, cases 1 to 7 therein). We can also work out the non-perturbative (as well as non-local) corrections to \eqref{kellerson2} without influencing the flux quantization condition from 
\eqref{kellerson1} or, in the heterotic dual, the anomaly cancellation condition. The latter result is consistent with the Adler-Bell-Jackiw (ABJ) theorem for anomaly cancellation \cite{ABJ}, namely, there are no higher loop corrections to the anomaly cancellation condition. Turning this around, our splitting in 
\eqref{lilcaduxox}, can be viewed as a proof that ABJ theorem should work in a time-dependent background also.
 
There is however one issue that might still be puzzling regarding the splitting \eqref{lilcaduxox} when we compare with the footnotes \ref{honikjul3} and \ref{hemramal}: could there be similar {\it merging} at higher orders in $\left(k_i, k_{n_i}, l, {g_s\over {\rm HH}_o}\right)$ discussed in the aforementioned footnotes? Clearly the merging at higher orders would be unwelcome and problematic due to the various issues discussed above and therefore the splitting in \eqref{lilcaduxox} at lowest orders should continue as we go to higher orders. How do we guarantee this? The answer lies in the similar analysis of \cite{wittenfluxes} when we compute the membrane anomalies from path-integral: as long as \eqref{kellerson1} is valid, the membrane anomalies cancel even for time-dependent backgrounds. In fact such an analysis would also tell us that the relative sign of the $c_2$ term in \eqref{kellerson1} would cease to matter and both 
$c_1 {\bf G}^{\rm global}_4 + c_2 \hat{\mathbb{Y}}_4$ and 
$c_1 {\bf G}^{\rm global}_4 - c_2 \hat{\mathbb{Y}}_4$ will be in integral cohomology classes. 
This way the splitting \eqref{kellerson1} and \eqref{kellerson2} continues to prevail even if we go to higher orders in 
$\left(k_i, k_{n_i}, l, {g_s\over {\rm HH}_o}\right)$. 
 
\vskip.2in

\noindent {\it Case 1: ${\bf G}_{\rm MNPQ}$ flux components}

\vskip.2in 

\noindent The ${\bf G}_{\rm MNPQ}$ flux components may be split into global and local components following \eqref{sessence70}. For the global piece, the flux quantization scheme from \eqref{kellerson1} will give us:
\bg\label{kellerson3}
\int_{{\bf \Sigma}_4} dy^{[{\rm MNPQ}]}~{\bf G}^{\rm global}_{\rm MNPQ} - {c_2\over c_1}\int_{{\bf \Sigma}_4} dy^{[{\rm MNPQ}]}~{\rm tr}
\Big(\mathbb{R}_{\rm tot} \wedge \mathbb{R}_{\rm tot}\Big)_{\rm MNPQ} = 
{\hat{\rm N}_5\over c_1}, \nd 
where $dy^{[{\rm MNPQ}]} \equiv dy^{\rm M} \wedge .... \wedge dy^{\rm Q}$ and
we identify the middle term with $\hat{\mathbb{Y}}_4$, such that the sign of this will be determined by the signs of $c_2$ and the Pontryagin terms. The quantization condition however {\it does not} depend on the relative sign, since $\hat{\mathbb{Y}}_4$ is integral \cite{wittenfluxes}: one may as well make the replacement $-c_2 \to \pm c_2$ in \eqref{kellerson3}.  The curvature form $\mathbb{R}_{\rm tot}$ is defined from 
\eqref{candace} and \eqref{elfadogg22}, and we can combine everything together to express the $g_s$ scaling of \eqref{kellerson3} in the following way:

{\footnotesize
\bg\label{kellerson4}
&& \sum_{k_1}\int_{{\bf \Sigma}_4} dy^{[{\rm MNPQ}]}~{\cal G}^{(k_1)}_{\rm MNPQ} \left({g_s\over {\rm HH}_o}\right)^{l_{\rm MN}^{\rm PQ} + {2k_1\over 3}}\\
&=& 
{c_2\over c_1} \sum_{k_2, k_3}  \int_{{\bf \Sigma}_4}dy^{[{\rm MNPQ}]}~{\rm tr}\left({\bf R}^{(k_2)}_{[{\rm MN}]}{\bf R}^{(k_3)}_{[{\rm PQ}]}\right) 
 \left({g_s\over {\rm HH}_o}\right)^{2{\rm dom}\left({\gamma\over 3} -2, 0\right) + {2\over 3}(k_2 + k_3)} 
+ {1\over c_1} \sum_{k_4} \hat{\rm N}_5^{(k_4)} 
\left({g_s\over {\rm HH}_o}\right)^{l + {2k_4\over 3}}, \nonumber \nd} 
where $k_i \in {\mathbb{Z}\over 2}$, and we used \eqref{sessence70}, \eqref{elfadogg22} and \eqref{benson} to fix the $g_s$ scalings of the flux components, the curvature terms and the dynamical M5-branes. The $g_s$ scalings of all the three terms would match when:
\bg\label{kellerson5}
l_{\rm MN}^{\rm PQ} + {2k_1\over 3} = 
2{\rm dom}\left({\gamma\over 3} -2, 0\right) + {2\over 3}(k_2 + k_3) = l + {2k_4\over 3}, \nd
where  $l_{\rm MN}^{\rm PQ} = 1$ from \eqref{collateral} and $\gamma = 6$ from \eqref{makibhalu3}. When $k_ i = 0$, we see that there is a mis-match of the dominant scalings: the flux term scales as 
$\left({g_s\over {\rm HH}_o}\right)^{+1}$, whereas the curvature term scales as 
$\left({g_s\over {\rm HH}_o}\right)^{0}$. We expect $l \ge 0$, and therefore if we take ${\bf G}_{\rm MNPQ}$ to be completely localized fluxes, as in the second term of \eqref{sessence70},  and $l = 0$ for this case from \eqref{benson}, the quantization condition yields:
\bg\label{kellerson6}
\pm c_2 \int_{{\bf \Sigma}_4}dy^{[{\rm MNPQ}]}~{\rm tr}\left({\bf R}^{(k_2)}_{[{\rm MN}]}
{\bf R}^{(k_3)}_{[{\rm PQ}]}\right) 
= \hat{\rm N}_5^{(k_4)}, \nd
where $k_4 = k_2 + k_3$ so the matching may be performed at every order in $g_s$ in the absence of global ${\bf G}_{\rm MNPQ}$ fluxes, and $\pm$ denotes the possibility of choosing either signs in \eqref{kellerson3}. In principle, the sign of the LHS will depend on the sign of the $k$-th order first Pontryagin class defined over a four-cycle in ${\cal M}_4 \times {\cal M}_2$ (once the value of $c_2$ is fixed); and the sign of the RHS will be determined whether we have five-branes or anti-five-branes. We can resolve any sign ambiguity by resorting to {\it positive} signs on both sides of \eqref{kellerson6}\footnote{For the case with positive $c_2$, if the sign of the LHS is negative then we need $\int {\rm tr} ~{\bf R} \wedge {\bf R} < 0$. 
A simple non-compact example is the Atiyah-Hitchin (AH) space (for a computation of the first Pontryagin class, see for example \cite{savstern}). We can geometrically join copies of AH spaces to construct a compact four-manifold (much like joining copies of Taub-NUT spaces to construct a K3 space). However since the relative sign is not important, we can stick with 
$\int {\rm tr}~{\bf R} \wedge {\bf R} > 0$ to satisfy \eqref{kellerson6} with positive $c_2$.}. Additionally, we see that the individual $\hat{\rm N}_5^{(k)}$ are not required to be integer as long as 
$\hat{\rm N}_5$ from \eqref{benson} is an integer.

On the other hand, if we consider the choice of the dominant scalings from \eqref{automoon} and 
\eqref{renaissance2}, without worrying too much about the values for $l_{0i}^{j{\rm M}}$, then the $g_s$ scaling of the flux components ${\bf G}_{\rm MNPQ}$ go as $\left({g_s\over {\rm HH}_o}\right)^0$, exactly as the dominant scaling of the first Pontryagin class. This means that the flux quantization condition becomes:
\bg\label{tagjapmey}
\int_{{\bf \Sigma}_4} dy^{[{\rm MNPQ}]}~{\cal G}^{(k_1)}_{\rm MNPQ} - c_2 \int_{{\bf \Sigma}_4}dy^{[{\rm MNPQ}]}~{\rm tr}\left({\bf R}^{(k_2)}_{[{\rm MN}]}
{\bf R}^{(k_1 - k_2)}_{[{\rm PQ}]}\right) 
= \hat{\rm N}_5^{(k_1)}, \nd
where $k_1 > k_2$ and we used $k_1 = k_2 + k_3 = k_4$ compared to what we had in \eqref{kellerson6}. The existence of {\it global} flux components for the case 
\eqref{automoon} (or even \eqref{renaissance2}, although the choice $l_{0i}^{j{\rm M}} = -3$ seems problematic), might suggest a preference of \eqref{automoon} over \eqref{collateral}, but as we shall show later, this is {\it not} the case. In fact \eqref{collateral} will still be the {\it preferred} choice for our case. The other two options in \eqref{pughmey}, where the dominant scalings of the G-flux components are 
${2\over 3}$ and ${1\over 3}$, again do not allow global fluxes and the condition \eqref{kellerson6} 
continues to provide the quantization condition there too. Note however that, according to footnote 
\ref{hetdual}, these G-flux components are eliminated when we are at the IIA orientifold point and therefore there is no pressure to map them to the heterotic side. Away from the orientifold point these fluxes survive and our above analysis provides the necessary quantization procedure. 

The {\it local} fluxes, in the vein of \eqref{sessence70}, are now balanced by the quantum terms as in 
\eqref{kellerson2}. In the first reference of \cite{desitter2}, the values for $\theta_{nl}$ in the schematic 
diagram \eqref{lilcaduxox} have been worked out carefully for all the relevant cases related to 
\eqref{collateral} (see cases 1 to 7 in section 4.2 therein). In the following we will briefly mention the choices for $\theta_{nl}$ when we go to the other possibilities from \eqref{pughmey}. 

For $l_{\rm MN}^{\rm PQ} = 1$, the first reference from \cite{desitter2} suggests that $\theta_{nl} = {14\over 3}$. Since both ${\bf G}_{\rm MNPQ}$ and it's dual form\footnote{Recall that the dual form scales as 
$l_{\rm MN}^{\rm PQ} - 2$, and therefore appears in the same way as in \eqref{botsuga4}.} 
contribute as $+{7\over 3}\left(l_{61}, l_{62}, l_{64}\right)$ to \eqref{fahingsha5} (see \eqref{botsuga4}), there are non-trivial quantum terms from $\ast\mathbb{Y}_7$ participating at this level.  When $l_{\rm MN}^{\rm PQ} = 0$, the quantum terms scale as $\theta_{nl} = {8\over 3}$ corresponding to \eqref{automoon}. For the remaining two cases in \eqref{pughmey} when $l_{\rm MN}^{\rm PQ} = {2\over 3}$ and $l_{\rm MN}^{\rm PQ} = {1\over 3}$, $\theta_{nl}$ takes values $4$ and ${10\over 3}$ respectively. There are still non-trivial quantum terms because the dual forms contribute respectively as $+2\left(l_{61}, l_{62}, l_{64}\right)$ and $+{5 \over 3}\left(l_{61}, l_{62}, l_{64}\right)$ to 
\eqref{fahingsha5}.

\vskip.2in

\noindent {\it Case 2: ${\bf G}_{{\rm MN}ab}$ components}

\vskip.2in

\noindent The case for ${\bf G}_{{\rm MN}ab}$ flux components is important because they cannot exist as global fluxes otherwise they would  imply the existence of the three-form fluxes ${\bf C}_{{\rm MN}a},
{\bf C}_{{\rm MN}b}$ and ${\bf C}_{{\rm M}ab}$. The last one T-dualizes to the metric cross-term   
${\bf g}_{3{\rm M}}$ in the IIB side, thus ruining the de Sitter structure altogether. To avoid such catastrophic
consequences, these flux components only exist as localized fluxes so that they appear as gauge fluxes on the IIB seven-branes. Nevertheless one may express the quantization condition for the {\it global} fluxes 
in the same way as \eqref{kellerson3}, namely:
\bg\label{kellerson30}
\int_{{\bf \Sigma}^{(1)}_4} dy^{[{\rm MN}ab]}~{\bf G}^{\rm global}_{{\rm MN}ab} \pm {c_2\over c_1}\int_{{\bf \Sigma}^{(1)}_4} dy^{[{\rm MN}ab]}~{\rm tr}
\Big(\mathbb{R}_{\rm tot} \wedge \mathbb{R}_{\rm tot}\Big)_{{\rm MN}ab} = 
{\hat{\rm N}_{5c}\over c_1}, \nd 
where we have used $\hat{\rm N}_{5c}$ to represent the number of five-branes for this case, $\pm$ to resolve any relative sign ambiguity \cite{wittenfluxes}, and 
${\bf \Sigma}^{(1)}_4 \equiv {\cal C}_2 \times {\mathbb{T}^2\over {\cal G}}$ where  ${\cal C}_2 \in 
{\cal M}_4 \times {\cal M}_2$ to denote a two-cycle inside the six-manifold. The $g_s$ scalings of the various terms
in \eqref{kellerson30} take the form:
\bg\label{kellerson40}
&& \sum_{k_1}\int_{{\bf \Sigma}^{(1)}_4} dy^{[{\rm MN}ab]}~{\cal G}^{(k_1)}_{{\rm MN}ab} \left({g_s\over {\rm HH}_o}\right)^{l_{\rm MN}^{ab} + {2k_1\over 3}} - {1\over c_1} \sum_{k_4} \hat{\rm N}_{5c}^{(k_4)} 
\left({g_s\over {\rm HH}_o}\right)^{l_c + {2k_4\over 3}}  \\
&=& \pm
{c_2\over c_1} \sum_{k_2, k_3}  \int_{{\bf \Sigma}^{(1)}_4}dy^{[{\rm MN}ab]}~{\rm tr}\left({\bf R}^{(k_2)}_{[{\rm M}a]}{\bf R}^{(k_3)}_{[{\rm N}b]}\right) 
 \left({g_s\over {\rm HH}_o}\right)^{2{\rm dom}\left({\gamma\over 3} - 1, 1\right) 
+ {2\over 3}(k_2 + k_3)}\nonumber\\
&\pm& {c_2\over c_1} \sum_{k_2, k_3}  \int_{{\bf \Sigma}^{(1)}_4}dy^{[{\rm MN}ab]}~{\rm tr}\left({\bf R}^{(k_2)}_{[{\rm MN}]}{\bf R}^{(k_3)}_{[ab]}\right) 
 \left({g_s\over {\rm HH}_o}\right)^{{\rm dom}\left({\gamma\over 3} -2, 0\right) 
+ {\rm dom}\left({\gamma\over 3}, 2\right) + {2\over 3}(k_2 + k_3)}, \nonumber \nd
where as before $dy^{[{\rm MN}ab]}$ implies the wedge product $dy^{\rm M} \wedge ...\wedge dy^b$. There are four different dominant scalings at play now: dominant scaling $l_{\rm MN}^{ab}$ for the flux components, dominant scaling $l_c$ for the dynamical M5-branes and two dominant scalings for the curvature wedge products. They would all balance when:
\bg\label{abarlisa}
l_{\rm MN}^{ab} + {2k_1\over 3} = l_c + {2k_4\over 3} = 2{\rm dom}\left({\gamma\over 3} - 1, 1\right)
= {\rm dom}\left({\gamma\over 3} -2, 0\right) 
+ {\rm dom}\left({\gamma\over 3}, 2\right), \nd
where ${\rm dom}(a, b)$ chooses the dominant scaling between $\left({g_s\over {\rm HH}_o}\right)^a$ 
and $\left({g_s\over {\rm HH}_o}\right)^b$. 
Once we consider \eqref{collateral}, $l_{\rm MN}^{ab} = 1$, and with the choice $\gamma = 6$ from 
\eqref{makibhalu3}, we see that the Pontryagin term scales as $\left({g_s\over {\rm HH}_o}\right)^{+2}$,  leading again to a mis-match between the $g_s$ scalings of the flux and the curvature terms. This mis-match is a good sign for us because it tells us that ${\bf G}_{{\rm MN}ab}$ cannot be global fluxes, thus avoiding the catastrophe alluded to earlier\footnote{A question could be asked regarding the scenario when $k_1 \ge {3\over 2}$ in \eqref{abarlisa}. What happens then? The answer is that, if there are no global fluxes when $k_1 = 0$, it would be meaningless to impose the existence of global fluxes for $k_1 \ge {3\over 2}$. Additionally, at this order any global fluxes would have contributed as $+{4\over 3}(l_{69}, l_{70}, l_{71})$ to 
\eqref{fahingsha5} and therefore would have anyway failed to influence the Schwinger-Dyson's equations at lower orders.}. What about the $g_s$ scalings of $\hat{\rm N}_{5c}^{(k_4)}$? If the M5-branes dualize to small instantons in the heterotic side then, to avoid issues with anomalies (considering also the comments from footnotes \ref{hetdual} and \ref{mca66k}), the $g_s$ scalings should match the ones coming from the Pontryagin class. This means $l_c = 2$ in \eqref{kellerson40}, giving us:
\bg\label{japiskull}
\pm c_2 \int_{{\bf \Sigma}^{(1)}_4}dy^{[{\rm MN}ab]}\left[{\rm tr}\left({\bf R}^{(k_2)}_{[{\rm MN}]}{\bf R}^{(k_3)}_{[ab]}\right) + {\rm tr}\left({\bf R}^{(k_2)}_{[{\rm M}a]}{\bf R}^{(k_3)}_{[{\rm N}b]}\right)\right] = 
  \hat{\rm N}_{5c}^{(k_4)}, \nd
where  $k_4 = k_2 + k_3$ and $k_i \in {\mathbb{Z}\over 2}$. As discussed before the $\pm$ sign implies that we can take  positive signs on both sides of \eqref{japiskull}, but now $\hat{\rm N}_{5c}^{(k_4)}$ are not necessarily constrained to be integers as long as $\hat{\rm N}_{5c}$ is an integer in the temporal domain set by the TCC. Since $l_c = 2$ in \eqref{kellerson40}, the M5-branes are not static compared to the case studied for the ${\bf G}_{\rm MNPQ}$ flux components. 

Once we consider \eqref{automoon} (or \eqref{renaissance2}) and take $\gamma = 6$, we see that the dominant scalings of all the four terms in \eqref{abarlisa} match precisely. Unfortunately this is {\it not} good because it would mean that ${\bf G}_{{\rm MN}ab}$ can be {\it global} fluxes. The catastrophe that we were worried about earlier, seems to come true now. Of course we can always impose that ${\bf G}_{{\rm MN}ab}$ $-$ despite scaling as $\left({g_s\over {\rm HH}_o}\right)^{+2}$ $-$ can be regarded as {\it localized} fluxes but there appears no strong theoretical reason to do so other than to avoid the aforementioned 
catastrophe. This provides a good reason to choose \eqref{collateral} over \eqref{automoon} (or even 
\eqref{renaissance2}). The two other cases in \eqref{pughmey}, namely $l_{\rm MN}^{ab} = {4\over 3}$ and
$l_{\rm MN}^{ab} = {5\over 3}$, cannot be global fluxes and therefore \eqref{japiskull} should be satisfied for both the cases. Of course as mentioned earlier these components cannot participate at the lowest order Schwinger-Dyson's equations for the metric.

For the localized fluxes, when $l_{\rm MN}^{ab} = 1$, it is easy to see that the dual seven-forms scale as
$l_{\rm MN}^{ab} - 6$, which means that the RHS of \eqref{kellerson2} scales as 
$\theta_{nl} - l_{\rm MN}^{ab} + {4\over 3}$, where $\theta_{nl}$ is given by \eqref{botsuga4}. Comparing both sides of \eqref{kellerson2}, gives us $\theta_{nl} = {2\over 3}$. This is barely enough to provide the kinetic term, so we see that non-perturbative effects are necessary. Incorporating the contributions from the BBS instantons \eqref{beverwickmey}, $\theta_{nl}$ becomes $\theta_{nl} = {8\over 3}$ which would easily balance both sides in \eqref{kellerson2}. When $l_{\rm MN}^{ab} = 2$ from \eqref{automoon} and 
\eqref{renaissance2}, $\theta_{nl} = {8\over 3}$ which could in principle be enough because the dual fluxes contribute as $+{4\over 3}$ to \eqref{fahingsha5} (see \eqref{botsuga4}). BBS instantons would change this to $\theta_{nl} = {14\over 3}$, as sub-leading contributions. However, the fact that these fluxes can exist as global fluxes, make the choices \eqref{automoon} and \eqref{renaissance2} less attractive for us as mentioned earlier. Finally, for the other two cases $l_{\rm MN}^{ab} = {4\over 3}$ and
$l_{\rm MN}^{ab} = {5\over 3}$ we see that the quantum terms scale as $\theta_{nl} = {4\over 3}$ and 
$\theta_{nl} = 2$ respectively which are barely enough because they contribute to \eqref{fahingsha5} as
$+{2\over 3}$ and $+1$ and therefore contributions from BBS instantons can change the quantum contributions to $\theta_{nl} = {10\over 3}$ and $\theta_{nl} = 4$ respectively. 

\vskip.2in

\noindent {\it Case 3: ${\bf G}_{{\rm MNP}a}$ components}

\vskip.2in

\noindent The case for the flux components ${\bf G}_{{\rm MNP}a}$ is interesting because for all the choices in \eqref{pughmey} they scale in the same way as $\left({g_s\over {\rm HH}_o}\right)^{+1}$. The $+1$ scaling is necessary for the system to satisfy anomaly cancellation condition \eqref{lindmonaco}, and therefore the quantization condition on the global fluxes takes the form:
\bg\label{ashfyre}
\int_{{\bf \Sigma}^{(2)}_4} dy^{[{\rm MNP}a]}~{\bf G}^{\rm global}_{{\rm MNP}a} \pm {c_2\over c_1}\int_{{\bf \Sigma}^{(2)}_4} dy^{[{\rm MNP}a]}~{\rm tr}
\Big(\mathbb{R}_{\rm tot} \wedge \mathbb{R}_{\rm tot}\Big)_{{\rm MNP}a} = 
{\hat{\rm N}_{5e}\over c_1}, \nd 
where $\hat{\rm N}_{5e}$ is the number of M5-branes with ${\bf \Sigma}^{(2)}_4 = {\cal C}_3 \times 
{\bf S}^1$, where ${\cal C}_3 \in {\cal M}_4 \times {\cal M}_2$, being a three-cycle in the six-manifold and ${\bf S}^1 \in {\mathbb{T}^2\over {\cal G}}$ being a one-cycle in the toroidal manifold. For both cases such odd cycles are possible because the metric on the internal eight-manifold is a non-K\"ahler one, and therefore allows all possible cycles, odd and even\footnote{There is again a subtlety here that needs some explanation. Existence of odd-cycles, especially in the toroidal space, does not necessarily imply vanishing Euler characteristics. Consider M5-branes wrapping a three-cycle $\hat{\cal C}_3$ (orthogonal to ${\cal C}_3$ above). The five-manifold ${\bf \Sigma}^{(2)}_5$ to which ${\bf \Sigma}^{(2)}_4$ is the boundary now spans $\hat{\cal C}_3 \times {\mathbb{T}^2\over {\cal G}}$, which boils down to finding the one-cycle boundary in the toroidal space. At the IIA orientifold point such one-cycle may be easily inferred from the ${\cal G}$ action (see footnote \ref{hetdual}), but the Euler characteristics remains non-vanishing. In any case, the anomaly cancellation condition \eqref{lindmonaco} allows non-trivial fluxes on the eight-manifold as long as $\int {\bf X}_8$ over the eight-manifold is non-vanishing (we are assuming flux products to be positive definite). For a 
non-K\"ahler manifold, the integral of ${\bf X}_8$ is not necessarily the Euler characteristics (as it was when the eight-manifold was a Calabi-Yau four-fold), so the existence of fluxes no longer depends on the Euler characteristics of the internal eight-manifold.}. The $g_s$ scalings of the various terms in \eqref{ashfyre} now becomes:

{\footnotesize
\bg\label{lazladi}
&& \sum_{k_1}\int_{{\bf \Sigma}^{(2)}_4} dy^{[{\rm MNP}a]}~{\cal G}^{(k_1)}_{{\rm MNP}a} \left({g_s\over {\rm HH}_o}\right)^{l_{\rm MN}^{{\rm P}a} + {2k_1\over 3}}
- {1\over c_1} \sum_{k_4} \hat{\rm N}_{5e}^{(k_4)} 
\left({g_s\over {\rm HH}_o}\right)^{l_e + {2k_4\over 3}}\\
&=& 
{c_2\over c_1} \sum_{k_2, k_3}  \int_{{\bf \Sigma}^{(2)}_4}dy^{[{\rm MNP}a]}~{\rm tr}\left({\bf R}^{(k_2)}_{[{\rm MN}]}{\bf R}^{(k_3)}_{[{\rm P}a]}\right) 
 \left({g_s\over {\rm HH}_o}\right)^{{\rm dom}\left({\gamma\over 3} -2, 0\right) 
 + {\rm dom}\left({\gamma\over 3} -1, 1\right) + {2\over 3}(k_2 + k_3)} , \nonumber \nd}    
 where for $\gamma = 6$ in \eqref{makibhalu3}, the wedge product of the curvature term scales as 
 $\left({g_s\over {\rm HH}_o}\right)^{+1}$, implying that $l_e = 1$ in the expression for the dynamical M5-branes. Thus yet again the M5-branes do not have a static part. Since $l_{\rm MN}^{{\rm P}a} = 1$ from all the choices in \eqref{pughmey}, global fluxes {\it do} exist now. This is good because existence of global flux components of the form ${\bf G}_{{\rm MNP}a}$ would imply the existence of NS-NS and RR three-forms
 $\left({\bf H}_3\right)_{\rm MNP}$  and  $\left({\bf F}_3\right)_{\rm MNP}$ respectively in the dual IIB side. Such three-forms would contribute to the IIB super-potential, but now, since they cannot be time-independent, the IIB super-potential would necessarily develop temporal dependences as emphasized also in \cite{heliudson}. The flux quantization condition is valid at all orders in 
 $\left(k_i, {g_s\over {\rm HH}_o}\right)$ because we expect:
 \bg\label{rook}
 l_{\rm MN}^{{\rm P}a} + {2k_1\over 3} = l_e + {2k_4\over 3} = 
 {\rm dom}\left({\gamma\over 3} -2, 0\right) 
 + {\rm dom}\left({\gamma\over 3} -1, 1\right) + {2\over 3}(k_2 + k_3), \nd
 for $\left(l_e, l_{\rm MN}^{{\rm P}a}, \gamma\right) = (1, 1, 6)$ and $k_i \in {\mathbb{Z}\over 2}$. Note that if we had taken $\gamma = 5$ or any values other than 6 this would not have been the case. In fact for 
 $\gamma = 5$, we see that the curvature term scales as 
 $\left({g_s\over {\rm HH}_o}\right)^{1\over 3}$, so would have differed from the $g_s$ scalings of the flux components thus prohibiting them to appear as global fluxes. While this scenario may not be inconsistent, the very existence of global fluxes implies the existence of a non-trivial super-potential in the IIB side that is helpful to dynamically stabilize at least the complex structure moduli (for some discussion on the dynamical moduli stabilization, the readers may refer to \cite{coherbeta, desitter2}).
 
The localized fluxes can now be balanced easily by the quantum terms. The dual seven-forms scale as 
$l_{\rm MN}^{{\rm P}a} - 4$, and therefore the RHS of \eqref{kellerson2} scales as 
$\theta_{nl} - l_{\rm MN}^{{\rm P}a} - {2\over 3}$ where $\theta_{nl}$ is defined in \eqref{botsuga4}. Balancing the $g_s$ scalings on both sides of \eqref{kellerson2} gives us $\theta_{nl} = {8\over 3}$. As the fluxes contribute as $+{4\over 3}$, the scaling of $\theta_{nl}$ tells us what possible perturbative contributions on the RHS of \eqref{kellerson2} are. Non-perturbative effects from BBS instantons will change the scaling to $\theta_{nl} = {14\over 3}$ providing enough sub-dominant quantum contributions. The non-local contributions would be further sub-dominant. Again, at every order in $g_s$, the localized fluxes are properly balanced over the four-cycle ${\bf \Sigma}^{(2)}_4$.
 
\vskip.2in

\noindent {\it Case summary: Flux quantizations and flux equations as selection principles}

\vskip.2in

\noindent Having laid out all the flux equations, Bianchi identities as well as the flux quantization rules for
the time-dependent background, we see the emergence of a few selection principles. These principles would help us to not only decide the forms of the fluxes  ${\rm G}_1$ to $G_{17}$ in the Bianchi identities 
\eqref{chloefern} and \eqref{20lis18} but also discern the most relevant dominant scalings from 
\eqref{pughmey}. They may be listed as follows. {\Su One}, we see that the following pair of flux components: $({\rm G}_4, {\rm G}_{15}), 
({\rm G}_5, {\rm G}_{16}), ({\rm G}_8, {\rm G}_{17}), ({\rm G}_1, {\rm G}_{9})$ and 
$({\rm G}_3, {\rm G}_{11})$ are necessarily {\it localized} fluxes. Despite that most do not contribute at the lowest orders Schwinger-Dyson's equations except $({\rm G}_3, {\rm G}_{11})$, although both 
$({\rm G}_1, {\rm G}_{3})$ can be consistently put to zero. {\Su Two}, the flux components 
$({\rm G}_6, {\rm G}_7, {\rm G}_{10})$ are {\it global} fluxes, although ${\rm G}_{10}$ can have localized pieces. In fact the global pieces of ${\rm G}_{10}$ satisfy the flux quantization conditions \eqref{ashfyre}. For the two latter components $-$ which are necessarily global $-$ the precise forms are given by 
\eqref{kurisada} and \eqref{HiGaToK} respectively, although in the limit $x^{11} \to 0$, the flux components
${\rm G}_7$ consistently decouple. The remaining flux components $({\rm G}_2, {\rm G}_{12}, {\rm G}_{13},
{\rm G}_{14})$ can be global or local but they do not contribute to the lowest orders Schwinger-Dyson's equations for the metric components. {\Su Three}, while the anomaly cancellation condition allows all the five choices in \eqref{pughmey}, the presence of localized fluxes and dynamical membranes seem to rule out the choice with $l_{0i}^{j{\rm M}} = -3$. On the other hand, out of the remaining four in \eqref{pughmey}, the flux quantization conditions appear to prefer the three over \eqref{automoon}. Finally, from the $g_s$ scalings and EOMs, it seems that the choice \eqref{collateral} is the most economical one, as the other two in the remaining three would forbid {\it any} of the flux components to appear at the lowest orders (this may be easily inferred from plugging in the scalings from \eqref{pughmey} in  \eqref{botsuga4}).  Therefore we can conclude that in the M-theory uplift of the IIB de Sitter space, irrespective of its embeddings, as long as the system consistently satisfies (a) flux EOMs, (b) anomaly cancellation, (c) Bianchi identities, and (d) flux quantization conditions, the internal and the external fluxes have to be necessarily time-dependent. The other flux components ${\bf G}_{0{\rm ABC}}$ may be put to zero without violating any of the 
aforementioned conditions. 

\section{de Sitter state in M-theory, toy cosmology, and exclusion principles \label{detourm}}

Our detailed analysis of the flux EOMs should have convinced the readers that the system is not only 
quantum mechanically consistent, but is also under theoretical control where precise computations may be performed. We would now like to analyze the situations from the point of view of having a Glauber-Sudarshan state, but before we go into that let us study two examples which would tell us how we can extend our earlier analysis to investigate new directions. In the process we will be able to answer questions related to not only the {\it early} time physics, but also find new selection principles that will help us exclude many cosmologies.

\subsection{A detour towards $4d$ de Sitter state from M-theory and EFT \label{eva}}

While discerning the requisite conditions we also see, in retrospect, why the M-theory uplift was {\it necessary} in the first place. The type IIB coupling is at a constant coupling point, and is generically fixed at 
${\cal O}(1)$ value, {\it i.e.} $g_b = {\cal O}(1)$ (we took vanishing axio-dilaton so $g_b = 1$ appears to be the natural choice). This means IIB is at strong coupling, and because of that we cannot study this background with available techniques (even S-duality doesn't help here). The duality to M-theory allows a temporal domain
where the IIA coupling $g_s < 1$ (see derivations in section \ref{sec2.1}) and as discussed here, detailed and precise computations may be performed. This temporal domain is also surprisingly related to TCC 
\cite{tcc} which here may be considered as an added bonus. 

The advantage that we gain by dualizing the IIB solution to M-theory is not just restricted to having a 
controlled laboratory for all the requisite computations. The M-theory uplift also helps us to interpret the computations as consequences of having a Glauber-Sudarshan state. In this matter, the M-theory uplift is not just powerful, but necessary. As shown in \cite{coherbeta}, the flux equations that we studied in the earlier sub-sections appear from the Schwinger-Dyson's equations. In fact the Schwinger-Dyson's equations split up in two parts. One part is directly related to the EOMs for both the metric and the flux components (recall that the metric and the flux components appear from the expectation values of the metric and the flux operators over the Glauber-Sudarshan state). The other part  relates the expectation values to the Faddeev-Popov ghosts. The latter is harder to work out, but fortunately the first part contains enough informations to precisely lay out the corresponding EOMs in the presence of hierarchically controlled quantum corrections. The perturbative parts of the quantum corrections appear from 
\eqref{fahingsha5}, but can make it even more general by allowing cross-terms in the internal eight-manifold 
(which could in-principle happen if we allow non-trivial fibrations). The quantum series then changes to:

{\footnotesize
\bg\label{fahingsha6}
\mathbb{Q}_{\rm T}^{(\{l_i\}, n_i)} &= & \left[{\bf g}^{-1}\right] \prod_{i = 0}^4 \left[\partial\right]^{n_i} 
\prod_{{\rm k} = 1}^{65} \left({\bf R}_{\rm A_k B_k C_k D_k}\right)^{l_{\rm k}} \prod_{{\rm r} = 66}^{105} 
\left({\bf G}_{\rm A_r B_r C_r D_r}\right)^{l_{\rm r}}\nonumber\\
& = & {\bf g}^{m_i m'_i}.... {\bf g}^{j_k j'_k} 
\{\partial_m^{n_1}\} \{\partial_\alpha^{n_2}\} \{\partial_a^{n_3}\}\{\partial_i^{n_4}\}\{\partial_0^{n_0}\}
\left({\bf R}_{mnpq}\right)^{l_1} \left({\bf R}_{abab}\right)^{l_2}\left({\bf R}_{pqab}\right)^{l_3}\left({\bf R}_{\alpha a b \beta}\right)^{l_4} \nonumber\\
&\times& \left({\bf R}_{\alpha\beta mn}\right)^{l_5}\left({\bf R}_{\alpha\beta\alpha\beta}\right)^{l_6}
\left({\bf R}_{ijij}\right)^{l_7}\left({\bf R}_{ijmn}\right)^{l_8}\left({\bf R}_{iajb}\right)^{l_9}
\left({\bf R}_{i\alpha j \beta}\right)^{l_{10}}\left({\bf R}_{0mnp}\right)^{l_{11}}
\nonumber\\
& \times & \left({\bf R}_{0m0n}\right)^{l_{12}}\left({\bf R}_{0i0j}\right)^{l_{13}}\left({\bf R}_{0a0b}\right)^{l_{14}}\left({\bf R}_{0\alpha 0\beta}\right)^{l_{15}}
\left({\bf R}_{0\alpha\beta m}\right)^{l_{16}}\left({\bf R}_{0abm}\right)^{l_{17}}\left({\bf R}_{0ijm}\right)^{l_{18}}
\nonumber\\
& \times & \left({\bf R}_{mnp\alpha}\right)^{l_{19}}\left({\bf R}_{m\alpha ab}\right)^{l_{20}}
\left({\bf R}_{m\alpha\alpha\beta}\right)^{l_{21}}\left({\bf R}_{m\alpha ij}\right)^{l_{22}}
\left({\bf R}_{0mn \alpha}\right)^{l_{23}}\left({\bf R}_{0m0\alpha}\right)^{l_{24}}
\left({\bf R}_{0\alpha\beta\alpha}\right)^{l_{25}}
\nonumber\\
&\times& \left({\bf R}_{0ab \alpha}\right)^{l_{26}}\left({\bf R}_{0ij\alpha}\right)^{l_{27}}
\left({\bf R}_{mnpi}\right)^{l_{28}}\left({\bf R}_{mni0}\right)^{l_{29}}
\left({\bf R}_{mn i\alpha}\right)^{l_{30}}\left({\bf R}_{0m0i}\right)^{l_{31}}
\left({\bf R}_{mijk}\right)^{l_{32}}\nonumber\\
&\times& \left({\bf R}_{m\beta i \alpha}\right)^{l_{33}}\left({\bf R}_{abmi}\right)^{l_{34}}
\left({\bf R}_{ijk0}\right)^{l_{35}}\left({\bf R}_{\alpha 0i0}\right)^{l_{36}}
\left({\bf R}_{\alpha\beta i 0}\right)^{l_{37}}\left({\bf R}_{ab0i}\right)^{l_{38}}
\left({\bf R}_{\alpha ijk}\right)^{l_{39}}\nonumber\\
&\times& \left({\bf R}_{ab i \alpha}\right)^{l_{40}}\left({\bf R}_{\alpha\beta i \alpha}\right)^{l_{41}}
\left({\bf R}_{mnpa}\right)^{l_{42}}\left({\bf R}_{mna0}\right)^{l_{43}}
\left({\bf R}_{mna i}\right)^{l_{44}}\left({\bf R}_{mna\alpha}\right)^{l_{45}}
\left({\bf R}_{m0a0}\right)^{l_{46}}\nonumber\\
&\times& \left({\bf R}_{maij}\right)^{l_{47}}\left({\bf R}_{ma\alpha\beta i}\right)^{l_{48}}
\left({\bf R}_{maba}\right)^{l_{49}}\left({\bf R}_{aij0}\right)^{l_{50}}
\left({\bf R}_{a0 i 0}\right)^{l_{51}}\left({\bf R}_{a\alpha\beta 0}\right)^{l_{52}}
\left({\bf R}_{a 0\alpha 0}\right)^{l_{53}}\nonumber\\
&\times& \left({\bf R}_{ab a0}\right)^{l_{54}}\left({\bf R}_{aijk}\right)^{l_{55}}
\left({\bf R}_{a\alpha ij}\right)^{l_{56}}\left({\bf R}_{a\alpha\alpha\beta}\right)^{l_{57}}
\left({\bf R}_{ab a\alpha}\right)^{l_{58}}\left({\bf R}_{abia}\right)^{l_{59}} 
\left({\bf R}_{\alpha\beta ia}\right)^{l_{60}}\nonumber\\
&\times& \left({\bf R}_{m\alpha i0}\right)^{l_{61}}\left({\bf R}_{mai0}\right)^{l_{62}}
\left({\bf R}_{ma\alpha 0}\right)^{l_{63}}\left({\bf R}_{mia\alpha}\right)^{l_{64}} 
\left({\bf R}_{a\alpha i0}\right)^{l_{65}}
\left({\bf G}_{mnpq}\right)^{l_{66}} \left({\bf G}_{mnp\alpha}\right)^{l_{67}}\nonumber\\
&\times & \left({\bf G}_{mnpa}\right)^{l_{68}}\left({\bf G}_{mn\alpha\beta}\right)^{l_{69}}
\left({\bf G}_{mn\alpha a}\right)^{l_{70}}
\left({\bf G}_{m\alpha\beta a}\right)^{l_{71}}\left({\bf G}_{0ijm}\right)^{l_{72}}
\left({\bf G}_{0ij\alpha}\right)^{l_{73}}
\left({\bf G}_{mnab}\right)^{l_{74}}\nonumber\\
&\times& \left({\bf G}_{ab\alpha\beta}\right)^{l_{75}}
\left({\bf G}_{m\alpha ab}\right)^{l_{76}}
 \left({\bf G}_{mnpi}\right)^{l_{77}}
\left({\bf G}_{m\alpha\beta i}\right)^{l_{78}}\left({\bf G}_{mn\alpha i}\right)^{l_{79}} 
\left({\bf G}_{mnai}\right)^{l_{80}}
\left({\bf G}_{mabi}\right)^{l_{81}}\nonumber\\
&\times &\left({\bf G}_{a\alpha\beta i}\right)^{l_{82}}
\left({\bf G}_{\alpha ab i}\right)^{l_{83}} \left({\bf G}_{ma\alpha i}\right)^{84} 
\left({\bf G}_{mn ij}\right)^{l_{85}}\left({\bf G}_{m\alpha ij}\right)^{l_{86}} 
\left({\bf G}_{\alpha\beta ij}\right)^{l_{87}}
\left({\bf G}_{maij}\right)^{l_{88}}\nonumber\\
&\times& \left({\bf G}_{\alpha a ij}\right)^{l_{89}}
\left({\bf G}_{ab ij}\right)^{l_{90}} \left({\bf G}_{0ija}\right)^{l_{91}}
\left({\bf G}_{0mnp}\right)^{l_{92}}
\left({\bf G}_{0mn\alpha}\right)^{l_{93}} 
\left({\bf G}_{0m\alpha\beta}\right)^{l_{94}}\left({\bf G}_{0mab}\right)^{l_{95}}\nonumber\\
&\times& 
\left({\bf G}_{0\alpha ab}\right)^{l_{96}}\left({\bf G}_{0mna}\right)^{l_{97}}
\left({\bf G}_{0m\alpha a}\right)^{l_{98}} \left({\bf G}_{0\alpha\beta a}\right)^{l_{99}}
\left({\bf G}_{0mni}\right)^{l_{100}}
\left({\bf G}_{0m\alpha i}\right)^{l_{101}} \left({\bf G}_{0\alpha\beta i}\right)^{l_{102}}
\nonumber\\
&\times& 
\left({\bf G}_{0mia}\right)^{l_{103}}
\left({\bf G}_{0\alpha ia}\right)^{l_{104}}\left({\bf G}_{0ab i}\right)^{l_{105}}, \nd}
which should be compared to \eqref{fahingsha5}. With a little effort, by specifying the precise 
fibrations, we can also work out the $g_s$ scaling much like the one from \eqref{botsuga4}. We will however 
refrain from doing this here because there isn't anything new to be learned by going towards more genericity. Instead we will study a {\it different} duality of IIB to M-theory that can allow us to address the four-dimensional de Sitter space directly from eleven-dimensions. In the process we can compare the 
consequence of viewing the de Sitter space as a {\it vacuum} or as a Glauber-Sudarshan state. 

To study the consequence of a new duality from IIB to M-theory, it would be better to first follow the old duality (that we have been using so far) but for a slightly different type IIB background. The background that
we have in mind may be represented by the following metric ans\"atze:

{\footnotesize
\bg\label{villvaleur}
 ds^2 = {a^2(t)\over {\rm H}^2(y)}\left(-dt^2 + g_{ij} dx^i dx^j + g_{33} dx^3dx^3\right) + 
 {\rm H}^2(y)\left({\rm F}_1(t) g_{\alpha\beta} dy^\alpha dy^\beta + {\rm F}_2(t) g_{mn} dy^m dy^n\right), \nd}
 which is almost similar to what we had before but now there are two main differences: {\Su one}, the coefficient $a^2(t)$ is not related to any de Sitter embedding studied in section \ref{2.1.1}, and {\Su two}, 
 ${\rm F}_i(t)$ do not satisfy the condition ${\rm F}_1(t) {\rm F}^2_2(t) = 1$. The precise connection of 
 $\left({\rm F}_i(t), a^2(t)\right)$ to four-dimensional de Sitter space in M-theory will be spelled out soon.
 Despite this, \eqref{villvaleur} is still a cosmological ans\"atze for a solution in IIB, much like \eqref{morbiusn} earlier, albeit with temporally varying Newton's constant.  The IIA string coupling will take the form similar to 
 \eqref{slivia} except that $\left(\Lambda t^2\right)^{n/2}$ therein will be replaced by $a(t)$. In other words:
 \bg\label{salvatonmay}
 {\rm H}_o({\bf x}, y) \equiv {g_b \over \sqrt{g_{33}({\bf x})}}, ~~~~ \Rightarrow ~~~~
 {g_s\over {\rm HH}_o} = {1\over a(t)}, \nd
 which as before may be used to define the temporal coordinates in M-theory. Here $g_b$ is the IIB coupling. Note two things: {\Su one}, $g_b = {\cal O}(1)$ is again at constant coupling point and therefore makes the IIB background 
 \eqref{villvaleur} strongly coupled, and {\Su two}, the late time is always {\it weakly} coupled in the IIA side no matter what choice of $a(t)$ we take, as long as we demand expanding cosmologies. The M-theory uplift 
 of \eqref{villvaleur} then becomes:
  \bg\label{matagre}
 ds^2 &= & g_s^{-8/3} {\rm H}^2_o\left(-dt^2 + g_{ij} dx^idx^j\right) 
 + g_s^{4/3}\left[(dx^{11})^2 + {(dx^3)^2\over g_b^2}\right]\nonumber\\
 &&~~~~~~ + 
 g_s^{-2/3} {\rm H}^2(y)\left({\rm F}_1(t) g_{\alpha\beta} dy^\alpha dy^\beta + {\rm F}_2(t) g_{mn} dy^m dy^n\right) , \nd
which goes on to show that the eleven-dimensional torus is a square one locally although globally we can allow ${\mathbb{T}^2\over {\cal G}}$. We note that, no matter how $a(t)$ is defined, the form of the metric 
\eqref{matagre} remains similar\footnote{As before we can use ${\rm F}_i(t) = 
{\rm F}_i\left({g_s\over {\rm HH}_o}\right)$, with the condition that ${\rm F}_i(t) \to 1$ as 
${g_s\over {\rm HH}_o} \to 0$. This means, no matter how $a(t)$ is defined, the form of ${\rm F}_i$ can always be ${\rm F}_i = \sum_{k_{n_i}} {\rm F}_i^{(k_{n_i})}\left({g_s\over {\rm HH}_o}\right)^{k_{n_i}}$ consistent with how we defined earlier in the paragraph following \eqref{cora2}. Clearly ${\rm F}_i^{(0)} = 1$ would keep the late time cosmology well-defined. \label{ryantel}} 
to what we had in \eqref{evader} with $g_s$ taking the form 
\eqref{salvatonmay}. This means that the fluxes required to support a background \eqref{villvaleur} in the IIB side may again be derived from the M-theory uplift. As shown in the first reference of \cite{desitter2}, the analysis does not depend on what conditions ${\rm F}_i(t)$ satisfy (in fact therein an example with 
${\rm F}_1 {\rm F}^2_2 = {\cal O}(g_s^2)$ was discussed in great details). After the dust settles we expect 
time-dependent internal fluxes ${\bf G}_{\rm MNPQ}, {\bf G}_{{\rm MNP}a}$ and ${\bf G}_{{\rm MN}ab}$ as well as time-dependent space-time fluxes ${\bf G}_{0ij{\rm M}}$ to be defined in a similar way as before (other localized flux components could also exist). Additionally we expect $\dot{g}_s$ to be given by positive powers of $g_s$, much like \eqref{senrem} before.

Nothing we said so far is new although presented  in a somewhat generalized way. To see the point of our computations, let us then dualize the IIB background \eqref{villvaleur} to M-theory in a slightly different way. Instead of choosing the toroidal direction to be $(x^3, x^{11})$, let us take the directions to be $(y_o, x^{11})$ where 
$y_o \in {\cal M}_4$. In fact to facilitate this we can rewrite the internal four-manifold ${\cal M}_4$ as 
${\cal M}_4 = {\cal M}_3 \times {\bf S}^1$ locally, where $y_o$ will be the coordinate parametrizing 
${\bf S}^1$. If $g_{y_oy_o}$ denotes the metric along ${\bf S}^1$, the type IIA coupling may be determined via:
\bg\label{salvatonmay2}
 {\rm H}_o({\bf x}, y) \equiv {g_b \over \sqrt{g_{y_oy_o}({\bf x})}}, ~~~~ \Rightarrow ~~~~
 {g_s {\rm H}\over {\rm H}_o} = {1\over \sqrt{{\rm F}_2(t)}}, \nd 
 which may now be compared to \eqref{salvatonmay}. Despite some similarities, there is an unfortunate disadvantage now. The technical advantage that we got from going to the M-theory uplift 
 \eqref{matagre} $-$ because the IIA string coupling \eqref{salvatonmay} can be made {\it small} in the temporal domain where $a(t) > 1$ $-$ now doesn't seem to help. This is because, for example, at late time when we expect  ${\rm F}_2 \to 1$, the IIA coupling ${g_s{\rm H}\over {\rm H}_o} \to 1$ so becomes strongly coupled.
 Additionally, $g_s$ is no longer related to $a(t)$, so our earlier manipulation of rewriting the metric as
 \eqref{matagre} cannot help. In fact the topology in M-theory is no longer ${\bf R}^{2, 1} \times {\cal M}_4
 \times {\cal M}_2 \times {\mathbb{T}^2\over {\cal G}}$. Instead we expect the topology to become
 ${\bf R}^{3, 1} \times {\cal M}_3
 \times {\cal M}_2 \times {\hat{\mathbb{T}}^2\over {\cal G}}$ where $\hat{\mathbb{T}}^2$ signifies the 
 $(y_o, x^{11})$ torus. What kind of metrics can be put on ${\bf R}^{3, 1}$ and on the internal {\it seven}-manifold will be ascertained soon.
 
 Despite these aforementioned hurdles, let us still push on. We have already identified the M-theory torus 
 $\hat{\mathbb{T}}^2$, and since M-theory and IIB are related by shrinking this torus, we can follow the duality chain to write the M-theory metric. There is a caveat though: since the internal space in the IIB side is threaded by fluxes, the duality chain will lead to a complicated metric for the internal space in M-theory where there could be a non-trivial mixing between the temporal and the internal-space coordinates.  We can simplify this situation by assuming that the IIB fluxes have no legs along the ${\bf S}^1$ direction. While this may not be always possible, we can nevertheless use such simplifications to generate possible solutions. The M-theory uplift of \eqref{villvaleur} now becomes:

{\footnotesize
\bg\label{tamtam} 
 ds^2 &= & {a^2(t){\rm F}^{1/3}_2(t) \over \left({\rm H}^2 {\rm H}_o\right)^{2/3}}\left(-dt^2 + g_{ij}dx^i dx^j + 
 g_{33} dx^3 dx^3\right)\\
 & + & {\rm F}^{1/3}_2(t)\left({{\rm H}^4\over {\rm H}_o}\right)^{2/3}\left({\rm F}_2(t) g_{mn} dy^m dy^n 
 + {\rm F}_1(t) g_{\alpha\beta} dy^\alpha dy^\beta\right) + 
 {1 \over {\rm F}_2^{2/3}(t)}\left({{\rm H}_o\over {\rm H}}\right)^{4/3}  
 \left[(dx^{11})^2 + {dy_o^2 \over g_b^2}\right], \nonumber \nd}
where $(y^m, y^\alpha) \in ({\cal M}_3, {\cal M}_2)$ and we have highlighted the temporal dependences. 
The toroidal structure is a consequence of the duality chasing, and in general could be more involved (for example the global topology is ${\hat{\mathbb{T}}^2\over {\cal G}}$). One may also easily note that in the IIB metric \eqref{villvaleur}, if $a^2(t)$ was related to any of the de Sitter embeddings (studied in section \ref{2.1.1}), \eqref{tamtam} {\it could not} contain a de Sitter space. Also, if 
we had imposed ${\rm F}_1{\rm F}_2^2 = 1$, the volume of the internal seven-manifold {\it could not} be time-independent. This means we can now impose the following conditions on $a^2(t)$ and ${\rm F}_i(t)$:
\bg\label{tt2}
a^2(t) {\rm F}^{1/3}_2(t) = a^2_{\bf dS}(t), ~~~~~~~ {\rm F}_1(t) {\rm F}_2^{5/3}(t) = 1, \nd
that would put a de Sitter metric along ${\bf R}^{3, 1}$ and a temporally varying metric (but with a time-independent Newton's constant)  on the internal seven-manifold ${\bf \Sigma}_7 \equiv {\cal M}_4 \times {\cal M}_2 \times {\hat{\mathbb{T}}^2\over {\cal G}}$. This is almost what we were aiming for: a four-dimensional de Sitter space with a seven-dimensional internal space but unfortunately, as alluded to earlier,  both the IIB and the M-theory uplift are at strong couplings.

One of the strong coupling issue for the M-theory uplift \eqref{tamtam} is that we can no longer suppress the non-perturbative contributions of the form ${\rm exp}\left(-{1\over g_s^2}\right)$. The perturbative series 
of the form \eqref{fahingsha5} is also a concern now. Additionally, in terms of $g_s$ defined in 
\eqref{salvatonmay2}, the metric \eqref{tamtam} may be rewritten as (using ${\hat{\rm H}}^3 \equiv 
{\rm H}^2 {\rm H}_o$):
\bg\label{tamtam2} 
 ds^2 &= & {a^2_{\bf dS}(t) \over {\hat{\rm H}}^2}\left(-dt^2 + g_{ij}dx^i dx^j + 
 g_{33} dx^3 dx^3\right)\\
 & + & g_s^{-2/3} {\rm H}^2 \left({\rm F}_2(t) g_{mn} dy^m dy^n 
 + {\rm F}_1(t) g_{\alpha\beta} dy^\alpha dy^\beta\right) + 
g_s^{4/3}
 \left[(dx^{11})^2 + {dy_o^2 \over g_b^2}\right], \nonumber \nd
which is almost similar to the form of the M-theory uplift \eqref{matagre}, defined therein using $g_s$ from 
\eqref{salvatonmay}, but with a few key differences. {\Su One}, which is already mentioned above, the metric \eqref{matagre} is split into $2+1$ dimensional space-time and an eight-dimensional internal space, whereas the split  in 
\eqref{tamtam2} is 
$3+1$ dimensional space-time and seven-dimensional internal space. {\Su Two}, since 
$a^2_{\bf dS}$ does not appear in the definition of $g_s$ in \eqref{salvatonmay2}, the temporal dependences in the metric 
\eqref{tamtam2} cannot be expressed by $g_s$ alone, as compared to \eqref{matagre}. This means, the 
advantages we got by the $g_s$ expansions in the quantum series \eqref{fahingsha5}, cannot be replicated for the metric choice \eqref{tamtam2}. This is unfortunate, because the corresponding IIB metric takes the form:

{\footnotesize
\bg\label{villvaleur2}
 ds^2 = {a^2_{\bf dS}(t)\over {\rm H}^2(y){\rm F}^{1/3}_2(t)}\left(-dt^2 + g_{ij} dx^i dx^j + g_{33} dx^3dx^3\right) + 
 {\rm H}^2(y)\left({\rm F}_1(t) g_{\alpha\beta} dy^\alpha dy^\beta + {\rm F}_2(t) g_{mn} dy^m dy^n\right), \nonumber\\ \nd}
 which {\it differs} from the de Sitter case by the presence of ${\rm F}^{1/3}_2(t)$ and from the fact that 
 ${\rm F}_1(t) {\rm F}^2_2(t) \ne 1$ (see \eqref{tt2}).  The latter would imply time-dependent Newton's constant in the IIB side, so the dual IIB cosmology appears to be leaning more towards a modified gravity framework than anything else (although the M-theory example is de Sitter). 
 
 The above conclusions, both in IIB and in M-theory, are a bit unusual and the predominance of strong couplings on both sides appear to dampen any hopes of doing concrete computations on either sides of the picture. Fortunately all is not lost and there is some light at the end of the tunnel when we look at the string coupling \eqref{salvatonmay}, because:
\bg\label{djinn1}
{g_s\over {\rm HH}_o} \equiv {{\rm F}_2 \sqrt{{\rm F}_1}\over a_{\bf dS}} ~ < ~ 1, \nd
which {\it is} possible because we can find a temporal domain wherein the requirement \eqref{djinn1} may be met\footnote{Unless it is the {\it static patch}, wherein all the three couplings: $g_b$, \eqref{salvatonmay2} and \eqref{djinn1}, are strong at late time. This is the familiar problem with the static patch that we also encountered earlier in section \ref{sec2.1}.}. 
This implies that the physics of the background \eqref{tamtam2} can only be inferred from the 
background \eqref{matagre}, and {\it not} from \eqref{villvaleur2}. Since we know that the M-theory background \eqref{matagre} can only be supported by time-dependent fluxes of the form 
${\bf G}_{\rm MNPQ}$, ${\bf G}_{{\rm MNP}a}$, ${\bf G}_{{\rm MN}ab}$ and ${\bf G}_{0ij{\rm M}}$ (keeping other components zero), duality chasing this to \eqref{tamtam}, shows that time-dependent fluxes are also necessary to support the de Sitter configuration \eqref{tamtam2}. 

The existence of time-dependences for the M-theory background \eqref{tamtam2}, similar to what we had for a background like \eqref{matagre}, means that all issues associated with the existence of an Wilsonian 
effective action with time-dependent frequencies re-appear now. Since the dual background 
\eqref{matagre} is realized as a Glauber-Sudarshan state, the background \eqref{tamtam2} should also be realized as a Glauber-Sudarshan state (and not as a {\it vacuum}). Such conclusion doesn't depend on the fact that \eqref{tamtam2}, and even \eqref{villvaleur2}, are at strong couplings. The Glauber-Sudarshan state for \eqref{tamtam2} is dual to the Glauber-Sudarshan state for \eqref{matagre}. 

There is however one advantage that we can have from the background \eqref{tamtam2} that is not present in the M-theory uplift \eqref{matagre}. This has to do with the {\it early} time physics. Recall that, from the expression of $g_s$ in \eqref{salvatonmay}, computational control only appears in the regime where 
$a(t) > 1$, {\it i.e.} at {\it late} time, making ${g_s\over {\rm HH}_o} < 1$. At early time, while the background 
\eqref{matagre} is unfortunately useless, the background \eqref{tamtam2} might become  useful. To see this
let us go to the following regime where:
\bg\label{pyncmay}
a^2_{\bf dS}(t) = {1\over \Lambda t^2}, ~~~~~ {g_s{\rm H}\over {\rm H}_o} = {1\over \sqrt{{\rm F}_2(t)}} ~< ~1, 
\nd
which is the {\it early} time in the flat-slicing coordinates of de Sitter. Since ${\rm F}_2(t) \to 1$ at 
late time ({\it i.e.} at $t \to 0$), we expect that it grows at early times ({\it i.e.} for $t < -{1\over \sqrt{\Lambda}}$), thus making ${g_s{\rm H}\over {\rm H}_o} < 1$. We are now in a surprisingly advantageous position: early time physics, which was so hard to decipher using the background \eqref{matagre}, can now be made amenable to study. There is however a minor technical disadvantage: the background \eqref{tamtam2} has a seven-dimensional internal space, so doesn't have the advantages\footnote{For example related to the computations of anomaly cancellations, flux quantizations, ${\bf X}_8$ polynomial {\it et cetera}. See however \cite{Cribiori}. Recently 
\cite{evaS} studied an example of a de Sitter space in M-theory with a seven-dimensional internal space. One of the ingredient used therein is the Casimir energy. It will be interesting to see if there is any connection between the background \eqref{tamtam2} and the constructions in \cite{Cribiori, evaS}.} related to the eight-dimensional internal space that we encountered in \eqref{matagre}. On the other hand, seven-dimensional manifolds have been well studied in the literature, so one might be able to extract useful informations from there. In any case, a new window of opportunity seems to have opened up for us related to the early time physics, so it is time to venture along that direction. More details on this will be presented elsewhere.

\subsection{Existence of EFT for a toy cosmology and exclusion principles \label{toyeft}}

Towards the end of section \ref{sec2.2} an interesting cosmology was studied in \eqref{morbiusn} where it
was pointed out that the existence of EFT implies the condition \eqref{senrem}, {\it i.e.} the temporal derivative of the dual IIA string coupling $g_s$ should be given by {\it positive} powers of $g_s$ to allow for the 
validity of an EFT description. This is a curious condition and one might want to figure out what exactly 
goes wrong if the aforementioned condition is not met. To this end, let us assume that the temporal derivative of $g_s$ is given in terms of a {\it negative} power of $g_s$. In other words we demand:
\bg\label{oniston}
{\partial \over \partial t}\left({g_s\over {\rm HH}_o}\right) = \left({g_s\over {\rm HH}_o}\right)^{-\vert\sigma\vert}, \nd
where $\vert\sigma\vert$ could be any number, integer, fractional or even irrational. 
The question that we want to ask is how the quantum terms, EOMs, anomaly cancellation, flux quantizations, and Bianchi 
identities behave once \eqref{oniston} is imposed. The answer, not surprisingly, lies in the behavior of the curvature and flux tensors once
\eqref{oniston} is taken into account. As one can easily show, the $g_s$ scalings for most (but not all) of the curvature tensors develop $\vert\sigma\vert$ dependences. Such $\vert\sigma\vert$ dependences affect the 
curvature two-forms, and changes it from \eqref{elfadogg2} to the following:

{\footnotesize
\bg\label{stowaway}
\mathbb{R}(x, y, w^a; g_s) &=& \sum_{l = 0}^\infty {\bf R}^{(l)}_{[{\rm MN}]}(x, y, w^a)\left({g_s\over {\rm HH}_o}\right)^{{l\over 3} +{\rm dom}\left({\gamma\over 3} - 2, - 2\vert\sigma\vert\right)} dy^{\rm M} \wedge dy^{\rm N}\\
& + & \sum_{l = 0}^\infty
\left[{\bf R}^{(l)}_{[ab]}\left({g_s\over {\rm HH}_o}\right)^{{l\over 3} +{\rm dom}\left({\gamma\over 3}, 2 - 2\vert\sigma\vert\right)} dw^{a} 
+ {\bf R}^{(l)}_{[{\rm M}b]}\left({g_s\over {\rm HH}_o}\right)^{{l\over 3} +{\rm dom}\left({\gamma\over 3} - 1, 1 - 2\vert\sigma\vert\right)}dy^{\rm M}\right]\wedge dw^{b}, \nonumber \nd}
where we have used the metric \eqref{makibhalu3} with $\gamma > 5$ arbitrary. Taking 
$\gamma = 6$, one can easily see that the dominant contributions come from the $\vert\sigma\vert$ 
factors in each of the curvature forms above. This changes the ${\bf X}_8$ polynomial from 
\eqref{thetrial} to the following: 
\bg\label{herman96}
{\bf X}_8(x, y, w^a; g_s) = \sum_{\{l_i\}}\widetilde{\bf X}_{(8, \vert\sigma\vert)}^{(l_1,...,l_4)}(x, y, w^a) 
\left({g_s\over {\rm HH}_o}\right)^{{1\over 3}(l_1 + l_2 + l_3 + l_4 + 6 - 24\vert\sigma\vert)}, \nd
with $l_i \in \mathbb{Z}$. One can work out the other ${\bf X}_8$ polynomials, but for our present computation \eqref{herman96} will suffice. Plugging this in the anomaly cancellation condition 
\eqref{lindmonaco} one may easily show that the dominant scalings of the internal G-flux components become:
\bg\label{gemmaton}
l_{\rm MN}^{{\rm P}a} = 1 - 4\vert\sigma\vert, ~~~~~ 
l_{\rm MN}^{ab} + l_{\rm MN}^{\rm PQ} = 2 - 8\vert\sigma\vert, \nd
where $({\rm M, N}) \in {\cal M}_4 \times {\cal M}_2, (a, b) \in {\mathbb{T}^2\over {\cal G}}$. One may also check that the dominant scaling for ${\bf G}_{{\rm MNP}a}$ satisfy the flux quantization conditions
\eqref{lazladi} and \eqref{rook} at every order implying that these components can behave as 
{\it global} fluxes. We also want the dominant scaling to be {\it positive} definite  so that the flux components remain finite at late time (See discussion towards the end of section \ref{mrain} and the footnote \ref{casin125}. The point is, although from \eqref{botsuga4} the bound is 
$l_{\rm MN}^{{\rm P}a} > -{1\over 3}$, the 
negative powers of $g_s$ will Borel sum to 
${\rm exp}\left(-{1\over g_s^{1/3}}\right)$ which vanishes at late time \cite{desitter2}).  This means $\vert\sigma\vert < {1\over 4}$ which, in terms of the ${1\over 3}$-modings
can be expressed as $\vert\sigma\vert < {0.75\over 3}$. In other words:
\bg\label{collette}
\vert\sigma\vert ~\le ~ {0.74999....9\over 3} ~ < ~ {1\over 4}. \nd
There are at least two problems associated with the choice \eqref{gemmaton}. 
{\Su One}, the $g_s$ scaling of ${\bf G}_{{\rm MNP}a}$
would imply that the corresponding three-form ${\bf C}_{{\rm NP}a}$ scales as $1 - 4\vert\sigma\vert$. This means ${\bf G}_{0{\rm NP}a}$ should at least scale\footnote{Recall that ${\bf G}_4 \ne d{\bf C}_3$, and the quantum terms play important role in designing the Bianchi identities.} 
as $l_{\rm 0N}^{{\rm P}a} = -5\vert\sigma\vert$. Unfortunately such scaling would imply that the flux components blow-up at late time and therefore cannot be put to zero, unless of course we make $\vert\sigma\vert = 0$ (see also footnote \ref{casin125}). On the other hand if we consider the EFT bounds from \eqref{perfum} and simply take the negative scaling without worrying too much about late time
behavior, then from \eqref{renaroy} we see that as long as $\vert\sigma\vert < {0.8\over 3}$, such negative scaling do not create problems in the quantum scaling \eqref{botsuga4}. Since we are already bounded by \eqref{collette}, the negative scaling for ${\bf G}_{{\rm 0NP}a}$ may still be tolerable modulo the late time issue. {\Su Two}, in the Schwinger-Dyson's equations for the space-time metric components ${\bf g}_{\mu\nu}$, the Einstein's tensor scales as 
$\left({g_s\over {\rm HH}_o}\right)^{-2 - 2\vert\sigma\vert + 2k/3}$, and the flux components ${\bf g}_{\mu\nu} {\bf G}_{{\rm MNP}a}{\bf G}^{{\rm MNP}a}$ 
scale as $\left({g_s\over {\rm HH}_o}\right)^{- 8\vert\sigma\vert + 2k'/3}$
where $k$ and $k'$ may be thought of as combinations of the modings of the paritcipating metric and flux components with $(k, k') \in \left({\mathbb{Z}\over 2}, {\mathbb{Z}\over 2}\right)$. The
two sides clearly cannot match at the lowest orders, which is consistent with the fact that ${\bf G}_{{\rm MNP}a}$ flux components do not contribute to the lowest order Einstein's equations. However at higher orders in $(k, k')$ with $k > k'$, we expect:
\bg\label{mca91} 
x_1 \equiv 2(k - k') = 6\Big(1 - 3\vert\sigma\vert\Big). \nd
Unfortunately not all values of $x_1 \in \mathbb{Z}$ can solve \eqref{mca91}.
Interestingly choosing $\vert\sigma\vert = {\mathbb{Z}\over 3}$ fixes 
$\vert\sigma\vert$ to be $\vert\sigma\vert = 0$, because the next possible values for $\vert\sigma\vert$, {\it i.e.} $\vert\sigma\vert \ge {1\over 3}$, break the $g_s$ and ${\rm M}_p$ hierarchies in the quantum series leading to the breakdown of the EFT \cite{necshort}. Alternatively, comparing this to \eqref{collette}, the only consistent solution is $\vert\sigma\vert = 0$. 
In general the number of solutions for \eqref{mca91}
may be listed as follows:
\bg\label{mcadosa1}
\left(x_1, \vert\sigma\vert\right) = \left\{ \left(0, {1\over 3}\right), \left(1, {5 \over 18}\right), \left(2, {2 \over 9}\right), 
\left(3, {1\over 6}\right), \left(4, {1\over 9}\right), \left(5, {1\over 18}\right), \left(6, 0\right)\right\}, \nd
out of which the first one is already eliminated, and the next three are eliminated on the ground that 
$\vert\sigma\vert < {1\over 7}$ (as we shall demonstrate soon in \eqref{fearstreet}). The only non-trivial ones appear to be 
$ \left(4, {1\over 9}\right)$, and  $\left(5, {1\over 18}\right)$ if we ignore the $(6, 0)$ choice. 
One may try to change this conclusion by making the following replacement: $\left({2k\over 3}, {2k'\over 3}\right) \to \left({2k\over 3} + {2k \vert\sigma\vert}, {2k'\over 3} +  
{2k'\vert\sigma\vert}\right)$ with $(k, k') \in \left({\mathbb{Z}\over 2}, {\mathbb{Z}\over 2}\right)$ as before. 
This will convert \eqref{mca91} to:
\bg\label{mcasin125}
{k - k' \over 3} - 1 = - 3 \vert\sigma\vert - (k - k') \vert\sigma\vert, \nd
where $k > k'$ because to the lowest orders the flux components ${\bf G}_{{\rm MNP}a}$ do not participate in the Einstein's equations (they would do so only if $\vert\sigma\vert = {1\over 3}$, thus not only contradicting \eqref{collette} but also creating issues with $g_s$ and ${\rm M}_p$ hierarchies). The RHS of \eqref{mcasin125} is always negative definite, whereas the LHS can be negative only for $k - k' < {3}$. For a random choice of $\vert\sigma\vert$, \eqref{mcasin125} has only a few solutions because:
\bg\label{jantalon}
x_2 \equiv 2(k - k') = 6\left({1 - 3\vert\sigma\vert\over 1 + 3\vert\sigma\vert}\right), \nd
with the LHS being multiples of integers, {\it i.e.} $2(k - k') \in {\mathbb{Z}}$ whereas the RHS is completely random. Using the above relation, and as before, one would be able to {count} the allowed choices for $\vert\sigma\vert$. For example, it is easy to see that the only allowed solutions for \eqref{jantalon} are:
\bg\label{mcadosa2}
\left(x_2, \vert\sigma\vert\right) = \left\{\left(0, {1\over 3}\right), \left(1, {5 \over 21}\right), \left(2, {1\over 6}\right), 
\left(3, {1\over 8}\right), \left(4, {1\over 15}\right), \left(5, {1\over 33}\right), \left(6, 0\right)\right\}, \nd
which should be compared to \eqref{mcadosa1}. Again the first three are eliminated on the ground that 
$\vert\sigma\vert < {1\over 7}$, so that now there are three non-trivial ones. Comparing \eqref{mcadosa1} and \eqref{mcadosa2} we see that out of the seven allowed solutions, the only over-lapping one is  
$\left(6, 0\right)$ which gives $\vert\sigma\vert = 0$. Of course, since we had used two different modings, the above analysis doesn't really fix a unique value for $\vert\sigma\vert$. 

Let us now see where does the bound $\vert\sigma\vert < {1\over 7}$ appear from our considerations. From the anomaly cancellation condition \eqref{gemmaton} we see that the sum of the two dominant scalings for 
${\bf G}_{{\rm MN}ab}$ and ${\bf G}_{\rm MNPQ}$ is $2 - 8\vert\sigma\vert$. Let us first fix the dominant couplings to be $l_{\rm MN}^{ab} = l_{\rm MN}^{\rm PQ} = 1 - 4\vert\sigma\vert$. This is clearly problematic for ${\bf G}_{{\rm MN}ab}$, because these flux components cannot appear at the lowest order in the space-time EOMs: the Einstein tensor scales as $\left({g_s\over {\rm HH}_o}\right)^{-2 - 2\vert\sigma\vert + 2k/3}$, and the flux contributions ${\bf g}_{\mu\nu}{\bf G}_{{\rm MN}ab}{\bf G}^{{\rm MN}ab}$ scale as
$\left({g_s\over {\rm HH}_o}\right)^{-2 - 8\vert\sigma\vert + 2k'/3}$. Clearly at lowest order, where 
$k = k' = 0$, the only way the $g_s$ scalings match is when $\vert\sigma\vert = 0$. At higher orders {\it i.e.} with $k > k'$, the condition becomes $k - k' = - 9\vert\sigma\vert$ which would lead to a contradiction unless $k = k'$ and $\vert\sigma\vert = 0$. The only other alternative is to assign $l_{\rm MN}^{ab} = 1 - \vert\sigma\vert$ and $l_{\rm MN}^{\rm PQ} = 1 - 7\vert\sigma\vert$ which, from the positivity of the dominant couplings, puts the bound:
\bg\label{fearstreet}
\vert\sigma\vert ~ \le ~ {0.428571.... - \epsilon\over 3} ~ < ~ {1\over 7}, \nd
where $\epsilon \to 0$. Such a bound helps us to eliminate some of the choices from \eqref{mcadosa1} and \eqref{mcadosa2}. 
Interestingly now, the condition for balancing the Einstein's tensor with the flux products 
${\bf g}_{\mu\nu} {\bf G}_{\rm MNPQ} {\bf G}^{\rm MNPQ}$ imposes the following condition:
\bg\label{mcadosa3}
x_3 \equiv 2(k - k') = 12\left(1 - 3\vert\sigma\vert\right), \nd
where the RHS is double the one from \eqref{mcadosa1}. This means that the number of allowed solutions would be slightly more than what we had earlier, because the extra factor of 2 would insert a new solution between two consecutive ones in \eqref{mcadosa1}, namely:
\bg\label{mcadosa4}
\left(x_3, \vert\sigma\vert\right) &= & \bigg\{\left(0, {1\over 3}\right), \left(1, {11 \over 36}\right), \left(2, {5 \over 18}\right), 
\left(3, {1\over 4}\right), \left(4, {2 \over 9}\right), \left(5, {7\over 36}\right), \left(6, {1\over 6}\right), \nonumber\\
&& \left(7, {5 \over 36}\right), \left(8, {1 \over 9}\right), \left(9, {1 \over 12}\right), 
\left(10, {1\over 18}\right), \left(11, {1 \over 36}\right), \left(12, {0}\right)\bigg\}, \nd
where we see that the first row is completely eliminated and in the second row, the overlapping ones with 
\eqref{mcadosa1} are expectedly $\left(8, {1\over 9}\right), \left(10, {1\over 18}\right)$ and 
$\left(12, 0\right)$ for $x_3 = 2x_1$. 

To see whether we can select fewer than three from the above set, we have to analyze different flux components. Let us then take the flux components ${\bf G}_{0{\rm MNP}}$, which would at least scale as
$\left({g_s\over {\rm HH}_o}\right)^{-8\vert\sigma\vert}$. Our earlier considerations of keeping {\it positive} dominant scaling should imply $\vert\sigma\vert = 0$, but since the flux components have one-legs along the temporal direction (whose metric blow-up at late time), we will ignore the late time issue for the time-being. Balancing the Einstein tensor with the flux products ${\bf g}_{\mu\nu} {\bf G}_{\rm 0MNP}{\bf G}^{\rm 0MNP}$ then gives us the following relation:
\bg\label{helenstan}
x_4 \equiv 2(k - k') = 6\left(2 - 7\vert\sigma\vert\right), \nd
where the RHS is now different from \eqref{mca91} as well as \eqref{mcadosa3}. The difference appears from our choice of the dominant scalings for the flux components ${\bf G}_{\rm MNPQ}$ and from there assigning the same dominant scalings for the three-form flux components ${\bf C}_{\rm MNP}$. Had we chosen ${\bf G}_{{\rm MNP}a}$ to fix the dominant scalings, we would have run into the problem with derivatives along ${\mathbb{T}^2\over {\cal G}}$ and the $\gamma$ factor from \eqref{makibhalu3}. As mentioned earlier, such subtleties appear because ${\bf G}_4$ is not exactly $d{\bf C}_3$, so the scalings for the three-form fluxes are a bit tricky to assign. Nevertheless, since derivatives along the internal six-directions do not introduce extra factors, our procedure to get the dominant scalings from ${\bf G}_{\rm MNPQ}$ and from there get \eqref{helenstan} may be consistently justified. After the dust settles, the number of solutions of \eqref{helenstan} may be listed as follows:  
\bg\label{mcadosa5}
\left(x_4, \vert\sigma\vert\right) &= & \bigg\{\left(0, {2 \over 7}\right), \left(1, {11 \over 42}\right), \left(2, {5 \over 21}\right), 
\left(3, {3\over 14}\right), \left(4, {4 \over 21}\right), \left(5, {1\over 6}\right), \left(6, {1\over 7}\right), \nonumber\\
&& \left(7, {5 \over 42}\right), \left(8, {2 \over 21}\right), \left(9, {1 \over 14}\right), 
\left(10, {1\over 21}\right), \left(11, {1 \over 42}\right), \left(12, {0}\right)\bigg\}, \nd
where as before, the first row is completely eliminated. In the second row, by comparing it to 
\eqref{mcadosa4} and \eqref{mcadosa1}, we see that the only overlapping choice is the one with 
$\vert\sigma\vert = 0$. Such a choice also makes the dominant scalings for the flux components 
${\bf G}_{\rm 0MNP}$ to be well-defined at late time. Note that contributions from other flux components cannot change this conclusion because they can at most have partial overlaps with the three set 
\eqref{mcadosa1}, \eqref{mcadosa3} and \eqref{mcadosa4}, but since the intersection of the set 
\eqref{mcadosa3} and \eqref{mcadosa4} is just one element, the only consistent choice appears to be
$\vert\sigma\vert = 0$. 

The above analysis pretty much kills any hope of keeping non-zero $\vert\sigma\vert$ in
\eqref{oniston}, although there is no constraint if the derivative of $g_s$ is given by {\it positive} powers of 
$g_s$, {\it i.e.} as $\left({g_s\over {\rm HH}_o}\right)^{+\vert\sigma\vert}$, as long as $\vert\sigma\vert \in
{\mathbb{Z}\over 3}$. We will still push on to see how the quantum terms behave once \eqref{oniston} is 
imposed. Of course imposing \eqref{oniston} would imply some changes to \eqref{botsuga4} as the $g_s$ scalings of the curvatures and the flux components themselves change. To see how the changes may be 
quantified, let us rearrange the sixty curvature products in \eqref{fahingsha5} in the following way:

{\footnotesize
\bg\label{fahingsha7}
\prod_{{\rm k} = 1}^{60} \left({\bf R}_{\rm A_k B_k C_k D_k}\right)^{l_{\rm k}} 
& = & \prod_{i = 1}^{9} {\bf R}_i(x, y, w^a; g_s) \nonumber\\
& = & \left({\bf R}_{a0b0}\right)^{l_1} \left({\bf R}_{abab}\right)^{l_2}\left({\bf R}_{pqab}\right)^{l_3}\left({\bf R}_{\alpha a b \beta}\right)^{l_4}\left({\bf R}_{abij}\right)^{l_5}\left({\bf R}_{\alpha\beta\alpha\beta}\right)^{l_6}
\left({\bf R}_{ijij}\right)^{l_7}\left({\bf R}_{ijmn}\right)^{l_8}\nonumber\\
&\times&
\left({\bf R}_{\alpha\beta mn}\right)^{l_9}
\left({\bf R}_{i\alpha j \beta}\right)^{l_{10}}\left({\bf R}_{0\alpha 0\beta}\right)^{l_{11}}
\left({\bf R}_{0m0n}\right)^{l_{12}}\left({\bf R}_{0i0j}\right)^{l_{13}}\left({\bf R}_{mnpq}\right)^{l_{14}}\left({\bf R}_{0mnp}\right)^{l_{15}}\nonumber\\
& \times & 
\left({\bf R}_{0\alpha\beta m}\right)^{l_{16}}\left({\bf R}_{0abm}\right)^{l_{17}}\left({\bf R}_{0ijm}\right)^{l_{18}}
\left({\bf R}_{0mn \alpha}\right)^{l_{19}}
\left({\bf R}_{0\alpha\beta\alpha}\right)^{l_{20}}
 \left({\bf R}_{0ab \alpha}\right)^{l_{21}}\left({\bf R}_{0ij\alpha}\right)^{l_{22}} \nonumber\\
&\times& \left({\bf R}_{mnp\alpha}\right)^{l_{23}}\left({\bf R}_{m\alpha ab}\right)^{l_{24}}
\left({\bf R}_{m\alpha\alpha\beta}\right)^{l_{25}}\left({\bf R}_{m\alpha ij}\right)^{l_{26}}
\left({\bf R}_{0m0\alpha}\right)^{l_{27}}
\left({\bf R}_{mnai}\right)^{l_{28}}\left({\bf R}_{\alpha\beta ai}\right)^{l_{29}}
\nonumber\\
&\times&
\left({\bf R}_{a0i0}\right)^{l_{30}}\left({\bf R}_{aijk}\right)^{l_{31}}
\left({\bf R}_{abai}\right)^{l_{32}}\left({\bf R}_{m\beta i \alpha}\right)^{l_{33}}\left({\bf R}_{abmi}\right)^{l_{34}}
\left({\bf R}_{\alpha 0i0}\right)^{l_{35}}\left({\bf R}_{\alpha ijk}\right)^{l_{36}}
\nonumber\\
&\times& \left({\bf R}_{ab i \alpha}\right)^{l_{37}}\left({\bf R}_{\alpha\beta i \alpha}\right)^{l_{38}}
\left({\bf R}_{mni\alpha}\right)^{l_{39}}\left({\bf R}_{mnpi}\right)^{l_{40}}\left({\bf R}_{0m0i}\right)^{l_{41}}
\left({\bf R}_{mijk}\right)^{l_{42}}\left({\bf R}_{ijk0}\right)^{l_{43}}\nonumber\\
&\times& 
\left({\bf R}_{\alpha\beta i 0}\right)^{l_{44}}\left({\bf R}_{ab0i}\right)^{l_{45}}
\left({\bf R}_{mni0}\right)^{l_{46}}\left({\bf R}_{maij}\right)^{l_{47}}\left({\bf R}_{ma\alpha\beta}\right)^{l_{48}}
\left({\bf R}_{maba}\right)^{l_{49}}\left({\bf R}_{mnpa}\right)^{l_{50}}\nonumber\\
&\times& 
\left({\bf R}_{a 0\alpha 0}\right)^{l_{51}}
\left({\bf R}_{mna\alpha}\right)^{l_{52}}
\left({\bf R}_{a\alpha ij}\right)^{l_{53}}\left({\bf R}_{a\alpha\alpha\beta}\right)^{l_{54}}
\left({\bf R}_{ab a\alpha}\right)^{l_{55}}\left({\bf R}_{m0a0}\right)^{l_{56}}
\left({\bf R}_{aij0}\right)^{l_{57}}\nonumber\\
&\times& 
\left({\bf R}_{a\alpha\beta 0}\right)^{l_{58}}
\left({\bf R}_{ab a0}\right)^{l_{59}}
\left({\bf R}_{mna0}\right)^{l_{60}}, \nd}
which incorporates some rearrangements from what we had earlier. These rearrangements are done to simply avoid cluttering of symbols in some of the computations that we are about to perform. Note that we will not be using the perturbative series \eqref{fahingsha6}, although a similar analysis could be performed using \eqref{fahingsha6}.

We will divide the sixty curvature products into a group of nine depending on how we assign the 
$g_s$ dependences once \eqref{oniston} is taken into account. There would also be contributions from the flux components, which are the products of forty terms from \eqref{fahingsha5}. If the $g_s$ scaling of the curvature products from \eqref{fahingsha7} is denoted by $\theta_{nl}({\bf R})$, and the corresponding 
$g_s$ scaling of the flux products by $\theta_{nl}({\bf G})$, then clearly $\theta_{nl}$ from \eqref{botsuga4}
is the sum $\theta_{nl}({\bf R}) + \theta_{nl}({\bf G}) \equiv \theta_{nl}$. As one would expect
$\theta_{nl}({\bf R})$ is not exactly the first line of \eqref{botsuga4}: there are five sub-divisions of the curvature products in \eqref{botsuga4}, whereas now we will have nine sub-divisions. It is not too hard to work out the precise $g_s$ scaling of the curvature products, which may be written as:

{\footnotesize
\bg\label{botsuga5}
\theta_{nl}({\bf R}) & = & {\rm dom}\left({\gamma\over 3} - {4\over 3}, {2\over 3} - 2\vert\sigma\vert\right)
\sum_{r = 1}^5 l_r+ {\rm dom}\left({2\gamma\over 3} - {4\over 3}, {2\over 3} - 2\vert\sigma\vert\right)
\sum_{k = 6}^{14} l_k + \left({2\over 3} - \vert\sigma\vert\right)\sum_{i = 15}^{22} l_i 
+ {2\over 3} \sum_{j = 23}^{27} l_j \nonumber\\
& + &  \left({2\over 3} + {\gamma \over 3}\right)\sum_{p = 28}^{32} l_p 
+ {5\over 3} \sum_{q = 33}^{42} l_q + \left({5\over 3} - \vert\sigma\vert\right) \sum_{s = 43}^{46} l_s
+  \left({\gamma\over 3} - {1\over 3}\right)\sum_{t = 47}^{56} l_t + 
\left({\gamma\over 3} - {1\over 3} - \vert\sigma\vert\right)\sum_{u = 57}^{60} l_u, \nonumber\\  \nd}
where we see how the $\vert\sigma\vert$ dependence influence the original scaling \eqref{botsuga4}, {\it i.e.} the first line of \eqref{botsuga4}. In a similar vein we expect the flux contributions to also get 
$\vert\sigma\vert$ dependences, and we shall discuss this briefly soon. Meanwhile, let us find out how the 
$\vert\sigma\vert$ dependence would influence other results. To this end, let us look at the lowest order Schwinger-Dyson's equations corresponding to the metric components ${\bf g}_{\mu\nu}$. As shown in 
\cite{coherbeta, desitter2}, these equations may be consistently solved by including non-perturbative instanton contributions. In terms of $g_s$ scalings, the Einstein tensor for the space-time components would scale as 
$\left({g_s\over {\rm HH}_o}\right)^{-2 - 2\vert\sigma\vert + 2k/3}$ as mentioned earlier, whereas the BBS instanton contributions to the energy-momentum tensor would scale as \cite{coherbeta, desitter2}:
\bg\label{shayfeliciti}
{\rm scale}\left(\mathbb{T}_{\mu\nu}\right) =  \sum_l lb_l \left({g_s\over {\rm HH}_o}\right)^{-2 + \theta_{nl} - {8\over 3} + {2k'\over 3}} {\rm exp}\left[-l\left({g_s\over {\rm HH}_o}\right)^{-2 + \theta_{nl} + {2k'\over 3}}\right], \nd
where $b_l$ is just a numerical constant and $l \in \mathbb{Z}$. The above scaling may be easily derived from the non-perturbative action that we discussed in section \ref{npeffects}, and one can easily see that for 
large values of $l$, the series \eqref{shayfeliciti} is convergent. Comparing the $g_s$ scaling with the one from the Einstein tensor, we get for $k = k' = 0$:
\bg\label{swoodley}
\theta_{nl} \equiv \theta_{nl}({\bf R}) + \theta_{nl}({\bf G}) = {8\over 3} - 2\vert\sigma\vert, \nd
where $\theta_{nl}({\bf R})$ is given in \eqref{botsuga5} and $\theta_{nl}({\bf G})$ may be worked out in a similar way. For the time being we will only take $\theta_{nl}({\bf R})$ into account. Taking the full expression 
\eqref{botsuga5} for $\theta_{nl}({\bf R})$ will complicate our analysis, so we can break this in small parts to see the consequences. We will also take $\gamma = 6$ from \eqref{makibhalu3}, so that the dominant parts would be the ones with $\vert\sigma\vert$ in them. Let us start with the first fourteen terms (the first five and the next nine terms all scale in the same way) from \eqref{botsuga5}, and define $n_1$ as $n_1 \equiv \sum_{r = 1}^{14} l_r \in \mathbb{Z}$ as $l_r \in \mathbb{Z}$. It is not too hard to see that this would lead to:
\bg\label{fejones}
\vert\sigma\vert = {n_1 - 4 \over 3(n_1 - 1)}, \nd
by balancing the $g_s$ scalings of the Einstein tensor with the non-perturbative BBS instantons. Since 
$n_1 \in \mathbb{Z}$ is an increasing function, we see that $\vert\sigma\vert$ would increase monotonically 
till it hits ${1\over 3}$ when $n_1 \to \infty$. On the other hand, it becomes negative for $n_1 < 4$ and blows-up for $n_1 = 1$. Since $\vert\sigma\vert \ge 0$, it is clear that $n_1 \ge 4$. We should also take into account the positivity bound $\vert\sigma\vert < {1\over 7}$ from \eqref{fearstreet}, which would restrict $n_1$ to lie between $4 \le n_1 \le 6$. Putting everything together, this implies that the only allowed solutions are:
\bg\label{joimasara}
\left(n_1, \vert\sigma\vert\right) = \left\{\left(4, 0\right), \left(5, {1\over 12}\right), \left(6, {2\over 15}\right)\right\}, \nd
with $\vert\sigma\vert = 0$ overlapping with all the three earlier sets \eqref{mcadosa1}, \eqref{mcadosa4}
and \eqref{mcadosa5}, although $\vert\sigma\vert = {1\over 12}$ overlaps with \eqref{mcadosa4} also. The
point however is this: introducing quantum effects $-$ which are absolutely essential here, compared to the classical contributions $-$ severely restricts the allowed choices of $\vert\sigma\vert$. In fact, as we expand 
to take other terms from \eqref{botsuga5} into account, we will see that only $\vert\sigma\vert = 0$ is consistently realized\footnote{Note that even if we had been {\it less} restrictive and taken $\vert\sigma\vert < {1\over 4}$ from \eqref{collette}, the allowed choices for $\vert\sigma\vert$ are again severely restricted from $4 \le n_1 \le 12$ and we get:
\bg\label{joimasara2}
\left(n_1, \vert\sigma\vert\right) = \left\{\left(4, 0\right), \left(5, {1\over 12}\right), \left(6, {2\over 15}\right),
\left(7, {1\over 6}\right), \left(8, {4\over 21}\right), \left(9, {5\over 24}\right), \left(10, {2\over 9}\right),
\left(11, {7\over 30}\right), \left(12, {8\over 33}\right)\right\}, \nonumber \nd
compared to {\it infinite} possible choices for $0 \le \vert\sigma\vert < {1\over 3}$ with 
$4 \le n_1 < \infty$. Of course, since the values of $\vert\sigma\vert$ lying between 
${1\over 7} < \vert\sigma\vert < {1\over 4}$ have numerous problems, the allowed cosmologies could only be a subset of \eqref{joimasara}, {\it i.e.} the first three terms of the above set.}.

Let us consider the next set of terms from \eqref{botsuga5}, namely the ones scaling as ${2\over 3} - 
\vert\sigma\vert$. There are eight such terms, and we will denote $n_2 \equiv \sum_{i = 15}^{22} l_i$ with 
$n_2 \in \mathbb{Z}$ because $l_i \in \mathbb{Z}$. We can again equate the $g_s$ scalings of the Einstein tensor with the non-perturbative BBS instantons from \eqref{shayfeliciti}. The result is:
\bg\label{jmasara}
\vert\sigma\vert = {2\over 3}\left({n_2 - 4\over n_2 - 2}\right), \nd
compared to \eqref{fejones} earlier. There is however already an issue with \eqref{jmasara}: for $n_2 \to
\infty$, we see that $\vert\sigma\vert \to {2\over 3}$ which is bigger than the EFT bound of ${1\over 3}$. In fact demanding the EFT bounds of ${1\over 3}$ or ${1\over 4}$, we see that $4 \le n_2 \le 5$, with:
\bg\label{miren2}
(n_2, \vert\sigma\vert) = \left\{\left(4, 0\right), \left(5, {2\over 9}\right)\right\}, \nd
thus only {\it two} possible values for $\vert\sigma\vert$: $\vert\sigma\vert = 0$ and $\vert\sigma\vert = {2\over 9}$. Of course the actual bound is $\vert\sigma\vert < {1\over 7}$ from \eqref{fearstreet}. Imposing this, the {\it only} possible solution turns out to be $\vert\sigma\vert = 0$ for $n_2 = 4$. 

The above analysis is an encouraging, albeit expected, sign that the quantum effects tend to tighten the choices for $\vert\sigma\vert$ severely leading to an almost unique value for $\vert\sigma\vert$ if we want to respect flux quantizations, anomaly cancellation, Bianchi identities and EOMs. We can ask if such restrictions continue to hold when we take the next five terms from \eqref{botsuga5} that scale as ${2\over 3}$. Defining $n_3 \equiv \sum_{j = 23}^{27} l_j$, and making the same aforementioned equivalence between the Einstein tensor and the non-perturbative terms \eqref{shayfeliciti}, we see that:
\bg\label{pascali}
\vert\sigma\vert = {4 - n_3\over 3}, \nd
compared to \eqref{fejones} and \eqref{jmasara} now. Clearly now $n_3 \le 4$, and one can easily see that the {\it only} possible solution satisfying any one of the bounds $\vert\sigma\vert < {1\over 3}$, or
$\vert\sigma\vert < {1\over 4}$ or $\vert\sigma\vert < {1\over 7}$, is $\vert\sigma\vert = 0$ with 
$n_3 = 4$. One may similarly play with the remaining set of quantum terms, or even combine the set of terms to cover everything from \eqref{botsuga5}, but the answer will not change: the quantum effects {\it only} allow $\vert\sigma\vert = 0$. On the other hand, and as alluded to earlier, flipping the sign of the exponent in \eqref{oniston}
leads to no restriction so long as $\sigma \in {\mathbb{Z}\over 3}$, the latter moding emanating from the 
${\mathbb{Z}\over 3}$ modings of the metric and the flux components in the construction.

What about the flux contributions? We have already dealt with some of the classical flux contributions wherein the choices for $\vert\sigma\vert$  were classified as
\eqref{mcadosa1}, \eqref{mcadosa2}, \eqref{mcadosa4} and \eqref{mcadosa5} with $\vert\sigma\vert = 0$ being the only overlapping one. The quantum, and especially the non-perturbative, contributions should tighten this a bit more as we shall see below. To avoid over-complicating the ensuing analysis, let us 
keep ${\bf G}_{0{\rm ABC}} = {\bf G}_{{\rm AB}ij} = {\bf G}_{0ija} = 0$, {\it i.e.} only allow the internal fluxes and the flux components ${\bf G}_{0ij{\rm M}}$, where $({\rm A, B}) \in {\bf R}^{2, 1} \times {\cal M}_4 \times {\cal M}_2 \times {\mathbb{T}^2\over {\cal G}}$ and $({\rm M, N}) \in {\cal M}_4 \times {\cal M}_2$ . How 
such a choice may be consistently realized should be clear from the flux EOMs that we discussed earlier. This means out of the forty flux components in \eqref{fahingsha5}, we will only take eleven components and, for computational efficiency, rearrange them in the following way:
\bg\label{sararent}
\prod_{p = 1}^{11} \left({\bf G}_{{\rm A}_p{\rm B}_p{\rm C}_p{\rm D}_p}\right)^{\hat{l}_p} = 
\left({\bf G}_{{\rm MN}ab}\right)^{l_1} \left({\bf G}_{{\rm MNP}a}\right)^{l_2}
\left({\bf G}_{{\rm MNPQ}}\right)^{l_3}\left({\bf G}_{0ij{\rm M}}\right)^{l_4}, \nd
where $(\hat{l}_i, l_i) \in (\mathbb{Z}, \mathbb{Z})$. As we saw earlier, the $\vert\sigma\vert$ dependence in \eqref{oniston} 
enters the $g_s$ scalings via the flux quantizations, anomaly cancellation and the Bianchi identities. This changes the $g_s$ scalings in \eqref{botsuga4}  to the following:
\bg\label{goldfinch}
\theta_{nl}({\bf G}) = {1\over 3}\left(1 - 3\vert\sigma\vert\right)\left(l_1 + 4 l_2 + 7l_3\right) + {l_4\over 3}, \nd
where the positivity of $\theta_{nl}({\bf G})$ implies $\vert\sigma\vert < {1\over 3}$, which is perfectly consistent with $\vert\sigma\vert < {1\over 7}$ imposed earlier. Let us then start with the $l_4$ term in 
\eqref{sararent}, which is related to the flux components ${\bf G}_{0ij{\rm M}}$. Note that these flux components are absolutely essential for the system to consistently fit with the dynamical M2-branes and the flux EOMs, as we saw earlier. Equating the $g_s$ scalings of the Einstein tensor with the $g_s$ scalings of the BBS instantons from \eqref{shayfeliciti}, we see that:
\bg\label{sararent2}
\vert\sigma\vert = {8 - l_4\over 6}, \nd
whose only consistent solution that keeps $\vert\sigma\vert < {1\over 7}$ is $\vert\sigma\vert = 0$ and 
$l_4 = 8$ (if we want to keep $\vert\sigma\vert < {1\over 3}$ or $\vert\sigma\vert < {1\over 4}$ one may also allow $(l_4, \vert\sigma\vert) = \left(7, {1\over 6}\right)$, but as mentioned earlier there are other issues if we 
take $\vert\sigma\vert \ge {1\over 7}$). This uniqueness\footnote{Interestingly, even at the classical level, where the $g_s$ scaling of the Einstein tensor is balanced by the $g_s$ scaling of 
${\bf g}_{\mu\nu} {\bf G}_{0ij{\rm M}} {\bf G}^{0ij{\rm M}}$, which in turn is 
$\left({g_s\over {\rm HH}_o}\right)^{-2 + 2k'/3}$, the condition becomes $\vert\sigma\vert = {l\over 6}$ with 
$l \equiv 2(k - k') \in \mathbb{Z}$. The only way $\vert\sigma\vert < {1\over 7}$, is with $l = 0$ giving us 
$\vert\sigma\vert = 0$.} should pretty much discourage us to proceed 
beyond $\vert\sigma\vert = 0$,  but we will still push on just to see what other flux components imply for 
$\vert\sigma\vert$.

To proceed, we will discuss individually the remaining three cases associated with the flux components 
${\bf G}_{{\rm MN}ab}, {\bf G}_{{\rm MNP}a}$ and ${\bf G}_{{\rm MNPQ}}$. Taking the first case with 
flux components ${\bf G}_{{\rm MN}ab}$ exponentiated by powers of $l_1$ with $l_1 \in \mathbb{Z}$, and equating the $g_s$ scalings of the Einstein tensor with the ones from the BBS instantons from \eqref{shayfeliciti}, we see that:
\bg\label{nikokid}
\vert\sigma\vert = {1\over 3}\left({l_1 - 8\over l_1 - 2}\right), \nd
where we restricted ourselves to the lowest orders with $k = k' = 0$ as before. Keeping $\vert\sigma\vert 
< {1\over 7}$ means that $8 \le l_1 \le 12$. This allows the following choices for $l_1$ and $\vert\sigma\vert$:
\bg\label{halan}
\left(l_1, \vert\sigma\vert\right) = \left\{\left(8, 0\right), \left(9, {1\over 21}\right), \left(10, {1\over 12}\right),
\left(11, {1\over 9}\right), \left(12, {2\over 15}\right)\right\}, \nd
which allows four more choices for $\vert\sigma\vert$ in addition to the one with $\vert\sigma\vert = 0$. This should be compared to the {\it classical} case where the $g_s$ scaling of the Einstein tensor was equated to the $g_s$ scaling of ${\bf g}_{\mu\nu} {\bf G}_{{\rm MN}ab}{\bf G}^{{\rm MN}ab}$. Since 
$l_{\rm MN}^{ab} = 1 - \vert\sigma\vert$, we see that classically {\it all} possible values for 
$\vert\sigma\vert < {1\over 7}$ is allowed. However non-perturbative effects restrict this severely to only {\it five}
possible choices out of which with further considerations, as we saw above, narrow them down to the unique value of $\vert\sigma\vert = 0$.    

For the second case we will take the flux components ${\bf G}_{{\rm MNP}a}$ that scale as 
$1 - 4\vert\sigma\vert$. We have already discussed in \eqref{mcadosa1} and \eqref{mcadosa2} the allowed choices for $\vert\sigma\vert$ when we equate the $g_s$ scaling of the Einstein tensor with the $g_s$ scaling coming from their classical energy-momentum tensor. Once non-perturbative effects from 
\eqref{shayfeliciti} are taken into account, we get the following value for $\vert\sigma\vert$:
\bg\label{fears1988}
\vert\sigma\vert = {2\over 3}\left({l_2 - 2\over 2l_2 - 1}\right), \nd
where $l_2 \in \mathbb{Z}$. As $l_2 \to \infty$, we see that $\vert\sigma\vert \to {1 \over 3}$ thus approaching  the EFT bound. Clearly then $2 \le l_2 \le 6$ if we want to keep $\vert\sigma\vert < {1\over 4}$ or $2 \le l_2 \le 3$ if we want to keep $\vert\sigma\vert < {1\over 7}$. The latter, which is in fact the more relevant one, gives us the following two choices:
\bg\label{fears1866}
(l_2, \vert\sigma\vert) = \left\{\left(2, 0\right), \left(3, {2\over 15}\right)\right\}, \nd
which again with further considerations narrow it down to $\vert\sigma\vert = 0$. The remaining flux components 
${\bf G}_{\rm MNPQ}$ cannot contribute at the lowest orders as may be seen from \eqref{goldfinch}. Therefore combining all the flux components together as in \eqref{sararent}, the only allowed choice becomes $\vert\sigma\vert = 0$. 

Our above discussions should convince the readers why $\vert\sigma\vert = 0$ is the only choice if we impose \eqref{oniston}. However before ending this section let us ask how such a choice fits in with the Wilsonian description of the system. Recall that, while deriving the Wilsonian effective action $-$ comprising of terms like \eqref{fahingsha5} plus the non-perturbative and the non-local terms $-$ we have taken an energy scale much below the massive KK modes or the massive stringy modes (which means that these states only propagate in the loops and integrating them out lead to the quantum series \eqref{fahingsha5} alongwith the non-perturbative and the non-local terms). What happens to this scenario if we take say
$\vert\sigma\vert = {2\over 15}$ instead of $\vert\sigma\vert = 0$? Note that $\vert\sigma\vert = {2\over 15}$ appears in \eqref{joimasara}, \eqref{halan}, and \eqref{fears1866}, so would be captured by at least a 
{\it subset} of the non-perturbative terms discussed above. Notwithstanding the issues that would plague 
the system with non-zero $\vert\sigma\vert$, the choice itself suffers from problems directly emanating from the
integration procedure. The key point, as always emphasized, is the existence of the Wilsonian effective action which appears from integrating out the heavy modes. Such integration does not care whether we impose \eqref{oniston} or not: the final answer would be a generic distribution of the {\it full} quantum series 
that includes the perturbative series \eqref{fahingsha5}, the non-perturbative terms, the non-local counter-terms, the topological terms, the fermionic terms, and additional terms from the branes and planes. As such, the choice 
$\vert\sigma\vert = {2\over 15}$ would lead to inconsistencies. Therefore the underlying Wilsonian procedure would {\it only} allow $\vert\sigma\vert = 0$ in \eqref{oniston}. (With positive powers of $g_s$, as mentioned earlier, solutions would exist as long as $\sigma \in {\mathbb{Z}\over 3}$.) In section 
\ref{bernardo}  we revisit the story again to show its surprising connection to the four-dimensional null energy condition from the IIB side.

\section{Properties of de Sitter space as a Glauber-Sudarshan state \label{GState}}

In the earlier sections we discussed how consistent time-{\it dependent} flux components may be switched on that would form a part of the Glauber-Sudarshan state. These fluxes satisfy EOMs which, when expressed using expectation values as in \eqref{pretroxie}, are basically the Schwinger-Dyson's equations as shown in great detail in \cite{coherbeta}. In defining these expectation values, we kept one subtlety under the rug (although this was only briefly dealt in \cite{coherbeta}). The subtlety has to do with our choice of the Glauber-Sudarshan state parametrized by $\bar{\sigma} \equiv (\bar{\alpha}, \bar{\beta})$ associated with the metric ($\bar{\alpha}$) and the flux ($\bar{\beta}$) components instead of using $\sigma \equiv (\alpha, \beta)$ to denote the metric and flux components respectively. The choice of $\bar{\sigma}$ instead of 
$\sigma$ is not just a change of notation\footnote{One should not confuse $\bar\sigma$ with the {\it complex conjugate} of $\sigma$. Instead $\bar\sigma$ should be viewed as a quantity different from $\sigma$, which will also be clear as we move along.}, but has a much deeper implications both for the  Glauber-Sudarshan state as well as the corresponding Agarwal-Tara state that governs the fluctuations over the Glauber-Sudarshan state. Note that the difference between $\sigma$ and $\bar{\sigma}$ is {\it not} related to the difference between the free and the interacting vacuum in M-theory, as both of them are associated with the interacting vacuum. The difference lies in some sense with the wave-function {\it renormalization} of the Glauber-Sudarshan state, as we shall elaborate in sub-section \ref{vac1}. This {\it renormalization} will have also have definite implications for the Agarwal-Tara state, as we shall see in sub-section \ref{vac2}.

Before moving ahead, let us clarify some standard properties of the Glauber-Sudarshan state for the free theory. Looking at this from the point-of-view of the configuration space, the wave-function of the vacuum state of the free theory, be it defined over some non-trivial solitonic configuration or otherwise,  may be expressed for a given ten-dimensional momentum ${\bf k}$ in the following way:
\bg\label{yiqiuma}
\Psi(f_{\bf k}) \equiv \sum_{g_{\bf k}} \delta\big(f_{\bf k} - g_{\bf k}\big) \Psi\left(g_{\bf k}\right), \nd
where $\left(f_{\bf k}, g_{\bf k}\right)$ denote the coordinates on the configuration space for a given spatial momentum ${\bf k}$, and the eleven-dimensional momentum $k \equiv ({\bf k}, k_0)$ is generically off-shell.
The delta function state in \eqref{yiqiuma}, when extended over all spatial momenta ${\bf k}$ leads to a {\it classical} field configuration in space. Such a field configuration is defined at a given instant of time 
$t$, and Feynman path-integral approach tells us how this evolves with time: every delta function state follows all possible paths in the configuration space {\it simultaneously} with temporal progression. All these paths, when extended over all the momenta ${\bf k}$, add up in such a way as to reproduce the standard Hamiltonian evolution of the vacuum state (wherein the probabilities do not change with time). 

Once we shift the vacuum state to create a coherent state in the free theory, a delta function state in the configuration space would still represent a classical field in ten-dimensional space (in fact the most {\it probable} one would represent the actual classical configuration that we are aiming for). The temporal evolution again spans all possible paths in the configuration space, but now the sum over the paths instead of giving us a stationary state, provides an oscillating state for any given momentum ${\bf k}$. Since now the probabilities would change with time, the most probably state would reproduce a temporally evolving classical configuration in both space and time\footnote{The fact that every delta function state in the configuration space $-$ contributing to the shifted vacuum state $-$ evolves simultaneously through all possible configurations (and in turn adds up) is important, otherwise the temporal evolution of the coherent state would not come about correctly. This simple underlying principle provides the necessary quantum {\it width} to the classical configuration.}. 

Such a clean picture is expected in free field theory and in quantum mechanics, but this is {\it not} what happens here because of the interactions. As we shall discuss below, interactions tend to influence the behavior of the Glauber-Sudarshan states drastically in at least three possible ways: {\Su one}, by changing the shifted-free-vacua for all spatial momenta ${\bf k}$ by a shifted-interacting-vacuum; 
{\Su two}, by changing the shapes and sizes of the states as they evolve temporally, and {\Su three}, by allowing them to reproduce the requisite classical configuration only in a finite temporal domain, beyond which the dominance of the strong type IIA coupling alters the dynamics. 
The latter case, and as mentioned earlier, is an issue because only when $g_s < 1$ we tend to have theoretical control over the perturbative, non-perturbative and non-local corrections in M-theory. For $g_s > 1$, it is not even clear if there is a simple way to introduce the quantum corrections using the curvature and the flux components.

\subsection{Temporal evolution of a Glauber-Sudarshan state \label{vac1}}

\noindent The distinction between free and interacting vacuum alluded to above is important. In the M-theory construction that we use here, there are no free vacua\footnote{M-theory doesn't have a coupling constant so there is no simple way to distinguish between the kinetic and the interaction parts of the Lagrangian (except via using ${\rm M}_p$ suppressions). However we can still define a {\it free} sector comprising of the kinetic terms for the metric and the fluxes (and similarly for the branes) by simply switching off the higher-order interactions. The vacua corresponding to these free sectors (for every momenta ${\bf k}$) will henceforth be described as the {\it free vacua}.}, so at the onset there is no meaning of the standard coherent states from quantum mechanics and free QFT. All we can talk about is an interacting vacuum that combines the free vacua using the full interacting Hamiltonian in M-theory. One can then displace the interacting vacuum using a {\it displacement operator}. In a free theory the displacement operator is typically written as $\mathbb{D}_0(\sigma, t)$, 
where\footnote{$\sigma$ or $\bar\sigma$ in this section should not be confused with $\sigma$ that we used in section \ref{toyeft} to denote the $g_s$ 
scalings.}  
$\sigma \equiv (\alpha, \beta)$ is related to coordinates on the configuration space associated with the metric and the G-flux configurations {\it i.e.} from
$\alpha$ and $\beta$ respectively. How exactly this is done is explained in equations (2.9), (2.17), (2.93) and (2.94) in the second reference of \cite{coherbeta}. What we now need is a bit more non-trivial: we want a displacement operator that displaces the interacting vacuum $\vert\Omega\rangle$ by an amount 
$\bar\sigma$ where $\bar\sigma$ differs from $\sigma$ by ${\cal O}(g_s)$ corrections. What exactly does this entail will be elaborated below. Our proposal for the displacement operator that shifts the interacting vacuum by an amount $\bar\sigma \equiv (\bar\alpha, \bar\beta)$ in the configuration space  may be 
written as \cite{coherbeta}:
\bg\label{amchase}
{\mathbb{D}(\bar\sigma, t) = \lim_{{\rm T} \to \infty(1 - i\epsilon)} \mathbb{D}_0(\bar\sigma, t) 
~{\rm exp} \left(i \int_{-{\rm T}}^t d^{11} x~{\bf H}_{\rm int}\right)} \nd
where $\mathbb{D}(\sigma, t)$ is {\it non-unitary} and ${\bf H}_{\rm int}$ is the full interacting Hamiltonian that involves the perturbative series 
\eqref{fahingsha5} as well as the non-perturbative and non-local terms (including the topological and the brane/plane terms). Such a choice of ${\bf H}_{\rm int}$ makes the analysis very complicated but, as we show below, we can still make precise predictions using \eqref{amchase}. Before proceeding however we should clear up a few questions that may arise regarding the choice of ${\rm T}$ and ${\bf H}_{\rm int}$. 
{\Su One}, how are we allowed to take 
${\rm T} \to \infty$ (in a slightly imaginary direction) when the dynamics is bounded by the temporal domain governed by the TCC?  The answer lies in the difference between the background solitonic configuration and the Glauber-Sudarshan state. The temporal coordinate ${\rm T}$ that is taken to infinity, albeit in a slightly imaginary direction, is associated with the solitonic vacuum. This is a supersymmetric Minkowski state\footnote{The reasons for choosing a {\it supersymmetric} solitonic vacuum have already been outlined in \cite{desitter2} 
and \cite{coherbeta}. The readers may find all the details there.}   
and there is no constraint on the temporal coordinate. On the other hand, once we construct the Glauber-Sudarshan state, the dynamics of this state remains well under control only in the temporal bound dictated by the TCC (as we saw in section \ref{sec2.1}). This temporal bound is controlled by $t$ in \eqref{amchase}. {\Su Two}, how are the flux EOMs etc., that we studied in the earlier sections, precisely related to the shifted interacting vacuum and the interacting Hamiltonian 
${\bf H}_{\rm int}$? We will answer this question in two steps. First, let us ignore the distinction between 
$\sigma$ and $\bar\sigma$. In that case, the Schwinger-Dyson's equations for the flux components may be written as\cite{coherbeta}:
\bg\label{407tegra}
\left\langle {\delta {\bf S}_{\rm tot} \over \delta {\rm C}^{\rm ABC}}\right\rangle_\sigma 
& = & {\delta {\bf S}^{(\sigma)}_{\rm tot} \over \delta \langle{\rm C}^{\rm ABC}\rangle_\sigma} +
\sum_{\sigma' \ne \sigma} \left\langle {\delta {\bf S}_{\rm tot} \over \delta {\rm C}^{\rm ABC}}
\right\rangle_{(\sigma'\vert\sigma)} \nonumber\\ 
& = & 
\left\langle {\delta {\bf S}_{\rm ghost} \over \delta {\rm C}^{\rm ABC}}\right\rangle_\sigma - 
\left\langle {\delta \over \delta {\rm C}^{\rm ABC}} {\rm log}\left(\mathbb{D}^\dagger(\sigma, t) 
\mathbb{D}(\sigma, t)\right)\right\rangle_\sigma, \nd 
with a similar expression for $\left\langle {\delta {\bf S}_{\rm tot} \over \delta {g}^{\rm AB}}
\right\rangle_\sigma$. The other functions appearing  in \eqref{407tegra} are defined as follows. The total action is ${\bf S}_{\rm tot} \equiv {\bf S}_{\rm kin} + {\bf S}_{\rm int}$, where the first part is the kinetic part and the second part involves all the interactions that we discussed above as ${\bf H}_{\rm int}$, and earlier as the 
quantum series \eqref{fahingsha5}, plus the non-perturbative, non-local and other contributions. The expectation values of metric and the flux components are related to the background metric and fluxes, {\it i.e.} $\langle {\rm C}_{\rm ABC}\rangle_\sigma \equiv {\bf C}_{\rm ABC}$ and 
$\langle {g}_{\rm AB}\rangle_\sigma \equiv {\bf g}_{\rm AB}$ appearing in our analysis of metric and fluxes in sections \ref{sec2.1}, \ref{sec2.2} and \ref{sec2.2.5}. The action ${\bf S}^{(\sigma)}_{\rm tot} \equiv 
{\bf S}_{\rm tot}\left(\langle {\rm C}_{\rm ABC}\rangle_\sigma, \langle {g}_{\rm AB}\rangle_\sigma\right) = 
{\bf S}_{\rm tot}({\bf C}_{\rm ABC}, {\bf g}_{\rm AB})$ which is expressed using the warped metric and the flux components\footnote{Note that we are writing the total action ${\bf S}_{\rm tot}$ in terms of the metric and the three-form fields instead of the curvatures and the four-form flux components. Since both metric and the three-form fields are not gauge invariant quantities, the action should involve Faddeev-Popov ghosts. These are precisely the ghosts that appear in \eqref{407tegra}.}. The bracket $(\sigma'\vert\sigma)$ involve intermediate states $\vert\sigma'\rangle$ which are summed over. In fact precisely because of these intermediate states, the expectation value of the curvature tensors have additional contributions as alluded to earlier in \eqref{vijuan2}. Finally, since 
$\mathbb{D}(\sigma, t)$ is non-unitary, the logarithmic term in the second line of \eqref{407tegra} does not vanish. 

The inclusion of propagating ghosts in the Schwinger-Dyson's equations in \eqref{407tegra}, while necessary, makes the analysis rather involved. However we can still make sense of the system if we note that the equations \eqref{407tegra}, including the ones with the metric, may be split into the following parts:
\bg\label{drakul}
&&{\delta {\bf S}^{(\sigma)}_{\rm tot} \over \delta \langle{\rm C}^{\rm ABC}\rangle_\sigma} = 0, ~~~~
{\delta {\bf S}^{(\sigma)}_{\rm tot} \over \delta \langle{g}^{\rm AB}\rangle_\sigma} = 0, 
\nonumber\\
&& \sum_{\sigma' \ne \sigma} \left\langle {\delta {\bf S}_{\rm tot} \over \delta {g}^{\rm AB}}
\right\rangle_{(\sigma'\vert\sigma)} = 
\left\langle {\delta {\bf S}_{\rm ghost} \over \delta {g}^{\rm AB}}\right\rangle_\sigma - 
\left\langle {\delta \over \delta {g}^{\rm AB}} {\rm log}\left(\mathbb{D}^\dagger(\sigma, t) 
\mathbb{D}(\sigma, t)\right)\right\rangle_\sigma\nonumber\\
&& \sum_{\sigma' \ne \sigma} \left\langle {\delta {\bf S}_{\rm tot} \over \delta {\rm C}^{\rm ABC}}
\right\rangle_{(\sigma'\vert\sigma)} = 
\left\langle {\delta {\bf S}_{\rm ghost} \over \delta {\rm C}^{\rm ABC}}\right\rangle_\sigma - 
\left\langle {\delta \over \delta {\rm C}^{\rm ABC}} {\rm log}\left(\mathbb{D}^\dagger(\sigma, t) 
\mathbb{D}(\sigma, t)\right)\right\rangle_\sigma, \nd
which differs slightly from the usual Schwinger-Dyson's equations (see the first reference in \cite{desitter2} for a more {\it standard} way of presenting the EOMs). Looking at \eqref{drakul}, one may easily note that the first equation is exactly the flux EOMs \eqref{mcacrisis} (including the non-perturbative contributions) that we studied in section \ref{sec2.2.5}. In a similar vein, from the second equation in \eqref{drakul}, we can recover the Einstein's equations for the metric components. 

The splitting of the Schwinger-Dyson's equations into the standard EOMs (for metric and flux components) and EOMs for the propagating ghosts is a useful way to show the consistencies of the Glauber-Sudarshan state because the first two equations in \eqref{drakul} capture the behavior of the most {\it probable} state in $\vert\sigma, t\rangle \equiv \mathbb{D}(\sigma, t) \vert\Omega\rangle$. However it makes the equations for ghosts harder to track because all the non-trivialities of say \eqref{407tegra}  (and the corresponding ones for the metric components) are transferred to the third and the fourth equations of \eqref{drakul}. Despite this the splitting in \eqref{drakul} is not without merit: we do know that the most probable state in the Glauber-Sudarshan wave-function is the closest one to the {\it classical} background and therefore should satisfy the first two equations in \eqref{drakul}. The ghosts may be dealt separately, but will not effect the outcome for the most probable state\footnote{Even in the first two equations of \eqref{drakul} both ghosts and gauge fixing may be inserted in. Due to the abelian nature one can choose them in a way that the ghosts decouple (much like what happens in QED). The gauge fixing will act appropriately to give us the correct propagators for the metric and the flux components. In any case we expect ${\bf S}^{(\sigma)}_{\rm tot}$ to be the effective action that provides the correct EOMs for $\langle g_{\rm AB}\rangle_\sigma$ and $\langle {\rm C}_{\rm ABC}\rangle_\sigma$. This also justifies how and why an alternative choice:
\bg\label{thiasin}
{\delta {\bf S}^{(\sigma)}_{\rm tot} \over \delta \langle{\rm C}^{\rm ABC}\rangle_\sigma}  =  
\left\langle {\delta {\bf S}_{\rm ghost} \over \delta {\rm C}^{\rm ABC}}\right\rangle_\sigma - 
\left\langle {\delta \over \delta {\rm C}^{\rm ABC}} {\rm log}\left(\mathbb{D}^\dagger(\sigma, t) 
\mathbb{D}(\sigma, t)\right)\right\rangle_\sigma - 
\sum_{\sigma' \ne \sigma} \left\langle {\delta {\bf S}_{\rm tot} \over \delta {\rm C}^{\rm ABC}}
\right\rangle_{(\sigma'\vert\sigma)}, \nonumber \nd
may not be the right way in which the background EOMs should appear. For example we should be able to reproduce \eqref{mcacrisis} from the Schwinger-Dyson's equations and clearly the above equation fails unless the splitting is as in \eqref{drakul}.\label{casin777}}. There is however a deeper question lurking behind: how do the choice 
$\sigma \to \bar\sigma$ effect the set of equations in \eqref{drakul}? The answer lies in the wave-function of the shifted interacting vacuum that takes the following form\footnote{The integration domain suggests that we can go to arbitrarily short distances. We can also rewrite the wave-function simply in terms of the allowed low energy modes, {\it i.e.} the long distance wave-lengths by appropriately renormalizing it. The fact that both leads to the same answer is clearly the consequence of the application of the Wilsonian viewpoint which we have been emphasizing all along. More importantly, the {\it absence} of any trans-Planckian effects (as shown in \cite{coherbeta} and also discussed later) is probably a stronger reason why this works so well.}:
\bg\label{csapphire}
\Psi^{(\bar{\sigma})}_{\Omega}\left(g_{\mu\nu}, t\right)  =  {\rm exp}\left[\int_{-\infty}^{+\infty} d^{10}{\bf k} ~{\rm log}
\left(\Psi^{(\bar{\sigma})}_{\bf k}\left(\widetilde{g}_{\mu\nu}({\bf k}), t\right)\right)\right], \nd
where the RHS is a combination of the shifted {\it free} vacua for every momenta ${\bf k}$ at any given instant $t$. The other parameters appearing in \eqref{csapphire} are the metric {\it coordinate} $g_{\mu\nu}$ 
in the configuration space and its corresponding Fourier transform $\widetilde{g}_{\mu\nu}$. We expect 
$\langle g_{\mu\nu}\rangle_{\bar\sigma} = {\bf g}_{\mu\nu}$, as we shall show below. 

The wave-function \eqref{csapphire} is a simplified version of a more complicated wave-function for the Glauber-Sudarshan state that takes into account the other metric and flux components. However, even in the simplified case of \eqref{csapphire}, one may see that the analytical advantage stems from 
\eqref{amchase} although the shift is by $\bar\sigma$ in the configuration space. This shift now effects the expectation value of the metric {\it operator} in the following way\footnote{With some abuse of notation, we 
express both the coordinate on the configuration space and the metric operator by the same symbol 
$g_{\mu\nu}$. The bold-faced metric component ${\bf g}_{\mu\nu}$ is reserved for the warped-metric 
in space-time. Which is which should be clear from the context.}:
 
 {\footnotesize
 \bg\label{kritins}
&& \langle {g}_{\mu\nu}(x, y, z)\rangle_{\bar\alpha} =   {\eta_{\mu\nu} \over \left(\Lambda\vert t\vert^2 
{\rm H}^2(y)\right)^{4/3}} = 
{\eta_{\mu\nu} \over {\rm H}^{8/3}(y)}\\ 
&&~~~~ +  
{\bf Re}\left[\int {d^{10} {\bf k}\over 2\omega^{(\psi)}_{\bf k}} ~
\left(\bar{\alpha}^{(\psi)}_{\mu\nu}({\bf k}, t) + {\cal O}\left({\bar{\alpha}^{(\psi)}_{\mu\nu}
\bar{\alpha}^{(\psi)\mu\nu}\over {\rm M}_p^{|c|}}\right) + 
{\cal O}\left({g_s^{|a|}\over {\rm M}_p^b}\right) 
+ {\cal O}\left[{\rm exp}\left(-{1\over g_s^{1/3}}\right)\right]\right)\psi_{\bf k}({\bf x}, y, z)\right], \nonumber \nd}
where we restricted to $\bar\alpha$ from $\bar\sigma \equiv (\bar\alpha, \bar\beta)$ with 
$\bar\alpha_{\rm AB}({\bf k}, t)$ and $\bar\beta_{\rm ABCD}({\bf k}, t)$; and $({\rm A, B}) \in {\bf R}^{2, 1} \times {\cal M}_4 \times {\cal M}_2 \times {\mathbb{T}^2\over {\cal G}}, y \in {\cal M}_4 \times {\cal M}_2$, correspond to the metric and flux components in the Fourier space; and $\psi_{\bf k}({\bf x}, y, z)$ is the spatial wave-function for a given momentum ${\bf k}$. The first term inside the momentum integrals in \eqref{kritins} comes from the quadratic piece of the action including the pole structure, and the ${\cal O}(g_s^{|a|})$ corrections appear from 
${\bf H}_{\rm int}$. For the second term onwards, the pole remains the same, but now the Fourier transform involve complicated functions
which when integrated over all momenta ${\bf k}$, provide the higher order $\bar\alpha$ factors as well as the $g_s$ corrections. In a similar way, the non-perturbative and the non-local counter-terms provide the exponential pieces for all the metric components. In section \ref{vac2} we will elaborate further on this in a more generic setting. For the present case, since we restrict ourselves to the space-time metric only,  
$\bar\alpha_{\rm AB}({\bf k}, t) \to \bar\alpha_{\mu\nu}({\bf k}, t)$. However, and here is the crucial 
observation, $\bar\alpha_{\mu\nu}({\bf k}, t)$ components are {\it not} the Fourier transforms in eqns. (2.9) and (2.17) in the second reference of \cite{coherbeta} (in a similar vein, $\bar\beta_{\rm ABCD}({\bf k}, t)$ 
components are not the Fourier transforms in eqns. (2.93) and (2.94) in the first reference of 
\cite{coherbeta}). We can then make the following 
identifications:

{\footnotesize
\bg\label{censusmay}
&&\bar\alpha_{\rm AB}({\bf k}, t) + {\cal O}\left({\bar{\alpha}_{\rm AB}
\bar{\alpha}^{{\rm AB}}\over {\rm M}_p^{|c|}}\right) + {\cal O}\left({g_s^{|a|}\over {\rm M}_p^b}\right) 
+ {\cal O}\left[{\rm exp}\left(-{1\over g_s^{1/3}}\right)\right] = \alpha_{\rm AB}({\bf k}, t) 
\equiv \widetilde{\bf g}_{\rm AB}({\bf k}, t) \\
&&\bar\beta_{\rm ABCD}({\bf k}, t) + {\cal O}\left({\bar{\beta}_{\rm ABCD}
\bar{\beta}^{{\rm ABCD}}\over {\rm M}_p^{|c|}}\right) + {\cal O}\left({g_s^{|a|}\over {\rm M}_p^b}\right) 
+ {\cal O}\left[{\rm exp}\left(-{1\over g_s^{1/3}}\right)\right] = \beta_{\rm ABCD}({\bf k}, t) 
\equiv \widetilde{\bf G}_{\rm ABCD}({\bf k}, t), \nonumber \nd}
where the RHS of both the equations are related to (2.9), (2.17) and (2.93), (2.94) respectively in the second reference of \cite{coherbeta} at least for the flat-slicing. In other words $\alpha_{\rm AB}({\bf k}, t) \equiv \alpha_{\rm AB}({\bf k}, \omega_{\bf k}){\rm exp}\left(-i\omega_{\bf k}t\right)$ and similarly for $\beta_{\rm ABCD}({\bf k}, t)$, but the temporal dependences of $(\bar\alpha_{\rm AB}({\bf k}, t), \bar\beta_{\rm ABCD}({\bf k}, t))$ are not so simple. We can easily extend the definitions of 
$\sigma \equiv (\alpha, \beta)$ to incorporate any embeddings of de Sitter studied in section \ref{sec2.1}. 
Note that the above identifications justify why ${g_s\over {\rm HH}_o} < 1$, otherwise we won't be able to 
control the perturbative series, or eliminate the non-perturbative ones. The latter easily decouples when 
${g_s\over {\rm HH}_o} \to 0$, as one would expect. In section \ref{vac2} we will revisit this computation to explain how various terms on the LHS of 
\eqref{censusmay} could arise. 

The subtlety with \eqref{censusmay} is that the relation is {\it not} linear, with the LHS being typically a polynomial in $\bar\sigma$. To express $\bar\sigma$ in terms of $\sigma$ will then require us to invert a polynomial equation. Fortunately the saving grace is the inverse powers of ${\rm M}_p^{|c|}$, so that for 
large ${\rm M}_p$ we can perturbatively express $\bar\sigma$ in powers of $\sigma$ and $g_s$. For 
${\rm M}_p \to \infty$ we can clearly decouple these terms and approximate \eqref{censusmay} to a linear equation, much like what we did in \cite{coherbeta}.

Despite the success in reproducing the precise de Sitter space-time there is something unusual about the Glauber-Sudarshan wave-function \eqref{csapphire}: it {\it does not} evolve as a coherent state! Evolution of the state may be understood as a simultaneous evolution of the delta function states in the wave-function
\eqref{yiqiuma}, where $\Psi(g_{\bf k})$ should now be constructed from $\mathbb{D}_0(\bar\sigma, 0)$. This should also be equivalent to the evolution expressed using Schr\"odinger formulation. In the language of the latter, 
this may be 
easily seen by evolving the state $\mathbb{D}_0(\bar\sigma, 0)\vert 0\rangle$ from $t = -{1\over \sqrt{\Lambda}}$, where $\Lambda$ is the four-dimensional cosmological constant (in IIB), to some temporal interval\footnote{We will henceforth be in the flat-slicing of de Sitter unless mentioned otherwise.} $t$:
\bg\label{momisto}
\vert\bar\sigma, t\rangle  = {\rm exp}\left[-i\int_{-1/\sqrt{\Lambda}}^t d^{11} x\Big({\bf H}_0 + 
{\bf H}_{\rm int}\Big)\right] \mathbb{D}_0(\bar\sigma, 0)\vert 0 \rangle, \nd
where ${\bf H}_0$ and ${\bf H}_{\rm int}$ are the free and the interacting parts of the M-theory Hamiltonian (in the operator formalism), the latter being the same interacting Hamiltonian appearing in \eqref{amchase}.  In fact precisely because of its presence, the evolution of the state does not correspond to the standard evolution of a coherent state. This is one of the key difference between our Glauber-Sudarshan state and the standard coherent state in quantum mechanics or quantum field theory. In the language of \eqref{momisto}, the difference between the two wave-functions may be expressed as 
$-\delta\Psi_\Omega^{(\sigma - \bar\sigma)}$, where:
\bg\label{privsoc}
\delta\Psi_\Omega^{(\sigma - \bar\sigma)} & = & \Psi^{(\sigma)}_\Omega - \Psi^{(\bar\sigma)}_\Omega\\
& = &{\rm exp}\left\{\int_{-\infty}^{+\infty} d^{10}{\bf k} ~{\rm log}\langle\widetilde{g}_{\mu\nu}({\bf k}) \vert 
{\rm exp}\left[-i\int_{-1/\sqrt{\Lambda}}^t d^{11} x~{\bf H}_0\right] \mathbb{D}_0(\sigma, 0)\vert 0 \rangle
\right\}\nonumber\\
&-&{\rm exp}\left\{\int_{-\infty}^{+\infty} d^{10}{\bf k} ~{\rm log}\langle\widetilde{g}_{\mu\nu}({\bf k}) \vert 
{\rm exp}\left[-i\int_{-1/\sqrt{\Lambda}}^t d^{11} x\Big({\bf H}_0 + 
{\bf H}_{\rm int}\Big)\right] \mathbb{D}_0(\bar\sigma, 0)\vert 0 \rangle\right\}, \nonumber \nd
which confirms what we mentioned earlier, namely, that the difference in the wave-function comes from the difference in $\bar\sigma$ and $\sigma$ and not due to the difference in interacting and free vacuum. Since 
$\bar\sigma$ and $\sigma$ differs perturbatively and non-perturbatively by powers of $g_s$ and 
${\rm M}_p$, the difference in the wave-functions could be attributed to {\it wave-function renormalization} in the Wilsonian sense. This is what one would have expected in an effective field theory description and it is comforting to see that such a picture in borne out of our construction. There is yet another upside to 
this: the temporal evolution of the state as in \eqref{momisto} implies that the number of gravitons and flux quanta (although we will ignore the latter to simplify the ensuing analysis), {\it changes} as the Glauber-Sudarshan state evolves in the temporal domain $-{1\over \sqrt{\Lambda}} \le t < 0$ in the flat-slicing. This change in the number of gravitons may easily be quantified, in the limit ${\rm M}_p \to \infty$, as:

{\footnotesize
\bg\label{ramchela}
{\rm N}^{(\psi)}(t) = \int_{-{\rm M}_p}^{+{\rm M}_p} d^{10}{\bf k} \left\vert \alpha^{(\psi)}_{\mu\nu}({\bf k}, \omega_{\bf k})
{\rm exp}\left(-i\omega^{(\psi)}_{\bf k}t\right) + {\cal O}\left({g_s^{|c|}\over {\rm M}_p^d}\right)
+  {\cal O}\left[{\rm exp}\left(-{1\over g_s^{1/3}}\right)\right] \right\vert^2, \nd}
where the super-script $\psi$ represents the spatial wave-function $\psi_{\bf k}({\bf x}, y, z)$ that appeared in \eqref{kritins},  $\omega_{\bf k}^{(\psi)}$ is the frequency associated with the mode ${\bf k}$, and $\alpha_{\mu\nu}^{(\psi)}$ is taken from the definition of $\sigma \equiv (\alpha, \beta)$ 
used above. As expected, the temporal dependence of the first term in \eqref{ramchela} appears from ${\bf H}_0$ part in the first term of \eqref{privsoc}.  The relative signs between the perturbative and non-perturbative corrections are governed by \eqref{censusmay}, and in fact the quantity that actually appears in \eqref{ramchela} is $\vert{\bar\alpha}_{\mu\nu}^{(\psi)}\vert^2$ from the definition of $\bar\sigma \equiv (\bar\alpha, \bar\beta)$. Note that the temporal evolution of $\bar\alpha_{\mu\nu}^{(\psi)}$ is {\it not} of the form 
${\rm exp}\left(-i\omega_{\bf k}^{(\psi)}t\right)$ $-$ this being the key difference between the Glauber-Sudarshan state and the usual coherent state as mentioned earlier $-$ because of the evolution \eqref{momisto} and therefore 
cannot be eliminated by the modding process in \eqref{ramchela}. This means ${\rm N}^{(\psi)}(t)$ is indeed a function that depends on time (for the standard coherent state the number of gravitons, ${\rm N}^{(\psi)}$, is fixed). Therefore as our Glauber-Sudarshan state evolves with respect to time, it gives out graviton and flux quanta, but the expectation values of the graviton and the flux operators reproduce precisely the de Sitter background as long as we are in the temporal domain governed by the TCC \cite{tcc}. 
For $t < -{1\over \sqrt{\Lambda}}$, ${g_s\over {\rm HH}_o} > 1$, and the series in \eqref{censusmay} cannot be summed up properly to reproduce the de Sitter space that we want. Within the temporal domain 
$-{1\over \sqrt{\Lambda}} \le t < 0$, the interacting part of the Hamiltonian ${\bf H}_{\rm int}$ is not only responsible for reproducing the background \eqref{makibhalu3} (or the simpler case of \eqref{evader}), but also for the change in the number of graviton and flux quanta. Note that this change in the number of quanta
is expected even at the {\it zeroth} order in $g_s$, and only for vanishing $g_s$ the standard picture of the coherent state is reproduced. For ${g_s\over {\rm HH}_o} < 1$, we can easily compute the precise change in, say, the number of gravitons as our Glauber-Sudarshan state evolves in the temporal domain dictated by TCC:
\bg\label{mauritius}
\Delta{\rm N}^{(\psi)}(t) = {\rm N}^{(\psi)}(-1/\sqrt{\Lambda}) - {\rm N}^{(\psi)}(0), \nd
where one can use \eqref{ramchela} to compute the individual pieces. The sign of $\Delta{\rm N}^{(\psi)}(t)$
is important: positive sign means that the Glauber-Sudarshan state loses  gravitons as it evolves towards late time, whereas negative sign means the addition of external quanta of gravitons  in the evolution process. Recently in \cite{dvali2} it was argued, mostly from four-dimensional perspective, that the Glauber-Sudarshan state would lose gravitons as it evolves. To see what happens for our case, let us do a small computation to the zeroth order in $g_s$ where at least $|a| = 0$ in \eqref{censusmay} (and ${\rm M}_p \to \infty$ so as to keep \eqref{censusmay} linear in $\bar\alpha$). We will also concentrate on a single mode ${\bf k}$ and assume that the relative sign for $|a| = 0$ remains positive. Plugging 
\eqref{ramchela} in \eqref{mauritius}, we find that as long as:
\bg\label{khalamay}
{\bf Re}\left[\alpha_{\mu\nu}^{(\psi)}({\bf k})~{\rm exp}\left({i\omega_{\bf k}^{(\psi)}\over \sqrt{\Lambda}}\right)
\right] ~ < ~ {\bf Re}\left[\alpha_{\mu\nu}^{(\psi)}({\bf k})\right], \nd
the Glauber-Sudarshan state would {\it lose} gravitons as it evolves in the allowed temporal domain. Taking real $\alpha_{\mu\nu}^{(\psi)}({\bf k})$ as in \cite{coherbeta}, the inequality \eqref{khalamay} is clearly 
satisfied, although the result does depend on the sign of the zeroth order terms (if any). Keeping the non-linear terms in \eqref{censusmay}, the simpler condition in \eqref{khalamay} will change. In any case, the precise corrections are not too hard to work out, but we will not pursue them here. We will instead study the 
more interesting topic of {\it fluctuations} over the Glauber-Sudarshan state using the so-called graviton (and flux) -added coherent state, {\it i.e.} the Agarwal-Tara state \cite{agarwal}.

\subsection{The Agarwal-Tara states and  fluctuations over de Sitter space \label{vac2}}

The Glauber-Sudarshan state responsible for the de Sitter space is in general quite non-trivial when we compare it with the standard coherent states in field theories. Even the temporal evolution doesn't follow the standard evolution of a coherent state, as we saw above. Despite that, the expectation values of the graviton and flux operators do reproduce the precise background related to de Sitter space-time. The question now is to study {\it fluctuations} over de Sitter space, {\it i.e.} fluctuations over the Glauber-Sudarshan state. As argued in \cite{coherbeta}, the fluctuations are given by graviton and flux-added Glauber-Sudarshan state, namely the Agarwal-Tara state \cite{agarwal}. 
The Agarwal-Tara state for a given momentum ${\bf k}$ is given by the following action:
\bg\label{ferrlisa}
\left\vert \Psi^{(c_1c_2)}_{\bf k}(t)\right\rangle = \left[c_1 + c_2 ~{\cal G}^{(\psi)}(a_{\bf k} + a^\dagger_{\bf k}; t)\right] \left\vert \Psi^{(\bar\sigma)}_{\bf k}(t) \right\rangle, \nd
where the operator ${\cal G}^{(\psi)}(a_{\bf k} + a^\dagger_{\bf k}; t)$ is defined in eq. (2.46) in the second reference of \cite{coherbeta}, and
$\left\vert \Psi^{(\bar\sigma)}_{\bf k}(t) \right\rangle$ is the simplified way to express the free vacuum wave-function in the configuration space as discussed in section \ref{vac1}. Note again the choice of $\bar\sigma$ compared to 
$\sigma$, although the definition from eq. (2.46) in the second reference of \cite{coherbeta} remains unaltered at least in the limit 
$a_{\rm eff}({\bf k}, t) \approx a_{\bf k}$ and $a^\dagger_{\rm eff}({\bf k}, t) \approx a^\dagger_{\bf k}$. For simplicity and for the brevity of the ensuing discussion, we will resort to this simplification. The relation 
\eqref{ferrlisa} is valid for all time $t$, and we get a pure coherent state for vanishing\footnote{Interestingly, for non-vanishing $c_2$ but vanishing $z_2$ in eq. (2.46) in reference 2 of \cite{coherbeta}, we get back the coherent state.}
$c_2$. 

There are a few subtleties that need to be pointed out before we move forward with our analysis. {\Su One}, 
The action ${\cal G}^{(\psi)}(a_{\bf k} + a^\dagger_{\bf k}; t)$ does what one would expect for an 
Agarwal-Tara state \cite{agarwal}, namely it adds (or removes) graviton and flux quanta (although for our discussion we will suppress the flux side of the story). However the action of ${\cal G}^{(\psi)}(a_{\bf k} + a^\dagger_{\bf k}; t)$ is more complicated compared to the original action in \cite{agarwal}. {\Su Two}, the action of ${\cal G}^{(\psi)}(a_{\bf k} + a^\dagger_{\bf k}; t)$ is on the shifted {\it free} vacuum for the mode 
${\bf k}$, instead of on the {\it interacting} vacuum. Clearly the latter should have been the expected action, but due to the relation \eqref{amchase}, the action of ${\cal G}^{(\psi)}(a_{\bf k} + a^\dagger_{\bf k}; t)$  can alternatively be acted individually on the free vacua for all momenta ${\bf k}$. While this surprising simplification helps tremendously to have quantitative control on the dynamics of the Glauber-Sudarshan state (as we saw in section \ref{vac1}), the complication, if any, lies elsewhere: finding the point $\bar\sigma$ in the configuration space because the relations connecting $\sigma$ (which we know from the Fourier transforms) to $\bar\sigma$ (which needs to be determined from \eqref{censusmay}) are in general non-linear. Nevertheless, as pointed out in section \ref{vac1}, quantitative predictions can be made. {\Su Three}, in the temporal evolution of the Glauber-Sudarshan state, the system loses graviton and flux quanta, so one would expect that the temporal evolution of the Agarwal-Tara state $-$ which is a graviton and flux-{\it added} Glauber-Sudarshan state $-$ to be naturally more complicated. Surprisingly however, a proper renormalization of the parameter involved in the analysis, much like what we did in \eqref{censusmay}, tells us that all the complications may be neatly packaged in certain Fourier transforms in such a way that the fluctuations over the de Sitter space appear in the requisite form. This may be seen explicitly from the following example. Let us start with the Agarwal-Tara state as in \eqref{ferrlisa}, with one simplification: we restrict ourselves to $\bar\sigma \to \bar\alpha$ so that it is only a graviton added coherent state. 
The expectation value of the metric operator in this state then takes the following form:

{\footnotesize
\bg\label{articoo}
\langle {g}_{\mu\nu}\rangle_{\Psi^{(c_1c_2)}} &\equiv&  
{\langle \Omega\vert \mathbb{D}^\dagger(\bar\alpha(t))\left(c_1^\ast + c_2^\ast {\cal G}^{\dagger(\psi)}(a, a^\dagger; t)\right) {\bf g}_{\mu\nu} \left(c_1 + c_2 {\cal G}^{(\psi)}(a, a^\dagger; t)\right)~\mathbb{D}(\bar\alpha(t)) \vert \Omega \rangle
\over 
\langle \Omega\vert \mathbb{D}^\dagger(\bar\alpha(t))\left(c_1^\ast + c_2^\ast {\cal G}^{\dagger(\psi)}(a, a^\dagger; t)\right) \left(c_1 + c_2 {\cal G}^{(\psi)}(a, a^\dagger; t)\right)~\mathbb{D}(\bar\alpha(t))\vert \Omega \rangle}, \nd}
where $(a, a^\dagger)$ are the annihilation and the creation operators for all modes ${\bf k}$. In the path-integral language the gauge fixing will provide a well-defined propagator for the graviton (the ghost may be decoupled as discussed in footnote \ref{casin777}).  
The RHS of \eqref{articoo} can actually be worked out using similar procedures laid out in eqns. (2.82), (2.83) and (2.85) in the second reference of \cite{coherbeta}, although now we will have to take more generic action than the ones taken in \cite{coherbeta} to elucidate the specific dependences on 
$\bar\alpha$ and $g_s$. This means we will modify eq. (2.82) in the second reference of \cite{coherbeta} in the following way:

{\footnotesize
\bg\label{salahrn}
&&{\rm Num}\left[\langle {g}_{\mu\nu}\rangle_{\Psi^{(c_1c_2)}}\right] 
=  \left(\prod_{k} \int d\left({\bf Re}~\widetilde{g}_{\mu\nu}(k)\right) d\left({\bf Im}~\widetilde{g}_{\mu\nu}(k)\right)\right)\\
&&~~~~~~~~~~~~~~~~~~~~~~~~~~\times
  {\rm exp}\left[{i\over V}\sum_{k} k^2 \vert \widetilde{g}_{\mu\nu}(k)\vert^2 + i\sum_{n, k} \gamma_n(k) 
  \vert \widetilde{g}_{\mu\nu}(k)\vert^{2(n+1)} + iS_{\rm sol} + .....\right] \nonumber\\
&&~~~~~~~~~~~~~~~~~~~~~~~~~~\times
 {\rm exp} \left[{2\over V} \sum_{k'} ~
\left({\bf Re}~\bar\alpha^{(\psi)}_{\mu\nu}(k')~{\bf Re}~\widetilde{g}^{\mu\nu}(k') + {\bf Im}~\bar\alpha^{(\psi)}_{\mu\nu}(k')~{\bf Im}~\widetilde{g}^{\mu\nu}(k')\right) + ...\right]
\nonumber\\
&&~~~~~~~~~~~~~~~~~ \times  {1\over V} \sum_{k''} ~\psi_{{\bf k}''}({\bf x}, y, z) e^{-ik''_0t}
\left({\bf Re}~\widetilde{g}_{\mu\nu}(k'') + i{\bf Im}~\widetilde{g}_{\mu\nu}(k'')\right)~{\rm exp}\left(-{1\over V}\sum_{k'} \vert\bar\alpha^{(\psi)}_{\mu\nu}(k')\vert^2\right) \nonumber\\ 
&&~~~~~~~~~~~~~~~~~ \times \Bigg\vert 1 + {c_2 \over V} \sum_{k'''}~C^{(\psi)}(k''')~
\left[\left({\bf Re}~\widetilde{g}^{\mu\nu}(k''') + i{\bf Im}~\widetilde{g}^{\mu\nu}(k''')\right)
\left({\bf Re}~\widetilde{g}_{\mu\nu}(k''') + i{\bf Im}~\widetilde{g}_{\mu\nu}(k''')\right)\right]^2\Bigg\vert^2, 
\nonumber \nd}
where we have added a perturbative series in powers of $n > 1$, ${\rm Num}[...]$ is the numerator in \eqref{articoo}, $S_{\rm sol}$ is the action of the 
background solitonic vacuum, and the dotted terms in the second line are the non-perturbative, non-local, fermionic and brane/plane terms (including the mixed ones). In the third line, we represent the action of the displacement operator $\mathbb{D}(\bar\alpha)$ and use \eqref{amchase} to express in the above form.The dotted terms therein represent the higher order interactions that will become useful as we shall see a bit later. We have also used $c_1 = 1$, and ${\cal G}^{(\psi)}(a, a^\dagger; t)$ is defined as a series 
in powers of $\widetilde{g}_{\mu\nu}\widetilde{g}^{\mu\nu}$ with coefficients $C_m^{(\psi)}$ (see details in the second reference of \cite{coherbeta}). Here we simplify this by taking $C_m^{(\psi)} \equiv C^{(\psi)}
\delta_{m2}$, which is the quadratic form appearing in the fifth line of \eqref{salahrn}. We can also 
express the denominator ${\rm Den}\left[\langle {g}_{\mu\nu}\rangle_{\Psi^{(c_1c_2)}}\right]$ as an integral similar to \eqref{salahrn} but without the fourth line. The coefficients $\gamma_n(k)$ 
appearing above provide some hierarchies as they are suppressed by powers of ${\rm M}_p$, but they also depend on the momenta because of all the derivative actions from \eqref{fahingsha5}. Since we are using Wilsonian action, the momenta are low (as the high energy degrees of freedom are integrated out and we are in a scale much below the KK or the massive stringy degrees of freedom). So low momenta and 
${\rm M}_p >> 1$, or more appropriately ${k^2\over {\rm M}_p^2} << 1$, should help us have some quantitative control on the path-integral from the first line of \eqref{salahrn}. (We will re-visit the issue of low momenta soon when we study the integrals over $k_0$.)  

Despite the possibility of the aforementioned simplifications, the path-integral in \eqref{salahrn} is still pretty hard to compute. In the absence of the third line, we expect the path-integral to vanish. However it is the shift in the interacting vacuum by the displacement operator $\mathbb{D}(\bar\alpha)$ that, on one hand, keeps the integral from vanishing, but on the other hand, makes the analysis pretty non-trivial. To make some sense of \eqref{salahrn}, let us resort to some additional simplifications: we will take only one mode $k$, and ignore the sum in the fourth line. However since the {\it sum} over the modes is absolutely essential to make any quantitative prediction using the path-integral, we will insert them in the end. We will also ignore the tensor indices and assume 
${\bf Re}~\widetilde{g}_{\mu\nu}(k) = {\bf Re}~\Phi(k) \equiv \Phi_k$, with ${\bf Im}~\widetilde{g}_{\mu\nu}(k) = {\bf Im}~\Phi_k = 0$. Similarly, ${\bf Re}~\bar\alpha_{\mu\nu}^{(\psi)} \equiv \bar\alpha_k$. We can now go to the Euclidean space-time where ${ik^2\over V} \to -\beta_k$, such that inverse $\beta_k$ will denote the pole structure. 

With these additional simplifications we are ready to work with the path-integral. There is one last notational change to avoid un-necessary clutter: ${c_2 C^{(\psi)}(k)\over V} \to {c}_k$ so the subscript will remind us that we are taking the mode $k$. The path-integral in \eqref{salahrn} simplifies to:

{\footnotesize
\bg\label{salahrn2}
{\rm Num}\left[\langle {g}_{\mu\nu}\rangle_{\Psi^{(c_1c_2)}}\right]_k &\equiv& 
\int d\Phi_k ~{\rm exp}\left(-\beta_k\Phi_k^2 + \bar\alpha_k \Phi_k + \sum_{n = 1}^\infty \gamma_{nk}\Phi_k^{2(n+1)} -\bar\alpha_k^2 + ...\right)
\Phi_k\left(1 + c_k \Phi_k^2\right)^2 \nonumber\\
& = & \int dy_k ~{\rm exp}\left[-\beta_k y_k^2 + \sum_{n = 1}^\infty\gamma_{nk}\left(y_k + {\bar\alpha_k\over 2\beta_k}\right)^{2(n+1)}- \bar\alpha_k^2\left(1 - {1\over 2\beta_k}\right) + ..\right] \nonumber\\
&&~~~~~~~~~~~~~~~~~~~~~~~~~~~~~~~~~~~~~~~\times \left(y_k + {\bar\alpha_k\over 2\beta_k}\right)\left[1 + c_k\left(y_k + {\bar\alpha_k\over 2\beta_k}\right)^2\right] \nonumber\\
&=& \prod_{n = 1}^\infty \sum_{m = 0}^\infty \int dy_k ~{\rm exp}\left[-\beta_k y_k^2 - \bar\alpha_k^2\left(1 - {1\over 2\beta_k}\right) + ...\right]{\gamma_{nk}^m\over m!}\left(y_k + {\bar\alpha_k\over 2\beta_k}\right)^{2m(n+1)}\nonumber\\
&&~~~~~~~~~~~~~~~~~~~~~~~~~~~~~~~~~~~~~~~\times \left(y_k + {\bar\alpha_k\over 2\beta_k}\right)\left[1 + c_k\left(y_k + {\bar\alpha_k\over 2\beta_k}\right)^2\right],\nd}
where in the second line we have defined $y_k \equiv \Phi_k - {\alpha_k\over 2\beta_k}$, and in the fourth line we have perturbatively expanded the interaction pieces assuming $\gamma_{nk} < \beta_k$. That this is possible may be justified from the ${\rm M}_p$ suppressions of the various interaction terms compared to the kinetic term. 
Note that in the third and the fifth lines we have the metric factor $y_k + {\alpha_k\over 2\beta_k}$, and because of that the path-integral in \eqref{salahrn2} can be split into two parts. The {\Su first} one is with 
${\bar\alpha_k\over 2\beta_k}$, that takes the form:

{\footnotesize
\bg\label{srein}
{\rm Num}^{(1)}\left[\langle {g}_{\mu\nu}\rangle_{\Psi^{(c_1c_2)}}\right]_k & \equiv & 
\prod_{n = 1}^\infty \sum_{m = 0}^\infty \int dy_k ~{\rm exp}\left[-\beta_k y_k^2 - \bar\alpha_k^2\left(1 - {1\over 2\beta_k}\right) + ...\right]{\gamma_{nk}^m\over m!}\left(y_k + {\bar\alpha_k\over 2\beta_k}\right)^{2m(n+1)}\nonumber\\
&\times & {\bar\alpha_k\over 2\beta_k}\left[1 + c_k\left(y_k + {\bar\alpha_k\over 2\beta_k}\right)^2\right] 
 =  {\bar\alpha_k\over 2\beta_k} ~
{\rm Den}\left[\langle {g}_{\mu\nu}\rangle_{\Psi^{(c_1c_2)}}\right]_k, \nd} 
which, as one would have expected, gives us simply ${\bar\alpha_k\over 2\beta_k}$ once we divide by the 
denominator ${\rm Den}\left[\langle {g}_{\mu\nu}\rangle_{\Psi^{(c_1c_2)}}\right]_k$. When $\gamma_{nk} = c_k = 0$, all the gaussian integrals could be performed, and due to the product structure with the momenta (coming from the products of all the gaussian functions), the final answer would be what we had in \cite{coherbeta}. Question is,
what happens now when we allow $\gamma_{nk} \ne 0, c_k \ne 0$? The answer is:

{\footnotesize
\bg\label{numden}
{{\rm Num}^{(1)}\left[\langle {g}_{\mu\nu}\rangle_{\Psi^{(c_1c_2)}}\right] \over 
 {\rm Den}\left[\langle {g}_{\mu\nu}\rangle_{\Psi^{(c_1c_2)}}\right]} &\equiv & 
 {\sum_{k'}{\rm Num}^{(1)}\left[\langle {g}_{\mu\nu}\rangle_{\Psi^{(c_1c_2)}}\right]_{k'}\left(\prod_{k \ne k'}{\rm Num}^{(1)}\left[\langle {g}_{\mu\nu}\rangle_{\Psi^{(c_1c_2)}}\right]_k\right)\psi_{{\bf k}'}({\bf x}, y, z)~
 e^{ik_0' t} \over 
 \prod_k {\rm Den}\left[\langle {g}_{\mu\nu}\rangle_{\Psi^{(c_1c_2)}}\right]_k}\nonumber\\
 &= & 
 \int d^{11}k~ {\bar\alpha_{\mu\nu}^{(\psi)}(k)\over 2\beta(k)} ~\psi_{\bf k}({\bf x}, y, z) ~e^{ik_0 t} = 
 \int d^{10} k ~\bar\alpha_{\mu\nu}^{(\psi)}({\bf k}, \omega_{\bf k})
 ~\psi_{\bf k}({\bf x}, y, z) ~e^{i\omega_{\bf k} t}, \nd}
where it is clear that ${\rm Den}\left[\langle {g}_{\mu\nu}\rangle_{\Psi^{(c_1c_2)}}\right]_k$ would be a complicated function of $k$. However since the full denominator is typically given by the product over all momenta $k$, we expect it as in \eqref{numden} above with
the products over all momenta coming from the product structure of the action, and the sum appearing from the graviton field itself in \eqref{salahrn}. In the first equality of \eqref{numden} it is understood that the $k'$ piece will come with an extra factor compared to the other pieces with $k \ne k'$. The extra pieces from each $k'$ can be summed over in a way shown in the second line above.  In the last equality, we have assumed that $\beta(k)$ has a simple pole and therefore the residue at the pole provides the form given above. This is however strictly true only when the $k_0$ integral ranges from $-\infty$ to $+\infty$. Because of our low energy Wilsonian form of the action, the range of $k_0$ cannot be that, so
 we can leave the integral over the eleven-dimensional momenta, thus giving a more off-shell description of the system. However if we take the limit of vanishing internal cycles and ${\rm M}_p \to \infty$, then the KK and the stringy modes may be made arbitrarily heavy and we can have access to a large range of momenta without violating\footnote{It is easy to maintain this condition. Let $k$ and ${\rm M}_p$ go to infinity as $k \to \epsilon^{-a}$ and ${\rm M}_p \to \epsilon^{-b}$ for $\epsilon \to 0$. As long as $b > a$, the ratio ${k^2\over {\rm M}_p^2} = \epsilon^{2(b - a)} << 1$.} the condition ${k^2\over {\rm M}_p^2} << 1$. In this case, the integral over $k_0$ can span the required range such that the residue at the requisite pole will reproduce the on-shell result of \eqref{numden}.   
 
The {\Su second} split comes from the $y_k$ piece in \eqref{salahrn2}. Such an integral would typically vanish if we do not shift the vacua by $\bar\alpha_k$, because of the gaussian action and also because the interactions are expressed in even powers of $y_k$ in \eqref{salahrn2}. The shift in $\bar\alpha_k$ creates the necessary odd powers of $y_k$, and we get:

{\footnotesize
\bg\label{srein2}
{\rm Num}^{(2)}\left[\langle {g}_{\mu\nu}\rangle_{\Psi^{(c_1c_2)}}\right]_k & = & 
\prod_{n = 1}^\infty \sum_{m = 0}^\infty \int dy_k ~{\rm exp}\left[-\beta_k y_k^2 - \bar\alpha_k^2\left(1 - {1\over 2\beta_k}\right) + ...\right]{\gamma_{nk}^m\over m!}\left(y_k + {\bar\alpha_k\over 2\beta_k}\right)^{2m(n+1)}\nonumber\\
&\times & y_k\left[1 + c_k\left(y_k + {\bar\alpha_k\over 2\beta_k}\right)^2\right] 
 =  \int dy_k ~y_k\mathbb{F}(y_k) + c_k\int dy_k~y_k\left(y_k + {\bar\alpha_k\over 2\beta_k}\right)^2
 \mathbb{F}(y_k), \nonumber\\  \nd} 
where $\mathbb{F}_k(y_k)$ is given by the integrand in the first line of \eqref{srein2}. As we see from the second line, the integral may be further split into two parts so that even for vanishing $c_k$ we have a non-vanishing result, provided $\gamma_{nk} \ne 0$. Incidentally, for vanishing $\gamma_{nk}$, even non-vanishing $c_k$ cannot save the integrals in \eqref{srein2} from vanishing. More so, for vanishing $\bar\alpha_k$, but non-vanishing $\gamma_{nk}$ or $c_k$, both the integrals vanish implying that the values of both the integrals should be proportional to the product $\gamma_{nk}^m \bar\alpha_k^{2p + 1}$ where 
$(m, p) \in (\mathbb{Z}, \mathbb{Z})$. The coefficients $\gamma_{nk}$ would typically be functions of the momenta (and are therefore suppressed by powers of ${\rm M}_p$). Thus keeping $\gamma_{nk}$ and 
$\bar\alpha_k$ both non-zero, the first integral in \eqref{srein2} gives:

{\footnotesize
\bg\label{crullatin} 
&&\int_{-\infty}^{+\infty} dy_k~y_k\mathbb{F}(y_k) =  \prod_{n = 1}^\infty \sum_{m = 1}^\infty 
\sum_{l = 0}^\infty {(2mn+2m)!(2l + 1)!!\over (2l+1)!(2mn + 2m - 2l - 1)!m!}~ {\gamma_{nk}^m\over (2\beta_k)^{l+1}} 
\sqrt{\pi\over \beta_k} \left({\bar\alpha_k\over 2\beta_k}\right)^{2(mn + m - l) - 1} \nonumber\\ 
& = &  \prod_{n = 1}^\infty \sum_{m = 0}^\infty 
\sum_{l = 0}^\infty{\left[2(m+1)(n+1)\right]! (2l+1)!!\over (2l+1)! \left[2m(n+1) + 2(n - l) + 1\right]! (m + 1)!}~
{\gamma_{nk}^m\over (2\beta_k)^{l}} 
\sqrt{\pi\over \beta_k} \left({\bar\alpha_k\over 2\beta_k}\right)^{2(mn + m - l)}
{\gamma_{nk}\over 2\beta_k} 
\left({\bar\alpha_k\over 2\beta_k}\right)^{2n+1}, \nonumber\\ \nd}
where the second line brings the integral in a suggestive form so it becomes useful when we compare it with the denominator (to be discussed shortly). In fact, since the functional form for $\gamma_{nk}$ is known from the quantum series \eqref{fahingsha5}, we see that the integral may be represented in powers of 
$\bar\alpha_k$ with the condition that $2(m + 1)(n+1) \ge 2l + 1$ from the second line. We also see that the denominator now takes the form:

{\footnotesize
\bg\label{crullatin2} 
\int_{-\infty}^{+\infty} dy_k~\mathbb{F}(y_k) = \prod_{n = 1}^\infty \sum_{m = 0}^\infty 
\sum_{l = 0}^\infty {(2mn+2m)!(2l - 1)!!\over (2l)!(2mn + 2m - 2l)!m!}~ {\gamma_{nk}^m\over (2\beta_k)^{l}} 
\sqrt{\pi\over \beta_k} \left({\bar\alpha_k\over 2\beta_k}\right)^{2(mn + m - l)}, \nd}
with the condition now being $m(n+1) \ge l$. When $m = 0$, we can only have $l = 0$, and from \eqref{crullatin2} we see that there is a term independent of $\gamma_{nk}$ and $\bar\alpha_k$. When $m = 1$, $l$ can be $l = 0, 1, 2$, and there are three terms all proportional to $\gamma_{nk}$ but with different powers of $\bar\alpha_k^2$. Thus for a fixed value of $n$, we can go to higher orders in $m$, and from there determine all the allowed choices of $l$. This way both the numerator and the denominator may be written down explicitly.

There are however a few caveats that we did not mention so far. {\Su First}, the interactions appearing in \eqref{salahrn} cannot be so simple. Even at the perturbative level, as should be clear from \eqref{fahingsha5}, the momentum modes of the graviton should mix non-trivially making \eqref{salahrn2} much more involved. Plus we have ignored the whole non-perturbative and the non-local (including the topological and the brane/plane) interactions. {\Su Secondly}, even for the simple case considered here, the numerator \eqref{crullatin} and the denominator \eqref{crullatin2} are not exactly proportional to each other for the relevant momentum $k$. We can then rewrite them as:
\bg\label{numden2}
&& {\rm Den}^{(a)}\left[\langle {g}_{\mu\nu}\rangle_{\Psi^{(c_1c_2)}}\right] = 
{\rm exp}\left[\int_{_\infty}^{+\infty} d^{11} k \left(\int_{-\infty}^{+\infty} dy_k ~\mathbb{F}(y_k)\right)\right]
\nonumber\\
&& {\rm Num}^{(2a)}\left[\langle {g}_{\mu\nu}\rangle_{\Psi^{(c_1c_2)}}\right] = 
\int_{-\infty}^{+\infty} d^{11}k'~\psi_{{\bf k}'}({\bf x}, y, z) ~e^{ik'_0 t} \int_{-\infty}^{+\infty} dy_{k'}~y_{k'} \mathbb{F}(y_{k'}) \nonumber\\
&& ~~~~~~~~~~~~~~~~~~~~~~~~~~~~~~~ \times {\rm exp}\left[\int_{_\infty}^{+\infty} d^{11} k~\Theta(k - k') \left(\int_{-\infty}^{+\infty} dy_k ~\mathbb{F}(y_k)\right)\right], \nd
where the super-script $a$ denotes the absence of $c_k$ interactions from \eqref{salahrn} in the computations, and $\Theta(k - k') = 0$ when $k = k'$ and is equal to 1 when $k \ne k'$. The above operation simply isolates the momentum $k'$ from the whole bunch of momenta because $k'$ would appear alongside $y_{k'}$ from our choice of the graviton field with spatial wave-function $\psi_{{\bf k}'}({\bf x}, y, z)$. In this language, the ratio between the numerator and the denominator would be proportional to powers of $\bar\alpha_k^2$, with the expansion parameter being controlled by $\gamma_{nk}$. {\Su Thirdly}, our analysis haven't so far revealed the sources of ${g_s^a\over {\rm M}_p^b}$ contributions that we predicted in \eqref{kritins}. One way the $g_s$ dependences would naturally appear is by replacing $\bar\alpha_k$ by $\alpha_k$ in \eqref{salahrn}. Since we know the forms of both $\gamma_{nk}$ and $\alpha_k$, the higher order terms of the type $\gamma_{nk}^m \alpha_k^{2p + 1}$ would serve as Fourier coefficients, and integrating them with $e^{ik_0 t}$ would reproduce the ${g_s^a\over {\rm M}_p^b}$ contributions. Unfortunately $\bar\alpha_k$ is {\it not} the same as $\alpha_k$, which is one of the crucial condition for getting the de Sitter background from the Glauber-Sudarshan state. Where do we then get the 
${g_s^a\over {\rm M}_p^b}$ terms? The answer lies in the {\it higher order} terms in the displacement operator, namely the dotted terms that appear in the third line of \eqref{salahrn}. Recall that the {\it linear} terms in the field $\widetilde{g}_{\mu\nu}(k)$, that are exponentiated in the definition of the displacement operator, are responsible in shifting the {\it free} vacua by an amount $\bar\alpha_k$ in the configuration space (we are using \eqref{amchase} to express everything in terms of the free vacua). But the higher order terms in the graviton fields, for example terms of the form\footnote{Here we quoted an example of the form $\kappa\widetilde{g}^3$. We could look for more general terms of the form $\underbracket[1pt][2pt]{\kappa_1\kappa_2......\kappa_q}_{q}\widetilde{g}^{2p + 1}$, where $p > 0$. These can produce the necessary {\it odd} powers of the fields that we need (note that if $\widetilde{g}_{\mu\nu}(k) \equiv \Phi_k$, then $\widetilde{g}^{\mu\nu}(k) \approx \Phi_k \left(g^{-2}\right)_{\rm sol} \ne \Phi^{-1}_k$ where $g_{\rm sol}$ is the metric of the solitonic background. Thus if we want to be more precise, $\kappa \widetilde{g}^3$ should actually be written as $\kappa \Phi^3 \left(g^{-4}\right)_{\rm sol}$. However since the solitonic background is fixed, presence of $g_{\rm sol}$ does not alter anything we said above).}
 $\kappa_{\mu\nu}(k) \widetilde{g}^{\mu\rho}(k)......
\widetilde{g}_{\alpha\sigma}(k) \widetilde{g}^{\sigma\nu}(k)$, do not influence the shift in the vacua but they do influence the interactions. Now recall that, all we need to get the ${g_s^a\over {\rm M}_p^b}$ dependences are the odd terms in $y_k$ {\it independent} of $\bar\alpha_k$ in the integrals of \eqref{numden2}. Putting everything together, we get:

{\footnotesize
\bg\label{lalski}
\langle g_{\mu\nu}\rangle_{\bar\alpha} = 
{\eta_{\mu\nu}\over {\rm H}^{8/3}(y)} + {{\rm Num}^{(1)}\left[\langle {g}_{\mu\nu}\rangle_{\Psi^{(c_1c_2)}}\right] \over 
 {\rm Den}\left[\langle {g}_{\mu\nu}\rangle_{\Psi^{(c_1c_2)}}\right]} + 
 {{\rm Num}^{(2a)}\left[\langle {g}_{\mu\nu}\rangle_{\Psi^{(c_1c_2)}}\right] \over 
 {\rm Den}^{(a)}\left[\langle {g}_{\mu\nu}\rangle_{\Psi^{(c_1c_2)}}\right]} = 
{\eta_{\mu\nu}\over \left(\Lambda\vert t\vert^2 {\rm H}^2\right)^{4/3}}, \nd}
 thus reproducing \eqref{kritins}, and also the identification \eqref{censusmay} that relates $\bar\alpha^{(\psi)}_{\mu\nu}$ with $\alpha_{\mu\nu}^{(\psi)}$. Here the first term comes from the solitonic background, the second term comes from \eqref{numden} and the third term comes from \eqref{numden2}. Once we incorporate the interactions governed by $c_2$ in \eqref{salahrn} the final result, as in \eqref{lalski}, can be neatly packaged in the following way:

{\footnotesize
\bg\label{kimbas}
\langle {g}_{\mu\nu}\rangle_{\Psi^{(c_1c_2)}} 
& = & {\eta_{\mu\nu}\over {\rm H}^{8/3}(y)} + {{\rm Num}^{(1)}\left[\langle {g}_{\mu\nu}\rangle_{\Psi^{(c_1c_2)}}\right] + {\rm Num}^{(2)}\left[\langle {g}_{\mu\nu}\rangle_{\Psi^{(c_1c_2)}}\right] \over 
 {\rm Den}\left[\langle {g}_{\mu\nu}\rangle_{\Psi^{(c_1c_2)}}\right]} \\
&= &  {\eta_{\mu\nu}\over \left(\Lambda\vert t\vert^2 {\rm H}^2\right)^{4/3}}
 + 
c_2 \int d^{11} k~f({\bf k}, k_0)~\left[\bar\alpha^{(\psi)}_{\mu\nu}({\bf k}, k_0) 
+ {\cal O}\left({\bar\alpha^{(\psi)}_{\mu\nu}\bar\alpha^{(\psi)\mu\nu}\over {\rm M}_p^{|c|}}\right)\right] 
\psi_{{\bf k}}({\bf x}, y, z)~e^{ik_0 t}, \nonumber \nd}
where we have taken $c_1 \equiv 1$ and $c_2 << 1$ in \eqref{articoo}. Note the appearance of powers of 
$\bar\alpha$ inside the square
bracket, somewhat similar to what we had in \eqref{kritins}, but {\it no} 
${\cal O}(g_s^a) \equiv {\cal O}\left(g_s^{|a|}\right) + 
{\cal O}\left(e^{-1/g_s^{1/3}}\right)$ pieces. In fact the first term, {\it i.e.} the {\it dual} de Sitter term, appears precisely from a similar manipulations that led to \eqref{lalski}. On the other hand in the second term, which appears from the $c_k$ parts of \eqref{srein2},  because of the presence of an additional piece proportional to $f({\bf k}, k_0)$, the integral over $dk_0$ cannot be automatically performed. This means the 
$e^{ik_0 t}$ part will hang around and not integrate to provide any pieces proportional to $g_s^a$, thus differing from a similar computation in \eqref{lalski}. 
 It can also be shown that, by performing the integrals in \eqref{srein2} more carefully in the presence of the $c_k$ terms, we recover the following form for $f({\bf k}, k_0)$:
\bg\label{HKPilla}
 f({\bf k}, k_0) \equiv  \sum_{m, n} {C^{(\psi)}_{m{\bf k}} b^{(m)}_n \bar{\alpha}^{2n}({\bf k}, k_0)
 \over \left({\bf k}^2 - k_0^2\right)^{n}}, \nd
 with $\bar\alpha^2 (k) \equiv \bar\alpha^{(\psi)}_{\mu\nu}(k)\bar\alpha^{(\psi)\mu\nu}(k)$;
 $C^{(\psi)}_{m{\bf k}}$ are the coefficients that appear in the definition eq. (2.46) in the second reference of \cite{coherbeta}, and $b^{(m)}_n$ are the coefficients that appear $-$ and in turn get automatically fixed $-$ from the path integral \eqref{srein2} (for a simple example the reader may refer to eq. (2.87) in the second reference of \cite{coherbeta}). 
 
 In fact this is all we need because once we dualize \eqref{kimbas} to IIB, this would provide the required fluctuations over the de Sitter space. The dualizing dictionary will involve a dimensional reduction and then multiplying the metric by requisite power of $g_s$. After the dust settles, the first term in \eqref{kimbas} will resemble the actual four-dimensional de Sitter space-time, and the second term will be the fluctuations. From our M-theory point of view let us express the fluctuation spectra as:
\bg\label{marieanne}
\delta g_{\mu\nu} \equiv \int d^{10}{\bf k} ~e_{\mu\nu}({\bf k})~ \nu_{\bf k}(t) ~\psi_{\bf k}({\bf x}, y, z), \nd
where $\nu_{\bf k}(t)$ is the {\it time-dependent} frequency spectrum. Depending on how we want to study, this fluctuation spectra could be {\it dual} to the ones over either the $\alpha$  or the Bunch-Davies vacua. In either cases, it may not be too hard to identify the precise coefficients $C^{(\psi)}_{m{\bf k}}$ in \eqref{HKPilla} of the 
Agarwal-Tara state by making the following identification:
\bg\label{brazpha}
 \int dk_0 f({\bf k}, k_0)\left[\bar\alpha^{(\psi)}_{\mu\nu}({\bf k}, k_0) 
+ {\cal O}\left({\bar\alpha^{(\psi)}_{\mu\nu}\bar\alpha^{(\psi)\mu\nu}\over {\rm M}_p^{|c|}}\right)\right] e^{ik_0 t} \equiv  e_{\mu\nu}({\bf k})~\nu_{\bf k}(t), \nd
which is in fact just the temporal Fourier transformation! Such a transformation would fix the form of 
$f({\bf k}, k_0)$ from the knowledge of the functional form of $\nu_{\bf k}(t)$. In the limit ${\rm M}_p >> 1$, we can ignore the ${\cal O}(\bar\alpha^2)$ terms, and using \eqref{censusmay} we can replace 
$\bar\alpha^{(\psi)}_{\mu\nu} \to \alpha^{(\psi)}_{\mu\nu}$.  One may then use \eqref{HKPilla} to determine the coefficients $C_{m{\bf k}}^{(\psi)}$, which in turn would fix the form of the Agarwal-Tara states for the dual of the Bunch-Davies or the $\alpha$ vacua. As an example let us consider the two allowed choices for $\nu_{\bf k}(t)$:
\bg\label{willsagao}
\nu_{\bf k}^{\rm BD}(t) \equiv   \frac{e^{-i{\bf k}t}}{\left(\Lambda \vert t \vert^2\right)^{1/3}\sqrt{2{\bf k}}}  \left(1 - \frac{i}{{\bf k}t}\right), ~~~~~
\nu_{\bf k}^{\alpha_o}(t) \equiv 2 {\bf Re}\left(\alpha_o \nu_{\bf k}^{\rm BD}\right), \nd 
where we use $\alpha_o$, instead of $\alpha$, to represent the modes for the $\alpha$-vacuum. We have also multiplied by a term proportional to $g_s^{-2/3}$ so that \eqref{willsagao} represents the correct frequencies from M-theory in the flat slicing of de Sitter. Taking the inverse Fourier transforms and plugging them in \eqref{brazpha}  in the limit ${\rm M}_p >> 1$ will reproduce the required form for 
$f({\bf k}, k_0)$. We can then compare this with \eqref{HKPilla} to get the required coefficients 
$C_{m{\bf k}}^{(\psi)}$ for either the Bunch-Davies or the $\alpha$ vacua respectively. In fact, through appropriate choices of these coefficients $C^{(\psi)}_{mk}$, we can reproduce any state for a quantum fluctuation on de Sitter.

There is one more thing that one might have noticed from the set of equations \eqref{kimbas} and 
\eqref{brazpha}: the fluctuations over our de Sitter Glauber-Sudarshan state are secretly linear combinations of the modes over the solitonic background (since both $\psi_{\bf k}({\bf x}, y, z)$ and 
$e^{ik_0t}$ appear from there). From this point of view there is no trans-Planckian problem as was also emphasized in \cite{coherbeta}. The time-dependent frequencies that we encounter from the fluctuations over a de Sitter {\it vacuum} are an artifact of the Fourier transforms over the solitonic vacuum. 

The aforementioned manipulations to determine the fluctuation spectra is clearly rich in details and many interesting dynamics may be pursued once we know the behavior of the Agarwal-Tara state. We will unfortunately not be able to elaborate the story further here and more details on the
 identifications of the Agarwal-Tara states corresponding to both the Bunch-Davies as well as the $\alpha$ vacua will be presented elsewhere.

\section{Energy conditions and requirements for an EFT description \label{bernardo}}

So far we have studied everything from M-theory point of view, which is clearly imposed on us because the IIB dynamics is typically at strong coupling with $g_b = 1$. While this throws a wrench on doing explicit computations in the IIB side, it doesn't however prohibit us to study the energy conditions that typically may be ascertained directly by manipulating the Einstein term without worrying about the quantum corrections. In particular, this will help us to make a surprising connection between the EFT conditions discussed in \eqref{senrem} and in section \ref{toyeft}, and the four-dimensional Null Energy Condition (NEC) in the IIB side. 

As discussed at the beginning of section \ref{sec2.2}, and also in section \ref{toyeft}, the condition 
\eqref{senrem} on the time dependence of a given flat FLRW scale factor is crucial for the existence of a $g_s$ hierarchy of all relevant curvature terms, which is a requirement for the existence of an EFT description with $g_s \ll 1$. In this section we show that condition (\ref{senrem}) is equivalent to the Null Energy Condition (NEC) on the effective energy-momentum tensor constructed from all possible corrections and fluxes. This implies that an effective description of cosmologies that violates the NEC might not be straightforwardly described by the extended formalism developed in the present work.

Energy conditions are useful to study acceptable metric ans\"atze in a given theory. For instance, the Gibbons-Maldacena-Nunez (GMN) no-go theorem \cite{GMN}, basically states that the possible sources in low energy supergravity (not including higher order corrections) cannot give rise to energy-momentum tensors that violates the Strong Energy Condition (SEC), which implies that it should hold, prohibiting compactifications to accelerating cosmologies. However, as discussed in \cite{Russo:2019fnk}, the generic metric considered has a time-independent internal space and we can relax the theorem if we consider a time-dependent internal space. In this case, extra terms appear in the RHS of the SEC inequality, such that it can be satisfied even with a four-dimensional accelerated solution. In \cite{Russo:2019fnk} the authors show that if we also impose the Dominant Energy Condition (DEC) on top of SEC, then any accelerating solution will have a singularity in the internal space.  

From this discussion, we see that it is important to be careful with the assumptions and to what metric we are imposing the energy conditions. In particular, in the variety of the no-go theorems discussed in the literature, the typical reasoning goes in the following way. A certain energy condition is shown to hold under some assumptions and then one argues that a given metric ans\"atze cannot be a solution to the theory if it violates that specific condition. In this line of reasoning, the two-derivative Einstein's equations are assumed in order to see whether the metric can satisfy the energy condition or not.

In our case, we want to try something slightly different. We already know from \cite{desitter2, coherbeta} that it should be possible to find solutions in type IIB theory whose metric contains a four-dimensional de Sitter space\footnote{In the language of the Glauber-Sudarshan wave-function in the dual M-theory, one may consider this to be the most probable state.}. Since the {\it proof of existence} of the solution is based on M-theory with a finite number of hierarchically controlled corrections (at least with the time-dependent fluxes case), the final equations that the type IIB metric (\ref{10dansatz1}) satisfy are not simply the two-derivative Einstein's equations. Regardless, one can still put all the corrections and sources to the RHS of the equation and work with an \emph{effective} ten-dimensional energy-momentum tensor ${\rm T}^{\text{eff}}_{\mu\nu}$ that supports the solution. What we want to do in this section is to write the energy conditions that 
${\rm T}^{\text{eff}}_{\mu\nu}$ should satisfy in order for a given metric ans\"atze to be a solution\footnote{One may also attempt to express the SEC \eqref{curvat} and NEC \eqref{nulle} in terms of the effective energy-momentum tensor ${\rm T}^{\rm eff}_{\mu\nu}$ but we will not do so here.}.

\subsection{Energy conditions for a general FLRW ans\"atze \label{flrow}}

As mentioned above, finding the energy conditions is not particularly sensitive to the explicit form of the quantum corrections. We will take advantage of this situation and re-visit the EFT condition that we discussed at the beginning of section \ref{sec2.2}. We argued therein that all the analysis of \cite{desitter2, coherbeta} is also valid for the metric ans\"atze \eqref{morbiusn}, which we quote again for simplicity:

{\footnotesize
\bg\label{10dansatz1}
    ds^2 = \frac{(\Lambda \vert \eta \vert^2)^n}{\mathrm{H}^2(y)}\left(-d\eta^2 + g_{ij}dx^i dx^j \right) 
    + \mathrm{H}^2(y)\left({\rm F}_1(\eta)g_{\alpha\beta}(y)dy^\alpha dy^\beta + {\rm F}_2(\eta) g_{mn}(y)dy^m dy^n\right), \nd}
where $(\alpha, \beta) \in {\cal M}_2$ and $(m, n) \in {\cal M}_4$, 
provided $1/n \geq -1$.   This condition appears from demanding that the derivative of the IIA string coupling $g_s$ with respect to the temporal coordinate is given in terms of {\it positive} powers of $g_s$. If this condition is violated, the system loses its EFT description as we demonstrated in section \ref{toyeft}. 
Note a small change of notation from the earlier sections: we have expressed \eqref{10dansatz1} in terms of  $\eta$ which will be the {\it conformal} time, as we shall reserve $t$ for the {\it cosmic} time\footnote{Although the preference of the cosmic time over the conformal time is purely a matter of convention here, the former is chosen in the earlier sections not only to facilitate computational efficiency but also to isolate the $g_s$ dependences.}.

It is interesting to compare such a bound on the evolution of the scale factor with the bounds from energy conditions. To this end, we will start with the well-known example of a FLRW metric and then connect it to 
\eqref{10dansatz1}. From the four-dimensional perspective, we will take the standard form of a 
FLRW metric:
\begin{equation}\label{FRLW}
    ds^2 = -dt^2 + a^2(t)\delta_{ij}dx^i dx^j,
\end{equation}
where $a^2(t)$ signifies the temporally varying scale factor. As it is written, \eqref{FRLW} is not expressed in terms of the conformal coordinates, but it is not necessary to bring it in that form to study the energy conditions. The metric \eqref{FRLW} satisfies the two-derivative Einstein's equations, and the Strong Energy Condition (SEC) for \eqref{FRLW} may be expressed as:
\begin{equation}\label{curvat}
    R_{\mu\nu}k^\mu k^\nu \geq 0, \quad k^2 < 0 
\end{equation}
where $R_{\mu\nu}$ is the curvature tensor. As mentioned above, the curvature tensors etc. from the metric 
\eqref{FRLW} is not yet connected to the curvature tensors from the metric \eqref{10dansatz1}. The inequality \eqref{curvat} may be re-expressed in terms of the Hubble parameter $H(\eta) \equiv \partial_\eta a(\eta)$  
as $\Dot{H} + H^2\leq 0$. Note that the Hubble parameter $H(t)$ should not be confused with the warp-factor ${\rm H}(y)$, the latter being a function of the internal six-dimensional coordinates only.  On the other hand, the Null Energy Condition (NEC) for the metric \eqref{FRLW} may be expressed as: 
\begin{equation}\label{nulle}
    R_{\mu\nu}l^\mu l^\nu \geq 0, \quad l^2 =0, 
\end{equation}
where $l^\mu$ is a light-like vector, thus satisfying the null condition $l^2 = 0$. This is where the difference with SEC appears which, in turn, is also reflected from the expression in terms of the Hubble parameter, namely: $\Dot{H}\leq 0$. From this discussion we see that a four-dimensional  de Sitter solution {\it violates} the four-dimensional SEC but {\it saturates} the four-dimensional NEC. The above conclusion may be further elaborated by choosing a power law scale factor, {\it i.e.} $a(t) \propto t^\gamma$. In the language of $\gamma$, we see that:
\begin{align}\label{secnec}
    \text{SEC}~ &\iff ~0<\gamma \leq 1,\\
    \text{NEC}~ &\iff  ~ \gamma\geq 0.
\end{align}
We can now try to relate the conditions \eqref{secnec} for the metric ans\"atze \eqref{10dansatz1}. For simplicity we will only look at the four-dimensional part of \eqref{10dansatz1}.  The scale-factor 
$a(t) \propto t^\gamma$ when expressed
in terms of the conformal time (which is used in \eqref{10dansatz1}), becomes $a(\eta) \propto \eta^{\gamma/(1-\gamma)} \equiv \eta^n $ with the proportionality factors given by appropriate powers of the cosmological constant $\Lambda$ to make this dimensionless. In the language of the conformal coordinates, the SEC and NEC from \eqref{secnec} become:
\begin{align}\label{secnec2}
    \text{SEC}~ &\iff ~~\frac{1}{n}\geq 0,\\
    \text{NEC}~ &\iff ~~ \frac{1}{n} \geq- 1, 
\end{align}
from which we see that the condition \eqref{senrem} on the time derivative of $g_s$ translates exactly to the NEC. This by itself is interesting, but the result should be taken with care, since the energy conditions we are discussing here are four-dimensional ones applied to a four-dimensional metric. Although it can be extended to arbitrary higher dimensions, such results do not apply directly to the metric (\ref{10dansatz1}), since this metric is not exactly of the form (\ref{FRLW}). It would be better to study the ten-dimensional energy conditions for the full type IIB ans\"atze (\ref{10dansatz1}) and see what comes out of them. This is what we turn to next. 

\subsection{Energy conditions for a type IIB ans\"atze and NEC \label{iibiib}}

The Gibbons-Maldacena-Nunez no-go theorem \cite{GMN} states that, in the absence of higher order corrections,  one cannot construct energy-momentum tensors that violate the SEC implying that  accelerating solutions with:
\bg\label{jomeshwar}
{\Ddot{a} \over a} = \Dot{H} +H^2 > 0, \nd
 are prohibited. Here we have expressed the temporal derivatives in terms of the cosmic time $t$. However, as discussed in \cite{Russo:2019fnk}, once we consider a generic metric with time-dependent internal space, we can relax the theorem and solutions could in-principle exist. In such a case, extra terms appear in the RHS of the inequality, such that it can be satisfied even with $\Dot{H} + H^2 > 0$. Additionally the authors of \cite{Russo:2019fnk} show that if we also impose the Dominant Energy Condition (DEC) on top of SEC, then any accelerating solution will have a late-time singularity in the internal space. 

To proceed, let us then start by revisiting the type IIB metric ans\"atze \eqref{10dansatz1}, but now express everything in terms of the {\it cosmic} time $t$. In the dual M-theory side, this is going to change all the $g_s$ scalings that we made, but we can easily resort back to the ones before by changing to the conformal coordinates. 
More concretely, we are interested in potential solutions with the following metric ans\"atze:
\begin{align}\label{metricansatz}
    ds^2 &= \frac{1}{\mathrm{H}^2(y)}(-dt^2 + g_{ij}dx^idx^j) + \mathrm{H}^2(y){\rm F}_1(t)g_{\alpha\beta}(y)dy^\alpha dy^\beta +\mathrm{H}^2(y) {\rm F}_2(t) g_{mn}(y)dy^m dy^n,\nonumber\\
    &= \frac{1}{\mathrm{H}^2(y)}ds^2_{\text{FLRW}} + \mathrm{H}^2(y){\rm F}_1(t)ds^2_{{\cal M}_2} 
    + \mathrm{H}^2(y){\rm F}_2(t)ds^2_{{\cal M}_4},
\end{align}
where $g_{ij}$ is such that the four-dimensional  metric is a FLRW universe and ${\rm F}_1 {\rm F}_2^2 =1$ to ensure a constant four-dimensional Planck mass (or time-independent Newton's constant). Note two things. {\Su One}, in rewriting \eqref{metricansatz}, we kept the form of ${\rm F}_i(t)$. In the conformal coordinates, as we saw in \eqref{10dansatz1}, they were taken to be ${\rm F}_i(\eta)$, whereas now we continue with ${\rm F}_i(t)$. These are scalar functions, so their specific form by changing $\eta \to t$ will not matter in the following discussion.  {\Su Two},  since most of the following results works for a general split on the number of dimensions, we will write $D \equiv (d - 1, 1) + \hat{d} + \tilde{d} = (3, 1) + 2+ 4$ and omit the numerical values of the dimensions whenever possible. We can also replace $(d -1, 1) \to d$ to avoid un-necessary clutter. To simplify the calculations, consider the conformally related metric:
\bg\label{generalmetric}
    d\bar{s}^2 &=& \bar{g}_{ac}dx^adx^c \equiv g_{\mu\nu}(x)dx^\mu dx^\nu + h_{ij}(x,y) dy^i dy^j\\
    & = & g_{\mu\nu} dx^\mu dx^\nu + \left(\delta^\alpha_i \delta^\beta_j \hat{g}_{\alpha\beta}(x, y) +
    \delta^m_i \delta^n_j \tilde{g}_{mn}(x, y)\right)dy^i dy^j \nonumber\\
    &=& g_{\mu\nu}(x)dx^\mu dx^\nu + {\rm H}^4(y){\rm F}_1(t)g_{\alpha\beta}(y)dy^\alpha dy^\beta 
    + {\rm H}^4(y){\rm F}_2(t)g_{mn}(y)dy^m dy^n, \nonumber \nd 
where $h_{ij}(x, y)$ is the metric of the internal space which are further sub-divided into $\hat{g}_{\alpha\beta}(x, y)$ and $\tilde{g}_{mn}(x, y)$, and in the third line we multiply the metric components 
in \eqref{metricansatz} by ${\rm H}^2$ to bring it to the required conformal form.  Note our convention: Greek indices are reserved for the $3+1$ dimensional space-time, while the Roman indices are for the internal six-manifold, {\it i.e.}
$(\mu, \nu) \in {\bf R}^{3, 1}$ and $(i, j) \in {\cal M}_4 \times {\cal M}_2$, although for the latter we can take any generic non-K\"ahler six-manifold.
The non-vanishing independent Christoffel symbols for a metric of the form (\ref{generalmetric}) are:
\begin{align}
    \Bar{\Gamma}^{\rho}_{ij} &= -\frac{1}{2}g^{\rho\sigma} \partial_{\sigma} h_{ij} = -\frac{1}{2}g^{\rho\sigma} \nabla_{\sigma} h_{ij} , \quad \Bar{\Gamma}^{\rho}_{\mu\nu} = \Gamma^{\rho}_{\mu\nu}(g),\\
    \Bar{\Gamma}^{i}_{\mu j} &= \frac{1}{2} h^{i k}\partial_\mu h_{k j} = \frac{1}{2} h^{i k}\nabla_\mu h_{k j} , \quad \Bar{\Gamma}^i_{jk} = \Gamma^i_{jk}(h),
\end{align}
where we denoted $\nabla_{\mu}$ as the covariant derivative constructed from $g_{\mu\nu}$. These are {\it not} the ten-dimensional covariant derivatives and nor are they expressed using the {\it torsional} connection. Recall that the torsion is given by the background three-form fluxes (especially the NS-NS three-form flux components), but their presence may be viewed as a contribution to the effective energy-momentum tensor $-$ much like what we did with \eqref{fahingsha5} $-$ so we can safely ignore them here.
The non-zero independent components of the Riemann tensor in this basis are:
\begin{align}
        \Bar{R}_{\mu \nu \tau}^{\;\;\;\;\;\sigma} &= R_{\mu\nu\tau}^{\;\;\;\;\;\sigma}(g) \nonumber\\
    \Bar{R}_{i j k}^{\;\;\;\;\;\sigma} &=  \nabla_{[i} \nabla^\sigma h_{j]k}\nonumber\\
    \Bar{R}_{i j\tau}^{\;\;\;\;\;\sigma} &= -\frac{1}{2}h^{kl}\nabla_\tau h_{l[i}\nabla^\sigma h_{j]k}\nonumber\\
    \Bar{R}_{i j k}^{\;\;\;\;\;l} &= R_{i j k}^{\;\;\;\;\;l}(h) -\frac{1}{2}h_{n[i}\nabla^{\rho}h_{j]k}\nabla_\rho h^{l n}
    \nonumber\\
 \Bar{R}_{i\mu j}^{\;\;\;\;\;\sigma} &= -\frac{1}{2}g^{\sigma\rho}\nabla_{\mu}\nabla_{\rho} h_{ij} +\frac{1}{4} g^{\sigma\rho} h^{kl}\nabla_{\mu} h_{jl}\nabla_\rho h_{i k},
\end{align}
where $\nabla_i$ with internal index is the six-dimensional covariant derivative constructed from $h_{ij}$, and again not involving any torsion. From these results, we can compute the components of the Ricci tensor in this basis, given by:
\begin{align}\label{Ricciforproductmanifold}
    \Bar{R}_{i\mu} &= h_{k[l}\nabla_{i]}\nabla_{\mu}h^{kl} \nonumber\\
    \Bar{R}_{\mu\nu} &= R_{\mu\nu}(g) -\frac{1}{4}\nabla_\mu h^{ij} \nabla_\nu h_{ij} - \frac{1}{2}h^{ij}\nabla_\mu \nabla_\nu h_{ij} \nonumber\\
    \Bar{R}_{ij} &= R_{ij}(h) -\frac{1}{2}g^{\rho\sigma}\nabla_\rho \nabla_\sigma h_{ij} -\frac{1}{2}h_{ik}\nabla^{\rho}h_{jl}\nabla_{\rho}h^{kl} +\frac{1}{4}h_{kl}\nabla^\rho h_{ij}\nabla_\rho h^{kl}.
  \end{align}
 In fact this is all we need to work on the consequences from the energy condition, although one might worry that the actual metric ans\"atze \eqref{metricansatz} is slightly different from the metric we took to compute the Riemann tensors. The difference is due to the conformal factor,  so to this end let us define a metric $\tilde{g}_{ab}$ in the following way:
\begin{equation}
    \widetilde{g}_{ab}(x, y) = \Omega^2(x, y)\Bar{g}_{ab}(x, y),
\end{equation}
where $\Omega^2(x, y)$ is the conformal factor that depends on coordinates $x^\mu \in {\bf R}^{3, 1}$ and 
$y^i \in {\cal M}_4 \times {\cal M}_2$ generically. Such a conformal transformation gives a way to relate the Ricci tensors derived with the metric components $\Bar{g}_{ab}$, to the Ricci tensors with the metric components $\widetilde{g}_{ab}$, via:

{\footnotesize
\begin{align}\label{confRicci}
    \widetilde{R}_{ab} = \Bar{R}_{ab} -(D-2)\left(\Bar{\nabla}_a\Bar{\nabla}_b \ln \Omega + \bar{g}_{ab}\bar{g}^{cd}\bar{\nabla}_c \ln \Omega \bar{\nabla}_d \ln \Omega - \bar{\nabla}_a \ln \Omega \bar{\nabla}_b \ln \Omega\right) - \bar{g}_{ab}\bar{g}^{cd}\bar{\nabla}_c \bar{\nabla}_d \ln \Omega.
\end{align}}
where $\bar{\nabla}_a$ is the torsion-free covariant derivative computed using the metric components 
$\Bar{g}_{ab}$. To proceed we will need the explicit form for the conformal factor $\Omega(x, y)$. It is easy to determine this by looking at the metric \eqref{generalmetric}: the conformal factor $\Omega(x, y)$ should take the form $\Omega(y) = {\rm H}^{-1}(y)$ so that it remains independent of $x^\mu$ coordinates. This choice of the conformal factor reproduces \eqref{metricansatz}, which is no surprise of course, but the aforementioned manipulation is not without merit. The relation \eqref{confRicci} actually provides a much easier way to compute the Ricci tensors of a warped metric and is applicable even if the warp factor develops dependences on $x^\mu$. For the simple case, with the choice of $\Omega(y) = {\rm H}^{-1}(y)$, since we recover the metric \eqref{metricansatz}, we can rewrite $\widetilde{R}_{ab} \equiv R^{(D)}_{ab}$, and use
\eqref{Ricciforproductmanifold} in \eqref{confRicci} to get: 

{\footnotesize 
\begin{align}\label{dRiccitensor}
    R_{\mu\nu}^{(D)} &= R_{\mu\nu}
    ^{(D-\hat{d}-\tilde{d})}(g) - \hat{d}\left(\frac{1}{2} 
    \frac{\nabla_\mu\nabla_\nu {\rm F}_1}{{\rm F}_1} -\frac{1}{4}\frac{\nabla_\mu {\rm F}_1\nabla_\nu {\rm F}_1}{{\rm F}_1^2}\right) - \tilde{d}\left(\frac{1}{2} 
    \frac{\nabla_\mu\nabla_\nu {\rm F}_2}{{\rm F}_2} -\frac{1}{4}\frac{\nabla_\mu {\rm F}_2\nabla_\nu {\rm F}_2}{{\rm F}_2^2}\right) \nonumber\\
    &+ \left(\frac{\nabla^2\mathrm{H}}{\mathrm{H}} - (D-1)\frac{(\nabla \mathrm{H})^2}{\mathrm{H}}\right)g_{\mu\nu} \nonumber\\
    R_{\alpha\beta}^{(D)} &= R_{\alpha\beta}^{(\hat{d})}(\hat{g}) + \left(-\frac{1}{2}\frac{\nabla^2{\rm F}_1}{{\rm F}_1}-\frac{1}{4}(\hat{d}-2)\frac{(\nabla {\rm F}_1)^2}{{\rm F}_1^2}-\frac{\tilde{d}}{4}\frac{\nabla_\rho {\rm F}_1\nabla^\rho {\rm F}_2}{{\rm F}_1 {\rm F}_2}\right)\mathrm{H}^4(y){\rm F}_1(t)g_{\alpha\beta} +(D-2)\frac{\nabla_\alpha \nabla_\beta \mathrm{H}}{\mathrm{H}} \nonumber\\
    &+ \left(\frac{\nabla^2\mathrm{H}}{\mathrm{H}}-(D-1)\frac{(\nabla \mathrm{H})^2}{\mathrm{H}^2}\right)\hat{g}_{\alpha\beta} \nonumber \\
    R_{mn}^{(D)} &= R_{mn}^{(\tilde{d})}(\tilde{g}) + \left(-\frac{1}{2}\frac{\nabla^2{\rm F}_2}{{\rm F}_2}-\frac{1}{4}(\tilde{d}-2)\frac{(\nabla {\rm F}_2)^2}{{\rm F}_2^2}-\frac{\hat{d}}{4}\frac{\nabla_\rho {\rm F}_1\nabla^\rho {\rm F}_2}{{\rm F}_1 {\rm F}_2}\right)\mathrm{H}^4(y){\rm F}_2(t)g_{mn} +(D-2)\frac{\nabla_m \nabla_n \mathrm{H}}{\mathrm{H}}  \nonumber \\
    &+ \left(\frac{\nabla^2\mathrm{H}}{\mathrm{H}}-(D-1)\frac{(\nabla \mathrm{H})^2}{\mathrm{H}^2}\right)\tilde{g}_{mn} \nonumber\\
    R^{(D)}_{m\mu} &= -\frac{(D-2)}{2}\frac{\nabla_\mu {\rm F}_2}{{\rm F}_2}\frac{\nabla_m \mathrm{H}}{\mathrm{H}},
  ~~  R_{\alpha \mu}^{(D)} = -\frac{(D-2)}{2}\frac{\nabla_\mu {\rm F}_1}{{\rm F}_1}\frac{\nabla_\alpha \mathrm{H}}{\mathrm{H}}, ~~
    R^{(D)}_{\alpha m} = R_{\alpha m } +(D-2)\frac{\nabla_\alpha \nabla_m \mathrm{H}}{\mathrm{H}},
\end{align}}
where $\hat{g}_{\alpha\beta}$ and $\tilde{g}_{mn}$ are defined in \eqref{generalmetric}.
Note also that the various operators are defined on the space in which the functions they are applying to have support: for example in $\nabla^2 {\rm F}_i(x)$ the Laplacian operator is the one constructed from $g_{\mu\nu}(x)$, while in $\nabla^2 \mathrm{H}(y)$ the Laplacian is constructed from the metric of the internal space $h_{ij}$ in (\ref{generalmetric}). 

Now that we have the expression for the $D$ dimensional Ricci tensor, we can study the implications of the $D$ dimensional energy conditions. Assuming a flat $d$ dimensional FLRW metric for the external space ($D = d + \hat{d} +\tilde{d}$), the SEC gives: 

{\footnotesize
\begin{align}\label{rdoo}
    R^{(D)}_{00} &=
    -(d-1)(\Dot{H}+H^2) - {\hat{d}\over 2} \left(\frac{\ddot{{\rm F}}_1}{{\rm F}_1}-\frac{\dot{{\rm F}}_1^2}{2 F_1^2}\right) - {\tilde{d} \over 2} \left(\frac{\ddot{{\rm F}}_2}{{\rm F}_2}-\frac{\dot{{\rm F}}_2^2}{2 {\rm F}_2^2}\right)
    -\left(\frac{\nabla^2\mathrm{H}}{\mathrm{H}}-(D-1)\frac{(\nabla \mathrm{H})^2}{\mathrm{H}^2}\right)
    \nonumber\\
    &= -(d-1)(\Dot{H}+H^2) -\left(2\hat{d}-\frac{\tilde{d}}{4}\right)\frac{\dot{{\rm F}}_2^2}{{\rm F}_2^2} -\left(\frac{\tilde{d}}{2}-\hat{d}\right)\frac{\Ddot{{\rm F}}_2}{{\rm F}_2} -\left(\frac{\nabla^2\mathrm{H}}{\mathrm{H}}-(D-1)\frac{(\nabla \mathrm{H})^2}{\mathrm{H}^2}\right)    \geq 0,
\end{align}}
where we used the relation ${\rm F}_1{\rm F}_2^2=1$ in the second equality to write derivatives of ${\rm F}_1$ in terms of ${\rm F}_2$. As mentioned earlier, this keeps the four-dimensional Newton's constant time-independent. As a reminder, $H(t)$ is the Hubble's constant whereas ${\rm H}(y)$ is the warp-factor.
For the case we are interested in, the dimensional split becomes $(d, \hat{d}, \tilde{d}) = (4, 2, 4)$ and \eqref{rdoo} can be re-written as:
\begin{align}\label{rdoo2}
    R^{(D)}_{00} &=
    -3(\dot{H}+ H^2) -\frac{\ddot{{\rm F}}_1}{{\rm F}_1} +\frac{1}{2}\frac{\dot{{\rm F}}_1^2}{{\rm F}_1^2}-2\frac{\ddot{{\rm F}}_2}{{\rm F}_2} +\frac{\dot{{\rm F}}^2_2}{{\rm F}_2^2} -\left(\frac{\nabla^2\mathrm{H}}{\mathrm{H}}-(D-1)\frac{(\nabla \mathrm{H})^2}{\mathrm{H}^2}\right)\nonumber\\
    &= -3\left(\dot{H}+ H^2 +\frac{\dot{{\rm F}}_2^2}{{\rm F}_2^2}\right) - \left(\frac{\nabla^2\mathrm{H}}{\mathrm{H}}-(D-1)\frac{(\nabla \mathrm{H})^2}{\mathrm{H}^2}\right) \geq 0.
 \end{align}
 This is almost the form in which we can express the SEC for our metric ans\"atze \eqref{metricansatz}, although one can do one more manipulation to bring it in a suggestive format. The manipulation involves 
rewriting the last parenthesis (which involves the warp-factor ${\rm H}(y)$) as a total Laplacian, and doing that we get:
\begin{align}\label{rood}
    R_{00}^{(D)} &= 
    -3(\dot{H}+ H^2) -\frac{\ddot{{\rm F}}_1}{{\rm F}_1} +\frac{1}{2}\frac{\dot{{\rm F}}_1^2}{{\rm F}_1^2}-2\frac{\ddot{{\rm F}}_2}{{\rm F}_2} +\frac{\dot{{\rm F}}^2_2}{{\rm F}_2^2} + \frac{\mathrm{H}^{D-2}}{D-2}\nabla^2\mathrm{H}^{2-D} \nonumber\\
    &=-3\left(\dot{H}+ H^2 +\frac{\dot{{\rm F}}_2^2}{{\rm F}_2^2}\right) + \frac{\mathrm{H}^{D-2}}{D-2}\nabla^2\mathrm{H}^{ 2- D}\geq 0,
\end{align}
where note the change in the relative sign of the second term involving the warp-factor ${\rm H}(y)$. The coefficient of the first term with the Hubble parameter $H(t)$ is clearly $-(d - 1)$, and we can use this to 
rearrange the inequality \eqref{rood} to get:
\begin{align}\label{rood2}
    (d-1)(D-2)\mathrm{H}^{2-D}\left[\dot{H}+ H^2 +\frac{1}{d-1}\left(\frac{\ddot{{\rm F}}_1}{{\rm F}_1} -\frac{1}{2}\frac{\dot{{\rm F}}_1^2}{{\rm F}_1^2}+2\frac{\ddot{{\rm F}}_2}{{\rm F}_2} -\frac{\dot{{\rm F}}^2_2}{{\rm F}_2^2}\right)\right]= \nonumber\\
    (d-1)(D-2)\mathrm{H}^{ 2-D}\left(\dot{H}+ H^2 +\frac{\dot{{\rm F}}_2^2}{{\rm F}_2^2}\right) \leq \nabla^2 \mathrm{H}^{2 - D},
\end{align}
where recall that the first line is with generic ${\rm F}_1$ and ${\rm F}_2$, whereas the second line is after we impose the condition ${\rm F}_1 {\rm F}_2^2 = 1$. Either of those can be used to express the inequality.
For example, we can use the first line, and integrate over the internal six-manifold of dimension 
$\hat{d} + \tilde{d}$ to get the following inequality:

{\footnotesize
\begin{equation}\label{rood3}
    (d-1)(D-2)\frac{G_D}{G_d}\left[\dot{H}+ H^2 +\frac{1}{d-1}\left(\frac{\ddot{{\rm F}}_1}{{\rm F}_1} -\frac{1}{2}\frac{\dot{{\rm F}}_1^2}{{\rm F}_1^2}+2\frac{\ddot{{\rm F}}_2}{{\rm F}_2} -\frac{\dot{{\rm F}}^2_2}{{\rm F}_2^2}\right)\right]\leq \int d^{\hat{d}+ \tilde{d}}y\sqrt{h}\nabla^2\mathrm{H}^{2 - D} = 0,
\end{equation}}
where $h$ is the determinant of the internal metric \eqref{generalmetric}, and we have used the fact that the warp-factor ${\rm H}(y)$ is a smooth function with no singularities over the six-manifold. As such integrating over the total derivative vanishes, which is what appears on the RHS. One may equivalently use  
${\rm F}_1{\rm F}_2^2=1$ to rewrite \eqref{rood3} as:
\begin{equation}\label{rood4}
    (d-1)(D-2)\frac{G_D}{G_d}\left(\dot{H}+ H^2 +\frac{\dot{{\rm F}}_2^2}{{\rm F}_2^2}\right)\leq \int d^{\hat{d}+ \tilde{d}}y\sqrt{h}\nabla^2\mathrm{H}^{2-D} = 0,
\end{equation}
which is expectedly a more condensed version of \eqref{rood3}. Note that in both \eqref{rood3} and \eqref{rood4} we have taken advantage of the fact that both the Hubble parameter $H(t)$ and ${\rm F}_i(t)$ are purely a function of the cosmic time $t$, and therefore the LHS integrate to $d$-dimensional Newton's constant defined as:
\begin{equation}\label{ddimensionalGNewton}
    \int d^{\hat{d} + \tilde{d}}y \sqrt{h}\mathrm{H}^{2 - D} \equiv  \frac{G_D}{G_d},
\end{equation}
which is clearly taken to be time-independent here. Note that according to our definition of $h_{ij}$ in 
\eqref{generalmetric}, ${\rm det}~h_{ij} \equiv h = {\rm det}~\hat{g}_{\alpha\beta}~{\rm det}~\tilde{g}_{mn}$, and therefore will have ${\rm F}_1^2 {\rm F}_2^4$ dependence. Thus if ${\rm F}_1 {\rm F}_2^2 \ne 1$, the Newton's constant will become time-dependent as mentioned above. After the dust settles, the SEC reduces then to:
\begin{equation}\label{rood5}
    \dot{H}+ H^2 +\frac{\dot{{\rm F}}_2^2}{{\rm F}_2^2} +\frac{1}{d-1}\left(\frac{\ddot{{\rm F}}_1}{{\rm F}_1} -\frac{1}{2}\frac{\dot{{\rm F}}_1^2}{{\rm F}_1^2}+2\frac{\ddot{{\rm F}}_2}{{\rm F}_2} -\frac{\dot{{\rm F}}^2_2}{{\rm F}_2^2}\right) = 
    \dot{H}+ H^2 +\frac{\dot{{\rm F}}_2^2}{{\rm F}_2^2} \leq 0,
\end{equation}
where we have used that fact that $d > 1$ and $D > 2$ along-with the condition of positivity of the Newton's constant (the latter appears from the compactness of the internal manifold).  Since the last term in the LHS is positive (or zero\footnote{It is instructive to check whether $\dot{\rm F}_2$ remains finite at late time. Since we have used conformal time $\eta$ in \eqref{metricansatz} and also in the earlier sections, it will be useful to relate it to $g_s$ to see the late time behavior. To this end, we can use the relation $\eta \propto t^{1-\gamma}$, and the definition of $g_s$ from \eqref{slivia}. It is now easy to see that:
\bg\label{84c}
\dot{\rm F}_2 \equiv {d{\rm F}_2\over dt} \propto g_s {d{\rm F}_2\over dg_s}\cdot {dg_s\over d\eta},
\nonumber \nd
where we have used the fact that $g_s \propto \eta^{-n}$ from \eqref{slivia}. Since ${\rm F}_2 \equiv
\sum_{k_{n_2}} {\rm F}_2^{(k_{n_2})} g_s^{k_{n_2}}$, as defined just after \eqref{cora2} with $k_{n_2} \in 
\mathbb{Z}$, the dominant term in ${d{\rm F}_2\over dg_s}$ is a constant. Similarly, as proved in section \ref{toyeft}, ${dg_s\over d\eta} = g_s^{+|\sigma|}$, {\it i.e.} it is given by a {\it positive} power of $g_s$. Combining everything together, we see that ${d{\rm F}_2\over dt} \to 0$ at late time, and since 
${\rm F}_2$ does not vanish at late time, the ratio ${\dot{\rm F}_2\over {\rm F}_2}$ does vanish at late time. Additionally, in the temporal domain governed by the TCC \cite{tcc}, $g_s < 1$ and the ratio remains finite and small.}), {\it i.e.} $\frac{\dot{{\rm F}}_2^2}{{\rm F}_2^2} \ge 0$, we have:
\begin{equation}
    \dot{H}+ H^2 \equiv \frac{\Ddot{a}}{a} \leq 0, 
\end{equation}
which is the GMN no-go theorem \cite{GMN}. We see that even allowing the internal space to be time dependent, if the overall volume is constant, {\it i.e.} ${\rm F}_1{\rm F}_2^2=1$, then it is not possible to avoid the GMN no-go theorem, as the implications of the SEC are not modified. Since we know that there could be de Sitter solutions with metric (\ref{metricansatz}), the effective energy-momentum tensor made up from all the corrections should necessarily violate the SEC.

For the Null Energy Condition (NEC), the situation is slightly different. In addition to $R_{00}^{(D)}$ from 
\eqref{rdoo}, we will also need $R_{11}^{(D)}$ from the list of Ricci tensors in \eqref{dRiccitensor}. Taking 
$l^\mu \equiv \left(1, {1\over a}, 0, 0\right)$ in \eqref{nulle}, and plugging in the values of the Ricci tensors, we get: 

{\footnotesize
\begin{align}
    R_{00}^{(D)} + a^{-2}R_{11}^{(D)} &=
    -(d-2)\dot{H} -\frac{\ddot{{\rm F}}_1}{{\rm F}_1} +\frac{1}{2}\frac{\dot{{\rm F}}_1^2}{{\rm F}_1^2}-2\frac{\ddot{{\rm F}}_2}{{\rm F}_2} +\frac{\dot{{\rm F}}^2_2}{{\rm F}_2^2} + H\left(\frac{\dot{{\rm F}}_1}{{\rm F}_1} + 2\frac{\dot{{\rm F}}_2}{{\rm F}_2}\right)\nonumber\\
    &=-(d-2)\dot{H}- \left(2\hat{d}-\frac{\tilde{d}}{4}\right)\frac{\dot{{\rm F}}_2^2}{{\rm F}_2^2} -\left(\frac{\tilde{d}}{2}-\hat{d}\right)\frac{\Ddot{{\rm F}}_2}{{\rm F}_2} 
    = -2\dot{H}-3\frac{\dot{{\rm F}}_2^2}{{\rm F}_2^2} \geq 0,
\end{align}}
where in the second line we used ${\rm F}_1{\rm F}_2^2 = 1$ and $(d, \hat{d}, \tilde{d}) = (4, 2, 4)$. In the last inequality, the second term is positive-definite, so this would imply:
\begin{equation}\label{tegchin}
    \dot{H} ~\leq ~0,
\end{equation}
implying that there is no modification to the lower dimensional NEC. Therefore, we conclude that the naive analysis done in the first section is correct: for the metric (\ref{metricansatz}) with an FLRW scale factor given by $a(\eta) \propto \eta^n$ (where $\eta$ is the conformal time), the condition ${1\over n}\geq -1$ is equivalent to the NEC, the four-dimensional one. 

Before ending this section let us make a few observations. {\Su First}, the above condition is derived using the 
$(2, 4)$ split of the internal six-manifold. A natural question then is: what happens if we take other kinds of splitting, for example $(1, 5), (3, 3)$ or even $(a_1, ..., a_6)$ with $\sum a_i = 6$? It turns out the result 
\eqref{tegchin} is {\it not} sensitive to how we split the internal six-manifold: any other splittings will result in a similar conclusion. This will be demonstrated elsewhere. 
{\Su Secondly}, the result also appears to be insensitive to the full ten-dimensional NEC and depends only on the four-dimensional NEC. This is an intriguing condition and recently in
\cite{necshort} we proposed this as a conjecture (see also \cite{vikman} for earlier developments in the subject). Combining the results of this section and the ones from 
section \ref{toyeft},  we can now provide a {\it proof} of this. Note however that the {\it reason} for this intriguing connection between the existence of EFT and four-dimensional NEC is still a bit mysterious. Maybe the ideas put forward in \cite{vikman} could be one of the keys to unravel this, but this would require further investigation before we make any concrete conclusion. {\Su Thirdly}, there are quite a few cosmological implications stemming from our conjecture that have been summarized in \cite{necshort}. The readers may want to look at this reference, and at the works cited therein, for more details.


\section{Discussions and conclusions \label{disco}}

In this paper we have given further evidence to argue that four-dimensional EFT description with de Sitter isometries in IIB, that overcomes the no-go theorem and the swampland criteria, can only exist if the de Sitter space is viewed as a Glauber-Sudarshan state in string theory or in its dual M-theory. The latter is preferred because the IIB background is typically at strong coupling with $g_b = 1$. In the following let us summarize some of the main results of the paper. 

\vskip.1in

\noindent $\bullet$ We argue that the all the computations related to EOMs, Bianchi identities, flux quantizations {\it et cetera} are independent of the de Sitter slicings.  Additionally, we show that for all the slicings, except for the static patch (and other patches related to it), there exist  temporal domains where the dual IIA backgrounds are weakly coupled despite the corresponding IIB backgrounds being at strong couplings. We also show that in the static patch not only there is no such weakly coupled regime but also other issues make the quantum analysis harder to perform. 

\vskip.1in

\noindent $\bullet$ We generalize the perturbative quantum series studied in \cite{desitter2, coherbeta} to incorporate dependences on the de Sitter slicings, {\it i.e.} the quantum terms to depend on $(x^i, x^j) \in {\bf R}^2$. The generalized perturbative series is given by \eqref{fahingsha3} which scales as 
\eqref{botsuga2}. This allows a well-defined EFT description. However, once we allow additional dependences on the toroidal directions, the quantum series becomes \eqref{fahingsha4}, but the $g_s$ scaling in \eqref{botsuga3} shows that the EFT description breaks down. Compared to the other similar breakdowns studied in \cite{desitter2}, now it extends to both the flux as well as the 
gravitational sectors. 

\vskip.1in

\noindent $\bullet$ We devise a new ans\"atze in M-theory, for the metric and the flux components that allow dependences on all coordinates (except for the T-duality direction $x^3$), in \eqref{makibhalu3} that depends on a real parameter $\gamma$. The $g_s$ scaling of the quantum series \eqref{fahingsha4} with the background \eqref{makibhalu3}, shown in  
\eqref{botsuga3}, shows that EFT description is now valid as long as $\gamma \ge 5$. Interestingly, for 
$\gamma = 6$, all computations in the background \eqref{makibhalu3} reduces to the ones for the simpler background \eqref{evader}. This surprising simplification even extends to the topological sector, that we demonstrate in details.  

\vskip.1in

\noindent $\bullet$ We compute the non-local and non-perturbative contributions to the energy-momentum tensors to all orders in non-localities. This is an extension of a similar computation in \cite{coherbeta} done to the first order in non-locality. We show that, once the trans-series associated with the non-local contributions to the energy-momentum tensors is summed up, the final result becomes perfectly local and finite. We also show the convergence of the series with higher orders in non-localities, and extend the analysis to include non-perturbative contributions from the BBS \cite{BBS} and KKLT \cite{KKLT} type instantons. 

\vskip.1in

\noindent $\bullet$ We show that a generic FLRW cosmological solution of the form \eqref{morbiusn} is a valid solution, meaning that EFT is preserved, if and only if the condition \eqref{senrem} is satisfied. We also argue that if the condition \eqref{senrem} is not satisfied, {\it i.e.} if the alternative condition \eqref{oniston} is imposed, then the perturbative and non-perturbative quantum effects only allow $\vert\sigma\vert = 0$. On the other hand, if the sign of the exponent in \eqref{oniston} is reversed, EFT description remans valid as long as 
$\vert\sigma\vert \in {\mathbb{Z}\over 3}$. We justify these by showing detailed computations in section \ref{toyeft}. 

\vskip.1in

\noindent $\bullet$ We show that the condition \eqref{senrem} is also related to the four-dimensional NEC in IIB. We argue this from a full ten-dimensional computation in the IIB side. Somewhat surprisingly, our identification does not in any way imply NEC in the full ten or eleven-dimensions. This connection of EFT with four-dimensional NEC was recently conjectured in \cite{necshort} and here we provide the full proof.

\vskip.1in

\noindent $\bullet$ We compute all the possible ${\bf X}_8$ curvature forms for the choice of the background \eqref{makibhalu3}. We show that when $\gamma = 6$ in \eqref{makibhalu3}, these curvature forms reduce to the ones with the metric ans\"atze \eqref{evader}, justifying the extension of the aforementioned simplification to the topological sector. These ${\bf X}_8$ forms are the important ingredients in the flux EOMs.

\vskip.1in

\noindent $\bullet$ We show how one should incorporate the non-local and the non-perturbative effects in the flux EOMs. Again, as in  for the case with the energy-momentum tensors earlier, the trans-series associated with the non-local terms can be summed up to provide local and finite contributions. We also discuss how one may compute the $g_s$ scalings of the non-local and non-perturbative terms. These contributions were not discussed in \cite{coherbeta}. 

\vskip.1in

\noindent $\bullet$ We study the flux EOMs for all possible allowed flux components. We divide the fluxes into three categories: {\Su internal fluxes}, dealing with flux components that have legs only along the internal eight-manifold in M-theory; {\Su ${\bf G}_{\rm 0ABC}$ fluxes}, dealing with all the flux components that have one leg along the temporal direction, and {\Su external fluxes}, dealing with flux components that have at least one leg along the spatial ${\bf R}^2$ directions. As a consequence of our choice, the perturbative quantum term in \eqref{fahingsha5} becomes more involved, but we argue, by computing the $g_s$ scaling in \eqref{botsuga4}, that the EFT description still remains valid as long as 
$\gamma \ge 5$ in \eqref{makibhalu3}. 

\vskip.1in

\noindent $\bullet$ To solve the flux EOMs we devise a new over-bracket and under-bracket scheme that not only allows us to match the $g_s$ scalings over the rank seven and eight tensors in 
\eqref{mcacrisis}, but also succinctly provides the precise perturbative and non-perturbative (including the non-local) quantum terms contributing to \eqref{mcacrisis}. We show that, at higher orders in the choice of the parameters, the under-brackets and the over-brackets tend to merge together across all the rank seven and eight tensors of \eqref{mcacrisis}.

\vskip.1in

\noindent $\bullet$ We show how the flux EOMs and the anomaly cancellation conditions help us to fix the dominant scalings of all the flux components in \eqref{makibhalu3}. However, we show in 
\eqref{pughmey} that some ambiguity in the choice of the dominant scalings for the internal flux components still remains, but once we impose the flux quantization scheme, it appears to prefer the choice 
\eqref{collateral}. Interestingly, most of the other flux components {\it do not} participate at the lowest orders in the background EOMs (which we show later to be related to the Schwinger-Dyson's equations). Our choice of \eqref{collateral} not only justifies the choice we made in \cite{desitter2, coherbeta}, but also shows that the internal fluxes necessary to support a de Sitter state in string theory {\it cannot} be time-independent (in fact the choices in \eqref{pughmey} only allow time-dependent fluxes).

\vskip.1in

\noindent $\bullet$ We derive the flux quantization conditions for all the internal fluxes when both the flux components and the internal manifold are varying with respect to time. We show how the interplay between the dynamical ${\rm M5}$-branes, global and localized fluxes, as well as the perturbative and non-perturbative quantum terms,  conspire in an interesting way to reproduce the flux quantization condition. Somewhat surprisingly, we do recover Witten's flux quantization condition 
\cite{wittenfluxes} even though all the underlying degrees of freedom have temporal dependences.  We justify our results by dualizing to the heterotic side and showing that the anomaly cancellation condition do not receive any corrections thus consistently fitting with the Adler-Bell-Jackiw theorem \cite{ABJ}. 

\vskip.1in

\noindent $\bullet$ We find a new four-dimensional de Sitter solution directly in M-theory by dualizing a IIB cosmological solution. Both the IIB and the dual IIA are at strong couplings now so, while from the IIA side we do not have the advantage of the weakly-coupled late-time physics, the {\it early-time} physics could become weakly-coupled. This gives us a new window to investigate early-time physics which we did not have in our earlier works \cite{desitter2, coherbeta}. We show that there exists {\it another} M-theory dual (with weakly coupled IIA limit) that may be used to compute results for the four-dimensional de Sitter space. Using the M-theory/M-theory duality we argue that the fluxes corresponding to the de Sitter background cannot be time-independent. 

\vskip.1in

\noindent $\bullet$ We find the precise wave-function renormalization for the Glauber-Sudarshan state associated with the de Sitter space in the dual IIB side. We determine the temporal evolution of the Glauber-Sudarshan state and show how the number of graviton and the flux quanta change in the temporal domain governed by the TCC \cite{tcc} which, in turn, is related to the onset of strong coupling in the IIA side. Under specific conditions we argue how the graviton and the flux quanta could decrease as the state evolves temporally. We also provide a formula in \eqref{mauritius} to compute the precise change in the number of quanta. 

\vskip.1in

\noindent $\bullet$ We compute the fluctuation spectra over the de Sitter Glauber-Sudarshan state using the Agarwal-Tara state \cite{agarwal}.  We show how our analysis may be used not only to reproduce the fluctuation spectra over Bunch-Davies or $\alpha$ vacua but also, with appropriate choices of certain parameters in \eqref{articoo}, the fluctuation spectra over any arbitrary vacua.  

\vskip.2in

\noindent The above is a list of the main results, although there are quite a few minor ones that we did not mention. Despite the number of results $-$ and as the patient reader may have realized after coming this far $-$ we have barely touched the tip of the iceberg. Many computations associated for example with the S-matrices, fluctuation spectra, early-time physics, inflationary state, dark energy, dark matter, structure formations {\it et cetera} still need to be performed to complete the story to its full glory. Again, as with our protagonists in the Valley of the Kings almost a century ago, viewing de Sitter space-time as a Glauber-Sudarshan state might open up a window into wonderful things!

\section*{Acknowledgements:} 

\noindent We would like to thank Robert Brandenberger, Gia Dvali, Andrew Frey and Shahin Sheikh-Jabbari for helpful comments and Savdeep Sethi for a useful reference.
The work of HB, KD and MMF, is supported in part by the Natural
Science and Engineering Research Council of Canada (NSERC). The work of SB is supported in part by the NSERC (funding reference \# CITA 490888-16) through a CITA National Fellowship and by a McGill Space Institute fellowship.


\end{document}